\documentclass[11pt]{JHEP3}
\input epsf.tex
\epsfclipon

\usepackage{color}
\let\normalcolor\relax
\usepackage{epsfig}
\usepackage{epstopdf}
\usepackage{epsf}
\input{epsf.sty}
\usepackage{graphicx,amsmath,amssymb}
\usepackage{epsfig,multicol}
\usepackage{multirow}
\allowdisplaybreaks

\usepackage{bbm,bm,amsmath,amssymb}
 
\usepackage[normalem]{ulem}

\usepackage{comment}
\def\be{\begin{equation}}
\def\ee{\end{equation}}
\def\figs/B{B}
\def\bea{\begin{eqnarray}}
\def\eea{\end{eqnarray}}
\def\bg{\begin{eqnarray}}
\def\nd{\end{eqnarray}}
\def\sin{{\rm sin}}
\def\cos{{\rm cos}}
\def\tan{{\rm tan}}
\def\log{{\rm log}}

\title{de Sitter Vacua in the String Landscape}
\author{Keshav Dasgupta$^{1}$, Maxim Emelin$^{1}$, Mir Mehedi Faruk$^{1}$,  
	Radu Tatar$^{2}$\\
	\vskip.03in
	${}^1$ Department of Physics, McGill University, Montr\'{e}al, Qu\'{e}bec, H3A 2T8, Canada \\
	${}^2$ Department of Mathematical Sciences,
University of Liverpool,  Liverpool, L69 7ZL, United Kingdom \\
	{\tt keshav@hep.physics.mcgill.ca, maxim.emelin@mail.mcgill.ca}
	~~{\tt  mir.faruk@mail.mcgill.ca, Radu.Tatar@Liverpool.ac.uk}}

\date{\today}
\maketitle

\abstract{The late-time behavior of our universe is one of accelerated expansion, or that of a de Sitter space, and therefore motivates us to look for time-dependent backgrounds. 
Finding such backgrounds in string theory has always been a challenging problem.  An even harder problem is to find time-dependent backgrounds that allow positive dark energies. As a first step to handle such scenarios, we study a 
time-dependent background in type IIB theory, with four-dimensional de Sitter isometries, by uplifting it to M-theory and then realizing it as a coherent, or squeezed-coherent, state over an appropriate solitonic configuration.  While classically such a background does not solve the equations of motion, the corresponding Schwinger-Dyson equations reveal that there are deeper issues that may even prohibit a solution to exist at the quantum level, as long as the internal space remains time-independent.  A more generic analysis is then called for, where both the effective four-dimensional space-time, the internal space, and the background fluxes are all time-dependent. We study in details such a background by including perturbative and non-perturbative 
as well as local and non-local quantum terms. Our  analysis reveals a distinct possibility of the emergence of  a four-dimensional positive curvature space-time with de Sitter isometries and time-independent Newton's constant
 in the landscape of type IIB string theory. We argue how the no-go and the swampland criteria are avoided in generating such a background, and compare it with other possibilities involving backgrounds with time-dependent Newton constants. These time-varying Newton constant backgrounds typically lead to unavoidable late time singularities, amongst other issues.}

\begin{document}

\section{Introduction and summary}	
\label{sec:intro}

The late-time behavior of our universe is one of accelerated expansion, as is its very early-time behavior according to the inflationary paradigm. Both of these facts motivate the search for solutions that exhibit accelerated expansion within string theory. A natural starting point is to search for the maximally symmetric variant of such a solution, namely de Sitter space. However, explicit top-down construction of any scale-separated string compactifications is technically challenging at the present time. Existing proposed constructions, the most prominent of which is the KKLT scenario \cite{KKLT}, involve a subtle patchwork of ten-dimensional and four-dimensional phenomena coming from an interplay of supergravity degrees of freedom with stringy effects such as higher derivative corrections, brane instantons or other brane world-volume phenomena. How and whether all the ingredients of any particular construction come together to produce the desired solution is still a matter of some dispute \cite{bena, westphal} (see also \cite{kalloshevan} for possible resolutions).  Furthermore, in \cite{danielsson} it is argued that the string loop corrections to the cosmological constant will generically induce a time-dependence of the whole background. 

The lack of full top-down constructions along with the various objections to existing proposals has led to several conjectures regarding the effective potentials that arise in string compactifications, which rule out de Sitter vacua \cite{vafa1, vafa2, vafa22, brennan, Ooguri, swampland}. These swampland conjectures, if true, favor quintessence models over time-independent meta-stable de Sitter vacua. These conjectures, however, are themselves largely based on the known behavior of effective potentials in regimes of string theory where top-down calculations can be performed. They could therefore be missing out on some of the more intricate effects, such as the back-reaction of world-volume effects in the presence of anti-branes, which are supposed to be responsible for the uplift in the KKLT scenario, thus coming back full circle. A systematic investigation of the possible quantum corrections in string theory is therefore called for in order to make progress on these questions.

In \cite{nogo, nodS} the viability of de Sitter vacua in type IIB theory were studied from the perspective of its M-theory uplift. There, all the corrections that are built out of various higher order combinations of the curvatures and fluxes and their derivatives were considered, yielding constraints that the series of quantum corrections have to obey to result in positive 4-dimensional scalar curvature. An important consequence of the analysis in \cite{nodS} is that for a time-independent compactification ansatz to de Sitter space, the corrections that must be switched on to give a positive cosmological constant result in the appearance of an infinite tower of additional time-independent corrections, all without any clear relative suppression. This was interpreted to indicate a breakdown of an effective field theory description. In other words, even if a de Sitter compactification ansatz could be realized, the physics in that space would not be described by an effective field theory with finitely many fields\footnote{A possible caveat to this conclusion could be a new duality frame, which reorganizes this infinite tower of corrections into finitely many fields. However, the existing constructions of de Sitter vacua make no such claim, but are rather claimed to be meta-stable states within the same effective theory as some nearby Minkowski or AdS vacuum.}.

The goal of this paper is two-fold. First, we wish to check the robustness of the results of \cite{nodS} with respect to deformations of the de Sitter ansatz. To this end, we consider dipole-type and Kasner-type deformations, which break the de Sitter isometries explicitly at the level of the original ansatz, yet still retaining the positivity of the four-dimensional cosmological constant.  
We will find that these deformations do not affect the general structure of the quantum corrections studied in 
\cite{nodS} and the same breakdown of effective field theory occurs. 
Second, we consider a new ansatz for the internal space geometry as well as the background fluxes where all are time-dependent. We will find that at least for some, rather natural choices of time dependence the infinite tower of relatively unsuppressed corrections gets lifted, as these corrections acquire a time-dependence and become suppressed at late times, precisely when the type IIB description is expected to be valid.

In section \ref{isometries} we describe our general setup and discuss several ways of viewing de Sitter solutions in string theory, either as a coherent state in a flat or AdS background, or as a background geometry in its own right. The latter approach suffers problems, related to general properties of quantum field theory in de Sitter space as well as the breaking of supersymmetry. The coherent state view, on the other hand, justifies the quantum-corrected equation of motion based approach used here as well as in \cite{nogo, nodS}. We proceed to study the dipole and Kasner deformations to the de Sitter ansatz and show that the conclusions of \cite{nodS} hold in the presence of these deformations. 

In section \ref{timeft} we turn to the case of time-dependent fluxes and internal manifold. Here we improve on the classification scheme of \cite{nodS} for the quantum corrections and study the most general local and non-local corrections to M-theory that can be built out of derivatives or integrals of various contractions of the fluxes and curvatures. We determine the relative scalings of these corrections with the type IIA string coupling, which also tracks their time-dependence, and investigate the possibility of an infinite series of time-independent corrections, such as that found in \cite{nodS}. We study two main choices of time-dependence for the fluxes and internal geometry. One choice allows us to completely eliminate the series of time-independent corrections. This choice, however, results in a variable Newton's constant, and is unappealing for that reason. The other choice has a time-independent four-dimensional Newton's constant and allows us to lift the time-independence of all perturbative quantum corrections.
However there are still possible non-local time-independent corrections, which are only suppressed at small type IIB string coupling.

Having classified the corrections and determined their scalings, we investigate the quantum-corrected equations of motion at every order in the type IIA coupling in section \ref{qotomth}. We find that a solution with positive 4-dimensional curvature can be achieved, provided the leading quantum corrections satisfy inequalities similar to those found in \cite{nogo}. The leading order equations also determine the un-warped internal metric components, while the higher order corrections can be solved for in terms of the lower-order quantities so as to maintain the existence of the solution to all orders. We also derive the flux quantization and anomaly cancellation conditions, which provide consistency checks for our approach. Finally we check for tachyonic directions for the scalar fields in the effective 4-dimensional theory and comment on the relationship between our construction and the swampland criteria. We conclude with a summary and discussion of our results and future outlook.

\subsection{Organization and summary of the paper \label{dhurjoti}}

A more detailed organization and summary\footnote{Interested reader, who may not have the time to go through our paper, may read this section to familiarize himself/herself with the main results of our work. Needless to say, we have tried to summarize all of the key concepts in a hopefully comprehensive way.} of the paper is as follows. Although the paper broadly concentrates on two topics: one without time-dependences and one with time-dependences, the latter, however,  covers the majority of the contents. In terms of sectional distributions, section \ref{isometries} studies basically the time-independent cases and sections \ref{timeft} and \ref{qotomth} study in details the time-dependent cases. Therefore readers who want to see our results for the time-dependent cases, may directly jump to sections \ref{timeft} and \ref{qotomth}. In fact many of the conclusions about the time-independent cases, emerge as corollaries of the results for the time-dependent cases justifying the broader outlook of the 
scenario that we present here. However the time-independent cases, discussed in section \ref{isometries} and also in \cite{nogo, nodS}, are important in themselves because we present them using the unique perspective of coherent and squeezed states that hitherto, we believe, have not been emphasized in the literature\footnote{At least not in string theory. In quantum gravity such an idea is not new and has been first studied in \cite{dvali}. Here, and in more details in \cite{coherbeta}, we show how this could be realized in full string theory.}. 
This is basically the content of section \ref{dyson}. The point of view adopted in section 
\ref{dyson} allows us to view the four-dimensional de Sitter space, uplifted to M-theory, as a coherent or a squeezed coherent states over a given solitonic background. This solitonic background could as well be a supersymmetric one, helping us to cancel the zero point vacuum energies from the bosonic and the  fermionic fluctuations. The question that we ask in section \ref{dyson} is whether such a combined background, i.e the background with soliton plus the coherent state fluctuations, is a solution in M-theory. 
While classically such a background does not solve the equations of motions, the corresponding Schwinger-Dyson equations reveal that there are deeper issues that may even prohibit a solution to exist at the quantum level, at least in the realm of investigation here. These issues have some bearings on the loss of hierarchy between various scales and coupling constants involved in the problem, that do not seem to get alleviated even if we try to break the effective four-dimensional isometries while keeping the internal space time-independent. Therefore a more generic analysis is called for, where both the effective four-dimensional space-time as well as the internal space, including the background fluxes are all time-dependent. In sections \ref{timeft} and \ref{qotomth}, we study in details the possibility of the existence of solutions at the quantum level, and therefore also the existence of four-dimensional effective field theories, while still keeping the type IIB coupling constant under control.

However before we discuss in full details a generic class of time-dependent solutions, we answer two pertinent questions that could arise at this stage. In section \ref{difuli} we argue why, by generating time-dependences using dipole-type deformations, solutions would still fail to exist. And in section \ref{kasner}, we argue why, even if we change the isometries of the four-dimensional de Sitter space while still keeping the geometry of the six-dimensional internal space time-independent, solutions would again {\it not} exist. Thus the situation at hand is more subtle than previously thought. 

Sections \ref{timeft} and \ref{qotomth}  are the main parts of the paper where we take a time-dependent type IIB background \eqref{pyncmey}, i.e a background where the four-dimensional space has de Sitter isometries and the compact internal six-dimensional space has time dependent warp-factors (with time running from $-\infty < t \le 0$). The background fluxes are also time-dependent, but we keep type IIB coupling constant to be time-independent. As alluded to earlier, this is necessary to make sense of any computations that we perform here. However time-dependent fluxes on compact internal space raises new questions on flux quantizations and anomaly cancellations. In addition to that, the fact that type IIB theory has NS and RR three-form fluxes, five-form fluxes as well as axio-dilaton, all on a time-varying six-dimensional internal space, raise numerous additional questions that have hitherto never been studied before. 

This proliferation of the number of time-dependent fields does have a slightly simpler representations from the M-theory perspective. However the reader should be warned from early on: M-theory will be used as a 
trick or for book-keeping purpose to solve the type IIB problem. Use of M-theory {\it does not} imply looking for a de sitter space in M-theory. The de Sitter space that we want to study will always be in the type IIB side.
Having said this, uplifting our type IIB background \eqref{pyncmey} to M-theory, will allow us to switch on time-dependent metric \eqref{vegamey} and G-flux components \eqref{frostgiant}. The issues of flux quantizations and anomaly cancellations are unfortunately {\it not} alleviated by this uplifting, rather all the type IIB questions should now be answered from M-theory point of view. None of the subtleties that we encounter in the type IIB side disappear from the M-theory uplifting, but the {\it only} advantage that we get from M-theory is the sheer compactness of the number of fluxes: most of the type IIB fluxes are packaged neatly as G-flux components. This viewpoint at least provides us with a controlled laboratory to perform our computations.  In section \ref{temporal} we 
illustrate the ingredient that go in the uplifted type IIB background to M-theory. 

Unfortunately the subtleties do not end here. As discussed in section \ref{isometries}, time-dependences 
{\it and} quantum corrections go hand in hand, and both are necessary to get any solutions, as de Sitter space in string theory is a highly quantum system and not a classical one.
Quantum corrections can be of various kinds: perturbative and non-perturbative, local and non-local, so the question is how to organize them so that meaningful computations could be performed. The additional subtlety is from the inclusion of all possible corrections
as {\it a-priori} there is no way to justify that de Sitter space could appear from finite number of quantum corrections, unless of course there is some inherent 
{\it hierarchy}. This then brings us to the sticky issue of justifying the existence of an inherent hierarchy 
with respect to both $M_p$ and type IIA string coupling $g_s$. In section \ref{pertu} we bring forth all these subtleties under various subsections, and provide possible answers. 

The quantum corrections are computed near weak flux backgrounds, so a generic quantum term could be expressed solely as polynomial functions of the G-flux components contracted appropriately with warped inverse metric components in M-theory. In subsection \ref{Gng} we study generic polynomial functions of the G-flux components. Interestingly the type IIA coupling $g_s$ now becomes a function of time, and we can use this to our advantage to trade the temporal dependences with $g_s$ dependences. This way we can simply ask for $g_s$ dependences of the quantum terms. Additionally, throughout the paper we study two categories of time-dependent backgrounds: one with time-independent volume of the six-dimensional internal space \eqref{olokhi}, and the other with time-dependent volume of the internal space 
\eqref{ranjhita}. These are  respectively related to time-independent and time-dependent four-dimensional Newton's constants.   

In subsection \ref{Gng2}, we add multiple derivative with G-fluxes and study the $g_s$ scalings of the quantum terms for the two cases \eqref{olokhi} and \eqref{ranjhita}. For both cases we find that time-dependences of the G-fluxes give rise to a certain level of $g_s$ hierarchies. These hierarchies were missing for the time-independent cases studied in section \ref{isometries}, which in turn lead to the non-existences of four-dimensional EFTs in the type IIB side. More importantly however, existence of $g_s$ hierarchy for the case \eqref{ranjhita} requires some derivative constraints that we illustrate  in subsection \ref{Gng2}.   
 
G-fluxes are not the only ingredients in M-theory, there are metric and curvature components  that need to be inserted in the quantum terms. Clearly this will make the story much more complex, so to deal with this we first study the curvature terms by themselves and ask the question whether polynomial powers of the curvature terms can induce hierarchies to the two cases \eqref{olokhi} and \eqref{ranjhita}. Introducing polynomial powers of curvatures require careful manipulations of the Christoffel symbols, Riemann and Ricci tensors as well as the Ricci scalars. This calls for a study of curvature algebras and product of curvature tensors. In subsection \ref{Gng3} we study in details such algebras and the $g_s$ scalings of the various curvature tensors. The results are shown in {\bf Table \ref{firozasutaria}} for the two cases 
\eqref{ranjhita} and \eqref{olokhi}.
 
The answer that we get,  from subsections \ref{Gng3} and \ref{Gng4} (the latter being with the inclusion of  multiple derivatives), is rather surprising. The curvature polynomials, no matter how they are arranged, always have positive $g_s$ hierarchies. In other words $g_s$ scalings of any polynomial powers of the curvature tensors and their derivatives always have non-zero $g_s$ scalings, compared to the cases studied with polynomial powers of the G-flux components. Therefore it appears that temporal dependences of the metric components seem to naturally induce $g_s$ hierarchies to polynomials constructed out of derivatives of the metric tensors, compared to the polynomials with G-fluxes. 

The story does not end here, because we can now combine everything and ask for polynomials containing 
product of curvatures, G-fluxes and derivatives. For the two cases, \eqref{ranjhita} and \eqref{olokhi}, the most generic quantum terms may be expressed as \eqref{phingsha} and \eqref{phingsha2} respectively. 
Additionally, we are interested in late time physics, i.e $g_s \to 0$, so polynomial powers of $g_s$ are allowed whereas 
${\rm exp}\left(-{1\over g_s}\right)$ may be consistently ignored as they die off much faster than polynomials
in $g_s$. This way credence could be given to the $g_s$ expansions of all the variables in the M-theory uplift. Taking all these into account, the results of subsection \ref{Gng5} are interesting and instructive. For the two cases, \eqref{ranjhita} and \eqref{olokhi}, the $g_s$ scalings of the generic quantum terms become \eqref{miai} and 
\eqref{melamon2} respectively i.e $g_s^{\theta_k}$ and $g_s^{\theta'_k}$ respectively. 
Both can be made non-zero if we make the modes $k$ (as defined in the G-flux expansions \eqref{frostgiant} with $n = 0$ therein) to have the following lower bounds $k \ge {9\over 2}$ and $k \ge {3\over 2}$ respectively (although we will speculate on other lower bounds too). As soon as we switch-off $k$, say for the case 
\eqref{olokhi}, we see that the $g_s$ scaling becomes \eqref{kkkbkb2}, i.e $g_s^{\theta'_0}$, 
which allows relative {\it minus} signs. These minus signs tell us that for any given value of $\theta'_0$ in \eqref{kkkbkb2}, there are in fact an {\it infinite} number of states classified by \eqref{evabmey2} thus ruining the $g_s$ hierarchy altogether (similar argument works for the case \eqref{ranjhita}).

This loss of $g_s$ hierarchy for vanishing $k$ (i.e for the time-independent fluxes), for both cases \eqref{olokhi} and \eqref{ranjhita}, is a reminder that these theories may be in the {\it swampland}, thus giving some credence to the conjectures of \cite{vafa1}. In fact we see that a stronger condition emerges: as long as the fluxes and the internal space are time-independent, no amount of quantum corrections can save the day. These theories will have no EFT descriptions in four-dimensions with de Sitter isometries. Breaking isometries in any way do not help either as shown in section \ref{isometries}. 

On the other hand, switching on time-dependence miraculously saves the day by creating at least the necessary $g_s$ hierarchy. The $M_p$ hierarchy, on the other hand, requires a different level of investigation because it involves positive and negative powers of $M_p$. The negative powers of $M_p$ are easy to argue: they come together with the derivative expansions that we have entertained so far. The positive powers however require nested integrals. These nested integrals appear from the non-local counter-terms in M-theory and are elaborated in details in subsection \ref{Gng6}. The non-local counter-terms are expressed using non-locality functions $\mathbb{F}^{(r)}(y - y')$ which, at low energies, are sharply peaked functions so that eleven-dimensional supergravity description still remains valid at low energies. 
In the absence of time-dependences, these non-local counter-terms ruin the $M_p$ hierarchies as shown first in \cite{nodS}. Again, time-dependences help us here by decoupling these non-local 
counter-terms and thus restoring back the $M_p$ hierarchies for both \eqref{olokhi} and \eqref{ranjhita}.

Our next set of quantum terms are the topological ones that we discuss in subsection \ref{tufi}. These are constructed out of curvature forms and various other forms from the G-flux components. On the other hand, 
the non-topological interactions could also be built using Hodge star operations on them. These non-topological interactions couple to the G-fluxes and are related to the quantum terms \eqref{phingsha} and \eqref{phingsha2} for the two cases \eqref{ranjhita} and 
\eqref{olokhi} respectively. One could also construct {\it dual} forms and therefore also the corresponding quantum terms $-$ say for the case \eqref{olokhi} $-$ as we show in \eqref{phingsha33}. The quantum terms associated with these dual forms, namely \eqref{verasofmey},  and their $g_s$ scalings, appear in {\bf Table \ref{dualforms}}. From here one may easily check that the $g_s$ scalings of the quantum terms with dual variables, as in \eqref{phingsha33}, are exactly the same as that of \eqref{phingsha2}. In other words the scalings are as in \eqref{melamon2}.

Having tabulated all the possible quantum effects in the M-theory uplift, we now go to the detailed study of the equations of motions (EOMs) in section \ref{qotomth}. Our first topic is the study of all the Einstein's EOMs in subsection \ref{instoo} by incorporating the energy-momentum tensors from the G-fluxes and from the quantum terms that we tabulated in subsection \ref{pertu}. The internal eight-dimensional manifold is of the form \eqref{melisett} with ${\cal M}_4$ parametrized by coordinates ($m, n$); ${\cal M}_2$ parametrized by ($\alpha, \beta$) and ${\mathbb{T}^2\over {\cal G}}$ parametrized by ($a, b$). Shrinking the ($a, b$) torus to zero size will take us to the type IIB background in the standard way, although late time automatically does this to our M-theory background. 

Our approach in subsection \ref{instoo} is to study each and every Einstein's equations for the two cases
\eqref{olokhi} and \eqref{ranjhita}. The aim would be to extract out the salient features for the two cases from their EOMs, so that in the end we could assimilate everything to see under what conditions an ansatze like \eqref{vegamey}, or equivalently \eqref{pyncmey}, would be a solution to the EOMs. Clearly since the metric input in \eqref{vegamey} follows the decomposition \eqref{melisett}, we will have to concentrate on 
{\it five} different types of Einstein's EOMs: one for each of the four allowed orientations in \eqref{melisett}, namely EOMs along ($m, n$), ($\alpha, \beta$), ($a, b$) and ($\mu, \nu$) directions corresponding to 
${\cal M}_4$, 
${\cal M}_2$, ${\mathbb{T}^2/ {\cal G}}$ and $2+1$ dimensional space-time directions respectively. The fifth one is for the possible cross-term EOMs.

We start with the analysis along ($m, n$) directions by first concentrating on the case \eqref{olokhi}. The 
G-flux components take the form \eqref{ravali} because we want to narrow down our analysis to the late time scenario where $g_s \to 0$. The 
energy-momentum tensor associated with the G-flux is given in \eqref{nkinsky}, and the energy-momentum tensor for the quantum terms may be extracted from \eqref{neveC} by making $k_2 = 0$ therein. Incorporating everything, the zeroth order in $g_s$ gives us \eqref{misslemon2} where the RHS has the quantum pieces, classified by $\theta'_k = {2\over 3}$ in \eqref{melamon2}, and the G-flux pieces, 
captured by ${\cal G}^{(3/2)}_{MNab}$ components in \eqref{nkinsky}. 

All is good except for two caveats: One, \eqref{misslemon2} actually mixes the un-warped metric components $g_{mn}$ with $g_{\alpha\beta}$ and $g_{ab}$, so we will need more information to solve it. 
And two, $\theta'_k = {2\over 3}$  in \eqref{melamon2} doesn't actually capture any {\it quantum} pieces 
because $l_i$ appearing in \eqref{melamon2} can either be 1 or 2 depending on whether we choose curvature terms or the G-flux terms respectively. Thus $\theta'_k = {2\over 3}$ can at-most renormalize the existing classical terms. The real {\it perturbative} quantum terms appear when we go beyond the zeroth order in $g_s$, i.e to order $g_s^{1/3}$. The EOM is now given by \eqref{rambha3} with contributions to the G-flux energy-momentum tensor now appearing from higher order G-flux components; and the contribution to the quantum energy-momentum tensor now appearing from $\theta'_k = 1$ in \eqref{melamon2}. The latter is classified by \eqref{melamon4}. Going even beyond this order, i.e going to order $g_s^{2/3}$, the EOM is governed by 
\eqref{clovbalti} with higher order G-flux components and the quantum terms being classified by 
\eqref{mameye} for $\theta'_k = {4\over 3}$ in \eqref{melamon2}.

For the second case, i.e \eqref{ranjhita}, the story repeats in a similar fashion although specific details about fluxes etc. differ. The G-flux components are now given by \eqref{nickid}, and the energy-momentum constructed out of them takes the form \eqref{nkinsky2}. To the zeroth order in $g_s$, the only G-flux components that contribute to the flux energy-momentum tensor are constructed from 
${\cal G}^{(9/2)}_{\alpha\beta ab}$ giving rise to the EOM \eqref{spoonR}. The {\it quantum} terms contributing to the EOM is classified by $\theta_k = {2\over 3}$ in \eqref{miai} and, as before, simply renormalize the existing classical data. To the next order in $g_s$, i.e to order $g_s^{1/3}$, the EOM is 
\eqref{rambha4} with the quantum terms classified by $\theta_k = 1$ in \eqref{miai}. The story then progresses in a similar fashion as we ascend to higher orders in $g_s$.

The analysis for the other two directions, namely ($\alpha, \beta$) and ($a, b$) directions, has many new subtleties associated with the choice of the G-flux components, distributions of the quantum terms and the various orders of $g_s$. These are discussed in details in subsections \ref{kocu2} and \ref{kocu3} 
respectively. In fact the two cases, namely \eqref{olokhi} and \eqref{ranjhita}, have many distinguishing features that point towards the subtle differences between them  that appear from analyzing their behavior for the two directions ($\alpha, \beta$) and ($a, b$). For example, case \eqref{ranjhita}, allows a conformally Calabi-Yau four-fold with vanishing Euler characteristics, whereas the eight-manifold for the case 
\eqref{olokhi} is typically non-K\"ahler (not necessarily complex either) and has a non-zero Euler characteristics.
Despite that, the zeroth orders in $g_s$, do have certain similarities in their EOMs  to what we saw for the ($m, n$) cases, at least regarding the 
behaviors of the quantum and the flux terms.

All these similarities however {\it do not} survive when we analyze the EOMs for the space-time, i.e 
($\mu, \nu$), directions for the two cases \eqref{olokhi} and \eqref{ranjhita}. This is detailed in subsection
\ref{kocu4}. Let us first consider the case \eqref{olokhi}. The spatial and the temporal Einstein's tensors are given by \eqref{synecdoche} and \eqref{synecdoche2} respectively. These Einstein's tensors would now have to be balanced not only by the energy-momentum tensors of the G-fluxes and the quantum terms, but also by the energy-momentum tensors of (integer and fractional) M2 and $\overline{\rm M2}$-branes
(the latter are essential in our set-up as may be seen from subsection \ref{maryse}). The flux energy-momentum tensor has a somewhat standard form 
of \eqref{keener}, but the quantum terms are very different from what we had earlier. In fact this is where the first sign that something beyond perturbative quantum terms are needed appears. The quantum terms are now classified by $\theta'_k = {8\over 3}$ in \eqref{melamon2}, with some {\it non-perturbative} inputs, and therefore satisfies and equation of the form \eqref{oleport}. This equation has an important consequence: it allows quantum terms with 
eighth order in derivatives, implying quartic in curvatures and/or eighth orders in G-fluxes!  The EOM then takes the form \eqref{fleuve} which is an equation to zeroth order in $g_s$, therefore classical, yet it contains 
terms with quartic orders in curvatures and/or eight orders in G-fluxes\footnote{This has also been observed recently by Savdeep Sethi \cite{sethiudem}. We thank him for discussions on this and many other related issues.}. 

The story for the case \eqref{ranjhita} resonates somewhat with the case \eqref{olokhi}, but again the specific details differ both in terms of the choices of the fluxes, quantum terms and the branes. Due to vanishing Euler characteristics the construction involves either vanishing number of M2-branes or equal number of branes and anti-branes (so as to cancel global charges). Additionally, it appears that there are two possible classes of backgrounds allowed: one with a harmonic warp-factor \eqref{ouletL} and another with a non-harmonic warp-factor satisfying \eqref{lebanmey}. The EOM for the latter is given by \eqref{emilee2}, with the quantum terms now represented by \eqref{1851}. This again allows terms quartic in curvatures and/or eighth orders in G-fluxes, so the story remains somewhat similar to \eqref{olokhi}. However the EOMs alone do not allow us to choose one over the other, but the G-flux EOMs in subsection \ref{jutamaro} do suggest the latter to be the correct EOM.

The first appearance of non-perturbative terms in subsection \ref{kocu4} suggests that this might be more generic and therefore should affect all the other Einstein's equations for $(m, n), (\alpha, \beta)$ and 
$(a, b)$ directions. This turns out to be true once we analyze the effects of the BBS \cite{BBS} and 
KKLT \cite{KKLT} -type instantons, both in their localized and de-localized forms. These are discussed in 
subsection \ref{instachela}, but more detailed analysis including derivations etc. have appeared recently
in \cite{coherbeta}. The result that we get from the non-perturbative terms are interesting: there  are higher order contributions $-$ notably eight-order polynomial forms $-$ that contribute to {\it all} the Einstein's EOMs and not just the ones along the $(\mu, \nu)$ directions. This democratic appearance of the quantum terms does have important consequences influencing the later sections.

In the time-independent case, the classification of the Einstein's EOMs that we performed above should have sufficed. However time-dependences bring forth additional subtleties and therefore additional EOMs.
What really happens is that the temporal dependences of the various metric components induce cross-terms EOMs
despite the non-existence of cross-terms in the metric (i.e over and above \eqref{melisett}). This is elaborated in subsection \ref{cross}. One of the important consequence of these EOMs is that the temporal behavior of some of the internal metric components (specifically the ones for ${\cal M}_4$ and 
${\cal M}_2$ in \eqref{melisett}) may be identified with the quantum terms that, in turn, are classified by 
$\theta'_k = 2$ in \eqref{melamon2} for the case \eqref{olokhi} and $\theta_k = 3$ in \eqref{miai} for the case \eqref{ranjhita}, at least perturbatively. One could add in the non-perturbative corrections from subsection
\ref{instachela}, but we do not pursue it further here as the main conclusions do not change much.

All that remains now is to see if solutions would exist for all the EOMs classified above. The subsection
\ref{kocu5} deals with analyzing these EOMs for the two cases \eqref{olokhi} and \eqref{ranjhita}. The result for \eqref{olokhi} may be neatly presented as \eqref{musemey}, 
which should be compared to eq. (6.10) of \cite{nogo}. The zero on the RHS of \eqref{musemey} appears from integrating the Laplacian on the warp-factor
over the compact base ${\cal M}_4 \times {\cal M}_2$, and since the warp-factor is a smooth function, the integral vanishes. The smoothness of the warp-factor is of course guaranteed from the series of quantum corrections appearing in \eqref{muse777}. 
Clearly, in the absence of the quantum pieces, the system has no solution because the integral involves only positive definite functions and therefore the consistency will demand vanishing fluxes and vanishing 
cosmological constant $\Lambda$. Interestingly {\it negative} $\Lambda$ is allowed even if the quantum terms are absent, implying both Minkowski  and AdS spaces may be realized in a set-up like ours.  In the presence of the quantum pieces, the consistency condition here differs in a crucial way with the one presented in \cite{nogo}. The quantum terms in 
\cite{nogo} are classified by $2/3 \le \theta'_0 \le 8/3$ for the internal and the space-time directions with $\theta'_0$ defined in \eqref{kkkbkb2}. These have infinite number of solutions for both cases, from local and non-local quantum terms, implying that an expression like eq. (6.10) in \cite{nogo} does not have any solution at all and is in the swampland. However now the scenario has changed. The internal and the space-time quantum terms are now classified by $2/3 \le \theta'_k \le 8/3$ with $\theta'_k$ defined as in 
\eqref{melamon2}. These have {\it finite} number of solutions in both cases, and in fact the lower bounds on the quantum terms, as we saw earlier, do not contribute much. This means the actual higher order quantum
terms appear from $\theta'_k \le 8/3$
in \eqref{musemey}. These quantum terms appear with a relative {\it minus} sign in \eqref{musemey}, and therefore if we can demand that the dominant positive contributions come from  the space-time quantum terms, then surprisingly solutions would exist where there were none before!

For the case \eqref{ranjhita}, assimilating all the EOMs in the same vein as above, unfortunately does not lead to an elegant conclusion like above. Indeed, combining with the derivative constraint for this case, and the fact that the quantum terms have to satisfy a stronger constraint like \eqref{murderland} (with the replacement \eqref{lucind}) along-with the possibility of late-time singularity as in \eqref{walmart},
  nonetheless  show that the late time physics with a four-dimensional de Sitter space-time, i.e with \eqref{olokhi}, is a preferable scenario over the ones with time-varying Newton constants. We summarize the differences between the two cases in {\bf Table \ref{jutameyL}}.

The next set of equations are from the G-flux EOMs, which open up numerous new subtleties that we believe have hitherto not been discussed much in the literature. Section \ref{jutamaro} is dedicated in elaborating all these subtleties.
One of the most important set of subtleties are related to quantization of fluxes on four-cycles of the internal eight-manifold. As we discussed earlier, the G-flux components vary with respect to time (here, since the temporal behavior is traded with $g_s$, the G-flux components have $g_s$ dependences) over a four-cycle that {\it also} varies with time. How does one go about understanding flux quantization in such a scenario? The answer lies in the subtle relationship between the flux EOMs and the infinite series of the quantum terms, as we show in subsection \ref{bianchi}. In fact what we need here is the EOM for the {\it dual} seven-form flux components. This is where the detailed analysis of the subsection \ref{tufi} becomes relevant now. The dual flux EOM is given by \eqref{uniprixtagra}, which eventually leads us to the modified flux quantization condition \eqref{sakuras}. Plugging in the temporal behaviors of the fluxes and the quantum terms, the quantization procedure becomes \eqref{scarymag}. Note that, in the absence of time-dependences, 
\eqref{sakuras} does lead to the well-known flux quantization procedure \eqref{cyclemey} reproducing the results of \cite{wittenflux}. 

The equation \eqref{scarymag} is interesting in its own right. It tells us how a $g_s$ varying G-flux component should be related to a $g_s$ varying quantum term, even if the system does not have movable 
M5-branes. We take \eqref{scarymag} as our fundamental equation and show that, in a hopefully convincing way, how for {\it each and every} G-flux components the quantization procedure works in a time-varying scenario. We have tabulated the results in 
{\bf Table \ref{blanchek}}.

There are two other potential contributions to the flux quantization conditions that we only gave cursory attentions in subsection \ref{bianchi}. These are the number of dynamical M5-branes, denoted by  $N$, and the integrated 
four-form, denoted by the integral of $\mathbb{\hat Y}_4$, in \eqref{scarymag}.  Both these could have potential $g_s$ dependences and would therefore contribute to the flux quantization conditions. 

The second set of subtleties appear when we ask the following question: how is Gauss' law satisfied on a compact internal space with time-varying fluxes and almost static membranes? Answering this question will open up new interconnections between flux EOMs and the quantum terms \eqref{phingsha2} for the case 
\eqref{olokhi}. The Gauss' law is represented here by \eqref{marbrick} containg all the ingredients assimilated from subsections \ref{tufi} and \ref{bianchi}. On the outset \eqref{marbrick} looks like the standard anomaly cancellation condition one would get from 
\cite{BB, DRS}, however a closer inspection reveals a few subtleties. One, the flux integral is now time-dependent because the ${\bf G}_4$ fluxes do not have any time-independent parts. Two, we have an integral over the topological 8-form $\mathbb{Y}_8$, whose polynomial form appears in \eqref{ashf2}, instead of just ${\bf X}_8$ as in \cite{BB, DRS}. Three, there appears a {\it new} contribution coming from the integral of a {\it locally} exact form $d\ast_{11} \mathbb{Y}_4$ over ${\cal M}_8$ from the quantum corrections. And four, we have $n_b$, the number of static M2-branes, that is a time-independent factor. Thus 
\eqref{marbrick} is not just a single relation as in \cite{DRS}, rather it is now a mixture of time-dependent and time-independent pieces juxtaposed together. How do we disentangle the various parts of \eqref{marbrick} to form consistent anomaly cancellation conditions for our case?  

The answer to these questions appears in subsection \ref{anoma}. There are two set of equations that we need to consider. The first set appears from integral over ${\bf X}_8$ as in \eqref{poladom}. Compared to the time-independent case, this equation may be divided into two parts: one, that is related to the Euler characteristics of the eight-manifold \eqref{sun18c}, and two, this is a time-dependent factor. The time-independent piece would be related to the number of membranes as \eqref{haroldR}.  There are some subtleties associated with the identification of the Euler characteristics to the ${\bf X}_8$ integral that we clarify in subsection \ref{anoma}. A careful analysis however reveals no time-independent piece from the 
${\bf X}_8$ polynomial, implying that there should be an equal number of branes and anti-branes in our set-up. 

This means the  set of equations appearing from the time-dependent parts of \eqref{poladom}, coupled to the G-flux EOM, form a consistency condition as \eqref{technox1}. Under certain simplification this equation may be represented in component form as \eqref{cleota}, which is the fundamental equation on which we base our second set of anomaly cancellation condition. This appears in a compact form as \eqref{ramamey} for the 
two cases \eqref{olokhi} and \eqref{ranjhita}.
Using all the information, one could perform many consistency checks now, and we name a couple here. We can easily argue that the G-flux components appearing in our set-up are no longer {\it self-dual}. Recall that self-duality, defined over the internal eight-dimensional internal space, is an important condition to guarantee
{\it supersymmetry}. Our space is clearly non-supersymmetric and thus non self-duality is a natural outcome of the analysis. Additionally, we can now show that \eqref{lebanmey} is the correct EOM, justifying our choice for the second case \eqref{ranjhita} in subsection \ref{kocu4}. All these, and other checks, form the contents of subsection \ref{anoma}.

What we are lacking so far are the flux EOMs in the same vein as the Einstein's EOMs discussed earlier. This is rectified in subsection \ref{maryse} where we elaborate the EOMs for each and every G-flux components in full details. One of the interesting outcome of our approach is the precise determination of the warp factor $H(y)$ in terms of the M2 and the $\overline{\rm M2}$-branes in \eqref{evaB102}. In addition to that the supersymmetry breaking condition is given by the non-self-duality of the G-flux components as in  \eqref{evebe}. 

The analysis of subsections \ref{anoma} and \eqref{maryse} however leaves open the possibility of having {\it dynamical} membranes. Allowing dynamical membranes should stir up additional corrections to the G-flux components. 
The question then is: could this change the very outcome that we have been advertising so far? The answer, that we elucidate in subsection \ref{branuliat}, turns out to be surprisingly no, as the only G-flux components that seem to be effected are of the form 
${\bf G}_{M0ij}$. Here $y^M$ parametrize the coordinates of ${\cal M}_4$ and ${\cal M}_2$, the latter being absent for the case \eqref{ranjhita} because of the derivative constraint. Our analysis shows that the exact form for ${\bf G}_{M0ij}$ can in fact be derived as in \eqref{hokyra} and in the limit $g_s \to 0$, which is the later time scenario, the most dominant part of \eqref{kyratagra} is exactly the G-flux components that we have been considering thus far. This points to the robustness of our conclusions even in the presence of dynamical membranes. 

However dynamical membranes, which become dynamical D3-branes in the IIB side, now lead to the possibility of realizing {\it inflation} in our set-up! In fact, in the presence of seven-branes this could be mapped to either the D3-D7 \cite{DHHK} or the KKLMMT \cite{kklmmt} inflationary model, albeit now in the presence of dynamical branes, fluxes and geometry. There are however a few caveats on the way to the possible realization of the D3-D7 inflation. The first one is the range of time that we could have quantitative control on the dynamics of the system. This is \eqref{alibet}, and if $\Lambda$, the cosmological constant, is very small, \eqref{alibet} could in principle allow us to access a reasonably large interval of time. Inflation being the {\it early time} physics, one might be able to access certain levels of e-folds from our set-up. Alternatively, it could be that our $g_s$ expansions of all the variables, that worked so well for the late time physics, may not be good enough to access any significant parts of the inflationary evolution of the four-dimensional space-time in the IIB side. In that case the $g_s$ expansions need to be modified 
(see \cite{coherbeta}). 

The second caveat seems to be related to the motion of the dynamical M2-branes. How do we make the M2-branes move slowly enough so that inflationary dynamics may actually be realized in our set-up (provided of course we have a way to take care of all the issues pointed out in the first caveat)? We will also need seven-branes, so question is how are the seven-branes realized in the compactification that we study here from the M-theory side. The latter does have an elegant answer and in fact ties up one loose end that we kept under the rug so far, namely, how do we interpret the ${\bf G}_{MNab}$ flux components?  

It turns out the G-flux components ${\bf G}_{MNab}$ are not global fluxes, rather they are {\it localized} fluxes as expressed as \eqref{teskimey} using normalizable forms $\Omega_{ab}$. The two-forms
${\bf F}_{MN}$ that appear from these localized G-flux components can now be interpreted as gauge fluxes on the D7-branes. The D7-branes, on the other hand, appear from the T-dual of the {\it orbifold} points 
on the internal eight-manifold \eqref{melisett} in M-theory. All these points towards a possible F-theory realization of our set-up, which of course ties up to the F-theory realization of the D3-D7 inflationary model in \cite{DHHK} or in \cite{kklmmt}.

Other allowed G-flux components, for example ${\bf G}_{MNPa}$ and ${\bf G}_{MNPQ}$, could be viewed as global fluxes leading respectively to the three and five-form fluxes in the IIB side. Along with the quantum terms \eqref{phingsha2}, for the case \eqref{olokhi}, now allow us to express the G-flux components as 
\eqref{fasit} thus satisfying the Bianchi identities, anomalies as well as the EOMs all in one go. Additionally, un-wanted components like ${\bf G}_{0MNP}$ could be easily made to vanish as \eqref{pashaclub} using the freedom in the choice of three-form potential ${\bf C}_3$. 

All these and other details, that we carefully and meticulously derived in subsections \ref{branuliat}, 
\ref{anoma} and \ref{bianchi}, prepare us to embed D3-D7 inflationary model to study early-time physics. However, how 
{\it early} it could be, as we discussed above, is a matter of some debate now. Additionally other subtleties, again as pointed out above, suggest that a more careful study is called for here. We therefore leave this for future work, and instead concentrate on interpreting some of our results in the light of the swampland criteria 
in section \ref{nala}. 

One of the  important question is the {\it stability} of our background. From the discussions 
in subsection \ref{stabul} we can summarize our view of stability here. The classical EOMs, or the EOMs to the lowest order in 
$g_s$ (which for most cases are to zeroth order in $g_s$ with the exception of one where the lowest order is $g_s^2$), for all the components are \eqref{misslemon2}, \eqref{uanaban}, \eqref{buskaM} and 
\eqref{fleuve}. They involve the so-called quantum terms that, for all cases except the space-time ones, renormalize only the existing classical data. The non-perturbative effects contribute eight-order (in derivatives) polynomials. Together with the G-flux components they determine the type IIB metric with four-dimensional de Sitter space-time and the un-warped internal six-dimensional non-K\"ahler metric. The quantum effects on {\it this} background, to order-by-order in powers of $g_s$, are balanced against the G-flux components and the higher order terms of the metric coefficients, again to order by order in powers of 
$g_s$, in a way so as to preserve the form of the dual type IIB metric to the {\it lowest} order in $g_s$.  This is one of the essential criteria of stability here.

What about tachyonic instabilities? They require more involved analysis because they call for varying the quantum action to second orders in metric, and fluxes. With only metric variations, the quantum terms contributing to the tachyonic instabilities are classified by ${4\over 3} \le \theta'_k \le {22\over 3}$ in \eqref{melamon2} for the case \eqref{olokhi}. These terms should make the RHS of \eqref{harmonir} negative definite. There are also other variations possible. For example second variations with respect to the three-form potentials ${\bf C}_{MNP}$, or even mixed variations by including metric components. The criteria to make them negative definite are discussed in subsection \ref{stabul}.

Our final set of analysis is related to the swampland criteria \cite{vafa1} and the energy conditions, namely the null, strong and dominant energy conditions (see recent study in \cite{russot}). They are all elaborated in subsection \ref{jola}, and have roots in the exact expression for the cosmological constant $\Lambda$ from \eqref{hathway}. The cosmological constant that we get for our case has contributions mostly from the zeroth order in $g_s$ 
in \eqref{phingsha2} for the case \eqref{olokhi}. This means, although the full quantum potential 
\eqref{ducksoup}  (or \eqref{chuachu}) has $g_s$ dependence (or time dependence), the pieces contributing to the cosmological constant are basically the $g_s$ independent pieces (see footnote 
\ref{expanda} for an explanation). This implies that the cosmological constant  is truly a {\it constant} here and, since the Newton's constant is also time-independent, the late time cosmology is de Sitter and {\it not} quintessence. This is also evident from the fact that the swampland criteria, as expressed in \cite{vafa1}, are easily taken care of as we show in 
\eqref{lojjabot}, using one scalar field \eqref{katygorom}, and in \eqref{lina05}, using all the relevant scalar fields.

Interestingly, the null energy condition (NEC) could also be satisfied once we use the quantum corrected energy-momentum tensors. The NEC can be expressed as \eqref{helen} and one may easily infer from there that, in the {\it absence} of the quantum terms, \eqref{helen} cannot be satisfied. The traces contributing to \eqref{helen} can be made explicit as \eqref{montcross} and \eqref{crossmey}, and it is not hard to see that with the choice 
\eqref{sortega}, the NEC can be re-written as \eqref{kortega}. Therefore the burden of satisfying the NEC 
lies solely on the 
$2+1$ dimensional space-time quantum corrections $[\mathbb{C}^\mu_\mu]^{(0, 0)}$, and since they are classified by $\theta'_k = {8\over 3}$, this provides us with enough freedom to satisfy \eqref{helen}. In fact under special choice of the higher order polynomials, the {\it strong} and {\it dominant} energy conditions, as expressed in \eqref{hanasuit}, might also be satisfied. Interestingly, from the exact expression of 
$\Lambda$ in \eqref{hathway}, the burden of getting $\Lambda > 0$  also lies solely on 
the positivity of the quantum corrections, thus bringing us full-circle. We end with a short discussion on moduli stabilization.


%


\section{Backgrounds with de sitter isometries and beyond \label{isometries}}

The issue of generating a positive cosmological constant solution in supergravity or string theory has been a challenging problem for a long time. Despite this level of difficulty, a stage of reconciliation has been achieved: it is now known that there are no classical four-dimensional de Sitter solutions in string theory. Quantum corrections are then essential, and the general consensus so far has been that four-dimensional de Sitter vacua could be generated by including quantum corrections in the system, thus going beyond supergravity approximations. This is pretty much the content of the no-go theorems given first by Gibbons \cite{gibbons}, followed by a more elaborate version by Maldacena and Nunez \cite{malnun}. All these works discussed the inabilities of fluxes or branes to {\it uplift} any background solutions with zero or negative cosmological constant to  the ones with positive cosmological constants. Other stringy ingredients like anti-branes and orientifold planes were later shown in \cite{nogo} to be equally ineffective, thus paving way to the sole savior of the situation, namely, the quantum corrections. In fact the study in \cite{nogo} revealed an additional constraint on the quantum corrections themselves: the quantum corrections, as they appeared in  specific ways in \cite{nogo},  should sum up to some {\it negative} definite quantity to allow for positive cosmological constant solutions  to appear in four-dimensions.

Such a constraint on the quantum corrections should already be alarming as every pieces of the quantum corrections appearing in the constraint is an infinite series by itself. Thus it would only make sense if there exists some inherent hierarchies in the quantum series expansion. Recall that the analysis of \cite{nogo} was done in the M-theory uplift of the type IIB theory and therefore the hierarchies in question are the $g_s$ and the $M_p$ hierarchies, where $g_s$ is the type IIA coupling. The specific type IIB background that we want to obtain as a solution of the quantum corrected EOMs in M-theory is of the form:
\bg\label{betta}
ds^2=\frac{1}{\Lambda(t)\sqrt{h}}(-dt^2+ dx_1^2+ dx_2^2+ dx_3^2)+\sqrt{h}{j}_{mn}dy^m dy^n,
\nd
where $h(y)$ is the warp-factor and $\Lambda(t) \equiv \Lambda \vert t\vert^2$ was chosen in \cite{nogo}
to allow for a four-dimensional de Sitter space under a flat slicing with $-\infty \le t \le 0$. Note that the metric 
of the internal space $j_{mn}$ is time independent so that the four-dimensional Newton's constant $G_N$ can remain time independent\footnote{The precise nature of the Newton's constant depends on our choice of the classical (or solitonic) background. This will be elaborated in \eqref{redjohn}.}. 
This is not an essential requirement, and we shall study variant of this later in the paper, although we do expect $\dot{G}_N/G_N$ to be constrained by cosmological data.  

The question however is the existence of a metric of the form \eqref{betta}. In type IIB side this boils down to the question of the existence of both the space-time metric components $g_{\mu\nu}(y, t)$ as well as the internal metric components $g_{mn}(y)$. To analyze this we will have to go to the M-theory uplift of the type IIB background as alluded to earlier, because the IIB background is more cumbersome to handle. Again, questions may be raised against the specific procedure of the duality, as the M-theory uplifting requires us to first put the $x_3$ direction on a circle and then dualize this to M-theory to be eventually combined with the $x_{11}$ circle to form a torus ${\bf T}^2$. The special role played by $x_3$ (or any other chosen space direction)  then breaks the isometry in the type IIB side converting \eqref{betta} to a geometry that isn't quite a de Sitter space that we want to study. A simple way out of this is to actually go to the zero volume limit of the M-theory torus ${\bf T}^2$ and then slowly increase the type IIA coupling. The latter procedure is however subtle because the type IIA coupling is in fact proportional to:
\bg\label{montse}
 g_s ~\propto~ h^{1/4}\left(\Lambda\vert t\vert^2\right)^{1/2}, \nd
so it is only the early time physics that is strongly coupled\footnote{Recall $-\infty \le t \le 0$ because of the flat slicing of the de Sitter space, so $t \to -\infty$ will be early time.\label{cori}}. 
Thus the very early times, keeping one of the cycle of ${\bf T}^2$ to be of vanishing size, would effectively capture the type IIB background that we want. Existence or non-existence of a vacua of the form \eqref{betta} could be answered there, and we can then move to a more generic point in the moduli space. On the other hand, at late time, since $g_s\to 0$, this is more automatic. The warped eleven-dimensional radius vanishes (see \eqref{recutvi}), and so does the radius of the $x_3$ circle. Together they take us to type IIB.

The above procedure is effective computationally, and has been used in \cite{nodS} to study the four-dimensional EFT description with a background like \eqref{betta} that is four-dimensional dS space with a time-independent internal six-dimensional space. The time-independence of the internal space guarantees two things: one, the four-dimensional Newton's constant $G_N$ can remain time independent, and two, the four-dimensional de Sitter isometries remain unbroken. The latter however implies additional constraints, namely that the internal fluxes, required to support a geometry like \eqref{betta}, should also be time-independent. From our M-theory perspective, this implies switching on time-independent flux components 
$G_{mnpa}(y)$ with ($y^m, y^a$) denoting coordinates of the 6d base the ${\bf T}^2$ respectively.

\subsection{Coherent states and the Schwinger-Dyson equations \label{dyson}}

The more subtle aspect of the story is to ask whether there exists a four-dimensional EFT description with full de Sitter isometries. There are multiple ways to address the question, and one such procedure is to analyze the on-shell conditions. This has been used in \cite{nodS}, and one of the benefits of such a procedure is the order-by-order expansion of the on-shell degrees of freedom that renders the $g_s$ and $M_p$ hierarchies  transparent. In fact this may be all that we need, but questions can be raised about the existence of the quantum vacuum itself that these ``on-shell" computations do not capture. Thus indulging in a {\it slight} off-shell computations may shed more light on the question of four-dimensional EFT. In other words, let us assume that the background on which we will analyze the quantum theory may be written as:
\bg\label{betbab}
ds^2= {1\over \sqrt[3]{h_2^2(y, x_i)}}\left(-dt^2+ dx_1^2+ dx_2^2\right) + \sqrt[3]{h_1(y)}{g}^{(0)}_{MN}dy^M dy^N,
\nd 
where ${g}^{(0)}_{MN}$ is the metric of the internal eight-dimensional manifold and $h_2(y, x_i), h_1(y)$ are the warp-factors (which are in principle different from $h(y)$). Such a background  
requires fluxes to support it, especially when the internal four-fold has a non-vanishing Euler 
characteristics \cite{BB, DRS, becker1}\footnote{We will discuss moduli stabilization later.}. 
For the special case when $h_2(y, x_i)$ is independent of $x_i$  in a way that $h_2(y) = h_1(y)$, the background has been discussed in details in \cite{BB, DRS}. The internal manifold becomes a Calabi-Yau 
four-fold which may be expressed as a ${\bf T}^2$ fibration over a six-dimensional base.  On the other hand if the Euler characteristic vanishes, which could happen when the ${\bf T}^2$  fibration becomes a product over a Calabi-Yau three-fold base, no background fluxes are needed and $h_1 = h_2 = 1$. Such a background dualizes to type IIB on the Calabi-Yau three-fold. In general however we can take the internal manifold to be a ${\bf T}^2$ fibration over a generic six-dimensional base.

Now an observable that captures the off-shell behavior could be the 2-point function of the metric components of the 6d base, i.e:
\bg\label{sakura01}
\langle \Omega \vert T g_{mn}(y_1, t_1) g_{pq}(y_2, t_2) \vert \Omega\rangle, \nd
 where ($y_1, y_2$) are two different points on the internal space and $T$ is the time ordering. Of course when the internal metric is time-independent, the time ordering is irrelevant here but we will keep it to make sense of the above analysis. The important thing in \eqref{sakura01} is the vacuum $\vert \Omega\rangle$ 
 which is an {\it interacting} vacuum in M-theory. We do expect such a vacuum to exist for any generic background in M-theory, so we will assume that such a vacuum may be defined, at least heuristically,
 for the background \eqref{betbab} with non-zero fluxes. Whether more complicated vacuum could exist will be discussed later.

 A few words about notations. The metric {\it fluctuations} over the background \eqref{betbab} may be divided into six-dimensional components 
 $g_{mn}$, two-dimensional toroidal components $g_{ab} \equiv g^{(2)}_{ab}$ and the three-dimensional space-time components $g_{\mu\nu} \equiv g^{(3)}_{\mu\nu}$. In this language it is clear that the information of the interacting vacuum may be replaced by the following path integral definition:
 
 {\footnotesize
 \bg\label{paulpine}
 \langle \Omega \vert T g_{mn}(y_1, t_1) g_{pq}(y_2, t_2) \vert \Omega\rangle \equiv 
 \mathbb{Z}^{-1} \int [{\cal D}g] [{\cal D}G][{\cal D}C]~{\rm exp}\left[iS(g, g^{(2)}, g^{(3)}, C)\right] g_{mn}(y_1, t_1) g_{pq}(y_2, t_2), \nd}
 \noindent where $\mathbb{Z}$ is partition function of the theory and $S(g, g^{(2)}, g^{(3)}, C)$ is the total action of M-theory that has all the perturbative corrections in it. Again question may be raised on the validity of such an action, but here we do not make any attempt to address such an issue as we take for granted at least the existence of perturbative series of quantum corrections. These corrections may be expressed 
in terms of polynomial powers of the metric and the G-flux components at weak curvatures and at small values of fluxes (all in units of $M_p$). This in turn implies that a correlation function of the form \eqref{paulpine}
cannot be used to explore regions of strong curvatures and strong G-fluxes. Finally, the measure of the path integral is defined as:    
\bg\label{jane}
[{\cal D}G] \equiv  [{\cal D}g^{(2)}][{\cal D}g^{(3)}], \nd
with a similar definition for $[{\cal D}C]$ by splitting the fluxes accordingly. Other meaningful entries, like the fermions, degrees of freedom on M2 and M5 branes will have to be included both in the measure as well as in the action, but we do not specify them for the time being. 

There is however one issue that may be addressed at this stage and it has to do with the {\it classical} metric configuration that we can extract from the full quantum theory. The quantum to classical 
correspondence\footnote{Note that a {\it classical} background can arise from a quantum theory in at least two possible ways. The first one is from an expectation value, or more generically from a solitonic solution. Such a solution is as classical as it gets in a given situation. The second one is from a coherent state in the quantum field theory. Such a state {\it simulates} a classical background by (a) not spreading in the Hilbert space of the quantum theory, and by (b) solving the classical EOMs that come from the Lagrangian description of the theory. However the state is also quantum by having a finite width of the coherent-state wave-packet. This quantum-ness of the classical state do not change with respect to time, so if the width is small, it pretty much provides the required classical background. Clearly such a state is constructed out of an infinite collection of gravitons with all possible frequency ranges. An alternative of using a delta function state  doesn't work because this will immediate spread in the quantum Hilbert space. However one could also use a squeezed coherent state which, although starts as a better representation of a classical background, eventually does have a varying quantum width as the system evolves in time. The latter may not be a real issue if the quantum-ness of the squeezed coherent state is not prominent.} 
has some bearing on the existence of coherent states, so it would be interesting to ask where a metric like 
\eqref{betta}, or more appropriately its M-theory uplift, could arise from a coherent state description in the full M-theory (see \cite{dvaliji} for 4d point of view).  In this language, the coherent state is easy to write down:
\bg\label{noharp}
\vert\alpha^{\tiny\tiny\tiny{MN}} \rangle \equiv {\rm exp}\left(\int d^dk \widetilde{g}^{MN}(k) a^\dagger_{MN}(k)\right) \vert 0 \rangle,
\nd
where $d$ are the spatial directions and for $d + 1 = 3$ it is the metric $\widetilde{g}_{\mu\nu}(k)$ that
is related to the Fourier transform of the three-dimensional metric over the background \eqref{betbab}. 
More appropriately, $\widetilde{g}_{\mu\nu}(k)$ may be expressed as, for fixed values of $y$:
\bg\label{garbo}
\widetilde{g}_{\mu\nu}(k) = \int d^3 x \left[{1\over \left(\Lambda\vert t\vert^2\sqrt{h}\right)^{4/3}}-
{1\over h_2^{2/3}}\right]
\psi^\ast_k(x)\eta_{\mu\nu}, \nd
where $\psi_k(x)$ is the Schr\"odinger wavefunction\footnote{When $\psi_k(x) = e^{ik.x}$ then \eqref{garbo} will be a standard Fourier transform. Here $\psi_k(x)$ could be more generic and we take 
$k^2 \equiv \omega_k^2 - {\bf k}^2 = m_k^2$ such that for ${\bf k} = 0$, $m_0$ is non-zero and the modes are on-shell. If this is not the case, then $g_{\mu\nu}(y, t)$ will have to be a background and not a coherent state.} for the solitonic background \eqref{betbab} with $k \equiv ({\bf k}, \omega_k)$ on-shell. It is clear that 
when the coherent state fluctuations add to the warped space-time, it provides the necessary four-dimensional space appearing from \eqref{betta}, or its M-theory uplift. Simultaneously, for $d = 6$ one may also construct 
the Fourier transform of the base metric $\widetilde{g}_{mn}(k)$ in the following way:
\bg\label{kilmer} 
\widetilde{g}_{mn}(k) = \int d^6 y dt \sqrt{g^{(0)}_{\rm base}}\left({h^{1/3} j_{mn}\over \Lambda^{1/3}\vert t\vert^{2/3}}
 - h_1^{1/3}g^{(0)}_{mn}\Big\vert_{\rm base}\right) \chi^\ast_k(y, t), \nd
where $j_{mn}(y)$ is the type IIB metric in \eqref{betta}, which is {\it not} necessarily a Calabi-Yau manifold, 
and the subscript {\it base} denote the four-fold metric $g_{mn}^{(0)}$ restricted to the six-dimensional 
base\footnote{We have used $g^{(0)}_{\rm base}$, and also $g^{(0)}_{\rm fibre}$ in \eqref{oshombhov}, to express the volumes of the compact spaces appropriately. This is not necessary and can be absorbed in the definitions of the Schr\"odinger wavefunctions $\chi^{}_k(y, t)$ and $\zeta^{}_k(z, t)$ respectively. For the non-compact $2+1$ dimensional space $\psi^{}_k(x)$ captures all the information in \eqref{kilmer}.} . 
The Schr\"odinger wavefunction $\chi_k(y, t)$ can be evaluated from the internal space in \eqref{betbab} and is expectedly more non-trivial. In a similar vein, and using two-dimensional Schr\"odinger wavefuncton 
$\zeta_k(z, t)$ we can define, at a fixed $y$:
\bg\label{oshombhov}
\widetilde{g}_{ab}(k) = \int d^2z dt \sqrt{g^{(0)}_{\rm fibre}}\left(h^{1/3} \Lambda^{2/3}\vert t \vert^{4/3} 
\delta_{ab}- h_1^{1/3}g^{(0)}_{ab}\bigg\vert_{\rm fibre}\right) \zeta^\ast_k(z, t). \nd
The other parameters appearing in \eqref{noharp} are the creation operators 
 $a^\dagger_{MN}(k)$; and $\vert 0\rangle$, the free vacuum. However the coherent state that we want for our case should be described on an interacting vacuum $\vert \Omega\rangle$ in M-theory, which in-turn is related to the free vacuum $\vert 0\rangle$ in the following standard way:
\bg\label{valentine}
\vert \Omega(t) \rangle ~\propto ~ \lim_{T \to \infty(1-i\epsilon)}{\rm exp} \left(-i\int_{-T}^t d^{11}x {\bf H}_{\rm int}\right) \vert 0 \rangle, \nd
where ${\bf H}_{\rm int}$ is the interacting part of the M-theory Hamiltonian. The claim is that such a state, when constructed out of the interacting vacuum, should satisfy the classical supergravity EOM in the presence of background fluxes\footnote{Note that, since most of flux components are taken to be time-independent, it is better to view them as background values instead of appearing from  coherent state fluctuations of the quantized fluxes. Thus once coherent states like \eqref{garbo}, \eqref{kilmer} and 
\eqref{oshombhov} are constructed, we will require background fluxes also to be switched on simultaneously. Together the system should solve supergravity EOMs.}. This is because, if the state didn't solve the EOM, then it will only contribute to the path integral (i.e the quantum behavior) but not to the classical dynamics of the system.

Thus either interpretation, classical or quantum coherent state, brings us to the point wherein we have to justify that the background \eqref{betta}, or its M-theory uplift, solves the supergravity EOMs. 
To analyze this in the path-integral language that we started off with,   it would be instructive to first study the expectation value of $g_{mn}$ in the interacting vacuum $\vert\Omega \rangle$ of M-theory.  In fact any two arbitrary 
configurations of internal metric may be related by the following standard identity:

{\footnotesize
\bg\label{chorigsby}
 \int [{\cal D}g] [{\cal D}G][{\cal D}C]~{\rm exp}\left[iS(g, G, C)\right] g_{mn}(y_1, t_1)   = 
  \int [{\cal D}g'] [{\cal D}G][{\cal D}C]~{\rm exp}\left[iS(g', G, C)\right] g'_{mn}(y_1, t_1), \nd}
\noindent as the path integral involves integrating over all possible metric configurations. A similar argument like \eqref{chorigsby} may also be given for all other components of the M-theory metrics.  We will dwell on this a bit later. 
  
The above identity implies that, in the field space, background with $g'_{mn}$ components may approach arbitrarily close  to the background with $g_{mn}$ components.  In other words, let us assume:

{\footnotesize
\bg\label{vanlisbon}
g'_{mn}(y, t) = g_{mn}(y, t) + \epsilon_{mn}(y, t) \equiv g_{mn}(y, t) + \int d^6z dt' 
\sqrt{g^{\rm CY}_{\rm base}} ~\epsilon(z, t')\delta^6(y-z)\delta(t-t') \delta_{mn},  \nd}
\noindent from where we can view $\epsilon_{mn}(y, t)$ to be a small fluctuation of the metric $g_{mn}$ at all points in the internal space parametrized by $y^m$ provided $\epsilon(z, t')$ remains small 
everywhere\footnote{Due to an abuse of notation, we have denoted the small tensor-fluctuation as 
$\epsilon_{nm}$. This should not be confused with the anti-symmetric tensor!}. 

The metric component $g_{mn}$, appearing from a coherent fluctuation, provides the curvature invariants as well as other local properties of the internal six-dimensional compact space. More appropriately however 
it is the total metric that captures the curvature invariants etc. Therefore we will replace $g_{mn}(y, t)$ as:
\bg\label{melmay}
g_{mn}(y, t) ~\to ~ h_1^{1/3}g_{mn}^{(0)}\Big\vert_{\rm base} + g_{mn}(y, t). \nd 
Generically it is assumed that this internal metric, say in the type IIB side, be non-K\"ahler and may even be non-complex as there are time-independent fluxes that provide the necessary energy-momentum tensor to support such a geometry. To see how this comes about, we can plug in 
\eqref{vanlisbon}, with the modification \eqref{melmay}, in \eqref{chorigsby} to get the following 
equation (to avoid clutter we will denote the total metric components by the same symbol $g_{mn}$):

{\footnotesize
\bg\label{montswing}
 \langle \Omega\vert g_{mn}(y, t) {\bf Tr}~{\bf G}_{\rm cl}(z, t')\vert \Omega\rangle
 &\equiv&\mathbb{Z}^{-1} \int [{\cal D}g] [{\cal D}G][{\cal D}C]~e^{iS} \delta^{pq}\left[{G}_{pq}  
 - \left(\mathbb{G}^2\right)_{pq}\right]g_{mn}(y, t)\\
&=&  -i \delta^8(y-z)\delta(t-t') +  h^{1/3}(z, t') \delta^{pq} \sum_{\{\alpha_i\}} \langle\Omega\vert 
  \widetilde{\mathbb{C}}_{pq}^{(\alpha_i)}(z, t') g_{mn}(y, t)\vert\Omega\rangle,  \nonumber \nd}
\noindent  where the delta function is defined over the eight-dimensional internal space in M-theory, although we could have also restricted to the six-dimensional subspace. The other factors appearing in \eqref{montswing} may be defined in the following way.  The ${\bf Tr}~{\bf G}_{\rm cl}(z, t')$ piece is the trace of the classical part of the $g_{mn}$ EOM, and which is defined on the right hand side of the above equation with the Einstein tensor $G_{pq}$. Finally, 
 $\left(\mathbb{G}^2\right)_{pq}$ is the energy momentum tensor coming from the flux components \cite{nogo, nodS} and may be expressed as:
 \bg\label{patjane}
\left(\mathbb{G}^2\right)_{pq} \equiv  {1\over 12}\left(G_{pABC}G_q^{~~ABC} 
- {1\over 8}g_{pq} G_{PQRS}G^{PQRS}\right), \nd
where ($p, q$) denote the coordinates of the six-dimensional base and ($A, P$) etc denote the coordinates of the eight-dimensional internal space, including the space-time components (see \cite{nogo, nodS}). In this language we may also express 
\eqref{montswing} as an integral equation over the full eight-dimensional space.

The above equation is an example of a Schwinger-Dyson equation from our M-theory perspective, and thus balances the classical and the quantum pieces. Solution would exists if the right hand side, which incorporates the quantum pieces $\widetilde{\mathbb{C}}_{pq}^{(\alpha_i)}$, can be controlled. In the absence of any time dependences, the right hand side of \eqref{montswing} will simply be the sum over the quantum pieces exactly as we had in \cite{nodS}. The series would make sense if, from type IIA point of view, there is some hierarchy in terms of $g_s$ and $M_p$. The way we expressed it in \cite{nodS}, there were no apparent hierarchy visible and thus the right hand side of \eqref{montswing} could not be expressed as a controlled expansion in terms of a small parameter. This at least ruined a simple EFT description of the system and solution could not be found\footnote{One may also find the Schwinger-Dyson equation for the fluctuation 
$g_{mn}$ only by expanding the M-theory action over the background \eqref{betbab}. The outcome of such  an exercise will reveal similar issues with hierarchy. See \cite{coherbeta} for more recent developments.}.  

Another disconcerting thing of the above discussion may be seen from the two-point function 
 \eqref{sakura01}, with $g_{mn}$ now defined as in \eqref{melmay}, which when plugged in the corresponding Schwinger-Dyson equation would lead to term of the form:
\bg\label{tunney}
\sum_{\{\alpha_i\}} \langle\Omega\vert g_{mn}(y_1, t_1)
  \widetilde{\mathbb{C}}_{pq}^{(\alpha_i)}(z, t) g_{rs}(y_2, t_2)\vert\Omega\rangle, \nd
 which would only make sense if the quantum series $\widetilde{\mathbb{C}}_{pq}^{(\alpha_i)}$ could be terminated in some way. In the absence of any hierarchy between $g_s$ and $M_p$ this is clearly impossible, leading us to 
 the same conclusion that we had before, namely: a coherent state fluctuation over a solitonic background
 \eqref{betbab} doesn't seem to lead to a sustained classical configuration of the form \eqref{betta} in the type IIB side (or in its M-theory uplift). 

The small time-dependence that we inserted in the definition of the quantum pieces $\mathbb{C}_{pq}^{(i)}$
in $\widetilde{\mathbb{C}}_{pq}^{\alpha_i}$ (see also \eqref{londry})
is to not only allow for a well-defined propagation of modes but also to allow  for a well-defined time ordering in \eqref{tunney}.
The key difference between this definition and the one used in \cite{nodS} is the use of $g_{mn}$ instead of 
$j_{mn}$ from \eqref{betta} as the latter is completely time independent. Thus expressing \eqref{montswing}
in terms of {\it unwarped} metric and flux components of \cite{nogo, nodS}, will immediately reproduce the 
time-independent EOMs. However the problem with EFT still persists. A different linear combinations of the quantum pieces as defined in eq. (5.44) of \cite{nodS} doesn't seem to alleviate the problem either.

One could also address the problem using a background of the form $AdS_4 \times \mathbb{M}_6$ in the type IIB side, or more generically analyze the coherent state construction directly from type IIB side by taking a background solitonic solution of the form\footnote{We will not try to prove the existence of such a vacua.}: 
\bg\label{karin}
ds^2 = {1\over \sqrt{h(y, u)}}\left(-dt^2 + dx_1^2 + dx_2^2 + du^2\right) + \sqrt{h_1(y)} g_{mn} dy^m dy^n, \nd
where $u$ is the radial direction and the warp-factor $h(y, u)$ depends on both $y^m$ as well as $u$. In the limit when $h(y, u) = h_1(y)$, this background would be dual to the M-theory background 
\eqref{betbab} with $h_2(y) = h_1(y)$. On the other hand, when:
\bg\label{lumiere}
h(y, u) =  u^4 h_1(y), \nd 
the background becomes $AdS_4 \times \mathbb{M}_6$, where $\mathbb{M}_6$ is   circle fibration over a squashed Sasaki-Einstein manifold \cite{luest}. Such a background requires all type IIB fluxes switched 
on, including varying axio-dilaton \cite{luest}. These fluxes should additionally help us to stabilize some of the moduli of the internal space, much like the stabilization of the complex structure moduli with fluxes for the background \eqref{betbab}. However all moduli do not get stabilized this way, but for the case 
\eqref{betbab} since
the internal space was eventually expected to be time {\it dependent}, the coherent state construction could be extended to the full eleven-dimensions as \eqref{garbo}, \eqref{kilmer} and 
\eqref{oshombhov}. The final IIB background,  or the corresponding M-theory uplift,  then had enough ingredients for moduli stabilization {\it provided} an EFT could be constructed. In the present case, the scenario is subtle. With the choice of \eqref{karin}, and our requirement of keeping the internal space 
time {\it independent} may pose an issue regarding coherent state construction unless we are able to express the time independent internal-space also as some kind of coherent state\footnote{Such a state will require 
$\omega_k = 0$ with non-zero ${\bf k}$. Since this is only possible off-shell, there is no on-shell or standard coherent state description of a time-independent background. \label{gowest}}. 
Additionally, since type IIB theory doesn't have a Lagrangian, an interacting vacuum becomes harder to construct (that doesn't imply non-existence of course).  Nevertheless, since we are dealing with a similar background, now from the type IIB side, 
one should be able to study this from four-dimensional perspective for energy scales below the sizes of the internal cycles \cite{Kutasov:2015eba, Green:2011cn, hirano}. We believe the issue of EFT should be confronted from this angle now. 

In such a background a coherent state could be created that converts the $AdS_4$ geometry to a four-dimensional de Sitter background. Such a configuration should again solve the type IIB EOMs in the presence of the full quantum corrections.  However questions have been raised, for example in 
\cite{Sethi:2017phn}, whether such a background is a good starting point to analyze the quantum theory. Leaving the issues of interacting vacuum aside, what we want to see whether the quantum fluctuations may form  close-to-classical coherent states that solve EOMs. From the space-time point of view, the dependence of the space-time metric on $u$ is an advantage over \eqref{betbab}\footnote{In the sense that
\eqref{betbab} can allow $\partial_{x_i}h_2(y, x_i) = 0$.}. However
the non-existence of a simple EFT description from our earlier analysis showed that this is a much harder problem to analyze because forming any localized states in the full quantum theory will immediately back-react, both on four-dimensional space-time as well as on the six-dimensional internal space. If all goes well, this should convert \eqref{karin} to \eqref{betta}. As of now, this remains an open problem \cite{wittendS, gates}.

Finally, one could take the background itself to be of the form $dS_4 \times \mathbb{M}_6$, which is 
\eqref{betta} instead of \eqref{betbab} or \eqref{karin} and study quantum fluctuations over this background. An immediate issue with such a choice is  the non-supersymmetric nature of the background \eqref{betta}; and therefore the vacuum energies of the bosonic and the fermionic fluctuations over this background do not cancel. In other words, we encounter the divergent integral of the form:
\bg\label{andhadhun}
\rho \equiv {1\over 2} \int {d^{d-1}k \over (2\pi)^{d-1}}  \sum_l\left(\pm n_l\sqrt{{\bf k}^2 + m_{kl}^2}\right) = 
{1\over 8} \int {d^dk\over (2\pi)^d}  \sum_l\left(\pm {in_l m_{kl}^2\over k^2 - m_{kl}^2 + i\epsilon}\right), \nd
for $d$ space-time dimensions. Here $\rho$ is the vacuum energy density for $n_l$ species of fermions and bosons each with mass $m_{kl}$ for asymptotic momenta ${\bf k}$ (the $\pm$ sign denotes the bosonic and the fermionic states respectively). 
In the two earlier choices, namely \eqref{betbab} and \eqref{karin}, the backgrounds were supersymmetric and the vacuum energies of the bosonic and the fermionic fluctuations cancel.  The non-supersymmetric solution \eqref{betta} for each case was then required to appear from a
coherent state in the supersymmetric 
theory\footnote{At this point it might be useful to point out the sources that generate the cosmological constant in four-dimensions. In fact there are three sources that are in operation here: (1) the vacuum energy as expressed in \eqref{andhadhun}; (2) the background fluxes on the internal manifold; and (3)  the controlled perturbative and non-perturbative quantum corrections. It is of course the first one out of the three sources that lead to the cosmological constant {\it problem} as we know it. For the supersymmetric vacua of the form \eqref{betbab} and \eqref{karin}, the vacuum energy contributions cancel. The cosmological constant then appears from the fluxes and the quantum corrections provided the latter has a well defined hierarchy. However once we choose \eqref{betta} as a {\it classical} background, i.e not as a quantum coherent state, all the three sources now contribute to the cosmological constant leading us back to the issue that we barely managed to avoid using \eqref{betbab} and \eqref{karin}.}. 
Such a state is expected to break  supersymmetry but the underlying vacuum for each cases do not. Of course the issue of hierarchy plaguing  our analysis deterred us from finding a solution of the form \eqref{betta}. 

Another issue with this choice of the background is related to the vacuum configuration itself. One would expect the 
vacuum now to be a Bunch-Davies vacuum, but a recent work \cite{danielsson} suggests that such a vacuum may itself be unphysical. This unphysicality, as suggested in \cite{danielsson}, may be related to the swampland conjectures \cite{vafa1, vafa2, vafa22, brennan, Ooguri, swampland}.

Finally, the quantum fluctuations over the effective four-dimensional background cannot be governed by a time-independent Newton's constant $G_N$. In the previous two cases with \eqref{betbab} and 
\eqref{karin}, the fluctuations over an effective three or four-dimensional space-time, respectively didn't have time varying Newton's constant. However now, if we denote the effective four-dimensional fluctuation over the
background metric components in \eqref{betta}
as $\epsilon h_{\mu\nu}$, in other words  consider 
\bg\label{sstones}
g_{\mu\nu}({\bf x}, t, y)  = {\eta_{\mu\nu} + \epsilon h_{\mu\nu}({\bf x}, t) \over \Lambda(t) \sqrt{h(y)}}, \nd
 where $\epsilon$ is a small number and $h(y)$ is the warp-factor used earlier (not to be confused with the metric fluctuation $h_{\mu\nu}$), then the 
effective action for $h_{\mu\nu}$ turns out to be the one with a time-dependent Newton's constant $G_N$ as \cite{choi}:
\bg\label{notyell}
S_{eff} =  - \epsilon^2 \int {d^4 x \over G_N} \left({1\over 4} \partial_\mu h \partial^\mu h 
-{1\over 2} \partial_\mu  h^{\sigma \nu} \partial^\mu h_{\sigma\nu}\right) + {\cal O}\left(\epsilon^3\right), \nd
where the indices are raised or lowered by the flat metric $\eta_{\mu\nu}$, and $h \equiv h^\mu_\mu$ is the trace of the metric fluctuations. The four-dimensional Newton's constant appearing above 
is easy to infer from \eqref{betta} and takes the following form:
\bg\label{redjohn}
{1\over G_N} =  {e^{-2\phi_b} \over \Lambda(t) l_s^8} \int d^6y{\sqrt{{\rm det}~j}}~h(y) , \nd
where $l_s$ is the ten-dimensional string length and $\phi_b$ is the type IIB dilaton (which is taken to be a constant here). The time-dependence in $G_N$ appears solely from $\Lambda(t) = \Lambda\vert t\vert^2$
in this case, but would be completely time-independent for the other two case, \eqref{betbab} and 
\eqref{karin}. To keep \eqref{redjohn} time independent, 
one way out would be to take the type IIB dilaton,  or equivalently the type IIB coupling constant, to be time-dependent so as to cancel the $\Lambda(t)$ factor. However this will make the type IIB analysis even harder to tackle than what it is now. Another way would be to take the internal space itself to be time-dependent. This is a curious scenario that might have  potentials of generating interesting cosmologies. We will discuss this case soon.

From all the above discussions, the pertinent question now is to inquire about the scenario that would allow a four-dimensional background with positive cosmological constant that may or may not actually be a 
{\it constant}. One scenario, as suggested in \cite{nodS} and alluded to above, is to take the internal metric in the type IIB side to be time-dependent. The time-dependence is supposed to induce some hierarchy between $g_s$ and 
$M_p$ which, at the end of the day, should allow a consistent solution of the EOMs  to emerge out of the analysis presented in meticulous details in \cite{nogo, nodS}. Whether this is the case is the subject of the 
following sections.


\subsection{Breaking the isometries using dipole type deformations \label{difuli}}

Inserting time dependences to the components of the internal metric in say \eqref{betta} naturally breaks the four-dimensional de Sitter isometries. But does this always allow solutions to exist? This is the question that we want to investigate here. In the process we will also be able to see if changing the $x_3$ isometry 
any way affects the conclusions that we got in the previous section. 

To start, let us assume that the internal six-dimensional space in \eqref{betta} may be expressed locally as a 
${\bf S}^1$ fibration over a five-dimensional base, in a way that there may not be any global one-cycle. We can parametrize the local coordinate as $\psi$ such that the NS three-form flux ${\bf H}_3$ do not have any leg along that direction. It is therefore the RR three-form flux ${\bf F}_3$ that has a $\psi$ component. Under a dipole deformation \cite{dipole} the metric \eqref{betta} changes to the following:
\bg\label{naviali}
ds^2 & = & {1\over \sqrt{h} \Lambda(t)}\left[-dt^2 + dx_1^2 + dx_2^2 + 
{\Lambda(t) dx_3^2 \over g_{\psi\psi} \sin^2\theta + \Lambda(t) \cos^2\theta}\right] \nonumber\\
 &+& \sqrt{h} \left[ \widetilde{g}_{mn}dy^m dy^n + {g_{\psi\psi} \Lambda(t) d\psi^2 \over 
 \Lambda(t) \cos^2\theta + g_{\psi\psi} \sin^2\theta}\right], \nd
where $\theta$ quantifies the dipole deformation and $g_{\psi\psi}$ is the unwarped local metric along the $\psi$ 
direction. The background fluxes, appearing from the $G_{mnpa}$ components in M-theory \cite{nodS}, do not change much beyond ${\bf F}_3$ getting an extra factor of $\cos~\theta$. However there does appear an extra NS B-field component proportional to:
\bg\label{kttedar}
{\bf B} = {g_{\psi\psi} \tan~\theta \over \Lambda(t) \cos^2\theta + g_{\psi\psi} \sin^2\theta}~dx_3 \wedge 
d\psi, \nd
which is in principle responsible for generating the dipole deformation and in turn breaking the $x_3$ isometry of the original metric \eqref{betta}. This B-field cannot be gauged away, and its dependence on $t$ creates some subtleties. These subtleties are important in understanding the dynamics of dipole theories  but are irrelevant for the ensuing discussions. Hence 
we are not going to discuss them here. Instead we will use the metric 
\eqref{naviali} simply as a  springboard to discuss a different issue, namely the inherent time dependence and the existence of an EFT description. 

It turns out, the metric configuration \eqref{naviali} along-with the B-field \eqref{kttedar}, despite having time-dependences, suffer from the same hierarchy issue that plagued the background \eqref{betta}. This may be easily checked by actually working out the EOMs as in \cite{nogo, nodS}, or by observing that the metric 
\eqref{naviali} appears from \eqref{betta} by making a TsT transformation with the shift $s$ given by \cite{dipole}:\bg\label{bi77}
\left(\begin{matrix} \psi \cr x_3 \cr \end{matrix}\right) ~ \to ~ \left(\begin{matrix}\cos~\theta & 0 \cr \sin~\theta & ~~~\sec~\theta \cr
\end{matrix} \right) \left(\begin{matrix} \psi \cr x_3 \cr \end{matrix}\right). \nd
 The subtlety that we encountered earlier regarding the existence of an EFT description with a ten-dimensional metric of the form \eqref{betta} thus appears {\it not} to get alleviated by simply introducing time-dependences in the internal metric components, or by breaking the $x_3$ isometry as evident from \eqref{naviali}.  Of course this is not  a generic statement and we will demonstrate soon that introducing different time-dependences to the internal metric components than the ones in \eqref{naviali} might alleviate certain problems. 

There are a few cases related to the background \eqref{naviali} and \eqref{kttedar} that we want to discuss before moving ahead with a different class of time-dependent solutions. The first one has to do with the 
B-field \eqref{kttedar} whose time dependence goes with $\Lambda(t) = \Lambda\vert t\vert^2$. Clearly at early times, i.e when $t \to -\infty$, the dipole deformation is invisible with finite $g_{\psi\psi}$. When $t$ ranges between $-t_2 < t < -t_1$ with $\vert t_2\vert > \vert t_1 \vert$, if we can allow the metric component 
$g_{\psi\psi}$ to satisfy:
\bg\label{diharmonicv}
g_{\psi\psi}(y_0) \gg \Lambda\vert t_2 \vert^2 \cot^2\theta, ~~~~ \forall ~y_0 \in y, \nd
the B-field appears to have a vanishing field strength (which would be gauge equivalent to zero B-field), yet the isometry along the $x_3$ direction is not restored. In fact the radius of the $x_3$ circle becomes very small, taking us to the T-dual IIA or the full M-theory version. This clearly shows that the breaking of the 
$x_3$ isometry has nothing much to do with the loss of the $g_s$ and $M_p$ hierarchy. From our earlier analysis we now know that the problems lies deep in the quantum region and any classical manipulations 
will be unable to alleviate the issue.

The second one has to do with the metric \eqref{naviali} itself. What if we break all the spatial isometries 
by doing multiple dipole transformations simultaneously? Does this help us to regain the four-dimensional 
hierarchy for our case? The answer turns out to be unfortunately no as can be inferred from the appendix of 
the third paper listed in \cite{dipole}.

\subsection{Kasner-de Sitter  type solutions and EFT  description \label{kasner}}

The failure of getting an EFT description using dipole type deformations suggests that a more generic analysis is called for. We still however want to retain the time independence of the internal manifold in the type IIB side, so let us choose the following metric:
\bg\label{sally}
ds^2 = {1\over \Lambda(t) \sqrt{h}}\left[-dt^2 + e^{f_1(t, x_i)} dx_1^2 + e^{f_2(t, x_i)} dx_2^2 + e^{f_3(t, x_i)} dx_3^2\right] + \sqrt{h} g_{mn} dy^m dy^n, \nonumber\\ \nd
where $f_i(t, x_j)$ are some generic functions of $t$ and the spatial coordinates $x_i$, $h(y)$ is the warp-factor and $\Lambda(t) = \Lambda\vert t\vert^2$ as in \cite{nogo, nodS}. The choice of $f_i$ functions break isometries maximally and lead to more cumbersome set of EOMs that are harder to dis-entangle. A slightly simpler and economical choice would be to take these functions as just functions of time, i.e:
\bg\label{tabu}
f_i(t, x_j) \equiv f_i(t), \nd
with the assumption that $f_1, f_2$ and $f_3$ are unequal as any equality between them would bring us back to the issues that we faced earlier. The M-theory uplift of the type IIB background is simpler:
\bg\label{heidkalu}
ds^2=e^{2{A}(y,t)}\left[-dt^2+e^{f_1(t)}dx_1 ^2+e^{f_2(t)}dx_2^2\right]+e^{2{B}(y,t)}{g}_{mn}dy^m dy^n+
e^{2{C}(y,t)}(dx_3^2+dx_{11}^2), \nonumber\\
\nd
which looks almost similar to the M-theory uplift of the background \eqref{betta} studied in \cite{nogo} except for the $f_i(t)$ factors. The difference however lies in the choice of the various warp-factors, namely:
\bg\label{facedx}
&&{A}(y,t) \equiv -{1\over 3} \log\left[\Lambda^2(t) h(y)\right]+\frac{1}{6}f_3(t) \nonumber\\
&& {B}(y,t) \equiv - {1\over 6} \log\left[{\Lambda(t)\over h(y)}\right]+\frac{1}{6}f_3(t)\nonumber\\
&& {C}(y,t)\equiv  {1\over 6} \log\left[\Lambda^2(t) h(y)\right] -\frac{1}{3}f_3(t), 
\nd
where, in the absence of $f_3(t)$, these coefficients would have been exactly the ones encountered in 
\cite{nogo}. The difference now lies in the actual value of the warp-factor $h(y)$ and the function $f_3(t)$ as the other 
$f_i(t)$ functions only contribute to the space-time metric components. It is also interesting to note that the 
curvature tensors may also be expressed in terms of $A, B, C$ and $f_i(t)$ in the following suggestive way:
\bg\label{bshai}
\mathbb{R}_{MN} = \mathbb{R}_{MN}(A, B, C) + \delta\mathbb{R}_{MN}(f_i), \nd
where $\mathbb{R}_{MN}(A, B, C)$ is related to the curvature tensors computed in section ${\bf 5}$ of 
\cite{nogo}. For example let us consider the curvature tensor $\mathbb{R}_{mn}$. This may be divided into 
$\mathbb{R}_{mn}(A, B, C)$ 
which is written as:

{\footnotesize
\bg\label{pavillonss}
\mathbb{R}_{mn}(A, B, C) & = & {R}_{mn} + 3\left[2\partial_{(m}{A}\partial_{n)}{B}-\partial_m {A} \partial_n {A}
- {g}_{mn}\partial_k {A}\partial^k {B}\right]+ 4\left[\partial_m {B} \partial_n {B}- {g}_{mn}\partial_k{B}\partial^k {B}\right]\nonumber\\
&-& 3D_{(m}\partial_{n)}{A} -2D_{(m}\partial_{n)}{C} +2\left[2\partial_{(m}{C}\partial_{n)}{B}-\partial_m {C} \partial_n {C} - {g}_{mn}\partial_k {C}\partial^k {B}\right] \nonumber\\
&+&{g}_{mn}\Box {B}  -4D_{(m}\partial_{n)}{B} + 
e^{2({B}-{A})}\left[\ddot {{B}}+\dot {{A}} \dot {{B}} +6 \dot {{B}}^2 +2 \dot {{C}} \dot {{B}}\right] g_{mn},
\nd}
where $R_{mn}$ is the Ricci tensor for the unwarped metric $g_{mn}$ in \eqref{heidkalu}. Note that in this form the tensor $\mathbb{R}_{mn}(A, B, C)$ resembles exactly the Ricci tensor in \cite{nogo} with 
$A, B$ and $C$ defined without the $f_i(t)$ factors. On the other hand, the extra factor appearing 
in \eqref{bshai} takes the following form:
\bg\label{nerdirt}
\delta\mathbb{R}_{mn}(f_i) \equiv {1\over 2} e^{2(B-A)} \dot{B}\left(\dot{f}_1 + \dot{f}_2\right) g_{mn}, \nd
which has the required explicit dependence on the $f_i$ factors. Thus it appears that the Ricci curvature 
divides into two pieces: one, which depends on the $f_i$ factors implicitly via the $A, B$ and $C$ parameters, and two, which depends explicitly on the $f_i$ factors. Such a division works for most of the 
Ricci tensors except for $\mathbb{R}_{11}$ and $\mathbb{R}_{22}$. These two special cases take the following form:
\bg\label{cosua}
e^{f_i} \mathbb{R}_{ii} \equiv \mathbb{R}_{ii}(A, B, C) + \delta\mathbb{R}_{ii}(f_k), \nd
where $i = (1, 2)$ and the repeated indices are not summed over. The extra $e^{f_i}$ factor in the definition above is the main difference and therefore $\mathbb{R}_{ii}(A, B, C)$ takes the following form:
\bg\label{brooklyn}
\mathbb{R}_{ii}(A, B, C) &\equiv &
-\eta_{ii} e^{2({A}-{B}}
   \left[\Box {A} + 3\partial_m {A} \partial^m {A} 
  + 4\partial_m {A} \partial^m {B} + 2 \partial_m {A} \partial^m {C}\right] \nonumber\\
  && ~~~~~~~~~~ +  \eta_{ii} \left[\ddot{A} + \dot{A}\left(\dot{A} +  6\dot{B} + 2\dot{C}\right)\right],  \nd
which expectedly coincides in form with a similar expression in \cite{nogo}. The difference, as mentioned earlier, lies in the definitions of the parameters involved in either of the two expressions.  The other pieces in \eqref{cosua} are defined in the following way:
\bg\label{greenbook}
&&\delta\mathbb{R}_{11} \equiv {1\over 2} \ddot{f}_1 + \dot{f}_1\left(\dot{A} + 3\dot{B} + \dot{C} + 
{1\over 4} \dot{f}_1 + {1\over 4} \dot{f}_2\right) + {1\over 2} \dot{A} \dot{f}_2 \nonumber\\
&&\delta\mathbb{R}_{22} \equiv {1\over 2} \ddot{f}_2 + \dot{f}_2\left(\dot{A} + 3\dot{B} + \dot{C} + 
{1\over 4} \dot{f}_2 + {1\over 4} \dot{f}_1\right) + {1\over 2} \dot{A} \dot{f}_1, \nd  
which vanish when $f_1$ and $f_2$ are constants. Note that it is not possible to choose functional forms for 
$f_1$ and $f_2$ such that $\delta\mathbb{R}_{ii}$ are cancelled globally over all points in the internal manifold. Local cancellations obviously happen, but are irrelevant for the ensuing discussions. 

Finally the other two Ricci tensors, namely $\mathbb{R}_{00}$ and $\mathbb{R}_{ab}$ take the expected form \eqref{bshai} with $\mathbb{R}_{00}(A, B, C)$ and $\mathbb{R}_{ab}(A, B, C)$ expressed in terms of $A, B$ and $C$ in exactly the same way as they appeared in \cite{nogo}. Thus we only need to write the functional forms for $\delta\mathbb{R}_{00}(f_i)$ and $\delta\mathbb{R}_{ab}(f_i)$, and they appear as:
\bg\label{murder2}
&&\delta\mathbb{R}_{ab} \equiv  {1\over 2} \delta_{ab} e^{2(C - A)} \dot{C} \left(\dot{f}_1 + \dot{f}_2\right), \nonumber\\
&&\delta\mathbb{R}_{00} \equiv -{1\over 2}\left(\ddot{f}_1 + \ddot{f}_2\right) - {1\over 2} \dot{A} 
\left(\dot{f}_1 + \dot{f}_2\right) - {1\over 4}\left(\dot{f}_1^2 - \dot{f}_2^2\right). \nd
At this point it may be interesting to note that if we keep the sum of $f_1$ and $f_2$ fixed with respect to time, then most of the variations $\delta\mathbb{R}_{MN}$ vanish, except for $\delta\mathbb{R}_{ii}$ where their sum vanish. In other words:
\bg\label{jaanebhi}
\delta\mathbb{R}_{mn} = \delta\mathbb{R}_{ab} = \delta\mathbb{R}_{00} = 0, ~~~
\delta\mathbb{R}_{11} = - \delta\mathbb{R}_{22}, \nd
for $\dot{f}_1 + \dot{f}_2 = 0$, but keeping $f_3(t)$ as a generic function of time. The above analysis serves as a motivation to impose the following condition on $f_\mu$, with the assumption that $f_0 \equiv 0$:
\bg\label{lagaga}
\sum_{\mu= 0}^3 f_\mu(t)  = f_0(t) + f_1(t) + f_2(t) + f_3(t) \equiv 2\mathbb{F}(t), \nd
such that if $\dot{f}_3 = 2\dot{\mathbb{F}}$, then clearly we have our condition \eqref{jaanebhi}. (The factor of 2 in \eqref{lagaga} is for later convenience.) On the other hand if $\mathbb{F}(t)$ vanishes, then this could be related to the Kasner condition \cite{kasner} because we can tune $f_0$ to absorb any changes to $t$. Our type IIB metric is not quite the Kasner kind, so generically we cannot impose the vanishing of $\mathbb{F}(t)$ as we have sources. These sources are the fluxes, branes and planes in the IIB side, or fluxes and branes in the M-theory uplift. The quantum corrections should also contribute to the sources, so they should be taken together with the classical sources. The string coupling in the type IIA side is now:
\bg\label{robnahood}
g_s ~\propto~ \left(\Lambda\vert t\vert^2\right)^{1/2} h^{1/4}{\rm exp}\left(-{f_3(t)\over 2}\right), \nd
with no explicit dependence on $f_1$ and $f_2$, although implicitly $h$ would eventually depend on all the background parameters once we solve the EOMs either in the form of Schwinger-Dyson equations or as classical equations with quantum corrections. The hierarchy between $g_s$ and $M_p$ should govern whether the equations can be analyzed in a controlled laboratory or not.

It is instructive at this stage to point out the various scales involved in the problem. If 
$g_s^{(o)}$ denotes the constant of proportionality in \eqref{robnahood}, and 
$R_{11}$ denotes the scale of the eleven-dimensional  radius, then the actual radius of the eleven dimension, from our metric \eqref{heidkalu}, appears to be:
\bg\label{recutvi}
\mathbb{R}_{11} =  g_s^{2/3} R_{11} = e^C \left(g_s^{(o)}\right)^{2/3} R_{11} \equiv e^C l_{11}, \nd
where $C$ is given in \eqref{facedx} and $l_{11} \equiv \left(g_s^{(o)}\right)^{2/3} R_{11}$ is the eleven-dimensional Planck length. This Planck length, and {\it not} $\mathbb{R}_{11}$, governs the scale in the theory such that $M_p = {1\over l_{11}}$. It is important to note that we have {\it one} scale $M_p$ in the theory and one tunable parameter, which we will take it to be the type IIA coupling $g_s$. The latter is dynamical.

There is however something a little disconcerting about the type IIA coupling constant \eqref{robnahood} compared to what we had earlier in \eqref{montse}. The dependence of $g_s$ on $f_3$ puts a special preference for the $x_3$ direction for duality purpose over the $x_1$ or $x_2$ directions.
There appears to be no compelling reason for this choice and in fact we could have considered certain combinations of $x_i$ spatial directions $-$ as the T-duality direction $-$ thus making the expression for $g_s$ in \eqref{robnahood} more involved. One way out of this would be to allow all the $f_i(t)$ functions to appear in the definition of $g_s$ in \eqref{robnahood}. This will however require a change of basis, but the end result would still reflect a special preference for the new compact direction. Alternatively we could consider:
\bg\label{yvesroche}
 f_1(t) = f_2(t) \equiv f_3(t), \nd
which would be consistent with the fact that a Kasner-de Sitter solution quickly tends to isotropic de Sitter solution (see for example \cite{blanco}). Additionally the worry of a preferential choice of $g_s$ disappears 
with this. We will also see that the quantum behavior is much more succinct with the choice 
\eqref{yvesroche}. Whether a more generic choice can be entertained here will be discussed later.

As discussed earlier, the quantum behavior is captured here in few steps. First we construct the M-theory uplift of \eqref{sally}, i.e the background \eqref{heidkalu}, as a coherent state exactly as in \eqref{sakura01}. 
The classical background remains as \eqref{betbab}, and the Fourier components appearing therein now take the following form:
\bg\label{indigobk}
&& \widetilde{g}_{\mu\nu}(k) = \int d^3 x \left[ e^{2A(y_0, t) + f_\mu(t)}-{1\over h_2^{2/3}}\right]
\psi^\ast_k(x)\eta_{\mu\nu} \nonumber\\
&& \widetilde{g}_{ab}(k) = \int d^2z dt \sqrt{g^{(0)}_{\rm fibre}}\left(e^{2C(z, t)}
\delta_{ab}- h_1^{1/3}g^{(0)}_{ab}\bigg\vert_{\rm fibre}\right) \zeta^\ast_k(z, t)\nonumber\\
&& \widetilde{g}_{mn}(k) = \int d^6 y dt \sqrt{g^{(0)}_{\rm base}}\left(e^{2B(y, t)}g_{mn}
 - h_1^{1/3}g^{(0)}_{mn}\Big\vert_{\rm base}\right) \chi^\ast_k(y, t),  \nd 
 where ($\psi_k(x), \zeta_k(z, t), \chi_k(y, t)$) are the same Schr\"odinger wave-functions that we encountered earlier in \eqref{garbo}, \eqref{kilmer} and \eqref{oshombhov} while defining the coherent states there. The difference is only in the appearance of $A, B$ and $C$ from \eqref{facedx} which involves the Kasner function $f_3(t)$. Of course as before the correct vacuum will again be the interacting vacuum 
 $\vert\Omega \rangle$  defined in \eqref{valentine}. 
 
The second step is to realize the quantum behavior via solutions of the equations of motion with quantum corrections, or as Schwinger-Dyson equations. This is a necessary condition otherwise the coherent states would only contribute to the path integral but not to the classical states of the system. Thus looking for sustained classical states now brings us closer to the analysis that we performed in \cite{nodS}, and for that we will need the explicit expressions for the energy momentum tensors. 

The energy momentum contributions coming from the quantum terms can be essentially expressed as in \cite{nodS}, i.e we can write $\mathbb{T}_{MN}^Q$ as:
\bg\label{londry}
\mathbb{T}_{MN}^Q \equiv \sum_{\{\alpha_i\}} \widetilde{\mathbb{C}}_{MN}^{\alpha_i}(y, t)
= \sum_{i} \left(\Lambda\vert t\vert^2 \left(g^{(0)}_s\right)^2\right)^{\alpha_i} 
e^{-\alpha_i f_3} 
\hat{\mathbb{C}}^{(i)}_{MN} \equiv \sum_i g_s^{2\alpha_i} \mathbb{C}^{(i)}_{MN}, \nd
where $g_s$ is given in \eqref{robnahood}; and $\widetilde{\mathbb{C}}_{MN}^{\alpha_i}$ is the form in which the quantum pieces appeared in the Schwinger-Dyson equations \eqref{montswing} and \eqref{tunney}. The other two functions,
$\hat{\mathbb{C}}^{(i)}_{MN}$ and ${\mathbb{C}}^{(i)}_{MN}$, respectively depend explicitly and implicitly on the warp factor $h$, and are both time-neutral functions of the curvature tensors 
${\bf R}, {\bf R}_{MN}$ and ${\bf R}_{MNPQ}$ and  the ${\bf G}_{MNPQ}$ components. For the present analysis we will assume that all the G-flux components with lower indices are time-independent, except 
${\bf G}_{m012}$ which may be written as:
\bg\label{colemanbook}
{\bf G}_{m012}=\partial_{m}\left({\epsilon_{012}e^{\mathbb{F}} \over
 h\left(\Lambda\vert t\vert^2\right)^2}\right),
\nd
and appears from assuming slowly moving membranes. The epsilon tensor is raised and lowered by the un-warped metric, and $\mathbb{F}$ is as in \eqref{lagaga}.  When $\mathbb{F}$ vanishes or is a constant, \eqref{colemanbook} takes the same form as the ones we had in \cite{nogo} and \cite{nodS}. Combining this with \eqref{yvesroche}, we see that $f_\mu$ becomes constant and we are back to the background 
\eqref{betta}. 

It is also useful at this stage to make a distinction between warped and un-warped G-fluxes. If we take the G-flux components with all lower indices to be time independent (except for \eqref{colemanbook}), then the G-flux components with all upper indices will become time-dependent. If we extract the time dependences out, then we can define  un-warped G-fluxes that may be raised or lowered by the time-independent parts of the metric. Such a distinction is  not necessary but is nevertheless very useful to study the quantum effects. 

To proceed, let us switch on all possible components of the G-flux, including the ones with legs along the space-time directions. Of course caution needs to be exercised here because we don't want to change the type IIB geometry \eqref{sally} while descending from \eqref{heidkalu}. For example NS B-field with leg along $x_3$ direction will create a cross term in the type IIB metric. Such terms will complicate the geometry so, while we entertain all choices of G-flux, certain components will have to be put to zero when we make the duality map. With this in mind, the warped and the un-warped G-flux components may be related in the following way: 
\bg\label{clmehta}
&&{\bf G}^{0mab}={G}^{0mab}[\Lambda(t)]^{1/3}h^{-1/3}e^{\frac{2}{3}f_3}\nonumber\\
&&{\bf G}^{0mna}= {G}^{0mna}[\Lambda(t)]^{4/3}h^{-1/3}e^{-\frac{1}{3}f_3}\nonumber\\
&&{\bf G}^{mnpa}={G}^{mnpa}[\Lambda(t)]^{1/3}h^{-4/3}e^{-\frac{1}{3}f_3}\nonumber\\
&&{\bf G}^{0mnp}={G}^{0mnp}[\Lambda(t)]^{7/3}h^{-1/3}e^{-\frac{4}{3}f_3}\nonumber\\
&&{\bf G}^{mnab}={G}^{mnab}[\Lambda(t)]^{-2/3}h^{-4/3}e^{\frac{2}{3}f_3}\nonumber\\
&&{\bf G}^{mnpq}={G}^{mnpq}[\Lambda(t)]^{4/3}h^{-4/3}e^{-\frac{4}{3}f_3}\nonumber\\
&&{\bf G}^{\mu mab}={G}^{\mu mab}[\Lambda(t)]^{1/3}h^{-1/3}e^{-f_\mu}e^{\frac{2}{3}f_3}\nonumber\\
&&{\bf G}^{\mu \nu ab}={G}^{\mu\nu ab}[\Lambda(t)]^{4/3}h^{2/3}e^{-f_\mu-f_\nu}e^{\frac{2}{3}f_3}
\nonumber\\
&&{\bf G}^{\mu mnp}={G}^{\mu mnp}[\Lambda(t)]^{7/3}h^{-1/3}e^{-\frac{4}{3}f_3}e^{-f_\mu}\nonumber\\
&&{\bf G}^{\mu mna}={G}^{\mu mna}[\Lambda(t)]^{4/3}h^{-1/3}e^{-\frac{1}{3}f_3}e^{-f_\mu}\nonumber\\
&&{\bf G}^{\mu \nu ma}={G}^{\mu\nu mn}[\Lambda(t)]^{7/3}
h^{2/3}e^{-f_\mu-f_\nu}e^{-\frac{1}{3}f_3}\nonumber\\
&&{\bf G}^{012a}={G}^{012a}[\Lambda(t)]^{10/3}h^{5/3}e^{-(f_1+f_2)}e^{-\frac{1}{3}f_3}\nonumber\\
&&{\bf G}^{\mu \nu mn}={G}^{\mu\nu mn}[\Lambda(t)]^{10/3}
h^{2/3}e^{-f_\mu-f_\nu}e^{-\frac{4}{3}f_3}\nonumber\\
&&{\bf G}^{012m}={G}^{012m}
[\Lambda(t)]^{13/3}h^{5/3}e^{-\frac{4}{3}f_3}e^{-(f_1+f_2)},
\nd
where the repeated indices are not summed over and $f_\mu$ will satisfy the two conditions \eqref{lagaga} and \eqref{yvesroche}. Using these, the energy-momentum tensor for the G-flux and along the six-dimensional base can be written as:

{\footnotesize
\bg\label{kaun}
&&\mathbb{T}^G_{mn}= -\left(\frac{\partial_m h\partial_n h}{2h^2}- g_{mn}\frac{\partial_k h\partial^k h}{4h^2}\right)
+\frac{1}{4h}\left({G}_{mlka}{G}_{n}^{lka}-\frac{1}{6}g_{mn} {G}_{pkla}{G}^{pkla}\right)\\
&+&
\frac{e^{-f_3}\Lambda(t)}{12h}\left({G}_{mlkr}{G}_{n}^{lkr}-\frac{1}{8}g_{mn}{G}_{pklr}{G}^{pklr}\right)+
\frac{e^{f_3}}{4h\Lambda(t)}\left({G}_{mlab}{G}_{n}^{lab}-\frac{1}{8}g_{mn} {G}_{pkab}{G}^{pkab}\right)\nonumber \\
&+&e^{-f_3-f_\mu }\Lambda(t)^2\left({G}_{mpq\mu}{G}_{n}^{pq\mu}-\frac{1}{8}g_{mn} {G}_{pqr\mu}
{G}^{pqr\mu}\right) + e^{-f_\mu }\Lambda(t)\left({G}_{mp\mu a}{G}_{n}^{p\mu a}-\frac{1}{8}g_{mn} 
{G}_{pq\mu a}{G}^{pq\mu a}\right)\nonumber\\
&+&
e^{-(f_\mu+f_\nu)}
\Lambda(t)^2 h 
\left({G}_{ma\mu\nu}{G}_{n}^{a\mu \nu}-\frac{1}{8}g_{mn} {G}_{pa\mu\nu}{G}^{pa\mu\nu}\right)
+ e^{-(f_\mu+f_\nu)} e^{f_3}
\Lambda(t) h 
\left(-\frac{1}{8}g_{mn} {G}_{ab\mu\nu}{G}^{ab\mu\nu}\right)\nonumber\\
&+& e^{f_3}e^{-f_\mu}
\left({G}_{m\mu ab}{G}_{n}^{\mu ab}-\frac{1}{8}g_{mn} {G}_{p\mu ab}{G}^{p\mu ab}\right)
+
e^{-f_3}e^{-(f_\mu+f_\nu)}
\Lambda(t)^3 h 
\left({G}_{mp\mu\nu}{G}_{n}^{p\mu \nu}-\frac{1}{8}g_{mn} {G}_{pq\mu\nu}{G}^{pq\mu\nu}\right). \nonumber
\nd}
Looking at the above expression, it is clear that the condition \eqref{yvesroche} now pays off in the sense that we can write \eqref{kaun} completely 
in terms of positive or negative powers of $g_s^2$. Since all the un-warped components of the G-fluxes are time independent, the time dependence in the above expressions all come from these powers of $g_s^2$. 
Additionally, certain choices of the fluxes in \eqref{kaun} are redundant here. For example we can put:
\bg\label{starkhouse}
G_{MN ab} = 0 = G_{\mu mab}, \nd
where $M, N = (\mu, \nu)$ or $(m, n)$.
The former is because all the flux components are assumed to be functions of the six-dimensional base coordinates $y^m$ only; whereas the latter is proportional to $\partial_M(C_{N ab})$ and therefore leads to a NS B-field $B^{(2)}_{N 3}(y)$ in the type IIA side. Such a B-field will change the type IIB background 
by creating a cross-term in the space-time metric of \eqref{sally} which we want to avoid at this stage for simplicity. (These components will be inserted back in section \ref{temporal} where we will require more generic picture.) One may also see that:
\bg\label{tagrach}
G_{m a \mu\nu} = 0. \nd
This is because \eqref{tagrach} leads to either a NS two-form $B^{(2)}_{\mu\nu}(y)$ or a RR two-form 
$C^{(2)}_{\mu\nu}(y)$ in the type IIB side. In the presence of D3-branes or higher dimensional D-branes the former will lead to a non-commutative space-time at weak type IIB coupling, whereas the latter will lead to a non-commutative theory at strong type IIB coupling (i.e under a S-duality). Although the latter is not much of an issue here, to avoid complicated space-time geometry we can impose \eqref{tagrach}. Thus in the same vein we can also 
make:
\bg\label{redvi}
G_{mn\mu a} = 0, \nd
as this will lead to either a NS B-field $B^{(2)}_{n\mu}$ or a RR B-field $C^{(2)}_{n\mu}$. The former, again in the presence of space-filling D-branes, leads to a dipole deformation in space-time thus changing the 
metric exactly as \eqref{naviali} with $x_2$ or $x_1$ replacing the $x_3$ direction. The later would have similar behavior at strong coupling. Again to avoid complicating the type IIB geometry from \eqref{sally} within the approximation \eqref{tabu}, we will impose \eqref{redvi}. Thus the only fluxes that would contribute here are the three-form fluxes ${\bf H}_3$ and ${\bf F}_3$, all defined on the six-dimensional base, and certain components of the five-form fluxes appearing from the following G-flux components:
\bg\label{cutevi}
G_{mnpq}, ~~~G_{mnp\mu}, ~~~ G_{mp\mu \nu}, ~~~ G_{mnpa}. \nd
Out of the above choices, some of the components will again have to be put to zero if one wants the full de Sitter isometry as in \eqref{betta}. At this stage we will however assume \eqref{cutevi} as a judicious choice of components which, when combing with \eqref{kaun} and \eqref{yvesroche}, leads to the following energy-momentum tensor for the G-flux\footnote{The appearance of $g_s^2, g_s^4$ and $g_s^6$ doesn't imply anything quantum about \eqref{haribou}. This is all a tree-level result and it's abrupt truncation at $g_s^6$ 
for all values of $g_s$ confirms it's classical nature. In comparison, the series in \eqref{londry} has no apparent truncation for any values of $g_s$ and is therefore truly quantum.} :

{\footnotesize
 \bg\label{haribou}
\mathbb{T}^G_{mn}&=& -\left(\frac{\partial_m h\partial_n h}{2h^2}
- g_{mn}\frac{\partial_k h\partial^k h}{4h^2}\right)
+\frac{1}{4h}\left({G}_{mlka}{G}_{n}^{lka}-\frac{1}{6}g_{mn} {G}_{pkla}{G}^{pkla}\right)\nonumber\\
&&+
\frac{g_s^2}{12h\sqrt{h}}\left({G}_{mlkr}{G}_{n}^{lkr}-\frac{1}{8}g_{mn}{G}_{pklr}{G}^{pklr}\right)+
{g_s^4\over h}\left({G}_{mpq\mu}{G}_{n}^{pq\mu}-\frac{1}{8}g_{mn} {G}_{pqr\mu}
{G}^{pqr\mu}\right)\nonumber \\
&& + {g_s^6\over \sqrt{h}}
\left({G}_{mp\mu\nu}{G}_{n}^{p\mu \nu}-\frac{1}{8}g_{mn} {G}_{pq\mu\nu}{G}^{pq\mu\nu}\right).\nd}
We are almost there. All we need to complete the story is the expression for the Einstein tensor 
$\mathbb{G}_{mn}$. Following the same convention of decomposing a tensor into un-warped and 
warped pieces, gives us:
\bg\label{oshotme}
\mathbb{G}_{mn} = G_{mn} - {\partial_m h \partial_n h\over 2h^2} + g_{mn} \left(g_s^2 \sqrt{h} 
\mathbb{G}(t) - 6h \Lambda + {\partial_k h \partial^k h\over 4h^2}\right), \nd
where $G_{mn}$ is the Einstein tensor defined with the un-warped metric $g_{mn}$, and $\Lambda$ is a positive constant given earlier. The time dependence of the above expression is now captured by both $g_s^2$ and 
the function $\mathbb{G}(t)$. The latter takes the following form:
\bg\label{hammer}
\mathbb{G}(t) \equiv 
- \frac{e^{f_3}}{4}\sum_{i, j}\dot{f}_{(i} \dot{f}_{j)}  -{e^{f_3}\over 2} \sum_i \left(\overset{\text{..}}{f_i} 
-\frac{3 \dot{f}_i}{t} \right) = {3e^{f_3}\over 2} \left({3\dot{f}_3\over t} - \dot{f}_3^2 - \ddot{f}_3\right), \nd
where we have used \eqref{yvesroche} to simplify the expression. The remaining pieces in 
\eqref{oshotme} are all time independent as we had before, implying a Schwinger-Dyson equation of the form \eqref{montswing}:

{\footnotesize
 \bg\label{montaxi}
 \langle \Omega\vert g_{mn}(y, t) {\bf Tr}~{\bf G}_{\rm cl}(z, t')\vert \Omega\rangle
 =  -i \delta^8(y-z)\delta(t-t') +  h^{1/3}(z, t') \delta^{pq} \sum_{\{\alpha_i\}} \langle\Omega\vert 
  \widetilde{\mathbb{C}}_{pq}^{(\alpha_i)}(z, t') g_{mn}(y, t)\vert\Omega\rangle, \nonumber\\ \nd}
where $g_{mn}$ is field similar to \eqref{melmay} in the sense that it is the sum of the background field \eqref{betbab} and the fluctuation; and the quantum part is defined in \eqref{londry}. The 
${\bf G}_{\rm cl}$ part has the fluxes as in \eqref{patjane}. As emphasized earlier, such a sum is necessary to extract the full Schwinger-Dyson equations for our case.  

We have however been a bit sloppy here. 
The computation of \eqref{montaxi} was done using regular states inserted inside the path integral. In other words, for a time-ordered  correlation function of the form \eqref{sakura01} written as:

{\footnotesize
\bg\label{lisbonjane}
 \langle 0 \vert \mathbb{\bf I} ~ {\rm exp}\left(-i \int_{t_1}^T d^{11}x {\bf H}_{\rm int}\right) \mathbb{\bf I} 
~ g_{mn}(y_1, t_1) \mathbb{\bf I}~
 {\rm exp} \left(-i \int_{t_2}^{t_1} d^{11}x {\bf H}_{\rm int}\right) \mathbb{\bf I} ~g_{rs}(y_2, t_2) \mathbb{\bf I}~ 
{\rm exp} \left(-i \int_{-T}^{t_2} d^{11}x {\bf H}_{\rm int}\right) \mathbb{\bf I}\vert 0\rangle, \nonumber\\ \nd}
where $T$ is taken to infinity in a slightly imaginary direction, 
the identity operator $\mathbb{\bf I}$ inserted between each pieces in \eqref{lisbonjane}, is typically taken to be of the following standard form:
\bg\label{salonfemme}
\mathbb{\bf I} \equiv \int {\cal D} \left(g_{mn}\right) ~ \vert g_{mn}\rangle \langle g_{mn} \vert, \nd
with no summation over the repeated indices. However in the presence of a coherent state of the form 
\eqref{noharp}\footnote{The coherent state in \eqref{noharp} is expressed in coordinate independent way by integrating the spatial coordinates in the range $[-\infty, +\infty]$. If we don't restrict ourselves to this, we can allow coordinate dependence in the definition of the coherent state. The latter is what we will use here.}, 
one could instead entertain the following identity operator:
\bg\label{oshognosy}
\mathbb{\bf I} \equiv \int{\cal D}\left(g_{mn}\right) {\rm exp}\left(-\int d^6y ~g_{mn} g^{mn}\right) 
\vert \alpha^{mn} \rangle \langle \alpha^{mn}\vert, \nd
to be inserted in \eqref{lisbonjane} above. Such insertions convert the correlation function \eqref{lisbonjane}
to an appropriate path-integral representation, aptly called the coherent-state path-integral, with a somewhat non-relativistic action. However the quantum pieces continue to appear in the form 
\eqref{tunney} above, albeit with an overall suppression factor appearing from the gaussian piece in
\eqref{oshognosy}. Such a gaussian piece suppresses {\it all} the terms in \eqref{tunney} equally so doesn't alleviate the hierarchy problem that we face here. Going to the external legs amputated correlation function:
\bg\label{veralengt}
\langle g_{mn}(y_1, t_1)\vert \mathbb{\bf I} ~ {\rm exp}\left(-i \int_{t_2}^{t_1} d^{11}x {\bf H}_{\rm int}\right) \mathbb{\bf I}\vert g_{pq}(y_2, t_2)\rangle, \nd
with $\mathbb{\bf I}$ as in \eqref{oshognosy} introduces, in addition to the gaussian suppression factors as before, new ingredients like the overlap wave-function between the standard graviton state 
$\vert g_{mn} \rangle$ and the coherent state \eqref{noharp} of the form:
\bg\label{iching}
\Psi_{mn}(y, t) \equiv \langle \alpha^{mn}(y, t) \vert g_{mn}(y, t)\rangle . \nd
Such overlap wave-functions simply complicate the path-integral description of the system, but since they all appear equally in the time-neutral series of \eqref{sakura01}, they again fail to alleviate the hierarchy issues plaguing our scenario. 

The lesson that we learn from the above discussion is that at the quantum level the time-neutral series 
$\mathbb{C}_{pq}^{(i)}$ that we constructed in \cite{nodS} is responsible for breaking the hierarchy between $g_s$ and $M_p$, thus prohibiting a simple four-dimensional effective field theory. It therefore doesn't quite matter if we view our background \eqref{betta} as a time-dependent (and therefore non-supersymmetric) background, or as a non-supersymmetric coherent (or squeezed-coherent) state over a supersymmetric solitonic background. The issue lies deeper in the theory, and not on the various realizations (or avatars) of   
\eqref{betta}. 

The time-dependent equations reveal similar story. Equating the time-dependent pieces in the Einstein term \eqref{oshotme} with the sum of the energy momentum tensors from fluxes in \eqref{haribou} and the quantum series in \eqref{londry}, one may easily infer the following equation:

{\footnotesize
\bg\label{thesleuth}
g_{mn} g_s^2 \sqrt{h} \mathbb{G}(t) &=&
\sum_{\alpha_i \ne 0} g_s^{2\alpha_i} \mathbb{C}^{(i)}_{mn} +
{g_s^6\over \sqrt{h}}
\left({G}_{mp\mu\nu}{G}_{n}^{p\mu \nu}-\frac{1}{8}g_{mn} {G}_{pq\mu\nu}{G}^{pq\mu\nu}\right) \\
&+& \frac{g_s^2}{12h\sqrt{h}}\left({G}_{mlkr}{G}_{n}^{lkr}-\frac{1}{8}g_{mn}{G}_{pklr}{G}^{pklr}\right)+
{g_s^4\over h}\left({G}_{mpq\mu}{G}_{n}^{pq\mu}-\frac{1}{8}g_{mn} {G}_{pqr\mu}
{G}^{pqr\mu}\right). \nonumber \nd}
At this stage, one may compare terms of order $g_s^2, g_s^4, g_s^6$ and beyond from the flux and the quantum pieces with the LHS defined in terms of $\mathbb{G}(t)$ whose functional form appears in 
\eqref{hammer}.  Such an identification will be possible if $\mathbb{G}(t)$ can be expressed as:
\bg\label{vega}
\mathbb{G}(t) = \sum_k c_k  ~ h^{-k/2} g_s^{2k}, \nd
where $c_k$ are strictly constants to keep \eqref{vega} as function of time only. Such a choice of 
$\mathbb{G}(t)$ changes the energy-momentum tensor from the quantum pieces \eqref{londry} to the following:
\bg\label{anastar}
\mathbb{T}_{mn}^Q ~ \to ~ \widetilde{\mathbb{T}}_{mn}^Q = \sum_{\alpha_i = k+1} g_s^{2\alpha_i} 
\mathbb{C}_{mn}^{(i)} - \sum_k {c_k ~g_s^{2k+2} g_{mn} \over h^{(k-1)/2}}, \nd
 where the second term is again a time-neutral series multiplying powers of $g_s$. Such a series may be absorbed in the definition of $\mathbb{C}_{mn}^{(i)}$ which were originally an implicit function of the warp factor $h$. Thus augmenting the definition of the quantum energy-momentum tensor from \eqref{londry} to 
 \eqref{anastar}  essentially implies that $\mathbb{G}(t)$ in \eqref{hammer} may be put to zero without any loss of
 generalities\footnote{Another way to see this may be inferred from the time-dependent equation of motion 
 \eqref{thesleuth}. The equation is expressed in terms of powers of $g_s^2$ and therefore one could compare the powers on both sides of the equation. Looking at the $g_s^2$ part, we get:
 $$g_{mn} \sqrt{h} \mathbb{G}(t) =  \frac{1}{12h\sqrt{h}}\left({G}_{mlkr}{G}_{n}^{lkr}-\frac{1}{8}g_{mn}{G}_{pklr}{G}^{pklr}\right)   + \sum_{\{\alpha_i = 1\}}  \mathbb{C}^{(i)}_{mn}.$$ 
 The RHS of the above equation is completely expressed in terms of time-neutral functions, but the LHS has a time-dependent function $\mathbb{G}(t)$. For this to make sense we have to impose $\mathbb{G}(t) = 0$.
 \label{neve}.}. 
This gives two possible values for $f_3(t)$, namely:
 \bg\label{lailam}
 f_3(t) = 0, ~~~~~~ f_3(t) = f_{00} + \log\left(e_0 + {t^4\over t_0^4}\right), \nd
 where $f_{00}, e_0$ and $t_0$ are constants. For the flat slicing that we consider for \eqref{betta} and for 
 \eqref{heidkalu}, the latter form for $f_3(t)$ shows a logarithmically decreasing behavior. Comparing the various powers of $g_s^2$ lead to similar issues encountered for the time-independent EOMs. 
 
 Let us now look at the behavior along the fibre torus direction. Since we allow flux components along these
 directions, there would be non-zero energy momentum tensor.  We can express it in the following way:
 
{ \footnotesize
 \bg\label{killalan}
 \mathbb{T}^G_{ab}&&=\frac{e^{-f_3}\Lambda(t)}{12h}\left({G}_{amnp}{G}_{b}^{mnp}-\frac{1}{2} \delta_{ab} {G}_{mnpc}{G}^{mnpc}\right)+\frac{1}{4h}\left({G}_{acmn}{G}_{b}^{cmn}-\frac{1}{2} \delta_{ab} {G}_{mncd}
 {G}^{mncd}\right)\nonumber\\
&&-\delta_{ab} \frac{\Lambda^2(t)e^{-2f_3}}{4.4! h} {G}_{mnpq}{G}^{mnpq}
+ {1\over 4h}\Lambda(t) e^{-f_3}
\delta_{ab}g^{mn}{\partial_m h\partial_n h}
-\delta_{ab}\Lambda^3(t)e^{-2f_3}e^{-f_\mu} {G}_{\mu mpn}{G}^{\mu mpn}\nonumber\\
&&+
{e^{-f_3-f_\mu}}\Lambda^2(t)\left({G}_{a\mu mp}{G}_{b}^{\mu mp}-\frac{1}{2} \delta_{ab} {G}_{c\mu mp}
{G}^{c\mu mp}\right)+
{e^{-f_\mu}}\Lambda(t)\left({G}_{acm\mu}{G}_{b}^{cm\mu }-\frac{1}{2} \delta_{ab} {G}_{cdm\mu }
{G}^{cdm\mu }\right)\nonumber \\
&&-{1\over 2} e^{-2f_3}{e^{-f_\mu-f_\nu}}\Lambda^4(t)h
 \delta_{ab} {G}_{mn\mu\nu }{G}^{mn\mu\nu }
+e^{-f_3}{e^{-f_\mu-f_\nu}}\Lambda^3(t)h\left({G}_{am\mu\nu}{G}_{b}^{m\mu \nu}-\frac{1}{2} \delta_{ab} 
{G}_{cdm\mu }{G}^{cdm\mu }\right)\nonumber \\
&&+ 
{e^{-f_\mu-f_\nu}}\Lambda^2(t)h\left({G}_{ac\mu\nu}{G}_{b}^{c\mu \nu}-\frac{1}{2} \delta_{ab} 
{G}_{cd\mu\nu }{G}^{cd\nu\mu }\right), \nd}   
where the repeated indices are summed over. The expression \eqref{killalan} is 
similar to what we had in \eqref{kaun} for $\mathbb{T}_{mn}^G$ and therefore subjected to the same truncation  \eqref{haribou} that we applied therein. The truncation appears from \eqref{cutevi}, and it 
converts \eqref{killalan} to the following:
\bg\label{mojarlat}
 \mathbb{T}^G_{ab}&&=\frac{g_s^2}{12h\sqrt{h}}\left({G}_{amnp}{G}_{b}^{mnp}-\frac{1}{2} \delta_{ab} {G}_{mnpc}{G}^{mnpc}\right) -{g_s^8 \over 2h} 
 \delta_{ab} {G}_{mn\mu\nu }{G}^{mn\mu\nu }\\
&&-{g_s^4\over 4. 4! h^2}\delta_{ab}  {G}_{mnpq}{G}^{mnpq}
+ {g_s^2\over 4h\sqrt{h}}
\delta_{ab}g^{mn}{\partial_m h\partial_n h}
- {g_s^6\over h\sqrt{h}}\delta_{ab}{G}_{\mu mpn}{G}^{\mu mpn}, \nonumber \nd
where we have used $g_s^{(0)} = 1$ in \eqref{londry} and the simplifying condition \eqref{yvesroche}. The above formulation is classical despite the appearance of terms up to $g_s^8$. The reasoning remains the same: the abrupt truncation and the simple identification of $\Lambda(t)e^{-f_3}$ to $g_s$ spells out it's classical behavior.  In a similar vein, the Einstein tensor becomes:
\bg\label{alan256}
\mathbb{G}_{ab} = - {g_s^2 \delta_{ab} \over 2\sqrt{h}}\left(R + 9 h \Lambda - {g^{mn} \partial_m h 
\partial_n h\over 4h^2}\right) + {g_s^4 \delta_{ab}} ~\mathbb{H}(t). \nd
The similarity of \eqref{alan256} with the corresponding one in \cite{nogo}, modulo the $\mathbb{H}$ factor, shouldn't be a surprise. The function $\mathbb{H}(t)$ is a function only of time, and may be written as:

{\footnotesize
\bg\label{jessfranco}
e^{-f_3} \mathbb{H}(t) \equiv -{1\over 4} \sum_{i, j=1}^2 \dot{f}_{(i} \dot{f}_{j)}  
- {1\over 2} \sum_{i = 1}^2 \left(\ddot{f}_i + \dot{f}_i \dot{f}_3 + {2\dot{f}_i\over t}\right) 
-\ddot{f}_3 - {\dot{f}_3^2 \over 2} + {3\dot{f}_3\over t} = 2 \ddot{f}_3 + {9 \dot{f}_3^2 \over 4} 
- {7 \dot{f}_3 \over t}. \nd}
We are now in a situation encountered earlier, and therefore subject to the same course of action implemented therein. As done for \eqref{vega}, the function $\mathbb{H}(t)$ may be expressed as powers of $g_s^2$, but now with a different coefficient $b_k$. Such a series representation of $\mathbb{H}(t)$ shifts
the energy-momentum of the quantum terms to the following:
\bg\label{teamsol}
\mathbb{T}_{ab}^Q ~ \to ~ \widetilde{\mathbb{T}}_{ab}^Q = \sum_{\alpha_i = k+2} g_s^{2\alpha_i} 
\mathbb{C}_{mn}^{(i)} - \sum_k {b_k ~g_s^{2k+4} g_{mn} \over h^{k/2}}, \nd
spelling out an essentially similar story encountered before in \eqref{anastar} with $b_k$ replacing $c_k$ in 
\eqref{vega}. Since the additional pieces in energy-momentum tensor are all time-neutral functions multiplying powers of $g_s^2$, we can absorb them in the definition of $\mathbb{C}_{ab}^{(i)}$, thus making $\mathbb{H}(t)$ equivalent to zero as before\footnote{In a similar vein, as discussed in footnote 
\ref{neve}, we can express the time-dependent equation involving $\mathbb{H}(t)$ by comparing the $g_s^4$ coefficients from \eqref{mojarlat}, \eqref{alan256} and \eqref{londry}. This gives us:
$$\mathbb{H}(t) =  -{1 \over 96 h^2} ~{G}_{mnpq}{G}^{mnpq} 
+ {1\over 2} \sum_{\{\alpha_i = 2\}} \delta^{ab} \mathbb{C}^{(i)}_{ab}.$$
Again we see that the LHS is a function of time whereas the RHS is completely made of time-neutral pieces. 
Such as equation would make sense if we impose $\mathbb{H}(t) = 0$, leading to essentially the same conclusion.}.
Implementing this gives us:
\bg\label{greenbook}
f_3(t) = f_{11} +  {8\over 9}~\log\left[e_1 + \left({t\over t_0}\right)^{9/2}\right], \nd 
in addition to the trivial solution, with $f_{11}, e_1$ and $t_0$ as constants. Comparing \eqref{greenbook} to 
\eqref{lailam} it is easy to argue that under the following condition:
\bg\label{rinehart}
f_{00} = f_{11}, ~~~~ e_0 = e_1 = 0, \nd
we can have one function solving two differential equations. To see whether this continues to be the case we will have to study the energy-momentum tensors for fluxes and quantum corrections, including the Einstein tensors, along the $2+1$ dimensional space-time directions.

The story then unfolds in an expected way. The generic form of the energy-momentum tensor is again easy to spell out, and takes the following form:

{\footnotesize
\bg\label{blakel}
\mathbb{T}^G_{\mu\nu} &=& -\eta_{\mu\nu}e^{f_\mu}\left(\frac{1}{4! \Lambda(t) h^2}{G}_{mnpa}{G}^{mnpa}
+\frac{e^{-f_3}}{4.4!  h^2}{G}_{mnpq}{G}^{mnpq}+
\frac{e^{f_3}}{16h^2\Lambda^2(t)}{G}_{mnab}{G}^{mnab}+ 
g^{mn}\frac{\partial_m h\partial_n h}{4\Lambda(t) h^3}\right) \nonumber\\
&+& \Lambda(t)h^{-1}e^{-f_3}\left({G}_{\mu mnp}{G}_{b}^{ mnp}-\frac{1}{2} \eta_{\mu \nu}
 {G}_{\rho mnp }{G}^{\rho mnp}\right)
+ 
{e^{-f_\nu}}e^{+f_3}
\left({G}_{\mu \rho ab}{G}_{\nu}^{\rho ab}-\frac{1}{2} \eta_{\mu \nu} {G}_{\rho \sigma ab }
{G}^{\rho \sigma ab}\right)\nonumber\\
&+&
{h}^{-1}\left({G}_{\mu mna}{G}_{b}^{ mna}-\frac{1}{2} \eta_{\mu \nu} {G}_{\rho mna }
{G}^{\rho mna }\right)
+
e^{f_3} [\Lambda(t)]^{-1}h^{-1}
\left({G}_{\mu mab}{G}_{\nu}^{ mab}-\frac{1}{2} \eta_{\mu \nu} {G}_{\rho mab }{G}^{\rho mab}\right)
\\
&+&
{e^{-f_\nu}}e^{-f_3}\Lambda(t)^{2}
\left({G}_{\mu \rho mn}{G}_{\nu}^{\rho mn}-\frac{1}{2} \eta_{\mu \nu} {G}_{\rho \sigma mn }
{G}^{\rho \sigma mn}\right) + {e^{-f_\nu}}\Lambda(t)
\left({G}_{\mu \rho ma}{G}_{\nu}^{\rho ma}-\frac{1}{2} \eta_{\mu \nu} {G}_{\rho \sigma ma }
{G}^{\rho \sigma ma}\right), \nonumber \nd}
where all possible terms contribution to the tensor is shown with appropriate coefficients. Clearly many of these terms are irrelevant for us and therefore keeping only the terms that we actually need, \eqref{blakel} 
changes to:

{\footnotesize
\bg\label{khalamey}
\mathbb{T}^G_{\mu\nu} &=& -{\eta_{\mu\nu}e^{f_\mu - f_3}\over g_s^2} 
\left(\frac{1}{4! h\sqrt{h}}{G}_{mnpa}{G}^{mnpa}
+\frac{g_s^2}{4.4!  h^2}{G}_{mnpq}{G}^{mnpq}+
\frac{1}{16h g_s^2}{G}_{mnab}{G}^{mnab}+ 
\frac{\partial_m h\partial^m h}{4 h^2\sqrt{h}}\right) \nonumber\\
& + & {g_s^2\over h \sqrt{h}}\left({G}_{\mu mnp}{G}_{b}^{ mnp}-\frac{1}{2} \eta_{\mu \nu}
 {G}_{\rho mnp }{G}^{\rho mnp}\right) + {g_s^4 e^{f_3 - f_\nu}\over h}
 \left({G}_{\mu \rho mn}{G}_{\nu}^{\rho mn}-\frac{1}{2} \eta_{\mu \nu} {G}_{\rho \sigma mn }
{G}^{\rho \sigma mn}\right), \nonumber\\ \nd} 
with raising and lowering to be done with un-warped metric components. The $g_s^2$ factor contains all the time-dependences, and so every term is naturally divided into a time-dependent and a time-independent parts. Such a procedure can be adapted, as before, for the Einstein tensor 
$\mathbb{G}_{\mu\nu}$ that may be expressed as:
\bg\label{bhalomey}
\mathbb{G}_{\mu\nu} = - {\eta_{\mu\nu}e^{f_\mu - f_3}\over 2 \sqrt{h} g_s^2}\left(R + 3\Lambda h + 
{\partial_m h \partial^m h\over 2h^2} - {\square h\over h}\right) + e^{f_\mu} \mathbb{J}_{\mu\nu}(t), \nd
where no summation over the repeated indices is implied. We see that, modulo the isometry breaking terms and the function $\mathbb{J}_{\mu\nu}(t)$, the expression \eqref{bhalomey} is very similar to the one we had in \cite{nogo}. The function $\mathbb{J}_{\mu\nu}$, for various choices of $\mu$ and $\nu$, takes the form:
\bg\label{kmeybmey}
\mathbb{J}_{00}(t) &=&  -\frac{\dot{f_1}}{t}-\frac{ \dot{f}_2}{t}-\frac{ \dot{f}_3}{t}
+\frac{1}{4}\dot{f}_1 \dot{f}_2+ \frac{1}{4}\dot{f}_2\dot{f}_3+ \frac{1}{4}\dot{f}_1 \dot{f}_3 = 
3 \dot{f}_3 \left({\dot{f}_3\over 4} - {1\over t}\right) \\
\mathbb{J}_{11}(t) & = & -\frac{ \dot{f_2}}{t}-\frac{ \dot{f_3}}{t}
+ \frac{1}{4}\dot{f_2}\dot{f_3}+\frac{1}{4}\dot{f_2}{}^2+\frac{1}{4}\dot{f_3}{}^2
   +\frac{1}{2}\overset{\text{..}}{f_2}+ \frac{1}{2}\overset{\text{..}}{f_3} = \ddot{f}_3 + \dot{f}_3
   \left({3\dot{f}_3\over 4} - {2\over t}\right)\nonumber\\
\mathbb{J}_{22}(t) & = & -   \frac{ \dot{f_1}}{t}-\frac{ \dot{f_3}}{t}
+ \frac{1}{4}\dot{f_1}\dot{f_3}+\frac{1}{4}\dot{f_1}{}^2+\frac{1}{4}\dot{f_3}{}^2
   +\frac{1}{2}\overset{\text{..}}{f_1}+ \frac{1}{2}\overset{\text{..}}{f_3} = \ddot{f}_3 + \dot{f}_3
   \left({3\dot{f}_3\over 4} - {2\over t}\right), \nonumber \nd
where we have again used \eqref{yvesroche} to express the RHS of the three equations. Note that in this limit the last two equations behave in a similar way, but differ from the first equation. We could also use the freedom to shift the quantum energy-momentum tensor \eqref{londry} to make $\mathbb{J}_{\mu\nu}(t) = 0$, 
similar to what we did in \eqref{anastar} and \eqref{teamsol}. Implementing this, the vanishing of 
$\mathbb{J}_{00}(t)$ produces:
\bg\label{sthrower}
f_3(t) = f_{22} + 4~\log\left({t\over t_0}\right), \nd
with constant $f_{22}$. The functional form for $f_3$ is similar to \eqref{lailam} and \eqref{greenbook}
in the limit \eqref{rinehart}, if we identify $f_{22}$ with $f_{11}$ and $f_{00}$. Interestingly, if we now put 
$\mathbb{J}_{11}(t)$ or $\mathbb{J}_{22}(t)$ to zero, we get:
\bg\label{lromay}
f_3(t) = f_{33} + {4\over 3}~\log\left(e_2 + {t^3\over t_0^3}\right), \nd
which becomes identical to \eqref{sthrower} if $f_{33} = f_{22}$ and vanishing $e_2$. Therefore comparing 
\eqref{lailam}, \eqref{greenbook}, \eqref{sthrower} and \eqref{lromay}, and imposing the condition
\eqref{rinehart} augmented by the additional identifications of $f_{22} $ and $f_{33}$, we find that the following function:
\bg\label{kalikut}
e^{f_1(t)} = e^{f_2(t)} = e^{f_3(t)} \equiv  e^f \left({t\over t_0}\right)^4, \nd
solves {\it all} the equations simultaneously despite the fact that there are more equations than the number of unknowns in the problem. Such a unique solution for an over-determined system of equations should convey some special feature that should also resonate with the fact that the quantum energy-momentum tensor can be shifted to absorb changes appearing from the $f_i(t)$ terms. To see this, let us plug in 
\eqref{kalikut} to \eqref{sally}. Since the internal space remains time-independent, we can only study the four-dimensional metric, which transforms to:
\bg\label{norhor}
ds^2 = {1\over \Lambda \vert t\vert^2}\left[-dt^2 + e^{f}\left({t\over t_0}\right)^4 \left(dx_1^2 + dx_2^2 
+ dx_3^2\right)\right] ~ \to ~ {1\over \Lambda  \vert t'\vert^2}\big(-dt'^2 + dx_1^2 +  dx_2^2 + dx_3^2\big) ,
 \nonumber\\ \nd
where the RHS is surprisingly similar to the four-dimensional metric that we considered earlier in 
\eqref{betta} as well as in \cite{nogo, nodS} provided we use $t'$ instead of $t$. They are related by:
\bg\label{nicobun}
t' ~ = - {e^{-f/2} t_0^2\over t}, ~~~~~~ {\rm and} ~~~~ 0 ~ \le ~ \vert t'\vert ~ \le + \infty, \nd
which only changes the de Sitter slicing. Note that $t_0$ takes care of the dimension in \eqref{nicobun}, but 
$e^f$ is a redundant coefficient which could have been originally absorbed in the definition of the space coordinates $x_i$. However despite certain novelty being attributed to  \eqref{kalikut}, the end result is not surprising. As alluded to above, our ability to shift the quantum energy-momentum tensor as \eqref{anastar} and \eqref{teamsol} has, in a certain sense, predestined the behavior of the $f_i(t)$ functions, However what is intriguing is the choice \eqref{yvesroche}. Is there a specific reason for this? 

To see this let us go back to \eqref{kmeybmey} and consider the functional forms for $\mathbb{J}_{11}(t)$ 
and $\mathbb{J}_{22}(t)$ without incorporating \eqref{yvesroche}. Since both the functions are equated to zero, subtracting them leads to the following differential equation:
\bg\label{chandra}
{d\over dt}~\log\left(\dot{f}_2 - \dot{f}_1\right) = {2\over t}  - \mathbb{F}(t), \nd
where $\mathbb{F}(t)$ is defined in \eqref{lagaga}. Integrating this equation from some initial time $T$ to the 
present time $t$, we get the following relation between $\dot{f}_2(t)$ and $\dot{f}_1(t)$:
\bg\label{tagramey}
\dot{f}_2(t) - \dot{f}_1(t)  = {ct^2\over T^2} ~{\rm exp}\left(- \int_T^t \mathbb{F}(t') dt'\right), \nd
where $c$ is the difference between $\dot{f}_2(T)$ and $\dot{f}_1(T)$ that controls the initial behavior. We will assume that it is a finite number so that the metric remains finite at $T$. Thus for the original de Sitter 
slicing: $ -\infty \le t \le 0$, we can take $T$ to be a large negative integer, and fix the initial condition such that $c$ is an arbitrarily small number. In this limit we see that taking:
\bg\label{thenightof}
f_2(t) = f_1(t) + {\rm constant}, \nd
is not inconsistent with the dynamical evolution of the system.
The constant is irrelevant for the dynamics and therefore the above computation at least puts some credence to the choice \eqref{yvesroche} related to $f_1(t)$ and $f_2(t)$, provided of course that 
the exponential factor involving $\mathbb{F}(t)$ in \eqref{tagramey} do not introduce extra large factors that could change the result. The choice \eqref{yvesroche} guarantees this to some extent, and if we compare 
\eqref{thenightof} to the first equation in \eqref{kmeybmey}, we get:
\bg\label{louise}
\dot{f}_3(t) = {\dot{f}(t)\over 2}\left({8 - t \dot{f}(t)\over t\dot{f}(t)  - 2}\right), \nd
where $f(t)$ is identified with either $f_1(t)$ or $f_2(t)$. The above equation doesn't immediately allows us to choose $f_3(t)$ to be equal to $f(t)$, but we can make:
\bg\label{chukka}
f_3(t) \equiv f(t) + \sum_{n > 0} h_n \left({t\over T}\right)^n, \nd
where $h_n$ are constants and $T$ is, as before, some initial time that serves as a scale here. This is different from the scale $t_0$ that we used earlier because $t_0$ can be a finite integer, whereas we will typically take $T$ to be a large number. Such a choice guarantees $f_3(t)$ to be a function close to $f(t)$ in the following sense:
\bg\label{louisnun}
f_3(t) = 4 ~\log\left({t\over t_0}\right) + {\cal O}\left({t\over T}\right), \nd
which, in the limit of large $T$ and finite scale $t_0$, reproduces \eqref{yvesroche} and thus the condition 
\eqref{kalikut}. 

We have hopefully tied up most of the loose ends although one question still remains: can we allow more generic conditions than \eqref{yvesroche} or \eqref{thenightof} and \eqref{louisnun}? Our preliminary investigation reveals that imposing more generic conditions do not allow for an analytical solution to exist, but a numerical solution could still exist. However such a generic case is not very useful to study the quantum behavior because we cannot express the energy-momentum tensor from the quantum pieces in a simplified form as 
in \eqref{londry}. Besides, a generic choice for $f_i(t)$, if it exists, will not alleviate the hierarchy issues that we faced earlier (and also in \cite{nodS}), implying that indulging in a more convoluted exercise fails to reveal new physics.

\section{Time-dependent backgrounds, fluxes and quantum effects \label{timeft}}

In \cite{nogo} and \cite{nodS} it was argued how a four-dimensional effective field theory description 
was harder to get with full de Sitter isometries and time-independent internal space. One would presume that deviating away from these conditions might alleviate these problems. However,
in sections \ref{difuli} and  \ref{kasner} we argued how it is still difficult to get a four-dimensional effective field theory when deviations from a pure de Sitter isometric backgrounds are introduced via dipole deformations or via  time-dependent isometry breaking factors, keeping the internal space time independent. Therefore it appears that the lessons we learnt so far may be tabulated as:

\vskip.1in

\noindent $\bullet$ Breaking the four-dimensional de Sitter isometries in type IIB theory by introducing four-dimensional isometry breaking factors do not help.

\vskip.1in

\noindent $\bullet$ Keeping the metric components of the internal space in type IIB theory time independent by introducing 
time-independent warp factors do not help.

\vskip.1in

\noindent $\bullet$ Keeping most of the background G-flux components 
time-independent\footnote{Except the one with components along space-time directions, for example 
\eqref{colemanbook} with arbitrary choice for $\mathbb{F}(t)$.}, in the M-theory uplift of the type IIB background, do not help.

\vskip.1in

\noindent Thus what should help is when we take {\it all} parameters in the type IIB theory time dependent. This implies taking not only the metric of the internal space time dependent, but also the fluxes threading through both the internal space as well as the four-dimensional space-time. This is a hard exercise and therefore to make sense of our computations,  we want to keep the type IIB coupling constant under some control.  One way is to take it as a slowly varying function of time. However this will not allow us to access all periods of cosmological evolution of the system because beyond certain range of time periods the coupling constant is bound to become large, thus inducing non-perturbative corrections. We want to avoid such scenarios, so as a first trial we shall take vanishing axion and the type IIB coupling constant to be a time-independent small parameter. Everything else will however have to become time dependent. With this in mind, let us take the following 
ansatze for the type IIB metric: 

{\footnotesize
\bg\label{pyncmey}
ds^2=\frac{1}{\Lambda(t)\sqrt{h}}(-dt^2+dx_1^2
+dx_2^2+dx_3^2)+ \sqrt{h} \Big(F_1(t){g}_{\alpha\beta}(y)dy^\alpha dy^\beta
+ F_2(t){g}_{mn}(y)dy^m dy^n\Big), \nd}
with $\alpha, \beta=4, 5$ and $m,n=6, 7, 8, 9$. This division of the metric components is not natural but is nevertheless useful. For example if we want to keep the volume of the internal space time independent
we can make the functions $F_i(t)$ to take the following form:
\bg\label{olokhi}
F_1(t) \equiv \omega^2(t), ~~~~~ F_2(t) \equiv  {1\over \omega(t)}, \nd
where $\omega(t)$ is another arbitrary function of time. Note that with this choice of the  metric the internal space is a strict product of a four-dimensional manifold ${\cal M}_4$ and a two-dimensional manifold 
${\cal M}_2$, implying that metric components like $g_{\alpha n}$ will be taken to zero. Generalization of this is easy to achieve simply by switching on $g_{\alpha n}$, so we will not discuss it much here. The division is also reflected in the M-theory uplift of \eqref{pyncmey}, which takes the form:

{\footnotesize
\bg\label{vegamey}
ds^2=e^{2{A}(y,t)}(-dt^2+dx_1 ^2+dx_2^2)+e^{2{B_1}(y,t)}
{g}_{\alpha \beta}dy^\alpha dy^\beta+
e^{2{B_2}(y,t)}{g}_{mn}dy^m dy^n + e^{2{C}(y,t)} g_{ab} dx^a dx^b, \nonumber\\  \nd}
where ($a, b$) are the coordinates of a square two-torus parametrized by coordinates $x_3$ and $x_{11}$. The internal eight-manifold in M-theory therefore takes the following form: 
\bg\label{melisett} 
{\cal M}_8 \equiv {\cal M}_4 \times \left({\cal M}_2 \times {\mathbb{T}^2\over {\cal G}}\right), \nd
where locally ${\cal G} = 1$ as clear from the metric \eqref{vegamey} and ${\cal M}_2$ is a local 2-cycle with ${\cal G}$ acting on $\mathbb{T}^2$. Globally however, as before, we don't want the manifold ${\cal M}_8$ to have a vanishing Euler characteristics, so ${\cal G}$ will have to be some symmetry group of the internal subspace. In terms of the metric \eqref{vegamey} this is invisible, so we can continue using the local metric (modulo subtleties with charge-neutral\footnote{Charge neutrality 
so as to maintain the conditions of vanishing axion and constant dilaton imposed earlier. This may be easily achieved by taking ${\cal G} = \mathbb{Z}_2, \mathbb{Z}_3, \mathbb{Z}_4$, and $\mathbb{Z}_6$, 
 which are essentially the branches of F-theory at constant couplings \cite{sendas}.  This converts
 ${\cal M}_2 \times {\mathbb{T}^2\over {\cal G}} \to {{\cal M}_4^{(2)}\over {\cal G}}$ globally with 
 non-K\"ahler ${{\cal M}_4^{(2)}\over {\cal G}}$, but we keep 
 ${\cal M}_2$ for local computations in EOMs.
  \label{plaza2019}} 
configuration of seven-branes that will be elaborated later). 
The various warp-factors appearing in \eqref{vegamey} may now be expressed as:  
\bg\label{mrglass}
&&e^{2A}= \left[\Lambda(t)\right]^{-{4\over 3}}\left[h(y)\right]^{-\frac{2}{3}}, ~~~ 
e^{2C}= \left[\Lambda(t)\right]^{\frac{2}{3}}\left[h(y)\right]^{\frac{1}{3}}\nonumber\\
&&e^{2B_1}=F_1(t) \left[\Lambda(t)\right]^{-\frac{1}{3}}\left[h(y)\right]^{\frac{1}{3}}, ~~~ 
e^{2B_2}=F_2(t)\left[\Lambda(t)\right]^{-\frac{1}{3}}\left[h(y)\right]^{\frac{1}{3}}, \nd
\noindent where all the parameters appearing above have been defined earlier. The way we have expressed the warp-factors, they appear to be functions of ($y^\alpha, y^m$) and $t$, but not functions of the space-time 
coordinates or of the fibre torus. If we relax the T-duality rules, we could even allow the warp-factors to be functions of the fibre torus, but then the analysis will get more involved. We want to avoid this, and also avoid complicating the space-time geometry by introducing isometry breaking factors. 

Our aim now is to express the solution \eqref{vegamey} as a coherent state over the same  solitonic background \eqref{betbab} that we used earlier. The coherent state formalism should be similar to 
\eqref{sakura01}, implying that the Fourier components are similar to what we had in \eqref{indigobk}
except for three changes. One, the $f_\mu(t)$ factor in the first term of \eqref{indigobk} vanishes. Two, the Fourier coefficient $\widetilde{g}_{mn}(y, t)$ now splits into two pieces:
\bg\label{sandcity}
&& \widetilde{g}_{\alpha\beta}(k) = \int d^2 y dt \sqrt{g^{(0, 2)}_{\rm base}}\left(e^{2B_1(y, t)}g_{\alpha\beta}
 - h_1^{1/3}g^{(0)}_{\alpha\beta}\Big\vert_{\rm base}\right) \eta^\ast_k(y, t) \nonumber\\
&&\widetilde{g}_{mn}(k) = \int d^4 y dt \sqrt{g^{(0, 4)}_{\rm base}}\left(e^{2B_2(y, t)}g_{mn}
 - h_1^{1/3}g^{(0)}_{mn}\Big\vert_{\rm base}\right) \xi^\ast_k(y, t), \nd
where $\xi_k(y, t)$ and $\eta_k(y, t)$ now replace the Schr\"odinger wave-function $\chi_k(y, t)$ in 
\eqref{indigobk}; and $g_{\rm base}^{(0, p)}$ denote the classical metric of a $p$-dimensional internal space  
in \eqref{betbab}. Finally, three: the $A, B_i$ and $C$ factors used in \eqref{indigobk} and \eqref{sandcity} should now be taken from \eqref{mrglass}. Note that the type IIA coupling will again resort back to 
\eqref{montse} that we had earlier.

 
 \subsection{Structure of the warp-factors and the background G-fluxes \label{temporal}}
 
There is also an alternative possibility of  viewing the solution \eqref{pyncmey} itself as the background  
(instead of being a coherent state over some solitonic background) and study fluctuation over 
this, as in \eqref{sstones}. These fluctuations couple with a Newton's constant given as in \eqref{redjohn}. There are of course problems associated with such a viewpoint, mostly as a consequence 
of being a non-supersymmetric vacuum  that we emphasized earlier. However if we assume that such issues may be alleviated at a deeper level, the cosmological framework that arises from this set-up should at least make sense with what we expect in four-dimensions.  In particular we can ask whether the Newton's constant $G_N$ may be kept time-independent for either vanilla de Sitter space or for fluctuations of the form \eqref{sstones} over de Sitter space. 
Comparing with \eqref{redjohn},
 it appears that there are at least two class of relations that $F_1(t)$ and $F_2(t)$ in \eqref{vegamey} satisfy, that may be written together as:
\bg\label{ranjhita2}
F_1(t) F_2^2(t) \equiv e_0 + {e_1g_s^2\over \sqrt{h}}, \nd
with specific choices for ($e_0, e_1$). For example, the choice ($1, 0$) i.e \eqref{olokhi} corresponds to vanilla de Sitter, whereas the choice ($0, 1$) corresponds to fluctuations of the form \eqref{sstones} over de Sitter. More elaborate generalizations are possible, but we will not indulge on them here\footnote{Note that the second condition on the warp-factors $F_i(t)$  implies that the fluctuations $\epsilon h_{\mu\nu}$ over the background \eqref{pyncmey} couple with a Newton's constant that is time-independent. However for the computation of EFT one may  view this simply as a constraint on the warp-factors $F_i(t)$. This choice therefore should not be viewed as giving an EFT on a flat space. The other possibility where $F_1(t) F_2^2(t) = 1$ (one choice being \eqref{olokhi}) will also be discussed simultaneously wherever we implement \eqref{ranjhita}.}. 
Here we have absorbed the constant type IIB coupling in the definition of $h$ to avoid introducing extra factors and used the IIA coupling $g_s$ to express the RHS. Note that the choice:
\bg\label{ranjhita}
F_1(t) F_2^2(t) = {g_s^2\over \sqrt{h}}, \nd 
is {\it not} the volume-preserving choice 
\eqref{olokhi}. The latter would have give us a time-independent overall volume of the internal space. The former i.e \eqref{ranjhita} would give a time-dependent Newton's constant if applied to vanilla de Sitter, so one may view the two cases from \eqref{ranjhita2} as representative of time-independent (i.e ($e_0, e_1$) = ($1, 0$)) and time-dependent (i.e ($e_0, e_1$) = ($0, 1$)) cases for vanilla de Sitter. Interestingly the choice \eqref{olokhi} resonates well with the condition prescribed for the Newton's constant in  \cite{russot} (see eq. (2.3) in \cite{russot}), so it will be interesting to compare the result of our investigations with the ones in 
\cite{russot}. We will discuss this later. 

The functional form for $F_1(t)$ and $F_2(t)$ are still undetermined and the two cases, namely 
\eqref{olokhi} and \eqref{ranjhita}, differ by having either a constant or $g_s^2$ on the RHS. For either of these two cases, we can start by defining $F_2(t)$ in the following way:
\bg\label{bobby}
F_2(t) & = & \sum_{k, n \ge  0} c_{kn} \left({g_s^2\over \sqrt{h}}\right)^{\Delta k} 
{\rm exp}\left(-{~n h^{\Delta/4} \over g^{\Delta}_s}\right) \nonumber\\
&=& c_{00} + \sum_{k > 0} c_{k0} \left({g_s^2\over \sqrt{h}}\right)^{\Delta k} 
+ \sum_{n > 0} c_{0n} ~{\rm exp}\left(-{~n h^{\Delta/4} \over g_s^\Delta}\right) + {\rm cross~terms}, \nd
where if $c_{00}$ vanishes then there is no time-independent piece: and $c_{kn}$ are integers with 
($k, n$) $ \in \left({\mathbb{Z} \over 2}, \mathbb{Z}\right)$. We have also inserted a constant parameter
$\Delta$ whose value will be determined later.
The above expansion is defined for small $g_s$ in type IIA, and we have assimilated the negative powers of $g_s$ 
as a non-perturbative sum. The latter is motivated from a resurgent sum of powers of inverse 
$g_s$ at weak IIA coupling so that all ($k, n$)-dependent terms in \eqref{bobby} are small.  However since  the type IIA coupling depends on both time and the coordinates of the internal space in the type IIB side, care is needed to interpret what is weak and what is strong coupling here. At a given point $y_0$ in the internal space, the time interval: 
\bg\label{dimplek}
\vert t \vert^2  < ~ {1\over {\Lambda\sqrt{h(y_0)}}}, \nd
should be related to weakly coupled interactions in the type IIA side. For small cosmological constant 
$\Lambda$ and small internal warp-factor at any point in the internal space, \eqref{dimplek} scans a reasonably wide range of time interval provided we can argue for the smallness of both $\Lambda$ and $h(y)$.  The smallness of $\Lambda$, in appropriate units, should be viewed as an experimental fact, whereas the smallness of $h(y)$ at all points $y^m$ in the internal space is  more non-trivial to establish.  We can take this as a requirement and arrange the fluxes etc to suit the equations of motion, but whether this can indeed hold needs to be seen. In any case as long as $h(y) < 1$ and $\Lambda <<1$, 
\eqref{dimplek} will assert a wide range of time interval for weakly coupled interactions. With this in mind, 
we can express $F_1(t)$ as:
\bg\label{jenlop}
F_1(t) \equiv  \left({g_s^2\over \sqrt{h}}\right) F_2^{-2}(t) =   \sum_{k, n >  0} b_{kn} 
\left({g_s^{2}\over \sqrt{h}}\right)^{\Delta k + 1} {\rm exp}\left(-{~n h^{\Delta/4} \over g^{\Delta}_s}\right), \nd
where $b_{kn}$ are constant coefficients that may be related to the $c_{kn}$ coefficients (for $k > 0, n > 0$)
in \eqref{bobby} at weak coupling. The way we have expressed \eqref{jenlop}, comparing to \eqref{bobby}
implies $b_{0n} = b_{1/2, n} = 0$ for $k = 0$ and $k = 1/2$ respectively.  
Similarly the single and double time derivatives of $F_2(t)$ may be expressed as:
\bg\label{scarjo} 
 {\dot{F}_2\over \sqrt{\Lambda}} &=&  \sum_{k, n \ge  0} c_{kn} \left[2k \Delta
\left({g_s^{2}\over \sqrt{h}}\right)^{\Delta k-1/2} + n\Delta  \left({g_s^{2}\over 
\sqrt{h}}\right)^{\Delta k- {\Delta\over 2} - {1\over 2}}\right] 
{\rm exp}\left(-{~n h^{\Delta/4} \over g^\Delta_s}\right) \nonumber\\
 {\ddot{F}_2\over {\Lambda}} &=&  \sum_{k, n \ge  0} c_{kn} \left[2k\Delta (2k\Delta-1) 
\left({g_s^{2}\over \sqrt{h}}\right)^{\Delta k-1} + n^2 \Delta^2 
\left({g_s^{2}\over \sqrt{h}}\right)^{\Delta k- \Delta - 1}\right]{\rm exp}\left(-{~n h^{\Delta/4} \over g^\Delta_s}\right)\nonumber\\
&+&  \sum_{k, n \ge  0} c_{kn} \left[n\Delta (4k\Delta-\Delta - 1)\left({g_s^{2}\over \sqrt{h}}\right)^{\Delta k- \Delta/2 -1}\right] 
{\rm exp}\left(-{~n h^{\Delta/4} \over g^\Delta_s}\right),  \nd
which shows that the time derivatives of $F_2(t)$ may also be expressed in terms of integer powers of
$g_s$. Needless to say, a similar conclusion also extends to the single and double time derivatives of 
$F_1(t)$ with the replacement of $c_{kn}$ by $b_{kn}$ in \eqref{scarjo}.  

The above discussion pretty much sums up the requirements that we want to impose on the warp-factors so that they solve the equations of motion. It is now time to dwell on the main ingredients, namely the G-fluxes. In our earlier attempt to study the Kasner-de Sitter type background, we had kept the G-flux components with all lower indices to be completely time independent. This made the G-flux components with all upper indices to be time dependent solely from the time-dependent warp-factors (see 
\eqref{clmehta} for details). Our present analysis will differ from this in one important respect: we will now keep the G-flux components with all lower indices to be inherently time dependent. In other words we take the following configuration:
\bg\label{frostgiant}
{\bf G}_{MNPQ}(y, t) = \sum_{k, n \ge 0} {\cal G}^{(k, n)}_{MNPQ}(y) 
\left({g_s^{2}\over \sqrt{h}}\right)^{\Delta k}
{\rm exp}\left(-{~n h^{\Delta/4} \over g^\Delta_s}\right), \nd
with the tensorial coefficient ${\cal G}_{MNPQ}^{(k, n)}$ for various choices of $k \in {\mathbb{Z}\over 2}$
and $n \in \mathbb{Z}$ being functions of the internal coordinates $y^m$. Such an expansion guarantees
that the flux components are expressed in terms of all positive and negative integer powers of 
$g^\Delta_s$. 
 There could also be a similar expansion for the potential $C_{MNP}$, but we only use the field strength here as these are the relevant variables for our case.  Note also the similarity of the expansion with \eqref{bobby} and \eqref{scarjo}. This is intentional as such time dependences should borne out of the time-dependent warp-factors for the internal space, and they in turn will be related to each other via the equations of motion to be satisfied by the corresponding coherent states. All these will be illustrated below, but before we proceed it may be worthwhile to isolate the time dependences of the G-flux components with all upper indices from the time dependent warp-factors much in the vein of \eqref{clmehta}. 

The necessity $-$ or more appropriately the usefulness $-$ of such an approach is two-fold. One: isolating the time dependences this way will emphasize the contributions of the warp-factors towards the temporal behavior of the fluxes more succinctly; and two: the time-independent cases would follow simply from the aforementioned expansion by switching off the un-related terms thus forming a single setup to study both time-dependent and time-independent cases.  With these in mind, we can isolate the time dependences in the following way: 
\bg\label{choshma}
&&{\bf G}^{012\alpha}= {G}^{012\alpha}
[\Lambda(t)]^{13/3}h^{5/3}F_1^{-1}\nonumber\\
&&{\bf G}^{012m}= {G}^{012m}
[\Lambda(t)]^{13/3}h^{5/3}F_2^{-1} \nonumber\\
&&{\bf G}^{\alpha \beta \gamma\delta}= {G}^{\alpha \beta \gamma\delta}[\Lambda(t)]^{4/3}h^{-4/3}
F_1^{-4}  \nonumber\\
&&{\bf G}^{\alpha \beta \gamma a}= {G}^{\alpha \beta \gamma a}[\Lambda(t)]^{1/3}h^{-4/3} 
F_1^{-3}\nonumber\\
&&{\bf G}^{mnpa}= {G}^{mnpa}[\Lambda(t)]^{1/3}h^{-4/3}F_2^{-3}\nonumber\\
&&{\bf G}^{mnpq}= {G}^{mnpq}[\Lambda(t)]^{4/3}h^{-4/3}F_2 ^{-4} \nonumber\\
&&{\bf G}^{\alpha \beta ab}= {G}^{\alpha\beta ab}[\Lambda(t)]^{-2/3}h^{-4/3}F_1 ^{-2} \nonumber\\
&&{\bf G}^{mnab}= {G}^{mnab}[\Lambda(t)]^{-2/3}h^{-4/3}F_2^{-2}\nonumber\\
&&{\bf G}^{mnp\alpha}= {G}^{mnp\alpha}[\Lambda(t)]^{4/3}h^{-4/3}F_2^{-3} F_1^{-1}\nonumber\\
&&{\bf G}^{mn\alpha a}= {G}^{mn\alpha a}[\Lambda(t)]^{1/3}h^{-4/3}F_2^{-2} F_1^{-1} \nonumber\\
&&{\bf G}^{m\alpha \beta a}= {G}^{m\alpha \beta a}[\Lambda(t)]^{1/3}h^{-4/3}F_1^{-2} F_2^{-1} \nonumber\\
&&{\bf G}^{mn\alpha \beta}= {G}^{mn\alpha \beta}[\Lambda(t)]^{4/3}h^{-4/3}F_2^{-2} F_1^{-2}\nonumber\\
&&{\bf G}^{m\alpha \beta \gamma}= 
{G}^{m\alpha \beta \gamma}[\Lambda(t)]^{4/3}h^{-4/3}F_2^{-1} F_1^{-3}\nonumber\\
&&{\bf G}^{m\alpha ab}= {G}^{mnab}[\Lambda(t)]^{-2/3}h^{-4/3}F_1^{-1} F_2^{-1}, \nd
where the division of the coordinates follow the prescription \eqref{melisett} namely, ($m, n, p$) denote coordinates of ${\cal M}_4$; ($\alpha, \beta$) denote coordinates of ${\cal M}_2$; ($a, b$) denote coordinates of  $\mathbb{T}^2/ {\cal G}$; and ($\mu, \nu$) denote coordinates of the $2+1$ dimensional space-time. It should be clear from \eqref{choshma} that the flux components with all upper indices, i.e 
$G^{MNPQ}(y, t)$ are functions of ($y^m, t$) and may be  got from \eqref{frostgiant} by raising the indices using the un-warped metric components $g_{\alpha\beta}(y), g_{mn}(y)$ and $g_{ab}(y)$ from 
\eqref{vegamey}. Additionally we can also switch on flux components with at most two legs along the space-time directions. These may be tabulated as:
\bg\label{mortaleng}
&&{\bf G}^{\mu \nu ab}= {G}^{\mu\nu ab}[\Lambda(t)]^{4/3}
h^{2/3}\nonumber\\
&&{\bf G}^{\mu \nu \alpha a}= {G}^{\mu\nu \alpha a}[\Lambda(t)]^{7/3}h^{2/3}F_1^{-1}\nonumber\\
&&{\bf G}^{\mu \alpha ab}= {G}^{\mu \alpha ab}[\Lambda(t)]^{1/3}
h^{-1/3}F_1^{-1}\nonumber\\
&&{\bf G}^{\mu \nu ma}= {G}^{\mu\nu mn}[\Lambda(t)]^{7/3}h^{2/3}F_2^{-1}\nonumber\\
&&{\bf G}^{\mu \nu \alpha\beta}= {G}^{\mu\nu \alpha\beta}[\Lambda(t)]^{10/3}
h^{2/3}F_1^{-2}\nonumber\\
&&{\bf G}^{\mu \alpha\beta\gamma}= {G}^{\mu \alpha\beta\gamma}[\Lambda(t)]^{7/3}h^{-1/3}F_1^{-3}
\nonumber\\
&&{\bf G}^{\mu \alpha\beta a}= {G}^{\mu \alpha\beta a}[\Lambda(t)]^{4/3}h^{-1/3}F_1^{-2}\nonumber\\
&&{\bf G}^{\mu mab}= {G}^{\mu mab}[\Lambda(t)]^{1/3}
h^{-1/3}F_2^{-1}\nonumber\\
&&{\bf G}^{\mu \nu mn}= {G}^{\mu\nu mn}[\Lambda(t)]^{10/3}h^{2/3}F_2^{-2}\nonumber\\
&&{\bf G}^{\mu mna}= {G}^{\mu mna}[\Lambda(t)]^{4/3}h^{-1/3}F_2^{-2}\nonumber\\
&&{\bf G}^{\mu mnp}= {G}^{\mu mnp}[\Lambda(t)]^{7/3}h^{-1/3}F_2^{-3}\nonumber\\
&&{\bf G}^{\mu \nu m\alpha}= {G}^{\mu\nu m\alpha}[\Lambda(t)]^{10/3}h^{2/3}F_2^{-1}F_1^{-1}\nonumber\\
&&{\bf G}^{\mu m\alpha a}={G}^{\mu m\alpha a}[\Lambda(t)]^{4/3}h^{-1/3}F_1^{-1}F_2^{-1}\nonumber\\
&&{\bf G}^{\mu mn\alpha}= {G}^{\mu mn\alpha}[\Lambda(t)]^{7/3}h^{-1/3}F_2^{-2} F_1^{-1}\nonumber\\
&&{\bf G}^{\mu m\alpha\beta}= {G}^{\mu m\alpha\beta}[\Lambda(t)]^{7/3}h^{-1/3}F_2^{-1} F_1^{-2}.
\nd
Fortunately we will not be required to keep all the flux components in our computations. 
Some of the G-flux components, such as \eqref{starkhouse}, \eqref{tagrach} and \eqref{redvi}, have to be put  to zero to keep the type IIB solution \eqref{pyncmey} as it is (otherwise cross-terms may develop). However since we saw in section \ref{difuli} that dipole deformations do not change any physics, components like 
${\bf G}_{MNab}$ should now be considered together\footnote{This is more subtle than it appears from first sight.  What configurations of ${\bf G}_{MNab}$ can be allowed here will become clearer as we move along.}. 
Additionally, we do not want to break the de Sitter like isometries apparent from our metric \eqref{pyncmey}, so as a first exercise we put to zero G-flux components with at most two legs in the space-time directions. After the dust settles, the components relevant for us are:
\bg\label{jayanti}
&& {\bf G}_{012m}, ~~~ {\bf G}_{012\alpha}, ~~~{\bf G}_{mnpa}, ~~~ {\bf G}_{mn\alpha a}, ~~~ 
{\bf G}_{mnab} \nonumber\\
&& {\bf G}_{m\alpha \beta a}, ~~~ {\bf G}_{mnpq}, ~~~ {\bf G}_{mnp\alpha}, ~~~ 
{\bf G}_{mn\alpha \beta},~~~ {\bf G}_{\alpha\beta ab}, ~~~ {\bf G}_{m\alpha a b},  \nd
whose upper indices may be extracted from \eqref{choshma}. Of course once a specific solution is constructed using the flux components \eqref{jayanti}, the freedom to construct new solutions by making dipole type deformations clearly exists. None of these new solutions constructed this way violate any of the no go conditions provided the existence of the original solutions is guaranteed. The latter however is an important requirement and in the following sections we will try to see if there is any possibility that the quantum corrections and the classical equations of motion conspire to generate solutions.

\subsection{Perturbative and non-perturbative quantum corrections \label{pertu}}

We have been a bit sloppy in describing the time-dependent warp-factors $F_1(t)$ and $F_2(t)$ in 
\eqref{jenlop} and \eqref{bobby} respectively, so it is now time to revisit them.  There are a two cases to consider with time-independent Newton's constant. First one is with vanishing $c_{00}$ for $F_2(t)$ in 
\eqref{bobby}. For this case $F_1(t)$ becomes:
\bg\label{masshley}
{1\over F_1(t)} = \sum c_{kn} c_{k'n'} \left({g_s^2\over \sqrt{h}}\right)^{\Delta k+ \Delta k'-1} 
{\rm exp}\left[- {(n+n')h^{\Delta/4}\over g^\Delta_s}\right], \nd
where ($k, k'$) $=$ (${\mathbb{Z}\over 2}, {\mathbb{Z}\over 2}$) and 
($n, n'$) $=$ (${\mathbb{Z}}, {\mathbb{Z}}$), and we see that we can equate the inverse of the RHS to the 
perturbative series \eqref{jenlop} because of the following limit:
\bg\label{sammuk}
\lim_{g_s\to 0}~ {1\over g_s^{2n\Delta}} {\rm exp}\left(-{1\over g^\Delta_s}\right) = 0, \nd
for any finite value of $n$, implying that for small $g_s$, both $F_1(t)$ and $F_2(t)$ may be expressed as  perturbative series. The difference however is that $F_2(t)$ does not have a time-independent piece whereas $F_1(t)$ does have a time-independent piece for $k = k' = {1\over 2}$. 

The second case is when we consider non-zero $c_{00}$, and we take $c_{00} = 1$ without loss of generalities. Clearly $F_2(t)$ now has a time-independent piece, but now $F_1(t)$ takes the following form:
\bg\label{hangup}
F_1(t) = {g_s^2\over \sqrt{h}} - 2 \sum_{k, n >  0} c_{kn} \left({g_s^{2}\over \sqrt{h}}\right)^{\Delta k+1} 
{\rm exp}\left(-{~n h^{\Delta/4} \over g^\Delta_s}\right) +  {\cal O}\left(g_s^{4\Delta k + 4}
e^{-2nh^{\Delta/4}/g^\Delta_s}\right), \nd
where the higher order terms appearing from going beyond quadratic orders for the series sum. We see that \eqref{hangup} do not have a time-independent piece, and in fact this could be equated to the perturbative $b_{nk}$ coefficients in \eqref{bobby} as alluded to earlier. 

Thus it appears that, demanding the fluctuation condition \eqref{ranjhita}, allows both $F_1(t)$ and $F_2(t)$ to have a perturbative series but selectively precludes a time-neutral piece in one over the other. This case may be rectified if the demand like \eqref{ranjhita} on Newton's constant is eliminated, wherein the perturbative series for both $F_1(t)$ and $F_2(t)$ may now be unconstrained. For the time being we will 
take $c_{00} = 1$ in the definition of $F_2(t)$, implying the following relations for the time derivatives of 
$F_1(t)$:
\bg\label{olsen}
&&{\dot F}_1 = {2g_s\over h^{1/4} F_2^2}\left({\Lambda}^{1/2} - {g_s\over h^{1/4}} 
.{\partial \over \partial t} {\rm log}~F_2\right) ~\propto ~ g_s\Big(1 + {\cal O}(g^\Delta_s)\Big)  \nonumber\\
&&{\ddot F}_1 = {2\Lambda\over F_2^2} - {4g_s {\Lambda^{1/2}}\over h^{1/4}F_2^3} 
-{4g_s \Lambda^{1/2} \dot{F}_2 \over h^{1/4} F_2^3} - {2g_s^2 \ddot{F}_2 \over h^{1/2} F_2^3} + 
{6g_s^2 \dot{F}^2_2 \over h^{1/2} F_2^4} ~ \propto ~ 1 + {\cal O}(g^\Delta_s), \nd
showing that both $\dot{F}_1$ as well as $\ddot{F}_1$ have perturbative expansions in powers of 
$g_s$ because $1/F_2^n$ has perturbative expansion in terms of $g_s$ for all values of $n$. However 
$1/F_1^n$ does not have any perturbative expansion in terms of $g_s$ for $g_s \to 0$, but could have once 
accompanied by other factors that go as positive powers of $g_s$. For example the power of $g_s$ that appears from a generic combination of $F_i(t)$ and their time derivatives may be written as:
\bg\label{evgreen} 
{g_s^m F_2^r \dot{F}^n_1 \dot{F}^p_2 \ddot{F}^l_1 \ddot{F}^q_2 \over F_1^k} ~ \sim ~ 
g_s^{m+n -2k} \Big(1 + {\cal O}(g^\Delta_s)\Big), \nd
where we only isolate the $g_s$ factor but do not show the perturbative series in the bracket. The latter could be easily ascertained from \eqref{scarjo} and \eqref{olsen}. The above analysis shows that as long as 
\bg\label{paradis}
k \le {m+n\over 2}, \nd
any series containing terms like \eqref{olsen} will have a perturbative $g_s$ expansion in the type IIA side. 
Our analysis also shows the irrelevancy of the other powers controlled by $r, p, l$ and $q$ as they are 
always proportional to $1 + {\cal O}(g^\Delta_s)$ and therefore already perturbative. 

\subsubsection{Product of G-fluxes and $g_s$ expansions \label{Gng}}

Let us now come to the other set of quantum corrections that contribute to the energy-momentum tensor, namely the ones that were written as \eqref{londry} involving the time-neutral series 
$\mathbb{C}_{MN}^{(i)}$. This is where we encounter more subtleties.  Let us illustrate this with an example.
Consider the following series:
\bg\label{kuttnerm}
\mathbb{Q}_1 \equiv \sum_k c_k \left({{\bf G}^{mnpq} {\bf G}_{mn}^{~~~ab} {\bf G}_{abpq}
\over M_p^3}\right)^k, \nd
where $c_k$ are numerical constants, ${\bf G}_{MNPQ}$ are the {\it warped} G-fluxes 
 and $M_p$ is the Planck scale in M-theory. This is an infinite series and clearly every term is time-neutral if we take the type IIB metric to be \eqref{betta}, or its M-theory uplift, as shown in \cite{nodS}.  Question is:  what happens now once we take the metric to be \eqref{vegamey}, supported by the warped G-fluxes of the form  \eqref{frostgiant} whose components may be separated into un-warped pieces as in \eqref{choshma} and  \eqref{mortaleng}? Plugging the flux and the metric ansatze \eqref{frostgiant} and \eqref{vegamey} respectively in \eqref{kuttnerm}, we get:
 
 {\footnotesize  
 \bg\label{valplant} 
\mathbb{Q}_1 = \sum_k c_k \left[\sum_{\{u_i\} \ge 0} {\left({\cal G}^{(u_1, u_2)}\right)^{mnpq} 
\left({\cal G}^{(u_3, u_4)}\right)_{mn}^{~~~~ab}\left({\cal G}^{(u_5, u_6)}\right)_{abpq} \over M_p^3 F_2^4 h^2}
\left({g_s^2\over \sqrt{h}}\right)^{\zeta^s \Delta u_{2s-1}} {\rm exp}
\left(-{\zeta^r u_{2r}h^{\Delta/4}\over g^\Delta_s}\right)
\right]^k, \nonumber\\ \nd}
where the indices are raised and lowered by the un-warped metric with ($m, n$) being the coordinates of 
${\cal M}_4$ and ($a, b$) being the coordinates of $\mathbb{T}^2/{\cal G}$. We have also used $\zeta^s$ to denote the sum with both $u_{2s-1}$ as well as $u_{2s}$ with:
\bg\label{fryesis}
\zeta^1 = \zeta^2 = \zeta^3 = 1, ~~~~ \zeta^0 = \zeta^k = 0 ~~ \forall ~k \ge 4, \nd
such that depending on the value of $u_i$ the series \eqref{valplant} may or may not have a time-neutral piece. (The repeated indices are summed over.) From the way we constructed the series, 
it should be clear that
$u_{2s-1} \in {\mathbb{Z}\over 2}$ and $u_{2s} \in \mathbb{Z}$, implying that if these parameters start from zero as denoted in \eqref{valplant}, $\mathbb{Q}_1$ will take the form:
\bg\label{calvani}
\mathbb{Q}_1 = \sum_k c_k \left[{\left({\cal G}^{(0, 0)}\right)^{mnpq} 
\left({\cal G}^{(0, 0)}\right)_{mn}^{~~~~ab}\left({\cal G}^{(0, 0)}\right)_{abpq} \over h^2 M_p^3} + 
{\cal O}(g^\Delta_s, e^{-1/g^\Delta_s})\right]^k, \nd
with the $g_s$ independent term will be the time-neutral piece exactly as we had in \cite{nodS}. Presence of such a term will create the same hierarchy problem that we encountered in \cite{nogo, nodS}, so our attempt here would be to somehow eliminate such a term. This is easily achieved by imposing, as a first trial:
\bg\label{prisc}
{\cal G}^{(0, 0)}_{MNPQ}(y) = 0, \nd
which in turn will eliminate all time-neutral pieces that have ${\bf G}_{MNPQ}$ in them (we will however see that this can be modified). The puzzle however is that the condition \eqref{prisc} does not preclude terms that were not originally time neutral with the IIB metric \eqref{betta}, but could now become time-neutral if one chooses the IIB metric \eqref{pyncmey} or it's M-theory uplift \eqref{vegamey}. To see whether this could happen then calls for a more careful analysis. 

To begin, let us first concentrate on quantum series constructed exclusively from product of G-fluxes with no extra derivatives. The G-flux may be represented from \eqref{frostgiant}, by including the condition 
\eqref{prisc}, in the following way:
\bg\label{hh4u} 
{\bf G}_{MNPQ} &= & g_s^{2\Delta k} \left[ {\cal G}_1(y) + G_1(y, g^\Delta_s)\right]_{MNPQ} 
+ e^{-1/g^\Delta_s}
\left[{\cal G}_2(y) + G_2(y, e^{-1/g^\Delta_s})\right]_{MNPQ} \nonumber\\
&+&  g^{2\Delta k}_s e^{-1/g^\Delta_s}
\left[{\cal G}_3 + G_3\left(y, g^\Delta_s, e^{-1/g^\Delta_s}\right)\right]_{MNPQ}, \nd
where $k \in {\mathbb{Z}\over 2}$; and 
$G_i(y, g^\Delta_s, e^{-1/g^\Delta_s})$ and  ${\cal G}_i(y)$ for $i = 1,..., 3$ may be read up from ${\cal G}^{(q, n)}$ appearing in \eqref{frostgiant} with or without including the $g_s$ pieces respectively. Note that, compared to \eqref{frostgiant}, the smallest power of $g_s$ for the G-flux is $2\Delta k$ whose range of values will be ascertained below\footnote{An erroneous way to proceed would be to expand 
${\rm exp}\left(-{1\over g^\Delta_s}\right)$ as powers of $1/g^\Delta_s$ to extract $g_s^{2\Delta k}$ from the series with $k \in {\mathbb{Z}\over 2}$,
Such an expansion is not valid at any stage of the expansion in the $g_s << 1$ limit that we are working on. \label{error}}. 
Clearly, once we pull out $g_s^{2\Delta k}$, the series still has a perturbative expansion thanks to the weak coupling limit \eqref{sammuk}.

With this we are now ready to write terms made exclusively with product of G-fluxes. We require two kinds of terms: one, with no free Lorentz indices, and two, with two free Lorentz indices. The one with no free Lorentz indices may be expressed as\footnote{One subtlety that we should keep track of is the fact that the G-fluxes are anti-symmetric whereas the metric components are symmetric in their respective indices.}:
\bg\label{ronan}
{\bf g}^{MM'} {\bf g}^{NN'}......{\bf g}^{DD'} {\bf G}_{MQPR} {\bf G}_{NUHG}..... {\bf G}_{ABCD}  
\equiv \left[{\bf g}^{-1}\right]^{2m} \left[{\bf G}\right]^m, \nd 
where $m$ is the number of G-flux components and  ${\bf g}_{MN}$ is the warped M-theory metric components. The indices $M, N, ..$ cover the coordinates of the eight dimensional internal space
\eqref{melisett}, and the RHS of \eqref{ronan} is the shortened way of expressing the product of the G-fluxes contracted by the metric indices. The power of the inverse metric is ascertained from the fact that the $4m$ components of the G-flux may be completely contracted by $2m$ inverse metric components. These $2m$ inverse metric components may be divided into $l_1$ inverse metric components from 
$\mathbb{T}^2/{\cal G}$; $l_2$ metric components from ${\cal M}_2$ and $l_3$ metric components from 
${\cal M}_4$ of the internal space \eqref{melisett}. Using this, the leading order $g_s$ dependence of 
\eqref{ronan} may be written as:
\bg\label{beanie}
\left[{\bf g}^{-1}\right]^{2m} \left[{\bf G}\right]^m ~ \sim ~ g_s^{2\Delta km - 2\left(2l_1 + 2l_2 - l_3\right)/3}
\left(1 + {\cal O}\left(g_s, e^{-1/g_s}\right)\right), \nd
where we have used the perturbative series for $F_1(t)$ and $F_2(t)$ given in \eqref{hangup} and 
\eqref{bobby} respectively to express their $g_s$ dependences. At this stage it is useful to note that the 
sum of the ($l_1, l_2, l_3$) factors should be equal to $2m$, i.e $l_1 + l_2 + l_3 = 2m$ so that \eqref{ronan} remains Lorentz invariant. This reproduces our first condition:
\bg\label{greta}
\left({6\Delta k - 8\over 3}\right)m + 2 l_3 \ge 0,  \nd
with the equality leading to the time-neutral case. 
Clearly for $\Delta k \ge {3\over 2}$ there is no constraint as $l_3 \ge 0$. In fact if $m > 1$, $l_3$ must satisfy $l_3 > 1$, otherwise it will be difficult to have Lorentz invariant terms. For $\Delta k \ge {1\over 2}$, we will at least require 
$l_3 \ge {5m\over 6}$, which means for $m = 3$ we require $l_3 = 4$. This is of course consistent with the simplest case \eqref{kuttnerm}. Thus for ${1\over 2}  \le \Delta k < {3\over 2}$ we can avoid the time-neutral series by constraining $l_3$. However if $\Delta k \ge {3\over 2}$, there seems to be no time-neutral series that can appear from any combinations of pure G-fluxes.  

Similarly for the case with two free Lorentz indices with $m$ G-flux components we now require 
$2m-1$ number of inverse metric components. The reasoning for this is simple to state. The generic energy-momentum tensor, for either G-fluxes $G$ or quantum terms $Q$, may be written as:
\bg\label{ryacone}
\mathbb{T}_{MN}^{(G, Q)} \equiv -{2\over \sqrt{{\bf g}_{11}}} {\delta S_{\rm eff} \over \delta {\bf g}^{MN}}, \nd
where $S_{\rm eff}$ is the effective action at any given scale. Such a procedure either {\it removes} an inverse metric component or {\it adds} an inverse-of-an-inverse metric component. In either case, the number of inverse metric components reduces by one.  
The $g_s$ expansion then remains similar to the 
RHS of \eqref{beanie} but $l_i$ satisfy $l_1 + l_2 + l_3 = 2m -1$. This gives rise to the following constraint:
\bg\label{foxvelvet}  
\left({6\Delta k - 8\over 3}\right)m + {4\over 3} + 2l_3 \ge 0, \nd
which may be compared to \eqref{greta}. For $\Delta k = {1\over 2}$, $l_3$ should at least satisfy $l_3 \ge {5m - 4 \over 6}$, implying that for $m = 3$, $l_3 \ge 2$. In general $l_3 \ge 1$ even for $m = 1$, although with $m = 1$ there 
doesn't appear any simple time-neutral term possible. Again we see that if $\Delta k \ge {3\over 2}$, there is no constraint on $l_3$, and it appears impossible to construct time-neutral series with two free Lorentz indices. Interestingly for certain G-flux components,
we will see that this is not always an essential condition and thus may be relaxed.

We can also discuss the case when $F_1(t)$ and $F_2(t)$ have inverses that are perturbatively expandable  as powers of $g_s$. Clearly for such a case, \eqref{ranjhita} cannot be satisfied and therefore the Newton's constant has to be defined using \eqref{olokhi}.  Nevertheless, one may see that the quantum terms with zero and two free Lorentz indices with only G-fluxes go as $g_s^{k_1}$ and $g_s^{k_2}$ respectively, where $k_1$ and $k_2$ are bounded by the following inequalities:
\bg\label{marilynm}
&&k_1 \equiv \left({6\Delta k + 4\over 3}\right)m - 2l_1 \ge 0 \nonumber\\
&& k_2 \equiv \left({6\Delta k + 4\over 3}\right)m - {2\over 3} - 2l_1 \ge 0, \nd
where we see that the constraints on $l_1$ are stronger than what we had for $l_3$ in \eqref{greta} and 
\eqref{foxvelvet} above. However since $l_1$ captures the metric for the toroidal fibre 
$\mathbb{T}^2/{\cal G}$, we expect $l_1$ to be small and satisfy the inequalities 
\eqref{marilynm}. In fact since $l_1 < 2m$, so if $\Delta k \ge {3\over 2}$ both the inequalities in 
\eqref{marilynm} are easily satisfied. Interestingly when $k = 0$, if we take $m = 3p$ for the scenario with zero Lorentz indices and $m = 3q + 2$ with two free Lorentz indices, we have:
\bg\label{hkapilla}
&&l_1 = 2p, ~~~ l_2 + l_3 = 4p, ~~~ m = 3p \nonumber\\
&&l_1 = 2q+1, ~~~ l_2 + l_3 = 4q + 2, ~~~ m = 3q + 2. \nd
where the combination $l_2 + l_3$ appears because ${\cal M}_6$ is not sub-divided into ${\cal M}_2$ and 
${\cal M}_4$. 
Thus we see that for ($p, q$) $\in$ ($\mathbb{Z}, \mathbb{Z}$) there are infinite possible solutions all giving rise to time-neutral series of the form \eqref{kuttnerm}\footnote{The example in \eqref{kuttnerm} is made of 
$m = 3$ so $p = 1$. Therefore $l_1 = 2, l_2 + l_3 = 4$ with zero free Lorentz indices.}. This justifies the claims made in \cite{nodS} regarding a class of time-neutral quantum series.

\subsubsection{G-fluxes with multiple derivatives \label{Gng2}}

Let us now consider the case where there are derivatives along with G-fluxes, all contracted in two possible ways: one with zero Lorentz indices and two, with two free Lorentz indices.  To illustrate this case, let us start with a simple example from \cite{nodS} that has no free Lorentz indices:
\bg\label{laurab}
\mathbb{Q}_2 \equiv \sum_k b_k \left({\square^2{\bf G}_{mnab} {\bf G}^{mnab} \over M_p^6}\right)^k, \nd
where $\square$ is the covariant derivative defined on the six-dimensional base 
${\cal M}_2 \times {\cal M}_4$ with the warped metric. With time-independent G-flux, and without any 
$F_i(t)$ factors in the metric, \eqref{laurab} is clearly time-neutral because every term in \eqref{laurab} is time-neutral. But now,
taking the G-flux as in \eqref{hh4u}, with ($m, n$) being the coordinates of ${\cal M}_4$, $\mathbb{Q}_2$ 
yields:

{\footnotesize
\bg\label{rupimey}
\mathbb{Q}_2 = \sum_k b_k\left[\sum_{\{u_i\} \ge 0}{\square^2\left({\cal G}^{(u_1, u_2)}\right)_{mnab}
\left({\cal G}^{(u_3, u_4)}\right)^{mnab}\over F_2^4 h^2 M_p^6}
\left({g_s^2\over \sqrt{h}}\right)^{\Delta(u_1+u_3)}
{\rm exp}\left(-{(u_2 + u_4)h^{\Delta/4}\over g^\Delta_s}\right)\right]^k,  \nd}
where the $g_s$ independent piece will lead to the same issue that we faced in \cite{nodS}, which in turn may be alleviated by imposing \eqref{prisc} as before. However the issue plaguing earlier, namely the possibility of generating {\it new} time-neutral series, now requires a careful assessment of terms of the form 
\eqref{laurab} taking the $g_s$ dependent G-flux \eqref{hh4u} into account. Therefore, the kind of term that we want to consider will be of the form:

{\footnotesize
\bg\label{leamornar}
{\bf g}^{MM'} {\bf g}^{M_1M_1'}...{\bf g}^{DD'} \partial_{M_1}\partial_{M_2} ... \partial_{M_n}
\left({\bf G}_{MQPR} {\bf G}_{NUHG}... {\bf G}_{ABCD}\right)  
\equiv \left[{\bf g}^{-1}\right]^{2m + {n\over 2}} \left[{\bf \partial}\right]^n\left[{\bf G}\right]^m, \nd}
where the RHS is a shortened symbolic expression for the derivative expressions. Clearly with only four derivative, contracted appropriately, will reproduce the terms in the series \eqref{laurab}. Interestingly the form of the $g_s$ expansion is exactly similar to the expression on the RHS of \eqref{beanie} i.e 
$g_s^{k_3}$, except now 
$l_i$ satisfy $l_1 + l_2 + l_3 = 2m + {n\over 2}$. This implies:
\bg\label{agathaC}
\vert k_3\vert \equiv \left\vert \left({6\Delta k - 8\over 3}\right)m - {2n\over 3} + 2l_3\right\vert \ge 0 \nd
where the equality would lead to the time-neutral series. On the other hand, since $n$ appears with a relative {\it minus} sign, sufficiently large $n$ will reverse the power of $k_3$ in $g^{k_3}_s$ and make it negative. Such 
a scenario should make sense if all the inverse powers of $g_s$ can be rearranged as:
\bg\label{kanna}
\sum_k {\alpha_k h^{\Delta k/4}\over g_s^{2\Delta k}} = 
\sum_l \beta_l  ~{\rm exp}\left(-{n_l h^{\Delta/4}\over g^\Delta_s}\right), \nd
with the integer $\alpha_k$ being related to the integers ($\beta_l. n_l$). The equality \eqref{kanna} is the 
consequence of summing the series in appropriate way, and should in principle be possible if non-perturbatively the series has to make sense\footnote{In other words at every order in $k$, terms on the LHS of \eqref{kanna} blow-up, yet the sum on the RHS remains perfectly finite. Thus the representation on the LHS is never the right way to study inverse $g_s$ expansion near $g_s \to 0$. The correct expression will always be the RHS of \eqref{kanna}.}. 
Assuming this to be the case, the puzzle however is more acute. What happens if we take a particular value of $n$ for a given $m$, i.e $n$ number of derivatives, such that 
$k_3$ vanishes? In fact all we require is for $n$ to take the following value:
\bg\label{duimey}
n = 3l_3 + \left(3\Delta k - 4\right)m, \nd 
to create a new class of time-neutral series with $m$ G-fluxes and $n$ derivatives. One might rewrite 
\eqref{kanna} in a slightly different way that puts the relative minus sign elsewhere as:
\bg\label{bhalosam}
\left({6\Delta k + 4\over 3}\right)m + {n\over 3} - 2\left(l_1 + l_2\right) \ge 0, \nd
which simply transfers the puzzle now on the values of $l_1$ and $l_2$ instead of on the number of derivatives. This doesn't appear to alleviate the issue because increasing $n$ also increases the metric components. However since $l_1$ and $l_2$ denote the metric components along 
$\mathbb{T}^2/{\cal G}$ and ${\cal M}_2$ respectively, and if we assume that the G-flux components are functions of the base ${\cal M}_4$ {\it only}, then increasing the number of derivatives will simply increase $l_3$ without changing $l_1$ and $l_2$! This way the constraint \eqref{bhalosam} may be easily satisfied without invoking any extra constraint on $k$. In fact even if we allow for two free Lorentz indices, the change 
from \eqref{bhalosam} is minimal:
\bg\label{bhalosammie}
\left({6\Delta k + 4\over 3}\right)m + {n\over 3} - {2\over 3} - 2\left(l_1 + l_2\right) \ge 0, \nd
since $n \ge 2$ in most cases.
Thus again with more derivatives, there would be no constraint on $k$. For small number of derivatives, 
we expect $l_1 + l_2 < 2m$. Therefore for $\Delta k \ge {3\over 2}$, $\left({6\Delta k + 4\over 3}\right)m > 4m$ implying that this 
would dominate over the term $-2(l_1 + l_2)$ making the LHS of both \eqref{bhalosam} as well as 
\eqref{bhalosammie} always positive definite. This brings us to similar conclusion that we had earlier, namely with $\Delta k \ge {3\over 2}$, arbitrary flux products with arbitrary number of derivatives do not lead to time-neutral    
series provided the G-fluxes are functions of the coordinates of the ${\cal M}_4$ base only. For $F_1$ and 
$F_2$ satisfying \eqref{olokhi} instead of \eqref{ranjhita}, the constraint equations for zero and two free Lorentz indices become respectively:
\bg\label{evabb} 
&&\left({6\Delta k + 4\over 3}\right)m + {n\over 3} - 2l_1 \ge 0 \nonumber\\
&&\left({6\Delta k + 4\over 3}\right)m + {n\over 3} -{2\over 3} - 2l_1 \ge 0, \nd
which are readily satisfied by imposing similar conditions on the G-fluxes and on $k$, because increasing 
$n$ does not affect $l_1$ and so $\Delta k \ge {3\over 2}$ still controls the positivity of the LHS of both the inequalities 
in \eqref{evabb}. We will however soon see that the condition can be relaxed. Again for $k = 0$, we expect the following two cases:
\bg\label{gaadka}
&&m = 3p_1 + p_2, ~~~ n = 2p_2, ~~~ l_1 = 2p_1 + p_2, ~~~ l_2 + l_3 = 4p_1 + 2p_2 \nonumber\\
&&m = 3q_1 + q_2 + 2, ~~~ n = 2q_2, ~~~ l_1 = 2q_1 + q_2 + 1, ~~~ l_2 + l_3 = 4q_1 + 2q_2 + 2, \nd
with zero and two free Lorentz indices respectively. Clearly since we expect 
($p_i, q_i$) $\in$ ($\mathbb{Z}, \mathbb{Z}$), there are infinitely many possible solutions each of which leading to a series like \eqref{rupimey}, and therefore justifying another class of time-neutral quantum series advertised in \cite{nodS}\footnote{In fact the term in \eqref{rupimey} is for $m = 2, n = 4$, therefore $p_1 = 0, p_2 = 2, l_1 = 2, l_2 + l_3 = 4$ with zero free Lorentz indices.}.

\subsubsection{Curvature algebra and product of curvatures \label{Gng3}}

Our next set of exercises will be to take quantum pieces with products of curvatures and curvature polynomials. In standard GR, curvatures may be represented by Riemann tensor, Ricci tensor and Ricci 
scalar. Since now multiple components will occur simultaneously, we will have to tread carefully. To simplify the ensuing analysis we will develop a curvature algebra which will also help us to facilitate computations.

One of the main element that governs all the curvature tensors is of course the metric of the internal space. For us, all we need is to actually see how everything scales with respect to $g_s$.  In view of that it will be easier to express everything as powers of $g_s$. For example, we can write the metric components as:
\bg\label{strawjane}
[{\bf g}] \equiv {\bf g}_{MN} &=& \left(g_s^{4/3} g_{ab}, g_s^{4/3} g_{\alpha\beta}, g_s^{-2/3} g_{mn}\right) 
\otimes \left(1 + {\cal O}(g^\Delta_s, e^{-1/g^\Delta_s})\right) \nonumber\\
&\equiv& \left(g_s^{4/3}, g_s^{4/3}, g_s^{-2/3}\right) \otimes \left(1 + {\cal O}(g^\Delta_s, 
e^{-1/g^\Delta_s})\right)_{MN} 
\rightarrow \left(g_s^{4/3}, g_s^{-2/3}\right), \nd
where the RHS of the second line of \eqref{strawjane} tells us how the terms in the metric scale as powers of 
$g_s$ as ${\cal O}(g^\Delta_s, e^{-1/g^\Delta_s})$ corrections are irrelevant to the analysis that we want to perform here. This means, in the same vein, we can express the Christoffel symbol in the following way:
\bg\label{tavader}
\Gamma^M_{NP} \equiv [{\bf g}^{-1}] \partial [{\bf g}] & = & \left[\left(g_s^{-4/3}, g_s^{2/3}\right) \times 
\left(g_s^{4/3}, g_s^{-2/3}\right)\right] \otimes \left(1 + {\cal O}(\partial, g^\Delta_s, 
e^{-1/g^\Delta_s})\right)^M_{NP} \nonumber\\
& = & \left(1, g_s^{-2}, g_s^2\right) \otimes \left(1 + {\cal O}(\partial, g^\Delta_s, 
e^{-1/g^\Delta_s})\right)^M_{NP} 
\rightarrow  \left(1, g_s^{-2}, g_s^2\right), \nd
where again the extreme RHS of the second line denotes the overall scaling of the terms of the Christoffel 
symbol.  Note that the derivative action in the definition of the Christoffel symbol does not act on 
$g_s/\sqrt{h}$ and therefore directly goes in ${\cal O}(\partial, g^\Delta_s, e^{-1/g^\Delta_s})$ implying that it would act on $y^M$ dependent pieces where $y^M$ are in general the coordinates of eight-dimensional internal space in M-theory\footnote{More precisely, defining $h(y) = H^4(y)$, it is easy to infer
that $\partial_0\left({g_s\over H}\right) = \sqrt{\Lambda}$ and 
$\partial_n\left({g_s\over H}\right) = 0$. To avoid clutter, we will ignore the $H(y)$ and simply denote the terms with $g_s$ scalings, unless mentioned otherwise.}.

The identity element in \eqref{tavader} is related to those terms in the Christoffel symbol where the $g_s$ scaling of $[{\bf g}^{-1}]$ cancels with the $g_s$ scaling of $\partial [{\bf g}]$. This happens when we deal with the metric components of the individual sub-spaces of the eight manifold, namely ${\cal M}_2$, ${\cal M}_4$ or $\mathbb{T}^2/{\cal G}$. Similarly the other powers of $g_s$ may also be explained by looking at various contributions to the Christoffel symbol. For us of course only the $g_s$ scaling matters for the time being. 
 
As expected, the Christoffel symbols now combine together to create the curvature tensors, namely the Riemann tensor, Ricci tensor and the Ricci scalar. Our symbolic manipulation should again work for these cases. For example the Riemann tensor with one upper index may be expressed in this language, in the following way:

{\footnotesize
\bg\label{lizplease}
{\bf R}^M_{~~NPQ}  & = & \partial_{[N} \Gamma^M_{P]Q} + \Gamma^M_{[N|S|} \Gamma^S_{P]Q} \\
& \equiv& \left(1, g_s^{-2}, g_s^2\right) \otimes \left(1 + {\cal O}(\partial^2, g^\Delta_s, 
e^{-1/g^\Delta_s})\right)^M_{NPQ} 
+ \left(1, g_s^{-2}, g_s^2, g_s^{-4}, g_s^4 \right) \otimes 
\left(1 + {\cal O}(\partial, g^\Delta_s, e^{-1/g^\Delta_s})\right)^2 \Big\vert^M_{NPQ}, \nonumber \nd} 
where in the first line $|S|$ implies that the index $S$ do not participate in the anti-symmetric operation 
of its neighboring indices (here it is between indices $N$ and $P$). The above form of the Riemann tensor implies that, in terms of $g_s$ scalings we can simply express this as:
\bg\label{chukkam}
{\bf R}^M_{~~NPQ} \equiv \left(1, g_s^2, g_s^{-2}, g_s^4, g_s^{-4}\right), \nd
which is got by combining the exponents of $g_s$ from the two terms without worrying about the 
${\cal O}(g^\Delta_s, e^{-1/g^\Delta_s})$ contributions. Such a shortened form captures the main message and is clearly much more economical to use, but does miss out in distinguishing various components that scale in the same way with $g_s$. This is not an immediate concern, so we will continue with this formalism unless a more sophisticated analysis is called for.
Similarly the Riemann tensor with all lower indices may be expressed as:
\bg\label{haramik}
{\bf R}_{MNPQ} &=& {\bf g}_{ML} {\bf R}^L_{~~{NPQ}} \equiv \left(g_s^{-2/3}, g_s^{4/3}, g_s^{-8/3}, 
g_s^{10/3}, g_s^{-14/3}, g_s^{16/3}\right) \\
&=& \left(g_s^{4/3}, g_s^{-2/3}\right) \otimes \left(1, g_s^2, g_s^{-2}\right) + 
\left(g_s^{4/3}, g_s^{-2/3}\right) \otimes \left(1, g_s^2, g_s^{-2}, g_s^4, g_s^{-4}\right), \nonumber \nd
where the second line shows how the scaling exponents came about by taking products of various terms. 
It is interesting to note that although the Riemann tensor with one upper index has a $g_s$ independent piece, the Riemann tensor with all lower indices do not seem to have any such piece. Additionally a specific component of Riemann tensor, since it is constructed out of derivatives and products of Christoffel symbols, has  at least four terms with leading $g_s$ exponents\footnote{This implies that each of these four terms have a leading $g_s$ exponent followed by higher powers of $g^\Delta_s$ and 
$e^{-1/g^\Delta_s}$.} and therefore may be expressed as:

{\footnotesize
\bg\label{kapilla}
{\bf R}_{MNPQ} = \sum_{i = 1}^4 g_s^{a_i} \left[\mathbb{R}_i(y) 
+  {\cal R}_i(y, g^\Delta_s, e^{-1/g^\Delta_s})\right]_{MNPQ}  = g_s^{a_k} \left[\mathbb{R}_k 
+ {\cal O}(y, g^\Delta_s, e^{-1/g^\Delta_s})
\right]_{MNPQ}, \nd}
where $a_k = {\rm min}(a_1, a_2, a_3, a_4)$ will govern the $g_s$ expansion for the particular Riemann 
tensor. Of course many of the above $g_s$ powers cannot be realized because of the absence of certain cross-terms in the metric. If we ignore these subtleties for the time being, the curvature tensors take the following form:
\bg\label{xjanex}
&&{\bf R}_{MNPQ} = \left(g_s^{-14/3}, g_s^{-8/3}, g_s^{-2/3}, g_s^{4/3}, g_s^{10/3}, g_s^{16/3}\right)
\nonumber\\
&& {\bf R}_{MP} = {\bf g}^{NQ} {\bf R}_{MNPQ} = \left(1, g_s^{-6}, g_s^{-4}, g_s^{-2}, g_s^2, g_s^4, g_s^6
\right)\nonumber\\
&& {\bf R} = {\bf g}^{MP} {\bf R}_{MP} = \left(g_s^{-22/3}, g_s^{-16/3}, g_s^{-10/3}, g_s^{-4/3}, g_s^{2/3}, g_s^{8/3}, g_s^{14/3}, g_s^{20/3}\right).\nd
All the above $g_s$ scalings got using the curvature algebra assume the
generic scenario where the  
metric components are functions of all the coordinates of the four manifold and, as mentioned earlier,  cross-terms exist.  However the former cannot be imposed in the flux sector if we want to avoid time-neutral series with derivatives on fluxes. Extending this to the metric components, we can assume that the un-warped metric components and the warp-factors are all functions of the coordinates 
$y^m$ of ${\cal M}_4$ implying that the curvature polynomials will also be functions of $y^m$. 

The latter condition, i.e the presumption that all metric cross-terms exist, again cannot be realized in our case because of the way we expressed the metric \eqref{vegamey} and the four-manifold \eqref{melisett}. Thus a more careful considerations of the scalings of the various tensor components are called for. Imposing the two constraints: (a) metric components and the curvature tensors are functions of ${\cal M}_4$ only; and (b) only cross-terms satisfying the division \eqref{melisett} are allowed, the various curvature tensors scale in the following way:

{\footnotesize
\bg\label{brooklynns}
&&{\bf R}_{mnpq} = g_s^{-2/3}, ~~~  {\bf R}_{abab} = g_s^{10/3}, ~~~ {\bf R}_{abmn} = {\bf R}_{ambn} = 
g_s^{4/3}, ~~~ {\bf R}_{\alpha a b\beta} = g_s^{10/3} \nonumber\\
&& {\bf R}_{mn \alpha\beta} = g_s^{4/3}, ~~~ {\bf R}_{\alpha \beta \alpha\beta} = g_s^{10/3}, ~~~{\bf R}_{\alpha mnp} = {\bf R}_{\alpha a np} = {\bf R}_{abc \alpha } = {\bf R}_{a mnp} =   
{\bf R}_{a \alpha \beta n} = 0, \nd}
where we do not show the ${\cal O}(g^\Delta_s, e^{-1/g^\Delta_s})$ corrections that accompany all the curvature tensors. Although the above set of tensors and their scalings are considerably simpler than what one would have expected from a generic set-up of \eqref{xjanex}, the generic scalings are nevertheless useful because they do not rely on the way we express the four-manifold. For our case, since we are searching for a specific cosmological solution with a specific internal space geometry, we will stick with \eqref{brooklynns} for now and look for quantum series with zero and two free Lorentz indices. A zero free Lorentz index quantum term now takes the following form:
\bg\label{hkpgkm}
\mathbb{Q}_3 &=& {\bf g}^{m_im'_i}....{\bf g}^{\beta_q\beta'_q}\prod_{\{i\}=1}^{\{l_i\}}{\bf R}_{m_in_ip_iq_i}
{\bf R}_{a_jb_ja_jb_j}{\bf R}_{p_kq_ka_kb_k}{\bf R}_{\alpha_l a_lb_l\beta_l} {\bf R}_{\alpha_p \beta_p m_p n_p}{\bf R}_{\alpha_q\beta_q\alpha_q\beta_q} \nonumber\\
&\equiv&
\left[{\bf g}^{-1}\right]^{L_1 + L_2 + L_3} \prod_{\{i\} = 1}^{\{l_i\}} \left[\mathbb{R}_i\right], \nd
where the set ${\{i\}}$ denotes the set of $i, j, k ...p$ integers that determines the product of all the available Riemann tensors with each set of Riemann tensors (and its various permutations for a given set of indices) occur $l_i, l_j, l_k, ...l_p$ times. The second line is a symbolic way to represent this using inverse metric components. It is clear that:
\bg\label{chukkamukha}
L_1 = 2l_2 + l_3 + l_4, ~~~~ L_2 = 2l_6 + l_4 + l_5, ~~~~ L_3 = 2l_1 + l_3 + l_5, \nd
with the assumption that $l_1, ..., l_6$ occur in the same order in which the curvature tensors appear in  the quantum piece $\mathbb{Q}_3$. 
In other words ${\bf R}_{mnpq}$ occurs $l_1$ times, ${\bf R}_{abab}$ occurs $l_2$ times, and so on\footnote{An underlying assumption is that the Riemann tensors are contracted in appropriate ways so that there is no need to explicitly insert the curvature scalar ${\bf R}$ or the Ricci tensor 
${\bf R}_{MN}$ in the expression \eqref{hkpgkm} for $\mathbb{Q}_3$. This way we can also avoid differentiating between symmetric or anti-symmetric Ricci tensors, namely ${\bf R}_{(MN)}$ or ${\bf R}_{[MN]}$ respectively.}. Similarly, $L_1, L_2$ and $L_3$ denote the number of inverse metric components along $\mathbb{T}^2/{\cal G}, {\cal M}_2$ and 
${\cal M}_4$ respectively\footnote{The inverse metric components that we are using here have components ${\bf g}^{ab}, {\bf g}^{\alpha\beta}$ and ${\bf g}^{mn}$, and in later sections we will use other space-time components like ${\bf g}^{ij}$ and ${\bf g}^{00}$. In this language the symbolic representation of the inverse metric components in \eqref{hkpgkm}, i.e the symbol $\left[{\bf g}^{-1}\right]^{L_1 + L_2 + L_3}$ 
may be expressed in the following way: 
$$\left[{\bf g}^{-1}\right]^{L_1 + L_2 + L_3} \equiv \left({\bf g}^{ab}\right)^{L_1}
\left({\bf g}^{\alpha\beta}\right)^{L_2}\left({\bf g}^{mn}\right)^{L_3}\equiv \prod_{i, j, k}^{L_{\tiny{1,2,3}}}
{\bf g}^{a_i b_i} {\bf g}^{\alpha_j\beta_j}{\bf g}^{m_kn_k}$$
in other words, $\left({\bf g}^{MN}\right)^{L_k}$ is defined as the following product 
$\left({\bf g}^{MN}\right)^{L_k} \equiv \prod_{i = 1}^{L_k} {\bf g}^{M_iN_i}$ where ($M. N$) = ($a, b$), ($\alpha, \beta$) or ($m, n$). More generic representations, that include space-time metrics in addition to the internal space metrics, appear 
in \eqref{ncresp} and in \eqref{tagchink}.}.
Using this formalism, and plugging in the appropriate $g_s$ scalings, it is easy to infer that:
\bg\label{lenaeng}
\mathbb{Q}_3 \equiv \left[{\bf g}^{-1}\right]^{L_1 + L_2 + L_3} \prod_{\{i\} = 1}^{\{l_i\}} \left[\mathbb{R}_i\right] = g_s^{2(l_1 + l_2 + l_3 + l_4 + l_5 + l_6)/3} \left(1 + {\cal O}(g^\Delta_s, e^{-1/g^\Delta_s})\right), \nd
implying that the quantum piece $\mathbb{Q}_3$ can {\it never} be time-neutral. Such a conclusion is interesting in the light of our earlier discussions with G-fluxes. Therein we had to impose some minimal 
$g_s$ scalings for the G-flux components to avoid time-neutral series. Here we see that the curvature terms 
avoid the time-neutrality without any imposition of extra constraints. 
This is good, but one would like to infer what happens when $F_i(t)$ are not constrained by \eqref{ranjhita} but follow \eqref{olokhi}. For such a case the scaling turns out to be:
\bg\label{tlizsq}
\mathbb{Q}'_3  = g_s^{2(l_1 + l_2 + l_3 + l_4 + l_5 + l_6)/3} \left(1 + {\cal O}(g^\Delta_s, 
e^{-1/g^\Delta_s})\right), \nd
which is exactly the same scaling as in \eqref{lenaeng} despite that fact that now the metric components have different $g_s$ scalings. The conclusion then remains the same as above: there can be no time-neutral series with zero Lorentz index with only curvature tensors. 

What happens when we have two free Lorentz indices? The answer here is simple as the only changes that can occur are in the values of $L_1, L_2$ and $L_3$. This is again easy to quantify: if we want free ($a, b$) Lorentz indices, all we need is to take ($L_1', L_2, L_3$) metric components where $L'_1 = L_1 - 1$, with $L_1$ being the value quoted in \eqref{chukkamukha}. Thus generically we need $L_j' = L_j - 1$ with $j$ defining the three possible class of metric choices. Putting everything together, the $g_s$ scaling 
may be expressed as $g_s^{\kappa}$ where $\kappa$ takes the following two values:
\bg\label{eyeswide}
\kappa \equiv {2\over 3} \sum_{i = 1}^6 l_i + {4\over 3}, ~~~~~
\kappa \equiv {2\over 3} \sum_{i = 1}^6 l_i - {2\over 3}, \nd
where the first one corresponds to indices along $\mathbb{T}^2/{\cal G}$ and ${\cal M}_2$ and the second one corresponds to indices along ${\cal M}_4$. Note that since at least one of the $l_i \ge 1$, 
$\kappa \ge 0$ where the strict inequality is for the first case. For the second case there is a possibility for 
$\kappa = 0$ when $l_1 = 1$, implying that the Ricci tensor ${\bf R}_{mn}$ is actually time-neutral with or without $F_i(t)$ being constrained by \eqref{ranjhita} as was also evident from our curvature algebra 
\eqref{xjanex}. This will not be an issue as we will discuss later. 

Let us now elaborate the quantum series with product of curvature tensors and derivatives. As with the 
G-fluxes we will consider the case where the derivatives are only along the ${\cal M}_4$ direction i.e all components of the metric are functions of the internal ${\cal M}_4$ coordinates. The quantum terms now take the form:

{\footnotesize
\bg\label{fidelio}
\mathbb{Q}_4 &=& {\bf g}^{m_im'_i}....{\bf g}^{\beta_q\beta'_q}\partial_{m_r}....\partial_{m_s}
\left(\prod_{\{i\}=1}^{\{l_i\}}{\bf R}_{m_in_ip_iq_i}
{\bf R}_{a_jb_ja_jb_j}{\bf R}_{p_kq_ka_kb_k}{\bf R}_{\alpha_l a_lb_l\beta_l} {\bf R}_{\alpha_p \beta_p m_p n_p}{\bf R}_{\alpha_q\beta_q\alpha_q\beta_q}\right) \nonumber\\
&\equiv&
\left[{\bf g}^{-1}\right]^{L_1 + L_2 + \hat{L}_3} \left[{\bf \partial}\right]^n \prod_{\{i\} = 1}^{\{l_i\}} \left[\mathbb{R}_i\right], \nd}
where $L_1$ and $L_2$ are as given in \eqref{chukkamukha} and $\hat{L}_3 = L_3 + {n\over 2}$ where $n$ is the number of derivatives. It is now easy to derive the following $g_s$ scalings with zero free Lorentz index: 
\bg\label{EWS} 
\mathbb{Q}_4 = g_s^{2(l_1 + l_2 + l_3 + l_4 + l_5 + l_6 + n/2)/3}\left(1 + {\cal O}(g^\Delta_s, 
e^{-1/g^\Delta_s})\right), \nd
showing that there are no time-neutral series possible with curvature tensors and derivatives without imposing any additional constraints. The above scaling remains unchanged even if $F_i(t)$ satisfy volume preserving condition \eqref{olokhi}.  On the other hand, if we demand two free Lorentz indices, the change is again minimal in the sense that the two $\kappa$ values quoted in \eqref{eyeswide} unequivocally change
by:
\bg\label{stguillaume}
\kappa \rightarrow \kappa + {n\over 3}, \nd
which is always positive because we expect at least one of the $l_i \ge 1$ and $n > 1$. Thus with derivatives there appears no possibilities of having time-neutral series whether or not $F_i(t)$ are constrained by \eqref{ranjhita}.   

\subsubsection{Adding space-time curvatures with derivatives \label{Gng4}}

Another aspect of the curvatures that is going to change our results is the inclusion of space-time curvature contributions. So far we have steered clear of space-time effects, namely fluxes and metric components along the space-time directions, but now it is time to include them in our quantum series. The space-time metric in M-theory scales as ${\bf g}_{\mu\nu} \sim g_s^{-8/3}$ which is different from all the metric scalings in the internal space. The $g_s$ scalings of the curvature tensors with legs along the spatial directions 
are easy to illustrate:
\bg\label{thierry}
{\bf R}_{ijij} = g_s^{-14/3}, ~~~~ {\bf R}_{ijmn}  = g_s^{-8/3}, ~~~~ {\bf R}_{iajb}  = g_s^{-2/3}, ~~~~ 
{\bf R}_{i\alpha j \beta}  = g_s^{-2/3}, \nd
with other spatial components vanishing. Compared to \eqref{brooklynns}, the spatial curvature tensors 
have predominantly negative powers of $g_s$ scalings. 

The curvature tensors with at least one temporal direction is bit more involved because of the time dependences of the various warp-factors creating numerous cross-terms. Nevertheless the $g_s$ scalings can be determined uniquely for each of the curvature tensors. For the present case we have the following tensor components:
\bg\label{tagramee}
&& {\bf R}_{0mnp} = g_s^{-5/3}, ~~~~{\bf R}_{0m0n} = g_s^{-8/3}, ~~~~{\bf R}_{0i0j} = g_s^{-14/3}, ~~~~
{\bf R}_{0a0b} = g_s^{-2/3} \nonumber\\
&& {\bf R}_{0\alpha 0 \beta} = g_s^{-2/3}, ~~~~~ {\bf R}_{0\alpha\beta m} = g_s^{1/3}, ~~~~~ 
{\bf R}_{0abm} = g_s^{1/3}, ~~~~~~ {\bf R}_{0ijm} = g_s^{-11/3}, \nd
including various possible permutations of each components. The $g_s$ powers are again predominantly negative, and the scalings are computed taken all the earlier considerations of the dependence of the metric components only on the coordinates of ${\cal M}_4$. Of course, as before, we have not specified the 
${\cal O}(g^\Delta_s, e^{-1/g^\Delta_s})$ corrections that accompany each of the curvature tensors listed in \eqref{thierry} and \eqref{tagramee}. 

With the curvature scalings at our disposal, let us work out the quantum terms with product of the curvature tensors. Comparing with \eqref{brooklynns}, \eqref{thierry} and \eqref{tagramee} we see that there are 
18 distinct curvature tensors excluding the allowed permutations of the indices of the individual tensors. Therefore to write the full quantum terms, we resort to some short-hand techniques. We define:
\bg\label{lilprince7}
\left({\bf R}_{MNPQ}\right)^{l_i} \equiv \prod_{k = 1}^{l_i} {\bf R}_{M_k N_k P_k Q_k}, \nd
where the subscript denote the various possible permutations and products of the curvature tensor for a give set of indices. Using this notation we can express the quantum piece, appearing from the curvature 
tensors only, in the following way:
\bg\label{lil777}
\mathbb{Q}_5 & = & {\bf g}^{m_i m'_i}.... {\bf g}^{j_k j'_k} \left({\bf R}_{mnpq}\right)^{l_1} \left({\bf R}_{abab}\right)^{l_2}\left({\bf R}_{pqab}\right)^{l_3}\left({\bf R}_{\alpha a b \beta}\right)^{l_4}\left({\bf R}_{\alpha\beta mn}\right)^{l_5}\left({\bf R}_{\alpha\beta\alpha\beta}\right)^{l_6} \nonumber\\
&\times&\left({\bf R}_{ijij}\right)^{l_7}\left({\bf R}_{ijmn}\right)^{l_8}\left({\bf R}_{iajb}\right)^{l_9}
\left({\bf R}_{i\alpha j \beta}\right)^{l_{10}}\left({\bf R}_{0mnp}\right)^{l_{11}}
\left({\bf R}_{0m0n}\right)^{l_{12}}\left({\bf R}_{0i0j}\right)^{l_{13}}\nonumber\\
& \times & \left({\bf R}_{0a0b}\right)^{l_{14}}\left({\bf R}_{0\alpha 0\beta}\right)^{l_{15}}
\left({\bf R}_{0\alpha\beta m}\right)^{l_{16}}\left({\bf R}_{0abm}\right)^{l_{17}}\left({\bf R}_{0ijm}\right)^{l_{18}},
\nd
where the components of the warped inverse metric are used to contract the indices of the curvature tensors in a suitable way (extra care needs to be implemented to contract the indices because of the anti-symmetry of the first two and the last two indices of a given curvature tensor). In a compact notation, \eqref{lil777} may be written as:
\bg\label{ncresp}
\mathbb{Q}_5 \equiv \left[{\bf g}^{-1}\right]^{E_1 + E_2 + E_3 + E_4 + E_5} \prod_{i = 1}^{18} 
\left({\bf R}_{MNPQ}\right)^{l_i}, \nd
where the term in the bracket is defined in terms of individual components in \eqref{lilprince7} and thus should be expanded accordingly. The powers of the inverse metric components $E_i$ are linear functions 
of $l_i$ and may be expressed as:
\bg\label{teenbar}
&&E_1 = 2l_7 + l_8 + l_9 + l_{10} + l_{13} + l_{18}\\
&&E_2 = {l_{11}\over 2} + l_{12} + l_{13} + l_{14} + l_{15} + {l_{16}\over 2} + {l_{17}\over 2} + {l_{18}\over 2} \nonumber\\
&& E_3 = 2l_1 + l_3 + l_5 + l_8 + {3l_{11}\over 2} + l_{12} + {l_{16}\over 2}  + {l_{17}\over 2}
+ {l_{18}\over 2}\nonumber\\
&&E_4 = 2l_2 + l_3 + l_4 + l_9 + l_{14} + l_{17}, ~~
E_5 = l_4 + l_5 + 2l_6 + l_{10} + l_{15} + l_{16}, \nonumber \nd
where $E_1, E_2, ...., E_5$ count the metric components along ($i, j$), ($0, 0$), ($m, n$), ($a, b$), and 
($\alpha, \beta$) respectively. Since we are only after the $g_s$ scalings, such a counting of the metric components would make sense. Therefore using the $g_s$ scalings of the metric components as well as the curvature tensors from \eqref{brooklynns}, \eqref{thierry} and \eqref{tagramee}, it is easy to see that the $g_s$ scaling of $\mathbb{Q}_5$ becomes:
\bg\label{asha}
\mathbb{Q}_5 = g_s^{2\left(l_1 + l_2 + l_3 + l_4 + ....... + l_{17} + l_{18}\right)/3}\left(1 
+ {\cal O}(g^\Delta_s, e^{-1/g^\Delta_s})\right), \nd 
which is a generalization of similar scaling for the part of the product of the curvature tensors in 
\eqref{lenaeng}. The conclusion then is also the same, namely, there is no time-neutral series possible with product of curvature tensors only.   

With multiple derivatives we can also work out the quantum terms. Since the derivatives are going to act only on the internal ${\cal M}_4$ coordinates, the correction to the $g_s$ scaling is easy to ascertain. The derivative action will only change $E_3$ in \eqref{teenbar} to $E_3 \to E_3 + {n\over 2}$ where $n$ is the number of derivatives. This implies:
\bg\label{autumn}
\mathbb{Q}_6 &\equiv& \left[{\bf g}^{-1}\right]^{E_1 + E_2 + E_3 + E_4 + E_5 + n/2}\left[{\bf \partial}\right]^n
\left( \prod_{i = 1}^{18} 
\left({\bf R}_{MNPQ}\right)^{l_i}\right) \nonumber\\
&=& g_s^{2\left(l_1 + l_2 + l_3 + l_4 + ....... + l_{17} + l_{18} + n/2
\right)/3}\left(1 + {\cal O}(g^\Delta_s, e^{-1/g^\Delta_s})\right) , \nd
with no possibility of any time-neutral series.  This is expectedly similar to what we had in \eqref{EWS}, and thus justifying the genericity of the arguments presented earlier. 

With two free Lorentz indices the story should again be similar to what we had earlier, but now, because of the possibility of multiple indices, things would be slightly involved. For example if we want free ($i, j$) Lorentz indices we convert $E_1$ to $E_1 - 1$ and keep other $E_i$ unchanged. We can quantify such changes by using a simple formalism. Let $k = (k_1, k_2$) such that $k$ identifies the subscript in $E_k$ and ($k_1, k_2$) identify the Lorentz indices. For example if $k = 1$ then $k_1 \equiv x_i$ and $k_2 \equiv 
x_j$. Using this let us define $E_k(w, z)$ as:
\bg\label{Amoon}
E_k(w, z) \equiv E_k - \delta_{wk_1} \delta_{zk_2}, \nd
with $E_k$ as in \eqref{teenbar}. The above form easily gives us the required exponent. For example 
$E_k(m, n) = E_k$ for $k \ne 3$ and $E_3(m, n) = E_3 - 1$. With this, the quantum terms with two free Lorentz indices will simply be:  
\bg\label{bandorka}
\mathbb{Q}_7(w, z) &\equiv& \left[{\bf g}^{-1}\right]^{\sum_k E_k(w, z) + n/2}\left[{\bf \partial}\right]^n
\left( \prod_{i = 1}^{18} \left({\bf R}_{MNPQ}\right)^{l_i}\right), \nd
where the choice of ($w, z$) specify which two Lorentz indices we want to keep free. Note that some care needs to be imposed in interpreting the results  as the derivation of the curvature tensors did not have cross-terms. So indices like $w = a, z = m$ has no meaning here. After the dust settles, the $g_s$ scaling for 
\eqref{bandorka} may be expressed as $g_s^{\chi}$ where $\chi$ takes the following {\it three} values:
\bg\label{baccha}
\chi \equiv {2\over 3} \sum_{i = 1}^{18} l_i + {n\over 3} -{8\over 3}, ~~~~
\chi \equiv {2\over 3} \sum_{i = 1}^{18} l_i + {n\over 3} -{2\over 3}, ~~~~
\chi \equiv {2\over 3} \sum_{i = 1}^{18} l_i + {n\over 3} + {4\over 3}, \nd
where the first one corresponds to two free Lorentz indices ($i, j$) and ($0, 0$); the second one corresponds to two free Lorentz indices along ${\cal M}_4$, i.e ($m,, n$); and the third one corresponds to two free Lorentz indices along ${\cal M}_2$ and $\mathbb{T}^2/{\cal G}$ i.e ($\alpha, \beta$) and ($a, b$) respectively. Note that the relative {\it minus} signs for the first two values of $\chi$ shows the possibility of time-neutral terms. For the first case, looking at $E_2$ in \eqref{teenbar}, and imposing:
\bg\label{angul}
l_{12} = l_{13} = l_{14} = l_{15} = 1, ~~~ n = 0, \nd
with all other $l_i$ vanishing gives us $\chi = 0$. This
exactly leads to a quantum term that appears  from the contraction ${\bf g}^{AB} {\bf R}_{0A0B}$ with 
($A, B$) spanning the four allowed choices, namely, ($i, j$), ($m, n$), ($a, b$) and ($\alpha, \beta$), as:
\bg\label{ryaconn}
\left({\bf g}^{00}\right)^3 {\bf g}^{\alpha\beta}{\bf g}^{ab}{\bf g}^{ij}{\bf g}^{mn}{\bf R}_{0m0n}
{\bf R}_{0i0j}{\bf R}_{0a0b}{\bf R}_{0\alpha 0\beta}~ \in ~ \left({\bf g}^{00}{\bf R}_{00}\right)^4 {\bf g}_{00},  \nd
where the LHS is the time-neutral piece in the expansion of the complete term given in the RHS, which for brevity be called the time-neutral ${\bf R}_{00}$ term. 
 In a similar vein, one can argue for time-neutral ${\bf R}_{ij}$ for the first case and time-neutral 
 ${\bf R}_{mn}$ for the second case. In fact the space-time terms appear from expanding 
 $\left({\bf g}^{\mu\nu}{\bf R}_{\mu\nu}\right)^4 {\bf g}_{MN}$ with ($M, N$) spanning ($0, 0$), and ($i, j$) indices; whereas the ($m, n$) term simply appears for ${\bf R}_{mn}$. Finally, the third case tells us that there are no time-neutral terms possible with either 
($a, b$) or ($\alpha, \beta$) indices.

The case with $F_i(t)$ satisfying \eqref{olokhi} with the inverses having perturbative expansions should in principle be redone in the light of the new $g_s$ scalings to the curvature tensors. At this stage, one might even generalize the story from \eqref{ranjhita2} to:
\bg\label{ranjhita3}
F_1(t) F_2^2(t) = \left({g_s^2\over \sqrt{h}}\right)^{\gamma\over 2}, \nd
with $\vert\gamma\vert \in \mathbb{Z}$ such that $\gamma = 0, 2$ correspond to \eqref{olokhi} and 
\eqref{ranjhita} respectively. Although most others values of $\gamma$ are not useful for us, it is
nevertheless interesting to speculate the fate of our background for generic choice of $\gamma$. Incidentally, the only scalings that are affected are: 
\bg\label{joihep}
&&{\bf R}_{\alpha\beta\alpha\beta} = g_s^{2\gamma -2/3} = g_s^{-2/3}, ~~
{\bf R}_{mn \alpha\beta} = g_s^{\gamma-2/3} = g_s^{-2/3}, ~~
{\bf R}_{\alpha ab \beta} = g_s^{\gamma + 4/3} = g_s^{4/3}\nonumber\\
&& {\bf R}_{ij \alpha\beta} = g_s^{\gamma - 8/3} = g_s^{-8/3}, ~~
 {\bf R}_{0 \alpha\beta m} = g_s^{\gamma - 5/3} = g_s^{-5/3}, ~~ 
 {\bf R}_{0 \alpha 0 \beta} = g_s^{\gamma - 8/3} = g_s^{-8/3}, \nd
where on the extreme RHS of every equation we have put $\gamma = 0$ to relate the result for 
\eqref{olokhi}. All these affected components 
 have legs along ${\cal M}_2$ but are functions of ${\cal M}_4$ only. Once the derivative constraints are removed for the case \eqref{olokhi}, i.e make them functions of ${\cal M}_2$ also,  the scalings \eqref{joihep} again work perfectly as shown in 
 {\bf Table \ref{firozasutaria}}.
 Putting these curvatures together and introducing $n$ derivatives, lead to exactly the same $g_s$ scalings for the quantum terms that we had
 in above for both zero and two free Lorentz indices for {\it any} choice of $\gamma$.  No extra conditions are needed and thus we share the same conclusion of the non-existence of time-neutral series with curvatures and multiple derivatives as before. 

\begin{table}[tb]  
 \begin{center}
\renewcommand{\arraystretch}{1.5}
\begin{tabular}{|c||c||c|}\hline Riemann tensors for \eqref{ranjhita} & ${g_s}$ scalings & 
Riemann tensors for \eqref{olokhi} \\ \hline\hline
${\bf R}_{mnpq}$ & $-{2\over 3}$& ${\bf R}_{mnpq}, {\bf R}_{mnp\alpha}, {\bf R}_{mn\alpha\beta}, {\bf R}_{m\alpha\alpha\beta}, {\bf R}_{\alpha\beta\alpha\beta}$ \\ \hline    
${\bf R}_{mnab}, {\bf R}_{mn\alpha\beta}$ & $~~{4 \over 3}$ & ${\bf R}_{mnab}, {\bf R}_{m\alpha ab}, 
{\bf R}_{\alpha\beta ab}$ \\ \hline    
${\bf R}_{abab}, {\bf R}_{ab\alpha\beta}, {\bf R}_{\alpha\beta\alpha\beta}$ & $~~{10 \over 3}$ & 
${\bf R}_{abab}$ \\ \hline    
${\bf R}_{mnp0}$ & $-{5\over 3}$ & ${\bf R}_{mnp0}, {\bf R}_{mn\alpha 0}, {\bf R}_{m\alpha\beta 0}, 
{\bf R}_{0\alpha\alpha\beta}$ \\ \hline    
${\bf R}_{mnij}, {\bf R}_{0m0n}$ & $-{8\over 3}$ & ${\bf R}_{mnij}, {\bf R}_{m\alpha ij}, {\bf R}_{\alpha\beta ij}, {\bf R}_{0m0n}, {\bf R}_{0\alpha 0\beta}, {\bf R}_{0m 0\alpha}$ \\ \hline    
${\bf R}_{m0 ij}$ & $-{11 \over 3}$ &${\bf R}_{m0ij}, {\bf R}_{\alpha 0 ij}$ \\ \hline    
${\bf R}_{ijij}, {\bf R}_{0i0j}$ & $-{14\over 3}$ & ${\bf R}_{ijij}, {\bf R}_{0i0j}$ \\ \hline    
${\bf R}_{0mab}, {\bf R}_{0m\alpha\beta}$ & $~~{1\over 3}$ & ${\bf R}_{0mab}, {\bf R}_{0\alpha ab}$ \\ \hline    
${\bf R}_{abij}, {\bf R}_{0a0b}, {\bf R}_{\alpha\beta ij}, {\bf R}_{0\alpha 0\beta}$ & $-{2\over 3}$& 
 ${\bf R}_{abij}, {\bf R}_{0a0b}$ \\ \hline    
  \end{tabular}
\renewcommand{\arraystretch}{1}
\end{center}
 \caption[]{The ${g_s}$ scalings of the various curvature tensors associated with the two cases 
 {\eqref{olokhi}} and {\eqref{ranjhita}}. These curvature tensors form the essential ingredients of the quantum terms \eqref{phingsha2} and \eqref{phingsha} respectively.  For the case \eqref{ranjhita} they depend only on the coordinates of ${\cal M}_4$, whereas for the case \eqref{olokhi}, they depend on the coordinates of
 ${\cal M}_4 \times {\cal M}_2$.
 The numbers in the middle column, say for example $-{2\over 3}$, should be understood as $\left({g_s\over H}\right)^{-2/3}$ where $H^4(y) \equiv h(y)$ is the warp-factor appearing in \eqref{pyncmey} and \eqref{vegamey}.} 
  \label{firozasutaria}
 \end{table}

\subsubsection{Product of curvatures, G-fluxes and derivatives \label{Gng5}} 

In the previous sub-sections we demonstrated how, by choosing G-fluxes  and curvature tensors and combining them independently with multiple derivatives, they do not lead to time-neutral quantum terms. Various cases were elaborated exhaustively by allowing $F_1(t)$ and $F_2(t)$ to satisfy either 
\eqref{ranjhita} or a variant of \eqref{olokhi} where each of their inverses have perturbative expansions in terms of $g_s$. It is now time to combine all of these together to write quantum terms as a combinations of G-fluxes, curvature tensors and their covariant derivatives. 

Our starting point is of course the G-flux ansatze \eqref{frostgiant} where we will assume that 
$\Delta k \ge {3\over 2}$, so as to comply with earlier constraints (although for certain cases we will see that 
$\Delta k \ge {1\over 2}$ suffice. More so, these constraints could be further generalized as we shall see later). However compared to what we analyzed before, we will now have to take individual components of G-fluxes carefully. The components that we want to consider are listed in 
\eqref{jayanti}. This way, when we consider the individual components of the curvature tensors in 
\eqref{brooklynns}, \eqref{thierry} and \eqref{tagramee} we will be able to quantify the behave of the quantum terms more accurately. 

To start, it is instructive then to specify the product of individual components of G-flux using a notation similar to \eqref{lilprince7} for the product of curvature tensors. This means, we define:
\bg\label{pachar}
\left({\bf G}_{MNPQ}\right)^{l_i} \equiv \prod_{k = 1}^{l_i} {\bf G}_{M_k N_k P_k Q_k}, \nd
the difference now being the complete anti-symmetry of the indices as compared to pair-wise anti-symmetry of the indices for the curvature tensors. Other than this, the two definitions, \eqref{pachar} and 
\eqref{lilprince7}, are similar in spirit.

Therefore combining the pieces of the curvature tensors and derivatives as in \eqref{autumn} and using the definition \eqref{pachar} to insert in the G-fluxes listed from \eqref{jayanti}, we get the following representation of the quantum terms (ignoring temporal, or $g_s$, derivatives for the time-being):
\bg\label{phingsha} 
\mathbb{Q}_{\rm T} & = & {\bf g}^{m_i m'_i}{\bf g}^{m_l m'_l}.... {\bf g}^{j_k j'_k} 
\partial_{m_1}\partial_{m_2}.....\partial_{m_n}
\left({\bf R}_{mnpq}\right)^{l_1} \left({\bf R}_{abab}\right)^{l_2}\left({\bf R}_{pqab}\right)^{l_3}\left({\bf R}_{\alpha a b \beta}\right)^{l_4} \nonumber\\
&\times& \left({\bf R}_{\alpha\beta mn}\right)^{l_5}\left({\bf R}_{\alpha\beta\alpha\beta}\right)^{l_6}
\left({\bf R}_{ijij}\right)^{l_7}\left({\bf R}_{ijmn}\right)^{l_8}\left({\bf R}_{iajb}\right)^{l_9}
\left({\bf R}_{i\alpha j \beta}\right)^{l_{10}}\left({\bf R}_{0mnp}\right)^{l_{11}}
\nonumber\\
& \times & \left({\bf R}_{0m0n}\right)^{l_{12}}\left({\bf R}_{0i0j}\right)^{l_{13}}\left({\bf R}_{0a0b}\right)^{l_{14}}\left({\bf R}_{0\alpha 0\beta}\right)^{l_{15}}
\left({\bf R}_{0\alpha\beta m}\right)^{l_{16}}\left({\bf R}_{0abm}\right)^{l_{17}}\left({\bf R}_{0ijm}\right)^{l_{18}}
\nonumber\\
&\times& \left({\bf G}_{mnpq}\right)^{l_{19}}\left({\bf G}_{mnp\alpha}\right)^{l_{20}}
\left({\bf G}_{mnpa}\right)^{l_{21}}\left({\bf G}_{mn\alpha\beta}\right)^{l_{22}}
\left({\bf G}_{mn\alpha a}\right)^{l_{23}}\left({\bf G}_{m\alpha\beta a}\right)^{l_{24}}\nonumber\\
&\times& \left({\bf G}_{0ijm}\right)^{l_{25}} \left({\bf G}_{0ij\alpha}\right)^{l_{26}}
\left({\bf G}_{mnab}\right)^{l_{27}}\left({\bf G}_{ab\alpha\beta}\right)^{l_{28}}
\left({\bf G}_{m\alpha ab}\right)^{l_{29}}
\nd
where we have inserted in all the available pieces of G-flux and the curvature tensors. 
Each of the pieces, either from the G-fluxes or curvatures,  will have  additional components. For example
 ${\bf R}_{mnpq}$ will have 36 components (excluding the permutations), and so on. Additionally each of the components are raised to $l_i$ powers giving rise to an elaborate set of terms. Note that we can now take advantage of the underlying anti-symmetries of the curvatures to contract some of the Riemann tensors to create 
anti-symmetric Ricci tensors of the form ${\bf R}_{[MN]}$. Of course the Ricci scalar ${\bf R}$ would also participate in the game as before. We can also express \eqref{phingsha} in a condensed form as:
\bg\label{tagchink}
\mathbb{Q}_{\rm T} &\equiv& \left[{\bf g}^{-1}\right]^{H_1 + H_2 + H_3 + H_4 + H_5 + n/2}
\left[{\bf \partial}\right]^n
\left( \prod_{i = 1}^{18} \left({\bf R}_{MNPQ}\right)^{l_i} \prod_{k = 19}^{29} \left({\bf G}_{RSTU}\right)^{l_k}\right), \nd 
which for a given choice of $\{ l_i\}$ determines a specific quantum term with the functional form for 
$H_k(l_j)$ to be determined soon. 
Since any such term has zero free Lorentz index, one may take arbitrary linear combinations of powers of this term. Such combinations lead to a complicated structure of the quantum series. Note that a term like \eqref{tagchink} is suppressed by 
$M_p^{\sigma}$ where:
\bg\label{sigme}
\sigma \equiv  \sigma(\{l_i\}, n) = n + 2\sum_{i = 1}^{18} l_i + \sum_{k = 19}^{29} l_k. \nd
The above quantum terms \eqref{phingsha} are generic enough but they could also have powers of metric components along-with the G-fluxes and curvature tensors\footnote{Taking advantage of the underlying pair-wise anti-symmetry of the curvature tensors and full anti-symmetry of the G-fluxes, two other possibilities exist for \eqref{phingsha} once we remove the derivatives. One: we can suitably 
contract the indices using eleven-dimensional epsilon tensor (i.e the eleven-dimensional Levi-Civita tensor and {\it not} tensor density); and two: we can suitable contract the indices using eleven-dimensional Gamma matrices. Since they don't change the $g_s$ scalings \eqref{miai} and 
\eqref{melamon2}, we will discuss them in the next section.}.
However since these metric components will not change the 
values of $H_k$ functions, we don't specify them here. Additionally all the derivatives should be replaced by covariant derivatives, but since we are taking the fluxes and curvatures, these extra pieces will appear from suitable combinations of these components.  One may then express the quantum potential 
as\footnote{Note that while writing \eqref{ducksoup} we have ignored constant coefficients. These may be easily inserted back, and different choices of these coefficients will specify different theories. These coefficients will have to be determined using the microscopic behavior of the theory, and it should be no surprise to find some (or many) of these coefficients vanishing, affecting the overall dynamics of the theory. However since here we only want to specify the generic behavior we will refrain from specifying the coefficients and assume them to be non-zero.} :
\bg\label{ducksoup}
 \mathbb{V}_Q \equiv \sum_{\{l_i\}, n}
 \int d^8 y \sqrt{{\bf g}_8} \left({\mathbb{Q}_T^{(\{l_i\}, n)} \over M_p^{\sigma(\{l_i\}, n)-8}}\right),
 \nd
 where the superscript on $\mathbb{Q}_T$ denotes the specific choice of $l_i$ and $n$ in \eqref{phingsha}
 with $\sigma$ as in \eqref{sigme} to make it dimensionless (see also \eqref{mcgillmey}). 
 The factor of determinant of the 
 eight-dimensional  warped metric is same for all terms in the potential \eqref{ducksoup}, so we will not count it's $g_s$ contribution in the following, unless mentioned otherwise\footnote{In any case the determinant will only contribute $g_s^{-2/3 + \gamma}$ to the overall scaling with $\gamma$ defined 
 in \eqref{ranjhita3}. Since this does not effect any of the conclusions, we will avoid inserting it in our analysis, unless mentioned otherwise. Note however that \eqref{ducksoup} is still {\it not} the most generic ansatze that we can make for the potential. Once we allow KK modes, i.e allow dependence on the 
 ($x_3, x_{11}$) directions, \eqref{ducksoup} can generalize to 
 \eqref{chuachu}.}. 
 However once we go to the non-local contributions to the potential, this determinant will occur multiple times, and then they {\it will} contribute to the $g_s$ scaling of the potential. 
 
How about other extra components of G-fluxes and curvature tensors that do not appear in the data specifying the background informations? For example various cross-terms in the metric would give rise to extra curvature tensors. Similarly cross-terms in the G-fluxes would contribute extra flux components   in \eqref{phingsha}. This is where the Wilsonian viewpoint becomes immensely useful. The quantum terms are indeed specified by all components of fluxes, derivatives and curvature tensors appearing from fluctuations over a given background, but we can {\it integrate} out the components that are not necessary to specify the background data. Such integrating out modes\footnote{This will involve by first giving small masses to the modes and then integrating them out in the path-integral sense. An example will be presented later assuming de Sitter as a coherent or a squeezed coherent state over a solitonic background. On the other hand, if the de Sitter space is taken as the {\it vacuum} configuration, subtleties associated with red-shifting modes appear. See \cite{petite, coherbeta} for details on this.}
 will result in an infinite series of quantum terms of the form 
\eqref{phingsha}, thus justifying our approach of expressing the quantum series with arbitrary values 
for $l_i$. With this in mind, the $H_k$ functions may be expressed in terms of the following linear combinations of $l_i$:   
\bg\label{lengchink}
H_1 &=& E_1 + l_{25} + l_{26}, ~~ H_2 = E_2 +   {l_{25}\over 2} + {l_{26}\over 2} \nonumber\\
 H_4 &=& E_4 + {l_{21}\over 2} + {l_{23}\over 2} + {l_{24}\over 2} + 
l_{27} + l_{28} + l_{29} \nonumber\\ 
 H_5 &=&  E_5 + {l_{20}\over 2} + l_{22} + {l_{23}\over 2} + l_{24}
+ {l_{26}\over 2} + l_{28} + {l_{29}\over 2}\nonumber\\
H_3 &=& E_3 + 2l_{19} + {3l_{20}\over 2}
 +{3l_{21}\over 2} + l_{22} + l_{23} + {l_{24}\over 2} + 
{l_{25}\over 2} + l_{27} + {l_{29}\over 2} + {n\over 2}, \nd
where $E_1, ..., E_5$ functions, which are themselves expressed as linear combinations of $l_i$,  are defined in \eqref{teenbar}; and $(H_1, ..., H_5)$ denote inverse metric components along 
$(i, j)$, $(0, 0)$, $(m, n)$, $(a, b)$ and $(\alpha, \beta)$ respectively. The story now proceeds in exactly the same way as outlined in the previous section.  The $g_s$ scaling of the quantum piece with zero free Lorentz index may be expressed as:
\bg\label{japmeye}
\mathbb{Q}_{\rm T} & \equiv& g_s^{\theta_k} \left(1 + {\cal O}(g^\Delta_s, e^{-1/g^\Delta_s})\right)
 \nonumber\\
&\equiv& \left[{\bf g}^{-1}\right]^{H_1 + H_2 + H_3 + H_4 + H_5 + n/2}
\left[{\bf \partial}\right]^n
\left( \prod_{i = 1}^{18} \left({\bf R}_{MNPQ}\right)^{l_i} \prod_{k = 19}^{29} \left({\bf G}_{RSTU}\right)^{l_k}\right), \nd
where $\theta_k$ is the scaling parameter that may now be computed by combining all the information that we have assimilated together, namely from the G-flux scaling in \eqref{frostgiant} to the curvature scalings in 
\eqref{tagramee}. The result is:
\bg\label{miai}
\theta_k & = & {2\over 3} \sum_{i = 1}^{18} l_i + {n\over 3} + {l_{25}\over 3} - {2l_{26}\over 3}
+ \left(2\Delta k + {4\over 3}\right)l_{19} + \left(2\Delta k + {1\over 3}\right)(l_{20} + l_{21}) \nonumber\\
&+& \left(2\Delta k - {2\over 3}\right)\left(l_{22} + l_{23} + l_{27}\right)  + \left(2\Delta k - {8\over 3}\right) 
l_{28}  +
\left(2\Delta k - {5\over 3}\right) \left(l_{24} + l_{29}\right), \nd 
where $k$ specifies the minimum $g_s$ scaling of the G-flux components in \eqref{frostgiant}. We expect this to be positive definite if we want the quantum terms in \eqref{phingsha} to have no time-neutral pieces. Unfortunately the relative minus signs in \eqref{miai} are worrisome, so is there way to demonstrate the positivity of \eqref{miai}? First, it is easy to see that if $\Delta k > {4\over 3}$ most of the terms, except the one with 
$l_{26}$, become positive definite\footnote{If $\Delta k = {4\over 3}$ then the coefficient of $l_{28}$ vanishes, implying that we can insert an arbitrary number of ${\bf G}_{ab\alpha\beta}$ components 
{\it without} changing the scaling. This will create a hierarchy issue similar to what we encountered in 
\cite{nodS}.}. This is where our earlier analysis comes in handy, as we have already argued that 
$\Delta k \ge {3\over 2}$  therein! Secondly, if $l_{26}$ vanishes then we are out of water. Can we make 
$l_{26} = 0$ here? Looking at \eqref{phingsha}, we see that $l_{26}$ appears with ${\bf G}_{0ij\alpha}$. It is clear from \cite{nogo, nodS} that: 
\bg\label{rogra}
{\bf G}_{0ij\alpha} = -\partial_\alpha\left({\epsilon_{0ij}\over h(y) \Lambda^2\vert t\vert^4}\right) = 0, \nd
because we have assumed in the earlier sections that all quantities are functions of the ${\cal M}_4$ coordinates, and are thus independent of $y^\alpha$. With these, we now see that $\theta_k > 0$ and therefore $F_i(t)$ satisfying \eqref{ranjhita}, there are no time-neutral series altogether.  

What happens when $F_i(t)$ satisfy the volume-preserving condition \eqref{olokhi}? The analysis becomes a bit more tricky because the metric components along ($\alpha, \beta$) directions scale differently and so do the curvature tensors. The new scalings of the curvature tensors are now \eqref{joihep}. After the dust settles, the scaling of the quantum terms \eqref{phingsha} can be expressed as $g_s^{\theta'_k}$, with additional ${\cal O}(g^\Delta_s, e^{-1/g^\Delta_s})$ corrections, where $\theta'_k$ now takes the following value:
\bg\label{melamon}
\theta'_k & = & {2\over 3} \sum_{i = 1}^{18} l_i + {n\over 3} + {1\over 3}\left(l_{25} + l_{26}\right) + 
\left(2\Delta k +{4\over 3}\right)\left(l_{19} + l_{20} + l_{22}\right) \nonumber\\
& & +  \left(2\Delta k +{1\over 3}\right)\left(l_{21} + l_{23} + l_{24}\right) + 
\left(2\Delta k -{2\over 3}\right)\left(l_{27} + l_{28} + l_{29}\right). \nd
Here we now notice a few important differences from \eqref{miai}; one, the coefficient of $l_{26}$ is positive, so the constraint \eqref{rogra} is not necessary; and two, we only require $\Delta k > {1\over 3}$ for 
 $\theta'_k$ to be a positive definite quantity\footnote{As will be clearer later, this condition is exactly equivalent to the condition $\Delta k \ge {1\over 2}$. Again imposing $\Delta k = {1\over 3}$ would make the coefficients of ($l_{27}, l_{28}, l_{29}$) vanish, implying the possibility of introducing an infinite possible combinations of ${\bf G}_{mnab}, {\bf G}_{ab\alpha\beta}$ and ${\bf G}_{m\alpha ab}$ components without changing $\theta'_k$ in \eqref{melamon}. As mentioned above, this will create similar problem as in 
 \cite{nodS}.}. 
 In addition to that we can relax the derivative constraint, which was originally along ${\cal M}_4$, to the full six dimensional internal manifold ${\cal M}_4 \times {\cal M}_2$ because now both the metric components along ($m, n$) and ($\alpha, \beta$) scale as $g_s^{-2/3}$. (This will lead to some subtleties that we will deal a bit later.) In other words, if there are $n_1$ derivatives along ${\cal M}_4$ and $n_2$ derivatives along ${\cal M}_2$, then $n$ in \eqref{melamon} can be replaced for the two cases, \eqref{olokhi} and \eqref{ranjhita}, respectively by:
 \bg\label{hkpGKM}
 n ~ \to ~ n_1 + n_2, ~~~~~ n ~ \to ~ n_1 - 2 n_2, \nd
 where the relative minus sign for the second case, i.e for background satisfying \eqref{ranjhita}, requires 
 $n_2 = 0$ to preserve the positivity of $\theta_k$ in \eqref{miai}. Interestingly for $k = 0$, the condition becomes:
 
 {\footnotesize
 \bg\label{kkkbkb}
 \theta'_0  =  {2\over 3} \sum_{i = 1}^{18} l_i + {n\over 3} + 
 {1\over 3}\left( l_{21} + l_{23} + l_{24} + l_{25} + l_{26}\right)  
 + {4\over 3}\left(l_{19} + l_{20} + l_{22}\right) -{2\over 3}\left(l_{27} + l_{28} + l_{29}\right), \nd} 
which by construction cannot always be positive definite. In fact the above scaling corresponds precisely to the scalings that we advocated in \cite{nodS} with time-independent internal space and time-independent G-flux. Of course there were no derivative constraints therein so we could even retain $l_{26}$ which, in turn, also allows us to retain $l_{27}, l_{28}$ and $l_{29}$, i.e G-fluxes with two indices along ($a, b$) directions.
Since this is important, let us clarify it in some details. To start, we define a scalar function along a compact direction $z$ as 
\bg\label{salom}
\Phi(z) = \sum_k {\phi}(k) e^{ikz}, \nd
with $k = {l\over R}$ where $l \in \mathbb{Z}$ and $R$ is the radius of the $z$-circle. 
Additionally, we impose ${\phi}^\ast(k) = {\phi}(-k)$ so that $\Phi(z)$ remains real. Using this, we can define a three-form:
\bg\label{montan}
{\bf C}_{MN3}(y^m, y^\alpha, x_{11}) \equiv \mathbb{C}_{MN3}(y^m, y^\alpha) \otimes  \Phi(x_{11}), \nd
where $(M, N)$ span coordinates of ${\cal M}_4 \times {\cal M}_2$ and $(x_3, x_{11})$ are the periodic coordinates of $\mathbb{T}^2/{\cal G}$ such that $\Phi(x_{11})$ is the zero-form on the torus that is not projected out by the ${\cal G}$ action. This also implies that 
the G-flux components are taken to be functions of all the coordinates\footnote{As we saw before, they are also functions of ($g^\Delta_s, e^{-1/g^\Delta_s}$) which we suppress to avoid clutter. However one question could be raised on the validity of the usage of the $g_s$ scaling \eqref{miai} or \eqref{melamon} for the case with no derivative constraints. The answer lies in the fact that the relative minus signs in 
\eqref{miai} or \eqref{melamon} mostly appear from the G-flux components whose $g_s$ scalings are not bounded by any derivative constraints. Since these relative minus signs are crucial in the discussion, the derivative bounds play no role in our choice. It is also important to note that ${\bf C}_{MN3}$ in \eqref{montan} is expressed in a series with 
${\rm exp}\left({ilx_{11}\over \mathbb{R}_{11}}\right)$, which for $l = 0$ becomes a constant. Thus to the lowest order, ${\bf C}_{MN3}$ is only a function of ($y^m, y^\alpha$) for the case \eqref{olokhi} and $y^m$ for the case \eqref{ranjhita}.} 
of the eight manifold except $x_3$, so components like 
${\bf G}_{MNab} \equiv {1\over 3!}\partial_{[11}{\bf C}_{MN3]}$ would lead to, in addition to other possible fields,  a RR field 
$\mathbb{C}^{(2)}_{MN}(y^m, y^\alpha)$ in the type IIB side. For  $l \ge 1$, we get KK modes 
$l/\mathbb{R}_{11}$, with $\mathbb{R}_{11}$ being the warped radius of the eleventh direction (which in turn will be related to $g_s$ as shown in \eqref{recutvi}). As $\mathbb{R}_{11}$ increases, the modes 
\eqref{montan} become lighter and we can no longer integrate them out! These light degrees of freedom now contribute to $l_{27}, l_{28}$ and $l_{29}$ in \eqref{phingsha} and therefore, from \cite{nodS}, 
time-neutrality for $\theta'_0$ now happens when:
\bg\label{evabmey}
l_{27} + l_{28} + l_{29} + {3l_{21}\over 2} = {n\over 2} + \sum_{i =1}^{18} l_{i} + 
2\sum_{j =19}^{22} l_{j} + {1\over 2} \sum_{k = 23}^{26} l_{k}, \nd
with $n$ being the number of derivatives that satisfy the first relation in \eqref{hkpGKM}. Since the $l_i$'s have no additional constraints, \eqref{evabmey} constitutes one relation between thirty variables, and as such will have infinite number of solutions, leading to the breakdown of an EFT 
description\footnote{One would expect that such breakdown implies the vanishing of the $M_p$ hierarchy simultaneously. This is subtle and will be elaborated more in section \ref{Gng6} (see also \cite{petite}). However assuming this to be the case, such a train of thought then is particularly consistent with the swampland conjecture as presented in \cite{vafa1}. In particular the swampland distance conjecture should be associated to the distance in the field space where the KK modes in \eqref{salom} and \eqref{montan} start becoming light. Note that one can potentially develop a similar story with three-form field components along $x_3$ direction as in \eqref{montan}. In such a picture, as the $x_3$ circle increases, the KK modes become lighter and start creating the same issues as above. However the $x_3$ dependences ruin the Busher's duality employed to convert the type IIB background to type IIA in the first place, although this may not be a big issue if we follow the T-duality rules proposed in \cite{mooreH}. Interestingly, ${\bf C}_{MN, 11}$ will dualize to a NS two-form field in the IIB side. \label{sirisayz}}.
A particular set of choice for the $l_i$ numbers, lets call them
$\{l_i, s\}$ such that for integer choice of $s$ (that is related to the KK states) we can allow different choices for 
$\{l_i\} = (l_1, l_2, ..., l_{29})$, satisfying 
\eqref{evabmey} would constitute a time-neutral quantum term of the form \eqref{phingsha}. 
Each of these quantum terms may in turn be arranged together as:
\bg\label{chuachu}
\mathbb{Q}_{{\rm T}\{i\}}^{(0)} \equiv \sum_{k_1, k_2,..} C_{k_1k_2...k_\infty} \prod_{s = 1}^\infty 
\left({\mathbb{Q}_{{\rm T}, \{l_i, s\}}\over M_p^{\sigma(\{l_i, s\})}}\right)^{k_s}, \nd
where the superscript denote time-neutrality and the subscript $\{i\} = $ ($1, 2, ..., 29$). The power of $M_p$ can be read off from \eqref{sigme} for a given choice of $\{l_i, s\}$ and furnish the inverse powers of
$M_p$ in the quantum series to keep them dimensionless. The series \eqref{chuachu} thus constitute the infinite class of time-neutral quantum pieces elaborated in \cite{nodS}.

The above construction gives a satisfying answer to the question of the non-existence of an EFT description in the set-up with time-independent fluxes in \cite{nodS}, although one question could be raised at this point. Since $\mathbb{R}_{11} \to 0$ decouples all the degrees of freedom coming from the KK states of
${\bf G}_{MNab}$, and clearly the vanishing of the warped eleven-dimensional radius is also a necessary condition to go to type IIB, couldn't we just decouple all the dangerous states and study the resulting EFT? The answer to this question lies in the three scaling behaviors that we derived earlier, namely \eqref{miai}, \eqref{melamon} and \eqref{kkkbkb}. For \eqref{miai} and \eqref{melamon}, whether or not we switch on 
($l_{27}, l_{28}, l_{29}$), they are {\it always} positive definite and therefore cannot create time-neutral series anywhere in the moduli space of M-theory. This is clearly not the case for \eqref{kkkbkb}, which does create an infinite class of time-neutral series as in \eqref{evabmey}. Thus although $g_s \to 0$ provides a false aura of a healthy EFT with $\theta'_0$ scaling in \eqref{kkkbkb}, it quickly disappears as we go away from this limit: a property not shared by \eqref{miai} and \eqref{melamon} for \eqref{ranjhita} and 
\eqref{olokhi} respectively. 

All the three scalings discussed above, namely \eqref{miai}, \eqref{melamon} and \eqref{kkkbkb} are related to special choices of $\gamma$ in \eqref{ranjhita3}. If we make an arbitrary choice of $\gamma$ then the 
$g_s$ scaling of the quantum term \eqref{phingsha} becomes $g_s^{\theta(k, \gamma)}$, where 
$\theta(k, \gamma)$ is:

{\footnotesize
\bg\label{dcmika}
\theta(k, \gamma)  & = & {2\over 3} \sum_{i = 1}^{18} l_i + {n\over 3} + {l_{25}\over 3} 
+ \left(2\Delta k + {4\over 3}\right)l_{19} + \left(2\Delta k + {1\over 3}\right)l_{21} + 
\left(2\Delta k - {2\over 3}\right)l_{27}\nonumber\\ 
&+& \left(2\Delta k + {4\over 3} - {\gamma\over 2}\right) l_{20}
+ \left(2\Delta k + {4\over 3} - {\gamma}\right)l_{22}
+ \left(2\Delta k + {1\over 3} - {\gamma\over 2}\right)l_{23}
+ \left(2\Delta k + {1\over 3} - {\gamma}\right)l_{24} \nonumber\\
&+& \left({1\over 3} - {\gamma\over 2}\right)l_{26} 
+ \left(2\Delta k - {2\over 3} - {\gamma}\right)l_{28}
+ \left(2\Delta k - {2\over 3} - {\gamma\over 2}\right)l_{29}, \nd}
where the first line is generic to all choices of $\gamma$, but the second and the third lines specifically depend on what value $\gamma$ takes. Plugging in  $\gamma = 0, 2$ one may easily derive 
\eqref{olokhi} and \eqref{ranjhita} respectively. It should also be clear that ${3\gamma + 2\over 3}$ is the largest attainable value with a relative minus sign, implying that it is only the coefficient of 
$l_{28}$ that can determine the lower bound on $k$ to avoid time-neutral series. For the present case, this happens when:
\bg\label{etochinki}
\Delta k > {1\over 3} + {\gamma\over 2}, \nd
from where one may easily derive the two earlier bounds we had. As $\gamma$ increases the lower  bound on $k$ increases. Since $\Delta k$ determines the {lower} power of $g_s$ for G-flux in 
\eqref{frostgiant} or \eqref{hh4u}, it implies that the lower power\footnote{As we will soon see, and alluded to earlier, \eqref{etochinki} does not provide the lowest bound for {\it all} the G-flux components. For some components one could go below \eqref{etochinki}.}
 is bigger for bigger $\gamma$. On the other hand $\gamma$ from \eqref{ranjhita3} also tells us the deviation of the four-dimensional Newton's constant from its standard {\it constant} value. Consequently, a more un-natural choice for Newton's constant
is directly proportional to a more un-natural choice of the $g_s$ dependence (or temporal dependence) of the G-flux components. Additionally, for $\gamma \ge 1$, the coefficient of $l_{26}$ starts becoming negative thus making \eqref{etochinki} prone to creating time-neutral series. The only way out appears from imposing \eqref{rogra}. Thus for $\gamma \ge 1$ the fields can only be functions of the ${\cal M}_4$ coordinates to avoid the breakdown of a EFT description of the system. This second level of un-naturalness prompts us to ask whether this is the reason why nature chooses the simplest value of
$\gamma = 0$ in \eqref{ranjhita3} and \eqref{dcmika}. We will speculate on this 
interesting possibility in section \ref{qotomth}.   

Let us pause for a moment to absorb the consequence of the two lessons that we learnt from generic choice of $\gamma$ in \eqref{dcmika}. One, larger $\gamma$ makes $k$ larger from \eqref{etochinki}, and two, 
larger $\gamma$ also makes the coefficient of $l_{26}$ negative. Thus $\gamma = 0$ and $\gamma > 0$ 
share different physics: $\gamma = 0$ no longer requires any derivative constraints so we can assume that all fields are functions of ${\cal M}_4 \times {\cal M}_2$; whereas $\gamma > 0$ has derivative constraint because of \eqref{rogra}. For both cases however we will  keep the fields independent of 
${\mathbb{T}^2/ {\cal G}}$ (even this could be relaxed for time-independent internal space, thus allowing the KK modes to appear in the discussion). Relaxing the derivative constraints for $\gamma = 0$
will create new components of curvature tensors, as well as new derivatives, that should modify 
\eqref{phingsha} to the following: 
\bg\label{phingsha2} 
\mathbb{Q}_{\rm T} & = & {\bf g}^{m_i m'_i}{\bf g}^{m_l m'_l}.... {\bf g}^{j_k j'_k} 
\partial_{m_1}..\partial_{m_{n_1}}\partial_{\alpha_1}..\partial_{\alpha_{n_2}}
\left({\bf R}_{mnpq}\right)^{l_1} \left({\bf R}_{abab}\right)^{l_2}\left({\bf R}_{pqab}\right)^{l_3}\left({\bf R}_{\alpha a b \beta}\right)^{l_4} \nonumber\\
&\times& \left({\bf R}_{\alpha\beta mn}\right)^{l_5}\left({\bf R}_{\alpha\beta\alpha\beta}\right)^{l_6}
\left({\bf R}_{ijij}\right)^{l_7}\left({\bf R}_{ijmn}\right)^{l_8}\left({\bf R}_{iajb}\right)^{l_9}
\left({\bf R}_{i\alpha j \beta}\right)^{l_{10}}\left({\bf R}_{0mnp}\right)^{l_{11}}
\nonumber\\
& \times & \left({\bf R}_{0m0n}\right)^{l_{12}}\left({\bf R}_{0i0j}\right)^{l_{13}}\left({\bf R}_{0a0b}\right)^{l_{14}}\left({\bf R}_{0\alpha 0\beta}\right)^{l_{15}}
\left({\bf R}_{0\alpha\beta m}\right)^{l_{16}}\left({\bf R}_{0abm}\right)^{l_{17}}\left({\bf R}_{0ijm}\right)^{l_{18}}
\nonumber\\
& \times & \left({\bf R}_{mnp\alpha}\right)^{l_{19}}\left({\bf R}_{m\alpha ab}\right)^{l_{20}}
\left({\bf R}_{m\alpha\alpha\beta}\right)^{l_{21}}\left({\bf R}_{m\alpha ij}\right)^{l_{22}}
\left({\bf R}_{0mn \alpha}\right)^{l_{23}}\left({\bf R}_{0m0\alpha}\right)^{l_{24}}
\left({\bf R}_{0\alpha\beta\alpha}\right)^{l_{25}}
\nonumber\\
&\times& \left({\bf R}_{0ab \alpha}\right)^{l_{26}}\left({\bf R}_{0ij\alpha}\right)^{l_{27}}
\left({\bf G}_{mnpq}\right)^{l_{28}}\left({\bf G}_{mnp\alpha}\right)^{l_{29}}
\left({\bf G}_{mnpa}\right)^{l_{30}}\left({\bf G}_{mn\alpha\beta}\right)^{l_{31}}
\left({\bf G}_{mn\alpha a}\right)^{l_{32}}\nonumber\\
&\times&\left({\bf G}_{m\alpha\beta a}\right)^{l_{33}}\left({\bf G}_{0ijm}\right)^{l_{34}} 
\left({\bf G}_{0ij\alpha}\right)^{l_{35}}
\left({\bf G}_{mnab}\right)^{l_{36}}\left({\bf G}_{ab\alpha\beta}\right)^{l_{37}}
\left({\bf G}_{m\alpha ab}\right)^{l_{38}},
\nd
where ($n_1, n_2$) are the number of derivatives along ${\cal M}_4$ and ${\cal M}_2$ directions respectively (again ignoring the $g_s$, or temporal, derivatives). Compared to \eqref{phingsha}, there are now nine extra pieces of curvature tensors, totalling to 38 total pieces of fluxes and curvature tensors. Each of these will have the required copies because of the $l_i$ factors, in addition to the internal permutations as mentioned earlier. Such a quantum term has a $M_p$ suppression of the form $M_p^\sigma$, where:
\bg\label{kamagni2}
\sigma \equiv \sigma(\{l_i\}, n_1, n_2) = n_1 + n_2 + 2\sum_{i = 1}^{27} l_i + \sum_{k = 28}^{38}l_k, \nd
which may be compared to \eqref{sigme}: the changes coming from new derivatives and new curvature tensors. By construction then every such quantum term, or a class of quantum terms that solves \eqref{kamagni2} for a fixed $\sigma$, has a distinct $M_p$ suppression going with it. 
We also expect both $H_i$ in
\eqref{lengchink} and $E_i$ in \eqref{teenbar} to change to $\widetilde{H}_i$ and $\widetilde{E}_i$ respectively. 
The change in the latter may be quantified as:
\bg\label{teenbar2}
&& \widetilde{E}_5 = E_5 + {l_{20}\over 2} + {3l_{21}\over 2} + {l_{22}\over 2} + {l_{23}\over 2} +
{l_{24}\over 2} + {3l_{25}\over 2} + {l_{27}\over 2}\nonumber\\
&&\widetilde{E}_1 = E_1 + l_{22} + l_{27}, ~~~~ \widetilde{E}_2 = E_2 + {l_{23}\over 2} + {l_{24}} 
+ {l_{25}\over 2} + {l_{26}\over 2} + {l_{27}\over 2} \nonumber\\
&&\widetilde{E}_3 = E_3 + {3l_{19}\over 2} + {l_{20}\over 2} + {l_{21}\over 2} + {l_{22}\over 2}
+ {l_{23}} + {l_{24}\over 2}, ~~~\widetilde{E}_4 = E_4 + {l_{20}} + {l_{26}}, \nd
with $E_n$ as defined in \eqref{teenbar}. The change in \eqref{lengchink} is now easy to determine: all the subscript would shift by $+9$ in addition to an extra contribution to 
$\widetilde{H}_5$ coming from the derivatives. The overall change is:
\bg\label{lengchink2}  
\widetilde{H}_1 &=& \widetilde{E}_1 + l_{34} + l_{35}, ~~ 
\widetilde{H}_2 = \widetilde{E}_2 +   {l_{34}\over 2} 
+ {l_{35}\over 2} \nonumber\\
 \widetilde{H}_4 &=&\widetilde{E}_4 + {l_{30}\over 2} + {l_{32}\over 2} + {l_{33}\over 2} + 
l_{36} + l_{37} + l_{38} \nonumber\\ 
 \widetilde{H}_5 &=&  \widetilde{E}_5 + {l_{29}\over 2} + l_{31} + {l_{32}\over 2} + l_{33}
+ {l_{35}\over 2} + l_{37} + {l_{38}\over 2} + {n_2\over 2} \nonumber\\
\widetilde{H}_3 &=& \widetilde{E}_3 + 2l_{28} + {3l_{29}\over 2}
 +{3l_{30}\over 2} + l_{31} + l_{23} + {l_{33}\over 2} + 
{l_{34}\over 2} + l_{36} + {l_{38}\over 2} + {n_1\over 2}, \nd
which expectedly takes the form similar to \eqref{lengchink}, with minor differences. One may also see that the quantum term in \eqref{phingsha2} scale with respect to $g_s$ as $g_s^{\theta'_k}$, with 
additional ${\cal O}(g_s^\Delta, e^{-1/g_s^\Delta})$ corrections, 
\bg\label{melamon2}
\theta'_k & = & {2\over 3} \sum_{i = 1}^{27} l_i + {n_1 + n_2\over 3} + {1\over 3}\left(l_{34} + l_{35}\right) + 
\left(2\Delta k +{4\over 3}\right)\left(l_{28} + l_{29} + l_{31}\right) \nonumber\\
& & +  \left(2\Delta k +{1\over 3}\right)\left(l_{30} + l_{32} + l_{33}\right) + 
\left(2\Delta k -{2\over 3}\right)\left(l_{36} + l_{37} + l_{38}\right), \nd
where the only change from \eqref{melamon} is from 2/3 curvature contributions from the additional Riemann tensors and 1/3 derivative contributions from the derivatives along ${\cal M}_2$ directions. Interestingly, $n_0$ temporal derivatives to \eqref{phingsha2} will simply change $n_1 + n_2$ to 
$n_0 + n_1 + n_2$. This $n_0$ factor can be absorbed by shifting $n_1$ and/or $n_2$, so we won't worry about them anymore. Note that these additional contributions do not change the sign and therefore the story remains unaltered from what we had earlier. When $k = 0$, we can further relax the derivative contributions to involve derivatives along $\mathbb{T}^2/{\cal G}$ directions.  This will involve more curvature tensors and additional 
$n_3$ derivatives 
with ($a, b$) indices. The extra curvature components will again add $+2/3$ to \eqref{melamon2} whereas the derivatives will add $-{2n_3/3}$. If $l_i^{(p)}$ denote the proliferation of each $l_i$ components due to the relaxation of the derivative constraints in \eqref{phingsha2}, then \eqref{kkkbkb} changes to:

 {\footnotesize
 \bg\label{kkkbkb2}
 \theta'_0  =  {2\over 3} \sum_p \sum_{i = 1}^{27} l^{(p)}_i + {n_1 + n_2 \over 3} - {2n_3\over 3} + 
 {1\over 3}\left( l_{30} + \sum_{p = 1}^{4} l_{31 + p}\right)  
 + {4\over 3}\left(l_{28} + l_{29} + l_{31}\right) -{2\over 3}\sum_{q = 1}^3 l_{35 + q}, \nd} 
 which as noted above differs from \eqref{kkkbkb} by the appearance of another set of relative minus signs from the derivatives along the toroidal direction. This makes it prone to creating additional time neutral series from $\theta'_0 = 0$. The condition for this to happen now becomes:   
\bg\label{evabmey2}
l_{36} + l_{37} + l_{38} + n_3 + {3l_{30}\over 2}  = {n_1 + n_2 \over 2} 
+ \sum_p \sum_{i =1}^{27} l_{i}^{(p)} + 
2\sum_{j =28}^{31} l_{j} + {1\over 2} \sum_{q = 32}^{35} l_{q}, \nd
which can be compared to \eqref{evabmey} and
again has more issues as expected leading to the problems with an effective field theory description pointed out in \cite{nodS}. Interestingly, although the proliferation of curvature tensors do not change much of the story, the proliferation of derivatives along $\mathbb{T}^2/{\cal G}$ tends to worsen the problem.

Interestingly, the three $g_s$ scalings that we discussed in \eqref{miai}, \eqref{melamon} and \eqref{melamon2} hint towards further generalization of the G-flux ansatze \eqref{frostgiant}. To see this, let us first consider \eqref{miai}. Due to the relative {\it plus} signs for the coefficients $l_{19}, l_{20}$ and $l_{21}$, the lowest value for $k$ could even be zero, i.e $k \ge 0$, implying that there are $g_s$ independent pieces associated with the G-flux components ${\bf G}_{mnpq}, {\bf G}_{mnp\alpha}$ and ${\bf G}_{mnpa}$. On the other hand, the $g_s$ scalings for the G-flux components ${\bf G}_{mn\alpha\beta}, {\bf G}_{mn\alpha a}$ 
and ${\bf G}_{mnab}$, now due to the relative {\it minus} signs, will only allow $\Delta k \ge {1\over 2}$ and therefore no time-independent pieces.  This is smaller than what we had before, but now we see that the $g_s$ scaling \eqref{miai} does allow this. The original bound of $\Delta k \ge {3\over 2}$ is now satisfied by
the G-flux component ${\bf G}_{\alpha\beta ab}$, whereas the remaining two G-flux components 
${\bf G}_{m\alpha\beta a}$ and ${\bf G}_{m\alpha ab}$ are bounded by $\Delta k \ge 1$. Of course, demanding the lower bound of $\Delta k \ge {3\over 2}$ for all the G-flux components is an easy way to take care of all the other bounds, but \eqref{miai} reveals the possibility of relaxing this condition. 

For the other two $g_s$ scalings, \eqref{melamon} and \eqref{melamon2}, all the G-flux components, except the three components ${\bf G}_{mnab}, {\bf G}_{\alpha\beta ab}$ and ${\bf G}_{m\alpha ab}$, allow time independent components because $k \ge 0$ for these components. The remaining three components cannot have time-independent pieces as they are all bounded by $\Delta k \ge {1\over 2}$. Again, as before, imposing $\Delta k \ge {1\over 2}$ is an easy way to take care of all the lower bounds, but now 
\eqref{melamon} or \eqref{melamon2} allows the possibility of relaxing this condition. It should however be noted that all these new possibilities should be verified against the G-flux EOMs as well as from the 
flux quantization conditions, before we could successfully use them to predict the dynamics of the system. All of these will be elaborated in section \ref{jutamaro} (see also \cite{petite}). For the time being however, a reliable way to
proceed would be to follow the universal lower bounds $\Delta k \ge {3\over 2}$ for \eqref{miai} and $\Delta k \ge {1\over 2}$ for \eqref{melamon2} (and \eqref{melamon}), before a more refined set of bounds are brought in.

With two free Lorentz indices we need to again discuss the two cases pertaining to \eqref{ranjhita} and
\eqref{olokhi}. The second case can be further fine-tuned to discuss the scenario advocated in 
\cite{nodS}, as we have done so far. The story for either of these cases remain simple as before. For 
\eqref{ranjhita}, it is easy to see that the $g_s$ scaling changes from \eqref{miai} to the following three values\footnote{Although $l_i > 0$ always, $H_i$ from \eqref{lengchink} or $E_i$ from \eqref{teenbar}, when two free Lorentz indices are allowed, can take integer values starting from $-1$, i.e $H_i \ge -1$ and 
$E_i \ge -1$. Similar criteria emerge from \eqref{lengchink2} and \eqref{teenbar2}. The negative value implies inserting a metric component, i.e the inverse of an inverse metric component, in either cases.}:
\bg\label{teenangul}
\theta_k ~\rightarrow~\left(\theta_k - {8\over 3}, \theta_k - {2\over 3}, \theta_k + {4\over 3}\right), \nd
where the first one corresponds to free Lorentz indices along ($i, j$) and ($0, 0$) directions; the second one corresponds to free Lorentz indices along ${\cal M}_4$ i.e along ($m, n$) directions and the third one corresponds to free Lorentz indices along $\mathbb{T}^2/{\cal G}$ and ${\cal M}_2$ i.e along ($a, b$) and ($\alpha, \beta$) directions respectively. On the other hand, $\theta'_k$ also changes from \eqref{melamon2}
in the aforementioned way:
\bg\label{charaangul}
\theta'_k ~\rightarrow~\left(\theta'_k - {8\over 3}, \theta'_k - {2\over 3}, \theta'_k + {4\over 3}\right), \nd
for both $\Delta k > {1\over 3}$ and $k = 0$, with the difference being the second one now corresponds to both ($m, n$) as well as ($\alpha, \beta$) directions as a consequence of identical scalings for the metric components along these directions for the case \eqref{olokhi} and \cite{nodS}. 
 
Let us now elaborate the scaling behavior in bit more details. For the case \eqref{melamon2} with
$\Delta k > {1 \over 3}$ we first note that switching on any components of G-fluxes or curvature tensors, $\theta'_k \ge 1/3$ and therefore makes every term in \eqref{melamon2} positive definite, thus ruling out all time-neutral series with zero Lorentz indices along directions $(i, j), (0, 0), (m, n), (a, b)$ and $(\alpha, \beta)$. With two Lorentz indices, there are no time-neutral series at least along the $(a, b)$ directions as is evident from both 
\eqref{teenangul} and \eqref{charaangul}. Along $(m, n)$ and $(\alpha, \beta)$ directions, for 
\eqref{melamon2}, there are a few cases. Since every Riemann tensor contribute an overall factor of $2/3$ to 
$\theta'_k$, it is easy to see that we need at most one of:
\bg\label{essences}
\left(l_1, l_5, l_8, l_{11}, l_{12}\right), ~~~ {\rm and}~~~\left(l_4, l_5, l_6, l_{10}, l_{15}, l_{16}\right),  \nd
for $(m, n)$ and $(\alpha, \beta)$ indices respectively,
to cancel the factor of $2/3$ in \eqref{charaangul}. In fact it is easy to see that we can only get two time-neutral pieces of the form ${\bf R}_{mn}$ and ${\bf R}_{\alpha\beta}$, using combinations of curvature tensors. Using G-fluxes, naively either of the three choices $l_{34} = 2$,  $l_{35} = 2$ and
$l_{34} = l_{35} = 1$ can cancel the $2/3$ factor in \eqref{charaangul}.   These are all easily eliminated as they imply either $\widetilde{H}_2, \widetilde{H}_5$ or $\widetilde{H}_3$ in \eqref{lengchink2} to be half-integers\footnote{Subtleties with half-integers will be discussed later.}. If we take $k = 1$ in 
\eqref{melamon2}, then the only other choices are associated with integer values for 
$(l_{36}, l_{37}, l_{38})$. Taking $l_{36} = 2$, $l_{37} = 2$ or 
$l_{38} = 2$ always make $\widetilde{H}_4 = 2$ and depending on the choices 
$(\widetilde{H}_3, \widetilde{H}_5) = (0, 1)$ or $(1, 0)$ 
from \eqref{lengchink2} respectively give rise to the following two set of 
tensors\footnote{Other possibilities include ${\bf g}_{mn} ~{\bf g}^{kl} \Lambda^{(1j)}_{kl}$ and 
${\bf g}_{\alpha\beta} ~{\bf g}^{\rho\sigma} \Lambda^{(2j)}_{\rho\sigma}$ that appear from expressing
$\widetilde{H}_3 = 1$ alternatively as $\widetilde{H}_3 \equiv 2 + (- 1)$ and $\widetilde{H}_5  = 1$ 
as $\widetilde{H}_5  \equiv 2 + (- 1)$ respectively where the minus signs denote inverse of the inverse metric components. Additionally, choices like 
${\bf g}_{mn} {\bf g}^{\alpha\beta} \Lambda^{(22)}_{\alpha\beta}$ etc. are also allowed.
All these manipulation don't change $\theta_k$ or $\theta'_k$.\label{aaloo}}: 
\bg\label{kimB}
&&\Lambda^{(11)}_{mn} \equiv {{\bf g}^{bd} {\bf g}^{ac}{\bf g}^{\alpha\beta}{\bf G}_{m\alpha ab}  
{\bf G}_{n\beta cd}\over M_p^2}, ~~~~
\Lambda^{(12)}_{mn} \equiv {{\bf g}^{bd} {\bf g}^{ac}{\bf g}^{lq}{\bf G}_{m l ab}  
{\bf G}_{nq cd}\over M_p^2} \nonumber\\
&& \Lambda^{(21)}_{\alpha\beta} \equiv {{\bf g}^{bd} {\bf g}^{ac}{\bf g}^{mn}{\bf G}_{m\alpha ab}  
{\bf G}_{n\beta cd}\over M_p^2}, ~~~~
\Lambda^{(22)}_{\alpha\beta} \equiv {{\bf g}^{bd} {\bf g}^{ac}{\bf g}^{\gamma\sigma}
{\bf G}_{\alpha\gamma ab}  
{\bf G}_{\beta\sigma cd}\over M_p^2},
\nd
as the sole examples of time-neutral rank two tensors along $(m, n)$ and $(\alpha, \beta)$ directions. The other choice with $l_{36} = l_{37} = 1$ is eliminated by the anti-symmetry of the G-fluxes. Similarly for 
$ n \ge 1$, there are no additional time-neutral quantum terms with the required indices. Clearly if we demand $\Delta k \ge {3\over 2}$, both the examples in \eqref{kimB} are no longer allowed.  In fact with 
$\Delta k \ge {3\over 2}$, we also eliminate any time-neutral rank two tensors from G-fluxes 
using \eqref{miai}. 

Along space-time directions the scenario is more delicate. With $\Delta k \ge {3\over 2}$ the only contributions from G-fluxes may appear from $(l_{34}, l_{35})$ taking integer values in \eqref{melamon2}. Taking $l_{34} = 8$ requires us to pick $\widetilde{H}_1 = 7, \widetilde{H}_2 = 4, \widetilde{H}_3 = 3$ from \eqref{lengchink2}. The other choice of $l_{35} = 8$ is similar to the first one because of the identical scalings of the metric components along $(m, n)$ and 
$(\alpha, \beta)$ directions. After the dust settles, the generic quantum term along the space-time directions
appears to be:
\bg\label{thurmanu}
\Lambda^{(3)}_{\mu_a\mu_{a+1}} \equiv M_p^{-8}\prod_{k = 1}^8 \prod_{n = 1}^4
{\bf G}_{\mu_k\nu_k\rho_km_k} {\bf g}^{m_{2n-1} m_{2n}} 
{\bf g}^{\mu_{2n-1} \mu_{2n}} {\bf g}^{\nu_{2n-1} \nu_{2n}} {\bf g}^{\rho_{2n-1} \rho_{2n}}{\bf g}_{\mu_a\mu_{a+1}}, \nd
where assuming $1 \le a \le 8$ and $\mu_a \in (0, i, j)$ is any one of the three space-time directions in M-theory, \eqref{thurmanu} creates two kind of terms: $\Lambda^{(3)}_{00}$ and $\Lambda^{(3)}_{ij}$. Exactly similar set of terms appear from \eqref{miai} (although $l_{26} = 0$ there). It turns out, since 
${\bf G}_{\mu\nu\rho m}$ takes the value similar to \eqref{rogra}, (but now the derivative is with respect to $y^m$ and consequently non-zero), \eqref{thurmanu} is just a function that may be expressed in terms of the warp-factor $h(y)$. Even more generically if we take $l_{34} = 2p$ and $n = 2q$ such that $p+q = 4$
in \eqref{melamon2}, then \eqref{lengchink2} implies $\widetilde{H}_1 = 2p -1, \widetilde{H}_2 = p$ and $\widetilde{H}_3 = 4$, with 
\eqref{thurmanu} becoming: 
\bg\label{mojarmet}
\Lambda^{(p, q)}_{\mu\nu} \equiv \partial_{m_1} \partial_{m_2}....\partial_{m_{2q}}
\left(\prod_{k=q+1}^{2p+ 2q}{\partial_{m_k} h \over h^{2p+2}}\right) 
{\eta_{\mu\nu}\over M_p^8}\prod_{r, s} {g}^{m_r m_s}, \nd
where we have expressed everything in terms of regular derivatives and inverse {\it unwarped} metric 
$g^{mn}$ so that 
\eqref{mojarmet} doesn't have to involve covariant derivatives. In fact the way we have written  the quantum terms in \eqref{phingsha2}, all informations of the internal metrics etc are contained in the definitions of the curvature tensors and the inverse metric components, and not in the derivatives. In this sense 
\eqref{mojarmet} has all the information in the warp-factor $h(y)$, and since $p+q = 4$, the allowed terns are $(p, q) = (4, 0), (3, 1), (2, 2), (1, 3)$, all being time-neutral by construction; and all suppressed by 
$M_p^8$. This $M_p$ suppression remains unchanged even if we add curvature tensors contributions to 
\eqref{mojarmet}.  The curvature tensors, at least those that could contribute to the space-time directions, are limited to only four tensors at a time because time-neutrality implies:
\bg\label{umakim}
2\sum_{i = 1}^{27} l_i + n_1 + n_2  + l_{34} = 8, \nd
thus $l_i \le 4$, and where many of the 27 $l_i$'s appearing in \eqref{phingsha2} are irrelevant to 
\eqref{umakim}. An example of such a term with only curvature tensors can be taken for 
$l_8 = l_9 = l_{10} = l_{13} = 1$  in \eqref{phingsha2} which allows us to choose 
$E_1 = 3, E_2 = E_3 = E_4 = E_5 =1$ from \eqref{lengchink} or \eqref{teenbar}. This gives:
\bg\label{monox}
\Lambda^{(4)}_{ij} \equiv M_p^{-8}{\bf R}_{i_1aj_1b} {\bf R}_{i_2\alpha j_2\beta} {\bf R}_{i_30 i0} 
{\bf R}_{i_4 mj n}
{\bf g}^{ab} {\bf g}^{\alpha\beta}{\bf g}^{mn}{\bf g}^{i_1i_2}{\bf g}^{i_3i_4}{\bf g}^{j_1j_2}{\bf g}^{00}, \nd
which is interestingly not just expressed in terms of the warp-factor $h(y)$ but also in terms of the 
temporal and spatial derivatives of the internal metric components. One can also mix  three curvature tensors and two derivatives or two curvature tensors and four derivatives etc satisfying \eqref{umakim} appropriately to generate additional terms. 
All these quantum terms are finite in number and they are all suppressed by $M_p^8$ 
(with $\Delta k > {1\over 3}$, the finiteness of quantum terms still remain and can be easily constructed). As we saw earlier, there are {\it no} time-neutral contributions that can come from \eqref{melamon2}, so the $M_p^8$ suppression cannot change. In fact exactly similar story could be constructed with \eqref{miai}, so we will not discuss this case separately here.



\subsubsection{Non-local counter-terms in M-theory and in type IIB \label{Gng6}}

The next set of quantum corrections are a bit unusual from standard quantum field theory, or even supergravity, point of view and are typically christened as non-local counter-terms. Such an umbrella term encompass a broad category of  quantum terms in M-theory, for which a detailed analysis is clearly beyond the scope of our work here. As such we will suffice ourselves here with some rudimentary exploration of the subject in the context of M-theory.

Our starting point would be to take the generic quantum terms in \eqref{phingsha} and \eqref{phingsha2}
and construct non-local interactions from them, as we believe that the non-local interactions should still contain powers of curvature tensors, G-fluxes and their covariant-derivatives. To proceed, let us denote  the specific quantum term of 
\eqref{phingsha} or \eqref{phingsha2} alternatively using the symbol $\mathbb{Q}_{\rm T}^{(\{l_i\}, n)}$ so that specific choice of the ($l_i, n$) integers, the former representing the powers of curvature tensors and G-fluxes and the latter representing the number of derivatives, allow us to specify one quantum term. It is clear that:
\bg\label{mcgillmey}
\left(\mathbb{Q}_{\rm T}^{(\{l_i\}, n)}\right) \otimes  \left(\mathbb{Q}_{\rm T}^{(\{l_j\}, m)}\right) 
  \equiv \mathbb{Q}_{\rm T}^{(\{l_i + l_j\}, n+ m)}, \nd
which may be easily derived using the explicit expression from either \eqref{phingsha} or 
\eqref{phingsha2}. The equality 
\eqref{mcgillmey} tells us that  an arbitrary product of any two elements in the set of all the quantum pieces 
labelled by $\left\{\mathbb{Q}_{\rm T}^{(\{l_k+l_m\}, n)}\right\}$ is also an element of the set. This is almost like giving a group structure to the set, except that the set doesn't have an inverse. The elements of the set may even be further generalized by introducing the following notation:

{\footnotesize
\bg\label{chukkaM}
t^{i_1i_2.....i_{2q}} &\equiv& \epsilon^{i_1i_2......i_{2q}} + 
c_1\left[ \left(g^{i_1i_3} g^{i_2i_4} - g^{i_1i_4}g^{i_2i_3}\right)....\left(g^{i_{2q-3} i_{2q-1}}g^{i_{2q-2}i_{2q}}
- g^{i_{2q-3} i_{2q}}g^{i_{2q-2}i_{2q-3}}\right) + ...\right]  \nonumber\\
&+&  {\rm permutations}, \nd}
where $c_1$ is a constant and the permutations are between other products of metrics to generate full anti-symmetry, and 
$\epsilon^{i_1i_2....i_{2q}}$ is the Levi-Civita tensor and {\it not} a tensor density. As such, with all it's indices lowered, it may be defined with the 
 square root of determinant of metrics and therefore scales in exactly the same way as the product of inverse metrics. However because of the total anti-symmetry of the Levi-Civita tensor (or of the anti-symmetric products of metrics), we cannot have too many of these terms at a given order.
This implies that, if we remove all the derivatives in say \eqref{phingsha}, and taking $q= 4$ in \eqref{chukkaM}, it is easy to get terms like:
\bg\label{karfish}
&&\mathbb{Q}_1 \equiv M_p^{-2}t^{i_1 i_2 ...... i_8} {\bf G}_{i_1i_2i_3i_4} {\bf G}_{i_5i_6i_7i_8}\nonumber\\
&& \mathbb{Q}_2 \equiv M_p^{-8}t^{i_1i_2......i_8} t^{j_1j_2..... j_8}{\bf R}_{i_1i_2j_1j_2} {\bf R}_{i_3 i_4 j_3 j_4} {\bf R}_{i_5 i_6 j_5 j_6} {\bf R}_{i_7 i_8 j_7 j_8}, \nd
with $i_k$ denoting coordinates of the internal eight-manifold, and $\mathbb{Q}_2$ can be identified with the famous $t_8 t_8 {\bf R}^4$ coupling in string theory \cite{witteng}.
It should be clear that the $g_s$ scalings of 
$\mathbb{Q}_1$ and $\mathbb{Q}_2$  are identical to the $g_s$ scalings of 
$\mathbb{Q}_{\rm T}^{(0,.., l_{19} = 2, .., 0; 0)}$ and $\mathbb{Q}_{\rm T}^{(l_1 = 4, 0,.., 0; 0)}$ respectively in 
\eqref{phingsha}.  Other combinations with curvature tensors and G-fluxes are clearly possible, and their $g_s$ scalings would be identical to the $g_s$ scalings of corresponding terms in 
$\mathbb{Q}_{\rm T}^{(l_i, n = 0)}$ at the  same order in curvature tensors and G-fluxes. This story could be elaborated to the same extent as earlier sections\footnote{Beyond the possible generalization to 
$\sum_k d_k \mathbb{Q}_1^k$ and $\sum_l f_l \mathbb{Q}_2^l$ with integer ($d_k, f_l$).}, 
but since we are only concerned with the $g_s$ scalings, we will not indulge in further discussions of the topic here.

Thus combining \eqref{karfish}, with their possible generalizations, and with the set of terms of the form \eqref{phingsha} or \eqref{phingsha2}, we have pretty much all the local perturbative and non-perturbative quantum terms (more details on the latter in subsection \ref{instachela}) at hand. The non-local quantum terms, which we label as non-local counter-terms, are a different class of objects which could nevertheless be related to the local terms \eqref{karfish}, 
\eqref{phingsha} and \eqref{phingsha2}. For example we could easily construct the following non-local counter-terms\footnote{See also \cite{calagni}
for operators of the form \eqref{koliman} and their possible connection to Witten's open string field theory.
This fascinating subject deserves more attention, but unfortunately any elaboration here will stray us from the main course of this paper.}:
\bg\label{koliman}
\mathbb{W}^{(\{l_i\}, n)} = 
\left(\sum_{q = 1}^{\infty} {C_q M_p^{2q}\over \square^q}\right)\mathbb{Q}_{\rm T}^{(\{l_i\}, n)}, \nd
where $\square$ is defined over the eight-manifold ${\cal M}_2 \times {\cal M}_4 \times 
\mathbb{T}^2/{\cal G}$ and $C_q$ could in general be function of the $y \equiv (y^m, y^\alpha, y^a)$ and of $(g^\Delta_s, e^{-1/g^\Delta_s})$ but only implicitly on $M_p$ (to first approximation) as we want to isolate the $M_p$ dependence as in \eqref{koliman}. The $g_s$ scalings appear from the quantum pieces 
$\mathbb{Q}_{\rm T}^{(\{l_i\}, n)}$ and also $C_q$. The inverse $\square$ operators may be combined together to create 
operators of the form ${\rm exp}\left({M_p^2\over \square}\right), {\rm sin}\left({M_p^2\over \square}\right)$ etc generating different levels of non-locality. All these operator actions may in turn be re-expressed as 
integrals which are much easier to handle. To elaborate this, let us first define the non-locality function 
$\mathbb{F}^{(r)}(y - y') \equiv \mathbb{F}^{(\{l_i\}, n; r)}(y - y')$ that is a function of two points $(y, y')$ on the eight-manifold, with $r$ denoting the level of non-locality. By construction the non-locality function should  become a sharply peaked function at low energies so that the low energy physics of M-theory could still be governed by 
local counter-terms, and hence by eleven-dimensional supergravity. On the other hand, the short distance 
behavior of this function could be complicated, revealing the full non-local structure of the 
system\footnote{We should keep in mind that the standard time-independent supergravity compactification of M-theory at the level of \cite{BB} is not in the swampland, and the full non-compact M-theory (or string theory) is also not in the swampland. This means, if $M_p^8 V^{(1)}_8$ denotes the {\it warped} volume of the eight-manifold in \eqref{vegamey}, in the limit $M_p^8 {V}^{(1)}_8 \to \infty$ the non-locality function should 
decouple. It should also decouple in compactifications of the form studied in \cite{BB}. To determine a generic function satisfying the above criteria,  let us denote a scale 
${\cal V}_8 \equiv V^{(1)}_8 - V^{(2)}_8$, where $V^{(2)}_8$ is the warped volume of the eight-manifold 
in \eqref{betbab}. In our coherent state language, ${\cal V}_8$ is easy to motivate: it is the volume of the eight-manifold coherent-state over the vacuum solitonic configuration \eqref{betbab}. Using this we can conjecture the following form of the non-locality function:
$$\mathbb{F}^{(r)}(y - y') \equiv 
\sum_{\{n\}} C^{(r)}_{\{n\}}~ \mathbb{H}_{\{n\}}\left({y - y'\over {\cal V}^{1/8}_8}\right)
{\rm exp}\left[-{(y - y')^2\over {\cal V}^{1/4}_8}\right]$$
\noindent where $\mathbb{H}_{\{n\}}\left({y - y'\over {\cal V}^{1/8}_8}\right)$ are the eight-dimensional
Hermite polynomials and $C^{(r)}_{\{n\}}$ are constants. Due to the orthogonality properties of the Hermite polynomials, any arbitrary non-locality function may be constructed with appropriate choices of the 
constant coefficients $C^{(r)}_{\{n\}}$. From the above series, the way one would proceed would be to first determine the non-locality function for a given scenario and then use it to construct the quantum terms as in \eqref{pmoran}. 
We can also perform a few checks to see if the function satisfies the required properties. In the standard supergravity limit of \cite{BB}, $V^{(1)}_8 = V^{(2)}_8$ and therefore $\mathbb{F}^{(r)}(y - y')$ is arbitrarily small at all points and thus decouples as a delta function. In the large volume limit where ${\cal V}_8 \to \infty$ for a fixed choice of $M_p$, all the Hermite polynomials, as well as the Gaussian piece, are constants and therefore 
$\mathbb{F}^{(r)}(y - y')$ is a constant. In the zero volume limit, i.e in the limit ${\cal V}_8 \to 0$ (which is not the supergravity limit of \cite{BB} where ${\cal V}_8 = 0$ but $V^{(1)}_8 = V^{(2)}_8 \ne 0$), the Gaussian part and the Hermite polynomials appear to become arbitrarily small and arbitrarily large respectively. However now 
$(y - y')^2$ is also very small, as any two points are very close to each other. The exponential term in general always dominates over any polynomial powers, but since we are summing over {\it all} Hermite polynomials, this could be subtle. Thus depending on how the series above is arranged, there could be a non-trivial 
$\mathbb{F}^{(r)}(y - y')$. In the limit $g_s \to 0$, $V^{(1)}_8 \to \infty$, so again the Gaussian function becomes arbitrarily small. On the other hand, in the limit where both the size of the torus 
${\mathbb{T}^2\over {\cal G}}$  and $g_s$ vanish, $V^{(1)}_8$ can remain finite but 
$V_8^{(2)} \to 0$. Since $(y - y')^2$ is no longer required to vanish, $\mathbb{F}^{(r)}(y - y')$ contributes 
non-trivially to the late time physics in the type IIB side. 
Away from these limits, $\mathbb{F}^{(r)}(y - y')$ generically remains a non-trivial function. Note that the 
{\it implicit} dependence on $M_p$ (as well as on $g_s$) of $\mathbb{F}^{(r)}(y - y')$ in the above series is only to the first approximation. One can always add another series, to the existing one, that explicitly depends on $M_p$ and satisfy all the criteria mentioned above. In fact our analysis reveals that 
$\mathbb{F}^{(r)}(y - y')$ has {\it four} possible contributions that depend (a) implicitly on $g_s$ and $M_p$ (as shown here), (b) implicitly on $g_s$ but explicitly on $M_p$, 
(c) explicitly on $g_s$ but implicitly on $M_p$ (an example will be presented soon), and finally (d) explicitly on $g_s$ and $M_p$. Needless to say, all of these should satisfy the criteria that went in the construction of (a). In fact, as long as these are taken care of, and with the possibility that a function  
$\mathbb{F}^{(r)}(y - y')$ could in principle exist (at least with the contribution (a) above), we will not be required to specify the actual form of $\mathbb{F}^{(r)}(y - y')$. See also \cite{petite}.
\label{chascage}}.
Using this function, let us define our first level of non-locality with zero free Lorentz indices using \eqref{phingsha} for example as:
\bg\label{pmoran}
\mathbb{W}^{(1)}(y) \equiv \mathbb{W}^{(\{l_i\}, n; 1)} 
= \int d^8 y' \sqrt{{\bf g}_8} \left({ \mathbb{F}^{(1)}(y - y') 
\mathbb{Q}_{\rm T}^{(\{l_i\}, n)} (y') \over M_p^{\sigma(\{l_i\}, n) - 8}}\right),  
 \nd
where the power of $M_p$ appearing above, i.e $\sigma(\{l_i\}, n)$ is defined in \eqref{sigme}, and the integral captures the first level of non-locality as advertised before. By construction $\mathbb{W}^{(1)}$ is dimensionless, and the non-locality appears from knowing the precise functional form for 
$\mathbb{F}^{(1)}(y - y')$,  which fortunately we won't need to specify. Suffice is to say that the $g_s$ dependence mostly appears from the quantum terms $\mathbb{Q}^{(\{l_i\}, n)}$ defined in \eqref{phingsha} and \eqref{karfish} (see footnote \ref{chascage}). We can also sum over all allowed choices of $(\{l_i\}, n)$ and, using the semi-group structure 
\eqref{mcgillmey}, the linear representation of the sum pretty much captures the generic picture. It should be clear that the $r$-th level of non-locality may be iteratively constructed from:
\bg\label{karalura} 
\mathbb{W}^{(r)}(y) &=& M_p^8 \int d^8y' \sqrt{{\bf g}_8(y')} ~\mathbb{F}^{(r)}(y - y') \mathbb{W}^{(r-1)}(y')\\
& = & M_p^{16} \int d^8 y' \sqrt{{\bf g}_8(y')}~\mathbb{F}^{(r)}(y - y') 
\int d^8y'' \sqrt{{\bf g}_8(y'')} ~\mathbb{F}^{(r-1)}(y' - y'') \mathbb{W}^{(r-2)}(y''), \nonumber
 \nd  
thus forming a series of nested integrals that capture the full non-locality of the system, for a given choice of 
$(\{l_i\}, n)$.
Clearly as $r$ increases the non-locality becomes more prominent  and starts coinciding with the non-locality generated from the operator action \eqref{koliman}. One expects: 
\bg\label{conner}
\sum_{\{l_i\}, n}\sum_{r = 1}^\infty b_r \mathbb{W}^{(r)}(y) = \sum_{\{l_i\}, n} f_{\{l_i\}, n} 
\mathbb{W}^{(\{l_i\}, n)}(y), \nd
with constants $b_r$ and $f_{\{l_i\}, n}$, as we can absorb all $y$-dependent factors in $\mathbb{F}^{(r)}(y)$ of \eqref{karalura} and
$C_q(y)$ of \eqref{koliman} respectively. Such a relation would not only justify the two forms of non-localities 
\eqref{koliman} and \eqref{karalura} as one and the same thing,  but would also help us relate $C_q(y)$ functions with the $\mathbb{F}^{(r)}(y)$ functions. A formal proof of \eqref{conner} is still lacking, despite evidences pointing towards the veracity of the conjecture. However since we will mostly concentrate on the non-localities of the form \eqref{karalura}, 
the exact equivalence depicted in \eqref{conner} will not be used here, and therefore the proof of 
\eqref{conner} will be relegated to future work. We do note that,  $\mathbb{W}^{(\infty)}(y)$ should be 
related to the $q \to \infty$ value of \eqref{koliman} when appropriately summed over ($\{l_i\}, n$) factors therein as,
at a given level of non-locality, the $M_p$ suppression changes from \eqref{sigme} or \eqref{kamagni2} to:
\bg\label{kamagni}
\sigma(\{l_i\}, n; r) \equiv \sigma_r = \sigma(\{l_i\}, n) - 8r, \nd
and therefore has both positive and negative values\footnote{Other contributions to 
$ \mathbb{F}^{(r)}(y - y')$ that explicitly depend on $M_p$, as discussed in footnote \ref{chascage}, could allow additional non-local contributions with different $M_p$ scalings. However for us all we need is a set of non-local quantum terms that have $M_p$ scalings as in \eqref{kamagni}. Additional contributions can and will exist, but they will not change the outcome.}. 
These additional positive and negative suppressions of the quantum terms were responsible for the loss of $M_p$ hierarchy as discussed in \cite{nodS}. Here our aim would be to see how the conclusions of 
\cite{nodS} may be avoided. 

To inquire how the $g_s$ scaling appears now, we will have to work out the non-localities order by order in 
$r$. We first work out the lowest level of non-locality from \eqref{pmoran}. Using the metric ansatze 
\eqref{vegamey} with the warp-factor as defined in \eqref{mrglass}, the non-local quantum piece 
\eqref{pmoran} yields:

{\footnotesize
\bg\label{telmaish}
\mathbb{W}^{(1)}(y) & = & \int d^8y' F_1(t) F_2^2(t) g_s^{-2/3} h^{3/2}
\sqrt{\left({\rm det}~g_{\alpha\beta}\right)\left({\rm det}~g_{mn}\right)\left({\rm det}~g_{ab}\right)}
 \left({ \mathbb{F}^{(1)}(y - y') 
\mathbb{Q}_{\rm T}^{(\{l_i\}, n)} (y') \over M_p^{\sigma(\{l_i\}, n) - 8}}\right)\nonumber\\
& = & \int d^8y' \left(e_0 g_s^{-2/3} + {e_1g_s^{4/3}\over \sqrt{h}}\right) {\bf V}_8(y')
 \left[{ \mathbb{F}^{(1)}(y - y') g_s^{\Theta_k}
\left(\widetilde{\mathbb{Q}}_{\rm T}^{(\{l_i\}, n)} (y') + {\cal O}(y', g^\Delta_s, e^{-1/g^\Delta_s})\right)\over M_p^{\sigma(\{l_i\}, n) - 8}}\right], \nonumber\\ \nd}
where in the second line we have used the relation \eqref{ranjhita2} to express the $g_s$ scalings of both the volume-preserving (i.e \eqref{olokhi} with $(e_0, e_1) = (1, 0)$),  and the fluctuating (i.e \eqref{ranjhita} with $(e_0, e_1) = (0, 1)$) cases (special care needs to be used to define the quantum pieces for the two cases \eqref{ranjhita} and \eqref{olokhi} as the former uses \eqref{phingsha} and the latter uses 
\eqref{phingsha2}. Modulo this subtlety, everything else remains identical.). The $g_s$ scalings of all the quantum terms in \eqref{phingsha} and 
\eqref{phingsha2} are expressed using $\Theta_k \equiv \Theta_k(\{l_i\}, n)$ which would cover for the two cases, \eqref{melamon2} related to \eqref{olokhi} and \eqref{miai} related to \eqref{ranjhita}. The 
$\widetilde{\mathbb{Q}}_{\rm T}^{(\{l_i\}, n)}(y')$ represent the spatial parts of the quantum terms 
\eqref{phingsha} and \eqref{phingsha2}
that do not depend on $e^{-1/g^\Delta_s}$. 
 Finally ${\bf V}_8(y')$ is defined as:
\bg\label{zaroori}
{\bf V}_8(y') \equiv h^{3/2}(y') \sqrt{\left({\rm det}~g_{\alpha\beta}\right)\left({\rm det}~g_{mn}\right)\left({\rm det}~g_{ab}\right)}, \nd
which would contribute to the warped volume of the internal space when integrated over the eight-manifold. 
All the metric components depend on coordinates of the eight-manifold generically, but there are certain 
constraints that restricted the dependences to certain sub-space of the internal manifold. Such constraints will help us evaluate the quantum terms in \eqref{telmaish} for the two cases, \eqref{olokhi} and 
\eqref{ranjhita}, and also compare our results with the generic case discussed in \cite{nodS}. 

To start, let us first consider the simplified case where $h(y) = h(y_0)\equiv h_0$ where $y_0$ is a chosen special point inside the eight-manifold. Such a choice allows us to choose the same string coupling $g_s$ at every order of the non-locality. 
All other variables, for example the metric components, remain functions of $y$ coordinates. Under such a simplification the $g_s$ scaling of the $r$-th level of non-locality becomes:
\bg\label{pdesimey} 
\mathbb{W}^{(r)}(y_{r+1}) = {1\over M_p^{\sigma_r}}\left(e_0 g_s^{-2/3} + {e_1g_s^{4/3}\over
 \sqrt{h_0}}\right)^r g_s^{\Theta_k}\mathbb{G}_8(y_{r+1}),  \nd 
which is defined for a given choice of $(\{l_i\}, n)$, and we have made a judicious coordinate choice of 
$y_{r+1}$ to label the non-local quantum term with zero Lorentz index\footnote{We take $y_0 = 0$ to comply with our choice of coordinates.}. The power of $M_p$ suppression may be read out from \eqref{kamagni} for the given choice of $(\{l_i\}, n)$, and the functional form for 
$\mathbb{G}_8(y_{r+1})$ may be expressed in terms of the nested integrals in the following way:
\bg\label{rambha}
\mathbb{G}_8(y_{r+1}) \equiv \prod_{q = 0}^{r -1} \int d^8y_{r-q} {\bf V}_8(y_{r-q}) \mathbb{F}^{(r-q)}(y_{r-q} - y_{r-q-1}) \left(\widetilde{\mathbb{Q}}_{\rm T}^{(\{l_i\}, n)}(y_1) 
+ {\cal O}(y_1, g^\Delta_s, e^{-1/g^\Delta_s})\right), 
\nonumber\\ \nd
with ${\bf V}_8(y_{r-q})$ being taken from \eqref{zaroori} with the constant choice of the warp-factor 
$h_0$. The nested integrals are expressed in terms of the ${\bf V}_8(y')$ and $\mathbb{F}^{(r)}(y - y')$, and this may help us to distinguish between the two choices, \eqref{olokhi} and \eqref{ranjhita}; and also between the generic case discussed in \cite{nodS}. By construction \eqref{rambha} will always be finite because the integrals are over finite domains, and the non-locality functions $\mathbb{F}^{(r)}(y - y')$ are chosen to be normalizable functions.

\vskip.1in

\noindent {\it Case 1: $F_1(t)$ and $F_2(t)$ satisfying the fluctuation condition \eqref{ranjhita}}

\vskip.1in

\noindent First, let us consider the choice \eqref{ranjhita} where the inverse of $F_2(t)$ has a perturbative expansion but the inverse of $F_1(t)$ does not. This means $e_0 = 0$ and $e_1 = 1$ in \eqref{pdesimey}. Additionally because of the derivative constraint there, all variables were taken to be functions of the coordinates 
of ${\cal M}_4$, and were thus independent of both ${\cal M}_2$ and $\mathbb{T}^2/{\cal G}$ coordinates. We will however take the warp-factor $h(y^m) = h_0$ as before to avoid changing the string coupling $g_s$ to any order in non-locality. Similarly, the non-locality functions will be taken to be functions of ${\cal M}_4$   
only. Putting everything together, \eqref{pdesimey} changes to:
\bg\label{susmita}
\mathbb{W}_1^{(r)}(y_{r+1}) = \left({\mathbb{G}_4(y_{r+1})g_s^{4r/3 + \theta_k} \over M_p^{\sigma_r}
\sqrt{h_0}}\right) \mathbb{V}^r_{\mathbb{T}^2}\mathbb{V}^r_{{\cal M}_2}, \nd
where the volume elements are defined as: $\mathbb{V}_{\mathbb{T}^2} = \int d^2y^a 
\sqrt{{\rm det}~g_{ab}}$ for the volume of the subspace $\mathbb{T}^2/{\cal G}$  
and $\mathbb{V}_{{\cal M}_2} = \int d^2y^\alpha 
\sqrt{{\rm det}~g_{\alpha\beta}}$ for the volume of the subspace ${\cal M}_2$. The metric components
$g_{ab}$ and $g_{\alpha\beta}$ are the un-warped metric coefficients that appear in \eqref{vegamey}. Note that the $r$-th level of non-locality requires these volume elements to be raised to the $r$-th powers, as evident from \eqref{rambha} above. The $g_s$ scaling for 
a choice of $(\{l_i\}, n)$ has the expected $\theta_k$ dependence from \eqref{miai}, but the non-locality adds another $+4r/3$ piece to it. This means that, there are no additional time-neutral pieces generated by non-locality here as $\theta_k$ from \eqref{miai} doesn't have any time-neutral solutions with 
$\Delta k \ge {3\over 2}$. Finally, the $\mathbb{G}_4(y^m_{r+1})$ factor has the following nested integral representation as \eqref{rambha}:
\bg\label{susen}
\mathbb{G}_4(y_{r+1}) \equiv \prod_{q = 0}^{r-1}\int d^4y_{r-q}\sqrt{g_4}~
\mathbb{F}^{(r-q)}(y_{r-q} - y_{r-q-1}) \left(\widetilde{\mathbb{Q}}_{\rm T}^{(\{l_i\}, n)}(y_1) 
+ {\cal O}(y_1, g^\Delta_s, e^{-1/g^\Delta_s})\right), \nonumber\\ \nd
where $g_4 = {\rm det}~g_{mn}$ with the integral defined over the subspace ${\cal M}_4$; and we have absorbed the factor of $h_0^{3/2}$ in the definition of $g_4$. The function 
$\mathbb{G}_4(y)$ captures the additional ${\cal O}(g^\Delta_s, e^{-1/g^\Delta_s})$ corrections and thus responsible for the perturbative and non-perturbative series in $g_s$. This is as what one would have expected, although a question might be raised on the dependence of the non-locality function $\mathbb{F}^{(r)}(y-y')$ only on 
${\cal M}_4$ coordinates. This may be justified, beyond declaring it as an imposed condition, by looking at 
\eqref{koliman} in the limit $q = 0$. In this limit $\mathbb{W}^{(\{l_i\}, n)}$, i.e for $q = 0$,  becomes a local function and therefore the derivative constraints will imply that the coefficients $C_0(y)$ will have to be a function of 
${\cal M}_4$ coordinates. Similarly taking $q = 1$, $\square\mathbb{W}^{(\{l_i\}, n)}$ becomes a local function and therefore $C_1(y)$ will have to be function of ${\cal M}_4$ coordinates. Following this chain of logic, $C_q$ for any $q$ becomes a function of ${\cal M}_4$ coordinates. Therefore at this stage, using the identification 
\eqref{conner}, the functions $\mathbb{F}^{(r)}(y - y')$ should only depend on the coordinates of 
${\cal M}_4$, justifying the integral representation \eqref{susen}.

All the above conclusions are good, and they get even better once we allow quantum  terms with two free Lorentz indices. The story evolves in the same way as above, so we will suffice ourselves in elaborating the $g_s$ scalings of the various terms. Looking at \eqref{teenangul}, and comparing it with \eqref{susmita}, 
the $g_s$ scaling become $g_s^{{\tilde\theta}_k}$, where:
\bg\label{akiasa}
{\tilde\theta}_k = \left(\theta_k + {4\over 3}(r - 2), \theta_k + {2\over 3}(2r - 1), \theta_k + {4\over 3}(r +1)
\right), \nd
with the first one corresponding to free Lorentz indices along ($i, j$) and ($0, 0$) directions; the second one corresponds to  free Lorentz indices along ${\cal M}_4$, i.e along ($m, n$) directions and the third one corresponds to free Lorentz indices along $\mathbb{T}^2/{\cal G}$ and ${\cal M}_2$ i.e along ($a, b$) and ($\alpha, \beta$) directions respectively. From \eqref{akiasa} we see that even with the lowest level of non-locality i.e with $r = 1$, there are no additional time-neutral series along $(m, n), (a, b)$ and 
$(\alpha, \beta)$ directions. Even more interestingly, since at the end we have to go to type IIB from M-theory, we can take the following limit\footnote{Note that this is an imposed condition on the un-warped 
volume of $\mathbb{T}^2/{\cal G}$. The warped eleven-dimensional radius $\mathbb{R}_{11}$ is related to $g_s$ via \eqref{recutvi}, so automatically goes to zero when $g_s \to 0$ at late time, i.e when $t \to 0$ in our choice of flat-slicing (see footnote \ref{cori}). The condition \eqref{tarfox} then provides a type IIB description at all time.} for fixed $M_p$:
\bg\label{tarfox}
\mathbb{V}_{\mathbb{T}^2} ~ \rightarrow~ 0, \nd
any additional time-neutral series along the $(i, j)$ and ($0, 0$) directions are heavily suppressed by 
powers of  $\mathbb{V}_{\mathbb{T}^2}$, which in turn should also be the case with zero free Lorentz index 
in \eqref{susmita}. 

\vskip.1in

\noindent {\it Case 2: $F_1(t)$ and $F_2(t)$ satisfying the volume-preserving condition \eqref{olokhi}}

\vskip.1in

\noindent The story that we elaborated for case 1 pretty much sums up all the procedure that we need for the present case where both $F_1(t)$ and $F_2(t)$ have perturbative expansions, including their inverses. However there are now a few crucial differences that will alter our story in an interesting way. First, the derivative constraints are weakened from case 1 in a way that we no longer restrict the derivatives to be along 
${\cal M}_4$ only. We do however want the functions to be independent of the ($x_3, x_{11}$) directions so that components like ${\bf G}_{MNab}$ do not complicate our analysis by switching on KK modes for 
($l_{36}, l_{37}, 
l_{38}$) in \eqref{phingsha2}\footnote{This is however only true if the ${\bf G}_{MNab}$ flux components are global fluxes. Once we entertain other possibilities, things do get more involved. See discussions later. \label{westgate}}. 
Therefore now we can allow all curvature tensors and G-fluxes to be functions of ${\cal M}_2 \times {\cal M}_4$, implying that, in the type IIB side, all curvature tensors and fluxes would  be functions of the six-dimensional internal space. This is good because the derivative constraint for 
case 1 was a tad bit un-natural in the light of the genericity that we want to impose on the quantum corrections.  The $r$-th level of non-locality may now be read from \eqref{pdesimey} by using $e_0 = 1$ and $e_1 = 0$ and using the quantum terms from \eqref{phingsha2}.  We will use the same approximation for the warp-factor, namely $h(y) = h_0$ to avoid changing  $g_s$ to any order in the non-locality. Putting everything together, \eqref{pdesimey} for the present case becomes:
\bg\label{lengmey}
\mathbb{W}_2^{(r)}(y_{r+1}) = \left({\mathbb{G}_6(y_{r+1})g_s^{-2r/3 + \theta'_k} \over M_p^{\sigma_r}
\sqrt{h_0}}\right) \mathbb{V}^r_{\mathbb{T}^2}. \nd
Compared to \eqref{susmita} there are a few key differences. First, there is no volume element 
$\mathbb{V}_{{\cal M}_2}$ appearing anymore because this goes inside $\mathbb{G}_4(y)$, as defined 
in \eqref{susen} to construct $\mathbb{G}_6(y)$. In other words, $\mathbb{G}_6(y)$ takes the following form:
\bg\label{chinmey}
\mathbb{G}_6(y_{r+1}) \equiv \prod_{q = 0}^{r-1}\int d^6y_{r-q}\sqrt{g_6}~
\mathbb{F}^{(r-q)}(y_{r-q} - y_{r-q-1}) \left(\widetilde{\mathbb{Q}}_{\rm T}^{(\{l_i\}, n)}(y_1) 
+ {\cal O}(y_1, g^\Delta_s, e^{-1/g^\Delta_s})\right), \nonumber\\ \nd
where again we have absorbed a factor of $h_0^{3/2}$ in the definition of $g_6$ and 
$\widetilde{\mathbb{Q}}_{\rm T}^{(\{l_i\}, n)}(y_1)$ being extracted from \eqref{phingsha2}.
 The second key difference, which is important, is the $g_s$ scaling. Using the original $g_s$ scaling 
 \eqref{melamon2} with zero Lorentz index for the quantum terms associated with the case \eqref{olokhi}, we now see that the $r$-th
order of non-locality now adds a factor of $-2r/3$ to the original scaling in the local case.  Recall that 
$\theta'_k$ as defined in \eqref{melamon2} for $\Delta k > {1 \over 3}$  {\it did not} have any time-neutral series, but now it appears that the non-locality would in fact help to create more time-neutral series. This is actually not an issue because for a fixed choice of $r$, $\Delta k > {1\over 3}$ will still allow only finite number of terms in \eqref{phingsha2}. Additionally, explicit dependence of $\mathbb{F}^{(r)}(y - y')$ on 
$g_s$ can also effect the $g_s$ scalings a bit. 
With two free Lorentz indices, the $g_s$ scaling now appears to be $g_s^{{\tilde\theta}'_k}$, where:
\bg\label{tamtam}
{\tilde\theta}'_k = \left(\theta'_k - {2\over 3}\left(r + 4\right), \theta'_k - {2\over 3}\left(r + 1\right), 
\theta'_k - {2\over 3}
\left(r -2\right)\right). \nd
In addition to the difference with the scaling behavior in \eqref{akiasa}, there are a few other differences.
The first one is in the ordering of the scaling behavior as it appears in \eqref{tamtam}. 
The first term in \eqref{tamtam} corresponds to free Lorentz indices along ($i, j$) and ($0, 0$) directions; but the second term corresponds to  free Lorentz indices along ${\cal M}_4$ as well as ${\cal M}_2$, i.e along ($m, n$) and ($\alpha, \beta$) directions respectively. The third term now corresponds to free Lorentz indices along 
$\mathbb{T}^2/{\cal G}$ i.e along ($a, b$) direction. 

The second difference between \eqref{akiasa} and \eqref{tamtam} appears from the value of $r$, i.e from the level of non-locality. While in \eqref{akiasa} increasing $r$ makes all the three terms there positive definite, in \eqref{tamtam} the effect is opposite. Increasing $r$ in \eqref{tamtam} actually creates  more relative minus signs, but as we saw earlier for a given $r$ there are still {\it finite} number of terms\footnote{For both the cases discussed here, note that $\theta'_k$ and $\theta_k$ from \eqref{tamtam} and \eqref{akiasa} respectively depend on $r$. This means $l_i = l_i(r)$ and $n = n(r)$, which in turn implies that the $M_p$ scaling \eqref{kamagni} also develops implicit $r$ dependence from  ($l_i, n$) in addition to the explicit $r$ dependent $-8r$ piece. This could of course change depending on the other contributions to the non-locality function that effect the $g_s$ and the $M_p$ scalings as elucidated in footnote \ref{chascage}.}. 
Additionally, the degree of non-locality is also suppressed by powers of $\mathbb{V}_{\mathbb{T}^2}$, as may be inferred from \eqref{lengmey}, and in the limit when the volume $\mathbb{V}_{\mathbb{T}^2}$ vanishes, for a fixed choice of $M_p$, all the additional series also decouple completely. The vanishing of 
$\mathbb{V}_{\mathbb{T}^2}$ is an essential requirement for our M-theory construction to connect it to type IIB theory.

\vskip.1in

\noindent {\it Case 3: Time-independent internal space with $F_1(t)= F_2(t) = 1$}

\vskip.1in

\noindent The non-localities discussed above do change the number of quantum terms for a given choice of $r$ in both \eqref{susmita} and \eqref{lengmey}, although finiteness of the number of quantum terms for given $g_s$ remains unchanged. 
The question is what happens when the internal space is time independent i.e when
$F_1(t) = F_2(t) = 1$? We expect the story to progress more or less in the same vein as above, and in fact most of the details remain somewhat similar to case 2 above, but with one crucial difference. Since 
${\bf G}_{MNab}$ features prominently in the discussion concerning this case, as evidenced from 
\eqref{salom} and \eqref{montan}, which in turn are responsible for the time-neutrality condition 
\eqref{evabmey2} with zero free Lorentz indices, all curvature tensors and G-fluxes in the theory need to be functions of 
${\cal M}_4 \times {\cal M}_2 \times \mathbb{T}^2/{\cal G}$ coordinates except the $x_3$ direction. 
In addition, there is as such no derivative condition imposed from the dynamics, the non-locality function 
$\mathbb{F}^{(r)}(y - y')$ could in principle be function of $x_3$ also by allowing components like
${\bf C}_{MN, 11}(y, x_3)$ in addition to \eqref{montan} (see footnote \ref{sirisayz}). The $r$-th level of non-locality then becomes:
\bg\label{japumey}
\mathbb{W}_3^{(r)}(y_{r+1}) =  {\mathbb{G}_8(y_{r+1})g_s^{-2r/3 + \theta'_0} \over M_p^{\sigma_r}
\sqrt{h_0}}. \nd
where $\theta'_0$ is as given in \eqref{kkkbkb2}, which already allows time-neutral series because 
there are relative minus signs due to the presence of ($l_{36}, l_{37}, l_{38}$) as well as $n_3$. We now see that the $r$-th level of non-locality creates additional relative minus signs that, for a given choice of $r$,  help in generating a different set of time-neutral series here. Similar picture emerges with two free Lorentz indices, as one may easily derive. Note also the absence of volume components like $\mathbb{V}_{\mathbb{T}^2}$ or $\mathbb{V}_{{\cal M}_2}$ as these factors appear in the nested integral \eqref{rambha} that defines $\mathbb{G}_8(y)$. It should be clear that 
in the limit of vanishing volume \eqref{tarfox}, the quantum term \eqref{japumey} doesn't have to decouple, thus paving way to the non-local counter-terms as advertised in \cite{nodS} (see footnote 25 of 
\cite{nodS} and the example cited in there). 

The discussion in \cite{nodS} can now be made more quantitative. In fact we will see that it is not just the time-neutral series that creates problem with a four-dimensional EFT description\footnote{Recall that one reason for looking at time-neutral series here, and also in \cite{nodS}, is to ensure that even for vanishing $g_s$ the un-controlled quantum terms with no $M_p$ hierarchies survive. This way the late time physics in the IIB side still retains the problems plaguing the non-zero $g_s$ case. Other reasons for looking at the time-neutral series will become clearer once we discuss the EOMs in section \ref{qotomth}.}, 
rather the problem persists
even when we go beyond zeroth order in $g_s$. To see this,  let us take the $g_s$ scaling \eqref{kkkbkb2} and make $n_3 = 0$ therein. We can also define the following new parameters:
\bg\label{mcteer}
&&\mathbb{N}_1 \equiv 2 \sum_{i = 1}^{27} l_i + n_1 + n_2 + l_{30} + \sum_{u = 1}^4 l_{31+u} \nonumber\\
&& \mathbb{N}_2 \equiv l_{28}  + l_{29} + l_{31}, ~~~ \mathbb{N}_3 \equiv l_{36}  + l_{37} + l_{38}, \nd    
where we have avoided the proliferation $p$ in $l_i^{(p)}$ and simply represented it as $l_i$ in 
\eqref{kkkbkb2} (we have also absorbed $n_0$ temporal derivatives by shifting $n_1$ and/or $n_2$, as alluded to earlier). In this language $\theta'_0$ can be rewritten as:
\bg\label{kkkbkb0}
3\theta'_0 \equiv \mathbb{N}_1 + 4\mathbb{N}_2 - 2\mathbb{N}_3. \nd
Clearly since $\mathbb{N}_i \in \mathbb{Z}$, and are positive definite, fixing a value for $\theta'_0$ will 
allow us to have an infinite number of choices for ($\mathbb{N}_1, \mathbb{N}_2, \mathbb{N}_3$). This in turn will provide an infinite number of choices for $l_i$ exponents in \eqref{phingsha2}. Let us call this value of $\theta'_0$ as $\Theta_0$, and therefore all the quantum terms in \eqref{phingsha2} that 
go as $g_s^{\Theta_0}$ have {\it different} $M_p$ scalings given by the $l_i$ choices in 
\eqref{kamagni2}. 

The question that we want to ask here is how many terms are allowed for a given power of $M_p$ once the  $g_s$ scaling has been fixed.  Clearly \eqref{kamagni2} can only allow a finite number of terms for a fixed 
$\sigma = 3\Sigma_0 \ge 0$, where the factor of 3 is for later convenience. However now we have the non-local terms whose $g_s$ and $M_p$ scalings, at least for the case where 
$\mathbb{F}^{(r)}(y - y')$ depends implicitly on $g_s$ and $M_p$, appear in \eqref{japumey} and 
\eqref{kamagni} respectively. How many terms do they allow for 
fixed (${\Theta}_0, 3{\Sigma}_0$)? The answer lies in the solutions to the following two equations:
\bg\label{euroshi}
&& \mathbb{N}_1 +  \mathbb{N}_2 +  \mathbb{N}_3  - 8r = 3\Sigma_0 \nonumber\\
&&  \mathbb{N}_1 + 4 \mathbb{N}_2 - 2 \mathbb{N}_3 - 2r = 3\Theta_0, \nd
where $ \mathbb{N}_i$ are defined in \eqref{mcteer} and $r$ is the level of non-locality that we consider here. The first equation in \eqref{euroshi} appears from \eqref{kamagni} while the second equation appears from the $g_s$ scaling in \eqref{japumey}. Subtracting these equations gives us:
\bg\label{keyslond}
 \mathbb{N}_2 -  \mathbb{N}_3 + 2r = \Theta_0 - {\Sigma_0}, \nd
 where again, the relative minus sign on the LHS of \eqref{keyslond} is important. Assuming, for simplicity, the RHS to be a positive definite integer, there are clearly an infinite number of choices for the triplet
 ($\mathbb{N}_2, \mathbb{N}_3, r$). For any given choice of the triplet, there is an integer solution for 
 $\mathbb{N}_1$, provided:
 \bg\label{ryanthai}
 \mathbb{N}_2 \le {1\over 4}\left(3\Theta_0 + 2r + 2\mathbb{N}_3\right), \nd
which is always possible if $\mathbb{N}_2$ is smaller than ${r\over 2}$. In general, \eqref{ryanthai} is a more relaxed condition and therefore easier to satisfy.  

The upshot of the above analysis can be easily summarized. The key set of relations are \eqref{euroshi}, where for $r = 0$ we get back the perturbative conditions discussed above. For vanishing $r$, the second equation in \eqref{euroshi} will provide an infinite set of quantum terms, all scaling as $g_s^{\Theta_0}$.  However they have different $M_p$ scalings as seen by plugging in the values of $\mathbb{N}_i$ in the first equation in \eqref{euroshi}. In this set, let us choose a quantum term (or a set of quantum terms) 
that scales as 
${g_s^{\Theta_0}\over M_p^{3\Sigma_0}}$. This is a meaningful exercise because the first equation in \eqref{euroshi} has finite number of solutions. 

We now go to $r = 1$, i.e to the first level of non-locality. The second equation in \eqref{euroshi} will now provide yet another infinite set of quantum terms, all scalings as 
$g_s^{\Theta_0}$ again, while the first equation in \eqref{euroshi} provides the $M_p$ scalings of these terms. It is clear that a quantum term (or a set of quantum terms) scaling as 
${g_s^{\Theta_0}\over M_p^{3\Sigma_0}}$ may be easily extracted from this new set. The process can then be continued for $r \ge 2$, with similar results. The conclusion is that, for a given order in
${g_s^{\Theta_0}\over M_p^{3\Sigma_0}}$, there are in fact an infinite number of quantum terms possible. This is precisely how an effective field theory description is ruined\footnote{We should compare this 
{breakdown} of EFT, which we will call as the {\it swampland breakdown}, to the well known breakdown of 
EFT, which we will call as the {\it standard breakdown}. In the standard breakdown of EFT, which may be seen in our expansion for $r = 0$ to any order in $g_s$, the series of operators that are suppressed by powers of $M_p$ become uncontrolled in the limit when the operators themselves are of order $M_p$. For example to order $g_s^{\Theta_0}$, the series of quantum terms are expected to appear from 
\eqref{phingsha2} (say for the case \eqref{olokhi}), and are variously suppressed by powers of $M_p$ from \eqref{kamagni2}, or from \eqref{kamagni} with $r = 0$. The quantum terms are in turn expressed as powers of curvature tensors and G-flux components, and when they take values of order $M_p$, the quantum series become uncontrolled. This is an example of a standard breakdown of EFT in the sense that such a breakdown 
does not preclude a UV completion of the theory. The {\it swampland breakdown} on the other hand, is slightly different as we saw above. In this case, when we take $r \ge 0$, we see that to order 
${g_s^{\Theta_0}\over M_p^{3\Sigma_0}}$ there are in fact an infinite number of operators, a finite subset of each being collected from every choice of $r \ge 0$. The operators themselves are no longer required to take values of order $M_p$ now, but the very fact that we have an infinite number of them prohibits a UV completion of the model. Of course a middle ground would be when the dynamics of the system themselves force the fields, and hence the operators, to take values of order $M_p$. In this case the standard breakdown of EFT will coincide with the swampland breakdown. More details on this appear in 
\cite{petite, maxpaper}. }. 

We can also make $\Theta_0 = 0$, and take all choices of $r \ge 0$. For any given value of $r$, there are  infinite choices of $\mathbb{N}_i$, all scaling differently with $M_p$. Summing up the series for all choices of $r$ then reproduces the time-neutral series of eq. (5.58) in \cite{nodS}. Our analysis above shows that going beyond time-neutral series do not alleviate the problems with the EFT description in lower dimensions. The only way an EFT description is possible if the coefficient of $\mathbb{N}_3$ in the second equation of \eqref{euroshi} becomes positive and $r$ in both the equations of \eqref{euroshi} decouples. As shown above, both are realized when time dependences are switched on. 

Before ending this section let us ask what happens in the standard supergravity case. As elaborated in 
footnote \ref{chascage}, the time-independent supergravity compactifications remain unaffected because 
${\cal V}_8 = 0$, as a time-independent background cannot have a coherent state representation 
(see the discussion in footnote \ref{gowest})
and therefore must coincide with the solitonic solution. 



\vskip.1in

\noindent {\it Case 4: Non-locality in time for the various choices of $F_i(t)$}

\vskip.1in

\noindent  The next case that we want to elaborate is a rather curious one, because it involves non-locality in both (internal) space and time. The temporal non-locality would only make sense as an integral condition. In other words we can take the non-locality function $\mathbb{F}^{(r)}(y-y', t- t')$ to be functions of both 
$(y, t)$ as well as ($y', t'$). However since we have identified any temporal dependence with 
${g_s^2 \over \sqrt{h}}$ (see \eqref{montse}), the non-locality function should now have both $y, y'$ and $g_s$ dependence, i.e explicit in $g_s$ but implicit in $M_p$. 
Therefore, much in the same vein as before, we can assign the following generic form for the non-locality function: 
\bg\label{oshpach}
\mathbb{F}^{(r)}(y - y', g_s) \equiv \sum_{l_a, l_b} f^{(r)}_{l_al_b}(y - y') 
\left({g_s^2\over \sqrt{h}}\right)^{\Delta l_a}
{\rm exp}\left(-{l_bh^{\Delta/4}\over g^\Delta_s}\right), \nd
where $(l_a, l_b) \in (\mathbb{Z}/2, \mathbb{Z})$, the warp-factor $h = h(y - y')$ and $f^{(r)}_{l_al_b}(y-y')$ to be a highly peaked function at low energies (we can identify it with the example in footnote \ref{chascage}). 
We can also resort to the simplification $h(y - y') = h_0$ to keep the $g_s$ itself unaltered to all order in the non-locality, as we have done before. Plugging this in \eqref{pdesimey} and \eqref{rambha} results in a complicated nested integral form, which would then have to be integrated over time to make sense of the result. In other words, we want:
\bg\label{advisor} 
\mathbb{U}^{(r)}(y_{r+1}, g_s(t)) \equiv \int_{-\infty}^t 
{dt' \sqrt{{\bf g}_{00}}\over M_p^{\sigma_r}}\left(e_0 g_s^{-2/3}(t') + {e_1g_s^{4/3}(t')\over
 \sqrt{h_0}}\right)^r g_s^{\Theta_k}(t')\mathbb{G}_8(y_{r+1}, g_s(t')),  \nonumber\\
 \nd 
where the three cases discussed above are described by assigning different values to the triplet 
($e_0, e_1, \Theta_k$) i.e  $(0, 1, \theta_k), (1, 0, \theta'_k)$ and $(1, 0, \theta'_0)$ with $\theta_k, \theta'_k$ and $\theta'_0$ as defined in \eqref{miai}, \eqref{melamon2} and \eqref{kkkbkb2} respectively. The $g_s(t')$
dependence of $\mathbb{G}_8(y_{r+1}, g_s(t'))$ may be determined by plugging in \eqref{oshpach} in
\eqref{rambha}. 

The concern however is the integral \eqref{advisor} itself. Since $g_s$, as defined in \eqref{montse} depends on time itself, so when $t \to -\infty$, $g_s \to +\infty$. The representation 
\eqref{oshpach} is not a suitable description at strong coupling. because \eqref{oshpach} is only defined perturbatively when $g_s \to 0$. We can do a change of variable $t \to 1/t$, or $g_s \to 1/g_s$ to study the strong coupling regime. In either formalism, it then appears that the relevant integral will be:
\bg\label{flower}
\int_0^{g_s} d{g}'_s ~{g'}^{\Delta q_1}_s {\rm exp}\left(-{q_2\over {g'}^\Delta_s}\right) & = & 
q_2^{q_1+{1\over \Delta}} \Gamma\left(-q_1 - {1\over \Delta}, {q_2\over g^\Delta_s}\right)\\
& = & {1\over q_2}\left(g_s^{q_1 + 1 + {1\over \Delta}} + {\cal O}(g_s^{q_1 + 2 + {1\over \Delta}})\right) 
{\rm exp}\left(-{q_2\over g^\Delta_s} + 
{\cal O}(g_s^{2\Delta})\right), \nonumber\\ \nd  
with $g_s < 1$ so that the expansion on the second line could be justified. The perturbative expansion then 
tells us that for any choice $q_1$ in the $g_s$ expansion, non-locality to any order only adds a 
$1 + {1\over \Delta}$ factor, and therefore doesn't alter any of our earlier conclusions regarding $g_s$ scalings. Additionally, the decoupling effect for vanishing volume as in \eqref{tarfox} still persists, so no new subtleties appear at this stage.  

\vskip.1in

\noindent {\it Case 5: Non-locality with volume independences and further generalization}

\vskip.1in

\noindent The volume condition \eqref{tarfox} pretty much saves the day by decoupling the effects of non-localities for the two cases \eqref{ranjhita} and \eqref{olokhi} discussed above. One could ask if this decoupling may also be applied to the time-independent case where $F_1(t) = F_2(t) = 1$. What if we assume that all the background fluxes and the metric components are made independent of the 
toroidal directions? Clearly now the KK modes are no longer important, but the relative minus signs in 
\eqref{kkkbkb} and \eqref{kkkbkb2} still survive. However the non-local counter-terms decouple. Couldn't this get us an EFT description according to the criteria that we presented above?

The answer turns out to be unfortunately not, as the breakdown of EFT that we discussed above is actually not sensitive to the toroidal volume despite featuring prominently in the above discussions. To see this let us express the generic form for the most dominant contribution to the non-locality function 
$\mathbb{F}^{(r)}(y - y')$ as:
\bg\label{spivibhalo}
\mathbb{F}^{(r)}(y - y') = \left(M_p^6 \mathbb{V}_6\right)^{\sigma_a(r) - r} \left(M_p^2 
\mathbb{V}_{\mathbb{T}^2}\right)^{\sigma_b(r) - r} 
\left({g^2_s\over \sqrt{h}}\right)^{{r\over 3} \pm {1\over 2} \vert\sigma_c(r)\vert} f(y - y'), \nd
where $\sigma_{a, b, c}(r) \in \mathbb{Z}$ and their dependences on the level of non-locality $r$ stems from our generic consideration. The function $f(y - y')$ do not depend on $g_s$ or $M_p$.
It is of course understood that the non-locality function would decouple for the
case ${\cal V}_8 = 0$ as discussed in footnote \ref{chascage}. The above form \eqref{spivibhalo} is not the most essential way to express the effects of non-locality, as one could instead express the full non-local quantum terms using the $M_p$ and $g_s$ scalings. This has been elaborated in  
\cite{petite, maxpaper}, so here we simply provide brief arguments. 

We can deal with various cases, and the first one would be to consider the scenario where the quantum terms keep the $g_s$ scalings unchanged but change the $M_p$ scalings.  Also, we will illustrate this and the subsequent ones for the 
situation with time-independent Newton's constant, i.e \eqref{olokhi}, unless mentioned otherwise. 
An additional crucial change is as alluded in footnote \ref{westgate}, namely, we now switch on 
$n_3$ derivatives, i.e dependences on the toroidal directions. The story now gets more involved, and the complete details may be extracted from \cite{petite, maxpaper}. A short summary is that,
to any order in ${g_s^a\over M_p^b}$ there are only {\it finite} number of terms allowed, and EFT description becomes possible. Once $k = 0$, the relative minus sign of $n_3$ in \eqref{kkkbkb2}, immediately rules out  finite number of terms to any order in 
 ${g_s^a\over M_p^b}$, thus also ruling out not only an EFT description, but also a UV completion. On the other hand, the scenario with the quantum terms that keep both $g_s$ and $M_p$ scalings unchanged i.e 
 $\sigma_a(r)  = \sigma_b(r) = \sigma_c(r) = 0$, simply renormalizes the terms to any order in 
  ${g_s^a\over M_p^b}$, therefore we only see the {\it standard} breakdown of EFT.
  
The next case would be to consider the quantum terms that change both the $g_s$ and the $M_p$ scalings. Of course once we change the $g_s$ scalings, the $M_p$ scalings would change automatically, so here we want to first deal with  $\sigma_b = 0$, so that the additional $M_p$ and $g_s$ scalings come from 
$\pm\vert\sigma_a\vert$ and $\pm\vert\sigma_c\vert$ respectively. Solving 
$\theta'_k = \vert\sigma_c\vert + \left({8\over 3}, {2\over 3}, -{4\over 3}\right) + {\mathbb{Z}\over 3}$ 
in \eqref{melamon2} should reproduce the number of quantum terms for the space-time; ${\cal M}_4$ and ${\cal M}_2$; and the toroidal directions respectively. Clearly for $r = 0$, they are all finite in number as before, but have different $M_p$ scalings as expected. Once we go to say $r = 1$, and also switch on $n_3$, although the $g_s$ scaling changes because 
$\sigma_c(1) \ne \sigma_c(0)$, the number of quantum terms remain countably finite. The $M_p$ scalings would again either shift to the right or left depending on the sign of $\vert\sigma_a\vert$. If we ask how many terms are there for a fixed value of 
$\sigma_r + 8r - \vert \sigma_a(r)\vert$, we see that there are still {\it finite} despite the relative minus signs from $n_3$ and other factors. Thus again to order  ${g_s^a\over M_p^b}$ there are only finite number of terms and EFT description becomes possible. Once $k = 0$, the finiteness goes away, and the system goes to the swampland.

\subsubsection{Topological quantum terms, curvature forms and fluxes \label{tufi}}

So far we have dealt with the non-topological quantum terms in terms of curvatures and G-flux components that would contribute to the energy-momentum tensor. However there are also EOMs associated with the 
G-fluxes that would demand contributions from the quantum terms \eqref{phingsha2}, and 
\eqref{phingsha} for the cases \eqref{olokhi} and \eqref{ranjhita} respectively. Interesting, once we look at the fluxes, we will have to study both the standard four-form G-fluxes and their dual, the seven-form, flux components. Thus we need to see how the $g_s$ scalings \eqref{melamon2} and \eqref{miai}, respectively for the two cases, would change.   
In addition to that there would also be topological terms that we will have to determine. In the following let us first analyze the topological terms. 

The topological contributions, as the name suggest, would appear from topological forms that are constructed using the Riemann tensors and the G-flux components by taking advantages of their anti-symmetries. They may be expressed as\footnote{G-flux could also contribute as a four-form by itself, or as a three-form by contracting ${\bf G}_{MNPQ}$ with a vielbein. The latter generically does not contribute because of the tracelessness condition whereas the former is already taken into account in the supergravity action.}:
\bg\label{ashf1}
&&\mathbb{R} \equiv {\bf R}_{MN}^{a_ob_o} {\bf M}_{a_ob_o}~ dy^M \wedge dy^N, ~~~~
\mathbb{G} \equiv {\bf G}_{MN}^{a_ob_o} {\bf M}_{a_ob_o} ~dy^M \wedge dy^N \nonumber\\
&&{\bf R}_{MN}^{a_ob_o} \equiv {\bf R}_{MNPQ}~ e^{a_o P} e^{b_o Q}, ~~~~~~~ 
{\bf G}_{MN}^{a_ob_o} \equiv {\bf G}_{MNPQ}~ e^{a_o P} e^{b_o Q}, \nd
where ${\bf M}_{a_ob_o}$ are the holonomy matrices on the compact manifold over which we will be taking traces. These are just like the generator matrices, for example as the ones appearing like 
${\bf A}_\mu^a {\bf T}^a$, in the definition of a gauge field one-form. Using \eqref{ashf1}, we can construct various higher order forms, one example being the following eight-form:
\bg\label{ashf2}
\mathbb{Y}_8 \equiv c_1 {\rm tr}~\mathbb{R}^4  + c_2 \left({\rm tr}~\mathbb{R}^2\right)^2 + 
c_3 \left({\rm tr}~\mathbb{R}^2\right) \left({\rm tr}~\mathbb{G}^2\right) + c_4 {\rm tr}~\mathbb{G}^4
+ c_5 \left({\rm tr}~\mathbb{G}^2\right)^2, \nd
where we have assumed that the holonomy matrices are traceless. For various choices of the $c_i$ coefficients, we can generate certain sub eight-forms. For example with:
\bg\label{ashf3}
c_1 = {1\over 3\cdot 2^{10}\cdot \pi^4}, ~~~ c_2 = -{1\over 12\cdot 2^{10}\cdot \pi^4}, ~~~ c_3 = c_4 = c_5 = 0, \nd
we have our ${\bf X}_8$ polynomial which is important to cancel anomalies as we shall see later. However now with non-zero ($c_3, c_4, c_5, ..$) more non-trivial polynomials (for example replacing $\mathbb{R}$ by 
$\mathbb{R} + \mathbb{G}$ in ${\bf X}_8$) may be constructed which, in a packaged form, is given as \eqref{ashf2}. In fact polynomials like \eqref{ashf2}, or their most generic form, open up the possibility of constructing 
topological and non-topological interactions in M-theory of the following form:
\bg\label{ashf4}
{\bf C}_3 \wedge \mathbb{Y}_8, ~~~~~~ 
{\bf G}_4 \wedge \ast_{11} \mathbb{Y}_4, \nd
where ${\bf C}_3$ is the M-theory three-form and the Hodge star is with respect to the full eleven-dimensional {\it warped} metric (as such it will be a function of $g_s$). The way we have expressed the non-topological piece, should allow us to extract this from the generalized quantum terms \eqref{phingsha2} and 
\eqref{phingsha} for \eqref{olokhi} and \eqref{ranjhita} respectively. For example the non-topological piece in \eqref{ashf4} may be expressed as:
\bg\label{aliceL2}
\int {\bf G}_4 \wedge \ast_{11} \mathbb{Y}_4 &\equiv &  \int d^{11} y \sqrt{-{\bf g}_{11}} 
\sum_{\{l_i\}, n_1, n_2}
\mathbb{Q}_{\rm T}\left(\{l_i\}, n_1, n_2\right)\\
&=& \int d^{11} y \sqrt{-{\bf g}_{11}} 
\left({\bf G}_4\right)_{M_1M_2M_3M_4} \left(\mathbb{Y}_4\right)_{N_1N_2N_3N_4} {\bf g}^{M_1N_1}
{\bf g}^{M_2N_2}{\bf g}^{M_3N_3}{\bf g}^{M_4N_4},\nonumber \nd
where we have used the warped metric both as inverses as well as in the definition of the determinant, and
the quantum terms $\mathbb{Q}_{\rm T}\left(\{l_i\}, n_1, n_2\right) $
are defined as in \eqref{phingsha2} for the case \eqref{olokhi} (changing the quantum terms to \eqref{phingsha} will provide information for the case \eqref{ranjhita}). The above relation could be used for identifying the $\mathbb{Y}_4$ tensor from the quantum series \eqref{phingsha2} or \eqref{phingsha}
for the two cases \eqref{olokhi} and \eqref{ranjhita} respectively.    
We can then ask the $g_s$ scalings of the following two kinds of quantum terms:
\bg\label{ashf5}
\left({\bf G}_4\right)_{012M} \left({\bf Y}_4\right)^{012M}, ~~~
\left({\bf G}_4\right)_{MNPQ} \left({\bf Y}_4\right)^{MNPQ}, \nd
where ($M, N, P$) are the coordinates of the eight-manifold. The $g_s$ scalings of these two interactions may be easily worked out by extracting a $\left({\bf C}_3\right)_{012}$ and a $\left({\bf C}_3\right)_{MNP}$
out of either \eqref{phingsha2} or \eqref{phingsha}. Since $\left({\bf G}_4\right)_{012M}$ and 
$\left({\bf G}_4\right)_{MNPQ}$ scale as $\left({g_s\over H}\right)^{-4}$ and 
$\left({g_s\over H}\right)^{2\Delta k}$ respectively, it is easy to infer the $g_s$ scalings of 
$\left({\bf Y}_4\right)^{012M}$ and $\left({\bf Y}_4\right)^{MNPQ}$ respectively as:
\bg\label{elroyale}
\theta'_k ~ \to ~ \theta'_k + 4, ~~~~~ \theta'_k ~ \to ~ \theta'_k - 2\Delta k, \nd
with $\theta'_k$ as given in \eqref{melamon2}. A similar scaling would work if we replace $\theta'_k$ with 
$\theta_k$ from \eqref{miai}, as one would expect. On the other hand, $\mathbb{Y}_8$ should be
topological. To see this let us first fix the time to $t = t_0$ in the M-theory metric \eqref{vegamey} and, for simplicity, switch off the G-fluxes. Plugging in the metric ansatze \eqref{vegamey} at the fixed time, with the choice \eqref{ashf3}, in \eqref{ashf2} then shows that at any $t = t_0 + \delta t$, \eqref{ashf2} may in general have $\delta t$ dependence in addition to a piece that depends on $t_0$. Since the temporal behavior is traded with $g_s$, \eqref{ashf2} will develop $g_s$ dependence. Additionally, because of the underlying non-K\"ahlerity of the internal eight-manifold (at least for the case \eqref{olokhi}), the  
integral of ${\bf X}_8$ is not exactly the Euler characteristics of the 
eight-manifold\footnote{We thank Savdeep Sethi for discussions on this point.}. 
Switching on the G-fluxes, the integral of $\mathbb{Y}_8$ should also have a 
$g_s$ dependent pieces. Together all of these would complicate the anomaly cancellation procedure that we have known for the time-independent case, implying a careful study is required in the time-dependent case. More details on this appears in section \ref{anoma}.

There are other topological contributions possible once we go to the {\it dual} formalism. Here duality implies a generalized form of electric-magnetic duality, much like the Montonen-Olive one \cite{monto}. 
To implement it here, at least at the level of perturbative and non-perturbative expansions that we have entertained so far, all we need is to express the flux contributions by their dual variables. The dual of a four-form flux is a 
seven-form flux, and therefore if we can express \eqref{phingsha2} and \eqref{phingsha} using the dual variables, we should be able to determine their $g_s$ scalings as well. This rather convoluted re-telling of the same story has a deeper purpose:  the dual description will not only help us to determine the Bianchi identities later but also help us to ascertain the flux quantization conditions.  The dual seven-form 
${\bf G}_7 = \ast_{11} {\bf G}_4$, may be expressed in terms of components in the following standard way:

{\footnotesize
\bg\label{moonie}
{\bf G}_7 = {1\over 7!} {\bf G}_{P'Q'R'S'} \sqrt{-{\bf g}_{11}} 
{\bf g}^{P'P}{\bf g}^{Q'Q}{\bf g}^{R'R}{\bf g}^{S'S}\epsilon_{PQRSM_1M_2.....M_7} 
dy^{M_1} \wedge dy^{M_2} ....... \wedge dy^{M_7}, 
\nd}
where the metric components as well as the determinant are all defined in terms of the warped metric and 
$\epsilon_{PQ....M_7}$ is the eleven-dimensional Levi-Civita {\it symbol}. The above formula is an useful way to determine the $g_s$ scalings of every components of the dual form once the original $g_s$ scalings are known. This will also help us to determine the $g_s$ scalings of the quantum terms, relevant for the case \eqref{olokhi}, that may now be expressed in the following way:
\bg\label{phingsha33} 
\mathbb{Q}^{(2)}_{\rm T} & = & {\bf g}^{m_i m'_i}{\bf g}^{m_l m'_l}.... {\bf g}^{j_k j'_k} 
\partial_{m_1}..\partial_{m_{n_1}}\partial_{\alpha_1}..\partial_{\alpha_{n_2}}
\left({\bf R}_{mnpq}\right)^{l_1} \left({\bf R}_{abab}\right)^{l_2}\left({\bf R}_{pqab}\right)^{l_3}\left({\bf R}_{\alpha a b \beta}\right)^{l_4} \nonumber\\
&\times& \left({\bf R}_{\alpha\beta mn}\right)^{l_5}\left({\bf R}_{\alpha\beta\alpha\beta}\right)^{l_6}
\left({\bf R}_{ijij}\right)^{l_7}\left({\bf R}_{ijmn}\right)^{l_8}\left({\bf R}_{iajb}\right)^{l_9}
\left({\bf R}_{i\alpha j \beta}\right)^{l_{10}}\left({\bf R}_{0mnp}\right)^{l_{11}}\\
& \times & \left({\bf R}_{0m0n}\right)^{l_{12}}\left({\bf R}_{0i0j}\right)^{l_{13}}\left({\bf R}_{0a0b}\right)^{l_{14}}\left({\bf R}_{0\alpha 0\beta}\right)^{l_{15}}
\left({\bf R}_{0\alpha\beta m}\right)^{l_{16}}\left({\bf R}_{0abm}\right)^{l_{17}}\left({\bf R}_{0ijm}\right)^{l_{18}}
\nonumber\\
& \times & \left({\bf R}_{mnp\alpha}\right)^{l_{19}}\left({\bf R}_{m\alpha ab}\right)^{l_{20}}
\left({\bf R}_{m\alpha\alpha\beta}\right)^{l_{21}}\left({\bf R}_{m\alpha ij}\right)^{l_{22}}
\left({\bf R}_{0mn \alpha}\right)^{l_{23}}\left({\bf R}_{0m0\alpha}\right)^{l_{24}}
\left({\bf R}_{0\alpha\beta\alpha}\right)^{l_{25}}
\nonumber\\
&\times& \left({\bf R}_{0ab \alpha}\right)^{l_{26}}\left({\bf R}_{0ij\alpha}\right)^{l_{27}}
\left({\bf G}_{0ij\alpha\beta ab}\right)^{l_{28}}\left({\bf G}_{0ijq\alpha ab}\right)^{l_{29}}
\left({\bf G}_{0ijq\alpha\beta b}\right)^{l_{30}}\left({\bf G}_{0ij mn ab}\right)^{l_{31}}
\left({\bf G}_{0ijmn\alpha b}\right)^{l_{32}}\nonumber\\
&\times&\left({\bf G}_{0ijnpqb}\right)^{l_{33}}\left({\bf G}_{mnp \alpha\beta ab}\right)^{l_{34}} 
\left({\bf G}_{mnpq\alpha ab}\right)^{l_{35}}
\left({\bf G}_{0ij mn \alpha\beta}\right)^{l_{36}}\left({\bf G}_{0ijmnpq}\right)^{l_{37}}
\left({\bf G}_{0ijmnp\alpha}\right)^{l_{38}}, \nonumber
\nd
which should now be compared to \eqref{phingsha2} written in terms of the original variables. We could also re-express \eqref{phingsha}, relevant for the case \eqref{ranjhita}, in terms of the dual variables, but
since the story would be similar to what we have in \eqref{phingsha33} we will avoid this exercise. In fact 
making the following two-step processes to \eqref{phingsha33}, we can convert this to the case corresponding to \eqref{ranjhita}: one, make 
$n_2 = l_{19} = l_{20} = ... = l_{27} = 0$, and two, relabel $l_{28}, ..., l_{38}$ to $l_{19}, ..., l_{29}$. 
The 
$g_s$ scalings 
are easy to determine using the method employed in the earlier sections (see {\bf Table \ref{dualforms}} for details). 
Following these footsteps, one may easily verify that the $g_s$ scalings of the quantum terms in \eqref{phingsha33} are {\it exactly} the same as in 
\eqref{melamon2}. Needless to say, the $g_s$ scalings of the quantum terms corresponding to the case \eqref{ranjhita}, are also exactly the same as in \eqref{miai}. This shows that resorting to the dual variables 
{\it do not} change the $g_s$ scalings of the quantum terms, and is therefore reassuring to see that the expected equivalences between dual theories are respected at every order in the $g_s$ expansions.     

Resorting to the dual fluxes ${\bf G}_7$ allow us to define six-form potentials ${\bf C}_6$ such that 
${\bf G}_7 = d{\bf C}_6 + ...$, where the dotted terms depend on how the Bianchi identities appear in our
set-up. This will be elaborated later when we discuss the EOMs for fluxes. What we want to study here are  
the various forms of interactions, both topological and non-topological, that may appear when we consider quantum terms like \eqref{phingsha33}. Motivated by \eqref{ashf4}, we expect interactions like:
\bg\label{verasofmey}
{\bf C}_6 \wedge \mathbb{Y}_5, ~~~~~~ 
{\bf G}_7 \wedge \ast_{11} \mathbb{Y}_7, \nd
where $\mathbb{Y}_5$ and $\mathbb{Y}_7$ are five and seven-forms constructed out of the curvature and the flux forms like \eqref{ashf1}. However an odd form like $\mathbb{Y}_5$ cannot be constructed out of 
the two-forms from \eqref{ashf1}, so can only be expressed as:
\bg\label{cambelle}
\mathbb{Y}_5 \equiv {\bf \Lambda}_5 + d\mathbb{\hat Y}_4, \nd
where ${\bf \Lambda}_5$ is a highly localized form which would represent a M5-brane 
once wedged with ${\bf C}_6$. The other four-form $\mathbb{\hat Y}_4$ can be 
constructed\footnote{The two four-forms $\mathbb{Y}_4$ and $\mathbb{\hat Y}_4$ are definitely related to each other because they describe similar interactions in M-theory, albeit in the relative dual 
pictures. We will however not elaborate on their precise equivalence here.}  out of the 
curvature two-form and gauge form coming from localized G-fluxes. Finally, the second term in 
\eqref{verasofmey} contributes the following non-topological interaction:
\bg\label{aliceL}
\int {\bf G}_7 \wedge \ast_{11} \mathbb{Y}_7 &\equiv &  \int d^{11} y \sqrt{-{\bf g}_{11}} 
\sum_{\{l_i\}, n_1, n_2}
\mathbb{Q}^{(2)}_{\rm T}\left(\{l_i\}, n_1, n_2\right) \nonumber\\
&=& \int d^{11} y \sqrt{-{\bf g}_{11}} 
\left({\bf G}_7\right)_{M_1.....M_7} \left(\mathbb{Y}_7\right)_{N_1.....N_7} {\bf g}^{M_1N_1}......
{\bf g}^{M_7N_7}, \nd
which is similar to what we had in \eqref{aliceL2} earlier. Again, 
the metric components are all taken as the warped ones and therefore involve $g_s$ factors in them, and 
$\mathbb{Q}^{(2)}_{\rm T}\left(\{l_i\}, n_1, n_2\right)$ are the quantum terms as given in 
\eqref{phingsha33}. The conjectured equality \eqref{aliceL} is to be used to define the functional form for 
$\mathbb{Y}_7$ tensor, much like what we had in \eqref{aliceL2} earlier, and basically tells us that that 
$\mathbb{Y}_7$ is constructed out of products of tensors in such a way that it is an anti-symmetric tensor of rank 7.  

Another important thing to notice about \eqref{phingsha}, \eqref{phingsha2} and \eqref{phingsha33} is that they are {\it not} globally defined functions, despite the fact that they contain globally defined tensors like four-form fluxes and the curvature tensors. The fact that inverse metric components show up in the definition of the quantum terms, and since the metric components are defined only on patches over the compact eight-manifold, render these quantum terms mostly local. Now because the Hodge dual of the forms $\mathbb{Y}_4$ and $\mathbb{Y}_7$ are related to the quantum terms 
\eqref{phingsha2} and \eqref{phingsha33} via \eqref{aliceL2} and \eqref{aliceL} respectively, they cannot be globally defined forms. This is much like the form ${\bf X}_8 = d{\bf X}_7$, where ${\bf X}_7$ is not globally defined, and therefore the integral of ${\bf X}_8$ over a compact eight-manifold is non-zero.

In the following we will elaborate on all the background EOMs, both for the metric and the G-flux components, that would appear for our case once the effects of the quantum terms are included. The analysis that we presented above will be used once we study the G-flux EOMs and their constraints.

\begin{table}[tb]
 \begin{center}
\renewcommand{\arraystretch}{1.5}
\begin{tabular}{|c|c|c|c|}\hline Tensors & Dual Forms & ${g_s\over H}$ scaling for \eqref{olokhi} & 
${g_s\over H}$ scaling for \eqref{ranjhita} \\ \hline\hline
$\mathbb{Y}_7^{npq\alpha\beta ab}$ & ${\bf G}_{0ijm}$  & $\theta'_k$   & $\theta_k - 2$  \\  \hline
$\mathbb{Y}_7^{mnpq\beta ab}$ & ${\bf G}_{0ij\alpha}$  & $\theta'_k$   & $\theta_k$  \\  \hline
$\mathbb{Y}_7^{0ij\alpha\beta ab}$ & ${\bf G}_{mnpq}$  & $\theta'_k -2\Delta k + 2$   & $\theta_k - 2\Delta k$  \\  \hline
$\mathbb{Y}_7^{0ijq\beta ab}$ & ${\bf G}_{mnp\alpha}$  & $\theta'_k -2\Delta k + 2$   & $\theta_k - 2\Delta k+ 2$  \\  \hline
$\mathbb{Y}_7^{0ijq\alpha\beta b}$ & ${\bf G}_{mnpa}$  & $\theta'_k -2\Delta k + 4$   & $\theta_k - 2\Delta k+ 2$  \\  \hline
$\mathbb{Y}_7^{0ijpq ab}$ & ${\bf G}_{mn\alpha\beta}$  & $\theta'_k -2\Delta k + 2$   & $\theta_k - 2\Delta k+ 4$  \\  \hline
$\mathbb{Y}_7^{0ijpq\beta b}$ & ${\bf G}_{mn\alpha a}$  & $\theta'_k -2\Delta k + 4$   & $\theta_k - 2\Delta k+ 4$  \\  \hline
$\mathbb{Y}_7^{0ijnpq b}$ & ${\bf G}_{m\alpha\beta a}$  & $\theta'_k -2\Delta k + 4$   & $\theta_k - 2\Delta k+ 6$  \\  \hline
$\mathbb{Y}_7^{0ijpq \alpha\beta }$ & ${\bf G}_{mnab}$  & $\theta'_k -2\Delta k + 6$   & $\theta_k - 2\Delta k+ 4$  \\  \hline
$\mathbb{Y}_7^{0ijmnpq}$ & ${\bf G}_{\alpha\beta ab}$  & $\theta'_k -2\Delta k + 6$   & $\theta_k - 2\Delta k+ 8$  \\  \hline
$\mathbb{Y}_7^{0ijnpq \beta}$ & ${\bf G}_{m\alpha ab}$  & $\theta'_k -2\Delta k + 6$   & $\theta_k - 2\Delta k+ 6$  \\  \hline
  \end{tabular}
\renewcommand{\arraystretch}{1}
\end{center}
 \caption[]{The ${g_s\over H}$ scalings of the various components of the seven-form $\mathbb{Y}_7$ 
  represented for the two cases {\eqref{olokhi}} and {\eqref{ranjhita}}. We have taken $\Delta = {1\over 3}$, 
  $k \ge {3\over 2}$ for \eqref{olokhi}, and $k \ge {9\over 2}$ for \eqref{ranjhita}. The other two parameters, 
  $\theta'_k$ and $\theta_k$, are defined in 
  {\eqref{melamon2}} and {\eqref{miai}} respectively.}
  \label{dualforms}
 \end{table}

\section{Analysis of the quantum equations of motion and constraints \label{qotomth}}

We now have at our hands all the necessary ingredients to pursue the equations of motion and from there extract any constraints that may effect the dynamics of the system. Before moving ahead, and for book-keeping purpose, let us summarize what we have so far. The M-theory metric that is relevant for us is 
\eqref{vegamey} with the warp-factors appearing there are defined as in \eqref{mrglass}. The $F_i(t)$ factors appearing in the metric are defined either using the volume preserving condition \eqref{olokhi} or the 
fluctuating condition \eqref{ranjhita}. Although both these forms allow perturbative expansions for $F_i(t)$, the former even allows the inverses to have perturbative expansions. The G-flux components are expressed as in \eqref{frostgiant} except the space-time components ${\bf G}_{\mu\nu\rho M}$ with $y^M$ being the internal coordinates of the eight-manifold. Of course not all $y^M$ are allowed, and we will deal with individual cases as we go along.

\subsection{Einstein's equations and effective field theories \label{instoo}}

An important aspect of our discussion is the quantum terms as they will be solely responsible to change or alter the course of our analysis. These quantum terms that we will be concerned about right now are the ones that will contribute to the energy-momentum tensors. The other quantum terms that  will effect the EOMs for the G-fluxes will be dealt a little later. The former category of quantum terms appear with two free Lorentz indices and whether or not they could create time-neutral series will form the basis of our discussion here. Thus keeping everything in perspective, we can represent the quantum terms in the following way that is a slight variant from what we had in \eqref{londry} or in \cite{nodS}:
\bg\label{neveC}
\mathbb{T}^Q_{MN} \equiv \sum_{k_1, k_2} \mathbb{C}_{MN}^{(k_1, k_2)}(y, M_p)
\left({g_s^2\over \sqrt{h}}\right)^{\Delta k_1} {\rm exp}\left(-{k_2 h^{\Delta/4}\over g^\Delta_s}\right), \nd
where $(k_1, k_2) = (\mathbb{Z}/2, \mathbb{Z})$ with $(M, N)$ being either of $(m, n), (\alpha, \beta), (a, b),
(i, j)$ or $(0, 0)$. The pattern of representation of the quantum terms follow the same pattern of perturbative series expansions employed for the G-fluxes, and the $F_i$ parameters.  This is of course intentional and in some sense necessary if we want to balance all the EOMs. 

The way we have expressed \eqref{neveC}, the $g_s$ scalings have been explicitly extracted out. Without pulling out the $g_s$ scalings, \eqref{neveC} should be identified with either \eqref{phingsha}
or \eqref{phingsha2} depending on the choice \eqref{ranjhita} or \eqref{olokhi} respectively for the case when we allow two free Lorentz indices. The $g_s$ scalings should then coincide with either 
\eqref{teenangul} or
\eqref{charaangul} respectively. These scalings immediately imply:
\bg\label{pabetha}
\Delta = {1\over 3}, ~~~~ \left(k_1, k_2\right) \in \left({\mathbb{Z}\over 2}, {\mathbb{Z}}\right), \nd
for \eqref{neveC} and also for scalings of $F_2(t), F_1(t)$ and ${\bf G}_{MNPQ}$ in \eqref{bobby}, 
\eqref{jenlop} and \eqref{frostgiant} respectively\footnote{Another way to see this is as follows. The typical $g_s$ exponent of a quantum term in say \eqref{miai}, \eqref{melamon2}, \eqref{teenangul} or 
\eqref{charaangul}
goes as ${n_1 + n_2\over 3} + 2\Delta k n_2$ with all $n_i \in \mathbb{Z}$ in the following. Similarly the $g_s$ exponent of a G-flux component from \eqref{frostgiant} and \eqref{hh4u} goes as $2\Delta k n_3$. Clearly with $n_2 = n_3 = 0$, the $g_s$ exponents for $F_{1, 2}(t)$ should also go as ${n_4\over 3}$ and ${n_5\over 3}$ respectively. On the other hand, if $n_1 = 0$, then the $g_s$ exponent of the quantum term goes as $\left(2\Delta k + {1\over 3}\right)n_2$. We could ask for similar scalings for the $F_{1, 2}(t)$ terms, but then the $g_s$ exponent for the G-flux can only go as 
$2\Delta k n_3$ as this is the lowest allowed exponent from \eqref{hh4u}.  The simplest, and probably the most economical, way to resolve all this is to allow $\Delta$ and $k$ to follow the values as in 
\eqref{pabetha}. For the generic case in \eqref{dcmika} the $g_s$ exponent for a typical quantum term appears to be ${n_1\over 3} + \left(2\Delta k + {n_6\over 3} - {\gamma\over 2}\right)n_2$ with $\gamma$ 
defined in \eqref{ranjhita3}. On the other hand, the $g_s$ exponent in the G-flux component still remains
$2\Delta k n_3$ as before. Assuming $k \in {\mathbb{Z}\over 2}$, we now have scaling issue associated with $\left(\Delta, {\gamma\over 2}\right)$ instead of $\Delta$ before. Again the simplest way to resolve this 
would be to allow $\Delta = {1\over 3}$ as in \eqref{pabetha}, and $\gamma = {2n_7\over 3}$. Clearly 
$n_7 = 0, 3$ are the cases \eqref{olokhi} and \eqref{ranjhita} respectively. 
 \label{triumf}}.
 Eventually however it all boils down to the question whether $\mathbb{C}_{MN}^{(0, 0)}$ exists or not, and if it exists, whether there is a $M_p$ hierarchy or not\footnote{As cautioned in footnote \ref{error}, it will be erroneous to expand \eqref{neveC} in inverse powers of $g_s$ to extract $g_s$ independent pieces. For example if one does it, then \eqref{neveC} becomes: 
$$\mathbb{T}^Q_{MN} = \sum_{k_1, k_2, m} {(-1)^m \Delta^m k_2^m \mathbb{C}_{MN}^{(k_1, k_2)}\over m!} 
~g_s^{\Delta(2k_1-m)}
h^{\Delta(m-2k_1)/4}$$
\noindent implying that there are time-neutral pieces whenever $m = 2k_1$. Such an analysis suffers from the problem that for any values of $m > 2k_1$ in the above expansion, the terms are not well defined in the limit $g_s \to 0$. Since all our expansions solely rely on the $g_s << 1$ limit, or more appropriately the 
$g_s \to 0$ limit, the inverse $g_s$ expansions are not advisable as they will lead to erroneous conclusions.\label{tantana}}.  
For the case \eqref{ranjhita}, our study of the scaling behavior \eqref{teenangul} with $\theta_k$ defined as in \eqref{miai}, tells us that: 
\bg\label{cambell}
\mathbb{C}_{ab}^{(0, 0)} = \mathbb{C}_{\alpha\beta}^{(0, 0)} = 0, ~~~~ \mathbb{C}_{mn}^{(0, 0)} = 
{\bf R}_{mn}, {\bf g}_{mn}~ {\bf g}^{\alpha\beta} \Lambda^{(22)}_{\alpha\beta}, \nd
but no $\Lambda_{mn}^{(11)}$  or $\Lambda_{mn}^{(12)}$ terms from \eqref{kimB}. This is because 
\eqref{miai} requires $l_{28} = 2$, implying $H_5 = 2, H_4 = 2$ and $H_3 = -1$ from \eqref{lengchink}. 
This actually vanishes, in the light of both the derivative constraint and the preservation of the type IIB metric form \eqref{pyncmey} as long as we ignore {\it localized} fluxes. The latter will be useful soon. The other non-zero tensor is the Ricci tensor ${\bf R}_{mn}$ that is time-neutral but is {\it not} a quantum piece. Therefore putting these together, all terms except $\mathbb{C}_{\mu\nu}^{(0, 0)}$ vanish for the case \eqref{ranjhita}. The non-local counter-terms do not add any extra time-neutral series for this case.

For the case \eqref{olokhi} the scenario turns out to be a bit different from \eqref{cambell} because now the non-localities do contribute towards creating new time-neutral series as may be inferred from 
\eqref{lengmey} with zero Lorentz indices and \eqref{tamtam} for two free Lorentz indices. This means we should again be looking for $\mathbb{C}_{MN}^{(0, 0)}$, which now takes the following form:
\bg\label{bhishonmey}
&&\mathbb{C}_{ab}^{(0, 0)} = 0 + \sum_{\{l_i\}, n}\sum_{r = 1}^\infty M_p^{-\sigma_r} 
\mathbb{V}^r_{\mathbb{T}^2} \mathbb{G}_{ab}^{(\{l_i\}, n)}(y_{r+1})\delta\left(\theta'_k -{2\over 3}(r-2)\right)
\nonumber\\
&&\mathbb{C}_{\mu\nu}^{(0, 0)} = \sum_j {\bf C}_{\mu\nu}^{(j)} + \sum_{\{l_i\}, n}\sum_{r = 1}^\infty 
M_p^{-\sigma_r} 
\mathbb{V}^r_{\mathbb{T}^2} \mathbb{G}_{\mu\nu}^{(\{l_i\}, n)}(y_{r+1})\delta\left(\theta'_k -{2\over 3}(r+4)\right)
\nonumber\\
&&\mathbb{C}_{A_iB_i}^{(0, 0)} = \left\{{\bf R}_{A_iB_i}, \Lambda_{A_iB_i}^{(ij)}\right\} + \sum_{\{l_i\}, n}\sum_{r = 1}^\infty M_p^{-\sigma_r} 
\mathbb{V}^r_{\mathbb{T}^2} \mathbb{G}_{A_iB_i}^{(\{l_i\}, n)}(y_{r+1})
\delta\left(\theta'_k -{2\over 3}(r+1)\right),
 \nd
where $(A_1, B_1)$  and $(A_2, B_2)$ correspond to $(m, n)$ and $(\alpha, \beta)$ respectively with the superscript notation as in \eqref{kimB}, 
$\theta'_k$ is defined in \eqref{melamon2}, and the 
$\mathbb{G}_{MN}^{(\{l_i\}, n)}$ may be extracted from the functional form \eqref{chinmey} by taking care of the Lorentz indices. The $M_p$ power at any degree of non-locality is given in \eqref{kamagni} by using 
\eqref{kamagni2}. 
One may easily see that all the three quantum series $\mathbb{C}_{ab}$, $\mathbb{C}_{mn}$ and 
$\mathbb{C}_{\alpha\beta}$ are suppressed by powers of $\mathbb{V}_{\mathbb{T}^2}$ and in the limit of vanishing volume, i.e \eqref{tarfox}, they decouple. However what survive in this limit are the time-neutral series given by sum over all $j$ in ${\bf C}_{\mu\nu}^{(j)}$ because $\Lambda^{(ij)}_{A_iB_i} = 0$ and 
${\bf R}_{A_iB_i}$ are classical. Again, the vanishings of $\Lambda^{(ij)}_{A_iB_i}$, in the light of both the derivative constraint and the preservation of the type IIB metric form \eqref{pyncmey}, are allowed as long as the {\it localized} fluxes are ignored.
Interestingly, the sum over the time-neutral quantum terms ${\bf C}_{\mu\nu}^{(j)}$ is now {\it finite} in number and have well defined hierarchy as evident from \eqref{thurmanu}, \eqref{mojarmet}, \eqref{umakim} and \eqref{monox}. This amazing turn of events will help us to find solutions where originally there were none \cite{nodS}. 

\subsubsection{Einstein equation along $(m, n)$ directions \label{kocu1}}

With all the quantum terms at hand, let us now compute the equations of motion for all the fields and parameters in the theory. We will start by first addressing the Einstein's equations. Since there are multiple
components in the theory, let us narrow it down to the Einstein's equation along ($m, n$) directions. The Einstein tensor is given by:
\bg\label{zalo}
\mathbb{G}_{mn} &=& {\bf G}_{mn}-\frac{\partial_m h \partial_n h}{2h^2}+ {g}_{mn }
\left[ {3}ht\Lambda \dot{F_2}  -6h \Lambda F_2+ \frac{F_2}{F_1}\frac{\partial_\alpha h \partial^\alpha h}{4h^2}+\frac{\partial_k h \partial^k h}{4h^2}\right]\nonumber\\
&-&g_{mn}\left[\frac{3}{2}ht^2\Lambda \ddot{F}_2
-\frac{ht^2\Lambda \dot{F}_1^2{F}_2 }{4F_1^2}
+\frac{3ht^2\Lambda \dot{F}_2\dot{F}_1 }{2F_1}
-\frac{2ht\Lambda \dot{F}_1{F}_2 }{F_1}
+\frac{ht^2\Lambda \ddot{F}_1{F}_2 }{F_1}\right] \nonumber\\
&=& {\bf G}_{mn}-\frac{\partial_m h \partial_n h}{2h^2}+ {g}_{mn }
\left[ {3}h^{3/4}\Lambda^{1/2} g_s \dot{F_2}  -6h \Lambda F_2+ \frac{F_2}{F_1}\frac{\partial_\alpha h \partial^\alpha h}{4h^2}+\frac{\partial_k h \partial^k h}{4h^2}\right]\nonumber\\
&-&g_s g_{mn}\sqrt{h}\left[\frac{3}{2} g_s \ddot{F}_2
-\frac{g_s \dot{F}_1^2{F}_2 }{4F_1^2}
+\frac{3g_s \dot{F}_2\dot{F}_1 }{2F_1}
-\frac{2h^{1/4}\sqrt{\Lambda} \dot{F}_1{F}_2 }{F_1}
+\frac{g_s \ddot{F}_1{F}_2 }{F_1}\right], \nd
where $g_{mn}$ is the un-warped metric from \eqref{vegamey}, which is also the ingredient used in 
the un-warped Einstein tensor ${\bf G}_{mn}$. In the third and the fourth lines, we have replaced the time parameter by $g_s$. Such a $g_s$ expansion should also be reflected in the definitions of $F_i(t)$ and whose behaviors are governed by either \eqref{olokhi} or \eqref{ranjhita}. Both these cases will be discussed separately as we go along.  

The other ingredient to balance the Einstein's equation is the expression for the energy-momentum tensor. As we saw earlier in section \ref{kasner}, there are potentially two contributions to it. One coming from the quantum pieces in \eqref{neveC}, and the other from the G-fluxes. 
The energy-momentum tensor from the G-flux is now given by:

{\footnotesize
\bg\label{redford}
\mathbb{T}^G_{mn}&=& 
\frac{1}{4hF_2^{2}}\left({\bf G}_{mlka}{\bf G}_{n}^{~lka}-\frac{1}{6} g_{mn}{\bf G}_{pkla}{\bf G}^{pkla}\right) -\frac{\partial_m h \partial_n h}{2h^2}+{g}_{mn }\left(\frac{F_2}{F_1}\frac{\partial_\alpha h \partial^\alpha h}{4h^2}+\frac{\partial_{m'} h \partial^{m'} h}{4h^2}\right)\nonumber\\
&+& \frac{1}{2hF_1F_2}\left({\bf G}_{ml\alpha a}{\bf G}_{n}^{~l\alpha  a}-\frac{1}{4}g_{mn} {\bf G}_{pl\alpha a}
{\bf G}^{pl\alpha a}\right)+\frac{1}{4hF_1 ^2}\left({\bf G}_{m\alpha\beta a}{\bf G}_{n}^{~\alpha\beta a}-\frac{1}{2}
g_{mn}{\bf G}_{p\alpha \beta a}{\bf G}^{p\alpha \beta a}\right)\nonumber\\
&+&
\frac{\Lambda(t)}{12h F_2^3}\left({\bf G}_{mlkr}{\bf G}_{n}^{~lkr}
-\frac{1}{8} g_{mn}{\bf G}_{pklr}{\bf G}^{pklr}\right)
+\frac{\Lambda(t)}{4hF_2^2 F_1}\left({\bf G}_{mlk\alpha}{\bf G}_{n}^{~lk\alpha}-\frac{1}{6} g_{mn} {\bf G}_{pkl\alpha}{\bf G}^{pkl\alpha}\right)\nonumber \\
&+& \frac{\Lambda(t)}{4hF_2F_1^2}
\left({\bf G}_{ml\alpha\beta}{\bf G}_{n}^{~l\alpha\beta}-\frac{1}{4} g_{mn}{\bf G}_{pl\alpha\beta}
{\bf G}^{pl\alpha\beta}\right)+
\frac{1}{4h\Lambda(t)F_2}\left({\bf G}_{mlab}{\bf G}_{n}^{~lab}-\frac{1}{4} g_{mn} {\bf G}_{pkab}{\bf G}^{pkab}\right)
\nonumber\\
&+& \frac{1}{4h\Lambda(t)F_1}\left({\bf G}_{m\alpha ab}{\bf G}_{n}^{~\alpha ab}-\frac{1}{2} g_{mn} {\bf G}_{p\alpha ab}{\bf G}^{p\alpha ab}\right)
-\frac{F_2}{16h\Lambda(t)F_1^2}\left(g_{mn} {\bf G}_{\alpha\beta a b}{\bf G}^{\alpha \beta ab}\right),
\nd}
where one may notice that we have retained components like ${\bf G}_{MNab}$. This is to keep in mind 
the case alluded to in footnote \ref{westgate} that will become important soon, and for the cases pertaining to our earlier constraints, we will be dealing with them on an individual basis as we go along. The other ingredients appearing in \eqref{redford} are the $F_i(t)$ functions and the warp-factor $h(y)$. The $F_i(t)$ functions satisfy \eqref{olokhi} or \eqref{ranjhita} depending on what conditions  we want to impose on the Newton's constant for the vanilla de Sitter case; and $h(y)$ is the warp-factor that is not required to be kept as a constant. Our aim in the following would be to study the two cases, \eqref{olokhi} and \eqref{ranjhita}, and ask if solutions exist corresponding to the 
background \eqref{vegamey} or \eqref{pyncmey}. 

\vskip.1in

\noindent {\it Case 1: $F_1(t)$ and $F_2(t)$ satisfying the volume-preserving condition \eqref{olokhi}}

\vskip.1in

\noindent The functional form for $F_2(t)$ has always been fixed to \eqref{bobby} for either \eqref{olokhi} or
\eqref{ranjhita}. For our purpose however the full form of \eqref{bobby} is not useful since we will only be concerned with $g_s \to 0$ limit which incidentally is also the late time limit. For this case, since 
$e^{-1/g^\Delta_s}$ dies off faster than any powers of $g_s$, we can simplify \eqref{bobby} and write it as:
\bg\label{karishma}
F_2(t) = \sum_{k\in{\mathbb{Z}\over 2}} C_{k} \left({g_s\over H}\right)^{2\Delta k}, ~~~~
F_1(t) = F_2^{-2}(t) = \sum_{k\in{\mathbb{Z}\over 2}} \widetilde{C}_{k} \left({g_s\over H}\right)^{2\Delta k}, \nd
where $H(y) \equiv h^{1/4}(y)$ is used to avoid fractional powers of warp-factors and $C_k \equiv
c_{k0}$ in \eqref{bobby}. 
Note that we have expressed $F_1(t)$ in the same format as $F_2(t)$, but with coefficients given by $\widetilde{C}_{k}$. These coefficients\footnote{The $C_k$ and $\widetilde{C}_k$ coefficients are 
related by $\sum_{\{k_i\}} \widetilde{C}_{k_1} C_{k_2} C_{k_3} \left({g_s\over H}\right)^{2\Delta(k_1 + 
k_2 + k_3)} = 1$ from where \eqref{oochi} may be determined.} 
may be easily found from \eqref{olokhi}, and here we quote a few of them:
\bg\label{oochi}
&&\widetilde{C}_{0} = C_{0} \equiv 1, ~~~~
\widetilde{C}_{{1\over 2}} = - 2 C_{{1\over 2}}, ~~~~ \widetilde{C}_{{1}} =  3 C^2_{{1\over 2}} -2 C_{1}
\\
&& \widetilde{C}_{{3\over 2}} = - 2 C_{{3\over 2}} + 6 C_{{1\over 2}} C_{1} - 4 C^3_{{1\over 2}}, ~~
\widetilde{C}_{{2}} = - 2 C_{{2}} + 5C^4_{1\over 2} + 3 C^2_{1}+ 6 C_{{1\over 2}} C_{{3\over 2}} 
- 12 C^2_{{1\over 2}} C_{1}. \nonumber \nd
These constant coefficients will have to be determined by plugging the ansatze in the supergravity equations of motion in the presence of the quantum terms. To proceed, we will need time derivatives of $F_2(t)$ and $F_1(t)$. For $F_2(t)$, they are some variants of \eqref{scarjo}:
\bg\label{liftbandh}
\dot{F}_2(t) = 2\Delta\sqrt{\Lambda} \sum_{k\in{\mathbb{Z}\over 2}} k C_{k} 
\left({g_s\over H}\right)^{2\Delta k-1}, ~~~
\ddot{F}_2(t) = 2\Delta{\Lambda} \sum_{k\in{\mathbb{Z}\over 2}} k(2\Delta k-1) C_{k} 
\left({g_s\over H}\right)^{2\Delta k-2}, \nd
arising due to the simplification adopted in \eqref{karishma}, and $\Lambda$ is the cosmological constant 
that appears in \eqref{vegamey}.
If we want to work with \eqref{scarjo} we will have to retain  $e^{-1/g^\Delta_s}$ pieces, but cannot expand it in inverse powers of $g^\Delta_s$ as cautioned in 
footnotes \ref{error} and \ref{tantana}. The time derivatives of $F_1(t)$ have exactly the same form as 
\eqref{liftbandh} except the $C_k$'s are replaced by $\widetilde{C}_k$. Plugging these in \eqref{zalo} we can express $\mathbb{G}_{mn}$ in powers of $g_s$ in the following way:
\bg\label{9do11}
\mathbb{G}_{mn} & = & {\bf G}_{mn} + 3\Lambda H^4 g_{mn} \sum_k \left(3\Delta k - 2\Delta^2k^2 -2\right) C_k 
\left({g_s\over H}\right)^{2\Delta k} \nonumber\\
&+& \Delta^2 \Lambda H^4 g_{mn} \sum_{\{k_l\}} k_1 k_2 \widetilde{C}_{k_1}\widetilde{C}_{k_2}\prod_{i = 3}^7
C_{k_i}\left({g_s\over H}\right)^{2\Delta(k_1 + ...+ k_7)} - {8\partial_m H \partial_n H\over H^2} \nonumber\\
 &-&  2\Delta \Lambda H^4 g_{mn} \sum_{\{k_l\}} k_1(3\Delta k_2 + 2\Delta k_1 -3) \widetilde{C}_{k_1}\prod_{i = 2}^4
C_{k_i}\left({g_s\over H}\right)^{2\Delta(k_1+ k_2 + k_3 + k_4)} \nonumber\\
&+&  {4g_{mn}\over H^2}\left(\partial_lH\partial^l H +
\partial_\alpha H\partial^\alpha H \sum_{\{k_l\}}C_{k_1} C_{k_2} C_{k_3}
\left({g_s\over H}\right)^{2\Delta(k_1+ k_2 + k_3)}\right), \nd
where the braces $\{k_l\}$ denote sum over all the $k_l \in {\mathbb{Z}\over 2}$ values. It is interesting that
only ($k_1, k_2$) explicitly show up as coefficients which implies summing over all other permutations of 
$k_p$ for $p \ne 1, 2$. This will be important when we want to extract various powers of $g_s$ to balance the equations.   

Let us now consider the energy-momentum tensor for the G-fluxes. The full expression has been given in
\eqref{redford}. One may note that the last three terms therein are exactly the ones we have in 
\eqref{kimB} (see also footnote \ref{aaloo}). In the $g_s \to 0$ limit, we can represent the G-flux from 
\eqref{frostgiant} as:
\bg\label{ravali}
{\bf G}_{MNPQ} = \sum_{k\in {\mathbb{Z}\over 2}} {\cal G}_{MNPQ}^{(k)}(y)
 \left({g_s\over H}\right)^{2\Delta k}, \nd
where $H = h^{1/4}$ is as defined earlier, and we have used the fact that in the limit of $g_s \to 0$, 
$e^{-1/g^\Delta_s}$ dies-off faster than any powers of $g_s$. Plugging \eqref{ravali} and \eqref{karishma} in 
\eqref{redford}, we get:
\bg\label{nkinsky} 
\mathbb{T}_{mn}^G & = & \sum_{\{k_i\}} {\widetilde{C}_{k_1}\over 4 H^4}
\left({\cal G}^{(k_2)}_{mlka} {\cal G}_n^{(k_3)lka} - {1\over 6} g_{mn} {\cal G}_{plka}^{(k_2)} {\cal G}^{(k_3)plka}\right)
\left({g_s\over H}\right)^{2\Delta(k_1 + k_2 + k_3)}\nonumber\\  
&+&  \sum_{\{k_i\}} {{C}_{k_1}\over 2 H^4}
\left({\cal G}^{(k_2)}_{ml\alpha a} {\cal G}_n^{(k_3)l\alpha a} - {1\over 4} g_{mn} {\cal G}_{pl\alpha a}^{(k_2)} 
{\cal G}^{(k_3)pl\alpha a}\right)
\left({g_s\over H}\right)^{2\Delta(k_1 + k_2 + k_3)}\nonumber\\ 
& + &  \sum_{\{k_i\}} {C_{k_1}C_{k_2} C_{k_3} C_{k_4}\over 4 H^4}
\left({\cal G}^{(k_5)}_{m\alpha\beta a} {\cal G}_n^{(k_6)\alpha\beta a} 
- {1\over 2} g_{mn} {\cal G}_{p\alpha \beta a}^{(k_5)} {\cal G}^{(k_6)p\alpha \beta a}\right)
\left({g_s\over H}\right)^{2\Delta(k_1+ .... + k_6)}\nonumber\\  
&+&  \sum_{\{k_i\}} {\widetilde{C}_{k_1}\widetilde{C}_{k_2} C_{k_3}\over 12 H^4}
\left({\cal G}^{(k_4)}_{mlkr} {\cal G}_n^{(k_5)lkr} - {1\over 8} g_{mn} {\cal G}_{pklr}^{(k_4)} {\cal G}^{(k_5)pklr}\right)
\left({g_s\over H}\right)^{2\Delta(k_1 + ... + k_5 + 1/\Delta)}\nonumber\\ 
&+& \sum_{\{k_i\}} {1 \over 4 H^4}
\left({\cal G}^{(k_1)}_{mlk\alpha} {\cal G}_n^{(k_2)lk\alpha} - {1\over 6} g_{mn} {\cal G}_{plk\alpha}^{(k_1)} {\cal G}^{(k_2)plk\alpha}\right)
\left({g_s\over H}\right)^{2\Delta(k_1 + k_2 + 1/\Delta)}\nonumber\\  
&+&  \sum_{\{k_i\}} {{C}_{k_1}{C}_{k_2} C_{k_3}\over 4 H^4}
\left({\cal G}^{(k_4)}_{ml\alpha\beta} {\cal G}_n^{(k_5)l\alpha\beta} - {1\over 4} g_{mn} 
{\cal G}_{pl\alpha\beta}^{(k_4)} {\cal G}^{(k_5)pl\alpha\beta}\right)
\left({g_s\over H}\right)^{2\Delta(k_1 + ... + k_5 + 1/\Delta)}\nonumber\\ 
&+&  \sum_{\{k_i\}} {\widetilde{C}_{k_1} C_{k_2}\over 4 H^4}
\left({\cal G}^{(k_3)}_{mlab} {\cal G}_n^{(k_4)lab} - {1\over 4} g_{mn} {\cal G}_{pkab}^{(k_3)} {\cal G}^{(k_4)pkab}\right)\left({g_s\over H}\right)^{2\Delta(k_1 + ... + k_4 - 1/\Delta)}\nonumber\\ 
&+&  \sum_{\{k_i\}} {{C}_{k_1} C_{k_2}\over 4 H^4}
\left({\cal G}^{(k_3)}_{m\alpha ab} {\cal G}_n^{(k_4)\alpha ab} - {1\over 2} g_{mn} 
{\cal G}_{p\alpha ab}^{(k_3)} {\cal G}^{(k_4)p\alpha ab}\right)
\left({g_s\over H}\right)^{2\Delta(k_1 + ... + k_4 - 1/\Delta)}\nonumber\\ 
&-& {g_{mn}\over 16H^4} \sum_{\{k_i\}} C_{k_1}.......C_{k_5} {\cal G}^{(k_6)}_{\alpha\beta ab} 
{\cal G}^{(k_7)\alpha\beta ab}\left({g_s\over H}\right)^{2\Delta(k_1 + ... + k_7 - 1/\Delta)}
- {8\partial_m H \partial_n H\over H^2} \nonumber\\
&+&  {4g_{mn}\over H^2}\left(\partial_lH\partial^l H +
\partial_\alpha H\partial^\alpha H \sum_{\{k_l\}}C_{k_1} C_{k_2} C_{k_3}
\left({g_s\over H}\right)^{2\Delta(k_1+ k_2 + k_3)}\right), \nd
where note that we have retained components like ${\cal G}^{(k)}_{MNab}(y)$, which immediately implies that these components cannot be expressed as \eqref{montan} because for the limit $g_s \to 0$ only the constant zero form survives. We also want to avoid switching on components like ${\bf C}_{Mab}$ to avoid developing cross-terms in the type IIB background \eqref{montse}. Thus the only option is to view them as 
{\it localized} fluxes which, in fact, will also be very useful to resolve other subtle issues surrounding 
flux quantization etc. in the full M-theory framework. By construction then:
\bg\label{metromey}
{\cal G}^{(0)}_{MNPQ} = 0.\nd
With these at hand, we are now ready to discuss all the equations of motion for the system. Our first step would be to study the EOMs at zeroth order in $g_s$. Looking at \eqref{9do11}, \eqref{ravali} and \eqref{neveC}, it is easy to infer the following:
\bg\label{misslemon}
&&{\bf G}_{mn} - 6\Lambda H^4 g_{mn}  = \sum_{\{k_i\}}\Bigg[ {\widetilde{C}_{k_1} C_{k_2}\over 4 H^4}
\left({\cal G}^{(k_3)}_{mlab} {\cal G}_n^{(k_4)lab} - {1\over 4} g_{mn} {\cal G}_{pkab}^{(k_3)} {\cal G}^{(k_4)pkab}\right)\nonumber\\
&& + {{C}_{k_1} C_{k_2}\over 4 H^4}
\left({\cal G}^{(k_3)}_{m\alpha ab} {\cal G}_n^{(k_4)\alpha ab} - {1\over 2} g_{mn} 
{\cal G}_{p\alpha ab}^{(k_3)} {\cal G}^{(k_4)p\alpha ab}\right)\Bigg] \delta(k_1 + k_2 + k_3 + k_4 - 3)
\nonumber\\
&& -{g_{mn}\over 16H^4} \sum_{\{k_i\}} C_{k_1}C_{k_2}C_{k_3}C_{k_4}C_{k_5} 
{\cal G}^{(k_6)}_{\alpha\beta ab} 
{\cal G}^{(k_7)\alpha\beta ab}\delta(k_1 + k_2 + .. + k_7 - 3) + \mathbb{C}^{(0, 0)}_{mn}, \nd
where the delta function is simply used to fix the condition on $k_i$. Note that all $k_i \in \mathbb{Z}/2$, and 
both set of ($k_3, k_4$) as well as ($k_6, k_7$) cannot vanish, and take the minimum values of $1/2$,  because of \eqref{metromey}. On the other hand, \eqref{melamon2} tells us that simplest condition of 
$\Delta k \ge 1/2$ which, with the delta function constraint above, immediately implies $k_3 = k_4 = 3/2$ in the first two lines  and 
$k_6 = k_7 = 3/2$ in the last line of \eqref{misslemon} and the rest zero. We could also analyze this using the refined conditions on $k_i$ of the G-flux components discussed in the paragraphs between \eqref{evabmey2} and \eqref{teenangul}, but this will not change any of our analysis significantly. Therefore to avoid further complicating the matter at hand, we will henceforth only consider the simplest modings of the G-flux components, namely $\Delta k \ge  {1\over 2}$ and $\Delta k \ge {3\over 2}$ for the two cases, \eqref{olokhi} and 
\eqref{ranjhita} respectively, unless mentioned otherwise. Thus:
\bg\label{misslemon2}
&&{\bf G}_{mn} - 6\Lambda H^4 g_{mn}  = \mathbb{C}^{(0, 0)}_{mn} + {1 \over 4 H^4}
\left({\cal G}^{(3/2)}_{mlab} {\cal G}_n^{(3/2)lab} - {1\over 4} g_{mn} {\cal G}_{pkab}^{(3/2)} {\cal G}^{(3/2)pkab}\right)\nonumber\\
&& + {1 \over 4 H^4}
\left({\cal G}^{(3/2)}_{m\alpha ab} {\cal G}_n^{(3/2)\alpha ab} - {1\over 2} g_{mn} 
{\cal G}_{p\alpha ab}^{(3/2)} {\cal G}^{(3/2)p\alpha ab}\right)
-{g_{mn}\over 16H^4} 
{\cal G}^{(3/2)}_{\alpha\beta ab} 
{\cal G}^{(3/2)\alpha\beta ab}, \nd
which is actually a set of 10 equations with 31 unknowns. The RHS is completely fixed once we know the functional form for ${\cal G}^{(3/2)}_{MNPQ}(y)$ components.
All these fluxes appearing above are localized fluxes and according to \eqref{bhishonmey}, at the zeroth order 
in $g_s$, there are no local quantum terms, except classical ones, and contributions to 
$\mathbb{C}^{(0, 0)}_{mn}$ come mostly from the non-local counter-terms. These are suppressed by powers of the torus volume and therefore their contributions are negligible\footnote{In the case where we do not impose the derivative constraints, as discussed in case 5 of section \ref{Gng6}, the non-local quantum terms are still finite in number \cite{petite}, thus keeping the conclusions unchanged. In this paper, since we are imposing the derivative constraints, we will not discuss the generic scenario.}. 
This is one of the key difference between a similar equation appearing in 
\cite{nodS} (see eq (5.25) in \cite{nodS}). The number of terms appearing in $\mathbb{C}_{mn}^{(i)}$ 
in eq (5.25) of \cite{nodS} are the number of solutions of $\theta'_0 = {2\over 3}$ in \eqref{kkkbkb2}. Clearly there are an {\it infinite} number of solutions for \eqref{kkkbkb2} with no hierarchy, the latter because of the inclusion of the non-local counter-terms. This ruined an EFT description in \cite{nodS}.
 
Before moving ahead let us clarify few questions that may be asked at this point regarding the two scaling behavior \eqref{melamon2} for \eqref{olokhi}, and \eqref{kkkbkb2} for the time-independent case. First, in determining the $g_s$ scaling $g^{\theta'_k}_s$ or 
$g_s^{\theta'_0}$, what values of the metric and G-flux components should we insert in  
\eqref{phingsha2}?
Recall from \eqref{vegamey} and \eqref{mrglass} the metric components are expressed in terms of their 
$g_s$ scalings as:
\bg\label{3amigomey}
&&{\bf g}_{\mu\nu} = g_s^{-8/3} \eta_{\mu\nu}, ~~~ {\bf g}_{ab} = g_s^{4/3} \delta_{ab} \nonumber\\
&& {\bf g}_{\alpha\beta} = g_{\alpha\beta}\left[\left({g_s\over H}\right)^{-{2\over 3}} + 
\widetilde{C}_{1\over 2}\left({g_s\over H}\right)^{-{1\over 3}} + \widetilde{C}_1 + \widetilde{C}_{3\over 2}
\left({g_s\over H}\right)^{{1\over 3}} + ..... \right] H^{4/3} \nonumber\\
&& {\bf g}_{mn} = g_{mn}\left[\left({g_s\over H}\right)^{-{2\over 3}} + 
{C}_{1\over 2}\left({g_s\over H}\right)^{-{1\over 3}} + {C}_1 + {C}_{3\over 2}
\left({g_s\over H}\right)^{{1\over 3}} + ..... \right] H^{4/3}, \nd
where the $C_k$ and $\widetilde{C}_k$ are related by \eqref{oochi}. Notice that near $g_s \to 0$, both 
${\bf g}_{mn}$ and ${\bf g}_{\alpha\beta}$ are dominated by their first terms. In fact for either of these components our perturbative expansion doesn't make any sense because of the inverse $g_s$ factors. This then implies that to determine the quantum terms \eqref{phingsha2} all we need is to just take the dominant contributions that go as $g_s^{-2/3}$ for the two metric components. Thus the $g_s$ scaling in 
\eqref{neveC} will appear solely from the $l_i$ terms in \eqref{melamon2} and {\it not} from the lower order scalings of the metric\footnote{The lower order scalings of the metric components in \eqref{3amigomey} do not change the general conclusion that we inherit from \eqref{melamon2}. To see this let us rewrite the 
$g_s$ scaling of the metric component ${\bf g}_{\alpha\beta}$ as $\left({g_s\over H}\right)^{-{2\over 3}} F_1(t)$. A generic term in the expansion will appear as $c_\gamma g_s^{-2/3 + \vert\gamma\vert}$. The inverse of the metric component then become ${\bf g}^{\alpha\beta} =  \left({g_s\over H}\right)^{2\over 3} F_1^{-1} 
g^{\alpha\beta} =  \left({g_s\over H}\right)^{2\over 3} F^2_2 g^{\alpha\beta}$, leading to a typical term in the $g_s$ scaling of the inverse to be of the form $b_\gamma g_s^{+2/3 + \vert\gamma\vert}$. Note that the 
sign of the 
$\vert\gamma\vert$ exponent has {\it not} changed. 
Thus inverses of $F_1(t)$ or $F_2(t)$ do not contribute negative exponents of $g_s$ because of their perturbative expansions. The only issue could be from the temporal derivatives of the metric components, 
and we should only care about one and two derivatives only. The $n$ temporal derivatives yield a generic form of $g_s^{-2/3 + \vert \gamma\vert - n}$, which for $n = 1, 2$ becomes $g_s^{-5/3 + \vert\gamma\vert}$
and $g_s^{-8/3 + \vert\gamma\vert}$ respectively. Again the sign of the $\gamma$ term has not changed, and the derivative action could be thought of changing only the dominant piece,
implying no chance of generating any time-neutral series from the higher order terms in \eqref{3amigomey}. \label{ironman}}
in \eqref{3amigomey}. The G-flux components on the other hand, do have perturbative expansions near $g_s\to 0$, which is evident from the factor  of $k$ appearing in \eqref{melamon2} with 
$k \ge 3/2$. Secondly, if the metric components are solely governed by their dominant terms, can we make 
$F_1(t) = F_2(t) = 1$ in \eqref{vegamey} or \eqref{pyncmey}? This could probably be the simplest solution to the system, but appears to over-constrain the scenario.  It turns out that the perturbative expansions of the 
$F_i(t)$ factors are directly related to the perturbative expansions of the G-flux and the quantum terms. This will be demonstrated soon. Finally, if the perturbative expansion of the metric components do not make sense, how are we even justified to proceed in the way we did with say, \eqref{misslemon2}? The answer lies in the miraculous way that the inverse $g_s$ dependences from the metric factors cancel out in the full EOMs, leaving only the perturbative series like that for $F_i(t), \mathbb{T}^G_{MN}$  and 
$\mathbb{T}^Q_{MN}$ to balance each other\footnote{Alternatively, if the inverse $g_s$ pieces exist then one would have to balance powers of ${\rm exp}\left(-{1\over g_s^{1/3}}\right)$. Such terms vanish 
for $g_s \to 0$, which is tantamount to saying that only perturbative powers of $g_s$ need to be balanced in the equations of motion.}. 
This is the reason why we can analyze the system order-by-order in $g_s$ despite the presence of inverse $g_s$ pieces as in \eqref{3amigomey}.    
 
Coming back, taking a trace on both sides of \eqref{misslemon2} immediately tells us that the 
internal manifold ${\cal M}_4$ cannot be a Calabi-Yau manifold. It cannot generically also be a conformally Calabi-Yau, as the non-K\"ahlerity will be controlled by the localized fluxes as well as the cosmological constant $\Lambda$. At this stage one can also count the number of variables we have in the problem. They can be tabulated as:
\bg\label{lelyx}
H(y);~~ g_{mn}(y);~~ {\cal G}^{(3/2)}_{MNPQ}(y),~ {\cal G}^{(2)}_{MNPQ}(y),~ {\cal G}^{(5/2)}_{MNPQ}(y), .... \nd
with 10 components for $g_{mn}$, 1 from $H(y)$ and 70 components from any choice of $k$ in 
${\cal G}^{(k)}_{MNPQ}$ totalling to at least 81 independent functions for a given $k$. The $g_{mn}$ EOM connects the 
metric components with the warp-factor and G-fluxes, which we elucidated to zeroth order in $g_s$ in 
\eqref{misslemon2}. 
In fact a more precise connection of $g_{mn}$ to the fluxes and the quantum terms appears from the next order in $g_s$ i.e $g_s^{1/3}$. The relation becomes:
\bg\label{rambha3} 
 g_{mn} &=& {3\over 58 \mathbb{A}(y)} \mathbb{C}^{(1/2, 0)}_{mn}
+ {3\over 58\mathbb{A}(y)}\sum_{\{k_i\}}\Bigg[ {\widetilde{C}_{k_1} C_{k_2}\over 4 H^4}
\left({\cal G}^{(k_3)}_{mlab} {\cal G}_n^{(k_4)lab} - {1\over 4} g_{mn} {\cal G}_{pkab}^{(k_3)} {\cal G}^{(k_4)pkab}\right)\nonumber\\
&+&  {{C}_{k_1} C_{k_2}\over 4 H^4}
\left({\cal G}^{(k_3)}_{m\alpha ab} {\cal G}_n^{(k_4)\alpha ab} - {1\over 2} g_{mn} 
{\cal G}_{p\alpha ab}^{(k_3)} {\cal G}^{(k_4)p\alpha ab}\right)\Bigg] 
\delta\left(k_1 + k_2 + k_3 + k_4 - {7\over 2}\right), \nonumber\\ \nd
which is another set of 10 equations with at least 44 unknowns. These would imply the precise connection between the ${\cal M}_4$ metric, localized fluxes and the quantum terms. 
The function\footnote{The function \eqref{ramravali} can never be zero globally 
because the G-flux components appearing in \eqref{ramravali} cannot globally cancel the contributions from the warp-factor, as they are by definition localized fluxes. \label{boston405}} 
$\mathbb{A}(y)$ is again a function of the localized fluxes, and the 
warp-factor $H(y)$, as:

{\footnotesize
\bg\label{ramravali}
\mathbb{A}(y) \equiv 
 {3\over 928 H^4} \sum_{\{k_i\}} C_{k_1}C_{k_2}C_{k_3}C_{k_4}C_{k_5} 
{\cal G}^{(k_6)}_{\alpha\beta ab} 
{\cal G}^{(k_7)\alpha\beta ab}\delta\left(k_1 + k_2 + .. + k_7 - {7\over 2}\right)  - C_{1\over 2}\Lambda H^4,
\nd}
where for both \eqref{rambha3} as well as \eqref{ramravali} we have to make sure that 
$(k_3, k_4) \ge (3/2, 3/2)$ as well as $(k_6, k_7) \ge (3/2, 3/2)$ so as to comply with  \eqref{metromey} as well as the positivity of \eqref{melamon2}. 
More crucially, note the dependence of $g_{mn}$ 
on the quantum terms 
$\mathbb{C}_{mn}^{(1/2, 0)}$ from \eqref{neveC}. Since we are looking at $g_s^{1/3}$, this means the local
quantum terms of 
$\mathbb{C}^{(1/2, 0)}_{mn}$ should be extracted from \eqref{phingsha2} and \eqref{charaangul} with 
 $\theta'_k = 1$ in \eqref{melamon2}, i.e:
\bg\label{melamon3}
&&{2} \sum_{i = 1}^{27} l_i + n_1 + n_2 + l_{34} + l_{35} + 
2\left(k +{2}\right)\left(l_{28} + l_{29} + l_{31}\right) 
+  \left(2 k +{1}\right)\left(l_{30} + l_{32} + l_{33}\right) \nonumber\\
&&~~~~~~~~~+ 
2\left( k -{1}\right)\left(l_{36} + l_{37} + l_{38}\right) = 3, \nd
with $(l_i, n_j) \in (\mathbb{Z}, \mathbb{Z})$ as it appears in \eqref{phingsha2}.
Again since $k \ge 3/2$, we see that there are only a few quantum terms that can appear from 
\eqref{melamon3}. These quantum terms may be extracted from a sub-class of \eqref{melamon3} 
that satisfy:
\bg\label{melamon4}
{2} \sum_{i = 1}^{27} l_i + n_1 + n_2 + \sum_{i = 0}^4 l_{34 + i} = 3, \nd
with other $l_i$ not contributing. These clearly select a finite number of local quantum terms from 
\eqref{phingsha2}. The remaining contribution to $\mathbb{C}^{(1/2, 0)}_{mn}$ in 
\eqref{rambha3} come from the non-local counter-terms, implying that to order $g_s^0$ and 
$g_s^{1/3}$, contributions to the metric can only come from the  fluxes and curvature tensors satisfying 
\eqref{melamon3}  and a set of non-local counter-terms (that 
in turn are heavily suppressed prohibiting us to go beyond a certain level of non-locality). For example, the non-local contributions to $r$-th order come from:
\bg\label{kitagra}
\theta'_k = {2\over 3}(r+1), ~~~~~ \theta'_k = {2r\over 3}+1, \nd
for the two cases $\mathbb{C}^{(0, 0)}_{mn}$ and $\mathbb{C}^{(1/2, 0)}_{mn}$ respectively with 
$\theta'_k$ as in \eqref{melamon2}. 
Additionally 
\eqref{misslemon2} is expressed in terms of ${\cal G}^{(3/2)}_{MNPQ}(y)$ whereas \eqref{rambha3} is expressed in terms of ${\cal G}^{(3/2)}_{MNPQ}(y)$ and ${\cal G}^{(2)}_{MNPQ}(y)$ allowing us to express 
${\cal G}^{(2)}_{MNPQ}(y)$ in terms of ${\cal G}^{(3/2)}_{MNPQ}(y)$ and other variables in the 
problem, where $y = (y^m, y^\alpha)$ form the coordinates of ${\cal M}_4 \times {\cal M}_2$.

To elucidate the story further, let us go to the next order in $g_s$, namely $g_s^{2/3}$. Our aim would be to see if there are additional constraints on the metric itself, or new degrees of freedom appear. Combining \eqref{9do11}, \eqref{ravali} and \eqref{neveC}, we get:
\bg\label{clovbalti}
g_{mn} & = & {9 \over \mathbb{B}(y)}  \mathbb{C}_{mn}^{(1, 0)}
+ {9\over \mathbb{B}(y)}\sum_{\{k_i\}}\Bigg[ {\widetilde{C}_{k_1} C_{k_2}\over 4 H^4}
\left({\cal G}^{(k_3)}_{mlab} {\cal G}_n^{(k_4)lab} - {1\over 4} g_{mn} {\cal G}_{pkab}^{(k_3)} {\cal G}^{(k_4)pkab}\right)\\
&+&  {{C}_{k_1} C_{k_2}\over 4 H^4}
\left({\cal G}^{(k_3)}_{m\alpha ab} {\cal G}_n^{(k_4)\alpha ab} - {1\over 2} g_{mn} 
{\cal G}_{p\alpha ab}^{(k_3)} {\cal G}^{(k_4)p\alpha ab}\right)\Bigg] 
\delta\left(k_1 + k_2 + k_3 + k_4 - {4}\right), \nonumber \nd
which is somewhat similar to \eqref{rambha3} but differs in three respects: one, the quantum terms are different; two, the $k_i$ sum over to 4 instead of 7/2 leading to a set of 10 equations with at least 58 unknowns; and three, the denominator is given by $\mathbb{B}(y)$ 
instead of $\mathbb{A}(y)$. This is defined as:

{\footnotesize
\bg\label{melhic}
\mathbb{B}(y) \equiv 
 {9\over 16 H^4} \sum_{\{k_i\}} C_{k_1}C_{k_2}C_{k_3}C_{k_4}C_{k_5} 
{\cal G}^{(k_6)}_{\alpha\beta ab} 
{\cal G}^{(k_7)\alpha\beta ab}\delta\left(k_1 + k_2 + .. + k_7 - {4}\right)  - \alpha_a \Lambda H^4,
\nd}
which should again be compared to \eqref{ramravali} (the non-vanishing of this is guaranteed from a similar argument presented in footnote \ref{boston405}). 
These similarities however do not survive beyond 
$g_s^{5/3}$ and we will comment on it below. The constant $\alpha_a$ is given by the following expression:
\bg\label{modad}
\alpha_a \equiv 43 C^2_{1\over 2} - 61 C_1 - 13 C_{1\over 2}, \nd
with $C_k$ being the constant appearing in the functional form for $F_2(t)$ in \eqref{karishma} and 
\eqref{oochi} an should in principle be determined along-with the metric, warp-factor and the G-flux components. 

Looking at \eqref{clovbalti} and \eqref{rambha3} we see that a pattern is emerging: \eqref{clovbalti} is expressed in terms of G-fluxes of the form ${\cal G}_{MNPQ}^{(5/2)}(y), {\cal G}_{MNPQ}^{(2)}(y)$ and 
${\cal G}_{MNPQ}^{(3/2)}(y)$. Thus knowing the metric information $g_{mn}(y)$ will enable us to express 
${\cal G}_{MNPQ}^{(5/2)}(y)$ in terms of ${\cal G}_{MNPQ}^{(2)}(y)$, ${\cal G}_{MNPQ}^{(3/2)}(y)$ and the warp-factors, as the quantum term in \eqref{clovbalti} is given by $l_i$ in \eqref{phingsha2} satisfying:
\bg\label{mameye}
{2} \sum_{i = 1}^{27} l_i + n_1 + n_2 + l_{34} + l_{35} + 2(k-1) (l_{36} + l_{37} + l_{38}) = 4 + 2r, \nd
with $r = 0$ producing the local terms. Note that $k \le 2$ otherwise the terms would be classical, implying that the quantum terms to this order cannot be constructed out of ${\cal G}^{(5/2)}_{MNPQ}$ justifying the above pattern. 

The form of the Einstein's equations would remain similar till $g_s^{5/3}$. For $g_s^2$ onwards, other components in the energy-momentum tensor \eqref{ravali} would start participating because the 
$k_i \ge  3/2$ bound for the G-flux components would no longer be prohibitive. Thus for any given component of the G-flux, say for example ${\cal G}^{(k)}_{mnab}$, there are infinite number of
sub-components classified by $k$ for $k \ge 3/2$. So far we have only dealt with a few G-flux components and their corresponding sub-components (classified above by $k_i$), but more would appear as we go to 
order $g_s^2$ and beyond. In fact 70 new components of G-flux would appear for every choice of $k_i$, implying that at least 70 new degrees of freedom are added at every order in $g_s$ as we go.

\vskip.1in

\noindent {\it Case 2: $F_1(t)$ and $F_2(t)$ satisfying the fluctuation condition \eqref{ranjhita}}

\vskip.1in

\noindent In the above section we discussed in details how the EOMs for the internal space 
${\cal M}_4$ may be determined from fluxes and the quantum terms. In this section we would like to see how this changes once we impose \eqref{ranjhita3}  or \eqref{ranjhita} on the metric coefficients $F_1(t)$ and $F_2(t)$. One of the first important distinction is the derivative constraint that appears from looking at the generalized scaling 
\eqref{dcmika}. This could even prompt us to analyze the whole section using \eqref{ranjhita3} instead of the special case \eqref{ranjhita}. The generic picture is more technically involved, and since we will not be gaining new physics by looking at \eqref{ranjhita3}, we will suffice ourselves here with a detailed consequence of imposing the special case \eqref{ranjhita} on the background EOMs.  We will however revert to the generic picture whenever possible. 

As a start, let us work out the behavior of the metric coefficients $F_1(t)$ and $F_2(t)$. We will keep 
$F_2(t)$ as in \eqref{karishma}, but change $F_1(t)$ accordingly. 
For example, the generic form for $F_i(t)$ may be expressed as:
\bg\label{fakhi}
F_2(t) = \sum_k C_k \left({g_s\over H}\right)^{2\Delta k}, ~~~~ 
F_1(t) = \sum_k \widetilde{C}_k \left({g_s\over H}\right)^{2\Delta k + \gamma} \equiv
\sum_k \hat{C}_k \left({g_s\over H}\right)^{2\Delta k}, 
\nd
this is almost similar to \eqref{karishma}, if we define $\hat{C}_k \equiv \widetilde{C}_k \left({g_s\over H}\right)^\gamma$. Note that, in this form
the ($C_k, \widetilde{C}_k$) coefficients satisfy the same relation as \eqref{oochi}. However the metric along the ($\alpha, \beta$) direction becomes:
\bg\label{anjaK}
 {\bf g}_{\alpha\beta} = g_{\alpha\beta}\left[\left({g_s\over H}\right)^{-{2\over 3} + \gamma} + 
\widetilde{C}_{1\over 2}\left({g_s\over H}\right)^{-{1\over 3} + \gamma} + \widetilde{C}_1 
\left({g_s\over H}\right)^\gamma + \widetilde{C}_{3\over 2}
\left({g_s\over H}\right)^{{1\over 3} + \gamma} + ..... \right] H^{4/3}, \nd
with the other coefficients remaining the same as in \eqref{3amigomey}. Choosing 
$\gamma = 2$ would explain the metric choice that we took earlier in analyzing the $g_s$ scaling 
\eqref{miai}. Again, we could resort to the dominant scalings of the metric coefficient i.e $g_s^{-2/3 + \gamma}$, but compared to footnote \ref{ironman} the inverse will become $g_s^{+2/3 - \gamma}$ with the 
$\gamma$ exponent picking up a negative sign. This is because $F_1^{-1}$ does not have a perturbative expansion compared to the case explored in footnote \ref{ironman}. The resulting physics will change as evident from the scaling behavior \eqref{dcmika} and \eqref{miai}. 

The time derivatives of $F_2(t)$ will expectedly remain the same as in \eqref{oochi}, but the time derivatives of $F_1(t)$ will change. The change is easy to quantify:
\bg\label{futuci}
&&\dot{F}_1(t) =  \sqrt{\Lambda} \sum_{k\in {{\mathbb{Z}\over 2}}}\widetilde{C}_k (2\Delta k + \gamma) 
\left({g_s\over H}\right)^{2\Delta k + \gamma - 1} \nonumber\\
&& \ddot{F}_1(t) = {\Lambda} \sum_{k\in {{\mathbb{Z}\over 2}}}\widetilde{C}_k (2\Delta k + \gamma) 
(2\Delta k + \gamma - 1) 
\left({g_s\over H}\right)^{2\Delta k + \gamma - 2}, \nd
where the inverse powers of $g_s$ will be dealt carefully once we go to the relevant EOMs. These functional form can now be used to determine the Einstein tensor along the ($m, n$) directions. The result is:
\bg\label{9do12}
\mathbb{G}_{mn} & = & {\bf G}_{mn} + 3\Lambda H^4 g_{mn} \sum_k \left(3\Delta k - 2\Delta^2k^2 -2\right) C_k 
\left({g_s\over H}\right)^{2\Delta k} + {4g_{mn} \partial_lH\partial^l H \over H^2} \\
&+& {1\over 4} \Lambda H^4 g_{mn} \sum_{\{k_l\}} (2\Delta k_1+ \gamma)(2\Delta k_2 + \gamma)
 \widetilde{C}_{k_1}\widetilde{C}_{k_2}\prod_{i = 3}^7
C_{k_i}\left({g_s\over H}\right)^{2\Delta(k_1 + ...+ k_7)} - {8\partial_m H \partial_n H\over H^2} \nonumber\\
 &-& \Lambda H^4 g_{mn} \sum_{\{k_l\}} (2\Delta k_1 + \gamma) (3\Delta k_2 + 2\Delta k_1 + \gamma  -3) \widetilde{C}_{k_1}\prod_{i = 2}^4
C_{k_i}\left({g_s\over H}\right)^{2\Delta(k_1+ k_2 + k_3 + k_4)}, \nonumber
\nd
which in the limit $\gamma = 0$ does {\it not} reproduce all the terms of \eqref{9do11}. In particular terms with derivatives with respect to $\alpha$ are missing. This is of course expected because $\gamma = 0$ and $\gamma > 0$ share different physics. Note also that none of the $g_s$ 
scaling gets effected by the $\gamma$ factor, although the $\gamma$ factor does change the 
some of the coefficients of the terms in a standard way. In a similar vein, the energy-momentum tensor from the G-fluxes may be represented as:
\bg\label{nkinsky2} 
\mathbb{T}_{mn}^G & = & \sum_{\{k_i\}} {\widetilde{C}_{k_1}\over 4 H^4}
\left({\cal G}^{(k_2)}_{mlka} {\cal G}_n^{(k_3)lka} - {1\over 6} g_{mn} {\cal G}_{plka}^{(k_2)} {\cal G}^{(k_3)plka}\right)
\left({g_s\over H}\right)^{2\Delta(k_1 + k_2 + k_3)} + {4g_{mn}\partial_lH\partial^l H\over H^2}
\nonumber\\  
&+&  \sum_{\{k_i\}} {{C}_{k_1}\over 2 H^4}
\left({\cal G}^{(k_2)}_{ml\alpha a} {\cal G}_n^{(k_3)l\alpha a} - {1\over 4} g_{mn} {\cal G}_{pl\alpha a}^{(k_2)} 
{\cal G}^{(k_3)pl\alpha a}\right)
\left({g_s\over H}\right)^{2\Delta(k_1 + k_2 + k_3 -\gamma/2\Delta)}\nonumber\\ 
& + &  \sum_{\{k_i\}} {C_{k_1}C_{k_2} C_{k_3} C_{k_4}\over 4 H^4}
\left({\cal G}^{(k_5)}_{m\alpha\beta a} {\cal G}_n^{(k_6)\alpha\beta a} 
- {1\over 2} g_{mn} {\cal G}_{p\alpha \beta a}^{(k_5)} {\cal G}^{(k_6)p\alpha \beta a}\right)
\left({g_s\over H}\right)^{2\Delta(k_1+ .... + k_6 - \gamma/\Delta)}\nonumber\\  
&+&  \sum_{\{k_i\}} {\widetilde{C}_{k_1}\widetilde{C}_{k_2} C_{k_3}\over 12 H^4}
\left({\cal G}^{(k_4)}_{mlkr} {\cal G}_n^{(k_5)lkr} - {1\over 8} g_{mn} {\cal G}_{pklr}^{(k_4)} {\cal G}^{(k_5)pklr}\right)
\left({g_s\over H}\right)^{2\Delta(k_1 + ... + k_5 + 1/\Delta)}\nonumber\\ 
&+& \sum_{\{k_i\}} {1 \over 4 H^4}
\left({\cal G}^{(k_1)}_{mlk\alpha} {\cal G}_n^{(k_2)lk\alpha} - {1\over 6} g_{mn} {\cal G}_{plk\alpha}^{(k_1)} {\cal G}^{(k_2)plk\alpha}\right)
\left({g_s\over H}\right)^{2\Delta(k_1 + k_2 + 1/\Delta)}\nonumber\\  
&+&  \sum_{\{k_i\}} {{C}_{k_1}{C}_{k_2} C_{k_3}\over 4 H^4}
\left({\cal G}^{(k_4)}_{ml\alpha\beta} {\cal G}_n^{(k_5)l\alpha\beta} - {1\over 4} g_{mn} 
{\cal G}_{pl\alpha\beta}^{(k_4)} {\cal G}^{(k_5)pl\alpha\beta}\right)
\left({g_s\over H}\right)^{2\Delta(k_1 + ... + k_5 + 1/\Delta -\gamma/\Delta)}\nonumber\\ 
&+&  \sum_{\{k_i\}} {\widetilde{C}_{k_1} C_{k_2}\over 4 H^4}
\left({\cal G}^{(k_3)}_{mlab} {\cal G}_n^{(k_4)lab} - {1\over 4} g_{mn} {\cal G}_{pkab}^{(k_3)} {\cal G}^{(k_4)pkab}\right)\left({g_s\over H}\right)^{2\Delta(k_1 + ... + k_4 - 1/\Delta)}\nonumber\\ 
&+&  \sum_{\{k_i\}} {{C}_{k_1} C_{k_2}\over 4 H^4}
\left({\cal G}^{(k_3)}_{m\alpha ab} {\cal G}_n^{(k_4)\alpha ab} - {1\over 2} g_{mn} 
{\cal G}_{p\alpha ab}^{(k_3)} {\cal G}^{(k_4)p\alpha ab}\right)
\left({g_s\over H}\right)^{2\Delta(k_1 + ... + k_4 - 1/\Delta -\gamma/2\Delta)}\nonumber\\ 
&-& {g_{mn}\over 16H^4} \sum_{\{k_i\}} C_{k_1}... C_{k_5} {\cal G}^{(k_6)}_{\alpha\beta ab} 
{\cal G}^{(k_7)\alpha\beta ab}\left({g_s\over H}\right)^{2\Delta(k_1 + ... + k_7 - 1/\Delta -\gamma/\Delta)}
- {8\partial_m H \partial_n H\over H^2},  \nd
where we have used the G-flux ansatze \eqref{ravali} to express it in powers of $g_s$. The above expression is similar to what we had in \eqref{nkinsky} and putting $\gamma = 0$ we get back most of the terms therein. The difference remains the same: terms with derivative with respect to $\alpha$ are missing.

Let us now analyze the EOMs. As before, we balance the Einstein tensor \eqref{9do12} with 
the energy-momentum tensors \eqref{nkinsky2}, for the G-fluxes and \eqref{neveC}, for the quantum terms. 
We will however have to specify some values for $\gamma$ to equate \eqref{9do12} with the sum of
\eqref{nkinsky2} and \eqref{neveC}. Let us take $\gamma = 2$. Such a choice immediately implies, from 
\eqref{dcmika} and \eqref{etochinki},
 that the 
{\it lowest} mode of G-flux that we can take to avoid generating time-neutral series is $9/2$, i.e 
${\cal G}^{(9/2)}_{MNPQ}$. As mentioned earlier, other choices are possible, but here we will stick with the simplest modings of the G-flux components to avoid over-complicating the scenario. This would imply:
\bg\label{nickid}
{\bf G}_{MNPQ} =  {\cal G}^{(9/2)}_{MNPQ} \left({g_s\over H}\right)^3 
+ {\cal G}^{(5)}_{MNPQ} \left({g_s\over H}\right)^{10/3} + ...., \nd
where we put $\Delta = 1/3$ to illustrate the $g_s$ dependence more precisely. The expansion 
\eqref{nickid} is a bit unnatural in the light of the G-flux behavior for $\gamma = 0$, and in fact increasing 
$\gamma$ increases the lower bound from \eqref{etochinki}, but let us carry on to see how this effects the EOMs\footnote{A case could be made for the other kind of modings that appears to alleviate this issue. However we would still retain some G-flux components with lowest modes of $9/2$. For these the apparent unnaturalities still remain.}. 

We will analyze the EOMs to order by order in powers of $g_s^{1/3}$. The lowest order is the zeroth power in $g_s$. Interestingly, because we took $\gamma = 2$, the only flux component that can contribute at this order is ${\cal G}^{(9/2)}_{\alpha\beta ab}$. In other words:
\bg\label{spoonR}
{\bf G}_{mn} - 3\Lambda H^4 g_{mn} = \mathbb{C}_{mn}^{(0, 0)} 
- {g_{mn} \over 16 H^4} {\cal G}^{(9/2)}_{\alpha\beta ab} {\cal G}^{(9/2)\alpha\beta ab}, \nd
where $\mathbb{C}^{(0, 0)}_{mn}$ collects all the quantum terms classified by $\theta_k = 2/3$ in 
\eqref{miai}, where the choice of $\theta_k$ is governed by the scaling argument in \eqref{teenangul}. The
equation \eqref{spoonR} should be compared to \eqref{misslemon2}. The latter has more G-flux components with much lower modes, but the overall story remains somewhat similar, albeit a bit more natural. A degree of freedom counting tells us that we have 10 equations with at least 17 unknowns, thus considerably more constrained than \eqref{misslemon2}.
Note that the coefficient of $\Lambda$, lets call it $\sigma_o$, is smaller that what we had in 
\eqref{misslemon2}. This is because $\gamma$ contributes to the coefficient as:
\bg\label{woodley}
\sigma_o \equiv {3\over 4}\left(4\gamma - \gamma^2 - 8\right), \nd
showing that no real choice of $\gamma$ can make the cosmological constant term in \eqref{spoonR} to vanish.  

To the next order in $g_s$ the story evolves in a similar way to what we had in \eqref{rambha3}. The metric can be directly related to the G-flux component ${\cal G}^{(9/2)}_{\alpha\beta ab}$ and the quantum terms 
$\mathbb{C}^{(1/2, 0)}_{mn}$. The precise expression is:
\bg\label{rambha4}
g_{mn} = {144 H^8\over \Lambda} \left({\mathbb{C}_{mn}^{(1/2, 0)} \over 16 H^8 \mathbb{J}(y) 
+ 45 C_{1\over 2} {\cal G}^{(9/2)}_{\alpha\beta ab} {\cal G}_{}^{(9/2)\alpha\beta ab}}\right), \nd
where the quantum terms are classified, as before, by $\theta_k = 1$, with $\theta_k$ defined as in 
\eqref{miai}. The equation \eqref{rambha4}, as also in \eqref{rambha3}, mixes all the un-warped metric components with the G-flux component ${\cal G}^{(9/2)}_{\alpha\beta ab}$ as well as the $C_k$ and the 
$\widetilde{C}_k$ coefficients, so one would need other equations to disantangle everything. The 
$C_k$ and the $\widetilde{C}_k$ coefficients also appear in the definition of $\mathbb{J}(y)$ which takes the following form:
\bg\label{mutapa}
\mathbb{J}(y) & \equiv & -42 C_{1\over 2} + \sum_{\{k_i\}} \left(k_1 + 3\right) \left(k_2 + 3\right)
\widetilde{C}_{k_1} \widetilde{C}_{k_2} \prod_{1 = 3}^7 C_{k_i} \delta\left(k_1 + ... + k_7 - {1\over 2}\right)
\\
&& - 2 \sum_{\{k_i\}} \left(k_1 + 3\right) \left(3k_2 + 2k_1 - 3\right)
\widetilde{C}_{k_1} C_{k_2} C_{k_3} C_{k_4} \delta\left(k_1 + k_2 + k_3  + k_4 - {1\over 2}\right).
\nonumber \nd
One could now go to the next order, i.e $g_s^{2/3}$, and analyze the background in a similar way to 
\eqref{clovbalti}, using the same component of G-flux and quantum terms $\mathbb{C}^{(1, 0)}_{mn}$ classified by $\theta_k = 4/3$ in \eqref{miai}. Compared to our analysis for case 1, only a few new degrees of freedom are added at this stage: the coefficients of the individual quantum terms and the $C_{1\over 2}$
coefficient. Thus \eqref{mutapa} is again a set of 10 equations with at least 18 unknowns. Compared to case 1 above, it appears that we have more equations than the number of unknowns, so existence of solution is a question here. Assuming solution exists, we see from \eqref{spoonR} and \eqref{rambha4} that the metric on
${\cal M}_4$ has to be a non-K\"ahler metric (or a conformally $K3$). 
The story can then be developed further in a somewhat similar way, but 
we will not do so here, and instead go with the analysis of the two cases along $(\alpha, \beta)$ directions.

\subsubsection{Einstein equation along $(\alpha, \beta)$ directions \label{kocu2}}

Having discussed in details the Einstein's equation along ($m, n$) directions, it is time to analyze the story for the ($\alpha, \beta$) directions, namely the directions along ${\cal M}_2$. The analysis will proceed more or less in the same way as before, although specific details would differ. In fact these are the differences 
that we want to illustrate in this section. We will proceed by first studying the volume preserving case \eqref{olokhi} and then go for the fluctuation case \eqref{ranjhita}. However before moving to the specific cases in question, we want to elucidate the general picture starting with the Einstein tensor. This takes the form:

{\footnotesize
\bg\label{suscott}
\mathbb{G}_{\alpha\beta} & = & {\bf G}_{\alpha\beta} - {8\partial_\alpha H \partial_\beta H \over H^2} + 
4g_{\alpha\beta}\left[{1\over 4} g_s \sqrt{\Lambda} H^3 \dot{F}_1 - {3\over 2} \Lambda H^4 F_1 +
{\partial_\alpha H \partial^\alpha H \over H^2} 
+ {F_1\over F_2}\left({\partial_m H \partial^m H\over H^2}\right)\right]\\
&-&4g_{\alpha\beta}\left[{1\over 8} g_s^2 H^2  \ddot{F}_1 - {g_s^2 H^2 \dot{F}_1^2\over 16 F_1} + 
{g_s^2 H^2 \dot{F}_2^2 F_1\over 8F_2^2} + {g_s^2 H^2 \dot{F}_2 \dot{F}_1 \over 4F_2} +
{g_s\sqrt{\Lambda} H^3 \dot{F}_2 F_1 \over F_2} + {g_s^2 H^2 \ddot{F}_2 F_1 \over 2 F_2}\right], 
\nonumber \nd}
where $h(y) \equiv H^4(y)$ and ${\bf G}_{\alpha\beta}$ is defined with the un-warped metric 
$g_{\alpha\beta}$. The $g_s$ dependence appearing in \eqref{suscott} is clearly not the full story as 
other $g_s$ dependences hide in the definitions of $F_i(t)$. This will be illustrated for the two case 
\eqref{olokhi} and \eqref{ranjhita} soon. The Einstein tensor \eqref{suscott} will have to be equated to the 
sum of the energy-momentum tensors for the G-flux as well as for the quantum terms. The latter is given in 
\eqref{neveC} whereas the former takes the form:

{\footnotesize
\bg\label{monroemey}
\mathbb{T}^G_{\alpha\beta}&=& 
 \frac{F_1}{H^4F_2^{3}}\left(-\frac{1}{24} 
g_{\alpha\beta}{G}_{mnp a}{G}^{ mnp a}\right)
+
\frac{\Lambda(t)}{12H^4 F_2^3}\left({G}_{\alpha lkr}{G}_{\beta}^{lkr}-\frac{1}{2}g_{\alpha\beta}{G}_{\gamma klr}{G}^{\gamma klr}\right)\nonumber\\
&+&\frac{1}{4H^4F_2^{2}}\left({G}_{\alpha lka}{G}_{\beta}^{lka}-\frac{1}{2}g_{\alpha\beta}
{G}_{\gamma kla}{G}^{\gamma kla}\right)+
\frac{1}{2H^4F_1F_2}\left({G}_{\alpha l\gamma a}{G}_{\beta}^{l\gamma  a}-\frac{1}{4} g_{\alpha \beta} 
{G}_{\delta l\gamma a}{G}^{\delta l\gamma a}\right)
\nonumber\\
&+&
\frac{\Lambda(t)}{4H^4F_2F_1^2}
\left({G}_{\alpha \eta lr}{G}_{\beta}^{\eta lr}-\frac{1}{4}g_{\alpha\beta}{G}_{\kappa\eta lr}
{G}^{\kappa\eta lr}\right) -\frac{F_1\Lambda(t)}{12H^4F_2^4}\left(\frac{1}{8}g_{\alpha\beta}G_{mnpq}
{G}^{mnpq}\right) -\frac{8\partial_\alpha H \partial_\beta H}{H^2}\nonumber \\
&+&\frac{1}{4H^4\Lambda(t)F_2}
\left({G}_{\alpha lab}{G}_{\beta}^{lab}
-\frac{1}{2}g_{\alpha\beta} {G}_{\alpha kab}{G}^{\beta kab}\right)
+ \frac{1}{4H^4\Lambda(t)F_1}
\left({G}_{\alpha \gamma ab}{G}_{\beta}^{\gamma ab}-\frac{1}{4}g_{\alpha \beta} {G}_{\eta\kappa ab}
{G}^{\eta\kappa ab}\right)\nonumber \\
&-&  {F_1 \over H^4 \Lambda(t) F_2^2} \left({1\over 16} g_{\alpha\beta} G_{mnab} G^{mnab}\right)
 +4 {g}_{\alpha\beta }\left[\frac{\partial_{\gamma} H \partial^{\gamma} H}{H^2}
+\frac{F_1}{F_2}\left(\frac{\partial_m H \partial^m H}{H^2}\right)\right], 
\nd}
which captures the contributions to the energy-momentum tensor from the G-fluxes. Interestingly, as in 
\eqref{nkinsky} all components of G-flux contribute, in addition to the space-time components. We will have to keep in mind that some of the G-flux components, namely ${\bf G}_{MNab}$ will have to be localized fluxes to preserve the de Sitter isometries in the IIB side as before.  With these at hand, let us discuss the individual cases.


\vskip.1in

\noindent {\it Case 1: $F_1(t)$ and $F_2(t)$ satisfying the volume-preserving condition \eqref{olokhi}}

\vskip.1in

\noindent Our starting point would be express both \eqref{suscott} and \eqref{monroemey} using the $g_s$ expansions of $F_i(t)$ as in \eqref{karishma} and G-flux as in \eqref{ravali}. Using these the Einstein tensor 
becomes:
\bg\label{pentchira}
\mathbb{G}_{\alpha\beta} & = & {\bf G}_{\alpha\beta} - {8\partial_\alpha H \partial_\beta H \over H^2} 
+ \Lambda H^4 g_{\alpha\beta} \sum_{\{k_i\}} \left[2\Delta  k \widetilde{C}_k  - 6  \widetilde{C}_k 
- \Delta k(2\Delta k - 1) \widetilde{C}_k\right] \left({g_s\over H}\right)^{2\Delta k}
\nonumber\\
& + & 4g_{\alpha\beta}\left[{\partial_\alpha H \partial^\alpha H \over H^2} 
+ \left({\partial_m H \partial^m H\over H^2}\right)\sum_{\{k_i\}} \widetilde{C}_{k_1}\widetilde{C}_{k_2}{C}_{k_3} \left({g_s\over H}\right)^{2\Delta(k_1 + k_2 + k_2)}\right]\nonumber\\
&-& \Lambda \Delta H^4 g_{\alpha\beta}\sum_{\{k_i\}}\Big[ 2\Delta k_1 k_2 C_{k_1} C_{k_2} 
\widetilde{C}_{k_3} \widetilde{C}_{k_4} - \Delta k_1 k_2 \widetilde{C}_{k_1} \widetilde{C}_{k_2}
C_{k_3} C_{k_4}  + 4\Delta k_2 k_4 \widetilde{C}_{k_1} \widetilde{C}_{k_2} C_{k_3} C_{k_4}\nonumber\\
&+& 8 k_1 C_{k_1} \widetilde{C}_{k_2}\widetilde{C}_{k_3} C_{k_4} + 4k_1(2\Delta k_1 -1) C_{k_1}
\widetilde{C}_{k_2}\widetilde{C}_{k_3}C_{k_4}\Big]\left({g_s\over H}\right)^{2\Delta(k_1 + k_2 + k_3 + k_4)}, \nd
which in turn should be compared to \eqref{9do11}. Expectedly their precise structures are a bit different, but the generic form remains somewhat equivalent.  This is also reflected in the form of the energy-momentum tensor, which may be expressed as:
\bg\label{tagjap} 
\mathbb{T}_{\alpha\beta}^G & =& 
{1\over 4 H^4}\sum_{\{k_i\}} \widetilde{C}_{k_1}\left({\cal G}^{(k_2)}_{\alpha lka} 
{\cal G}^{(k_3)lka}_{\beta} -{1\over 2} g_{\alpha\beta} {\cal G}^{(k_2)}_{\gamma lka} 
{\cal G}^{(k_3)\gamma lka}\right) \left({g_s\over H}\right)^{2\Delta(k_1 + k_2 + k_3)}\\
&+& {1\over 2 H^4}\sum_{\{k_i\}} {C}_{k_1}\left({\cal G}^{(k_2)}_{\alpha l\gamma a} 
{\cal G}^{(k_3)l\gamma a}_{\beta} -{1\over 4} g_{\alpha\beta} {\cal G}^{(k_2)}_{\delta l\gamma a} 
{\cal G}^{(k_3)\delta l\gamma a}\right) \left({g_s\over H}\right)^{2\Delta(k_1 + k_2 + k_3)}
\nonumber\\
&-& {g_{\alpha\beta}\over 24 H^4} \sum_{\{k_i\}}\widetilde{C}_{k_1}
\widetilde{C}_{k_2} \widetilde{C}_{k_3} C_{k_4} {\cal G}^{(k_5)}_{mnpa} {\cal G}^{(k_6)mnpa} 
\left({g_s\over H}\right)^{2\Delta(k_1 + k_2 + k_3 + k_4 + k_5 + k_6)}\nonumber\\
&-& {g_{\alpha\beta} \over 96 H^4}\sum_{\{k_i\}}\widetilde{C}_{k_1}\widetilde{C}_{k_2} \widetilde{C}_{k_3}
 {\cal G}^{(k_4)}_{mnpq} 
{\cal G}^{(k_5)mnpq} \left({g_s\over H}\right)^{2\Delta(k_1 + k_2 + k_3 + k_4 + k_5 + 1/\Delta)}
\nonumber\\
&-& {g_{\alpha\beta} \over 16 H^4} \sum_{\{k_i\}}\widetilde{C}_{k_1}\widetilde{C}_{k_2}
 {\cal G}^{(k_3)}_{mnab} 
{\cal G}^{(k_4)mnab} \left({g_s\over H}\right)^{2\Delta(k_1 + k_2 + k_3 + k_4 - 1/\Delta)}
\nonumber\\
&+& {1\over 4 H^4}\sum_{\{k_i\}} \widetilde{C}_{k_1}{C}_{k_2} \left({\cal G}^{(k_3)}_{\alpha l ab} 
{\cal G}^{(k_4)lab}_{\beta} -{1\over 2} g_{\alpha\beta} {\cal G}^{(k_3)}_{\gamma lab} 
{\cal G}^{(k_4)\gamma lab}\right) \left({g_s\over H}\right)^{2\Delta(k_1 + k_2 + k_3 + k_4 - 1/\Delta)}
\nonumber\\
&+& {1\over 4 H^4}\sum_{\{k_i\}} {C}_{k_1}{C}_{k_2} \left({\cal G}^{(k_3)}_{\alpha \gamma ab} 
{\cal G}^{(k_4)\gamma ab}_{\beta} -{1\over 4} g_{\alpha\beta} {\cal G}^{(k_3)}_{\gamma \eta ab} 
{\cal G}^{(k_4)\gamma \eta ab}\right) \left({g_s\over H}\right)^{2\Delta(k_1 + k_2 + k_3 + k_4 - 1/\Delta)}
\nonumber\\
&+& {1\over 4 H^4}\sum_{\{k_i\}} {C}_{k_1}{C}_{k_2} C_{k_3}\left({\cal G}^{(k_4)}_{\alpha \eta l r} 
{\cal G}^{(k_5)\eta l r}_{\beta} -{1\over 4} g_{\alpha\beta} {\cal G}^{(k_4)}_{\gamma \eta kr} 
{\cal G}^{(k_5)\gamma \eta k r}\right) \left({g_s\over H}\right)^{2\Delta(k_1 + k_2 + k_3 + k_4 + k_5 + 1/\Delta)}
\nonumber\\
&+& {1\over 12 H^4}\sum_{\{k_i\}}\widetilde{C}_{k_1}\widetilde{C}_{k_2} C_{k_3}\left({\cal G}^{(k_4)}_{\alpha lkr} 
{\cal G}^{(k_5)lkr}_{\beta} -{1\over 2} g_{\alpha\beta} {\cal G}^{(k_4)}_{\gamma lkr} 
{\cal G}^{(k_5)\gamma klr}\right) \left({g_s\over H}\right)^{2\Delta(k_1 + k_2 + k_3 + k_4 + k_5 + 1/\Delta)}
\nonumber\\
&-& {8\partial_\alpha H \partial_\beta H \over H^2} + 
4g_{\alpha\beta}\left[{\partial_\gamma H \partial^\gamma H \over H^2} 
+ \left({\partial_m H \partial^m H\over H^2}\right)\sum_{\{k_i\}} \widetilde{C}_{k_1}\widetilde{C}_{k_2}{C}_{k_3} \left({g_s\over H}\right)^{2\Delta(k_1 + k_2 + k_2)}\right], \nonumber \nd
which should again be compared to \eqref{nkinsky} and we see that the relevant G-flux components and the warp-factors fall in their rightful places. Expectedly the last three terms of \eqref{tagjap} matches with the three equivalent terms in \eqref{pentchira}. To the zeroth order in $g_s$, the equation of motion becomes:
\bg\label{uanaban}
 {\bf G}_{\alpha\beta} - 6\Lambda H^4 g_{\alpha\beta} &=& \mathbb{C}_{\alpha\beta}^{(0, 0)} +
{1\over 4 H^4} \left({\cal G}^{(3/2)}_{\alpha \gamma ab} 
{\cal G}^{(3/2)\gamma ab}_{\beta} -{1\over 4} g_{\alpha\beta} {\cal G}^{(3/2)}_{\gamma \eta ab} 
{\cal G}^{(3/2)\gamma \eta ab}\right) \\
&+& {1\over 4 H^4} \left({\cal G}^{(3/2)}_{\alpha l ab} 
{\cal G}^{(3/2)lab}_{\beta} -{1\over 2} g_{\alpha\beta} {\cal G}^{(3/2)}_{\gamma lab} 
{\cal G}^{(3/2)\gamma lab}\right) - {g_{\alpha\beta}\over 16 H^4}{\cal G}^{(3/2)}_{mn ab} 
{\cal G}^{(3/2)mn ab}, \nonumber \nd
showing us that the internal space ${\cal M}_2$ again cannot be a Calabi-Yau manifold. The non-K\"ahlerity of ${\cal M}_2$ is generated by both G-fluxes and the cosmological constant. The G-fluxes entering in 
\eqref{uanaban} are the special ones that have legs along the ($a, b$) directions much like the ones entering in \eqref{misslemon2}. As mentioned earlier, these fluxes cannot be of the form \eqref{montan} and therefore will be treated as localized fluxes. However their $3/2$ modings are consistent and cannot be smaller than this compared to the possibilities with other G-flux components (in any case we stick with one set of modings for all components). The other ingredient is the quantum term 
$\mathbb{C}_{\alpha\beta}^{(0, 0)}$. More details on this will be discussed below.

The next order is $g_s^{1/3}$. We need to be careful here because some of the $k_i$ that determine the G-flux components are bounded below as $k_i \ge 3/2$. Others can take any, i.e zero and positive, values lying in $\mathbb{Z}/2$. Keeping this in mind, expanding to $g_s^{1/3}$ gives us:
\bg\label{ajanta}
g_{\alpha\beta} &=& {9\over 2 \mathbb{C}(y)} \mathbb{C}_{\alpha\beta}^{(1/2, 0)} + 
{9\over 8 H^4 \mathbb{C}(y)}\sum_{\{k_i\}}\Bigg[ \widetilde{C}_{k_1}{C}_{k_2} \left({\cal G}^{(k_3)}_{\alpha l ab} 
{\cal G}^{(k_4)lab}_{\beta} -{1\over 2} g_{\alpha\beta} {\cal G}^{(k_3)}_{\gamma lab} 
{\cal G}^{(k_4)\gamma lab}\right) \nonumber\\
&+& {C}_{k_1}{C}_{k_2} \left({\cal G}^{(k_3)}_{\alpha \gamma ab} 
{\cal G}^{(k_4)\gamma ab}_{\beta} -{1\over 4} g_{\alpha\beta} {\cal G}^{(k_3)}_{\gamma \eta ab} 
{\cal G}^{(k_4)\gamma \eta ab}\right)\Bigg]\delta\left(k_1 + k_2 + k_3 + k_4 - {7\over 2}\right)\nonumber\\
&-& {9g_{\alpha\beta}\over 32 H^4 \mathbb{C}(y)} \sum_{\{k_i\}} \bigg(\widetilde{C}_{k_1} \widetilde{C}_{k_2}
{\cal G}^{(k_3)}_{mnab}  {\cal G}^{(k_4)mnab}\bigg) \delta\left(k_1 + k_2 + k_3 + k_4 - {7\over 2}\right),
 \nd
where we note that $(k_3, k_4) \ge (3/2, 3/2)$ as alluded to above. The uniqueness of the lower bounds should again be apparent from the choice of the components appearing in the above equation. This means we are looking at G-flux components with $(k_3, k_4) = (3/2, 3/2), (3/2, 2)$ and $(2, 3/2)$. This, in turn, should be compared to the 
$(3/2, 3/2)$ distribution that we got in \eqref{uanaban}. The coefficient $\mathbb{C}(y)$ is defined as:
\bg\label{chotom}
\mathbb{C}(y) \equiv 50 \Lambda H^2(y) C_{\tiny{1\over 2}}, \nd
which is always a non-zero function because $H(y)$ is a non-vanishing real function globally. 
The other ingredient of \eqref{ajanta} are the quantum terms.
These are classified by $\mathbb{C}_{\alpha\beta}^{(1/2, 0)}$ and should be compared to the quantum terms classified by
$\mathbb{C}_{\alpha\beta}^{(0, 0)}$ in \eqref{uanaban}. Following \eqref{charaangul}, the latter would be classified by $\theta'_k = {2\over 3}$ whereas the former would be classified by 
$\theta'_k = 1$ in \eqref{melamon2}. 

The next order is $g_s^{2/3}$, and follows in exactly the same footsteps of the previous case, although details differ. The equation now becomes:
\bg\label{vanandmey}
g_{\alpha\beta} &=& {9\mathbb{C}^{(1, 0)}_{\alpha\beta}\over \mathbb{E}(y)} + 
{9\over 4 H^4\mathbb{E}(y)}\sum_{\{k_i\}} \bigg[\widetilde{C}_{k_1}{C}_{k_2} 
\left({\cal G}^{(k_3)}_{\alpha l ab} 
{\cal G}^{(k_4)lab}_{\beta} -{1\over 2} g_{\alpha\beta} {\cal G}^{(k_3)}_{\gamma lab} 
{\cal G}^{(k_4)\gamma lab}\right) \nonumber\\
&+& \left({\cal G}^{(k_3)}_{\alpha \gamma ab} 
{\cal G}^{(k_4)\gamma ab}_{\beta} -{1\over 4} g_{\alpha\beta} {\cal G}^{(k_3)}_{\gamma \eta ab} 
{\cal G}^{(k_4)\gamma \eta ab}\right)\bigg]\delta\left(k_1 + k_2 + k_3 + k_4 - 4\right) \nonumber\\
&-& {9g_{\alpha\beta}\over 16 H^4 \mathbb{E}(y)} \sum_{\{k_i\}} \bigg(\widetilde{C}_{k_1} \widetilde{C}_{k_2}
{\cal G}^{(k_3)}_{mnab}  {\cal G}^{(k_4)mnab}\bigg) \delta\left(k_1 + k_2 + k_3 + k_4 - {4}\right), \nd
in exactly the same format as in \eqref{clovbalti}. Again $k_3$ and $k_4$ are bounded as 
$(k_3, k_4) \ge (3/2, 3/2)$ so we have G-flux contributions  from ${\cal G}^{(3/2)}_{MNPQ}, 
{\cal G}^{(2)}_{MNPQ}$ and ${\cal G}^{(5/2)}_{MNPQ}$. 
In the same vein, the quantum terms are classified by an equation of the form \eqref{mameye} for local and non-local contributions. Finally the function $\mathbb{E}(y)$ appearing above is defined in the following way:
\bg\label{sweetdream}
\mathbb{E}(y) &\equiv& -\Lambda H^4(y)\left[47 \widetilde{C}_1 + 3 \mathbb{D}(y)\right]\\
\mathbb{D}(y) &\equiv &{2\over 3} \sum_{\{k_i\}}\Big[k_1 k_2 C_{k_1} C_{k_2} 
\widetilde{C}_{k_3} \widetilde{C}_{k_4} - {1\over 2} k_1 k_2 \widetilde{C}_{k_1} \widetilde{C}_{k_2}
C_{k_3} C_{k_4}  + 2 k_2 k_4 \widetilde{C}_{k_1} \widetilde{C}_{k_2} C_{k_3} C_{k_4}\nonumber\\
&+& 12 k_1 C_{k_1} \widetilde{C}_{k_2}\widetilde{C}_{k_3} C_{k_4} + 2k_1(2 k_1 -3) C_{k_1}
\widetilde{C}_{k_2}\widetilde{C}_{k_3}C_{k_4}\Big]\delta\left(k_1 + k_2 + k_3 + k_4 - 1\right), 
\nonumber \nd
where we expect both these functions to be non-vanishing globally. All the three EOMs that we listed above, namely \eqref{uanaban}, \eqref{ajanta} and \eqref{vanandmey}, are each a set of three equations with 
at least 31, 40 and 49 unknowns respectively.

\vskip.1in

\noindent {\it Case 2: $F_1(t)$ and $F_2(t)$ satisfying the fluctuation condition \eqref{ranjhita}}

\vskip.1in

\noindent The analysis of $(\alpha, \beta)$ directions will be a bit more subtle from what we encountered 
for case 1, part of the reason being the different modings of the G-flux components and part of the reason being the different scaling behavior of the quantum terms as evident from \eqref{teenangul}. Before we go into these discussion, let us present the Einstein tensor for this case:

{\footnotesize
\bg\label{pentchira2}
\mathbb{G}_{\alpha\beta} & = & {\bf G}_{\alpha\beta}  
+ \Lambda H^4 g_{\alpha\beta} \sum_{\{k_i\}} \left[(2\Delta  k + \gamma) \widetilde{C}_k  - 6  \widetilde{C}_k 
- {1\over 2}(2\Delta k + \gamma)(2\Delta k +\gamma  - 1) \widetilde{C}_k\right] 
\left({g_s\over H}\right)^{2\Delta k + \gamma} 
\nonumber\\
& + & 4g_{\alpha\beta}
\left({\partial_m H \partial^m H\over H^2}\right)\sum_{\{k_i\}} \widetilde{C}_{k_1}\widetilde{C}_{k_2}{C}_{k_3} \left({g_s\over H}\right)^{2\Delta(k_1 + k_2 + k_2) + \gamma}  - \Lambda \Delta H^4 g_{\alpha\beta}
\left({g_s\over H}\right)^{2\Delta(k_1 + k_2 + k_3 + k_4) + \gamma}\nonumber\\
&\times & \sum_{\{k_i\}}\Big[ 2\Delta k_1 k_2 C_{k_1} C_{k_2} 
\widetilde{C}_{k_3} \widetilde{C}_{k_4} -{1\over 4 \Delta} (2\Delta k_1+ \gamma)(2\Delta k_2 + \gamma) \widetilde{C}_{k_1} \widetilde{C}_{k_2}
C_{k_3} C_{k_4} \nonumber\\ 
&+& 2 (2\Delta k_2 + \gamma) k_4 \widetilde{C}_{k_1} \widetilde{C}_{k_2} C_{k_3} C_{k_4}
+ 8 k_1 C_{k_1} \widetilde{C}_{k_2}\widetilde{C}_{k_3} C_{k_4} + 4k_1(2\Delta k_1 -1) C_{k_1}
\widetilde{C}_{k_2}\widetilde{C}_{k_3}C_{k_4}\Big],  \nd}
which may be compared to \eqref{pentchira}. As before, the difference lies in the absence of $\alpha$ 
dependent terms and the appearance of the $\gamma$ factor at various places, including the $g_s$ scalings of most of the terms. We will eventually make $\gamma = 2$, but for the time being we shall carry on with the generic picture as far as possible. 

The energy-momentum tensor for the G-flux is much easier to compute. All we need is to ask how the $g_s$ scalings of each terms in \eqref{tagjap} could change. Taking this into account, the expression for the energy-momentum tensor becomes:

{\footnotesize
\bg\label{tagjap2} 
\mathbb{T}_{\alpha\beta}^G & =& 
{1\over 4 H^4}\sum_{\{k_i\}} \widetilde{C}_{k_1}\left({\cal G}^{(k_2)}_{\alpha lka} 
{\cal G}^{(k_3)lka}_{\beta} -{1\over 2} g_{\alpha\beta} {\cal G}^{(k_2)}_{\gamma lka} 
{\cal G}^{(k_3)\gamma lka}\right) \left({g_s\over H}\right)^{2\Delta(k_1 + k_2 + k_3)}\nonumber\\
&+& {1\over 2 H^4}\sum_{\{k_i\}} {C}_{k_1}\left({\cal G}^{(k_2)}_{\alpha l\gamma a} 
{\cal G}^{(k_3)l\gamma a}_{\beta} -{1\over 4} g_{\alpha\beta} {\cal G}^{(k_2)}_{\delta l\gamma a} 
{\cal G}^{(k_3)\delta l\gamma a}\right) \left({g_s\over H}\right)^{2\Delta(k_1 + k_2 + k_3 - \gamma/2\Delta)}
\nonumber\\
&-& {g_{\alpha\beta}\over 24 H^4} \sum_{\{k_i\}}\widetilde{C}_{k_1}
\widetilde{C}_{k_2} \widetilde{C}_{k_3} C_{k_4} {\cal G}^{(k_5)}_{mnpa} {\cal G}^{(k_6)mnpa} 
\left({g_s\over H}\right)^{2\Delta(k_1 + k_2 + k_3 + k_4 + k_5 + k_6 + \gamma/2\Delta)}\nonumber\\
&-& {1\over 96 H^4}g_{\alpha\beta} \sum_{\{k_i\}}\widetilde{C}_{k_1}\widetilde{C}_{k_2} \widetilde{C}_{k_3}
 {\cal G}^{(k_4)}_{mnpq} 
{\cal G}^{(k_5)mnpq} \left({g_s\over H}\right)^{2\Delta(k_1 + k_2 + k_3 + k_4 + k_5 + 1/\Delta 
+ \gamma/2\Delta)}
\nonumber\\
&-& {g_{\alpha\beta} \over 16 H^4} \sum_{\{k_i\}}\widetilde{C}_{k_1}\widetilde{C}_{k_2}
 {\cal G}^{(k_3)}_{mnab} 
{\cal G}^{(k_4)mnab} \left({g_s\over H}\right)^{2\Delta(k_1 + k_2 + k_3 + k_4 - 1/\Delta + \gamma/2\Delta)}
\nonumber\\
&+& {1\over 4 H^4}\sum_{\{k_i\}} \widetilde{C}_{k_1}{C}_{k_2} \left({\cal G}^{(k_3)}_{\alpha l ab} 
{\cal G}^{(k_4)lab}_{\beta} -{1\over 2} g_{\alpha\beta} {\cal G}^{(k_3)}_{\gamma lab} 
{\cal G}^{(k_4)\gamma lab}\right) \left({g_s\over H}\right)^{2\Delta(k_1 + k_2 + k_3 + k_4 - 1/\Delta)}
\nonumber\\
&+& {1\over 4 H^4}\sum_{\{k_i\}} {C}_{k_1}{C}_{k_2} \left({\cal G}^{(k_3)}_{\alpha \gamma ab} 
{\cal G}^{(k_4)\gamma ab}_{\beta} -{1\over 4} g_{\alpha\beta} {\cal G}^{(k_3)}_{\gamma \eta ab} 
{\cal G}^{(k_4)\gamma \eta ab}\right) \left({g_s\over H}\right)^{2\Delta(k_1 + k_2 + k_3 + k_4 - 1/\Delta
-\gamma/2\Delta)}
\nonumber\\
&+& {1\over 4 H^4}\sum_{\{k_i\}} {C}_{k_1}{C}_{k_2} C_{k_3}\left({\cal G}^{(k_4)}_{\alpha \eta l r} 
{\cal G}^{(k_5)\eta l r}_{\beta} -{1\over 4} g_{\alpha\beta} {\cal G}^{(k_4)}_{\gamma \eta kr} 
{\cal G}^{(k_5)\gamma \eta k r}\right) \left({g_s\over H}\right)^{2\Delta(k_1 + k_2 + k_3 + k_4 + k_5 + 1/\Delta - \gamma/\Delta)}
\nonumber\\
&+& {1\over 12 H^4}\sum_{\{k_i\}}\widetilde{C}_{k_1}\widetilde{C}_{k_2} C_{k_3}\left({\cal G}^{(k_4)}_{\alpha lkr} 
{\cal G}^{(k_5)lkr}_{\beta} -{1\over 2} g_{\alpha\beta} {\cal G}^{(k_4)}_{\gamma lkr} 
{\cal G}^{(k_5)\gamma klr}\right) \left({g_s\over H}\right)^{2\Delta(k_1 + k_2 + k_3 + k_4 + k_5 + 1/\Delta)}
\nonumber\\
&+& 
4g_{\alpha\beta}
\left({\partial_m H \partial^m H\over H^2}\right)\sum_{\{k_i\}} \widetilde{C}_{k_1}\widetilde{C}_{k_2}{C}_{k_3} \left({g_s\over H}\right)^{2\Delta(k_1 + k_2 + k_2)},  \nd}
where expectedly the last term matches with an equivalent term in \eqref{pentchira2}. Other terms could be compared to \eqref{tagjap}, and here we notice something interesting: to allow for a zeroth power of $g_s$, 
the sum of the two modings of the G-flux components, i.e the sum of the two $k_i$ values of the 
${\cal G}^{(k_i)}_{MNPQ}$ appearing in any term above, should at most be:
\bg\label{boatcasin}
k_i + k_j = {1\over \Delta}\left(1 + {\gamma\over 2}\right), \nd
where ($k_i, k_j$) are the modings appearing in the product of two G-flux components in \eqref{tagjap2}
that contribute to the energy-momentum tensor. With $\gamma = 2$ and $\Delta = 1/3$, this means the sum 
in \eqref{boatcasin} should at most be 6. This is unfortunately not possible in the light of \eqref{nickid} and 
\eqref{metromey}, where $k_i \ge 9/2$ for the G-flux components from \eqref{etochinki}, implying that to zeroth order in $g_s$, there are no G-flux contributions to the ($\alpha, \beta$) EOMs. 

What about the quantum terms \eqref{neveC}? Here we face another conundrum: according to the 
scalings of the quantum terms in \eqref{teenangul}, with two free Lorentz indices along ($\alpha, \beta$)
directions, the $g_s$ expansion should go as:
\bg\label{telmaro}
g_s^{\theta_k + 4/3} =  g_s^2, ~~ g_s^{7/3}, ...., \nd
with $\theta_k$ defined in \eqref{miai},
implying that there are no quantum terms to zeroth order in $g_s$. The {\it minimum} allowed power of 
$g_s$ is $g_s^2$ because terms with $\theta_k = 1/3$ vanishes due to the anti-symmetry of the G-fluxes. 
The non-local terms cannot contribute anything because it {\it adds} a factor of $+4r/3$ at $r$-th level of 
non-locality to \eqref{miai} as evident from \eqref{susmita} and \eqref{akiasa}. This means that at zeroth order in $g_s$, perturbatively even the quantum terms cannot contribute. Putting everything together, \eqref{pentchira2}, \eqref{tagjap2} and \eqref{neveC} with \eqref{teenangul}, gives us:
\bg\label{casinmey}
{\bf G}_{\alpha\beta} = 0, \nd
implying that the internal space ${\cal M}_2$ can be a conformally Calabi-Yau space\footnote{A more precise statement is that \eqref{casinmey} directly implies $R^{(4)} = 0$, i.e the Ricci scalar of ${\cal M}_4$
vanishes and we can take the metric $g_{mn}$ to be that of a $K3$ space. Imposing this on \eqref{casinmey} provides a source-free equation for the metric $g_{\alpha\beta}$ whose solution is a torus. This way the metric for ${\cal M}_4 \times {\cal M}_2$ can be conformal to $K3 \times {\bf T}^2$.}. 
This doesn't imply the metric to be that of a flat torus, because of the warp-factors. On the other hand since ${\cal M}_2$ can now have toroidal topology, it's Euler characteristics would vanish, implying the vanishing of the Euler characteristics of the full eight manifold. One might now worry whether non-zero fluxes could be allowed on a manifold with vanishing Euler number \cite{BB, DRS}. This is a pertinent question and we will analyze this in more details soon, but the short answer is the following. Since the fluxes involved are 
{\it time-dependent} the constraints discussed in \cite{BB, DRS} will have to be modified allowing fluxes to exist on the eight manifold with vanishing Euler number. These fluxes will have to be supported by quantum effects,  so there is no contradiction yet\footnote{Another possibility is to take the metric of ${\cal M}_2$ to be flat everywhere except at four points. Geometrically this is 
${\bf T}^2/{\bf Z}_2$ and therefore doesn't have a vanishing Euler characteristics. However quantum corrections would eventually make this into a smooth space with non-vanishing curvature, so will not be a solution to \eqref{casinmey}. Thus we will continue with 
$K3 \times {\bf T}^2$ as our un-warped background. This will eventually lead to some subtleties that we shall clarify in section \ref{branuliat}. \label{choolmaro}}.

To the next order in $g_s$, i.e $g_s^{1/3}$, there are no contributions from \eqref{pentchira2}, 
\eqref{tagjap2} and \eqref{teenangul}. In fact the next contributions only come from order $g_s^2$, and leads to the following EOM:
\bg\label{gordhov}
\mathbb{C}_{\alpha\beta}^{(3, 0)} + {1\over 4 H^4} \left({\cal G}^{(9/2)}_{\alpha \gamma ab} 
{\cal G}^{(9/2)\gamma ab}_{\beta} -{1\over 4} g_{\alpha\beta} {\cal G}^{(9/2)}_{\gamma \eta ab} 
{\cal G}^{(9/2)\gamma \eta ab}\right) + 4\Lambda H^4 g_{\alpha\beta} = 0, \nd
which is a set of 3 equations with at least 7 unknowns. Note that this is also the first time the quantum 
terms contribute to the EOM; and here they are classified by $\theta_k = 2/3$ with $\theta_k$ given as in 
\eqref{miai}. The above equation however is a bit puzzling in the light of \eqref{casinmey}. In terms of the un-warped metric $g_{\alpha\beta}$ we expect from \eqref{casinmey} that the internal space be Ricci flat. Putting $g_{\alpha\beta} = \delta_{\alpha\beta}$ then puts a constraint on the form of the quantum terms 
$\mathbb{C}^{(3, 0)}_{\alpha\beta}$ from \eqref{gordhov}. In particular \eqref{gordhov} tells us that the trace of the quantum terms has to be a negative definite function, i.e:
\bg\label{saraN}
\left[\mathbb{C}_\alpha^{\alpha}\right]^{(3, 0)} = -{1\over 8H^4} \left({\cal G}^{(9/2)}_{\alpha\beta ab}\right)^2
- 8\Lambda H^4. \nd
Whether such a constraint could be satisfied will be discussed later when we analyze all the EOMs together.
From here the story progresses in the usual way with the Einstein tensor \eqref{pentchira2} 
being balanced by the energy-momentum tensors \eqref{tagjap2} and \eqref{neveC}.

\subsubsection{Einstein equation along $(a, b)$ directions \label{kocu3}}

The story along the ($a, b$) directions, i.e directions along $\mathbb{T}^2/{\cal G}$ has a little more subtlety 
than what we encountered so far. Part of the reason being that the variables we took so far are independent of the toroidal direction. This was not the case in \cite{nodS}, which is of course reflected in the 
scaling expression \eqref{evabmey2}. The other main reason has to do with the quantum terms that we will discuss when we study the individual cases, \eqref{olokhi} and \eqref{ranjhita}, soon. For the immediate 
discussion, let us present the expression for the Einstein tensor:
\bg\label{dreambox}
\mathbb{G}_{ab}&=&\delta_{ab}\left(-\frac{R}{2} -9h \Lambda + \frac{4{g}^{\alpha\beta}\partial_\alpha
H\partial_\beta H}{H^2F_1}
+ \frac{4{g}^{mn}\partial_m H\partial_n H}{H^2F_2}\right)\left({g_s\over H}\right)^2  \nonumber\\
&+&\delta_{ab}H^4\left(\frac{\dot{F}_1^2}{4F_1^2}
+ \frac{3\dot{F}_1}{tF_1}
- \frac{\ddot{F_1}}{F_1}
-\frac{\dot{F}_2^2}{2F_2^2}+ \frac{6\dot{F}_2}{tF_2}
-\frac{2 \ddot{F}_2}{F_2}
-\frac{2\dot{F}_1\dot{F}_2}{F_1 F_2}\right)\left({g_s\over H}\right)^4,
\nd
where $R$ is the curvature scalar of the six-dimensional base ${\cal M}_4 \times {\cal M}_2$ and {\it not} the full eight-manifold. The reason is of course because we have assigned non-trivial metric to the six-dimensional base, whereas the metric of the toroidal space is governed by the warp-factors only. This is also the reason why $\delta_{ab}$ appears in \eqref{dreambox} above instead of a non-trivial metric 
$g_{ab}$. Inclusion of the latter would complicate the dynamics of the system, for example, by changing the coupling constant etc., so we will avoid it here\footnote{As discussed in footnote \ref{plaza2019}, both 
${\cal G}$ and the seven-branes' distributions have to be fixed so as to maintain zero axion and constant dilaton on a flat background, at all orders in $g_s$. The fact that this is possible will hopefully become apparent from our analysis presented in this section.}.
Note also the absence of $g_s$ independent terms in \eqref{dreambox}. This differs from 
\eqref{zalo} and \eqref{suscott}, both of which allow $g_s$ neutral terms in the definitions of the Einstein tensors. Similarly the energy-momentum tensor is given by:

{\footnotesize 
\bg\label{brazmey}
\mathbb{T}^G_{ab} &=&\frac{\Lambda(t)}{12H^4F_2^3}\left({G}_{amnp}{G}_{b}^{mnp}-\frac{1}{2} \delta_{ab} {G}_{mnpc}{G}^{mnpc}\right)+
\frac{\Lambda(t)}{4H^4F_2^2 F_1}\left({G}_{amn\alpha}{G}_{b}^{mn\alpha}-
\frac{1}{2} \delta_{ab} {G}_{mn\alpha c}{G}^{mn\alpha c}\right)\nonumber\\
&+&\frac{\Lambda(t)}{4H^4F_1^2  F_2}\left({G}_{am\alpha\beta}
{G}_{b}^{m\alpha\beta}-\frac{1}{2} \delta_{ab}{G}_{cm\alpha\beta}{G}^{cm\alpha\beta}\right)
+\frac{1}{2H^4F_1 F_2}\left({G}_{acm\rho}{G}_{b}^{cm\rho}-\frac{1}{4} \delta_{ab} 
{G}_{m\rho cd}{G}^{m\rho cd}\right)\nonumber\\
&+&
\frac{1}{4H^4F_2^2}\left({G}_{acmn}{G}_{b}^{cmn}-\frac{1}{4} \delta_{ab} {G}_{dcmn}{G}^{dcmn}\right)
+\frac{1}{4H^4F_1^2}\left({G}_{ac\alpha\beta}{G}_{b}^{c\alpha \beta}-\frac{1}{4} \delta_{ab} 
{G}_{cd \alpha\beta }{G}^{cd\alpha\beta }\right)
\nonumber\\
&-&\delta_{ab} \frac{\Lambda(t)^2}{4.4!
H^4 F_2^4} {G}_{mnpq}{G}^{mnpq}
-\delta_{ab} \frac{\Lambda(t)^2}{24 H^4 F_2^3F_1}{G}_{mnp\alpha}{G}^{mnp\alpha}
 -\delta_{ab} \frac{\Lambda(t)^2}{16 H^4 F_2^2F_1^2} {G}_{mn\alpha\beta}
 {G}^{mn\alpha\beta}\nonumber\\
&+&\frac{4\Lambda(t)}{H^2 F_1}\delta_{ab}{g}^{\alpha \beta}\partial_\alpha H\partial_\beta H
+\frac{4\Lambda(t)}{H^2 F_2}\delta_{ab}{g}^{mn}\partial_m H\partial_n H,
\nd}
where one may note the specific placement of $\Lambda(t) \equiv \left({g_s\over H}\right)^2$ which will determine the subsequent dynamics of the system once quantum terms are added to the system. In the following, we proceed with the various cases 
in consideration. 

\vskip.1in

\noindent {\it Case 1: $F_1(t)$ and $F_2(t)$ satisfying the volume-preserving condition \eqref{olokhi}}

\vskip.1in

\noindent Our starting point then is to study the volume preserving case, where now, as mentioned above, some subtleties will arise due to the specific forms of the Einstein and the energy-momentum tensors. The latter for both G-fluxes as well as the quantum terms. The former, i.e the Einstein tensor \eqref{dreambox}, takes the following form:

{\footnotesize
\bg\label{novafatai}
\mathbb{G}_{ab} & = & {4\delta_{ab}\over H^2} \sum_{\{k_i\}}\left(C_{k_1} C_{k_2} g^{\alpha\beta}
\partial_\alpha H \partial_\beta H +\widetilde{C}_{k_1} C_{k_2} g^{mn}
\partial_m H \partial_n H\right)
\left({g_s\over H}\right)^{2\Delta(k_1 + k_2 + 1/\Delta)}\\
&-&{\delta_{ab}\over 2}\left(R + 18 H^4 \Lambda\right)\left({g_s\over H}\right)^2 
+\Delta^2 \Lambda H^4 \delta_{ab} \sum_{\{k_i\}} k_1 k_2 \widetilde{C}_{k_1} \widetilde{C}_{k_2} 
C_{k_3}.... C_{k_6} \left({g_s\over H}\right)^{2\Delta(k_1 + ... + k_6 + 1/\Delta)}\nonumber\\
&+& 2\Delta\Lambda H^4\delta_{ab} \sum_{\{k_i\}}C_{k_1} C_{k_2} \widetilde{C}_{k_3}\Big(k_1\left(8 -
\Delta k_2 - 4\Delta k_1\right) + 2k_3\left(2 -\Delta k_3 - 2\Delta k_2\right)\Big)
\left({g_s\over H}\right)^{2\Delta(k_1 + k_2 + k_3 + 1/\Delta)},\nonumber
\nd}
where expectedly there are no terms to zeroth order in $g_s$. There is also no curvature term for the toroidal manifold, evident from the $\delta_{ab}$ factor appearing from \eqref{novafatai}, presence of which would have altered the coupling constant itself. Similarly, one may represent the energy momentum tensor
in the following way:
\bg\label{kristinaQ}
\mathbb{T}^G_{ab} & = & {1\over 12 H^4} \sum_{\{k_i\}} \widetilde{C}_{k_1}\widetilde{C}_{k_2} C_{k_3}
\left({\cal G}^{(k_4)}_{amnp}{\cal G}^{(k_5)mnp}_{b}-\frac{1}{2} \delta_{ab} {\cal G}^{(k_4)}_{mnpc}
{\cal G}^{(k_5)mnpc}\right)\left({g_s\over H}\right)^{2\Delta(k_1 + ... + k_5 + 1/\Delta)}
\nonumber\\
&+& {1\over 4 H^4} \sum_{\{k_i\}} 
\left({\cal G}^{(k_1)}_{amn\alpha}{\cal G}_b^{(k_2)mn\alpha}-\frac{1}{2} \delta_{ab} {\cal G}^{(k_1)}_{mn\alpha c}
{\cal G}^{(k_2)mn\alpha c}\right)\left({g_s\over H}\right)^{2\Delta(k_1 + k_2 + 1/\Delta)}
\nonumber\\
&+& {1\over 4 H^4} \sum_{\{k_i\}} {C}_{k_1}{C}_{k_2} C_{k_3}
\left({\cal G}^{(k_4)}_{am\alpha\beta}{\cal G}_b^{(k_5)m\alpha\beta}-
\frac{1}{2} \delta_{ab} {\cal G}^{(k_4)}_{cm\alpha\beta}
{\cal G}^{(k_5)cm\alpha\beta}\right)\left({g_s\over H}\right)^{2\Delta(k_1 + ...+ k_5 + 1/\Delta)}
\nonumber\\
&+& {1\over 4 H^4} \sum_{\{k_i\}} \widetilde{C}_{k_1}
\left({\cal G}^{(k_2)}_{acmn}{\cal G}_b^{(k_3)cmn}-\frac{1}{4} \delta_{ab} {\cal G}^{(k_2)}_{dcmn}
{\cal G}^{(k_3)dcmn}\right)\left({g_s\over H}\right)^{2\Delta(k_1 + k_2 + k_3)}
\nonumber\\
&+& {1\over 4 H^4} \sum_{\{k_i\}} {C}_{k_1}{C}_{k_2} C_{k_3} C_{k_4}
\left({\cal G}^{(k_5)}_{ac\alpha\beta}{\cal G}_b^{(k_6)c\alpha\beta}-\frac{1}{4} \delta_{ab} 
{\cal G}^{(k_5)}_{cd\alpha\beta}
{\cal G}^{(k_6)cd\alpha\beta}\right)\left({g_s\over H}\right)^{2\Delta(k_1 + ...+ k_6)}
\nonumber\\
&+& {1\over 2 H^4} \sum_{\{k_i\}} {C}_{k_1}
\left({\cal G}^{(k_2)}_{acm\rho}{\cal G}_b^{(k_3)cm\rho}-\frac{1}{4} \delta_{ab} {\cal G}^{(k_2)}_{cdm\rho}
{\cal G}^{(k_3)cdm\rho}\right)\left({g_s\over H}\right)^{2\Delta(k_1 + k_2 + k_3)}
\nonumber\\
&-& {\delta_{ab}\over 4\cdot 4! H^4} \sum_{\{k_i\}} \widetilde{C}_{k_1}\widetilde{C}_{k_2} 
{\cal G}^{(k_3)}_{mnpq}
{\cal G}^{(k_4)mnpq}\left({g_s\over H}\right)^{2\Delta(k_1 + k_2 + k_3 + k_4 + 2/\Delta)}
\nonumber\\
&-& {\delta_{ab}\over 4! H^4} \sum_{\{k_i\}} \widetilde{C}_{k_1}{C}_{k_2} 
{\cal G}^{(k_3)}_{mnp\alpha}
{\cal G}^{(k_4)mnp\alpha}\left({g_s\over H}\right)^{2\Delta(k_1 + k_2 + k_3 + k_4 + 2/\Delta)}
\nonumber\\
&-& {\delta_{ab}\over 16 H^4} \sum_{\{k_i\}} {C}_{k_1}{C}_{k_2} 
{\cal G}^{(k_3)}_{mn\alpha\beta}
{\cal G}^{(k_4)mn\alpha\beta}\left({g_s\over H}\right)^{2\Delta(k_1 + k_2 + k_3 + k_4 + 2/\Delta)}
\nonumber\\
&+& {4\delta_{ab}\over H^2} \sum_{\{k_i\}} \left({C}_{k_1}{C}_{k_2}\partial_\alpha H \partial^\alpha H 
+ \widetilde{C}_{k_1}{C}_{k_2}\partial_m H \partial^m H\right) 
\left({g_s\over H}\right)^{2\Delta(k_1 + k_2 + 1/\Delta)}, \nd
where as one would expect, the last line of this matches with the first line of the Einstein tensor 
\eqref{novafatai}. Note also the absence of terms to zeroth order in $g_s$ because of the condition 
\eqref{metromey}. This is consistent with what we expect from \eqref{novafatai}, but one may now question whether this also appears from the energy-momentum tensor for the quantum terms in \eqref{neveC}. From the look of \eqref{neveC} it appears that $k_1 = 0$ should be an allowed choice. However, as discussed earlier in \eqref{telmaro}, looking at 
\eqref{charaangul} we see that tensors with two free Lorentz indices along ($a, b$) direction scale as:
\bg\label{malinB}
g_s^{\theta'_k + 4/3} \equiv g_s^{5/3}, ~g_s^2, ~g_s^{7/3}, ~g_s^{8/3}, ~g_s^3, ...., \nd
as $\theta'_k$ defined in \eqref{melamon2} is bounded below by $\theta'_k \ge 1/3$. Now since the lowest 
value of $\theta'_k = 1/3$  corresponds to switching on either $(l_{36}, l_{37}, l_{38}) = (1, 0, 0), 
(0, 1, 0)$ or $(0, 0, 1)$ in \eqref{melamon2} $-$ and they vanish due to the antisymmetry of the G-flux components $-$ it then appears that  the lowest  allowed 
scaling of $g_s$ can only be $g_s^2$. This seems perfectly consistent with the scalings expected from 
\eqref{novafatai} and \eqref{kristinaQ}, resolving a possible conundrum in our 
construction\footnote{One may alternatively view the quantum energy-momentum tensor to be represented not as \eqref{neveC} but as the following shifted one near $g_s \to 0$: 
$$\mathbb{T}^Q_{ab} = \sum_{k\in \mathbb{Z}/2} \mathbb{C}_{ab}^{(k + 5/2, 0)}
\left({g_s\over H}\right)^{2\Delta(k + 5/2)}$$
\noindent which would reproduce the correct $g_s$ scalings from \eqref{phingsha2}. Such redefinition is possible because \eqref{neveC} is conjectured to be equivalent to \eqref{phingsha2}, the latter being the main focal point of our analysis. Interestingly however one could even resort to a more generic definition of the quantum energy-momentum tensor of the following form reproducing all the examples encountered so far (indices to be lowered using warped metric):
$$\big[\mathbb{T}^Q\big]^M_N = \sum_{k\in \mathbb{Z}/2} \big[\mathbb{C}^{(k_1, k_2)}\big]^M_N 
\left({g_s\over H}\right)^{2(k_1+1)/3} {\rm exp}\left(-{k_2 H^{1/3}\over g_s^{1/3}}\right).$$
\label{cbmoja}}.

Now that the quantum issues are clarified, we should look at the equations of motion to order $g_s^2$ by balancing the Einstein tensor in \eqref{novafatai} with the energy-momentum tensors in \eqref{kristinaQ} and \eqref{neveC}. This produces:

{\footnotesize
\bg\label{buskaM}
&&\left({R\over 2} + 9H^4 \Lambda\right)\delta_{ab}  =  -\mathbb{C}_{ab}^{(3, 0)} - {1\over 4 H^4} 
\Bigg[
\left({\cal G}^{(3/2)}_{acmn}{\cal G}_b^{(3/2)cmn}-\frac{1}{4} \delta_{ab} {\cal G}^{(3/2)}_{dcmn}
{\cal G}^{(3/2)dcmn}\right)\\
&+& 
\left({\cal G}^{(3/2)}_{ac\alpha\beta}{\cal G}_b^{(3/2)c\alpha\beta}-\frac{1}{4} \delta_{ab} 
{\cal G}^{(3/2)}_{cd\alpha\beta}
{\cal G}^{(3/2)cd\alpha\beta}\right)
+ 
2\left({\cal G}^{(3/2)}_{acm\rho}{\cal G}_b^{(3/2)cm\rho}-\frac{1}{4} \delta_{ab} {\cal G}^{(3/2)}_{cdm\rho}
{\cal G}^{(3/2)cdm\rho}\right)\Bigg], \nonumber \nd}
where the quantum terms manifest themselves as $\mathbb{C}_{ab}^{(3, 0)}$ instead of 
$\mathbb{C}_{ab}^{(0, 0)}$, the former being defined for $\theta'_k = 2/3$ in \eqref{melamon2} exactly as before. It is also interesting to note that, so far all the G-flux energy-momentum tensors appear from 
${\cal G}^{(k)}_{mnab}, {\cal G}^{(k)}_{m\alpha ab}$ and ${\cal G}^{(k)}_{\alpha\beta ab}$ for various choices of $k$ satisfying $k \ge 3/2$. 

The next order is $g_s^{7/3}$. Interestingly, the Einstein tensor \eqref{novafatai} cancels out to this order, leaving only the energy-momentum tensor of the G-flux to balance with the energy-momentum tensor of the quantum terms. This gives us:

{\footnotesize
\bg\label{gargere}
4H^4 \mathbb{C}_{ab}^{(7/2, 0)} & = & \sum_{\{k_i\}} \Bigg[\widetilde{C}_{k_1}
\left({\cal G}^{(k_2)}_{acmn}{\cal G}_b^{(k_3)cnm}-\frac{1}{4} \delta_{ab} {\cal G}^{(k_2)}_{dcmn}
{\cal G}^{(k_3)dcnm}\right)\\
&-& 2 C_{k_1}\left({\cal G}^{(k_2)}_{acm\rho}{\cal G}_b^{(k_3)cm\rho}-\frac{1}{4} \delta_{ab} {\cal G}^{(k_2)}_{cdm\rho}
{\cal G}^{(k_3)cdm\rho}\right)\Bigg]\delta\left(k_1 + k_2 + k_3 - {7\over 2}\right) \nonumber\\
&-& \sum_{\{k_i\}} {C}_{k_1}... C_{k_4}
\left({\cal G}^{(k_5)}_{ac\alpha\beta}{\cal G}_b^{(k_6)c\alpha\beta}-\frac{1}{4} \delta_{ab} 
{\cal G}^{(k_5)}_{cd\alpha\beta}{\cal G}^{(k_6)cd\alpha\beta}\right)\delta\left(k_1 + ...+ k_6 - {7\over 2}\right),
\nonumber \nd}
where the quantum terms on the LHS of the above equation is determined for $\theta'_k = 1$ in 
\eqref{melamon2}. This is similar to the choice of the quantum terms in \eqref{rambha3} and 
\eqref{ajanta}. In fact now the story follows the pattern laid out for higher order in $g_s$ as seen previously. For example, the 
next order in $g_s$, which is $g_s^{8/3}$, gives us the following equation:

{\footnotesize
\bg\label{hemra}
\delta_{ab} & = & {9\over \Lambda \mathbb{F}(y)} \mathbb{C}_{ab}^{(4, 0)} +
{9\over 4\Lambda H^4 \mathbb{F}(y)} \sum_{\{k_i\}} \Bigg[\widetilde{C}_{k_1}
\left({\cal G}^{(k_2)}_{acmn}{\cal G}_b^{(k_3)cmn}-\frac{1}{4} \delta_{ab} {\cal G}^{(k_2)}_{dcmn}
{\cal G}^{(k_3)dcmn}\right)\\
&+& 2 C_{k_1} \left({\cal G}^{(k_2)}_{acm\rho}{\cal G}_b^{(k_3)cm\rho}-\frac{1}{4} \delta_{ab} {\cal G}^{(k_2)}_{cdm\rho}
{\cal G}^{(k_3)cdm\rho}\right)\Bigg]\delta\left(k_1 + k_2 + k_3 - {4}\right) \nonumber\\
&+& {1\over 4} \sum_{\{k_i\}} {C}_{k_1}... C_{k_4}
\left({\cal G}^{(k_5)}_{ac\alpha\beta}{\cal G}_b^{(k_6)c\alpha\beta}-\frac{1}{4} \delta_{ab} 
{\cal G}^{(k_5)}_{cd\alpha\beta}{\cal G}^{(k_6)cd\alpha\beta}\right)\delta\left(k_1 + ...+ k_6 - {4}\right),
\nonumber \nd}
with the quantum terms being classified by $\theta'_k = 4/3$ as in \eqref{clovbalti} and \eqref{vanandmey}. 
This pattern of fluxes would change eventually as we go higher in $g_s$, and in fact for $g_s^4$ we will see new components entering  for both G-flux and the quantum  energy-momentum tensors. Finally, the function $\mathbb{F}(y)$ appearing in \eqref{hemra} is defined as:

{\footnotesize
\bg\label{shulay}
\mathbb{F}(y) \equiv H^4(y) C^2_{1\over 2} + 4 H^4(y) \sum_{\{k_i\}}C_{k_1} C_{k_2}
\widetilde{C}_{k_3}
\Big[k_1\left(24 -{k_2} - {4k_1}\right) + 2k_3\left(6 -{k_3} - {2 k_2}\right)
\Big] \delta(k_1 + k_2 + k_3 - 1), \nonumber\\ \nd}
which should be compared to \eqref{ramravali}, \eqref{melhic}, \eqref{chotom} and \eqref{sweetdream}. The structural similarities of all these functions are of course not a coincidence: they rely on the forms of the EOMs for the various directions analyzed above. 

\vskip.1in

\noindent {\it Case 2: $F_1(t)$ and $F_2(t)$ satisfying the fluctuation condition \eqref{ranjhita}}

\vskip.1in

\noindent The volume preserving case seems to work rather well, so now we want to see how the story changes once the $\gamma$ factor is introduced in. We expect changes at all fronts now: the Einstein tensor, the energy-momentum tensors for the G-flux and the quantum terms should all reflect the changes. 
The subtleties that we encountered with the quantum terms had a nicer resolution here so we will also have to see what happens now. As before we start with the Einstein tensor, that takes the following form:
\bg\label{novafatai2}
\mathbb{G}_{ab} & = & - {\delta_{ab}\over 2}\left(R + 18 H^4 \Lambda\right)\left({g_s\over H}\right)^2 + 
{4\delta_{ab}\over H^2} \sum_{\{k_i\}}\widetilde{C}_{k_1} C_{k_2} g^{mn}
\partial_m H \partial_n H
\left({g_s\over H}\right)^{2\Delta(k_1 + k_2 + 1/\Delta)}\nonumber\\
&+&{1 \over 4} \Lambda H^4 \delta_{ab} \sum_{\{k_i\}} 
(2\Delta k_1 + \gamma) (2\Delta k_2 + \gamma)  \widetilde{C}_{k_1} \widetilde{C}_{k_2} 
C_{k_3}.... C_{k_6} \left({g_s\over H}\right)^{2\Delta(k_1 + ... + k_6 + 1/\Delta)}\nonumber\\
&+&\Lambda H^4\delta_{ab} \sum_{\{k_i\}}
\bigg(2\Delta k_1\left(8 -
\Delta k_2 - 4\Delta k_1\right) + (2\Delta k_3 + \gamma)\left(4 - 2 \Delta k_3 -\gamma - 4\Delta k_2\right)
\bigg)
\nonumber\\
&\times& C_{k_1} C_{k_2} \widetilde{C}_{k_3}\left({g_s\over H}\right)^{2\Delta(k_1 + k_2 + k_3 + 1/\Delta)},
\nd
where interestingly none of the $g_s$ scalings get effected by the $\gamma$ term, but most of the 
individual terms do have $\gamma$ dependent coefficients. Similar, the energy-momentum tensor for the G-fluxes changes in an expected way:
\bg\label{kristinaQ2}
\mathbb{T}^G_{ab} & = & {1\over 12 H^4} \sum_{\{k_i\}} \widetilde{C}_{k_1}\widetilde{C}_{k_2} C_{k_3}
\left({\cal G}^{(k_4)}_{amnp}{\cal G}^{(k_5)mnp}_{b}-\frac{1}{2} \delta_{ab} {\cal G}^{(k_4)}_{mnpc}
{\cal G}^{(k_5)mnpc}\right)\left({g_s\over H}\right)^{2\Delta(k_1 + ... + k_5 + 1/\Delta)}
\nonumber\\
&+& {1\over 4 H^4} \sum_{\{k_i\}} 
\left({\cal G}^{(k_1)}_{amn\alpha}{\cal G}_b^{(k_2)mn\alpha}-\frac{1}{2} \delta_{ab} {\cal G}^{(k_1)}_{mn\alpha c}
{\cal G}^{(k_2)mn\alpha c}\right)\left({g_s\over H}\right)^{2\Delta(k_1 + k_2 + 1/\Delta -\gamma/2\Delta)}
\nonumber\\
&+& {1\over 4 H^4} \sum_{\{k_i\}} {C}_{k_1}{C}_{k_2} C_{k_3}
\left({\cal G}^{(k_4)}_{am\alpha\beta}{\cal G}_b^{(k_5)m\alpha\beta}-
\frac{1}{2} \delta_{ab} {\cal G}^{(k_4)}_{cm\alpha\beta}
{\cal G}^{(k_5)cm\alpha\beta}\right)\left({g_s\over H}\right)^{2\Delta(k_1 + ..+ k_5 + 1/\Delta 
-\gamma/\Delta)}
\nonumber\\
&+& {1\over 4 H^4} \sum_{\{k_i\}} \widetilde{C}_{k_1}
\left({\cal G}^{(k_2)}_{acmn}{\cal G}_b^{(k_3)cmn}-\frac{1}{4} \delta_{ab} {\cal G}^{(k_2)}_{dcmn}
{\cal G}^{(k_3)dcmn}\right)\left({g_s\over H}\right)^{2\Delta(k_1 + k_2 + k_3)}
\nonumber\\
&+& {1\over 4 H^4} \sum_{\{k_i\}} {C}_{k_1}{C}_{k_2} C_{k_3} C_{k_4}
\left({\cal G}^{(k_5)}_{ac\alpha\beta}{\cal G}_b^{(k_6)c\alpha\beta}-\frac{1}{4} \delta_{ab} 
{\cal G}^{(k_5)}_{cd\alpha\beta}
{\cal G}^{(k_6)cd\alpha\beta}\right)\left({g_s\over H}\right)^{2\Delta(k_1 + ...+ k_6 -\gamma/\Delta)}
\nonumber\\
&+& {1\over 2 H^4} \sum_{\{k_i\}} {C}_{k_1}
\left({\cal G}^{(k_2)}_{acm\rho}{\cal G}_b^{(k_3)cm\rho}-\frac{1}{4} \delta_{ab} {\cal G}^{(k_2)}_{cdm\rho}
{\cal G}^{(k_3)cdm\rho}\right)\left({g_s\over H}\right)^{2\Delta(k_1 + k_2 + k_3 -\gamma/2\Delta)}
\nonumber\\
&-& {\delta_{ab}\over 4\cdot 4! H^4} \sum_{\{k_i\}} \widetilde{C}_{k_1}\widetilde{C}_{k_2} 
{\cal G}^{(k_3)}_{mnpq}
{\cal G}^{(k_4)mnpq}\left({g_s\over H}\right)^{2\Delta(k_1 + k_2 + k_3 + k_4 + 2/\Delta)}
\nonumber\\
&-& {\delta_{ab}\over 4! H^4} \sum_{\{k_i\}} \widetilde{C}_{k_1}{C}_{k_2} 
{\cal G}^{(k_3)}_{mnp\alpha}
{\cal G}^{(k_4)mnp\alpha}\left({g_s\over H}\right)^{2\Delta(k_1 + k_2 + k_3 + k_4 + 2/\Delta 
-\gamma/2\Delta)}
\nonumber\\
&-& {\delta_{ab}\over 16 H^4} \sum_{\{k_i\}} {C}_{k_1}{C}_{k_2} 
{\cal G}^{(k_3)}_{mn\alpha\beta}
{\cal G}^{(k_4)mn\alpha\beta}\left({g_s\over H}\right)^{2\Delta(k_1 + k_2 + k_3 + k_4 + 2/\Delta 
-\gamma/\Delta)}
\nonumber\\
&+& {4\delta_{ab}\over H^2} \sum_{\{k_i\}} 
 \widetilde{C}_{k_1}{C}_{k_2}\partial_m H \partial^m H
\left({g_s\over H}\right)^{2\Delta(k_1 + k_2 + 1/\Delta)}, \nd
where taking $\gamma = 2$ we see that there are no zeroth order in $g_s$ possible because the lower bound on the moding $k_i$ of any G-flux component has to be  $k_i \ge 9/2$. The largest allowed 
suppression factor is $-\gamma/\Delta = -6$ for the component of G-flux ${\cal G}^{(9/2)}_{\alpha\beta ab}$
in \eqref{kristinaQ2}, implying that the lowest power of $g_s$ contribution to the EOM will be 
$g_s^2$.  This fits rather well with the $g_s$ scaling of the quantum terms in \eqref{teenangul}, which now has a similar form as \eqref{telmaro} and \eqref{malinB} with $\theta_k$ defined as in \eqref{miai}. Therefore combining \eqref{novafatai2} with \eqref{kristinaQ2}, \eqref{neveC} and \eqref{telmaro} we get, to order 
$g_s^2$, the following EOM:
\bg\label{stjean}
 \left({R\over 2} + 9 H^4 \Lambda\right)
\delta_{ab} + {1\over 4 H^4} 
\left({\cal G}^{(9/2)}_{ac\alpha\beta}{\cal G}_b^{(9/2)c\alpha\beta}-\frac{1}{4} \delta_{ab} 
{\cal G}^{(9/2)}_{cd\alpha\beta}{\cal G}^{(9/2)cd\alpha\beta}\right) +  \mathbb{C}_{ab}^{(3, 0)} = 0, \nd
which may now be compared to \eqref{buskaM}. The quantum terms appearing here is similar to what we had in \eqref{buskaM}, and is classified by $\theta_k = 2/3$ in \eqref{miai}. However the number of G-flux components contributing to \eqref{stjean} is much smaller; and \eqref{stjean} is a set of two equations with at least 7 unknowns. 

To the next order in $g_s$, i.e $g_s^{7/3}$, the Einstein tensor \eqref{novafatai2} does contribute compared to the scenario with \eqref{novafatai}. In fact both the energy-momentum tensors also contribute to this order. The resulting EOM becomes:

{\footnotesize
\bg\label{toolused}
\delta_{ab} = {1\over 4q\Lambda H^8} \sum_{\{k_i\}} {C}_{k_1}....C_{k_4}
\left({\cal G}^{(k_5)}_{ac\alpha\beta}{\cal G}_b^{(k_6)c\alpha\beta}-\frac{1}{4} \delta_{ab} 
{\cal G}^{(k_5)}_{cd\alpha\beta}
{\cal G}^{(k_6)cd\alpha\beta}\right)\delta\left(k_1 + ... + k_6 - {19\over 2}\right) + 
{\mathbb{C}_{ab}^{(7/2, 0)}\over q\Lambda H^4}, \nonumber\\ \nd}
where $q \equiv 4 - 10 C_{1\over 2}$, and one may use this equation to fix the form of the quantum terms classified by $\theta_k = 1$ in \eqref{miai} with the G-flux component appearing above\footnote{Compared
to the ($\alpha, \beta$) case the traces of \eqref{toolused} and \eqref{gargere} do not fix the signs of 
$\left[\mathbb{C}\right]^{(7/2, 0)}$ in both cases.}. 
Once we go to higher orders in $g_s$ new components of G-flux start contributing to the EOM as evident from the form of 
\eqref{kristinaQ2}, but the flatness of the toroidal direction $-$ i.e the metric choice $\delta_{ab}$ $-$ 
remains unchanged to any order in $g_s$. Since the story is expected to proceed in a similar vein as earlier, we will not discuss this further here, and instead go to the study of space-time components.

\subsubsection{Einstein equation along $(\mu, \nu)$ directions \label{kocu4}}

The structural similarities of the equations for all the previous cases have some bearings on the choices of G-flux components (at least to some low orders in $g_s$) that enter in the EOMs. The quantum terms are also similar, modulo the subtlety for $\mathbb{T}_{ab}^Q$ requiring some redefinition (see however the second formula in footnote 
\ref{cbmoja} for a universal definition of the quantum energy-momentum tensor).
 
The story for the space-time components will require additional subtleties that we will illustrate as we go along.  First, let us express the Einstein tensor along the two spatial directions in the following way:
\bg\label{sugarmey}
\mathbb{G}_{ij} &=& -\frac{\eta_{ij}}{\Lambda(t)}\Bigg(3\Lambda + \frac{{R}}{2H^4}
+\frac{4 {g}^{\alpha\beta}\partial_\alpha H
\partial_\beta H}{H^6 F_1}
+ \frac{4 {g}^{mn}\partial_m H\partial_n H}{H^6 F_2}
- \frac{\square_{(m)}H^4}{2H^8 F_1}\Bigg)
\\
&+& {\eta_{ij}\over \Lambda(t)}\Bigg(\frac{\square_{(\alpha)}H^4}{2 H^8 F_2}\Bigg)
+\eta_{ij}\left( \frac{\dot{F}_1^2}{4F_1^2}+ \frac{\dot{F}_1}{tF_1}- \frac{\ddot{F}_1}{F_1}
-\frac{\dot{F}_2^2}{2F_2^2} +\frac{2\dot{F}_2}{tF_2} - \frac{2\ddot{F}_2}{F_2}
- \frac{2\dot{F}_1\dot{F}_2}{F_1 F_2}\right), \nonumber
\nd
where, since we identified $\Lambda(t) = \left({g_s\over H}\right)^2$, the appearance of $\Lambda^{-1}(t)$
is a bit disconcerting for the late time physics where $t \to 0$ or $g_s \to 0$. We will not worry about this right now and carry on with the Einstein tensor along the temporal direction which, in turn,  takes the following form:
\bg\label{broshai}
\mathbb{G}_{00} &=& {\eta_{00}\over \Lambda(t)}\Bigg(\frac{\square_{(m)}H^4}{2 H^8 F_2}\Bigg) -\eta_{00}\left( \frac{\dot{F}_1^2}{4F_1^2}
- \frac{3\dot{F}_1}{tF_1}
+ \frac{3\dot{F}_2^2}{2F_2^2} -\frac{6\dot{F}_2}{tF_2}
+ \frac{2\dot{F}_1\dot{F}_2}{F_1 F_2}\right)\nonumber\\
&-& \frac{\eta_{00}}{\Lambda(t)}\Bigg(3\Lambda + \frac{{R}}{2H^4}
+\frac{4 {g}^{\alpha\beta}\partial_\alpha H
\partial_\beta H}{H^6 F_1}
+ \frac{4 {g}^{mn}\partial_m H\partial_n H}{H^6 F_2}
- \frac{\square_{(\alpha)}H^4}{2 H^8 F_1}\Bigg),
\nd
where the key difference from \eqref{sugarmey}, other than the appearance of $\eta_{00}$, is in the terms 
with derivatives on $F_i(t)$. Other than these, both the Einstein tensors are similar in terms of the appearance of the warp-factor $H(y)$ and the six-dimensional curvature scalar $R$.  In the similar vein, we can express the energy-momentum tensor for the G-flux in the following way:
\bg\label{thanda}
\mathbb{T}^G_{\mu\nu} &=&-{\eta_{\mu\nu}\over 8\Lambda(t) H^8}
\left(\frac{1}{3 F_2^3}{G}_{mnpa}{G}^{mnpa}
+\frac{1}{F_2^2F_1}{G}_{m\alpha pa}{G}^{m\alpha pa}
+
\frac{1}{F_1^2F_2}{G}_{\alpha\beta pa}{G}^{\alpha\beta pa}\right)\nonumber\\
&-&{\eta_{\mu\nu} \over 24 H^8}\left( \frac{1}{4 F_2^4}{G}_{mnpq}{G}^{mnpq}
+
\frac{1}{F_2^3F_1}{G}_{mnp\alpha}{G}^{mnp\alpha}+
\frac{1}{4F_2^2F_1^2}{G}_{mn\alpha\beta}{G}^{mn
\alpha\beta}\right)\nonumber\\
&-& {\eta_{\mu\nu}\over 8\Lambda^2(t) H^8}\left(\frac{1}{2 F_2^2}{G}_{mnab}{G}^{mnab}+ 
\frac{1}{F_2F_1}{G}_{m\alpha ab}{G}^{m\alpha ab}
+
\frac{1}{2 F_1^2}{G}_{\beta\alpha ab}{G}^{\beta\alpha ab}\right)\nonumber\\
&-& {4\eta_{\mu\nu} \over \Lambda(t)H^6}\Bigg(\frac{g^{mn}\partial_m H \partial_n H}{F_2}   
+ \frac{g^{\alpha\beta}\partial_\alpha H \partial_\beta H}{F_1}
\Bigg), \nd
where again expectedly the last two terms cancel with equivalent terms in both $G_{ij}$ and $G_{00}$ in 
\eqref{sugarmey} and \eqref{broshai} respectively. With these at our disposal, let us go to the individual cases now.


\vskip.1in

\noindent {\it Case 1: $F_1(t)$ and $F_2(t)$ satisfying the volume-preserving condition \eqref{olokhi}}

\vskip.1in

\noindent The inverse $\Lambda(t)$ factors appearing in the expressions of the Einstein tensors as well as the energy-momentum tensors for the G-fluxes are a case of worry in the late time limit that we want to analyze the background. Of course the existence of these factors are expected from the inverse 
$\Lambda(t)$ factor appearing in the type IIB metric \eqref{vegamey}, but since our construction involve 
finite values in the $g_s \to 0$ limit, we will need to tread carefully to interpret our answers. To analyze the story further, let us write the Einstein tensor along spatial direction first in the following way:

{\footnotesize
\bg\label{synecdoche}
\mathbb{G}_{ij} & = & -\eta_{ij} \left(3\Lambda + {R\over 2H^4}\right)\left({g_s\over H}\right)^{-2} + 
{\Lambda \eta_{ij}\over 9} \sum_{\{k_i\}} k_1 k_2 \widetilde{C}_{k_1} 
\widetilde{C}_{k_2}C_{k_3}...C_{k_6}
\left({g_s\over H}\right)^{2\Delta(k_1 + ...+k_6 -1/\Delta)}\nonumber\\
&-& {4\eta_{ij}\over H^6} \sum_{\{k_i\}}\Bigg[\left((\partial_\alpha H)^2 
- {\square_{(m)} H^4 \over 8 H^2}\right) C_{k_1} C_{k_2}  
+ \left((\partial_m H)^2 
 -{\square_{(\alpha)} H^4 \over 8 H^2}  \right) C_{k_1} \widetilde{C}_{k_2}\Bigg]
\left({g_s\over H}\right)^{2\Delta(k_1 + k_2 -1/\Delta)}\nonumber\\
&+& {2\Lambda\eta_{ij}\over 9}\sum_{\{k_i\}}
\Big[2k_3(3 - k_3 - 2k_2) + k_1(12 - 4 k_1 - k_2)\Big]C_{k_1} C_{k_2}
\widetilde{C}_{k_3}\left({g_s\over H}\right)^{2\Delta(k_1 + k_2 + k_3 -1/\Delta)}, \nd}
where we have defined $(\partial_\alpha H)^2  \equiv g^{\alpha\beta} \partial_\alpha H\partial_\beta H$ and 
the same for $(\partial_m H)^2 \equiv g^{mn} \partial_mH\partial_nH$ with un-warped metrics. It is also easy to read out the form of the $\mathbb{G}_{00}$ tensor:

{\footnotesize
\bg\label{synecdoche2}
\mathbb{G}_{00} & = & -\eta_{00} \left(3\Lambda + {R\over 2H^4}\right)\left({g_s\over H}\right)^{-2} -
{\Lambda \eta_{00}\over 9} \sum_{\{k_i\}} k_1 k_2 \widetilde{C}_{k_1} 
\widetilde{C}_{k_2}C_{k_3}...C_{k_6}
\left({g_s\over H}\right)^{2\Delta(k_1 + ...+k_6 -1/\Delta)}\nonumber\\
&-& {4\eta_{00}\over H^6} \sum_{\{k_i\}}\Bigg[\left((\partial_\alpha H)^2 
- {\square_{(m)} H^4 \over 8 H^2}\right) C_{k_1} C_{k_2}  
+ \left((\partial_m H)^2 
 -{\square_{(\alpha)} H^4 \over 8 H^2}  \right) C_{k_1} \widetilde{C}_{k_2}\Bigg]
\left({g_s\over H}\right)^{2\Delta(k_1 + k_2 -1/\Delta)}\nonumber\\
&+& {2\Lambda\eta_{00}\over 9}\sum_{\{k_i\}}
\Big[k_3(9 - 4k_2) + 3k_1(6  - k_2)\Big]C_{k_1} C_{k_2}
\widetilde{C}_{k_3}\left({g_s\over H}\right)^{2\Delta(k_1 + k_2 + k_3 -1/\Delta)}, \nd}
which differs from \eqref{synecdoche} in three respects: presence of $\eta_{00}$, sign of the second term, and a different coefficient of the last term. On the other hand, 
from the various terms of \eqref{synecdoche} and \eqref{synecdoche2}, it is easy to infer that the lowest 
power of $g_s$, which is $g_s^{-2}$, appears when $k_i = 0$. In the limit $g_s \to 0$, this blows up, so to 
extract finite terms we have to carefully analyze the other contributions to the EOMs.  

The other contributions to the EOM for the spatial components appear from the energy-momentum 
tensors of the G-flux and the quantum terms. The energy-momentum tensor for the G-fluxes for both spatial and temporal components may be expressed in the following way:

{\footnotesize
\bg\label{keener}
\mathbb{T}_{\mu\nu}^G & = & {\eta_{\mu\nu}\over 4 H^8}\sum_{\{k_i\}}\left({1\over 6}
\widetilde{C}_{k_1}\widetilde{C}_{k_2}
C_{k_3}{\cal G}^{(k_4)}_{mnap}{\cal G}^{(k_5)mnpa} - {1\over 2} {C}_{k_1}{C}_{k_2}
C_{k_3}{\cal G}^{(k_4)}_{\alpha\beta pa}{\cal G}^{(k_5)\alpha\beta pa}\right)
\left({g_s\over H}\right)^{2\Delta(k_1 + ...+k_5 - 1/\Delta)}\nonumber\\
&-& {\eta_{\mu\nu}\over 24 H^8}\sum_{\{k_i\}}\left({1\over 4}\widetilde{C}_{k_1}\widetilde{C}_{k_2}
{\cal G}^{(k_3)}_{mnpq}{\cal G}^{(k_4)mnpq} + 
\widetilde{C}_{k_1}{C}_{k_2}
{\cal G}^{(k_3)}_{mnp\alpha}{\cal G}^{(k_4)mnp\alpha} +
{1\over 4}{C}_{k_1}{C}_{k_2}
{\cal G}^{(k_3)}_{mn\alpha\beta}{\cal G}^{(k_4)mn\alpha\beta}\right)\nonumber\\
&\times& \left({g_s\over H}\right)^{2\Delta(k_1 + ..
+k_4)}
- {\eta_{\mu\nu}\over 8H^8}\sum_{\{k_i\}}\left({1\over 2} \widetilde{C}_{k_1}{\cal G}^{(k_2)}_{mnab}{\cal G}^{(k_3)mnab} + 
{C}_{k_1}{\cal G}^{(k_2)}_{m\alpha ab}{\cal G}^{(k_3)m\alpha ab}\right)
\left({g_s\over H}\right)^{2\Delta(k_1 + k_2 + k_3 - 2/\Delta)} \nonumber\\
&-&{\eta_{\mu\nu}\over H^6}\sum_{\{k_i\}}\left({1\over 8H^2} {\cal G}_{m\alpha p a}^{(k_1)} {\cal G}^{(k_2)m\alpha p a}
+ 4(\partial_\alpha H)^2 C_{k_1} C_{k_2} + 4(\partial_m H)^2 C_{k_1} \widetilde{C}_{k_2}\right)
\left({g_s\over H}\right)^{2\Delta(k_1 + k_2 - 1/\Delta)}\nonumber\\
&-& {\eta_{\mu\nu}\over 16 H^8}\sum_{\{k_i\}}C_{k_1}C_{k_2}C_{k_3}C_{k_4} {\cal G}^{(k_5)}_{\alpha\beta ab}
{\cal G}^{(k_6)\alpha\beta ab}\left({g_s\over H}\right)^{2\Delta(k_1 + k_2 + k_3 + k_4 + k_5 + k_6 - 2/\Delta)}, \nd}
where since some of the $k_i$, accompanying the G-flux components are bounded below as $k_i \ge 3/2$, 
we would get the $g_s^{-2}$ powers from the ${\cal G}^{(3/2)}_{mnab}, {\cal G}^{(3/2)}_{m\alpha ab}$ 
and ${\cal G}^{(3/2)}_{\alpha\beta ab}$ components. However this is puzzling in light of the quantum terms 
\eqref{neveC}. Our expression from \eqref{neveC} allows only $g_s^0$ as the lowest power of $g_s$ because 
the negative powers are assimilated to a series in $e^{-1/g_s}$. In the limit $g_s \to 0$ this dies off faster 
than any powers of $g_s$. Additionally as cautioned in footnote \ref{tantana} it is not advisable to expand 
$e^{-1/g_s}$ to any finite orders in inverse $g_s$. One way out of this would be to multiply the Einstein tensor 
\eqref{synecdoche}, the G-flux energy-momentum tensor \eqref{keener} and the quantum energy-momentum tensor \eqref{neveC} by $\left({g_s\over H}\right)^2$. This unfortunately will {\it not} solve the problem, because now the lowest power of \eqref{neveC} will be $g_s^2$ so cannot be used to balance the $g_s^0$ terms of \eqref{synecdoche} and \eqref{keener}. The quantum terms are  essential, to avoid over-constraining the system. Additionally, the $g_s$ scaling along the space-time direction is in fact:
\bg\label{ewatson}
g_s^{\theta'_k - 8/3} \equiv g_s^0, ~ g_s^{1/3}, ~ g_s^{2/3}, ~g_s, ~g_s^{4/3}, ......, \nd
as evident from \eqref{charaangul}, implying that the minimum value of $\theta'_k$ in \eqref{melamon2} is
$\theta'_k = 8/3$ to account for $g_s$ independent terms. All of these then imply the following way out: interpret the $g_s^{-2}$ as an M5-instanton wrapping the base ${\cal M}_4 \times {\cal M}_2$ such that it will contribute the extra $g_s^{-2}$ factor, thus redefining  the energy-momentum tensor for the quantum pieces as:
\bg\label{emilee}
\mathbb{T}^Q_{\mu\nu} \equiv \sum_{\{k\}} \mathbb{C}_{\mu\nu}^{(k, 0)} 
\left({g_s\over H}\right)^{2\Delta(k - 1/\Delta)}, \nd
instead of \eqref{neveC} for $(\mu, \nu)$ indices. Such a re-definition is similar to the re-definition we did for the $(a, b)$ case with perturbative terms (see also the second formula in footnote \ref{cbmoja}) and is consistent with the scalings employed in \cite{nogo} and 
\cite{nodS} (see eq (5.29) in \cite{nogo}). In section \ref{instachela} we will discuss in more details how these non-perturbative contributions arise.

There is yet another contribution that we have ignored so far and has to do with the energy-momentum tensor of an almost {\it static} set of membranes. These are related to static D3-branes (integer and fractional) in the type IIB side, and we can consider both branes and anti-branes in our picture. For simplicity, let us assume that we have 
$n_b$ number of coincident membranes at a point on the internal eight-dimensional manifold. These membranes are therefore stretched along the $2+1$ dimensional space-time\footnote{We will consider both integer and fractional M2-branes. The latter being M5-branes wrapped on 3-cycles. The M5-instanton wrapping ${\cal M}_4 \times {\cal M}_2$ already contributes to \eqref{emilee}.}. 
The analysis of the energy-momentum tensor proceeds in exactly the same way as given in \cite{nogo}, so we will suffice ourselves by simply quoting the answer:
\bg\label{casamey}
\mathbb{T}_{\mu\nu}^{(B)} \approx  - {\kappa^2 T_2 n_b \over H^8 \sqrt{g_6}}\left({g_s\over H}\right)^{-2}
\delta^8\left(y - Y\right) \eta_{\mu\nu}, \nd
where $T_2$ is the tension of the individual membranes, $\kappa$ is a constant related to $M_p$, 
$g_6$ is the determinant of the unwarped metric of the six-dimensional base ${\cal M}_4 \times {\cal M}_2$, and $n_b$ is the number of membranes located at $Y^M$ in the internal eight-manifold. 

With these definitions of the quantum energy-momentum tensor in \eqref{emilee} and the membrane energy-momentum tensor in \eqref{casamey}, we are ready to move ahead with the EOMs. First we multiply all the tensors with $\left({g_s\over H}\right)^2$ to get rid of any infinities arising in the 
$g_s\to 0$ limit. Secondly, we compare the zeroth order in $g_s$ for \eqref{synecdoche}, \eqref{keener} and 
\eqref{emilee}, to get the following EOM:
\bg\label{fleuve}
&&  6\Lambda + {R\over H^4} - {\square H^4\over H^8} + \left[\mathbb{C}^i_{i}\right]^{(0, 0)}   
- {2 \kappa^2 T_2 n_b \over H^8 \sqrt{g_6}} \delta^8(y - Y)\nonumber\\
&& = {1\over 8 H^8} \Big({\cal G}^{(3/2)}_{mnab}{\cal G}^{(3/2)mnab} +
2 {\cal G}^{(3/2)}_{m\alpha ab}{\cal G}^{(3/2)m\alpha ab} +  {\cal G}^{(3/2)}_{\alpha\beta ab}
{\cal G}^{(3/2)\alpha\beta ab}\Big),  
 \nd
showing how the same set of G-flux components appear again to balance the spatial equation of motion. 
We have also defined $\square \equiv \square_{(m)} + \square_{(\alpha)}$ to avoid clutter.
The equation \eqref{fleuve} is somewhat similar to what we had in eq (5.32) of \cite{nogo} with two crucial differences. One, the G-flux components are the set ${\cal G}^{(3/2)}_{mnab}, 
{\cal G}^{(3/2)}_{m\alpha ab}$ and ${\cal G}^{(3/2)}_{\alpha\beta ab}$ of {\it localized} fluxes and not the globally-defined time-independent flux component appearing in \cite{nogo}\footnote{If we consider the other set of modings for the G-flux components, described in the paragraphs between \eqref{evabmey2} and 
\eqref{teenangul} (see also \cite{petite}), we expect additional G-flux components to appear alongside the ones we have now. In fact this would repeat for all the cases studied so far, at least for the volume preserving case \eqref{olokhi}, and  the end result could then be compared to \cite{nogo}. Here we want to avoid these complications.}. 
Two, the quantum terms 
$\mathbb{C}_{\mu\nu}^{(0, 0)}$ are classified by, including non-localities:
\bg\label{oleport} 
2\sum_{i = 1}^{27}l_i + n_1 + n_2 + \sum_{i = 0}^4 l_{34 + i} = 8 + 2r, \nd
i.e with $\theta'_k = 8/3$ in \eqref{melamon2} ($l_i, n_i$ are defined in \eqref{phingsha2}), 
compared to $\theta'_0 = 8/3$ in 
\eqref{kkkbkb2}. The former, i.e \eqref{oleport}, has a large but {\it finite} number of solutions, whereas 
the latter has an {\it infinite} number of solutions with no $g_s$ or $M_p$ hierarchies \cite{petite}. In a similar vein one may work out the $\mathbb{G}_{00}$ EOM, but to this order the result \eqref{fleuve} will not change. 

The next order in $g_s$, i.e for $g_s^{1/3}$, one may easily find the EOMs by comparing terms of 
this order from \eqref{synecdoche}, \eqref{synecdoche2}, \eqref{keener} and \eqref{emilee} with no contributions from the membranes.  The G-flux components contributing now are of the form 
${\cal G}^{(3/2)}_{MNab}$ and 
${\cal G}^{(2)}_{MNab}$ with ($M, N$) spanning the coordinates of ${\cal M}_4 \times {\cal M}_2$. The 
quantum terms $\mathbb{C}_{ij}^{(1/2, 0)}$ are classified by $\theta'_k = 3$ in \eqref{melamon2}. Combining the two set of equations, one from the ($i, j$) components, and one from the ($0, 0$) components, we get:
\bg\label{urisis}
2\left[\mathbb{C}_0^0\right]^{(1/2, 0)} = \left[\mathbb{C}_i^i\right]^{(1/2, 0)}, \nd
where the quantum terms $\mathbb{C}^{(1/2, 0)}_{\mu\nu}$ are the specific linear combinations of all terms classified by $\theta'_k = 3$ for individual components in \eqref{melamon2}. According to the discussions 
around \eqref{3amigomey} these quantum terms are computed using the dominant scalings of the metric components ${\bf g}_{mn}$ and ${\bf g}_{\alpha\beta}$. Thus the LHS of \eqref{urisis} is fixed in terms of the 
known components of the metric and the G-fluxes in a way that their sum vanishes. Such an equation can be used to predict the relative coefficient of the various terms to the same order in curvatures and G-fluxes.

One can even go higher orders in $g_s$, say for example $g_s^{2/3}$ as we have done before, and compare the ($i, j$) and the ($0, 0$) EOMs. The quantum terms would be of the form 
$\mathbb{C}_{\mu\nu}^{(1, 0)}$ and are classified by $\theta'_k = 10/3$ in \eqref{melamon2}. These could be used to fix the higher order coefficients of $F_i(t)$ in terms of the quantum terms.  For example 
taking the traces of \eqref{synecdoche} and \eqref{synecdoche2} appropriately, we get:
\bg\label{ellemey}
C_{1\over 2}^2 = 3\left(2\left[\mathbb{C}_0^0\right]^{(1, 0)} - \left[\mathbb{C}_i^i\right]^{(1, 0)}\right), \nd
which tells us that it is only the constant pieces of the quantum terms \eqref{phingsha2} that are responsible in generating the $F_i(t)$ functions. Note that, to this order $C_1$ and $\widetilde{C}_1$ coefficients cancel out. To determine these, we have to go to the next order in $g_s$ where, in turn the $C_{3\over 2}$ and 
$\widetilde{C}_{3\over 2}$ pieces cancel out, leaving us with $C_1$ and $\widetilde{C}_1$. We will leave the evaluation of these coefficients for interested readers, and instead go to the discussion of the case with 
$\gamma$ switched on. 

\vskip.1in

\noindent {\it Case 2: $F_1(t)$ and $F_2(t)$ satisfying the fluctuation condition \eqref{ranjhita}}

\vskip.1in

\noindent The analysis along the space-time directions has a few subtleties that we clarified above. Additional subtleties arise when we switch on non-zero $\gamma$ from the fact that the internal eight-manifold has zero Euler characteristics. This implies that one cannot switch on either non-zero components of G-fluxes that are time-independent, or dynamical M2-branes at least in the supersymmetric limit 
\cite{BB, DRS}. Our study is for non-supersymmetric states, plus we take vanishing time-independent component of G-flux \eqref{metromey}, so the situation is a bit more subtle. Nevertheless the bound considered in \cite{BB, DRS} does not allow us to take static M2-branes\footnote{See however footnote 
\ref{choolmaro}.}. 
What happens for dynamical branes will be discussed later.  

We will start by elaborating the Einstein tensor for both spatial and temporal directions. The Einstein tensor for the two spatial directions may be expressed in the following way:

{\footnotesize
\bg\label{synecdoche3}
\mathbb{G}_{ij} & = & -\eta_{ij} \left(3\Lambda + {R\over 2H^4}\right)\left({g_s\over H}\right)^{-2} + 
{\Lambda \eta_{ij}\over 4} \sum_{\{k_i\}} (2\Delta k_1+\gamma)(2\Delta k_2 +\gamma) \widetilde{C}_{k_1} 
\widetilde{C}_{k_2}C_{k_3}...C_{k_6}
\left({g_s\over H}\right)^{2\Delta(k_1 + ...+k_6 -1/\Delta)}\nonumber\\
&+& {4\eta_{ij}\over H^6} \sum_{\{k_i\}}C_{k_1}\Bigg[C_{k_2}\left(  
{\square_{(m)} H^4 \over 8 H^2}\right) \left({g_s\over H}\right)^{2\Delta(k_1 + k_2 -1/\Delta 
- \gamma/2\Delta)} - \widetilde{C}_{k_2} (\partial_m H)^2 
\left({g_s\over H}\right)^{2\Delta(k_1 + k_2 -1/\Delta)}\Bigg]
\\
&+& {\Lambda\eta_{ij}\over 9}\sum_{\{k_i\}}
\Big[(2k_3 + 3\gamma)(6 - 2k_3 - 3\gamma- 4k_2) +2 k_1(12 - 4 k_1 - k_2)\Big]C_{k_1} C_{k_2}
\widetilde{C}_{k_3}\left({g_s\over H}\right)^{2\Delta(k_1 + k_2 + k_3 -1/\Delta)}, \nonumber \nd}
where we see that only one $g_s$ scaling is effected by the $\gamma$ factor, although quite a few coefficients do pick up $\gamma$ dependent factors. In addition to that, derivatives with respect to 
$\alpha$ are missing compared to \eqref{synecdoche}. Similar story also shows up for the Einstein tensor along the temporal directions in the following way:

{\footnotesize
\bg\label{synecdoche4}
\mathbb{G}_{00} & = & -\eta_{00} \left(3\Lambda + {R\over 2H^4}\right)\left({g_s\over H}\right)^{-2} -
{\Lambda \eta_{00}\over 4} \sum_{\{k_i\}} (2\Delta k_1+\gamma)(2\Delta k_2 +\gamma) \widetilde{C}_{k_1} 
\widetilde{C}_{k_2}C_{k_3}...C_{k_6}
\left({g_s\over H}\right)^{2\Delta(k_1 + ...+k_6 -1/\Delta)}\nonumber\\
&+& {4\eta_{00}\over H^6} \sum_{\{k_i\}}C_{k_1}\Bigg[C_{k_2}\left(  
{\square_{(m)} H^4 \over 8 H^2}\right) \left({g_s\over H}\right)^{2\Delta(k_1 + k_2 -1/\Delta 
- \gamma/2\Delta)} - \widetilde{C}_{k_2} (\partial_m H)^2 
\left({g_s\over H}\right)^{2\Delta(k_1 + k_2 -1/\Delta)}\Bigg]
\nonumber\\
&+& {\Lambda\eta_{00}\over 9} \sum_{\{k_i\}} \Big[(2k_3 + 3\gamma)(9 - 4k_2) + 6k_1(6- k_2)\Big]
C_{k_1} C_{k_2}
\widetilde{C}_{k_3}\left({g_s\over H}\right)^{2\Delta(k_1 + k_2 + k_3 -1/\Delta)}, \nd}
where again, as compared to \eqref{synecdoche2}, other than the last term and one relative sign difference, the two Einstein tensors are identical. In a similar vein, the energy-momentum tensor for the G-flux takes the following form:
\bg\label{keener2}
\mathbb{T}_{\mu\nu}^G & = & {\eta_{\mu\nu}\over 24 H^8}\sum_{\{k_i\}}\left(
\widetilde{C}_{k_1}\widetilde{C}_{k_2}
C_{k_3}{\cal G}^{(k_4)}_{mnap}{\cal G}^{(k_5)mnpa}\right)
\left({g_s\over H}\right)^{2\Delta(k_1 + ...+k_5 - 1/\Delta)}\nonumber\\
& -&  {\eta_{\mu\nu}\over 8 H^8}\sum_{\{k_i\}}\left( {C}_{k_1}{C}_{k_2}
C_{k_3}{\cal G}^{(k_4)}_{\alpha\beta pa}{\cal G}^{(k_5)\alpha\beta pa}\right)
\left({g_s\over H}\right)^{2\Delta(k_1 + ...+k_5 - 1/\Delta - \gamma/\Delta)}\nonumber\\
&-& {\eta_{\mu\nu}\over 96 H^8}\sum_{\{k_i\}}\left(\widetilde{C}_{k_1}\widetilde{C}_{k_2}
{\cal G}^{(k_3)}_{mnpq}{\cal G}^{(k_4)mnpq}\right)\left({g_s\over H}\right)^{2\Delta(k_1 + k_2 + k_3
+k_4)}\nonumber\\
&+& {\eta_{\mu\nu}\over 24  H^8}\sum_{\{k_i\}}\left(\widetilde{C}_{k_1}{C}_{k_2}
{\cal G}^{(k_3)}_{mnp\alpha}{\cal G}^{(k_4)mnp\alpha}\right)\left({g_s\over H}\right)^{2\Delta(k_1 + k_2 + k_3 +k_4 -\gamma/2\Delta)}\nonumber\\
& +& {\eta_{\mu\nu}\over 96 H^8}\sum_{\{k_i\}}\left(
{C}_{k_1}{C}_{k_2}
{\cal G}^{(k_3)}_{mn\alpha\beta}{\cal G}^{(k_4)mn\alpha\beta}\right)\left({g_s\over H}\right)^{2\Delta(k_1 + 
k_2 + k_3 +k_4 - \gamma/\Delta)}\nonumber\\
&- & {\eta_{\mu\nu}\over 16 H^8} \sum_{\{k_i\}}\left(\widetilde{C}_{k_1}{\cal G}^{(k_2)}_{mnab}{\cal G}^{(k_3)mnab}\right)\left({g_s\over H}\right)^{2\Delta(k_1 + k_2 + k_3 - 2/\Delta)}\nonumber\\
&-&  {\eta_{\mu\nu}\over 8 H^8}\sum_{\{k_i\}} \left({C}_{k_1}{\cal G}^{(k_2)}_{m\alpha ab}{\cal G}^{(k_3)m\alpha ab}\right)
\left({g_s\over H}\right)^{2\Delta(k_1 + k_2 + k_3 - 2/\Delta - \gamma/2\Delta)} \nonumber\\
&-&{\eta_{\mu\nu}\over 8 H^8}\sum_{\{k_i\}}\left({\cal G}_{m\alpha p a}^{(k_1)} {\cal G}^{(k_2)m\alpha p a}\right)
\left({g_s\over H}\right)^{2\Delta(k_1 + k_2 - 1/\Delta -\gamma/2\Delta)}\nonumber\\
&+&  {4\eta_{\mu\nu}\over H^6}\sum_{\{k_i\}}\left(g^{mn}\partial_m H \partial_n H\right) C_{k_1} \widetilde{C}_{k_2}
\left({g_s\over H}\right)^{2\Delta(k_1 + k_2 - 1/\Delta)}\\
&-& {\eta_{\mu\nu}\over 16 H^8}\sum_{\{k_i\}} C_{k_1}C_{k_2}C_{k_3}C_{k_4} {\cal G}^{(k_5)}_{\alpha\beta ab}
{\cal G}^{(k_6)\alpha\beta ab}\left({g_s\over H}\right)^{2\Delta(k_1 + k_2 + k_3 + k_4 + k_5 + k_6 - 
2/\Delta - \gamma/\Delta)}, \nonumber \nd
where the various shifts of the $g_s$ scalings due to the $\gamma$ are shown above. Taking $\gamma = 2$, we see that the issue regarding the lowest order $g_s$ scaling appear here too, albeit in a more severe way. When $\gamma = 0$, the lowest order scaling of the Einstein tensor from \eqref{synecdoche} 
is $g_s^{-2}$. For $\gamma > 0$, the lowest order scaling from \eqref{synecdoche3}
becomes $g_s^{-2\Delta \omega_1}$. On the other hand, the lowest order $g_s$ scaling that can emerge from the energy-momentum tensor \eqref{keener2} is  $g_s^{-2\Delta \omega_2}$, where:
\bg\label{gdwest}
\omega_1 \equiv {\gamma + 2\over 2\Delta},  ~~~~
\omega_2 \equiv {\gamma + 2 \over \Delta} - 9,
\nd
which for $\gamma = 2$ and $\Delta = {1\over 3}$ is $g_s^{-4}$ and $g_s^{-2}$ respectively\footnote{The factor of 9 in \eqref{gdwest} appears from the minimum moding of the G-flux components ${\cal G}_{\alpha\beta ab}^{(9/2)}$ that contributes to \eqref{keener2}.}, implying that there cannot be any contributions from the energy-momentum tensor \eqref{keener2} to this order. In fact increasing $\gamma$ only worsens the problem. 

Looking at the modified form of the energy-momentum tensor from the quantum terms in \eqref{emilee}, shows that it also does not contribute terms to order $g_s^{-4}$. Therefore one of the simplest way out of this could be to demand:
\bg\label{ouletL}
\square_{(m)} H^4(y) \equiv \square_{(m)} h(y) =  0, \nd
on ${\cal M}_4$ where the Laplacian is computed using the un-warped metric $g_{mn}(y)$. As we saw before, the manifold ${\cal M}_4$ is a compact four-dimensional manifold that supports a non-K\"ahler metric. Thus $H^4(y) = h(y)$ is a harmonic function on the compact non-K\"ahler manifold 
${\cal M}_4$. The manifold ${\cal M}_2$ is conformally a torus, and the full Ricci scalar of the six-dimensional space ${\cal M}_4 \times {\cal M}_2$ is then given by:
\bg\label{bratmey}
R = {1\over 8H^4} {\cal G}^{(9/2)}_{\alpha\beta ab} {\cal G}^{(9/2)\alpha\beta ab} 
- H^4 \left[\mathbb{C}_i^i\right]^{(0, 0)} - 4\Lambda H^4, \nd
which vanishes when we take the un-warped metric of the six-dimensional space to be that of 
$K3 \times {\bf T}^2$. Additionally, the quantum terms are again classified by $\theta_k = 8/3$ from \eqref{teenangul}, with 
$\theta_k$ defined as in \eqref{miai}. Comparing this to \eqref{fleuve}, we notice a few key differences:
the brane term is absent and so are some of the G-flux components. The warp-factor is harmonic so naturally decouples out of \eqref{fleuve}. The contribution from the cosmological constant term is smaller 
 because the coefficient of the $\Lambda$ term, i.e $\sigma_2 \Lambda$, changes to:
 \bg\label{ginason}
 \sigma_2 \equiv {1\over 4}\left(8\gamma - 3\gamma^2 - 12\right). \nd 
 To the next order in $g_s$, i.e $g_s^{1/3}$, surprisingly we get exactly the same relation \eqref{urisis} that we encountered earlier despite the presence of the $\gamma$ factor (which we take as $\gamma = 2$). We expect the other coefficient to appear in a way reminiscent of \eqref{ellemey} and the story follows the path laid out for case 1. 

 Before moving to the next sub-section, let us ask if there is an alternative to the choice \eqref{ouletL}. The choice \eqref{ouletL} tells us that the warp-factor $h(y)$ is simply a harmonic function on the non-K\"ahler manifold ${\cal M}_4$, and all information of the fluxes and the quantum corrections enter indirectly.  An alternative to this choice would be to modify further the definition of the quantum energy-momentum tensor 
\eqref{emilee} by changing the $g_s$ exponent from:
\bg\label{kirumol}
{1\over \Delta} ~ \to ~ {\gamma + 2 \over 2 \Delta}, \nd
which would equate the Laplacian of the warp-factor directly to the quantum corrections at zeroth order in $g_s$. The Einstein's equation can then be realized at second order in $g_s$ equating 
\eqref{synecdoche3} with \eqref{keener2} and the quantum terms. To see how this works out, let us rewrite the quantum corrections, using the input \eqref{kirumol}, in the following way:
\bg\label{emilee2}
\mathbb{T}^Q_{\mu\nu} \equiv \sum_{\{k\}} \mathbb{C}_{\mu\nu}^{(k, 0)} 
\left({g_s\over H}\right)^{2\Delta(k - 2/\Delta)}, \nd
instead of \eqref{emilee}, where we took $\gamma = 2$ (see section \ref{instachela} for more details). This extra $\left({g_s\over H}\right)^{-4}$ suppression tells us that the warp-factor $H^4$ is no longer needed to be a harmonic function as in 
\eqref{ouletL}, rather it can now satisfy the following equation:
\bg\label{1851}
\square_{(m)} H^4 = H^8 \left[\mathbb{C}^i_i\right]^{(0, 0)}, \nd
with the quantum terms being classified by $\theta_k = {8\over 3}$ in \eqref{miai}, and therefore involve a mixture of terms in fourth powers of curvature,  eighth powers of G-fluxes or a combination of both to the relevant powers. Note that there are no G-flux contributions to this order, as we noted earlier. However once we go to the next order, i.e to order $\left({g_s\over H}\right)^{-2}$, flux contributions get poured in and the equation becomes:
\bg\label{lebanmey}
{\square_{(m)} H^4\over H^8} = {1\over \gamma_o}\left(4\Lambda + {R\over H^4} - {1\over 8H^8} 
{\cal G}^{(9/2)}_{\alpha\beta ab} {\cal G}^{(9/2)\alpha\beta ab} + \left[\mathbb{C}^i_i\right]^{(3, 0)}\right), \nd
which has some surprising similarities with \eqref{fleuve}. The similarities being the appearances of equivalent forms of curvature, fluxes and quantum terms on the RHS. However there are also few crucial differences. One, the G-flux components are not as many as in \eqref{fleuve}. Two, the coefficient of the cosmological constant term is now 4 instead of 6 before. Three, the warp-factor $H^4$ satisfy a much simpler relation 
\eqref{1851} in addition to \eqref{lebanmey}. And four, the quantum terms are classified by 
$\theta_k = {14\over 3}$ with $\left[\mathbb{C}^i_i\right]^{(3, 0)}$ instead by $\theta_k = {8\over 3}$ 
with $\left[\mathbb{C}^i_i\right]^{(0, 0)}$ in \eqref{miai}. Finally, $\gamma_o$ is given by:
\bg\label{bus80mey}
\gamma_o \equiv \sum_{\{k_i\}} C_{k_1} C_{k_2} \delta\left(k_1 + k_2 - 3\right). \nd
The question now is which of the two descriptions is the correct one. Clearly we will need more constraints to distinguish one from the other, and in section \ref{anoma} we will see that the flux EOMs provide the required constraints to justify \eqref{lebanmey}, instead of \eqref{bratmey}, to be the correct EOM for this case.

\subsubsection{Non-perturbative effects  from instantons \label{instachela}}

In the analysis of the energy-momentum tensors from the quantum terms \eqref{emilee} and \eqref{emilee2}, we find $g_s^{-2}$ and even $g_s^{-4}$ dependences respectively. It was argued therein to come from the five-brane instantons. Question then is whether these instanton effects also modify the energy-momentum tensors along the other internal directions. In the following we will argue that they do {\it contribute} to the corresponding tensors. 

Our discussion here will be brief as most of the details have appeared in \cite{coherbeta}. The non-perturbative terms that we discuss in \cite{coherbeta}, appear from the non-local counter-terms that we encountered earlier in section \ref{Gng6}, so they are not new. Their contributions may be classified as 
BBS \cite{BBS} and KKLT \cite{KKLT} type instantons, both in localized and de-localized forms. 

The {\it localized} BBS instanton type contributions come from the non-local counter-terms when we restrict the nested integral structure to the internal sub-manifold ${\cal M}_4 \times {\cal M}_2$. The energy-momentum tensors then become \cite{coherbeta}:
\bg\label{yellstop1}
&& \mathbb{T}^{({\rm np}; 1)}_{ab}(z) = \sum_k b_k \left[g_s^{4/3} + {k \over g_s^2} \left(g_s^{\theta'_k + 4/3}\right)\right]
~{\rm exp} \left(-{k \over g_s^2} \cdot g_s^{\theta'_k}\right) \nonumber\\
&& \mathbb{T}^{({\rm np}; 1)}_{\mu\nu}(z) = \sum_k b_k \left[g_s^{-8/3} + {k \over g_s^2} \left(g_s^{\theta'_k - 8/3}\right)\right]
~{\rm exp} \left(-{k \over g_s^2} \cdot g_s^{\theta'_k}\right) \nonumber\\
&& \mathbb{T}^{({\rm np}; 1)}_{mn}(z) = \mathbb{T}^{({\rm np}; 1)}_{\alpha\beta}(z)  = \sum_k b_k \left[g_s^{-2/3} + {k \over g_s^2} \left(g_s^{\theta'_k - 2/3}\right)\right]
~{\rm exp} \left(-{k \over g_s^2} \cdot g_s^{\theta'_k}\right), \nd
where $b_k$ are constants, and $\theta'_k$ is as defined in \eqref{melamon2}. We will analyze only for 
the case \eqref{olokhi}, but a similar analysis may be extended to the case \eqref{ranjhita} too. Note that the first terms in each of the energy-momentum tensors correspond to the $g_s$ scalings of the corresponding metric components, so won't concern us here (see \cite{coherbeta} for details how they are cancelled by the counter-terms). The set of equations in \eqref{yellstop1} contributes to all the energy-momentum tensors for $\theta'_k = {8\over 3}$ in the sense that it not only contributes as $g_s^{-2}$ to \eqref{emilee}, but also to \eqref{neveC} for all choice of $(M, N)$ directions.  For example it contributes to $\mathbb{C}_{mn}^{(0, 0)}$  and 
$\mathbb{C}_{\alpha\beta}^{(0, 0)}$ as $g_s^0$; and to $\mathbb{C}_{ab}^{(3, 0)}$ as $g_s^2$. Going beyond the lowest orders, the contributions from \eqref{yellstop1} effect the higher order equations in a similar way. On the other hand, the {\it de-localized} BBS type instantons contribute to the energy-momentum tensors in the following way: 
\bg\label{yellstop11}
&& \mathbb{T}^{({\rm np}; 2)}_{ab}(z) = \sum_k b_k \left[g_s^{4/3} + {k \mathbb{V}_{{\bf T}^2}\over g_s^{2/3}} \left(g_s^{\theta'_k + 4/3}\right)\right]
~{\rm exp} \left(-{k\mathbb{V}_{{\bf T}^2} \over g_s^{2/3}} \cdot g_s^{\theta'_k}\right) \\
&& \mathbb{T}^{({\rm np}; 2)}_{\mu\nu}(z) = \sum_k b_k \left[g_s^{-8/3} + {k \mathbb{V}_{{\bf T}^2} \over g_s^{2/3}} \left(g_s^{\theta'_k - 8/3}\right)\right]
~{\rm exp} \left(-{k \mathbb{V}_{{\bf T}^2}\over g_s^{2/3}} \cdot g_s^{\theta'_k}\right) \nonumber\\
&& \mathbb{T}^{({\rm np}; 2)}_{mn}(z) = \mathbb{T}^{({\rm np}; 2)}_{\alpha\beta}(z)  = \sum_k b_k \left[g_s^{-2/3} + {k \mathbb{V}_{{\bf T}^2} \over g_s^{2/3}} \left(g_s^{\theta'_k - 2/3}\right)\right]
~{\rm exp} \left(-{k \mathbb{V}_{{\bf T}^2}\over g_s^{2/3}} \cdot g_s^{\theta'_k}\right), \nonumber \nd
where the de-localization is defined as de-localization over the toroidal volume $\mathbb{V}_{{\bf T}^2}$,
which is the volume of the sub-manifold ${\mathbb{T}^2\over {\cal G}}$. The lowest order contributions to all components of the energy-momentum tensors now come from $\theta'_k = {4\over 3}$, as compared to $\theta'_k = {8\over 3}$ for the localized case. 

Let us now take the second set of instanton contributions that we will label as localized and de-localized KKLT type instanton contributions. The {\it localized} KKLT type instantons contribute as M5-branes wrapped on the six-manifold ${\cal M}_4 \times {\mathbb{T}^2\over {\cal G}}$. These instantons dualize to D3-brane instantons wrapping ${\cal M}_4$ in the IIB side and were one of the essential ingredients that went in the KKLT construction \cite{KKLT}. Their contributions to the energy-momentum tensors may be expressed as:
\bg\label{yellstop101}
&& \mathbb{T}^{({\rm np}; 3)}_{ab}(z) = \sum_k b_k \left[g_s^{4/3} + {k} \left(g_s^{\theta'_k + 4/3}\right)\right]
~{\rm exp} \left(-{k}~g_s^{\theta'_k}\right) \nonumber\\
&& \mathbb{T}^{({\rm np}; 3)}_{\mu\nu}(z) = \sum_k b_k \left[g_s^{-8/3} + {k} \left(g_s^{\theta'_k - 8/3}\right)\right]
~{\rm exp} \left(-{k} ~g_s^{\theta'_k}\right) \nonumber\\
&& \mathbb{T}^{({\rm np}; 3)}_{mn}(z) = \mathbb{T}^{({\rm np}; 3)}_{\alpha\beta}(z)  = \sum_k b_k \left[g_s^{-2/3} + {k} \left(g_s^{\theta'_k - 2/3}\right)\right]
~{\rm exp} \left(-k~g_s^{\theta'_k}\right), \nd
which tells us that the lowest order contributions come from $\theta'_k = {2\over 3}$ in \eqref{melamon2}. Interestingly the exponential factor do not go as inverse $g_s^2$ for any values of $\theta'_k$, so the exponential factor may be expanded to higher orders in $g_s$. On the other hand, the {\it de-localized} KKLT instantons contribute as:
\bg\label{yellstop111}
&& \mathbb{T}^{({\rm np}; 4)}_{ab}(z) = \sum_k b_k \left[g_s^{4/3} + {k \mathbb{V}_{2}\over g_s^{2/3}} \left(g_s^{\theta'_k + 4/3}\right)\right]
~{\rm exp} \left(-{k\mathbb{V}_{2} \over g_s^{2/3}} \cdot g_s^{\theta'_k}\right) \nonumber\\
&& \mathbb{T}^{({\rm np}; 4)}_{\mu\nu}(z) = \sum_k b_k \left[g_s^{-8/3} + {k \mathbb{V}_{2} \over g_s^{2/3}} \left(g_s^{\theta'_k - 8/3}\right)\right]
~{\rm exp} \left(-{k \mathbb{V}_{2}\over g_s^{2/3}} \cdot g_s^{\theta'_k}\right) \nonumber\\
&& \mathbb{T}^{({\rm np}; 4)}_{mn}(z) = \mathbb{T}^{({\rm np}; 4)}_{\alpha\beta}(z)  = \sum_k b_k \left[g_s^{-2/3} + {k \mathbb{V}_{2} \over g_s^{2/3}} \left(g_s^{\theta'_k - 2/3}\right)\right]
~{\rm exp} \left(-{k \mathbb{V}_{2}\over g_s^{2/3}} \cdot g_s^{\theta'_k}\right),  \nd
which is essentially similar to \eqref{yellstop11}, at least when we consider the $g_s$ scalings. The volume factor $\mathbb{V}_2$ that appears here is the volume of the sub-manifold ${\cal M}_2$, about which we define our delocalization, and the lowest order $g_s$ contributions come from $\theta'_k = {4\over 3}$, as in \eqref{yellstop11}.

For the case \eqref{ranjhita}, the contributions to the quantum energy-momentum tensors from the localized BBS type instanton gas now should behave like the contributions from the localized KKLT type instanton gas, i.e \eqref{yellstop101}, with the only difference being that the $g_s$ scaling of 
$\mathbb{T}^{({\rm np}; 3)}_{\alpha\beta}(z)$ should now equal the $g_s$ scaling of $\mathbb{T}^{({\rm np}; 3)}_{ab}(z)$ instead that of $\mathbb{T}^{({\rm np}; 3)}_{mn}(z)$. In fact this equality should hold for all cases considered in the following, unless mentioned otherwise.
On the other hand, the de-localized BBS type instanton gas behaves like \eqref{yellstop11} except $\mathbb{V}_{{\bf T}^2}$ therein is replaced by 
$g_s^2 \mathbb{V}_{{\bf T}^2}$. In a similar vein, the localized KKLT type instanton gas does behave like the KKLT type instanton gas before, namely like \eqref{yellstop101}, although for de-localized KKLT type instanton gas we need to make the replacement $\mathbb{V}_2 \to g_s^2 \mathbb{V}_2$ in 
\eqref{yellstop111}. Therefore compared to the case \eqref{olokhi}, none of the instanton gas appears to have an inverse $g_s$ dependence. 

The inverse $g_s$ dependence could appear from a different direction. Let us consider the other important ingredient in our model, namely the seven-branes. For the case \eqref{olokhi}, the seven-branes wrap the four-manifold ${\cal M}_4$ in the IIB side. In M-theory it would be a configuration of uplifted six-branes wrapping ${\cal M}_4$ and filling the remaining space-time ${\bf R}^{2, 1}$. The gauge fluxes on the seven-branes typically appear from the G-flux components ${\bf G}_{MNab}$, where $(M, N) \in {\cal M}_4$, using the decomposition \eqref{teskimey} that we will discuss later. Here it will suffice to point out the presence of the {\it localized} two-form $\Omega_{ab}$ that is one of the important ingredient in the decomposition. This two-form is defined over the sub-manifold ${\cal M}_2 \times {\mathbb{T}^2\over {\cal G}}$, and is generically a function of $(y^\alpha, y^a)$, although here it could only be a function of $y^\alpha$. 

Such consideration do not go well for the case \eqref{ranjhita} because of the derivative constraint stemming from \eqref{rogra}. Thus if we impose \eqref{rogra}, then $\Omega_{ab} = \epsilon_{ab}$, which is not a localized flux. Alternatively, we can allow seven-branes to be oriented along 
${\cal M}_2 \times \mathbb{C}_2$, where $\mathbb{C}_2$ is a two-cycle in ${\cal M}_4$. In that case $\Omega_{ab}$ could be a two-form defined on $\mathbb{C}'_2 \times {\mathbb{T}^2\over {\cal G}}$ where 
$\mathbb{C}_2 \times \mathbb{C}'_2 \in {\cal M}_4$, locally. The localized two-form can then be expressed as:
\bg\label{thaigon}
\Omega_{ab} = \sum_{n > 0} {\rm B}_n ~{\rm exp}\left(-{y^{2n}_m M^{2n}_p\over g_s^{2n/3}}\right)
\epsilon_{ab}, \nd
where $y^m \in \mathbb{C}_2$, and $\Omega_{ab}$ is small, but non-zero, if $g_s < 1$, but vanishes rapidly if $g_s \to 0$. The usefulness of \eqref{thaigon} is felt once it is plugged inside the quantum series \eqref{phingsha}: $y^m$ derivatives will bring down inverse powers of $g_s^{2/3}$ from \eqref{teskimey}, if we keep $B_n = 0$ except $B_1$. The story is then similar to what is been done in \cite{coherbeta} for the case \eqref{olokhi}. From the non-perturbative analysis for the de-localized 
BBS type instanton gas, wrapping  
$\mathbb{C}_2 \times \mathbb{C}'_2 \times {\cal M}_2$, the contribution to the energy-momentum tensor becomes:
\bg\label{yellthai}
&& \mathbb{T}^{({\rm np}; 5)}_{mn}(z) = \sum_k b_k \left[g_s^{-2/3} + {k \mathbb{V}_{{\bf T}^2} \over g_s^{2(n_2 - 2)/3}} \left(g_s^{\theta_k - 2/3}\right)\right]
~{\rm exp} \left(-{k\mathbb{V}_{{\bf T}^2} \over g_s^{2(n_2 - 2)/3}} \cdot g_s^{\theta_k}\right) \\
&& \mathbb{T}^{({\rm np}; 5)}_{\mu\nu}(z) = \sum_k b_k \left[g_s^{-8/3} + {k \mathbb{V}_{{\bf T}^2} \over g_s^{2(n_2 - 2)/3}} \left(g_s^{\theta_k - 8/3}\right)\right]
~{\rm exp} \left(-{k \mathbb{V}_{{\bf T}^2}\over g_s^{2(n_2 - 2)/3}} \cdot g_s^{\theta_k}\right) \nonumber\\
&& \mathbb{T}^{({\rm np}; 5)}_{ab}(z) = \mathbb{T}^{({\rm np}; 5)}_{\alpha\beta}(z)  = \sum_k b_k \left[g_s^{4/3} + {k \mathbb{V}_{{\bf T}^2} \over g_s^{2(n_2 - 2)/3}} \left(g_s^{\theta_k + 4/3}\right)\right]
~{\rm exp} \left(-{k \mathbb{V}_{{\bf T}^2}\over g_s^{2(n_2 - 2)/3}} \cdot g_s^{\theta_k}\right), \nonumber \nd
where $\theta_k$ is given in \eqref{miai}. The $g_s$ scalings of the metric factors appearing as the first terms in each of the energy-momentum pieces in \eqref{yellthai} are eliminated \cite{coherbeta}, so do not concern us here. The factor $n_2$ is the number of derivatives along $\mathbb{C}_2$. The exponential factor {\it do not} go to zero fast as one might think, because the torus volume $\mathbb{V}_{{\bf T}^2}$ goes to zero, so there would exist a limit where \eqref{yellthai} could contribute to the Einstein EOMs. Indeed the limit may be seen from the derivative action that brings down a factor of $M_p^2$ from \eqref{thaigon}, but there is also an inverse $M_p$ factor in the quantum terms \eqref{phingsha} due to the presence of the derivative itself. Together they give a numerical factor like $(2n_2)!$ or $M_p^{n_2}$ depending on how the derivatives are acting. The exponential term in \eqref{yellthai} then is always suppressed by a factor
${\rm exp}\left(-M_p^{n_2} \mathbb{V}_{{\bf T}^2} g_s^{\bar{\theta}}\right)$, where $\bar{\theta}$ is typically a positive integer. For large values of $n_2$, this would die off faster, and so convergence of the series may be achieved (see also \cite{coherbeta}). The non-perturbative effects from delocalized KKLT type instanton gas, if oriented appropriately, would display similar behavior as \eqref{yellthai}. 

The energy-momentum tensor for the space-time part has the necessary inverse $g_s$ dependence to account for either $g_s^{-2}$ or $g_s^{-4}$ behavior in the EOM from the Einstein tensors \eqref{synecdoche3} and \eqref{synecdoche4} or the G-flux energy-momentum tensor \eqref{keener2}. It turns out, if we take the following value for $\theta_k$:
\bg\label{padekhai}
\theta_k = {2\over 3}\left(n_2 - 1 + {n\over 2}\right), ~~~~\left(n_2, n\right) \in \left(\mathbb{Z}, \mathbb{Z}\right), \nd
with $\theta_k$ as in \eqref{miai}, then it can explain all the $g_s$ dependences along the internal directions as well as the $g_s^{-2}$ behavior along the space-time directions for the quantum terms, with $n = 0$ being the non-perturbative contributions to the lowest order EOMs. For $n = 0$, clearly $n_2 \ge 3$ for this to make sense ($n_2 = 2$ is just the perturbative expansion of the exponential factor in \eqref{yellthai}). For any $n$, $\bar{\theta} = {1\over 3}(n + 2)$, so it is indeed a positive definite integer, implying that the suppression factor in the exponential terms in \eqref{yellthai} can only come from 
$-M_p^{n_2}$ (as for any given $n$, $n_2$ can be arbitrarily large). The $g_s^{-4}$ dependence of the non-perturbative terms, on the other hand, appears for the following choice of $\theta_k$ in \eqref{miai} and for all values of $(n_2, n)$:
\bg\label{metroaangul}
\theta_k = {2\over 3}\left(n_2 - 4 + {n\over 2}\right), ~~~~ \left(n_2, n\right) \in \left(\mathbb{Z}, \mathbb{Z}\right), \nd 
which contributes, at the lowest order, to the space-time EOM. Interestingly, the contribution is now for $n_2 \ge 5$. There are two ways to proceed here: one, we can assume \eqref{metroaangul} to be the universal behavior for all $n$ i.e $ \mathbb{T}^{({\rm np}; 5)}_{MN} = g^{l_a}_s + {g_s^{\theta_k + l_a} \over g_s^4}$, where $l_a = -{2\over 3}, {4\over 3}$ or $-{8\over 3}$ depending on which direction we consider (an example being the form \eqref{emilee2}). In that case the lowest order 
contribution comes from $\theta_k = {8\over 3}$, which is for $n_2 = 8$ and $n = 0$ in \eqref{metroaangul}. The next order would appear from $n = 6$, going as $g_s^{-2}$ in this language, and will contribute for $\theta_k = {14\over 3}$. Two, we can take \eqref{padekhai} to be the universal behavior for all $n$, and define the $g_s^{-4}$ behavior with $\theta_k = {2\over 3}(n_2 - 4)$ with no dependence on $n$ anymore. Such a choice fixes $n_2 \ge 5$, and therefore $n= 0$ case in \eqref{padekhai} will get contributions for 
$\theta_k = {8\over 3}$ for $\theta_k$ as in \eqref{miai}.  From here onwards, the story progresses in the usual way.

\subsubsection{Metric cross-terms and the $F_i(t)$ factors \label{cross}} 

So far we have studied the equations of motion without worrying about the cross-terms. To complicate the matter, cross-terms in the Einstein tensor {\it do} arise because of two reasons: one,  the internal metric has time-dependent factors (i.e the functions $F_i(t)$), and two, the warp-factor $H(y)$ is in general a function of all the coordinates of ${\cal M}_4 \times {\cal M}_2$. Thus at least we expect the following 
three cross-terms:
\bg\label{chutchor}
\mathbb{G}_{0n} = -2\left({\dot{F}_1 \over F_1} + {\dot{F}_2 \over F_2}\right) {\partial_n H\over H}, 
~~~ \mathbb{G}_{0\alpha} = -4\left({\dot{F}_2 \over F_2}\right) {\partial_\alpha H\over H}, ~~~
\mathbb{G}_{\alpha m} = -{8\partial_\alpha H \partial_m H \over H^2}, \nd
with other cross-components vanishing. For the Einstein tensors $\mathbb{G}_{0n}$ and 
$\mathbb{G}_{0\alpha}$, it is easy to argue that there are no corresponding energy-momentum tensors from the G-fluxes because we do not allow $G_{mn\mu\nu}$ and $G_{m\alpha\mu\nu}$ components. Allowing them would not only add new complications to the existing EOMs studied earlier, but also break the de-Sitter isometries in the type IIB side. We want to avoid the latter, so it appears that the Einstein tensors with the cross-terms along temporal direction  will have to be balanced solely by the quantum terms. If $y^M$ denote the coordinates of ${\cal M}_4 \times {\cal M}_2$, the energy-momentum tensor associated with the quantum cross-terms may be expressed in the $g_s \to 0$ limit as:
\bg\label{neveC2}
\mathbb{T}^Q_{0M} \equiv \sum_{\{k\}} \mathbb{C}_{0M}^{(k, 0)} 
\left({g_s\over H}\right)^{2\Delta(k - 1/2\Delta)}, \nd
where the specific choice of the $g_s$ scaling is to take care of $g_s^{-1}$ pieces that may arise from 
$\dot{F}_i(t)$ in \eqref{chutchor}. Taking for example the volume preserving case \eqref{olokhi}, it is easy to see where the $g_s^{-1}$ factor appear from. The Einstein tensors become:
\bg\label{madhubala}
&&\mathbb{G}_{0\alpha} = -8\Delta \sqrt{\Lambda} \left({\partial_\alpha H\over H}\right)
\sum_{\{k_i\}} k_1 C_{k_1} C_{k_2} \widetilde{C}_{k_3}
\left({g_s\over H}\right)^{2\Delta(k_1 + k_2 + k_3 - 1/2\Delta)} \nonumber\\
&& \mathbb{G}_{0n} = -4\Delta \sqrt{\Lambda} \left({\partial_n H\over H}\right)
\sum_{\{k_i\}} (k_1+ k_2) \widetilde{C}_{k_1} C_{k_2} {C}_{k_3}
\left({g_s\over H}\right)^{2\Delta(k_1 + k_2 + k_3 - 1/2\Delta)}, \nd
with the $g_s$ scaling showing the inverse factor, alluded to above, which we can easily get rid of by multiplying all the tensors in \eqref{madhubala} and \eqref{neveC2} by $g_s$. To zeroth order in $g_s$ there are no contributions from either \eqref{madhubala} or \eqref{neveC2}. To next order in $g_s$, i.e 
$g_s^{1/3}$, we get:
\bg\label{babycasin}
C_{1\over 2} = - {\mathbb{C}_{0\alpha}^{(1/2, 0)}\over 12\sqrt{\Lambda}} 
\left({\partial_\alpha H\over H}\right)^{-1} = {\mathbb{C}_{0n}^{(1/2, 0)}\over 6 \sqrt{\Lambda}} 
\left({\partial_n H\over H}\right)^{-1}, \nd
which should be compared to \eqref{ellemey}. The above set of Einstein tensors provide a much easier way to get the $C_k$ and $\widetilde{C}_k$ coefficients of the $F_i(t)$ functions. Expectedly, they are related to the quantum terms, so classically we can only see time-independent internal space. The latter has problems with EFT as we saw before and also in \cite{nogo, nodS}.  

Switching on the $\gamma$ factor to study the case \eqref{ranjhita} or \eqref{ranjhita3} eliminates 
$\mathbb{G}_{0\alpha}$ and $\mathbb{G}_{\alpha m}$ because of the derivative constraint. This only leaves $\mathbb{G}_{0n}$ which takes the following form:
\bg\label{seli}
\mathbb{G}_{0n} =  -4\Delta \sqrt{\Lambda} \left({\partial_n H\over H}\right)
\sum_{\{k_i\}} \left(k_1+ k_2 + {\gamma \over 2\Delta}\right) \widetilde{C}_{k_1} C_{k_2} {C}_{k_3}
\left({g_s\over H}\right)^{2\Delta(k_1 + k_2 + k_3 - 1/2\Delta)}, \nd
which now does allow a term to the zeroth order in $g_s$. By ignoring the $g_s^{-1}$ piece for the time being $-$ to be reconciled later using the same line of thought as before $-$ the zeroth order in $g_s$ yields the following relation for the quantum term:
\bg\label{phonemar}
\mathbb{C}_{0n}^{(0, 0)} = - 4\sqrt{\Lambda} \left({\partial_n H\over H}\right), \nd
which, once combined with \eqref{ouletL}, should determine the functional form of the quantum term when we take $\gamma = 2$. Going to the next order in $g_s$, i.e $g_s^{1/3}$, we get exactly the same relation that we have in \eqref{babycasin}, i.e:
\bg\label{marielou}
C_{1\over 2} = {\mathbb{C}_{0n}^{(1/2, 0)}\over 6 \sqrt{\Lambda}} 
\left({\partial_n H\over H}\right)^{-1}. \nd
All these appear to lead to some consistent formulation of the background data, although there is one puzzle that we have kept under the rug so far. This has to do with the computation of the quantum energy-momentum tensor \eqref{neveC2}. How do we interpret this term? If we follow the definition of the energy-momentum tensor in 
\eqref{ryacone}, then the {\it absence} of ${\bf g}_{0n}$ should tell us that one cannot construct the cross-term energy-momentum tensor at all. In fact even the formulation of the Einstein tensor comes under scrutiny now. 

The key point that we are missing here is the Wilsonian viewpoint that we already emphasized earlier (see the discussions between \eqref{ducksoup} and \eqref{lengchink}). The background that we consider should contain all the components of metric and fluxes and we integrate out all the ones that would potentially ruin the four-dimensional de Sitter isometries in the type IIB side. This amounts to integrating out specific components of metric and G-fluxes in the M-theory side, leading to an effective action. In the following, let us see how this works when we integrate out one component of the metric, say ${\bf g}_{0n}$. We define:
\bg\label{kabir}
{\rm exp}\left(-iS_{\rm eff}\right) &\equiv& \int {\cal D}{\bf g}_{0n} ~{\rm exp}\Bigg[-i\int d^{11} x 
\sqrt{{\bf g}_{11}({\bf g}_{0n})}\left({\bf R}^{(11)} - {\bf g}^{0n} \mathbb{T}^G_{0n} 
- {\bf g}^{0n} \mathbb{T}^Q_{0n} + ...\right)\Bigg], \nonumber\\  \nd
where the dots denote terms that are independent of ${\bf g}_{0n}$, and the bold-faced components are defined with respect to the warped metric. Since ${\bf g}_{0n}$ is a dummy variable, we can re-define this
to ${\bf g}'_{0n}$
without changing the effective action $S_{\rm eff}$. Taking 
${\bf g}'_{0n} = {\bf g}_{0n} + {\bf h}_{0n}$, where ${\bf h}_{0n}$ is a small shift of the metric component, does not change the measure. This leads us to:

{\footnotesize
\bg\label{kristian} 
{\rm exp}\left(-iS_{\rm eff}\right) &\equiv&  \int {\cal D}{\bf g}'_{0n} ~{\rm exp}\Bigg[-i\int d^{11} x 
\sqrt{{\bf g}_{11}({\bf g}'_{0n})}\left({\bf R}^{(11)}({\bf g}'_{0n}) - {\bf g}'^{0n} \mathbb{T}^G_{0n} 
- {\bf g}'^{0n} \mathbb{T}^Q_{0n} - \mu^2 {\bf g}'^{0n} {\bf g}'_{0n} + ...\right)\Bigg] \nonumber\\ 
& = & \int {\cal D}{\bf g}_{0n} ~{\rm exp}\Bigg[-i\int d^{11} x 
\sqrt{{\bf g}_{11}({\bf g}_{0n})}\left({\cal L}_0({\bf g}_{0n})
+  {\bf h}^{0n} \left(\mathbb{R}_{0n} - {1\over 2} {\bf g}_{0n} {\bf R} - 
\mathbb{T}^G_{0n} - \mathbb{T}^Q_{0n}\right)+ ...\right)\Bigg], \nonumber\\ \nd}
where in the second line we have expanded to first order in ${\bf h}_{0n}$ to express the factor involving
Ricci tensor. We have also inserted a small mass to the graviton so as to integrate this out.
Note that ${\bf g}_{0n}$ does show up with a coefficient ${\bf h}^{0n}$, and we have defined: 
\bg\label{lizshort}
\mathbb{R}_{0n} \equiv {\hat{\bf R}}_{0n}({\bf g}_{0n}) + {\bf R}_{0n}, \nd
where only ${\hat{\bf R}}_{0n}$ is a function of ${\bf g}_{0n}$. Therefore,
neither ${\bf R}_{0n}$ nor the energy-momentum tensors are functions of ${\bf g}_{0n}$. 
For the latter we could have divided into a piece that depends on ${\hat{\bf R}}_{0n}$, i.e indirectly on 
${\bf g}_{0n}$, and a piece independent of ${\bf g}_{0n}$;
but since we are eventually going to integrate out ${\bf g}_{0n}$, their presence or absence will not change much the generic 
quantum term \eqref{phingsha} or \eqref{phingsha2}. Finally, 
the Lagrangian
${\cal L}_0({\bf g}_{0n})$ is defined as:
\bg\label{anuradha}
{\cal L}_0({\bf g}_{0n}) = {\bf R}^{(11)}({\bf g}_{0n}) - {\bf g}^{0n} \mathbb{T}^G_{0n} 
- {\bf g}^{0n} \mathbb{T}^Q_{0n} - \mu^2 {\bf g}^{0n} {\bf g}_{0n}. \nd 
The above equation, \eqref{kristian}, combined with \eqref{anuradha}, is  a form of the Schwinger-Dyson equation for our case, but is presented in a slightly different way because we want to integrate out 
${\bf g}_{0n}$. Doing this leads us to the following two conclusions. One, we recover the terms with polynomial powers of $\left(\mathbb{T}^G_{0n}\right)^2$ and $\left(\mathbb{T}^Q_{0n}\right)^2$ (along-with the mixed terms). These are of course contained in \eqref{phingsha} and \eqref{phingsha2} according to 
\eqref{mcgillmey}: a consequence of the semi-group structure of the system. Two, ${\bf g}_{0n}$ appears inside the bracket multiplying ${\bf h}^{0n}$. This means, once we integrate out ${\bf g}_{0n}$, there would be terms with powers of ${\bf h}^{0n}$ accompanied with the combination of the Ricci curvature 
${\bf R}_{0n}$ and the energy-momentum tensors $\mathbb{T}^G_{0n}$ and $\mathbb{T}^Q_{0n}$,
{\it without} the ${\bf g}_{0n} {\bf R}$ piece.
We also expect the effective action $S_{\rm eff}$  to be independent of any arbitrary parameter like 
${\bf h}^{0n}$. Combining everything together it appears that if we demand at ``on-shell" the following two conditions: 
${\bf g}_{0n} = 0$ and
\bg\label{dipmarie}
{\bf R}_{0n} - \mathbb{T}^G_{0n}  - \mathbb{T}^Q_{0n} = 0, \nd
then there is a well defined effective action $S_{\rm eff}$, with 
 the latter reproducing the expected EOM for the cross-term. Notice that none of the terms in 
 \eqref{dipmarie} can depend on ${\bf g}_{0n}$, because of the procedure that we have adopted to derive the equations and the effective action. In retrospect this is of course consistent with what we have been considering so far. 

The short analysis presented above reveals one crucial fact: we can allow energy-momentum tensors of the form $\mathbb{T}^G_{0n}$ and $\mathbb{T}^Q_{0n}$ even if cross-components of the metric, like 
${\bf g}_{0n}$, do not appear in the background. The point is that it is not necessary for certain components of the metric (or G-flux) to physically appear as long as they appear inside quantum {\it loops}. The Wilsonian way of course guarantees this by allowing a small mass to these components that would facilitate their {\it off-shell} appearances. Such a line of thought does lead to consistent picture as we saw from all our earlier analysis, however one question still lingers: how do we  actually determine  the $g_s$ scalings for these cross-component energy-momentum tensors?   

This can be answered using a simple trick. For concreteness let us consider the quantum series 
\eqref{phingsha2} meant for the volume preserving case \eqref{olokhi}. Before we contract this completely with inverse metric components, let us insert a function $t_{0n}$ with the property $t^{0m}t_{0n} = 
\delta^m_n$ as $(t_{0n})^{l_{39}}$ in \eqref{phingsha2}, where $l_{39}$ can take values ($0, 1$) only.
 We can now put back all the inverse metric components to make it Lorentz invariant. We can also assume that $t_{0n}$ has no $g_s$ scaling, i.e it scales as $g_s^0$. The $g_s$ scaling of the modified 
\eqref{phingsha2} now becomes $\hat{\theta}'_k$ where:
\bg\label{HofSm}
\hat{\theta}'_k \equiv \theta'_k + \left({5\over 3} -{\gamma\over 2}\right)l_{39}, \nd
with $\theta'_k$ as defined in \eqref{melamon2} and we have inserted $\gamma$ just for the completeness sake (as $\gamma$ should have been inserted with $\theta_k$ in \eqref{miai}). To extract an expression with one free $0$ index and one free $n$ index, to account for the energy-momentum tensor $\mathbb{T}^Q_{0n}$, all we need is to {\it remove} one ${\bf g}^{00}$ and one ${\bf g}^{nn}$ metric components to create two free indices anywhere inside the modified quantum terms 
\eqref{phingsha2}. This will change the $g_s$ scaling from \eqref{HofSm} to $\widetilde{\theta}'_k$, where:  
\bg\label{kototagmey}
\widetilde{\theta}'_k \equiv \theta'_k + \left({5\over 3} - {\gamma\over 2}\right)l_{39} - {10\over 3}, \nd
with $\theta'_k$ as in \eqref{melamon2}. If we replace $\theta'_k$ in \eqref{kototagmey} by $\theta_k$ 
of \eqref{miai}, we get the result for \eqref{ranjhita}. Finally, contracting the resulting expression with 
$t^{0m}$ will give us the required expression for $\mathbb{T}^Q_{0m}$ with $g_s$ scaling as in 
\eqref{kototagmey} and $l_{39} = 1$. Clearly for vanishing $\gamma$ and concentrating on the perturbative terms for the time being, the $g_s$ scaling is $\theta'_k - 5/3$, whereas for $\gamma = 2$ we get $\theta_k - 8/3$ representing the two cases \eqref{olokhi} and 
\eqref{ranjhita} respectively. Our $g_s$ scaling for the quantum terms in \eqref{neveC2} for \eqref{olokhi} should be interpreted in the following way:
\bg\label{casin3}
g_s^{\theta'_k - 5/3} ~\equiv~ g_s^0, ~~g_s^{1/3}, ~~g_s^{2/3}, ~~ g_s, ....., \nd
so that the zeroth order terms are classified by $\theta'_k = 5/3$ in \eqref{melamon2}. Similarly for 
\eqref{ranjhita}, the zeroth order terms are classified by $\theta_k = 8/3$ in \eqref{miai}. As we saw above, the latter do contribute so that $\mathbb{C}_{0n}^{(0, 0)}$ are classified as above for the case 
\eqref{ranjhita}. However for the volume preserving case, i.e \eqref{olokhi}, the first non-trivial contributions come from $\mathbb{C}_{0n}^{(1/2, 0)}$ and $\mathbb{C}_{0\alpha}^{(1/2, 0)}$. They are classified by 
$\theta'_k = 2$ in \eqref{melamon2}. In a similar vein one could analyze the $\mathbb{G}_{\alpha m}$ 
equations for the volume preserving case \eqref{olokhi}, and even include the contributions from the non-perturbative sector as detailed in subsection \ref{instachela}. We will not elaborate them here and instead analyze the consequence of the perturbative and the non-perturbative corrections on de Sitter solution.

\subsubsection{de Sitter solution from the quantum constraints \label{kocu5}}

\noindent In the above sections we managed to assimilate all the possible quantum corrected  EOMs that can occur in the system. Many subtleties regarding the distribution of the quantum terms were noticed, but in the end the arrangement of the these terms reflected a certain level of consistencies that were expected in set-up like ours and also of our earlier works \cite{nogo, nodS} with one noticeable difference: the quantum terms could now be precisely classified using the scaling \eqref{melamon2} for \eqref{olokhi} and 
\eqref{miai} for \eqref{ranjhita}. Thus the issue of the existence of effective field theories could now be answered in the affirmative provided the EOMs themselves have solutions. In the following therefore we would like to analyze this for the two cases in question.

\vskip.1in

\noindent {\it Case 1: $F_1(t)$ and $F_2(t)$ satisfying the volume-preserving condition \eqref{olokhi}}

\vskip.1in
 
\noindent We start by analyzing the volume-preserving case \eqref{olokhi}, by first taking the traces of all the EOMs to lowest order in $g_s$ and try to find if certain consistency condition(s) could be generated. Our first equation is for the ($m, n$) directions. 
In the zeroth order in $g_s$, 
the equation is given in \eqref{misslemon2}, which is constructed using un-warped metric and G-flux components. Taking a trace of this equation yields:
\bg\label{lori}
R^{(4)} - 2R - 24 H^4 \Lambda = \left[\mathbb{C}_m^m\right]^{(0, 0)}  
-{1\over 4H^4} \Big({\cal G}^{(3/2)}_{m\alpha ab}{\cal G}^{(3/2)m\alpha ab} + 
{\cal G}^{(3/2)}_{\alpha\beta ab}{\cal G}^{(3/2)\alpha\beta ab}\Big), \nd
where $R^{(4)}$ is the Ricci scalar for the four-dimensional manifold ${\cal M}_4$ and $R$ remains the Ricci scalar of the full six-dimensional base ${\cal M}_4 \times {\cal M}_2$. As mentioned above, both are computed using un-warped metric components, including the traces unless mentioned otherwise.

The quantum terms $\left[\mathbb{C}_m^m\right]^{(0, 0)}$ are classified by $\theta'_k = 2/3$ in 
\eqref{melamon2} and one may easily see that with such a small value for $\theta'_k$ there are only a few classical terms mostly made of G-fluxes. The classical terms can only renormalize the existing terms that we have from the energy-momentum tensor for the G-fluxes. However as we saw in section \ref{instachela}, there are also {\it non-perturbative} contributions to the EOM. They are typically classified by $\theta'_k \le {8\over 3}$ from the BBS and KKLT type instantons, as shown in \eqref{yellstop1}, \eqref{yellstop11}, \eqref{yellstop101} and \eqref{yellstop111}. In fact an exactly similar story unfolds for the EOM along the ($\alpha, \beta$) directions. Taking the trace of \eqref{uanaban}, written for the zeroth order in 
$g_s$, we get:
\bg\label{maria}
R^{(2)} - R - 12 \Lambda H^4 = \left[\mathbb{C}_\alpha^\alpha\right]^{(0, 0)}  
+{1\over 8H^4} \Big({\cal G}^{(3/2)}_{\alpha\beta ab}{\cal G}^{(3/2)\alpha\beta ab} - 
{\cal G}^{(3/2)}_{mn ab}{\cal G}^{(3/2)mn ab}\Big), \nd
where $R^{(2)}$ is the un-warped curvature of ${\cal M}_2$, and since ${\cal M}_2$ is a non-K\"ahler 
two-dimensional space, this does not vanish. The quantum terms 
$\left[\mathbb{C}_\alpha^\alpha\right]^{(0, 0)}$ are again classified perturbatively by $\theta'_k = 2/3$ in 
\eqref{melamon2}, thus renormalizing the existing classical terms, and non-perturbatively by $\theta'_k \le {8\over 3}$. Compared to 
\eqref{lori}, the relative factors, signs and G-flux components differ but the main message of \eqref{maria} remains similar to \eqref{lori}. 

The next set of equations are a bit different from what we had so far and the differences appear mostly from the scalings of the quantum terms. For example looking at the EOM for the ($a, b$) direction, i.e.
\eqref{buskaM} appearing to order $g_s^2$ instead of the expected zeroth order in $g_s$, and taking the trace, we get:
\bg\label{emmaS}
R + 18 \Lambda H^4 = -\left[\mathbb{C}_a^a\right]^{(3, 0)}  
-{1\over 8H^4} \Big(2{\cal G}^{(3/2)}_{m\alpha ab}{\cal G}^{(3/2)m\alpha ab} +
{\cal G}^{(3/2)}_{mn ab}{\cal G}^{(3/2)mn ab} 
+ {\cal G}^{(3/2)}_{\alpha\beta ab}{\cal G}^{(3/2)\alpha\beta ab} 
\Big), \nonumber\\ \nd 
where now we see that the quantum terms have different modings than what we had in \eqref{lori} and
\eqref{maria}. However they are still classified perturbatively by $\theta'_k = 2/3$ in \eqref{melamon2}, and therefore renormalizing the existing classical terms, and non-perturbatively by $\theta'_k \le {8\over 3}$. This shared similarities between the three traces, \eqref{lori}, 
\eqref{maria} and \eqref{emmaS}, {\it do not} imply that the quantum effects are relatively unimportant because we haven't yet analyzed the space-time EOMs. All the EOMs are inter-related  so conclusions based on analyzing only parts of the story typically fail to reveal the true picture. 

This becomes clear once we look at the space-time EOMs. Looking at the zeroth order in $g_s$ 
in \eqref{fleuve} we notice that the quantum effects play an equally important role as above. To facilitate discussion, let us quote \eqref{fleuve} again:
\bg\label{fleuve2}
&&  6\Lambda + {R\over H^4} - {\square H^4\over H^8} + \left[\mathbb{C}^i_{i}\right]^{(0, 0)}   
- {2 \kappa^2 T_2 \left(n_b + \bar{n}_b\right) \over H^8 \sqrt{g_6}} \delta^8(y - Y)\nonumber\\
&& = {1\over 8 H^8} \Big({\cal G}^{(3/2)}_{mnab}{\cal G}^{(3/2)mnab} +
2 {\cal G}^{(3/2)}_{m\alpha ab}{\cal G}^{(3/2)m\alpha ab} +  {\cal G}^{(3/2)}_{\alpha\beta ab}
{\cal G}^{(3/2)\alpha\beta ab}\Big),  
 \nd
where $\square$ is now over the full six-dimensional space ${\cal M}_4 \times {\cal M}_2$, $(n_b, \bar{n}_b)$ are the number of M2 and $\overline{\rm M2}$-branes; and the quantum terms are again classified by $\theta'_k \le 8/3$ in \eqref{melamon2}, compared to similar cases for the three traces above. Such choices of $\theta'_k$ for the the metric EOMs will allow a large number of terms by choosing various combinations of $l_i$ in \eqref{phingsha2}, thus mixing curvature terms with the G-flux components. 

All the four equations above shows how the Ricci scalar $R$ may be related to the G-fluxes and the quantum terms. The quantum terms are shown to be classified by choosing appropriate values for 
$\theta'_k$ in \eqref{melamon2}, but there are additional non-local contributions to them. Fortunately, in the limit of vanishing ($a, b$) torus these contributions are negligible so may be avoided in the $g_s \to 0$ limit, i.e in the late time limit. Adding \eqref{lori} and \eqref{maria} we get:
\bg\label{lp777}
R + 18 H^4\Lambda &=& -{1\over 2} \left[\mathbb{C}_m^m \right]^{(0, 0)} 
-{1\over 2} \left[\mathbb{C}_\alpha^\alpha\right]^{(0, 0)} \\
&+& {1\over 16 H^4} 
\Big({\cal G}^{(3/2)}_{\alpha\beta ab}{\cal G}^{(3/2)\alpha\beta ab} + 
2 {\cal G}^{(3/2)}_{m\alpha ab}{\cal G}^{(3/2)m\alpha ab} + 
{\cal G}^{(3/2)}_{mnab}{\cal G}^{(3/2)mnab}\Big), \nonumber \nd
which, in the absence of the G-flux pieces, would be equivalent to a similar equation in \cite{nogo} for the time-independent internal space (see eq. (6.4) in \cite{nogo}). It is reassuring to see the emergence of familiar equations once we resort to the time-independent scenario. The time-dependences therefore not only add new fluxes to the time-independent equations, but also allows us to consider a controlled set of quantum corrections. Interestingly, now looking at \eqref{emmaS}, we notice that the LHS is identical to the 
LHS of \eqref{lp777}. In the absence of the G-flux pieces, we could have concluded that the quantum corrections in these two set of equations are related to each other; much like eq. (6.6) of \cite{nogo}. This is {\it not} the case now. The quantum corrections along ($a, b$) directions are not related in a simple way to the sum of the quantum corrections along ($m, n$) and ($\alpha, \beta$) directions. The G-fluxes interfere to make this a bit more involved. We could however add \eqref{lp777} and \eqref{emmaS} to get the following 
equation:
\bg\label{lp7fasa}
R + 18 H^4\Lambda &=& -{1\over 2} \left[\mathbb{C}_a^a \right]^{(3, 0)} 
-{1\over 4} \left[\mathbb{C}_m^m \right]^{(0, 0)} 
-{1\over 4} \left[\mathbb{C}_\alpha^\alpha\right]^{(0, 0)} \\
&-& {1\over 32 H^4} 
\Big({\cal G}^{(3/2)}_{\alpha\beta ab}{\cal G}^{(3/2)\alpha\beta ab} + 
2 {\cal G}^{(3/2)}_{m\alpha ab}{\cal G}^{(3/2)m\alpha ab} + 
{\cal G}^{(3/2)}_{mnab}{\cal G}^{(3/2)mnab}\Big), \nonumber \nd
combining all the quantum terms and the G-fluxes together. Note the difference in the moding of the 
($a, b$) quantum terms, but as mentioned earlier, they are all classified by 
$2/3 \le \theta'_k \le {8\over 3}$ in 
\eqref{melamon2}. Since $\theta'_k = 2/3$ is almost classical (one may easily see by choosing the appropriate $l_i$ in \eqref{phingsha2}), all they do here is to renormalize the existing classical pieces without introducing any higher order corrections. The more interesting parts are the $\theta'_k \le {8\over 3}$. They contribute non-trivial higher order quantum corrections to the all the Einstein EOMs, but more sigificantly, they are all countably {\it finite}.  This was clearly not the case in \cite{nogo, nodS}, where 
$\theta'_0 \le {8\over 3}$ in \eqref{kkkbkb2} would have led to an infinite number of quantum terms without any visible hierarchies. Switching on time-dependences have completely changed the scenario. On the other hand, subtracting \eqref{lp777} from \eqref{emmaS}, we get:

{\footnotesize
\bg\label{lulu}
\left[\mathbb{C}_m^m \right]^{(0, 0)} 
+ \left[\mathbb{C}_\alpha^\alpha\right]^{(0, 0)} -2 \left[\mathbb{C}_a^a \right]^{(3, 0)} 
= {3\over 8 H^4} 
\Big({\cal G}^{(3/2)}_{\alpha\beta ab}{\cal G}^{(3/2)\alpha\beta ab} + 
2 {\cal G}^{(3/2)}_{m\alpha ab}{\cal G}^{(3/2)m\alpha ab} + 
{\cal G}^{(3/2)}_{mnab}{\cal G}^{(3/2)mnab}\Big), \nonumber\\ \nd}
which instead would directly connect the quantum terms to the fluxes. Such an equation leads to the following interesting observation. All the quantum pieces appearing above are classified by 
${2\over 3} \le \theta'_k \le {8\over 3}$. The appearance of the flux factors on the RHS of \eqref{lulu} immediately confirms the fact that the sum of the higher order quantum terms in \eqref{lori} and \eqref{maria} is twice the higher order quantum terms in \eqref{emmaS}, and therefore exactly {\it cancels} out in the combination \eqref{lulu}. The remaining quantum pieces, classified by $\theta'_k = {2\over 3}$, only renormalize the existing classical
data, and therefore the RHS of \eqref{lulu} has only contributions from the G-flux components.

We can now use the curvature scalar, defined in terms of the quantum terms for the eight-dimensional manifold and the G-fluxes in \eqref{lp7fasa}, and plug this \eqref{fleuve2}. Doing this yields:

{\footnotesize
\bg\label{muse777}
- {\square H^4} & = & 12 \Lambda H^8 + {5\over 32} \Big({\cal G}^{(3/2)}_{mnab}{\cal G}^{(3/2)mnab} +
2 {\cal G}^{(3/2)}_{m\alpha ab}{\cal G}^{(3/2)m\alpha ab} + 
{\cal G}^{(3/2)}_{\alpha\beta ab}{\cal G}^{(3/2)\alpha\beta ab}\Big)\\
& + & {2 \kappa^2 T_2 \left(n_b + \bar{n}_b\right) \over \sqrt{g_6}} \delta^6(y - Y)
+ \left({1\over 2} \left[\mathbb{C}_a^a \right]^{(3, 0)} 
+{1\over 4} \left[\mathbb{C}_m^m \right]^{(0, 0)} 
+{1\over 4} \left[\mathbb{C}_\alpha^\alpha\right]^{(0, 0)} - H^4 \left[\mathbb{C}^i_{i}\right]^{(0, 0)}\right)H^4,   
 \nonumber \nd}
where we have made one change: the M2 and the $\overline{\rm M2}$-branes are now restricted to move on the six-dimensional base 
${\cal M}_4 \times {\cal M}_2$ only as this will facilitate as easier interpretation in the type IIB side.  Note also that the only {\it minus} sign appears from the quantum terms in the space-time directions. This equation is somewhat similar to eq. (6.8) in \cite{nogo}. The differences being in (a) the relative factors, (b) the choice of the G-flux components and (c) the dependence on the full eight-dimensional coordinates instead of only on the six-dimensional base here; but both equations share one similarity regarding the appearance of the relative minus sign. This is crucial because integrating \eqref{muse777} over the six-dimensional base gives us:

{\footnotesize
\bg\label{musemey}
&&12 \Lambda\int d^6 y \sqrt{g_6} H^8 + {5\over 32} \int d^6 y \sqrt{g_6}
\Big({\cal G}^{(3/2)}_{mnab}{\cal G}^{(3/2)mnab} +
2 {\cal G}^{(3/2)}_{m\alpha ab}{\cal G}^{(3/2)m\alpha ab} + 
{\cal G}^{(3/2)}_{mnab}{\cal G}^{(3/2)mnab}\Big)\nonumber\\
& & + 2 \kappa^2 T_2 (n_b + \bar{n}_b)
+ \int d^6 y \sqrt{g_6}\left({1\over 2} \left[\mathbb{C}_a^a \right]^{(3, 0)} 
+{1\over 4} \left[\mathbb{C}_m^m \right]^{(0, 0)} 
+{1\over 4} \left[\mathbb{C}_\alpha^\alpha\right]^{(0, 0)} - H^4 [\mathbb{C}^i_{i}]^{(0, 0)}\right)H^4
= 0,\nd}
which should be compared to eq. (6.10) of \cite{nogo}. The zero on the RHS appears from integrating 
$\square H^4$ over the compact base ${\cal M}_4 \times {\cal M}_2$, and since $H^4(y) \equiv h(y)$ is a smooth function, the integral vanishes. The smoothness of $H^4(y)$ is guaranteed from the series of quantum corrections appearing in \eqref{muse777}. 
Clearly, in the absence of the quantum pieces, the system has no solution because the integral involves only positive definite functions and therefore the consistency will demand vanishing fluxes and vanishing 
$\Lambda$. Interestingly {\it negative} $\Lambda$ is allowed even if the quantum terms are absent, implying both Minkowski  and AdS spaces may be realized in a set-up like ours. The recent swampland conjectures concerning AdS spaces may be overcome by introducing back the quantum corrections, but we don't want to discuss this here. In the presence of the quantum pieces, the consistency condition here differs in a crucial way with the one presented in \cite{nogo}. The quantum terms in 
\cite{nogo} are classified by $2/3 \le \theta'_0 \le 8/3$ for the internal and the space-time directions with $\theta'_0$ defined in \eqref{kkkbkb2}. These have infinite number of solutions for both cases, from the local and the non-local quantum terms,
implying that an expression like eq. (6.10) in \cite{nogo} does not have any solution at all and is in the swampland\footnote{The fact that {\it both} $g_s$ and $M_p$ hierarchies are lost  in this case may be seen from \cite{petite}.}. However now the scenario has changed. The internal and the space-time quantum terms are now classified by $2/3 \le \theta'_k  \le 8/3$  with $\theta'_k$ defined as in 
\eqref{melamon2}. These have {\it finite} number of solutions in both cases, and in fact the non-perturbative internal space quantum terms, as we saw in \eqref{lulu}, are related in a special way. This means, if  
${2\over 3} < \theta'_k \le {8\over 3}$ contributions to $\left[\mathbb{C}_a^a \right]^{(3, 0)}$ be 
$\mathbb{Q}_a^{({\rm np})}$, then by \eqref{lulu}, the non-perturbative contributions to 
$\left[\mathbb{C}_m^m \right]^{(0, 0)} + \left[\mathbb{C}_\alpha^\alpha \right]^{(0, 0)}$ would be 
$2\mathbb{Q}_a^{({\rm np})}$. On the other hand, let the quantum contributions for $\theta'_k \le {8\over 3}$ 
to $\left[\mathbb{C}_i^i \right]^{(0, 0)}$ be $\mathbb{Q}_i^{({\rm np})}$. 
 These quantum terms appear with an overall {\it minus} sign in \eqref{musemey}, and therefore as long as 
 $H^4 \mathbb{Q}_i^{({\rm np})}$ dominates over other values in \eqref{musemey},
then surprisingly solutions would exist where there were none before!

In section \ref{maryse} we will study the EOM for the G-flux components, and in \eqref{evaB102} we relate the warp-factor $H(y)$ to the number of M2 and $\overline{\rm M2}$-branes. The appearance of 
$n_b$ and $\bar{n}_b$, i.e the number of M2 and $\overline{\rm M2}$-branes respectively, in 
\eqref{muse777} and \eqref{evaB102} is important: there is a relative sign difference in \eqref{evaB102} compared to the one in \eqref{muse777} ($2\kappa^2 = 1$ in \eqref{evaB102}). The reason is simple. 
The flux EOM captures the charges, whereas the Einstein EOM captures the energy-momentum tensors 
of the branes and the anti-branes. The equation \eqref{evaB102} differs from an equivalent flux equation in 
\cite{nodS}, because of the absence of quantum and flux factors, but there exists a related equation, \eqref{evaBgon}, that is very similar to \eqref{muse777}. Once we subtract \eqref{muse777} from \eqref{evaBgon}, 
we can easily get the supersymmetry breaking condition \eqref{evebe}, as a non self-duality condition on the fluxes.

The details gathered so far will help us to determine the metric of the internal space in terms of the fluxes and the quantum corrections. For example, let us start by expressing the un-warped metric $g_{mn}$
using \eqref{rambha3} in the following way:

{\footnotesize
\bg\label{rambhafin}
g_{mn} = {3\over 58}\left[{\mathbb{C}^{(1/2, 0)}_{mn} + 
 {1\over 4H^4} \sum_{\{k_i\}}C_{k_2}\left({\widetilde{C}_{k_1}}
{\cal G}^{(k_3)}_{mlab} {\cal G}_n^{(k_4)lab} 
+{{C}_{k_1}}
{\cal G}^{(k_3)}_{m\alpha ab} {\cal G}_n^{(k_4)\alpha ab}\right) \over  \mathbb{A}(y)
+ {3 \over 928 H^4} \sum_{\{k_i\}}C_{k_2}\left({\widetilde{C}_{k_1}}
{\cal G}_{pkab}^{(k_3)} {\cal G}^{(k_4)pkab}
+ 2C_{k_1} {\cal G}_{p\alpha ab}^{(k_3)} {\cal G}^{(k_4)p\alpha ab}\right)}\right], \nd}
 where $\mathbb{A}(y)$ is defined in \eqref{ramravali} and $k_i$ satisfy $\sum_i k_i = 7/2$, with the constraint that $(k_3, k_4) \ge (3/2, 3/2)$. The $C_k$ and the $\widetilde{C}_k$ coefficients can be determined using the cross-term EOMs as we saw in section \ref{cross}. Finally, the quantum terms appearing above are governed by $1 \le \theta'_k \le {3}$ in \eqref{melamon2}. For the lower limit of $\theta'_k$, the quantum terms are mostly expressed as powers of G-flux components instead of curvature tensors as may be easily seen from \eqref{melamon3}. The higher powers of curvature tensors and G-flux components  start appearing for $\theta'_k \le 3$. This means the RHS of \eqref{rambhafin} is expressed mostly by powers of G-fluxes, curvature tensors and the ($C_k, \widetilde{C}_k$) coefficients (the latter are also 
 determined by fluxes for small values of $k$). In fact a somewhat similar story repeats for the metric 
 component $g_{\alpha\beta}$ also, which now takes the following form:
 
 {\footnotesize
\bg\label{rambhafin2}
g_{\alpha\beta} = {9\over 2}\left[{\mathbb{C}^{(1/2, 0)}_{\alpha\beta} + 
 {1\over 4H^4} \sum_{\{k_i\}}C_{k_2}\left({\widetilde{C}_{k_1}}
{\cal G}^{(k_3)}_{\alpha lab} {\cal G}_\beta^{(k_4)lab} 
+{{C}_{k_1}}
{\cal G}^{(k_3)}_{\alpha\gamma ab} {\cal G}_\beta^{(k_4)\gamma ab}\right) \over  \mathbb{C}(y)
+ {9 \over 32 H^4} \sum_{\{k_i\}}C_{k_2}\left(2{\widetilde{C}_{k_1}}
{\cal G}_{\gamma lab}^{(k_3)} {\cal G}^{(k_4)\gamma lab}
+ C_{k_1} {\cal G}_{\gamma\eta ab}^{(k_3)} {\cal G}^{(k_4)\gamma\eta ab}
+ \hat{C}_{k_{1,2}} {\cal G}_{mn ab}^{(k_3)} {\cal G}^{(k_4)mn ab} \right)}\right], \nonumber\\ \nd} 
as gathered from \eqref{ajanta}; 
 where $\mathbb{C}(y)$ defined as in \eqref{chotom} and 
 $\hat{C}_{k_{1, 2}} \equiv \widetilde{C}_{k_1} \widetilde{C}_{k_2}/ C_{k_2}$ with $k_i$ satisfying as before
 $\sum_i k_i = 7/2$ with the standard constraint $(k_3, k_4) \ge (3/2, 3/2)$. 
 The quantum terms are again classified by $ 1 \le \theta'_k \le 3$ in \eqref{melamon2}, and therefore are most populated by powers of G-flux components and curvature tensors. Both the metric components, 
 \eqref{rambhafin} and \eqref{rambhafin2} are non-K\"ahler, but the un-warped metric along the ($a, b$) directions is flat as expected\footnote{We can also make some general observations regarding the {\it sign} of the internal curvature term $R$ from \eqref{emmaS} and \eqref{lp777}. Let us first assume that the quantum terms in \eqref{emmaS} and \eqref{lp777} are zero. Then the only solution is with vanishing flux components ${\cal G}^{(3/2)}_{MNab}$ and $R = -18 \Lambda H^4$. It is also clear from \eqref{musemey}, for vanishing quantum terms and vanishing fluxes, $\Lambda = 0$ and therefore $R = 0$.  When the fluxes vanish, but all the quantum terms are non-zero, then the internal quantum terms must satisfy the relation 
 \eqref{lulu} with zero on the RHS. The consistency condition \eqref{musemey} allows positive 
 $\Lambda$ if the space-time quantum terms $[\mathbb{C}^i_{i}]^{(0, 0)}$ dominates over all other terms. 
  In this case $\Lambda > 0$ is allowed. However if the internal space quantum terms vanish (which still allows positive $\Lambda$ in \eqref{musemey}), then from \eqref{emmaS} and \eqref{lp777} the internal curvature scalar has to be {\it negative} i.e $R = -18\vert\Lambda\vert H^4$ with the warp-factor $H(y)$ 
  satisfying: $$\square H^4 = \left([\mathbb{C}^i_{i}]^{(0, 0)} - 12 \vert\Lambda\vert 
  -{2\kappa^2 T_2n_b\over H^8 \sqrt{g_6}} \delta^8(y-Y)\right) H^4$$
  \noindent where $n_b$ is the number of M2-branes, $T_2$ is the tension of a M2-brane  and $g_6$ is the determinant of the six-dimensional internal metric. The six-dimensional base of the eight-manifold now becomes a non-K\"ahler space with a negative Ricci scalar. Clearly for {\it vanishing}
  $[\mathbb{C}^i_{i}]^{(0, 0)}$, and vanishing fluxes, $\Lambda$ can only be negative from \eqref{musemey}
  if the internal quantum terms are all positive definite.  In this case either $R < 0$ or 
  $R < 18 H^4 \vert\Lambda\vert $. If the internal quantum terms are all negative definite, then there can be 
  $\Lambda > 0$ for vanishing fluxes and vanishing space-time quantum terms. In this case $R > 0$ or 
  $R > -18 H^4\vert\Lambda\vert $. In the same vein, other possible choices can be entertained. It would also be interesting to compare our results with \cite{saurya}.}.
 Thus solving for $h(y)$ from \eqref{muse777}, and ($C_k, \widetilde{C}_k$) 
 from the cross-term EOMs in section \ref{cross} (see for example \eqref{babycasin} and \eqref{ellemey}), we can pretty much determine the full background data provided information about the G-flux components are 
 provided. The latter will require us to solve the flux EOMs, that we shall discuss soon. 
 
The miracle that has happened here has its root in the time-dependence of the G-flux components and the 
internal space. The time dependences of the G-fluxes are responsible for changing the relative signs of 
the ($l_{36}, l_{37}, l_{38}$) terms in \eqref{kkkbkb2} to the $k$-dependent scaling \eqref{melamon2}. On the other hand, the time-dependences of the internal space i.e the existence of the $F_i(t)$ factors are related to the quantum terms. The quantum terms are classified by $\theta'_k$ in \eqref{melamon2}, thus bringing us back full-circle. This interdependency of the temporal behavior of fluxes and the metric components is solely responsible for the generation of a four-dimensional positive curvature space-time in the type IIB side with de Sitter isometries. Switching off time-dependences (or the quantum terms) will immediately ruin the picture and drag us back to the swampland.    

\vskip.1in

\noindent {\it Case 2: $F_1(t)$ and $F_2(t)$ satisfying the fluctuation condition \eqref{ranjhita}}

\vskip.1in

\noindent Our procedure to study the scenario corresponding to $\gamma > 0$ will essentially be the same: we will take the traces of the various EOMs and from there inquire whether solutions could be constructed. We first take the trace of the EOM along the ($m, n$) directions. The EOM is given in \eqref{spoonR} and is defined at the zeroth order in $g_s$. The trace yields:
\bg\label{marlock}
R = {1\over 8H^4}  {\cal G}^{(9/2)}_{\alpha\beta ab} {\cal G}^{(9/2)\alpha\beta ab} - {1\over 2} 
\left[\mathbb{C}_m^m\right]^{(0, 0)} - 6\Lambda H^4, \nd
where we have used the fact that the un-warped Ricci scalar of ${\cal M}_4$ vanishes, which in turn appears from looking at \eqref{casinmey}. In fact this led us to choose the un-warped geometry of the six-dimensional base to be that of $K3 \times {\bf T}^2$, implying that the cosmological constant $\Lambda$ in this set-up may be expressed as:
\bg\label{laceymey}
\Lambda = {1\over 48H^8}  {\cal G}^{(9/2)}_{\alpha\beta ab} {\cal G}^{(9/2)\alpha\beta ab} - {1\over 12H^4} 
\left[\mathbb{C}_m^m\right]^{(0, 0)}, \nd
which at the face value doesn't contradict anything because the quantum terms are classified perturbatively by 
$\theta_k = 2/3$ in
\eqref{miai} for $\gamma = 2$, and this allows us to choose  $l_{28} = 2$ renormalizing the classical flux piece such that the RHS of \eqref{laceymey} becomes a positive constant. Non-perturbatively, the contributions come from \eqref{padekhai} and can take values $\theta_k = {2\over 3}(n_2 - 1)$, but is exponentially suppressed for large values of $n_2$. 
However this puts a tighter constraint on the behavior of the G-flux component ${\cal G}^{(9/2)}_{\alpha\beta ab}$. An alternative to this would be to take $R^{(2)} \ne 0$ in \eqref{casinmey}. This however would be a bit difficult to argue because 
\eqref{casinmey} is a source-free equation (see also footnote \ref{choolmaro}).  It is also interesting to note that \eqref{saraN} provides a relation similar to \eqref{laceymey}, namely:
\bg\label{saraN2}
\Lambda = -{1\over 64 H^8} \left({\cal G}^{(9/2)}_{\alpha\beta ab}\right)^2 -{1\over 8H^4}
\left[\mathbb{C}_\alpha^{\alpha}\right]^{(3, 0)}, \nd
which again shows that there has to be a delicate cancellation to allow for the cosmological constant 
term to appear from the RHS. Of course again the quantum terms are classified by 
${2\over 3} \le \theta_k \le {2\over 3}(n_2 - 1)$ 
in \eqref{miai} so we haven't faced a contradiction yet.  However the fact that first term in \eqref{saraN2}
is negative definite shows that the quantum terms have to be negative definite also to reproduce the positive
$\Lambda$ from RHS.  
We will not worry about whether \eqref{saraN2} and \eqref{laceymey} could be mutually consistent, and instead proceed with analyzing the other equations of the system.

Our next equation is the equation along the ($a, b$) directions. There are some subtleties in the construction of the EOMs, that we explained earlier, and after the dust settles, the EOM to order $g_s^2$ (which is the lowest order now) is given by \eqref{stjean}. Taking the trace leads to:
\bg\label{saraN3}
\Lambda = -{1\over 144 H^8} \left({\cal G}^{(9/2)}_{\alpha\beta ab}\right)^2 -{1\over 18H^4}
\left[\mathbb{C}_a^a \right]^{(3, 0)}, \nd
which is an equation similar to \eqref{saraN2} above. The concern associated with this equation remains the same as before as the quantum terms are classified by ${2\over 3} \le \theta_k \le {2\over 3}(n_2 - 1)$ in \eqref{miai}. We should then go to the space-time EOM to see if any of our concerns could be lifted. As we saw before, there are two space-time EOMs given by \eqref{bratmey} and \eqref{lebanmey}, out of which \eqref{lebanmey} will be the correct EOM once we gather all the constraints from flux EOM in section \ref{anoma}.  For the time being there is no way to choose \eqref{bratmey} over \eqref{lebanmey}, so we shall put both to test now and see what comes out from our exercise. 

We then start with the first set of EOM, i.e \eqref{bratmey}. In this case the 
 story, like \eqref{saraN3}, also repeats for the EOM along the space-time direction as may be seen from \eqref{bratmey}, and we reproduce it here again for completeness:
\bg\label{bratmey3}
\Lambda = {1\over 32 H^8} {\cal G}^{(9/2)}_{\alpha\beta ab} {\cal G}^{(9/2)\alpha\beta ab} 
- {1\over 4} \left[\mathbb{C}_i^i\right]^{(0, 0)}. \nd
We now face an interesting situation. If the $g_s^{-4}$ behavior from \eqref{metroaangul} with $n = 0$, fixes $n_2 \ge 5$, then the quantum terms classified by ${2\over 3} \le \theta_k \le {2\over 3}(n_2 - 1)$ in 
\eqref{miai} may have completely different set of terms 
compared to the earlier cases where the quantum terms are classified for $n_2 \ge 2$.
None of these terms are as simple as the classical flux term appearing in \eqref{bratmey3}, and therefore to reproduce the constant $\Lambda$ factor, there needs to be strong constraints on all the quantum terms classified by ${2\over 3} \le \theta_k \le {2\over 3}(n_2 - 1)$ with $n_2 \ge 5$ in \eqref{miai}.  

There is also no integral constraint like the one in \eqref{musemey} for the volume preserving case 
\eqref{olokhi} because the warp-factor $h(y)$ is harmonic from \eqref{ouletL} (if we incorporate non-perturbative contributions from \eqref{metroaangul} with $n = 0$, then they have to integrate to zero over the eight-manifold).  Combining \eqref{saraN2} 
and \eqref{bratmey3} yields:
\bg\label{sabull}
\Lambda = -{1\over 12 H^4} \Big(\left[\mathbb{C}_\alpha^{\alpha}\right]^{(3, 0)}
+ H^4 \left[\mathbb{C}_i^i\right]^{(0, 0)}\Big), \nd
which relates $\Lambda$ directly to the quantum terms. Since $\Lambda > 0$, the quantum terms or their sum have to be a negative definite integer. Additionally, they have to be proportional to $H^4$ (at least from the first term in \eqref{sabull}) if 
\eqref{bratmey3} has to make sense. Also since the square of the flux piece appearing in the above equations is a positive quantity, we expect:
\bg\label{murderland} 
H^4 \left[\mathbb{C}_i^i\right]^{(0, 0)} ~>~ {1\over 3} \left[\mathbb{C}_m^m\right]^{(0, 0)} ~ > ~
{2\over 9} \left[\mathbb{C}_a^a\right]^{(3, 0)} ~ > ~
{1\over 2} \left[\mathbb{C}_\alpha^\alpha\right]^{(3, 0)}, \nd
as a possible hierarchy between all the quantum terms classified by appropriate values of $\theta_k$ 
in \eqref{padekhai} and \eqref{miai}. 
All these lead to some strong constraints that are unclear if they could be consistently satisfied. Let us then ask whether the correct EOM, namely \eqref{lebanmey}, could ease some of the tension here. Combining \eqref{1851} with \eqref{lebanmey}, we get:
\bg\label{jlcurtis}
\Lambda = {1\over 32 H^8}{\cal G}^{(9/2)}_{\alpha\beta ab} {\cal G}^{(9/2)\alpha\beta ab} 
- {1\over 4} \left(\left[\mathbb{C}_i^i\right]^{(3, 0)} - \left[\mathbb{C}_i^i\right]^{(0, 0)}\right), \nd
which is similar to \eqref{bratmey3}, and appears to not alleviate any of the issues that we faced above.  The only difference between \eqref{bratmey3} and \eqref{jlcurtis} is the quantum terms, so 
\eqref{murderland} would  remain as before with the sole replacement:
\bg\label{lucind}
\left[\mathbb{C}_i^i\right]^{(0, 0)} \longrightarrow \left[\mathbb{C}_i^i\right]^{(3, 0)} - \left[\mathbb{C}_i^i\right]^{(0, 0)},
\nd
leading to same sort of strong constraints as before, {\it unless} the higher order non-pertubative contributions cancel out in the difference. 
Furthermore switching on $\gamma$ leads to an unnatural derivative constraint that is harder to justify. The absence of M2-branes, due to the vanishing Euler characteristics\footnote{There is some subtlety regarding the interpretation of an Euler characteristics for a time-dependent background that we shall discuss later. It suffices to say that none of these so-called topological data remain completely time-independent or even topological unless they vanish. So the conclusion that we have here is not far from truth.}, is also an issue because M2-branes dualize to D3-branes in the type IIB side and account for the color degrees of freedom (although presence of $\overline{\rm M2}$-branes could easily alleviate this issue). Additionally, the late-time behavior, as may be inferred from \eqref{fakhi}, shows that:
\bg\label{walmart}
F_1(t) \to 0, ~~~ F_2(t) \to 1, \nd
thus the subspace ${\cal M}_2$ shrinks to zero size leading to singularities at late time. However since we are never at $g_s = 0$ point, the quantum EOMs do not show any signs of complications at this stage. Thus
although none of the arguments presented here is damning enough to discard the model with 
non-zero $\gamma$, the issues presented here nonetheless  show that the late time physics with a four-dimensional de Sitter space-time, i.e with \eqref{olokhi}, is a preferable scenario over the ones with time-varying Newton constants. In {\bf Table \ref{jutameyL}} we summarize the differences between the two choices \eqref{olokhi} and \eqref{ranjhita}.

\begin{table}[tb]
 \begin{center}
\renewcommand{\arraystretch}{1.5}
\begin{tabular}{|c||c|}\hline Time-independent Newton's constant & Time-dependent Newton's constant 
\\ \hline\hline
No derivative constraint on ${\cal M}_4 \times {\cal M}_2$ & Derivative constraint on ${\cal M}_2$\\ \hline
${\cal M}_4$: non-K\"ahler & ${\cal M}_4$: conformally $K3$ \\ \hline
${\cal M}_2$: non-K\"ahler & ${\cal M}_2$: conformally ${\bf T}^2$ \\ \hline
$\chi_8 \ne 0$ & $\chi_8 = 0$ \\ \hline
Allows static and dynamical M2-branes & Only dynamical M2-branes allowed \\ \hline
No late time singularities & Late time singularities \\ \hline
G-flux components with $k \ge {3\over 2}$ & G-flux components with $k \ge {9\over 2}$\\ \hline
  \end{tabular}
\renewcommand{\arraystretch}{1}
\end{center}
 \caption[]{The key differences between backgrounds with time-independent Newton's constant coming from 
 \eqref{olokhi} and time-dependent Newton's constant coming from \eqref{ranjhita}. The Euler characteristics of the eight-manifold \eqref{melisett} is denoted by $\chi_8$. The case with dynamical membranes will be discussed in subsection \ref{branuliat}.}
  \label{jutameyL}
 \end{table}

%


\subsection{Analysis of the G-flux quantizations and anomaly cancellations \label{jutamaro}}

The study of all the Einstein's equation performed above revealed a delicate interconnection between the metric components, the quantum terms and the G-flux components at every order in the $g_s$ expansions. 
However the story is far from over: there are also flux EOMs that would introduce yet another layer of interconnections and constraints. Some of the details have appeared in our earlier works 
\cite{nogo, nodS}, and here we would like to specifically concentrate on two aspects of this: flux quantization and anomaly cancellation. In the process we shall also be able to tie up few of the loose ends from the earlier sections.

\subsubsection{Bianchi identities and flux quantizations \label{bianchi}}

The concept of flux quantization is intimately tied up with the Bianchi identity. In the time-independent case this was analyzed in details by \cite{wittenflux}.  Let us first elaborate this using the dual forms 
${\bf G}_7$ discussed in section \ref{tufi}. In the absence of the quantum terms, i.e in the absence of 
$\mathbb{Y}_7$ from \eqref{verasofmey}, the M-theory action using the dual variables may be written as:
\bg\label{seshmeye}
\mathbb{S}_{11} \equiv c_1 \int {\bf G}_7 \wedge \ast_{11} {\bf G}_7 + N\int {\bf C}_6 \wedge \Lambda_5
+ c_2\int {\bf C}_6 \wedge d\mathbb{\hat Y}_4, \nd
where $N$ represents the number of M5-branes, $c_i$ are constants that are defined in terms of certain powers of $M_p$ that may be easily specified\footnote{For example $c_1 = M_p^9$ and 
$c_2 = M_p^6$, but the term with $c_2$ will involve other powers of $M_p$.}, $\Lambda_5$ is a localized 
five-form that captures the singularities of the M5-branes, 
the Hodge star is with respect to the warped eleven-dimensional metric and ${\bf C}_6$ appears from defining 
${\bf G}_7 = d{\bf C}_6 + ...$ where the dotted terms appears from M2 and M5-branes in appropriate ways. The EOM for ${\bf C}_6$ turns out to be:
\bg\label{kkd}
d\ast_{11} {\bf G}_7 = {1\over c_1} \left(N \Lambda_5 + c_2 d\mathbb{\hat Y}_4\right) \equiv d{\bf G}_4, \nd
where on the RHS we expressed the equation in terms of the four-form ${\bf G}_4$. The above equation represents the Bianchi identity in the absence of any extra contributions from the quantum terms. Integrating the above equation over a five-manifold ${\bf \Sigma}_5$ with boundary 
${\bf \Sigma}_4 = \partial {\bf \Sigma}_5$, we get:
\bg\label{cyclemey}
c_1\int_{{\bf \Sigma}_4} {\bf G}_4 = N + c_2 \int_{{\bf \Sigma}_4} \mathbb{\hat Y}_4, \nd
where the RHS is expressed in terms of $N$, the number of {\it static} M5-branes, and an integral of a 
four-form over the four-manifold ${\bf \Sigma}_4$. In deriving the above equation we have assumed that the integral of $\Lambda_5$ over the five-manifold ${\bf \Sigma}_5$ is identity. Now defining:
\bg\label{goleta}
c_1 = {1\over 2\pi}, ~~~ c_2 = -1, ~~~ \mathbb{\hat Y}_4 = 
{1\over 16\pi^2}\left({\rm tr}~\mathbb{F}\wedge \mathbb{F}
 - {1\over 2}{\rm tr}~\mathbb{R} \wedge \mathbb{R}\right) + N\Lambda_5, \nd 
 where the curvature form $\mathbb{R}$ is as defined in \eqref{ashf1} and the gauge two-form 
 $\mathbb{F}$ will appear from the flux-form $\mathbb{G}$, also defined in \eqref{ashf1}, once we view the G-flux components as {\it localized} fluxes (this will be elaborated soon). Therefore combining 
 \eqref{goleta} with \eqref{cyclemey}, we reproduce the G-flux quantization as expressed in \cite{wittenflux}. 
 
 The question now is what happens when the G-flux components become time-dependent? One easy way out would be to introduce moving M5-branes, as the other pieces appearing in \eqref{cyclemey} are topological.  These topological pieces could also have time dependences, but as we saw earlier, the time dependences of the G-flux and metric components are correlated to the quantum corrections which in turn are classified by $\theta'_k$ in \eqref{melamon2} or $\theta_k$ in \eqref{miai} for \eqref{olokhi} and 
 \eqref{ranjhita} respectively. This therefore calls for the quantum corrections to the Bianchi identities themselves. 
 
Introducing the quantum corrections here would imply switching on the Hodge dual of $\mathbb{Y}_7$, which in turn implies switching on the second interaction in \eqref{verasofmey}. Implementing this changes the Bianchi identity from \eqref{kkd} to the following:
\bg\label{uniprixtagra}
d\ast_{11} {\bf G}_7 
= {1\over c_1} \left(N \Lambda_5 + c_2 d\mathbb{\hat Y}_4  - c_3 d\ast_{11} \mathbb{Y}_7\right) 
\equiv d{\bf G}_4, \nd 
where $c_3$ is yet another constant defined in terms of powers of $M_p$. 
As discussed in \eqref{aliceL}, the $\mathbb{Y}_7$ interaction should be understood as coming from 
\eqref{phingsha33} and is therefore non-topological. It may not be globally defined because it involves metric  components on the compact space ${\cal M}_4 \times {\cal M}_2 \times {\mathbb{T}^2\over {\cal G}}$, 
although in the ensuing analysis it doesn't matter whether it is globally defined or not. 
 Integrating \eqref{uniprixtagra} in the same way as above, leaves us with the following flux quantization condition:
\bg\label{sakuras}
c_1\int_{{\bf \Sigma}_4} {\bf G}_4 = N + c_2 \int_{{\bf \Sigma}_4} \mathbb{\hat Y}_4
- c_3 \int_{{\bf \Sigma}_4} \ast_{11}\mathbb{Y}_7, \nd 
where $N$, the number of M5-branes, would be affected if $\Lambda_5$ itself becomes $g_s$ (i.e time) 
dependent. Recall that $\Lambda_5$ in \eqref{uniprixtagra} is like a delta function and therefore if there are moving M5-branes, it would pick up $g_s$ dependence. Similarly $\mathbb{\hat Y}_4$ would also pick up some $g_s$ dependence. However these are all classical, and what we are looking for is more on the quantum side that could account for all {\it higher order} $g_s$ dependence of the ${\bf G}_4$ flux-components ${\cal G}_{MNPQ}^{(k)}$ for all $k \ge 3/2$. To see how this would come about, let us express 
\eqref{sakuras} in terms of components in the following way:
\bg\label{scarymag}
&&c_1\sum_{k\in {\mathbb{Z}\over 2}} \int_{{\bf \Sigma}_4} {\cal G}^{(k)}_{N_8N_9N_{10}N_{11}} 
\left({g_s\over H}\right)^{2\Delta k}
dy^{N_8} \wedge ....\wedge dy^{N_{11}} = N + c_2 \int_{{\bf \Sigma}_4} \mathbb{\hat Y}_4\\
&&  -c_3 \sum_l \int_{{\bf \Sigma}_4} 
\sqrt{-{g}_{11}} \left(\mathbb{Y}^{(l)}_7\right)_{N_1'.... N_7'} {g}^{N'_1 N_1}...... 
{g}^{N'_7 N_7}\left({g_s\over H}\right)^{\hat{\theta}_l}
\epsilon_{N_1....N_7 N_8....N_{11}} dy^{N_8} \wedge ....\wedge dy^{N_{11}},
\nonumber  \nd
where the metric components are all the {\it un-warped} metric components (including the determinant), and the epsilon is the 
Levi-Civita symbol (i.e not a tensor). Note also that although the
 LHS has been expanded in the standard way as in \eqref{ravali}, the RHS needs some explanation. According to \eqref{aliceL}, the quantum terms \eqref{phingsha33} are expanded by first choosing a particular component from the set of allowed dual forms and then labelling the remaining pieces as the associated seven-form $\mathbb{Y}_7$ accompanying the dual component. This way 
$\mathbb{Y}_7$ is uniquely identified once the dual G-flux component is chosen. However we expect the dual G-flux component to have a similar expansion as \eqref{ravali}, albeit with different $g_s$ scalings. The corresponding $\mathbb{Y}_7$ form will then have the $g_s$ scalings as given in 
{\bf Table \ref{dualforms}}. The RHS of the \eqref{scarymag} therefore represents precisely these scalings that we will simply label as $\hat{\theta}_l$. For every choice of ${\cal G}^{(k)}_{MNPQ}$ on the LHS, the 
$g_s$ scalings of the corresponding seven-form $\mathbb{Y}^{(l)}_7$ should 
match-up\footnote{We have been a bit sloppy in defining $\hat{\theta}_l$. The actual $g_s$ scalings of every components of $\mathbb{Y}_7$ may be read from {\bf Table \ref{dualforms}}. However 
$\hat{\theta}_l$ will have an additional contribution from $\sqrt{-{\bf g}_{11}}$, where the determinant is now expressed in terms of the warped metric components. To avoid all these un-necessary complications we just define $\hat{\theta}_l$ once and for all in \eqref{scarymag} without worrying too much of its source.}.  In the following we will do a detailed check of this, although note that we will not consider the non-perturbative corrections to $\mathbb{Y}_7$ from section \ref{instachela} at this stage. With some efforts this may be accommodated in, but the analysis is cumbersome without revealing new physics, so will leave it for future studies.

Before delving into this note that if the M5-branes are static, then $N$ will appear with no $g_s$ factor 
accompanying it in \eqref{scarymag}. Thus if there are no time-neutral G-flux components we cannot allow static M5-branes, although M2-branes can still be allowed\footnote{This is a bit more subtle than one would think. Dynamical M2-branes would back-react on the background stirring up corrections to fluxes and the metric. This is however surprisingly tractable, and we will elaborate the story in subsection \ref{branuliat}.}.
There is however some subtlety that we are hiding under the rug here. Since the 
$\mathbb{Y}_7$ piece in the Bianchi identity \eqref{uniprixtagra} should always have $g_s$ dependence, the {\it static} quantities that can actually appear from the Bianchi identity may be combined as 
$\mathbb{S}_5$ where:
\bg\label{bonemey}
\mathbb{S}_5 \equiv N \Lambda_5 - {c_2 \over 32\pi^2} d\Big({\rm tr}~\mathbb{R} \wedge \mathbb{R}\Big), \nd
\noindent where the second term comes from the definition of $\mathbb{\hat Y}_4$ in \eqref{goleta}, 
and 
$\Lambda_5$ is the localized five-form. The gauge field $\mathbb{F}$ will in general have $g_s$ dependence, but here we will simply put it to zero. Now,  
clearly if the trace or $\mathbb{R}$ in \eqref{bonemey} has only $g_s$ dependent terms, then 
$N = 0$ as ${\bf G}_4$ has no $g_s$ independent piece. However if the trace or the curvature form allows a $g_s$ independent piece then we can cancel $\mathbb{S}_5$ locally by identifying $\Lambda_5$ with the 
trace part. The global condition: 
\bg\label{pasadena}
N = {c_2 \over 32\pi^2} \int_{\Sigma_4} {\rm tr}~\mathbb{R} \wedge \mathbb{R}, \nd 
over a specific four-cycle $\Sigma_4 \equiv \partial \Sigma_5$ is then automatic. However compared to 
\cite{wittenflux}, we now require the integral of the first Pontryagin class to be an 
integer\footnote{The sign will be determined from the sign of $c_2$.} as we cannot switch on time-independent 
G-flux components here. Thus time-dependences put some extra constraints that did not exist for the 
time-independent case. In general, since we will only be concerned about comparing the $g_s$ scalings,
the time-independent part of $N$ can be effectively taken to zero without altering the flux quantization condition \eqref{scarymag}. There is however no reason to make 
$c_2 = 0$ because $\mathbb{\hat Y}_4$ can have $g_s$ dependences. We will not worry too much about this as we want to match the $g_s$ scalings of the LHS to the $g_s$ scaling of the quantum terms on the 
RHS of \eqref{scarymag}. More details on this will appear in \cite{coherbeta2}.

\vskip.1in

\noindent{\it Case 1: ${\bf G}_{mnab}$ component}

\vskip.1in

\noindent We will start by taking $c_2 = 0$ in \eqref{scarymag} just for simplicity. This may be restored back at the end with appropriate $g_s$ scalings. Such a procedure will help us to compare the LHS and the RHS succinctly. Therefore for a given order in $k$ the matching becomes:
\bg\label{birjoani} 
c_1 \int_{{\bf \Sigma}^{(1)}_4} {\cal G}^{(k)}_{mnab} dy^m \wedge....\wedge dy^b = - c_3 
\int _{{\bf \Sigma}^{(1)}_4}
\sqrt{-{g}_{11}}\left({\mathbb{Y}}^{(k)}_7\right)^{0ijpq\alpha\beta} \epsilon_{0ijpq\alpha\beta mnab}
dy^m\wedge ... \wedge dy^b, \nonumber\\ \nd
where ${{\bf \Sigma}^{(1)}_4} = {\cal C}_2 \times {\mathbb{T}^2\over {\cal G}}$, and ${\cal C}_2$ is a two-cycle in ${\cal M}_4$. The LHS of \eqref{birjoani} scales as $\left({g_s\over H}\right)^{2\Delta k}$ with 
$k \ge 3/2$ for the case \eqref{olokhi} and $k \ge 9/2$ for the case \eqref{ranjhita}. The $g_s$ scaling on the RHS is $\left({g_s\over H}\right)^{\hat{\theta}_k}$ where $\hat{\theta}_k$ for \eqref{olokhi} becomes:
\bg\label{tl1}
\hat{\theta}_k = \theta'_k - 2\Delta k + 6 - {14\over 3} = \theta'_k - 2\Delta k + {4\over 3}, \nd
where the first three terms in the first equality appears from {\bf Table \ref{dualforms}} and $-{14\over 3}$ 
comes from $\sqrt{-{\bf g}_{11}}$ (note that the determinants in \eqref{birjoani} and \eqref{scarymag} have un-bolded metric components). For $k = 3/2$ the $g_s$ scaling of the LHS becomes $2\Delta k = 1$ whereas the $g_s$ scaling of the RHS becomes $\hat{\theta}_k = \theta'_k + {1\over 3}$ with 
$\theta'_k$ as in \eqref{melamon2}. 
This means when 
$\theta'_k = {2\over 3}$ the $g_s$ scalings on both sides of \eqref{birjoani} matches exactly.

For the case \eqref{ranjhita} there are two changes: the determinant changes to 
$\sqrt{-{\bf g}_{11}} \propto  g_s^{-{8/ 3}}$ and $k \ge {9\over 2}$. Putting the information from {\bf Table 
\ref{dualforms}}, we get:
\bg\label{tl2}
\hat{\theta}_k = \theta_k - 2\Delta k + 4 - {8 \over 3} = \theta_k - 2\Delta k + {4\over 3}, \nd
where $\theta_k$ is as in \eqref{miai}. The $g_s$ scaling of the LHS for $k = 9/2$ is 
$2\Delta k = 3$ whereas the $g_s$ scaling of the RHS becomes $\hat{\theta}_k = \theta_k - {5 \over 3}$, implying that when $\theta_k = {14\over 3}$ the $g_s$ scaling on both sides of \eqref{birjoani} match exactly. Comparing the two cases, we see that the quantization scheme for \eqref{olokhi} is a bit more natural. However we could also consider the moding scheme\footnote{As described earlier, we will follow the universal moding scheme of $k \ge 3/2$ for the case \eqref{olokhi}, and $k \ge 9/2$ for the case 
\eqref{ranjhita}. However in the following we will also briefly mention the changes in the quantization procedure with the second set of  moding scheme.}
for the G-flux components described in the paragraphs between \eqref{evabmey2} and \eqref{teenangul}. For \eqref{olokhi}, we expect no change, but for \eqref{ranjhita} the flux component will have a moding given by $k \ge {3/ 2}$.  Thus, if we take $k = 3/2$, we see that $\theta_k = {8\over 3}$.

\vskip.1in

\noindent{\it Case 2: ${\bf G}_{\alpha\beta ab}$ component}

\vskip.1in

\noindent Following the same procedure as before we can define the quantization scheme for the G-flux component
${\bf G}_{\alpha\beta ab}$ defined over a four-cycle ${{\bf \Sigma}^{(2)}_4} \equiv {\cal M}_2 \times 
{\mathbb{T}^2\over {\cal G}}$ in the following way:

{\footnotesize
\bg\label{harper1}
c_1 \int_{{\bf \Sigma}^{(2)}_4} {\cal G}^{(k)}_{\alpha\beta ab} dy^\alpha \wedge....\wedge dy^b = - c_3 
\int _{{\bf \Sigma}^{(2)}_4}
\sqrt{-{g}_{11}}\left({\mathbb{Y}}^{(k)}_7\right)^{0ijmnpq} \epsilon_{0ijmnpq\alpha\beta ab}
dy^\alpha\wedge ... \wedge dy^b, \nd}
where now the seven-form has different set of indices. Looking at {\bf Table \ref{dualforms}} it is easy to see that the $g_s$ scaling of this seven-form component remains the same as earlier and therefore then matching of the $g_s$ scalings on both LHS and RHS of \eqref{harper1} happens exactly when 
$\theta'_k = {2\over 3}$ with $\theta'_k$ defined as in \eqref{melamon2}. The matching of the higher order terms then follows automatically. 

On the other hand, for the case \eqref{ranjhita}, the analysis is not similar to what we had before because the $g_s$ scaling of the seven-form changes as should be evident from {\bf Table \ref{dualforms}}. In fact the scaling becomes:
\bg\label{chloeS}
\hat{\theta}_k = \theta_k - 2\Delta k + 8 - {8 \over 3} = \theta_k - 2\Delta k + {16\over 3}, \nd
implying that for $k = {9\over 2}$, we will require $\theta_k = {2\over 3}$ in \eqref{miai} to match the lowest powers of $g_s$ on both sides of \eqref{harper1}. Once matched at the lowest powers, all higher order $g_s$ scalings get matched automatically. Interestingly, with the second set of moding scheme, we expect no changes to the quantization procedure, at least for this component.

\vskip.1in

\noindent{\it Case 3: ${\bf G}_{m\alpha ab}$ component}

\vskip.1in

\noindent This is an interesting case where the four-cycle on which we define our flux component is chosen from a combination of two one-cycles, one each from ${\cal M}_4$ and ${\cal M}_2$ respectively, and combined with the existing two-cycle ${\mathbb{T}^2\over {\cal G}}$. The one-cycles are possible because neither ${\cal M}_4$ nor ${\cal M}_2$ are Calabi-Yau manifolds as we saw earlier. We will call this four-cycle as 
${{\bf \Sigma}^{(3)}_4}$ and the quantization condition becomes:
\bg\label{harper2}
c_1 \int_{{\bf \Sigma}^{(3)}_4} {\cal G}^{(k)}_{m\alpha ab} dy^m \wedge....\wedge dy^b = - c_3 
\int _{{\bf \Sigma}^{(3)}_4}
\sqrt{-{g}_{11}}\left({\mathbb{Y}}^{(k)}_7\right)^{0ijnpq\beta} \epsilon_{0ijnpq\beta m\alpha ab}
dy^m\wedge ... \wedge dy^b. \nonumber\\ \nd
The $g_s$ scaling of the RHS remains similar to what we had for the two cases above for \eqref{olokhi}. This means that choosing $\theta'_k = {2\over 3}$ we can match the lowest order $g_s$ scalings on both sides of \eqref{harper2}. The second set of moding scheme introduces no change, and the higher order terms, as expected, match automatically after that. 

The story for the case \eqref{ranjhita} is however a bit different because the $g_s$ scaling of the dual form 
appearing in \eqref{harper2} is different as can be seen from {\bf Table \ref{dualforms}}. In addition to that, 
since ${\cal M}_4$ and ${\cal M}_2$ are conformally CY,  {\it global} one-cycles are non-existent here. Nevertheless local one-cycles are possible and thus ${\bf \Sigma}^{(3)}_4$ could only be viewed as a local four-cycle, implying that a relation like \eqref{harper2} cannot quite capture the flux quantization scheme for this case. Locally however we can still give some meaning to an equation like \eqref{harper2}, and if we carry on with such a local quantization condition, it will  tell us that the 
$g_s$ scaling of the RHS of \eqref{harper2} becomes:
\bg\label{pateromey}
\hat{\theta}_k  = \theta_k - 2\Delta k + {10 \over 3}, \nd
where $k \ge{9\over 2}$. This means that the bound on $\theta_k$ from \eqref{miai} is now 
$\theta_k \ge {8\over 3}$, implying that the flux quantization scheme here pits the time variation of the 
integrated G-flux component with the integrated quantum terms classified by $\theta_k = {8\over 3}$ for 
the case \eqref{ranjhita} and $\theta'_k = {2\over 3}$ for the case \eqref{olokhi}. On the other hand with the second moding scheme, if we take
$k = 3$, then $\theta_k = {2\over 3}$, i.e same as $\theta'_k$.

\vskip.1in

\noindent{\it Case 4: ${\bf G}_{mnpq}$ component}

\vskip.1in

\noindent We now start with components of G-fluxes that do not contribute at lower order in $g_s$ scalings to the EOMs. This means the quantization scheme will involve even higher order quantum corrections that are captured by the dual seven-form. This may be seen from the following quantization condition:
\bg\label{lizzie12}
c_1 \int_{{\cal M}_4} {\cal G}^{(k)}_{mnpq} dy^m \wedge....\wedge dy^q = - c_3 
\int _{{\cal M}_4}
\sqrt{-{g}_{11}}\left({\mathbb{Y}}^{(k)}_7\right)^{0ij\alpha\beta ab} \epsilon_{0ij\alpha\beta abmnpq}
dy^m\wedge ... \wedge dy^q. \nonumber\\ \nd
where the four-cycle is clearly the manifold ${\cal M}_4$. Looking at {\bf Table \ref{dualforms}} one can easily work out the $g_s$ scaling of the RHS of \eqref{lizzie12}. Putting everything together, this gives us:
\bg\label{kristen}
\hat{\theta}_k  = \theta'_k - 2\Delta k - {8 \over 3}, \nd
with $\theta'_k$ as in \eqref{melamon2} and $k \ge {3\over 2}$. The $g_s$ scaling of the LHS of 
\eqref{lizzie12} remains the same, i.e $2\Delta k$, and therefore to match both sides of \eqref{lizzie12}, we need $\theta'_k \ge {14\over 3}$ in \eqref{melamon2}. Clearly for this value of $\theta'_k$ there are multiple terms which we can easily work out from \eqref{phingsha33}. However if we take $k \ge 0$, with the second moding scheme, then 
$\theta'_k = {8\over 3}$.

The case with \eqref{ranjhita} is also different. The $g_s$ scaling of the seven-form may be read from 
{\bf Table \ref{dualforms}}, Putting things together, the $g_s$ scaling of the RHS of \eqref{lizzie12} now becomes:
\bg\label{lizzie13}
\hat{\theta}_k  = \theta_k - 2\Delta k - {8 \over 3}, \nd
with $\theta_k$ as in \eqref{miai}, and therefore the only way to match both sides of \eqref{lizzie12} is to impose $\theta_k \ge {26\over 3}$ in \eqref{miai}. This is a large number and therefore will involve many quantum terms, making the quantization scheme a bit more complicated. Nevertheless, matching of both sides could be made succinctly. On the other hand, taking $k \ge 0$, gives us $\theta_k = {8\over 3}$.

\vskip.1in

\noindent{\it Case 5: ${\bf G}_{mnp\alpha}$ component}

\vskip.1in

\noindent Quantization of flux in this case requires us to find a three-cycle in ${\cal M}_4$ and a one-cycle in 
${\cal M}_2$. This is possible thanks to the non-K\"ahler nature of ${\cal M}_4$ and ${\cal M}_2$ for the case \eqref{olokhi}. The quantization scheme now becomes:
\bg\label{borden}
c_1 \int_{{\bf \Sigma}^{(4)}_4} {\cal G}^{(k)}_{mnp\alpha} dy^m \wedge....\wedge dy^\alpha = - c_3 
\int _{{\bf \Sigma}^{(4)}_4}\sqrt{-{g}_{11}}\left({\mathbb{Y}}^{(k)}_7\right)^{0ijq\beta ab} \epsilon_{0ijq\beta abmnp\alpha}
dy^m\wedge ... \wedge dy^\alpha, \nonumber\\ \nd
where ${{\bf \Sigma}^{(4)}_4}$ is the corresponding four-cycle. Now according to {\bf Table \ref{dualforms}},
the $g_s$ scaling of the dual seven-form remains exactly the same as what we had for the 
${\bf G}_{mnpq}$ component and therefore the analysis will proceed in the same way as before. The net result is that the $g_s$ of the RHS remains \eqref{kristen}, and therefore the $g_s$ scalings of both sides of
\eqref{borden} match when  $\theta'_k \ge {14\over 3}$ in \eqref{melamon2}. It also implies that taking 
$k \ge 0$, we get $\theta'_k = {8\over 3}$ as before.

For the case \eqref{ranjhita}, finding a globally defined four-cycle is not possible as both ${\cal M}_4$ and 
${\cal M}_2$ are conformally CY manifolds. Local construction is possible, but that weakens the flux quantization scheme here. Nevertheless if we proceed with a relation like \eqref{borden}, but now defined over a local four-cycle ${\bf \Sigma}^{(4)}_4$, we could still make some sense of \eqref{borden}, at least in identifying the $g_s$ scalings on both sides of the relation. This gives us:
\bg\label{lizzie14}
\hat{\theta}_k  = \theta_k - 2\Delta k - {2 \over 3}, \nd
with $\theta_k$ as defined in \eqref{miai} and $k \ge {9\over 2}$. Thus if $\theta_k \ge {20\over 3}$ 
we can in principle match both sides of \eqref{borden} for the case \eqref{ranjhita}. These bigger numbers, for both $\theta'_k$ and $\theta_k$, are somewhat consistent with the fact that the corresponding G-flux components do not contribute at lower values of the $g_s$ to the EOMs. On the other hand, with 
$k \ge 0$ we get $\theta_k = {2\over 3}$, a more manageable result.

\vskip.1in

\noindent{\it Case 6: ${\bf G}_{mn\alpha\beta}$ component}

\vskip.1in

\noindent This case is in many sense similar to the one studied for the ${\bf G}_{mnpq}$ component, because the $g_s$ scalings of the metric components, for the case \eqref{olokhi}, are similar. Both the metric components, ${\bf g}_{mn}$ and ${\bf g}_{\alpha\beta}$, scale as $g_s^{-2/3}$ and therefore it is no surprise that the $g_s$ scaling of the dual seven-form is again similar to what we had for the other component. However the flux quantization scheme involve the following components:
\bg\label{kerahintel}
c_1 \int_{{\bf \Sigma}^{(5)}_4} {\cal G}^{(k)}_{mn\alpha\beta} dy^m\wedge ....\wedge dy^\beta = - c_3 
\int _{{\bf \Sigma}^{(5)}_4}\sqrt{-{g}_{11}}\left({\mathbb{Y}}^{(k)}_7\right)^{0ijpqab} 
\epsilon_{0ijpqabmn\alpha\beta}
dy^m\wedge ... \wedge dy^\beta, \nonumber\\ \nd 
where ${\bf \Sigma}^{(5)}_4 \equiv {\cal C}_2 \times {\cal M}_2$, with ${\cal C}_2$ is the same 
two-cycle in ${\cal M}_4$ that we had chosen for the case with ${\bf G}_{mnab}$ component. The $g_s$ scaling of the RHS of \eqref{kerahintel} remains identical to \eqref{kristen} and therefore if 
$\theta'_k \ge {14\over 3}$ in \eqref{melamon2}, we can easily match both sides of \eqref{kerahintel}. As mentioned earlier, the higher order terms then match automatically. Again, with $k \ge 0$ with the second scheme, we get 
$\theta'_k = {8\over 3}$.

For the case \eqref{ranjhita}, we are in a better shape now because it is easy to find a two-cycle in 
${\cal M}_4$ when it is a conformally CY manifold. The four-cycle then becomes a product of the two-cycle in ${\cal M}_4$ and the conformally CY manifold ${\cal M}_2$ (which is topologically a torus). The $g_s$ scaling of the RHS of \eqref{kerahintel} becomes:
\bg\label{lizzie15}
\hat{\theta}_k  = \theta_k - 2\Delta k + {4 \over 3}, \nd
for $\theta_k$ as in \eqref{miai}. This implies that if $\theta_k \ge {14\over 3}$ we should be able to match the $g_s$ scalings of both sides of \eqref{kerahintel} for any order of $k \ge {9\over 2}$. Similarly for 
$k \ge {3\over 2}$, the matching happens with $\theta_k = {2\over 3}$.

\vskip.2in

\noindent{\it Case 7: ${\bf G}_{mnpa}, {\bf G}_{mn\alpha a}$ and ${\bf G}_{m\alpha\beta a}$ components}

\vskip.1in

\noindent The final three cases are to be defined on four-cycles that are to be constructed with one-cycles 
from ${\mathbb{T}^2\over {\cal G}}$ manifold. By definition such a one-cycle do not exist in   
 ${\mathbb{T}^2\over {\cal G}}$ for both cases \eqref{olokhi} and \eqref{ranjhita}. Previously the case with 
 \eqref{olokhi} did not suffer from any non-existence of global cycles, although the case with \eqref{ranjhita} 
 did have issues with the existence of global cycles. Now we see that for either case, global four-cycles are not possible, and we have to make sense of flux quantization with only local four-cycles. Although the non-existence of global cycles make the quantization scheme questionable, we can nevertheless compare the 
 $g_s$ scalings of flux integrals and the quantum terms using local four-cycles. Allowing this, we now have three set of equations:
 \bg\label{stewartk}
&&c_1 \int_{{\bf \Sigma}^{(6)}_4} {\cal G}^{(k)}_{mnpa} dy^m \wedge....\wedge dy^a = - c_3 
\int _{{\bf \Sigma}^{(6)}_4}\sqrt{-{g}_{11}}\left({\mathbb{Y}}^{(k)}_7\right)^{0ijq\alpha\beta b} 
\epsilon_{0ijq\alpha\beta bmnpa}
dy^m\wedge ... \wedge dy^a, \nonumber\\
&&c_1 \int_{{\bf \Sigma}^{(7)}_4} {\cal G}^{(k)}_{mn\alpha a} dy^m \wedge....\wedge dy^a = - c_3 
\int _{{\bf \Sigma}^{(7)}_4}\sqrt{-{g}_{11}}\left({\mathbb{Y}}^{(k)}_7\right)^{0ijpq\beta b} 
\epsilon_{0ijpq\beta bmn\alpha a}
dy^m\wedge ... \wedge dy^a, \nonumber\\ 
&&c_1 \int_{{\bf \Sigma}^{(8)}_4} {\cal G}^{(k)}_{m\alpha\beta a} dy^m \wedge....\wedge dy^a = - c_3 
\int _{{\bf \Sigma}^{(8)}_4}\sqrt{-{g}_{11}}\left({\mathbb{Y}}^{(k)}_7\right)^{0ijnpqb} 
\epsilon_{0ijnpqbmn\alpha a}
dy^m\wedge ... \wedge dy^a, \nonumber\\ \nd 
where the four-cycles  ${\bf \Sigma}^{(i)}_4$ for $i = 6, 7, 8$ respectively are 
${\cal C}_3 \times S^1_{(3)}, 
{\cal C}_2 \times S^1_{(2)} \times S^1_{(3)}$ and $S^1_{(1)} \times {\cal M}_2 \times S^1_{(3)}$, with the 
subscript denoting which one-cycle is meant. Clearly $S^1_{(1)}$  and $S^1_{(2)}$ are global one-cycles, 
but $S^1_{(3)}$ is not, as explained earlier. Therefore the set of equations \eqref{stewartk} can at most help us identify the $g_s$ scalings on both sides of the equalities, but would not serve as flux quantization conditions (as the four-cycles could shrink to zero sizes). From {\bf Table \ref{dualforms}} we can easily see that, for the case \eqref{olokhi}, the RHS of all the three equations scale in exactly the same way as:
\bg\label{lizzie16}
\hat{\theta}_k  = \theta'_k - 2\Delta k - {2 \over 3}, \nd 
with $\theta'_k$ as in \eqref{melamon2} and $k \ge {3\over 2}$. This means that if we take 
$\theta'_k \ge {8\over 3}$ we can match the $g_s$ scalings of both sides of each individual equalities for 
all $k \ge {3\over 2}$, and to any subsequent orders. With the second moding scheme, $k \ge 0$, and therefore $\theta'_k \ge {2\over 3}$.

The case for \eqref{ranjhita} is however not as uniform as above. The $g_s$ scalings of the dual seven-forms themselves are different as may be inferred from {\bf Table \ref{dualforms}}. This directly translates to the $g_s$ scalings of the RHS of the three equations in \eqref{stewartk} in the following way:
\bg\label{lizzie17}
\hat{\theta}_k  = \theta_k - 2\Delta k - {2 \over 3}, ~~~~ \hat{\theta}_k  = \theta_k - 2\Delta k + {4 \over 3}, ~~~~
\hat{\theta}_k  = \theta_k - 2\Delta k + {10 \over 3}, 
\nd 
with $\theta_k$ as in \eqref{miai} and $k \ge {9\over 2}$. Of course now none of the one-cycles are globally defined, and neither is the three-cycle ${\cal C}_3$, so the four-cycles in each of the three 
cases in \eqref{stewartk} are local in much weaker sense than what we had earlier. This means the 
flux-quantization conditions are even more weakly defined than before. Nevertheless we see that 
the above three scalings in \eqref{lizzie17} puts the following lower bounds on $\theta_k$:
\bg\label{lizzie18}
\theta_k \ge {20\over 3}, ~~~~~~~ \theta_k \ge {14\over 3}, ~~~~~~~ \theta_k \ge {8\over 3}, \nd
respectively for the three cases in \eqref{stewartk} for the $g_s$ scalings to match on both sides of the equalities. Once they match at the lowest orders, matchings at higher orders are almost automatic. Similarly with the second moding scheme we expect $k \ge 0$, $k \ge 3/2$ and $k \ge 3$ for the three components, giving us $\theta_k \ge {2\over 3}$  uniformly for all the three cases.

Our detailed analysis above should justify how flux quantizations should be understood in the case when the fluxes themselves are varying with respect to time, or alternatively, have $g_s$ dependences (as we packaged all temporal dependences as $g_s$ scalings). The original time-independent quantization scheme of \cite{wittenflux} where:
\bg\label{liftchinki}
\left[{{\bf G}_4 \over 2\pi}\right] - {p_1(y)\over 4} ~\in~ \mathbb{H}^4(y, \mathbb{Z}), \nd
doesn't quite work in the time-dependent case as ${\bf G}_4$ is always time-dependent (i.e $g_s$ dependent) in our set-up whereas $p_1(y)$, the first Pontryagin class, may not always be (i.e for some sub-manifold in the internal eight-manifold, $p_1(y)$ may be time, or $g_s$, independent). Therefore the combination on the LHS of \eqref{liftchinki} being in the fourth cohomology class 
$\mathbb{H}^4(y, \mathbb{Z})$ doesn't make much sense here, and  the quantization scheme now becomes little more involved as we showed above. 
In principle one would expect both the G-flux components as well as the four-cycles to vary with respect to time. However we have managed to rewrite the flux quantization condition in such a way that all $g_s$ dependences go in the definition of the fluxes,
and the cycles themselves are defined using un-warped metric components. Such a procedure then helped us to balance the $g_s$ dependences of the integrated flux components on given four-cycles with the $g_s$ dependences of the corresponding quantum corrections. We have tabulated the results in 
{\bf Table \ref{blanchek}}.

Despite the aforementioned computations, the story however is not complete.
There are two other potential contributions to the flux quantization conditions that we only gave cursory attentions. These are the number of dynamical M5-branes, denoted by  $N$, and the integrated 
four-form, denoted by the integral of $\mathbb{\hat Y}_4$, in \eqref{scarymag}.  Both these could have potential $g_s$ dependences and would therefore contribute to the flux quantization conditions. A more complete analysis will appear in \cite{coherbeta2}.

\begin{table}[tb]
 \begin{center}
\renewcommand{\arraystretch}{1.5}
\begin{tabular}{|c|c|c|c|c|c|}\hline Forms & Dual Forms & ${\hat\theta}_k$ for \eqref{olokhi} & 
${\hat\theta}_k$ for \eqref{ranjhita} & $\left[\theta'_k\right]_{\rm min}$ & $\left[\theta_k\right]_{\rm min}$ 
\\ \hline\hline
${\cal G}^{(k)}_{mnab}$  & $\left(\mathbb{Y}^{(l)}_7\right)^{0ijpq \alpha\beta }$ & $\theta'_k -2\Delta k + {4\over 3}$   & $\theta_k - 2\Delta k+ {4\over 3}$ & ${2\over 3}$ & ${14\over 3}$   \\  \hline
${\cal G}^{(k)}_{\alpha\beta ab}$  & $\left(\mathbb{Y}^{(l)}_7\right)^{0ijmnpq}$ & $\theta'_k -2\Delta k + {4\over 3}$   & $\theta_k - 2\Delta k+ {16 \over 3}$ & ${2\over 3}$ & ${2 \over 3}$   \\  \hline
${\cal G}^{(k)}_{m\alpha ab}$  & $\left(\mathbb{Y}^{(l)}_7\right)^{0ijnpq \beta }$ & $\theta'_k -2\Delta k + {4\over 3}$   & $\theta_k - 2\Delta k+ {10 \over 3}$ & ${2\over 3}$ &~ ${8\over 3}$~$\ast$   \\  \hline
${\cal G}^{(k)}_{mnpq}$  & $\left(\mathbb{Y}^{(l)}_7\right)^{0ij\alpha\beta ab}$ & $\theta'_k -2\Delta k - {8\over 3}$   & $\theta_k - 2\Delta k - {8\over 3}$ & ${14 \over 3}$ & ${26 \over 3}$   \\  \hline
${\cal G}^{(k)}_{mnp\alpha}$  & $\left(\mathbb{Y}^{(l)}_7\right)^{0ijq\beta ab}$ & $\theta'_k -2\Delta k - {8\over 3}$   & $\theta_k - 2\Delta k- {2 \over 3}$ & ${14\over 3}$ &~ ${20 \over 3}$~$\ast$   \\  \hline
${\cal G}^{(k)}_{mn\alpha\beta}$  & $\left(\mathbb{Y}^{(l)}_7\right)^{0ijpqab}$ & $\theta'_k -2\Delta k - 
{8\over 3}$   & $\theta_k - 2\Delta k+ {4\over 3}$ & ${14 \over 3}$ & ${14\over 3}$   \\  \hline
${\cal G}^{(k)}_{mnpa}$  & $\left(\mathbb{Y}^{(l)}_7\right)^{0ijq \alpha\beta b }$ & $\theta'_k -2\Delta k - {2\over 3}$   & $\theta_k - 2\Delta k- {2 \over 3}$ &~${8 \over 3}~\ast$ &~ ${20\over 3}$~$\ast$   \\  \hline
${\cal G}^{(k)}_{mn\alpha a}$  & $\left(\mathbb{Y}^{(l)}_7\right)^{0ijpq\beta b}$ & $\theta'_k -2\Delta k - {2\over 3}$   & $\theta_k - 2\Delta k+ {4\over 3}$ &~ ${8 \over 3}~\ast$ &~ ${14\over 3}~\ast$   \\  \hline
${\cal G}^{(k)}_{m\alpha\beta a}$  & $\left(\mathbb{Y}^{(l)}_7\right)^{0ijnpq\beta }$ & $\theta'_k -2\Delta k - {2\over 3}$   & $\theta_k - 2\Delta k+ {10 \over 3}$ &~ ${8 \over 3}~\ast$ &~ ${8 \over 3}~\ast$   \\  \hline
  \end{tabular}
\renewcommand{\arraystretch}{1}
\end{center}
 \caption[]{Flux quantization associated with \eqref{scarymag} keeping $N = c_2 = 0$. All the integrated 
 flux components scale as $g_s^{2\Delta k}$, and the $g_s$ scalings of the dual forms, that incorporate the quantum corrections, go as $g_s^{\hat{\theta}_k}$. These are tabulated above for the two cases 
 \eqref{olokhi} and \eqref{ranjhita}. The other two parameters, $\theta'_k$ and $\theta_k$, are defined in 
  {\eqref{melamon2}} and {\eqref{miai}} respectively. The symbol $\ast$ denotes the non-existence of global four-cycles.}
  \label{blanchek}
 \end{table}

\subsubsection{Anomaly cancellations and localized fluxes \label{anoma}}

In the above section we studied how the flux quantization conditions as well as the Bianchi identities go hand in hand, and how the $g_s$ scalings could be matched for every allowed G-flux components. The results are shown in {\bf Table \ref{blanchek}}.  It is time now to go to the next level of subtleties, namely the 
interpretation of the flux components that thread the internal manifold, and the cancellations of anomalies that arise from fluxes and branes on compact spaces.

We will start by defining the eleven-dimensional action much like how we described it in \eqref{seshmeye}, but now using the fundamental variables and not the dual ones. This means four-form G-flux components will appear instead of the seven-form dual flux components. In this language the action becomes:

{\footnotesize
\bg\label{weisz}
\mathbb{S}_{11} \equiv b_1\int {\bf G}_4 \wedge \ast_{11} {\bf G}_4 + b_2 \int {\bf C}_3 \wedge {\bf G}_4
\wedge {\bf G}_4 + b_3 \int {\bf C}_3 \wedge \mathbb{Y}_8 + b_4 \int {\bf G}_4 \wedge \ast_{11}
\mathbb{Y}_4 + n_b \int {\bf C}_3 \wedge {\bf \Lambda}_8, \nonumber\\ \nd}
where $b_i$ are all proportional to certain powers of $M_p$ (that may be easily fixed by derivative counting), $\mathbb{Y}_8$ is as defined in \eqref{ashf2} which contains the ${\bf X}_8$ polynomial, and $n_b$ is the number of static M2-branes. The other important ingredient of \eqref{weisz} is the  
$\ast_{11}\mathbb{Y}_4$ piece that captures the quantum corrections from either \eqref{phingsha} or 
\eqref{phingsha2} as elucidated in \eqref{aliceL2}. Such a term appearing in \eqref{weisz} leads to 
the non-topological interactions, and by construction $\ast_{11} \mathbb{Y}_4$ is not a globally defined 
function on a compact space. The EOM that arises from varying ${\bf C}_3$ now takes the following form:
\bg\label{marianc}
d\ast_{11} {\bf G}_4 = {1\over b_1} \Big(b_2~ {\bf G}_4 \wedge {\bf G}_4 + b_3~\mathbb{Y}_8 - 
b_4 ~d\ast_{11} \mathbb{Y}_4 + n_b {\bf \Lambda}_8\Big). \nd
Since both ${\bf G}_4$ and ${\bf G}_7 \equiv \ast_{11} {\bf G}_4$ are globally defined forms on the compact eight-manifold ${\cal M}_8$, as given in \eqref{melisett}, integrating the LHS of \eqref{marianc} over  
${\cal M}_8$ would automatically vanish. Doing this on the RHS then reproduces the following anomaly cancellation condition:
\bg\label{marbrick}
b_2 \int_{{\cal M}_8} {\bf G}_4 \wedge {\bf G}_4  + b_3 \int_{{\cal M}_8} \mathbb{Y}_8 
-b_4\int_{{\cal M}_8} d\ast_{11} \mathbb{Y}_4 + n_b = 0, \nd
where we have assumed that the integral of the localized form ${\bf \Lambda}_8$ over the eight-manifold 
is identity. This is true of course when the M2-branes are completely {\it static}. We will discuss more on this later. 

On the outset \eqref{marbrick} looks like the standard anomaly cancellation condition one would get from 
\cite{BB, DRS}, however a closer inspection reveals a few subtleties. One, the flux integral is now time-dependent because the ${\bf G}_4$ fluxes do not have any time-independent parts. Two, we have an integral over the topological 8-form $\mathbb{Y}_8$, whose polynomial form appears in \eqref{ashf2}, instead of just ${\bf X}_8$ as in \cite{BB, DRS}. Three, there appears a {\it new} contribution coming from the integral of a {\it locally} exact form $d\ast_{11} \mathbb{Y}_4$ over ${\cal M}_8$ from the quantum corrections. And four, we have $n_b$, the number of static M2-branes, that is a time-independent factor. Thus 
\eqref{marbrick} is not just a single relation as in \cite{DRS}, rather it is now a mixture of time-dependent and time-independent pieces juxtaposed together. How do we disentangle the various parts of \eqref{marbrick} to form consistent anomaly cancellation conditions for our case?  

\vskip.1in

\noindent{\it The ${\bf X}_8$ polynomial and Euler characteristics of the eight-manifold}

\vskip.1in

\noindent First let us look at the ${\bf X}_8$ part of $\mathbb{Y}_8$. As should be clear from \eqref{ashf2}, the choice 
\eqref{ashf3} allows us to construct the ${\bf X}_8$ polynomial from $\mathbb{Y}_8$. In the time-independent case, we expect (see the first reference in \cite{BB}):
\bg\label{nickaha}
\int_{{\cal M}_8} {\bf X}_8 = -{1\over 4!(2\pi)^4}\chi_8, \nd
where $\chi_8$ is the Euler-characteristics of the eight-manifold ${\cal M}_8$ when it has a Calabi-Yau metric on it. In fact, in the time-independent case 
\eqref{nickaha} makes sense, but if we now take the metric ansatze \eqref{vegamey} with the warp-factors as defined in \eqref{mrglass}, how does \eqref{nickaha} translates to the present case?

To answer this question let us look for the regime of validity of our $g_s$ expansions for all the parameters involved in our analysis. It is easy to see that as long as $0 \le \left({g_s\over H}\right)^2 < 1$ we have pretty much controlled quantum series expansions for all the parameters here. Clearly we {\it cannot} analyze the 
cases when $\left({g_s\over H}\right)^2 \ge 1$ because of the way we expressed the G-flux components 
in \eqref{frostgiant}, quantum terms in \eqref{neveC} etc. Thus $\left({g_s\over H}\right)^2 = 1$ 
forms a kind of {\it boundary}, below which all the analysis that we performed remains valid. Interestingly when 
$\left({g_s\over H}\right)^2 = 1$, the M-theory metric \eqref{vegamey} takes the following form:

{\footnotesize
\bg\label{sun18c}
ds^2 = H^{-8/3}\Big(-dt^2 + dx^2_1 + dx^2_2\Big) + H^{4/3} \Big(g_{\alpha\beta} dy^\alpha dy^\beta + 
g_{mn} dy^m dy^n + g_{ab} dy^a dy^b\Big), \nd}
where the metric components appearing above are all the un-warped ones and we have absorbed the 
$F_i\left(-{1\over \sqrt{\Lambda}}\right)$ in the definition of the internal coordinates ($\Lambda$ being the cosmological constant). 
We will not worry about the fluxes and the quantum corrections in this limit as they are any way not well defined according to our $g_s$ expansion
scheme. What we do want to point out is the similarity of the metric \eqref{sun18c} to the time-independent metric that we took in \eqref{betbab} (of course a redefinition of the internal space according to 
\eqref{melisett} is called for here). For the case \eqref{betbab} we had assumed that a time-dependent background like \eqref{vegamey} could appear from {\it coherent states} description of the form 
\eqref{garbo}, \eqref{kilmer} and \eqref{oshombhov}, now appropriately modified by introducing the $F_i(t)$ factors and the internal sub-division \eqref{melisett}, over the vacuum solution \eqref{betbab}. Our present scenario is somewhat similar to the one we encountered earlier, although we do not want to give a coherent state interpretation when comparing \eqref{vegamey} and \eqref{sun18c} just yet. What we can say is that as:
\bg\label{alibet}
-{1\over \sqrt{\Lambda}} ~ < ~ t ~ \le ~0, \nd
the metric \eqref{sun18c} slowly transforms into \eqref{vegamey}, implying that all temporal evolution should be defined for $t \equiv -{1\over\sqrt{\Lambda}} + \delta t$. Such a point of view does not rule out a coherent state formalism for our present background because we can still view the time-dependent evolution for 
$-\infty < t \le 0$ to be over a solitonic configuration of the form \eqref{sun18c}. Unfortunately the inaccessibility of the regimes $t \le -{1\over \sqrt{\Lambda}}$ prohibits us to provide a quantitative analysis 
of such a scenario.

What it does provide is a way to interpret the integral of ${\bf X}_8$ over the eight-manifold. Let us first consider the eight-manifold as given in \eqref{sun18c}. This is not a Calabi-Yau four-fold so the ${\bf X}_8$ integral will not necessarily capture the Euler characteristics of the internal eight-manifold ${\cal M}_8$ 
defined as in \eqref{melisett}.  Once we switch on a time interval $\delta t$, the warp-factors \eqref{mrglass} changes to the following:

{\footnotesize
\bg\label{klina01}
&& e^{2A} = \left(1 + {8\over 3} \sqrt{\Lambda} \delta t\right) H^{-8/3}, ~~
e^{2C} = \left(1 - {4\over 3} \sqrt{\Lambda} \delta t\right) H^{4/3}, ~~
\Lambda t^2 \equiv \left({g_s\over H}\right)^2 = 1 - 2\sqrt{\Lambda} \delta t\\
&&e^{2B_1} = F_1\left(-{1\over \sqrt{\Lambda}} + \delta t\right)
\left(1 + {2 \over 3} \sqrt{\Lambda} \delta t\right) H^{4/3}, ~~~
e^{2B_2} = F_2\left(-{1\over \sqrt{\Lambda}} + \delta t\right)
\left(1 + {2 \over 3} \sqrt{\Lambda} \delta t\right) H^{4/3}, \nonumber \nd}
where we see that the temporal evolution of the metric \eqref{sun18c} appears as additive pieces, each proportional to $\delta t$, to every metric components (including the space-time ones) up-to the $F_i$ factors. The $F_i$ factors do not change this observation because:
\bg\label{rosybab}
&& F_2\left(-{1\over \sqrt{\Lambda}} + \delta t\right) = 1 + \sum_k C_k\Big(1 - 2 \Delta \sqrt{\Lambda} 
\delta t\Big) \nonumber\\
&& F_1\left(-{1\over \sqrt{\Lambda}} + \delta t\right) = 
\left[1 + \sum_k \widetilde{C}_k\Big(1 - 2 \Delta \sqrt{\Lambda} 
\delta t\Big)\right]\Big(1 - \gamma \sqrt{\Lambda} \delta t\Big), \nd
where $\gamma = 0, 2$ are related to the two cases \eqref{olokhi} and \eqref{ranjhita} respectively. The other two set of parameters $C_k$ and $\widetilde{C}_k$ have been determined earlier in terms of the quantum corrections in section \ref{cross}. 

Therefore combining \eqref{klina01} and \eqref{rosybab}, the metric ansatze \eqref{vegamey} can actually be viewed as a perturbation over the initial metric configuration \eqref{sun18c}. In fact in this language, the 
late time cosmological evolution may be viewed as evolving from the metric configuration \eqref{sun18c}
via the warp-factors \eqref{klina01} and \eqref{rosybab}. It is also easy to replace $\delta t$ to a finite temporal value by iterating \eqref{klina01} and \eqref{rosybab} or by directly summing over binomial coefficients. All in all, our little exercise above tells us that:
\bg\label{poladom}
\int_{{\cal M}_8} {\bf X}_8 \equiv {1\over 3\cdot 2^9\cdot \pi^4} \int_{{\cal M}_8}\left({\rm tr}~\mathbb{R}^4 
- {1\over 4}\left({\rm tr}~\mathbb{R}^2\right)^2\right) = 
-{\omega_o\over 4!(2\pi)^4}\chi_8 +  g_o(\delta t), \nd
where $\mathbb{R}$ is the curvature two-form as it appears in \eqref{ashf1}, and $\omega_o$ measures the deviation from the Euler characteristics $\chi_8$. This could be integer or fraction depending on our choice of the eight-manifold. Note that the integral 
\eqref{poladom} splits into two pieces: 
$\omega_o\chi_8$, which is the piece independent of $\delta t$, is now only proportional to the Euler characteristics of 
the eight-manifold appearing in \eqref{sun18c}; and 
$g_o(\delta t)$ is a factor that depends on our temporal evolution parameter $\delta t$. The latter doesn't automatically vanish, at least not for the kind of background that we analyze here, and therefore should contribute to the anomaly cancellation condition \eqref{marbrick}. Exactly how this happens will be illustrated soon. 

The Euler characteristics $\chi_8$ can take either values, positive or negative, and both will be useful in analyzing the anomaly cancellation\footnote{Thus without loss of generalities we will take $\omega_o > 0$
in \eqref{poladom}.}. 
The case with vanishing Euler is interesting in its own way, but it appears not to be realized at least for the case \eqref{olokhi}. Question however is the robustness of the 
interpretation \eqref{poladom}. How is the split \eqref{poladom} understood in the full cosmological  setting?
This is where the coherent state interpretation becomes useful. If we assume that the cosmological evolution for $-{1\over \sqrt{\Lambda}} < t \le 0$ is via coherent states that evolve over a solitonic background like \eqref{sun18c} then $\chi_8$ appears to be related to the Euler characteristics of the vacuum eight-manifold. As we saw in section \ref{dyson}, study of non-supersymmetric backgrounds via such coherent states addresses many issues in a controlled setting that are hitherto difficult to  manage otherwise. This at least puts more confidence on our interpretation here. 

The above analysis, although correct, undermines the fact that $\chi_8 \equiv \chi_8(-1/\sqrt{\Lambda})$, and therefore is secretly a dynamical variable. It happens to pick up the Euler number at 
$t = -{1\over \sqrt{\Lambda}}$, but in general $\chi_8$ itself cannot be  independent of time (hence $g_s$), and for $ t \ne -{1\over \sqrt{\Lambda}}$ it may not even be the Euler characteristic of the eight-manifold. It is therefore instructive to work out the $g_s$ dependence of the polynomial ${\bf X}_8$. First however, we will need the 
curvature two-forms. They take the following values:

{\footnotesize
\bg\label{beachbb}
{\bf  R}^{a_o b_o}_{ab} &=&  g_s^2 ~{\rm R}^{a_o b_o}_{[ab]} + {\cal O}(F_1, F_2)\nonumber\\
&\equiv& {\bf R}_{abM'N'} ~e^{a_o M'} e^{b_o N'}  
+ {\bf R}_{aba'b'} ~e^{a_o a'} e^{b_o b'} + {\bf R}_{abM'0} ~e^{a_o M'} e^{b_o 0} 
+ {\bf R}_{abij} ~e^{a_o i} e^{b_o j}\nonumber\\
{\bf  R}^{a_o b_o}_{MN} &=& g_s^0~ {\rm R}^{a_o b_o}_{[MN]} + {\cal O}(F_1, F_2) \nonumber\\
&\equiv& {\bf R}_{MNM'N'} ~e^{a_o M'} e^{b_o N'}  
+ {\bf R}_{MNab} ~e^{a_o a} e^{b_o b} + {\bf R}_{MNM'0} ~e^{a_o M'} e^{b_o 0} 
+ {\bf R}_{MNij} ~e^{a_o i} e^{b_o j}, \nd}    
where $(M, N) \in {\cal M}_4 \times {\cal M}_2$, $(a, b) \in {\mathbb{T}^2\over {\cal G}}$ and $(a_o, b_o)$ are the locally $SO(11)$ indices. 
The $(F_1, F_2)$ dependence from \eqref{karishma} come from the definition of the vielbeins, and could involve factors like $\sqrt{F_i}$ etc. These can be expanded from the explicit expressions for the $F_i(t)$ in \eqref{karishma} because we are in the limit $g_s << 1$.   One may now express the curvature two-form in the following way:

{\footnotesize
\bg\label{beachmaa}
\mathbb{R} \equiv {\bf R}^{a_o b_o}_{MN} {\bf M}_{a_o b_0} ~dy^M \wedge dy^N + 
{\bf R}^{a_o b_o}_{ab} {\bf M}_{a_o b_o} ~dy^a \wedge dy^b + 
{\bf R}^{a_o b_o}_{Ma} {\bf M}_{a_o b_0} ~dy^M \wedge dy^a, \nd}
where the holonomy matrices ${\bf M}_{a_o b_o}$ are defined in such a way that taking traces of products of them provide some constraints. In fact they are not essential in the description of the curvature forms, and there exists definition of the two-forms without involving the holonomy matrices provided we define traces properly. Here the advantage of introducing ${\bf M}_{a_o b_o}$ is to impose the following conditions:
\bg\label{brianmaa}
&& {\rm tr}~{\bf M}_{a_o b_o} = 0, ~~ {\rm tr} \left({\bf M}_{a_o b_o} {\bf M}_{a_1 b_1}\right) \equiv 
\delta_{b_o a_1} ~\delta_{b_1 a_o} \nonumber\\
&& {\rm tr}\left({\bf M}_{a_o b_o}{\bf M}_{a_1 b_1}{\bf M}_{a_2 b_2}
{\bf M}_{a_3 b_3}\right) = \delta_{b_o a_1}~\delta_{b_1 a_2}~\delta_{b_2 a_3} ~\delta_{b_3 a_0}, 
\nd
using which we can easily define the traces of the curvature forms. What we need here are simply 
${\rm tr}~\mathbb{R}^4$  and ${\rm tr}~\mathbb{R}^2$ as they appear in the definition of ${\bf X}_8$
in \eqref{poladom}. Plugging all these in \eqref{poladom} gives us:
\bg\label{polameys}
\int_{{\cal M}_8} {\bf X}_8 = g_s^2 \int_{{\cal M}_8} \widetilde{\bf X}_8(y) + {\cal O}\left(F_1, F_2\right)
\equiv \sum_{k \ge 0} {\bf X}_8^{(k)}(y) \left({g_s\over H}\right)^{2(1 + \Delta k)}, \nd
which could be identified to the RHS of \eqref{poladom} once we take $g_s = 1 - \delta g_s$; and 
$\widetilde{\bf X}_8 \equiv {\bf X}^{(0)}_8$ is computed from ${\rm R}^{a_o b_o}$, without the $g_s$ and $F_i(t)$ factors. However identifications aside, the point of the above exercise is to show that the ${\bf X}_8$ polynomial is generically a $g_s$ dependent function. This will make the anomaly cancellation analysis a bit more non-trivial as we shall see below. 

The ${\bf X}_8$ polynomial that we computed above is defined over the eight-manifold ${\cal M}_8$ in 
\eqref{melisett}. 
Clearly other choices of ${\bf X}_8$ exist if we don't restrict ourselves exclusively to the eight-manifold. How do they behave? Before discussing this, let us resolve one puzzle associated with the above computation. The value of ${\bf X}_8$ rely on the existence of the Riemann tensor ${\bf R}_{MNab}$ where $(M, N) \in
{\cal M}_4 \times {\cal M}_2$. In our case this may be written as:
\bg\label{pola2mey}
{\bf R}_{MNab} = {\bf g}^{cd} \Big({\bf g}_{ac, M} {\bf g}_{bd, N} - {\bf g}_{ac, N} {\bf g}_{bd, M}\Big)
+ {\bf g}_{bd}\left({\bf g}_{ac, M}{\bf g}^{cd}_{~~, N} - {\bf g}_{ac, N}{\bf g}^{cd}_{~~, M}\right), \nd  
which relies on two things: existence of cross-term for the metric of ${\mathbb{T}^2\over {\cal G}}$, and the dependence of the metric on the coordinates $(M, N) \in {\cal M}_4 \times {\cal M}_2$. This {\it does not} vanish as explained in footnote \ref{plaza2019} simply because of the fact that charge-cancellation will require us to go to the constant coupling limit of F-theory \cite{sendas}. This implies that the local geometry of ${\cal M}_2 \times {\mathbb{T}^2\over {\cal G}}$ goes to ${{\cal M}_4^{(2)} \over {\cal G}}$ globally with 
${\cal G} \equiv \mathbb{Z}_2, \mathbb{Z}_3, \mathbb{Z}_4$, and $\mathbb{Z}_6$ \cite{sendas}. Taking
${\cal G}$ as $\mathbb{Z}_2$ for simplicity, this means that over all points of ${\cal M}_2$ the manifold is $\mathbb{T}^2$ except at {\it four} points where the metric actually becomes ${\mathbb{T}^2\over \mathbb{Z}_2}$. This means the choice $g_{ab} = \eta_{ab}$ that we took in section \ref{kocu3}, holds at all points over ${\cal M}_2$ and only develops a cross-term $g_{ab}$ at four points. Clearly then, $g_{ab} \equiv 
g_{ab}(y^\alpha)$, or more generically, $g_{ab} \equiv g_{ab}(y^m, y^\alpha)$, but  {\it cannot} be a function of
$y^a$ (dependence on $y^a = x^3$ will ruin the T-duality rules that connect IIB to M-theory). This precisely makes \eqref{pola2mey} non-zero resulting in non-zero contributions to  ${\bf X}_8$ from these points. Such a procedure intimately connects the non-zero Euler characteristics with the existence of non-trivial fluxes, a fact rather well known from earlier studies \cite{BB, DRS}.   

To determine the behavior of the other ${\bf X}_8$ polynomials we will be required to determine a few more curvature two-forms, in addition to demanding the existence of certain odd-dimensional cycles inside our eight-manifold \eqref{melisett}. The curvature forms that we require may be listed as:
\bg\label{polamaala}
{\bf  R}^{a_o b_o}_{0b} &=&  g_s^0 ~{\rm R}^{a_o b_o}_{[ab]} \equiv 
{\bf R}_{0b0a} ~e^{a_o 0} e^{b_o a} \nonumber\\
{\bf  R}^{a_o b_o}_{0M} &=&  g_s^{-1} ~{\rm R}^{a_o b_o}_{[0M]} + {\cal O}(F_1, F_2)\nonumber\\
&\equiv& {\bf R}_{0MPQ} ~e^{a_o P} e^{b_o Q}  
+ {\bf R}_{0Mij} ~e^{a_o i} e^{b_o j} + {\bf R}_{0Mab} ~e^{a_o a} e^{b_o b} 
+ {\bf R}_{0M0N} ~e^{a_o 0} e^{b_o N}\nonumber\\
{\bf  R}^{a_o b_o}_{ij} &=&  g_s^{-2} ~{\rm R}^{a_o b_o}_{[ij]} + {\cal O}(F_1, F_2)\nonumber\\
&\equiv& {\bf R}_{ijPQ} ~e^{a_o P} e^{b_o Q}  
+ {\bf R}_{iji'j'} ~e^{a_o i'} e^{b_o j'} + {\bf R}_{ijab} ~e^{a_o a} e^{b_o b} 
+ {\bf R}_{ij0N} ~e^{a_o 0} e^{b_o N}, \nd
where note that the first curvature two-form do not have any $F_i$ dependence, which stems from the fact that the $F_i(t)$ factors only effect the metric along ${\cal M}_4 \times {\cal M}_2$. We can now use 
\eqref{polamaala} to construct a more generic curvature form than the one given earlier in \eqref{beachmaa}, in the following way:

{\footnotesize
\bg\label{beachmaa2}
\mathbb{R}_{\rm tot} \equiv \mathbb{R} + {\bf R}^{a_o b_o}_{\mu M} {\bf M}_{a_o b_0} ~dx^\mu \wedge dy^M + 
{\bf R}^{a_o b_o}_{\mu\nu} {\bf M}_{a_o b_o} ~dx^\mu \wedge dx^\nu + {\bf R}^{a_o b_o}_{\mu b} {\bf M}_{a_o b_0} ~dx^\mu \wedge dy^b, \nd}
which is not just restricted to the eight-manifold \eqref{melisett}; and $(\mu, \nu) \in {\bf R}^{2, 1}$. 
Here $\mathbb{R}$ is the curvature form 
\eqref{beachmaa}. Using these we can construct three other polynomials by replacing $\mathbb{R}$ in 
\eqref{poladom} by $\mathbb{R}_{\rm tot}$, and restricting the polynomials to the following sub-manifolds:

{\footnotesize
\bg\label{polathai}
{\bf X}_8\Big\vert_{{\bf R}^{2, 1} \times \mathbb{C}_5} = g_s^{-3} \widetilde{\bf X}^{(1)}_8,~~~
{\bf X}_8\Big\vert_{{\bf R}^{2, 1} \times \mathbb{C}_4 \times {\bf S}^1} = g_s^{-2} \widetilde{\bf X}^{(2)}_8, ~~~
{\bf X}_8\Big\vert_{{\bf R}^{2, 1} \times \mathbb{C}_3 \times \mathbb{T}^2/{\cal G}} = 
g_s^{-1} \widetilde{\bf X}^{(3)}_8, \nd}
where $\mathbb{C}_3$, $\mathbb{C}_4$ and $\mathbb{C}_5$ are respectively three, four and five cycles in 
${\cal M}_4 \times {\cal M}_2$; and ${\bf S}^1$ is a one-cycle in ${\mathbb{T}^2\over {\cal G}}$. Globally none of these odd cycles might exist, but in  section \ref{maryse}, when we derive the G-flux EOMs, we will only be concerned with local odd-cycles, so global criteria of non-existence would not matter too much. 
There are also ${\cal O}(F_1, F_2)$ corrections that will accompany each of the polynomials in \eqref{polathai}. Additionally, all these eight-manifolds, other than the one of \eqref{polameys}, are non-compact, so there are no additional integral conditions on the corresponding ${\bf X}_8$ polynomials. 

\vskip.1in

\noindent{\it Anomaly cancellation conditions and time-dependent G-fluxes}

\vskip.1in

\noindent Let us now come to the anomaly cancellation conditions from \eqref{marbrick}. This equation should now naturally split into at least two parts: one, that is time-independent (i.e independent of $g_s$), and two, that depends on time, and hence on $g_s$. It appears that, out of the four set of pieces 
in \eqref{marbrick}, only one set is apparently time independent. It is the number $n_b$ and 
$\bar{n}_b$ of M2 and $\overline{\rm M2}$-branes. On the other hand,  if there exists a time-independent part of $\mathbb{Y}_8$ that is related to the Euler characteristics of the eight-manifold, as in \eqref{poladom}, and we take $\chi_8 > 0$, then we expect the first anomaly cancellation condition to become:
\bg\label{haroldR}
n_b  - \bar{n}_b =  {b_3 \over 4!(2\pi)^4}~\chi_8, \nd
where $b_3$ is the factor that depends on $\omega_o$ and $M_p$. Thus if $\chi_8$ remains a time-independent quantity, the Euler characteristics of the internal manifold \eqref{sun18c}, even for a 
non-K\"ahler eight-manifold,  governs the number of {\it static} M2-branes in our model in some sense\footnote{We expect this to change once we impose the second moding scheme introduced in the paragraphs between \eqref{evabmey2} and \eqref{teenangul}, at least for the case \eqref{olokhi}. This is because $k \ge 0$ for the G-flux components 
${\bf G}_{mnpa}$ and ${\bf G}_{q\alpha\beta b}$ and therefore enters the anomaly cancellation condition 
\eqref{marbrick} giving rise to the condition similar to the one encountered in the last reference of \cite{BB}. For the case \eqref{ranjhita}, the change is similar because $k \ge 0$ is now for the G-flux components ${\bf G}_{mnpq}, {\bf G}_{mnp\alpha}$ and ${\bf G}_{mnpa}$.}. 
Since the number of M2 and $\overline{\rm M2}$-branes have to be an integer, the equation \eqref{haroldR} puts an extra constraint on $b_3$ and the Euler characteristics of the eight-manifold itself, namely the combination on the RHS of \eqref{haroldR} should be an integer. Such a condition should be reminiscent of a similar condition in the second reference of 
\cite{BB}, and here we see that in a time-dependent background, \eqref{haroldR} is realized instead of the full anomaly cancellation condition with G-fluxes of \cite{DRS} (see also the last reference of \cite{BB}).

There are two assumptions that have inadvertently gone in that requires special attention. One, even if some part of $\mathbb{Y}_8$ remains time-independent, there is no apparent reason why the integral could be related to the Euler characteristics of the eight-manifold (although it could still be a topological quantity); and two, as we saw in \eqref{polameys} and \eqref{polathai} it is not clear there exists a $g_s$ independent 
polynomial ${\bf X}_8$ in our construction.  This means,  as
the integral of ${\bf X}_8$ becomes a $g_s$ {\it dependent} function, it would leave only the M2 and the $\overline{\rm M2}$-branes to be time-independent pieces. Therefore \eqref{haroldR} would  make sense if $b_3 = 0$, giving us $n_b = \bar{n}_b$, i.e equal number of M2 and anti-M2 branes. This will also become clear from \eqref{evaB102}, when we derive the G-flux EOMs
in section \ref{maryse}. 

For the time-dependent parts of \eqref{marbrick} there are a couple of subtleties. One, we need to tread carefully as various parts of the G-flux components have different $g_s$ scalings; and two, time-dependent contributions now come from both topological and non-topological parts of \eqref{marbrick}. In fact the non-topological piece, given in terms of $\ast_{11} \mathbb{Y}_4$, is solely time dependent as it is constructed out of the quantum terms \eqref{phingsha} or \eqref{phingsha2}  as shown in 
\eqref{aliceL2}. On the other hand, the topological part does have a time independent piece as seen from \eqref{poladom}. Combining everything together, our second anomaly cancellation condition may be expressed as:
\bg\label{technox1}
b_2 \int_{{\cal M}_8} {\bf G}_4 \wedge {\bf G}_4  + b_3 \int_{{\cal M}_8} \Big(\mathbb{Y}_8 
-{\bf X}_8\Big) 
-b_4\int_{{\cal M}_8} d\ast_{11} \mathbb{Y}_4 = -b_3\int_{{\cal M}_8} {\bf X}_8, \nd
which is in fact not a {\it single} condition, rather it is an infinite number of conditions on various components of the G-fluxes and the quantum terms. We have also divided the $\mathbb{Y}_8$ part into two parts, one that depends exclusively on curvature forms $\mathbb{R}$ from \eqref{beachmaa}, and the other that depends on G-fluxes and curvature forms. The first part scales as $g_s^2$ while the second part has more complicated $g_s$ dependent scalings that contributes only if $(c_3, c_4, c_5, ..)$ in \eqref{ashf2} take non-zero values. For the simplest case, $\mathbb{Y}_8$ could then be identified to ${\bf X}_8$ only, and this would scale as $g_s^2$ from \eqref{polameys} as mentioned above. The G-flux components 
${\cal G}^{(k)}_{MNPQ}, {\cal G}^{(k)}_{MNPa}$ and ${\cal G}^{(k)}_{MNab}$ all scale as 
$\left({g_s\over H}\right)^{2\Delta k}$, and for $k \ge {3\over 2}$, the minimum scaling for each of these components would be $\left({g_s\over H}\right)^{3\Delta} = {g_s\over H}$ for $\Delta = {1\over 3}$. This means that the quadratic term in G-flux components, namely the coefficient of $b_2$ part in \eqref{technox1}, scales as $g_s^2$, thus matching up exactly with the $g_s^2$ scaling from the ${\bf X}_8$ part 
in \eqref{polameys}. This matching is encouraging, but the G-flux components also have higher order terms 
for $k > {3\over 2}$. They could in principle be matched with the ${\cal O}(F_1, F_2)$ parts of \eqref{polameys}, but that cannot be the full story, as we also have the quantum terms, namely the coefficients of $b_4$ in \eqref{technox1}. How do the anomaly cancellation work when the $b_4$ and $b_3$ terms are switched on?
To see this, we plug in in the G-flux components and the quantum series in \eqref{technox1}, to get:

{\footnotesize
\bg\label{cleota}
&&b_2 \sum_{\{k_i\}}\int_{{\cal M}_8} {\cal G}^{(k_1)}_{N_1N_2N_3N_4} 
{\cal G}^{(k_2)}_{N_5N_6N_7N_8} \left({g_s\over H}\right)^{2\Delta(k_1 + k_2)} 
dy^{N_1} \wedge .... \wedge dy^{N_8}
+ b_3 \sum_{k_3}\int_{{\cal M}_8} {\bf X}_8^{(k_3)}\left({g_s\over H}\right)^{2(1+\Delta k_3)}\nonumber\\
&& = b_4 \sum_{k} \int_{{\cal M}_8} \partial_{N_8}\left( \sqrt{-g_{11}}
\left(\mathbb{Y}_4^{(k)}\right)_{M'_8...M'_{11}}
{g}^{M_8M'_8}...{g}^{M_{11}M'_{11}}\left({g_s\over H}\right)^{\tilde{\theta}_k}
\right)\epsilon_{N_1...N_7M_8...M_{11}} dy^{N_1} \wedge ....
\wedge dy^{N_{8}}, \nonumber\\ \nd}
where we see that the RHS is expressed in terms of a total derivative and un-warped metric components
and we took ${\bf X}_8^{(k_3)} \equiv \left({\bf X}_8^{(k_3)}\right)_{N_1...N_8}$ from \eqref{polameys}. 
Since ${\cal M}_8$ is a compact eight-manifold without a boundary, generically the RHS would vanish.
 However if $d\ast_{11}\mathbb{Y}_4$ is only a locally-exact form then there is a chance that it may not. 
 This could happen if some of the components entering $\ast_{11}\mathbb{Y}_4$, say the metric and the flux components, are  
{\it not} globally defined. This is like the 
${\bf X}_8$ form that is expressed as a locally-exact form $d{\bf X}_7$ where ${\bf X}_7$ is not a globally defined form on a compact eight-manifold. This could in principle make the RHS non-zero even in the absence of any boundary. If this is the case then the $g_s$ scaling $\tilde{\theta}_k$ appearing 
in \eqref{cleota} may be defined as:
\bg\label{alybrite}
\tilde{\theta}_k \equiv \theta'_k - {2\over 3}, ~~~~~ \tilde{\theta}_k \equiv \theta_k + {4 \over 3}, \nd
for the two cases, \eqref{olokhi} and \eqref{ranjhita} respectively where $\theta'_k$ and $\theta_k$ are defined as in 
\eqref{melamon2} and \eqref{miai} respectively. The anomaly cancellation condition then requires us to match the $g_s$ scalings on both sides of the equation \eqref{cleota}. This gives us:
\bg\label{ramamey}
&& \theta'_k = {2\over 3}\left(k_1 + k_2 +1\right) = {2\over 3}(4 + k_3) , ~~~~~ \left(k_1, k_2\right) \ge \left({3\over 2}, ~{3\over 2}\right) \nonumber\\
&&\theta_k = {2\over 3}\left(k_1 + k_2 - 2\right) = {2\over 3}(\hat{\gamma} + k_3), ~~~~~ \left(k_1, k_2\right) \ge \left({9\over 2}, ~
{9\over 2}\right),
\nd
as the set of anomaly cancellation conditions for the two cases \eqref{olokhi} and \eqref{ranjhita} respectively with $\hat{\gamma} \ge 4$. As a check one may see that, for the case \eqref{olokhi}, $k_1 + k_2 \ge 3$ and therefore $k_i > 0$ implying no time-{\it independent} G-flux components. Interestingly,  with locally-exact form $d\ast_{11}\mathbb{Y}_4$, inserting $k_1 = k_2 = {3\over 2}$, would imply $\theta'_k = {8\over 3}$ and therefore involves the same set of quantum terms that we had for example in \eqref{fleuve}, wherein the quantum terms were classified by \eqref{oleport}. This makes sense because the equation governing the
G-flux components is as in \eqref{marianc}, and therefore if we restrict the LHS of \eqref{marianc} to the G-flux components ${\bf G}_{0ijm}$ or ${\bf G}_{0ij\alpha}$, then it appears as though the LHS may be expressed in terms of 
$\square H^4$ exactly as in \eqref{fleuve}. This identification is a bit subtle, and we will clarify it in subsection \ref{maryse}. 
In fact, as we shall see there,  the similarity goes even deeper: \eqref{fleuve} has the same number of ingredients as \eqref{marianc}, for example there are M2-branes, fluxes and quantum corrections almost in one-to-one correspondence to \eqref{marianc}. 

There is however at least one crucial difference between \eqref{fleuve} and \eqref{marianc} apart from the appearance of the $b_3$ factor in the latter. The difference lies in the choice of the G-flux components themselves: \eqref{fleuve} is defined in terms of ${\cal G}^{(k)}_{MNab}$ components whereas \eqref{marianc} 
involves $\ast_8 {\cal G}^{(k)}_{MNab}$ components, with $\ast_8$ being the Hodge dual over the internal eight-manifold. For the time-independent case this observation has already been registered in \cite{nogo} (see 
eq. (7.11) therein), and now we see that such a case happens here too. It is easy to show that in general the G-flux components are no longer self-dual, where the self-duality is defined with respect to the internal eight-dimensional space. In fact presence of self-duality would have been a sign of supersymmetry, but since supersymmetry is broken, it is no surprise that we see non self-dual G-flux components. This will be elaborated in more details in section \ref{maryse}.

For the case \eqref{ranjhita} governed by $\theta_k$ in \eqref{miai}, there appears to be some mis-match if we compare to \eqref{bratmey}. On one hand, again assuming locally-exact form $d\ast_{11}\mathbb{Y}_4$ and taking $k_1 = k_2 = {9\over 2}$, we get $\theta_k = {14\over 3}$ from \eqref{ramamey}. On the other hand, \eqref{bratmey} tells us that the quantum terms are classified by $\theta_k = {8\over 3}$
in \eqref{bratmey}. This difference may be attributed to the multiple constraints appearing from 
\eqref{ouletL}, vanishing Ricci scalar for the six-dimensional base, and vanishing Euler characteristics for the eight-manifold, or even the vanishing of the RHS of \eqref{cleota} by instead taking a globally-exact form 
$d\ast_{11}\mathbb{Y}_4$; and therefore a simple comparison between the set of equations cannot be performed.  

However a more likely scenario is that  \eqref{bratmey} is {\it not} the correct EOM, and the correct EOM for this case is actually \eqref{lebanmey}. In fact the similarity of \eqref{lebanmey} with \eqref{fleuve}, and the fact that the quantum terms are classified by $\theta_k = {14\over 3}$ puts extra confidence in the 
\eqref{lebanmey} to be the correct EOM. Taking this to be the case, and comparing \eqref{lebanmey} and 
\eqref{cleota}, we again observe the non-existence of self-dual fluxes. The number of flux components in 
\eqref{cleota} do not match with the ones in \eqref{lebanmey}, but if we only allow components 
${\cal G}^{(9/2)}_{\alpha\beta ab}$ in \eqref{cleota} then the story would be exactly similar to what we had for the case \eqref{olokhi}, reassuring, in turn, the correctness of our procedure so far. Thus
we see that the anomaly cancellation provides useful consistency checks on our earlier EOMs derived using Einstein's equations\footnote{In retrospect this could in principle justify taking a locally-exact form
$d\ast_{11}\mathbb{Y}_4$, although if we take a globally-exact form, then it would not contribute to the RHS of 
\eqref{cleota} as an integral condition but would contribute {\it locally}. As we shall discuss in subsection 
\ref{maryse}, this is in concordance with the EOMs from the Einstein's equations. Interestingly however if we are not careful there might appear apparent mis-match between the two set of equations with the second moding schemes of the G-flux components described in the paragraphs between \eqref{evabmey2} and 
\eqref{teenangul}.  For example for the case \eqref{olokhi}, the second moding scheme implies $k \ge {3\over 2}$ only for the G-flux components ${\bf G}_{MNab}$, and $k \ge 0$ for the other components. Plugging this in \eqref{ramamey} gives us $\theta'_k = {5\over 3}$ which is different from the expected classification of $\theta'_k = {8\over 3}$ as well as a contradiction with $k_1 + k_2 = 3 + k_3$. This apparent mis-match is because the moding scheme only provides the lower bounds on $k$. Our analysis reveals that 
$(k_1, k_2) > (0, 0)$ and therefore the second moding scheme will only make sense if we take this into account.  This again justifies the choice of temporally varying degrees of freedom. \label{plazamey}}.  

\subsubsection{Flux equations along various directions \label{maryse}}

In the above section we discussed how anomaly cancellation conditions may be understood from the constraint equations.  The key equation appears to be \eqref{cleota} which helps us to pitch the higher order flux components to the higher order quantum terms. However an unsatisfactory, or more appropriately an incomplete, feature of the above analysis is that the condition \eqref{cleota} is an integral condition. Are there local conditions for the flux components?

These local conditions are of course the equations of motion for the flux components, which in turn would determine how these components have their spatial spread. The temporal evolutions of these flux components, i.e their $g_s$ behavior,  are already accounted for in the ansatze \eqref{ravali} for the fluxes, and therefore the EOMs will determine a second set of consistency conditions for them. In the following, we will determine these behavior for all the allowed components.

\vskip.1in

\noindent{\it Case 1: ${\bf G}_{0ijM}$ components}

\vskip.1in

\noindent The behavior for the ${\bf G}_{0ijM}$ is known both in terms of it's $g_s$ as well as $y^m$ dependences, where $(M, m) \in {\cal M}_4 \times {\cal M}_2$.
Typically this scales as $\left({g_s\over H}\right)^{-4}$, but a more complete derivation of it's behavior may be ascertained in the presence of {\it dynamical} branes. Thus dynamical branes allow an alternative derivation of this component, which we shall discuss in section \ref{branuliat}. The final answer is simple and is given by \eqref{kyratagra}, and here we shall ask what this would imply for the warp-factor $H(y)$ etc., once we express the EOM in the following way:
\bg\label{evaB10}
&& \partial_{N_8}\Big(\sqrt{-{\bf g}_{11}}~{\bf G}_{0ijM} ~{\bf g}^{00'} {\bf g}^{ii'} {\bf g}^{jj'} {\bf g}^{MM'}\Big)
\epsilon_{0'i'j'M'N_1 N_2 ...... N_7}\\
&& ~~~= b_4~\partial_{N_8}\Big(\sqrt{-{\bf g}_{11}} \left(\mathbb{Y}_4\right)_{0ijM} 
{\bf g}^{00'} {\bf g}^{ii'} {\bf g}^{jj'} {\bf g}^{MM'}\Big)\epsilon_{0'i'j'M'N_1.....N_7}\nonumber\\
&&~~~+ b_1~{\bf G}_{N_1.....N_4} {\bf G}_{N_5.....N_8} + \left(\mathbb{Y}_8\right)_{N_1......N_8} + 
T_2\Big(n_b \delta^8(y - y_1) - \bar{n}_b \delta^8(y - y_2)\Big) \epsilon_{N_1.....N_8}, \nonumber\nd
where $b_i$ are $g_s$ independent constants (but could depend on $M_p$), $n_b$ and ${\bar{n}_b}$ are the number of M2 and anti-M2 branes, $T_2$ is the brane tension (we take $2\kappa^2 = 1$), 
$N_i \in$ 8-manifold,
$\mathbb{Y}_8$ is defined in \eqref{ashf2} and the $\mathbb{Y}_4$ appears in \eqref{ashf4} and 
\eqref{aliceL2}. In the following we will be more interested in the ${\bf X}_8$ part of the $\mathbb{Y}_8$ polynomial, as it's $g_s$ scaling may be easily determined. In fact, as we did for the Einstein's equations in the earlier sections, it'll be useful to express \eqref{evaB10} in terms of the $g_s$ scalings of the various terms. This goes as:

{\footnotesize
\bg\label{evaB101}
&&-\square H^4\sum_{\{k_i\}} C_{k_1} \widetilde{C}_{k_2}\left({g_s\over H}\right)^{2\Delta(k_1 + k_2)}
+ {1\over \sqrt{g_8}}\sum_{\{k_i\}} \partial_{N_8}\Big(\sqrt{g_8} ~H^8 {\cal G}^{(k_3)}_{012M} g^{MN_8}\Big) 
C_{k'_1} \widetilde{C}_{k'_2} \left({g_s\over H}\right)^{2\Delta(k_3 + k'_1 + k'_2)} \nonumber\\
&& ~~~~= b_1 \sum_{\{k_i\}} {\cal G}^{(k_4)}_{N_1... N_4} \left(\ast_8{\cal G}^{(k_5)}\right)^{N_1...N_4}
\left({g_s\over H}\right)^{2\Delta(k_4 + k_5)} + {g_s^2\over \sqrt{g_8}} \left(\widetilde{\bf X}_8\right)_{N_1...N_8} \epsilon^{N_1...N_8} \nonumber\\
&& ~~~~+ {b_4\over \sqrt{g_8}}\sum_{\{k\}} \partial_{N_8}\Big(\sqrt{g_8}\left(\mathbb{Y}^{(k)}_4\right)^{012N_8}\Big)
\left({g_s\over H}\right)^{\theta'_k - 2/3} +  {T_2\over \sqrt{g_8}}
\Big(n_b \delta^8(y - y_1) - \bar{n}_b \delta^8(y - y_2)\Big), 
\nd}
where all raising and lowering are done using the {\it un-warped} metric components, including the Hodge star and the $\widetilde{\bf X}_8$ polynomial \eqref{polameys} defined over the internal eight manifold. We have also used
$\epsilon_{N_1 N_2...} \epsilon^{N'_1 N'_2...} = \delta^{N'_1}_{N_1} \delta^{N'_2}_{N_2}...
\delta^{N'_8}_{N_8}$. The flux components are defined as in \eqref{ravali} and \eqref{kyratagra} and 
$(C_k, \widetilde{C}_k)$ appear in \eqref{karishma}. 
The $g_s$ scalings in \eqref{evaB101} are discussed for the case \eqref{olokhi}, and we shall stick with this for this section unless mentioned otherwise. This means $\theta'_k$ that appears above is from \eqref{melamon2}. The $g_s^2$ scaling for ${\bf X}_8$ is interesting, and we could have gone with the higher order $g_s$ dependence as discussed before, but we will not do here. The lowest order in $g_s$, i.e the zeroth order in $g_s$, equation may be easily read off from \eqref{evaB101} as\footnote{One might worry that there could be potential contributions to \eqref{evaB102} from the fifth term of \eqref{evaB101}. The fifth term, i.e the quantum term, scales as $\theta'_k = {2\over 3}$ which can only get contributions from 
$l_{34} = 2$ or $l_{35} = 2$ in \eqref{melamon2}.  But this is exactly the first term, so we can think of this as simply changing the coefficient of the first term. As such this is harmless. The second term in \eqref{evaB101} does not contribute because $k_3 > 0$ from \eqref{kyratagra}. \label{kalaryan}} :
\bg\label{evaB102}
-\square H^4 = {T_2\over \sqrt{g_8}}\Big(n_b \delta^8(y - y_1) - \bar{n}_b \delta^8(y - y_2)\Big), \nd
which determines the warp-factor $H(y)$ completely in terms of the M2-branes and the anti-M2-branes. The lowest order $g_s$ scaling is determined from the fact that $(k_1, k_2, k'_1, k'_2) \ge 0, k_3 \ge {1\over 2}$ 
and $(k_4, k_5) \ge {3\over 2}$ in \eqref{evaB101}. Using $\Delta = {1\over 3}$, it is easy to see that the $k_i$ appearing above are further related to each other by:
\bg\label{evaBB}
k_1 + k_2 = k'_1 + k'_2 + k_3 = k_4 + k_5 = {3\over 2} \left(\theta'_k - {2\over 3}\right), \nd
where $\theta'_k$ is as in \eqref{melamon2}.  Note the absence of both ${\bf X}_8$ polynomial as well as any components of the G-flux. This is important: both the fluxes and curvatures have no $g_s$ {\it independent} parts here. If by any chance there would be $g_s$ independent flux components, or if the scaling analysis for the curvature forms reveal $g^0_s$ pieces, they would enter \eqref{evaBB}. For a discussion on this, one may refer to \cite{maxpaper}.  To order $g_s^2$, the flux equation may be written as:

{\footnotesize
\bg\label{evaBgon}
&&-\square H^4\sum_{\{k_i\}} C_{k_1} \widetilde{C}_{k_2}~\delta(k_1 + k_2 - 3)
+ {1\over \sqrt{g_8}}\sum_{\{k_i\}} \partial_{N_8}\Big(\sqrt{g_8} ~H^8 {\cal G}^{(k_3)}_{012M} g^{MN_8}\Big) 
C_{k_1} \widetilde{C}_{k_2} ~\delta(k_1 + k_2 + k_3 - 3) \nonumber\\
&& = b_1 \sum_{\{k_i\}} {\cal G}^{(3/2)}_{N_1... N_4} \left(\ast_8{\cal G}^{(3/2)}\right)^{N_1...N_4}
+ {1 \over \sqrt{g_8}} [\widetilde{\bf X}_8]
+ {b_4\over \sqrt{g_8}}\sum_{\{k\}} \partial_{N_8}\Big(\sqrt{g_8}\left(\mathbb{Y}^{(k)}_4\right)^{012N_8}\Big)
\delta\left(\theta'_k - {8\over 3}\right), 
\nd}
where $[\widetilde{\bf X}_8] \equiv (\widetilde{\bf X}_8)_{N_1...N_8} \epsilon^{N_1...N_8}$, is the index-free notation. Interestingly, the quantum terms are classified by $\theta'_k = {8\over 3}$  for $\theta'_k$ as in 
\eqref{melamon2}. One should now compare \eqref{evaBgon} to \eqref{muse777}: both these equations are determined in terms of $\square H^4$, square of the G-flux components ${\cal G}^{(3/2)}_{MNab}$, and the quantum terms that are classified by $\theta'_k \le {8\over 3}$. The inequality sign comes from the contributions of the non-perturbative terms, discussed in section \ref{instachela}, to the Einstein's EOMs. Comparing \eqref{evaBgon} with \eqref{fleuve}, we see that:
\bg\label{evebe}
\left\vert {\cal G}^{(3/2)}_{MNab} - (\ast {\cal G}^{(3/2)})_{MNab} \right\vert  ~ > ~ 0, \nd
showing that the relevant flux components that appear at the lowest order in $g_s$, thus contributing to both Einstein and G-flux EOMs, {\it cannot} be self-dual. This is a clear signal of supersymmetry 
breaking.  In the coherent state description of the de Sitter space, this breaking of supersymmety via non self-dual fluxes shows that the coherent state itself breaks supersymmetry whereas the vacuum remains supersymmetric \cite{coherbeta}.

\vskip.1in

\noindent{\it Case 2: ${\bf G}_{MNPQ}$ components}

\vskip.1in

\noindent The equations of motion for the ${\bf G}_{MNPQ}$ components where 
$(M, N) \in {\cal M}_4 \times {\cal M}_2$ are somewhat similar to what we discussed above. The precise 
EOM for these components may be expressed as:
\bg\label{evaB00}
&& \partial_{N_8}\Big(\sqrt{-{\bf g}_{11}}~{\bf G}_{MNPQ} ~{\bf g}^{MM'} {\bf g}^{NN'} {\bf g}^{PP'} 
{\bf g}^{QQ'}\Big)
\epsilon_{M'N'P'Q'N_1 N_2 ...... N_7}\nonumber\\
&&~~~= b_1~{\bf G}_{0ijQ} {\bf G}_{MNab}~\delta^{[0}_{[N_1} \delta^i_{N_2} \delta^j_{N_3}\delta^Q_{N_4}
\delta^M_{N_5} \delta^N_{N_6} \delta^a_{N_7} \delta^{b]}_{N_8]} + \left({\bf X}_8\right)_{N_1......N_8}\\
&& ~~~+ b_4~\partial_{N_8}\Big(\sqrt{-{\bf g}_{11}} \left(\mathbb{Y}_4\right)_{MNPQ} 
{\bf g}^{MM'} {\bf g}^{NN'} {\bf g}^{PP'} {\bf g}^{QQ'}\Big)\epsilon_{M'N'P'Q'N_1.....N_7}, \nonumber\nd
where all the metric components appearing above are the {\it warped} ones. Note the absence of the brane terms compared to \eqref{evaB10}: any branes wrapping internal cycles will break the de Sitter isometries, so they cannot contribute here. As before, what is now important is to match the various $g_s$ components from the above equation. This means we need to see how every term of \eqref{evaB00} scale with $g_s$. This is not hard, and the answer is:

{\footnotesize
\bg\label{evaB000}
&& \sum_{\{k_i\}}\partial_{N_8}\Big(\sqrt{{g}_{8}}~{\cal G}^{(k_1)}_{MNPQ} ~{g}^{MM'} {g}^{NN'} {g}^{PP'} 
{g}^{QQ'}\Big) C_{k_2} C_{k_3}\left({g_s\over H}\right)^{2\Delta(k_1 + k_2 + k_3) - 2}
\epsilon_{M'N'P'Q'N_1 N_2 ...... N_7}\nonumber\\
&&~~~= b_1\sum_{\{k_i\}}{\cal G}^{(k_4)}_{0ijQ} {\cal G}^{(k_5)}_{MNab}~
\left({g_s\over H}\right)^{2\Delta(k_4 + k_5) - 4}
\delta^{[0}_{[N_1} \delta^i_{N_2} \delta^j_{N_3}\delta^Q_{N_4}
\delta^M_{N_5} \delta^N_{N_6} \delta^a_{N_7} \delta^{b]}_{N_8]} + g_s^{-1} \left(\widetilde{X}^{(3)}_8\right)_{N_1......N_8}\nonumber\\
&& ~~~+ b_4 \sum_{\{k,k'\}}\partial_{N_8}\Big(\sqrt{{g}_{8}} (\mathbb{Y}^{(k)}_4)^{M'N'P'Q'} 
\Big)\left({g_s\over H}\right)^{\theta'_k - 2\Delta k' - 14/3} \epsilon_{M'N'P'Q'N_1.....N_7}, \nd}
where, other than the variations of the components of the fluxes, we now have $C_{k_1}$ and $C_{k_2}$ 
from \eqref{karishma} with no contributions from $\widetilde{C}_{k}$. The quartic curvature forms in 
the definition of ${\bf X}_8$ in \eqref{polathai} now scales as $g_s^{-1}$ if we ignore the contributions from \eqref{karishma}; and $\theta'_k$ is defined in \eqref{melamon2}. The $g_s$ scalings of all the other terms in \eqref{evaB000} can now be matched as:
\bg\label{briwhit}
2\Delta(k_1 + k_2 + k_3) - 2 = 2\Delta(k_4 + k_5) - 4 = \theta'_k - 2\Delta k' - {14\over 3}, \nd 
where $(k_1, k_5) \ge (3/2, 3/2)$ and $(k_2, k_3, k_4) \ge (0, 0, 0)$ with $k' \ge 3/2$ in the flux sector. 
We have also defined ${\cal G}^{(0)}_{0ijQ} \equiv  -\partial_Q\left({\epsilon_{0ij} \over H^4}\right)$ from 
\eqref{kyratagra}. Choosing $\Delta = {1\over 3}$, it is easy to see that the $k_i$ are related by 
$k_1 + k_2 + k_3 + 3 = k_4 + k_5$. The quantum terms are then classified by:
\bg\label{shonpapdi}
\theta'_k = {2\over 3}\left(1 + k' + k_4 + k_5\right) = {2\over 3}\left(4 + k' + k_1 + k_2 + k_3\right), \nd
telling us that if $k_1 = {3\over 2}$ and $k_2 = k _3 = 0$, then $k_5 = {3\over 2}$ is only achieved with 
$k_4 = 3$. This means, as will become clearer from section \ref{branuliat}, dynamical branes become necessary to realize this. This also means that 
the lowest value of $\theta'_k = {14\over 3}$ with $\theta'_k$ as in 
\eqref{melamon2}, implying that the $g_s$ scalings of each terms go as $g_s^{-1}$. Interestingly, this is also the $g_s$ scaling of the ${\bf X}_8$ term, and therefore the lowest order equation may be expressed as:
\bg\label{evaBladdu}
&& \partial_{N_8}\Big(\sqrt{{g}_{8}}~{\cal G}^{(3/2)}_{MNPQ} ~{g}^{MM'} {g}^{NN'} {g}^{PP'} 
{g}^{QQ'}\Big) 
\epsilon_{M'N'P'Q'N_1 N_2 ...... N_7}\nonumber\\
&&~~~= b_1 ~{\cal G}^{(k_1)}_{0ijQ} {\cal G}^{(k_2)}_{MNab}~
\delta\left(k_1 + k_2- {9\over 2}\right)
\delta^{[0}_{[N_1} \delta^i_{N_2} \delta^j_{N_3}\delta^Q_{N_4}
\delta^M_{N_5} \delta^N_{N_6} \delta^a_{N_7} \delta^{b]}_{N_8]} + \left(\widetilde{X}^{(3)}_8\right)_{N_1......N_8}\nonumber\\
&& ~~~+ b_4\sum_{\{k,k'\}} \partial_{N_8}\Big(\sqrt{{g}_{8}} (\mathbb{Y}^{(k)}_4)^{M'N'P'Q'} 
\Big)\delta\left(\theta'_k - 2\Delta k' - {11\over 3}\right) \epsilon_{M'N'P'Q'N_1.....N_7}, \nd 
where in the last term, once we choose $k' = {3\over 2}$, $\theta'_k = {14\over 3}$ as discussed above. Switching on the $C_{k_i}$ coefficients, for $k_i = {\mathbb{Z}\over 2}$,  in the first line of \eqref{evaBladdu},
and higher $k_j$ for the second line will allow us to go to higher order in $g_s$, for example $g_s^{-2/3}, 
g_s^{-1/3}, g_s^0$ etc. The way we have constructed, ${\bf X}_8$ in \eqref{polathai} do not contribute beyond $g_s^{-1}$, but could be made to do so by switching on $F_i(t)$ factors from \eqref{karishma} in ${\bf X}_8$. The story then progresses in the usual way.

On the other hand, in the absence of dynamical branes, 
 there is a possibility of going below 
$g_s^{-1}$. For example by choosing $k_4 = 0$ and $k_5 = {3\over 2}$ in \eqref{evaB000}, we can go as low as $\left({g_s\over H}\right)^{-3}$. We will however be required to impose 
${\cal G}^{(-\vert k\vert)}_{MNPQ} = 0$, to avoid awkward components from appearing in our equation. The equation governing such components may be expressed in the following way:
\bg\label{cambermar}
\sum_{\{k\}} \partial_{N_8}\Big(\sqrt{{g}_{8}} (\mathbb{Y}^{(k)}_4)^{MNPQ} \Big) \delta\left(\theta'_k - {8\over 3}\right) \epsilon_{MN...N_7}= {b_1\over b_4} ~\partial_Q'
\left({\epsilon_{0ij}\over H^4}\right) {\cal G}^{(3/2)}_{M'N'ab} ~\delta^{[0}_{[N_1}...\delta^{b]}_{N_8]}, \nonumber\\ \nd
with the quantum terms classified by $\theta'_k = {8\over 3}$ in \eqref{melamon2}. There is no integral constraint from \eqref{cambermar} because of the non-compactness of the $2+1$ dimensional space-time. Note that ${\cal G}^{(3/2)}_{MNab}$ appears on {\it both} sides of the equality because $\theta'_k = {8\over 3}$ involve at most eighth order in G-flux components of the form ${\cal G}^{(3/2)}_{MNab}$ and at most quartic order in curvature tensors. The $g_s$ scaling of ${\cal G}^{(k)}_{MNPQ}$ go as 
$2\Delta k + {4\over 3}$, so for $k = {3\over 2}$ we can have at most one such component. This is not a quantum term, so we will need to go to higher values of $\theta'_k$ to allow such components to appear on the LHS of \eqref{cambermar}. The analysis then progresses as above.

\vskip.1in

\noindent{\it Case 3: ${\bf G}_{MNab}$ components}

\vskip.1in

\noindent The next set of components are ${\bf G}_{MNab}$ where $(M, N) \in {\cal M}_4 \times {\cal M}_2$, and $(a, b) \in {\mathbb{T}^2\over {\cal G}}$. These are important components because they occur in the lowest order Einstein's EOMs as we saw in the earlier sections. In \eqref{cambermar} we discussed how these components may be related to each other through the higher order quantum terms, in the absence of the dynamical branes. In the following we will study a more direct way of generating the relation between these components via the following EOM:
\bg\label{evaBoo}
&& \partial_{N_8}\Big(\sqrt{-{\bf g}_{11}}~{\bf G}_{MNab} ~{\bf g}^{MM'} {\bf g}^{NN'} {\bf g}^{aa'} 
{\bf g}^{bb'}\Big)
\epsilon_{M'N'a'b'N_1 N_2 ...... N_7}\nonumber\\
&&~~~= b_1~{\bf G}_{0ijQ} {\bf G}_{MNPR}~\delta^{[0}_{[N_1} \delta^i_{N_2} \delta^j_{N_3}\delta^Q_{N_4}
\delta^M_{N_5} \delta^N_{N_6} \delta^P_{N_7} \delta^{R]}_{N_8]} + \left({\bf X}_8\right)_{N_1......N_8}\\
&& ~~~+ b_4~\partial_{N_8}\Big(\sqrt{-{\bf g}_{11}} \left(\mathbb{Y}_4\right)_{MNab} 
{\bf g}^{MM'} {\bf g}^{NN'} {\bf g}^{aa'} {\bf g}^{bb'}\Big)\epsilon_{M'N'a'b'N_1.....N_7}, \nonumber\nd 
which is very similar to \eqref{evaB00}, the only difference being the choice of the components.  The $g_s$ scalings of each of the terms in \eqref{evaBoo} are a bit different from what we had in \eqref{evaB00}, in the following way:

{\footnotesize
\bg\label{evaBooo}
&& \sum_{\{k_i\}}\partial_{N_8}\Big(\sqrt{{g}_{8}}~{\cal G}^{(k_1)}_{MNab} ~{g}^{MM'} {g}^{NN'} {g}^{aa'} 
{g}^{bb'}\Big) C_{k_2}\left({g_s\over H}\right)^{2\Delta(k_1 + k_2) - 6}
\epsilon_{M'N'P'Q'N_1 N_2 ...... N_7}\nonumber\\
&&~~~= b_1\sum_{\{k_i\}}{\cal G}^{(k_3)}_{0ijQ} {\cal G}^{(k_4)}_{MNPR}~
\left({g_s\over H}\right)^{2\Delta(k_3 + k_4) - 4}
\delta^{[0}_{[N_1} \delta^i_{N_2} \delta^j_{N_3}\delta^Q_{N_4}
\delta^M_{N_5} \delta^N_{N_6} \delta^a_{N_7} \delta^{b]}_{N_8]} + g_s^{-3} \left(\widetilde{X}^{(1)}_8\right)_{N_1......N_8}\nonumber\\
&& ~~~+ b_4 \sum_{\{k,k'\}}\partial_{N_8}\Big(\sqrt{{g}_{8}} (\mathbb{Y}^{(k)}_4)^{M'N'a'b'} 
\Big)\left({g_s\over H}\right)^{\theta'_k - 2\Delta k' - 14/3} \epsilon_{M'N'a'b'N_1.....N_7}, \nd} 
 where note the presence of only $C_{k_2}$ from \eqref{karishma}, compared to $(C_{k_2}, C_{k_3})$ 
 in \eqref{evaB000}. The $g_s$ scalings of both the first line as well as that for ${\bf X}_8$ are noticeably different from what we had in \eqref{evaB000}. The ${\bf X}_8$ scaling come from the curvature forms that we studied in \eqref{polathai}, and again, incorporation of the $F_i(t)$ factors from \eqref{karishma} will change the scaling a bit, but we will not discuss this here. The $g_s$ scalings of the other terms are related to each other via the following relation:
 \bg\label{bonfacha}
 2\Delta(k_1 + k_2) - 6 = 2\Delta(k_3 + k_4) - 4 = \theta'_k - 2\Delta k' - {14\over 3}, \nd
 telling us that $k_1 + k_2 = k_3 + k_4 + 3$ with $(k_1, k_4) \ge (3/2, 3/2)$ whereas $(k_2, k_3) \ge (0, 0)$. 
 The lowest order values are attained when $k_1 = {3\over 2}, k_2 = 0$. This scales as 
 $\left({g_s\over H}\right)^{-5}$. There are no such terms on the second line of \eqref{evaBooo}, so we go to the quantum terms. They scale as $\theta'_k = {2\over 3}$, leading to a term of the form\footnote{Recall that specifying a particular value of $\theta'_k$ in \eqref{melamon2}, we are in fact looking at the fully Lorentz {\it invariant} quantum series in \eqref{phingsha2}. For the G-flux EOMs, one can then {\it eliminate} one of the relevant G-flux components. This way the scaling becomes $\theta'_k - 2\Delta k'$, where $k' \ge {3\over 2}$, as advocated earlier in \eqref{elroyale}.}: 
 \bg\label{pierette}
 \left(\mathbb{Y}^{(3/2)}_4\right)^{M'N'a'b'} = b_5~{\cal G}^{(3/2)}_{MNab} ~{g}^{MM'} {g}^{NN'} {g}^{aa'} 
 g^{bb'}, \nd 
 up-to a possible constant $b_5$. This is in fact the first line itself, provided $b_5 b_4 = 1$, showing that the lowest order equation doesn't reveal anything new. 
 
 It is encouraging to see some consistency appearing from \eqref{evaBooo}. Let us now go to the next order, which is determined for $(k_1, k_2)$ in \eqref{evaBooo} taking the two set of values: $\left({3\over 2}, {1\over 2}\right)$ and 
 $(2, 0)$. The first line of \eqref{evaBooo} then scales as $\left({g_s\over H}\right)^{-14/3}$, implying that the 
 quantum terms should be classified by $\theta'_k = 1$ in \eqref{melamon2} with $k' = {3\over 2}$. This means:

{\footnotesize
\bg\label{pierettetag}
 \left(\mathbb{Y}^{(3/2)}_4\right)^{M'N'a'b'} = b_6~{\cal G}^{(2)}_{MNab} ~{g}^{MM'} {g}^{NN'} {g}^{aa'} 
 g^{bb'} + b_7 ~{\cal G}^{(3/2)}_{MPac} {\cal G}^{(3/2)}_{NP'bc'} ~g^{MM'} g^{NN'} g^{PP'} g^{aa'} g^{bb'} 
 g^{cc'} + .., \nonumber\\ \nd}
where the dotted terms are other possible contributions, from curvature or non-local counter-terms, and 
$b_i$ are $g_s$ independent constants. The first term in \eqref{pierettetag}, cancels with the $(2, 0)$ term from the first line of \eqref{evaBooo}. There are no contributions to this order from the second line in 
\eqref{evaBooo}. This means:
\bg\label{lunatimin}
C_{1\over 2} {\cal G}^{(3/2)}_{MNab} = {b_4 b_7\over b_1}~ {\cal G}^{(3/2)}_{MPac} {\cal G}^{(3/2)}_{NP'bc'}
g^{PP'} g^{cc'} + ...., \nd
where $C_{1\over 2}$ appears in the $g_s$-expansion of $F_2(t)$ in \eqref{karishma}. Its form is determined in \eqref{marielou} and here we see that it cannot be zero, implying that the $F_i(t)$ functions should be non-trivial functions of ${g_s\over H}$ for our analysis to make sense. The dotted terms in 
\eqref{lunatimin} could even have terms of the form:

{\footnotesize
\bg\label{lovibond}
\left(\mathbb{Y}_4^{(3/2)}\right)^{MNab} = \int d^6 y' \sqrt{{\bf g}_6(y')} ~\mathbb{F}^{(1)}(y - y') {\bf G}^{0ijM}(y') 
{\bf G}^{0klN}(y') {\bf R}_{ijkl}(y') {\bf R}^{[ab]}(y')
{\bf R}_{00}(y'), \nd}
which arises from the non-local counter-terms\footnote{An interesting non-local operator is of the form
${\cal O}_{\rm NL} \equiv \sum_n e_n \left({{\bf R} \over \square_y}\right)^n$, $n > 0$, where ${\bf R}$ is the warped 
Ricci scalar, and $\square_y$ is the Laplacian on the internal manifold ${\cal M}_4 \times {\cal M}_2$. Such an operator has no $M_p$ or $g_s$ dependence. However as shown in section \ref{Gng6}, their action is suppressed by the volume of ${\mathbb{T}^2\over {\cal G}}$, and in the limit the volume goes to zero, they decouple.} 
 where the non-locality is restricted to the internal six-manifold ${\cal M}_4 \times {\cal M}_2$, and the non-locality function $\mathbb{F}^{(1)}(y-x)$ has no explicit dependence on $g_s$. The G-flux components appearing above should be identified with 
${\cal G}^{(0)}_{0ijM}$, i.e the first term in \eqref{kyratagra}. 
 In the limit it $\mathbb{F}^{(1)}(y - y') \to {\delta^6(y - x)\over \sqrt{g_6}}$ with unwarped metric, the volume factor will scale as $\left({g_s\over H}\right)^{-2}$, making \eqref{lovibond} scale as $g_s^0$, thus matching up with the scaling of each term in \eqref{lunatimin}. 
 
Going to even higher order would imply taking the following values for $(k_1, k_2)$ in \eqref{evaBooo}: 
$\left({3\over 2}, {1}\right), \left(2, {1\over 2}\right)$ and $\left({5\over 2}, 0\right)$.
  This would mean that the first line in \eqref{evaBooo} scales as $\left({g_s\over H}\right)^{-13/3}$, with  no contributions from the second line. The quantum terms then scale as $\theta'_k = {4\over 3}$ in \eqref{melamon2}, and therefore:

{\footnotesize
\bg\label{gertagra}
\left(\mathbb{Y}_4^{(3/2)}\right)^{MNab} = b_8~{\cal G}^{(5/2)MNab} + b_9~{\cal G}^{(3/2)MNab}
\Big({\cal G}^{(3/2)}_{PQcd} {\cal G}^{(3/2)PQcd} + b_{10} {\cal G}^{(0)}_{0ijQ} {\cal G}^{(0)0ijQ}\Big) + ....
\nd}
where ${\cal G}^{(0)}_{0ijQ}$ should be identified with the first term in \eqref{kyratagra}. It is also clear that the first term of \eqref{gertagra} is exactly the same term that appears for $k_1 = {5\over 2},  k_2 = 0$ in the first line of \eqref{evaBooo} for $b_4 b_8 = 1$, implying:

{\footnotesize
\bg\label{duolipa}
\hskip-.25in C_1~{\cal G}^{(3/2)}_{MNab} + C_{1\over 2}~ {\cal G}^{(2)}_{MNab} = {b_4 b_9\over b_1}
\Big({\cal G}^{(3/2)}_{PQcd} {\cal G}^{(3/2)PQcd} + b_{10} {\cal G}^{(0)}_{0ijQ} {\cal G}^{(0)0ijQ} + b_{11} {\rm R} + b_{12}\square_y\Big){\cal G}^{(3/2)}_{MNab} + .. 
\nd}
where ${\rm R}$ is the $g_s$ independent part of the Ricci scalar, $\square_y$ is the Laplacian on the base manifold ${\cal M}_4 \times {\cal M}_2$, and  the dotted terms are from non-local counter-terms and curvature corrections. Again, the above equation shows that neither $C_1$, nor 
$C_{1\over 2}$ can be zero (the latter is already non-zero from \eqref{lunatimin}). 

The next order considers $(k_1, k_2)$ in \eqref{evaBooo} to take the following values: 
$\left({3\over 2}, {3\over 2}\right), \left({2}, {1}\right)$, $\left({5\over 2}, {1\over 2}\right)$ and 
$\left({3}, 0\right)$, making the $g_s$ scaling of the first line of \eqref{evaBooo} to be $\left({g_s\over H}\right)^{-4}$, still precluding the contributions from the second line. The quantum terms are classified by
$\theta'_k = {5\over 3}$ in \eqref{melamon2}, and one could construct the EOM similar to \eqref{duolipa}. Going beyond, the second line of \eqref{evaBooo} starts contributing, and the analysis becomes more non-trivial.

\vskip.1in

\noindent{\it Case 4: ${\bf G}_{MNPa}$ components}

\vskip.1in

\noindent The final set of components that we consider here are of the kind ${\bf G}_{MNPa}$ with 
$(M, N) \in {\cal M}_4 \times {\cal M}_2$ and $a \in {\mathbb{T}^2\over {\cal G}}$. These components are intresting because they reduce to the NS and RR three-form fluxes ${\bf H}_3$ and ${\bf F}_3$ respectively in the type IIB side.  The EOM governing these components may be written as:
\bg\label{evaoo}
&& \partial_{N_8}\Big(\sqrt{-{\bf g}_{11}}~{\bf G}_{MNPa} ~{\bf g}^{MM'} {\bf g}^{NN'} {\bf g}^{PP'} 
{\bf g}^{aa'}\Big)
\epsilon_{M'N'P'a'N_1 N_2 ...... N_7}\nonumber\\
&&~~~= b_1~{\bf G}_{0ijM} {\bf G}_{QRNb}~\delta^{[0}_{[N_1} \delta^i_{N_2} \delta^j_{N_3}\delta^M_{N_4}
\delta^Q_{N_5} \delta^R_{N_6} \delta^N_{N_7} \delta^{b]}_{N_8]} + \left({\bf X}_8\right)_{N_1......N_8}\\
&& ~~~+ b_4~\partial_{N_8}\Big(\sqrt{-{\bf g}_{11}} \left(\mathbb{Y}_4\right)_{MNPa} 
{\bf g}^{MM'} {\bf g}^{NN'} {\bf g}^{PP'} {\bf g}^{aa'}\Big)\epsilon_{M'N'P'a'N_1.....N_7}, \nonumber\nd 
with no contributions from the two-branes. As before we would like to know the $g_s$ scalings of each of the terms in \eqref{evaoo}. This may be expressed as:

{\footnotesize
\bg\label{evaooo}
&& \sum_{\{k_i\}}\partial_{N_8}\Big(\sqrt{{g}_{8}}~{\cal G}^{(k_1)}_{MNPa} ~{g}^{MM'} {g}^{NN'} {g}^{PP'} 
{g}^{aa'}\Big) \left({g_s\over H}\right)^{2\Delta k_1 -4}\epsilon_{M'N'P'a'N_1 N_2 ...... N_7}\nonumber\\
&&~~~ \times \left(a_{(mn\alpha)} + \left(a_{(mnp)}~C_{k_2} C_{k_3} \widetilde{C}_{k_4}  + a_{(m\alpha\beta)}~
\widetilde{C}_{k_2} \widetilde{C}_{k_3} \widetilde{C}_{k_4}\right) \left({g_s\over H}\right)^{2\Delta(k_2 + k_3 + k_4)}\right)\nonumber\\
&&~~~= b_1\sum_{\{k_i\}}{\cal G}^{(k_5)}_{0ijM} {\cal G}^{(k_6)}_{QRNb}~
\left({g_s\over H}\right)^{2\Delta(k_5 + k_6) - 4}
\delta^{[0}_{[N_1} \delta^i_{N_2} \delta^j_{N_3}\delta^M_{N_4}
\delta^Q_{N_5} \delta^R_{N_6} \delta^N_{N_7} \delta^{b]}_{N_8]} + g_s^{-2} \left(\widetilde{X}^{(2)}_8\right)_{N_1......N_8}\nonumber\\
&& ~~~+ b_4 \sum_{\{k,k'\}}\partial_{N_8}\Big(\sqrt{{g}_{8}} (\mathbb{Y}^{(k)}_4)^{M'N'P'a'} 
\Big)\left({g_s\over H}\right)^{\theta'_k - 2\Delta k' - 14/3} \epsilon_{M'N'P'a'N_1.....N_7}, \nd} 
where the coefficients $a_{(mn\alpha)}$ for example specify the coordinates to be located as 
$(m, n) \in {\cal M}_4$ and $\alpha \in {\cal M}_2$. Correspondingly, their $g_s$ scalings differ and have been shown above in \eqref{evaooo}. The $g_s$ scaling of $\widetilde{\bf X}_8^{(2)}$ polynomial is given in \eqref{polathai}.  
It is also easy to see that the $k_i$ are related  to each other via:
\bg\label{gerpole}
2\Delta(k_1 + k_2 + k_3 + k_4) - 4 = 2\Delta(k_5 + k_6) - 4 = \theta'_k - 2\Delta k' - {14\over 3}, \nd 
with $(k_2, k_3, k_4, k_5) \ge (0, 0, 0, 0)$ and 
$(k_1, k_6, k') \ge \left({3\over 2}, {3\over 2}, {3\over 2}\right)$. The lowest order equation\footnote{Again we see that if we take $k_1 = 0$ in \eqref{evaooo}, then  the last term, i.e the quantum term, scales as 
$\theta'_k = {2\over 3}$ which is only possible with $l_{30} = 2$ or $l_{32} = 2$ or $l_{33} = 2$ in 
\eqref{melamon2}. However this is exactly the first term in \eqref{evaooo} if we take $b_4 = 1$, implying again that it might be hard to maintain a time-independent component here.}
can be ascertained for $k_1 = k_6 = k' = {3\over 2}$ and the remaining $k_i = 0$, and therefore we are looking at terms that scale as $\left({g_s\over H}\right)^{-3}$. The quantum terms would be classified by $\theta'_k = {8\over 3}$ in \eqref{melamon2}. The EOM then is given by:

{\footnotesize
\bg\label{cbgerman}
&& \partial_{N_8}\Big(\sqrt{{g}_{8}}~{\cal G}^{(3/2)}_{MNPa} ~{g}^{MM'} {g}^{NN'} {g}^{PP'} 
{g}^{aa'}\Big)\left(a_{(mnp)} + a_{(mn\alpha)} + a_{(m\alpha\beta)}\right)~\epsilon_{M'...N_7}\nonumber\\
&& = b_1 ~{\cal G}^{(0)}_{0ijM} {\cal G}^{(3/2)}_{QRNb}~\delta^{[0}_{[N_1}....\delta^{b]}_{N_8]}
+ b_4~\partial_{N_8}\Big(\sqrt{{g}_{8}} (\mathbb{Y}^{(3/2)}_4)^{M'N'P'a'}\Big) \delta\left(\theta'_k - {8\over 3}\right)
\epsilon_{M'...N_7} , \nd}
which provides the spatial dependence of the G-flux components ${\cal G}^{(3/2)}_{MNPa}$ when 
${\cal G}^{(0)}_{0ijM}$ is the first term in \eqref{kyratagra}. From here one could go about finding the higher order EOMs $-$ the story remains somewhat similar to what we had earlier. One interesting case is when 
$k_1 = k_6 = k' = k_5 = {3\over 2}$ and $k_2 = .. = k_4 = {1\over 2}$. For this case ${\bf X}_8$ contributes along with the quantum terms which are classified by $\theta'_k = {11\over 3}$ in \eqref{melamon2}.

\subsubsection{Dynamical branes, fluxes and additional constraints \label{branuliat}}

The interconnections between the G-flux EOMs and the Einstein's EOMs, in particular the ones that match the quantum terms, although satisfying to a certain degree, do conceal an additional layer of subtleties that we kept hidden under the rug so far. 
These subtleties arise once we look at the M2 and M5-branes, especially the ones endowed with dynamical motions. To illustrate this, let us first discuss the static M2-branes ignoring, for the time being, the 
M5-branes\footnote{The M5-branes wrapped on three-cycles of the internal eight-manifold could be viewed as fractional M2-branes. If we ignore the subtleties associated with the KK modes from the wrapped directions, then the dynamics of these will be no different from the M2-branes. In this paper we will avoid distinguishing between the integer and the fractional M2-branes.}.  

\vskip.1in

\noindent{\it Dynamical membranes and G-fluxes}

\vskip.1in

\noindent The subtleties alluded to above arise when the dynamical motions of the membranes tend to stir up additional corrections to the G-flux components, in particular the ones with components along the $2+1$ space-time direction, for example ${\bf G}_{M0ij}$.  
Question then is: how robust is our earlier analysis that we did using the space-time flux components borrowed from \cite{nogo}? To see this, we will have to re-visit the dynamics of membranes more carefully now. For simplicity however we will only consider single membrane, and ignore M5-branes (as mentioned above). 
The action for a {\it single} membrane\footnote{Due to the condition $n_b = \bar{n}_b$ coming from say \eqref{evaB102}, there has to be another $\overline{\rm M2}$-brane somewhere at a point in the internal eight-manifold \eqref{melisett}. See also footnote \ref{kalaryan}.}  can be written as:
\bg\label{radimed}
\mathbb{S}_B = -{T_2\over 2}\int d^3\sigma \left\{\sqrt{-\gamma_{(2)}}\Big(\gamma^{\mu\nu}_{(2)} 
\partial_\mu X^M\partial_\nu X^N {\bf g}_{MN} - 1\Big) + {1\over 3} \epsilon^{\mu\nu\rho} 
\partial_\mu X^M\partial_\nu X^N\partial_\rho X^P {\bf C}_{MNP}\right\}, \nonumber\\ \nd
where $\gamma_{(2)\mu\nu}$ is the world-volume metric, $\epsilon_{\mu\nu\rho}$ is the Levi-Civita 
{\it symbol}, 
${\bf g}_{MN}$ is the warped metric in M-theory, 
$X^M$ are the coordinates of eleven-dimensional space-time and ${\bf C}_{MNP}$ is the three-form potential. The EOM for the world-volume metric easily relates it to the M-theory metric ${\bf g}_{MN}$ as the following pull-back:
\bg\label{neveluce}
\gamma_{(2)\mu\nu} = \partial_\mu X^M \partial_\nu X^N {\bf g}_{MN}, \nd
which means in the {\it static-gauge}, we will simply have $\gamma_{(2)\mu\nu} = {\bf g}_{\mu\nu}$, i.e the world-volume metric is the $2+1$ dimensional space-time metric. On the other hand, the EOM for the 
membrane motion takes the following condensed form:
\bg\label{kitite}
\square_{(\sigma)}X^P + \gamma^{\mu\nu}_{(2)}\partial_{\mu}X^M \partial_{\nu}X^N 
{\bf \Gamma}^P_{MN} - {\epsilon^{\mu\nu\rho}\over 3!\sqrt{-\gamma_{(2)}}}\partial_{\mu}X^Q
\partial_{\nu}X^N\partial_{\rho}X^R {\bf G}_{SQNR}{\bf g}^{SP} = 0, \nonumber\\ \nd
with $\square_{(\sigma)}$ forming the Laplacian\footnote{$\square_{(\sigma)} X^P = 
{1\over \sqrt{-\gamma_{(2)}}}\partial_\mu\Big(\sqrt{-\gamma_{(2)}} \gamma_{(2)}^{\mu\nu} \partial_\nu 
X^P\Big)$.} 
in $2+1$ dimension described using the world-volume 
metric $\gamma_{(2)\mu\nu}$, ${\bf \Gamma}^P_{MN}$ is the Christoffel symbol described using the warped metric ${\bf g}_{MN}$, and ${\bf G}_{SQNR}$ is the G-flux components that we have been using so far. In the static-gauge we expect $\square_{(\sigma)} X^P = 0$, and then the remaining two terms of 
\eqref{kitite}, simply gives us:
\bg\label{anwbeer}
{\bf G}_{0ijM} = -{3\over 2} \sqrt{-\gamma_{(2)}} ~{\bf g}^{\mu\nu} {\bf g}_{\mu\nu, M}, \nd
where we identify the world-volume metric to the $2+1$ dimensional space-time warped metric 
${\bf g}_{\mu\nu}$. Therefore plugging in the metric components from \eqref{vegamey} and \eqref{mrglass} we can reproduce the familiar results for ${\bf G}_{0ijm}$ and ${\bf G}_{0ij\alpha}$ in \cite{nogo, nodS}, including the Kasner one in \eqref{colemanbook} and the one for the case \eqref{ranjhita} in 
\eqref{rogra}. 

All we did above is very standard, but the keen reader must have already noticed the subtlety. The form 
\eqref{anwbeer} is {\it only} possible if there are static M2-branes. If the system doesn't have any static 
M2-branes, or the M2-branes are somehow absent, the result \eqref{anwbeer} doesn't follow naturally. 
For the case \eqref{ranjhita} all the parameters are independent of $y^\alpha$ so, at least at the face-value, 
\eqref{rogra} makes sense once we compare it with \eqref{anwbeer}. However since the Euler characteristics\footnote{Here it is just the integral of the ${\bf X}_8$ polynomial over the eight-manifold. See discussion earlier.} of the internal eight-manifold also vanishes, all static M2-branes are eliminated. How can we then justify the non-zero value of ${\bf G}_{0ijm}$ for the case \eqref{ranjhita}?
 
This is where the difference between time-independent (and also supersymmetric) and time-dependent cases becomes more prominent. In the time-independent supersymmetric case\footnote{For the time-independent non-supersymmetric case, as we saw earlier, it is hard to establish an EFT description in lower dimensions
with de Sitter isometries. Thus it doesn't make sense to talk about it here and we shall ignore this case altogether.}, vanishing Euler characteristics for a four-fold implies vanishing fluxes and branes \cite{BB, DRS}. This is clearly not the case for the time-dependent case where, as we saw above, G-flux components that are time-dependent (i.e $g_s$ dependent) are allowed. This means for vanishing Euler characteristics, {\it dynamical} M2-branes can be allowed too.  

Introducing dynamics open up a new class of subtleties that we have hitherto left unexplored. One of the first subtlety is that the world-volume metric is no longer the $2+1$ dimensional space-time metric. In fact 
$\gamma_{(2)00}$ becomes:
\bg\label{kviard}
\gamma_{(2)00} &=& {\bf g}_{00} + \dot{y}^m \dot{y}^n {\bf g}_{mn} 
+ \dot{y}^\alpha \dot{y}^\beta {\bf g}_{\alpha\beta}
+ \dot{y}^a \dot{y}^b {\bf g}_{ab} \\
& = & \left({g_s\over H}\right)^{-8/3}\left(g_{00} + \dot{y}^m \dot{y}^n {g}_{mn}\left({g_s\over H}\right)^2 
+ \dot{y}^\alpha \dot{y}^\beta {g}_{\alpha\beta}\left({g_s\over H}\right)^2 
+ \dot{y}^a \dot{y}^b {g}_{ab}\left({g_s\over H}\right)^4\right), \nonumber \nd
where the components are defined, for the case \eqref{olokhi}, using warped M-theory metric and therefore involve $g_s$ dependent 
terms. The other components of the metric may be taken to be the corresponding space-time metric if 
$y^M \equiv y^M(t)$. We can now quantify what is meant by slowly moving membrane by specifying the behavior of $y^M$ as:
\bg\label{lodger}
y^M({\bf x}, g_s) = \sum_{k\in {\mathbb{Z}\over 2}} y^M_{(k)}({\bf x}) \left({g_s\over H}\right)^{2\Delta k}, \nd
near $g_s \to 0$ and $y^M_{(k)}({\bf x})$ could in principle depend on the world-volume spatial
coordinates, but here we will take it to be a constant as in \eqref{kviard}. In this representation of 
$y^M$, slowly moving membrane means small $k$ at late times, i.e for $g_s << 1$. In the limit $k \to 0$, 
the membrane is truly static and when $g_s \to 0$, $y^M({\bf x}, 0) \to 0$. This is almost like the end point of a D3-D7 \cite{DHHK} or a KKLMMT \cite{kklmmt} inflationary model where, in IIB, a D3-brane (T-dual of our M2-brane) dissolves in the D7-brane (T-dual of an orbifold point in our eight-manifold) keeping a 
$\overline{\rm D3}$-brane (T-dual of a $\overline{\rm M2}$-brane) somewhere deep in a throat-like geometry\footnote{The original throat geometry appeared in \cite{kklmmt} where the internal manifold was a {\it compactified} warped-deformed conifold. Here instead we have a geometry of the form ${\cal M}_4 \times {\cal M}_2$ from \eqref{melisett} so the specific details regarding throat-like geometry will be different, although such a geometry is not an essential feature of the inflationary model. We could have also taken a resolved warped-deformed conifold wherein locally we could identify the resolved two-cycle with 
${\cal M}_2$ and the remaining geometry with ${\cal M}_4$. The underlying non-K\"ahlerity would fix the global structure.

A related question is whether the {\it back-reaction} from the $\overline{\rm D3}$, as advertized in \cite{bena}, is of any concern here. The work of \cite{bena} specifically used the deformed conifold geometry and multiple $\overline{\rm D3}$-branes. Our geometry is very different and we can use single $\overline{\rm D3}$-brane, so the concerns of \cite{bena} look irrelevant here.}. Additionally, 
the $y^M$ represent the eight scalar fields on the world-volume of the M2-brane, and once we dualize them to type IIB, only six scalar fields would remain. The Laplacian action on $y^M$ then yields:
\bg\label{rebwest}
\square_{(\sigma)}y^M & = & {2\Delta^2\Lambda\over \vert g_{00}\vert}
\sum_{k_3} {k_3(2k_3 -7)\over 1+ f_o}
\left({g_s\over H}\right)^{2\Delta(k_3 + 1)} y^M_{(k_3)} \nonumber\\
&-& {8\Lambda^2 \Delta^4\over \vert g_{00}\vert}
\sum_{\{k_i\}}{k_1k_2k_3(k_1+k_2) g_o\over (1+f_o)^{2}}
\left({g_s\over H}\right)^{2\Delta\left(k_1+k_2+k_3 + 1\right)}y^M_{(k_3)}, \nd
where note that both the terms are suppressed by positive powers of ${g_s\over H}$, $g_{00}$ is the un-warped metric component, $\Delta = {1\over 3}$ as chosen before and $\Lambda$ is the cosmological constant. We have also assumed no motion along the ($a, b$) directions and therefore $y^M$ above can either be $y^m$ or $y^\alpha$.  The remaining two factors, ($f_o, g_o$) are defined in the following way:
\bg\label{kepnerA}
&& f_o \equiv f_o(y) = 4\Lambda \Delta^2 \sum_{\{k_i\}}
g_o(k_1, k_2; y) k_1 k_2 \left({g_s\over H}\right)^{2\Delta(k_1 + k_2)}\nonumber\\
&& g_o \equiv g_o(k_1, k_2; y) = g^{00}\left(y^m_{(k_1)} y^n_{(k_2)} g_{mn}(y)  
+ y^\alpha_{(k_1)} y^\beta_{(k_2)} g_{\alpha\beta}(y)\right), \nd 
where the metric involved are all the un-warped ones. Note that, since $f_o$ is a series in positive powers in $g_s$, any series of the form $(1 + f_o)^{-\vert q\vert}$ for arbitrary $q$ will only contribute {\it positive} powers of ${g_s\over H}$ to the series \eqref{rebwest}. Thus the generic conclusion of $\square_{(\sigma)}$
being defined in terms of positive powers of ${g_s\over H}$, remains unchanged. In fact
this also persists for the second term in the EOM \eqref{kitite}.  To see this, 
let us take $M = \alpha$ in \eqref{lodger} for the case \eqref{olokhi}. We get:

{\footnotesize
\bg\label{angelD}
\gamma^{00}_{(2)} \partial_{0}X^P \partial_{0}X^Q {\bf \Gamma}^\alpha_{PQ} = 
{\vert g^{00}\vert \over 1 + f_o}\left({g_s\over H}\right)^{2/3} \left[\Gamma^\alpha_{00} + 
4\Delta^2\Lambda \sum_{\{k_i\}} k_1 k_2 h^\alpha_o(k_1, k_2; y) 
\left({g_s\over H}\right)^{2\Delta(k_1+k_2)}\right], \nd}
where $f_o$ is defined in \eqref{kepnerA}; and ${\bf \Gamma}^\alpha_{PQ}$ and
$\Gamma^\alpha_{00}$ are the Christoffel symbols defined with respect to the warped and the un-warped metrics respectively. The other factors, namely $\Delta$ and $\Lambda$, appearing above have already been defined with \eqref{rebwest}. Finally the factor $h_o(k_1, k_2; y)$ takes the following form:
\bg\label{mastronini}
h^\alpha_o(k_1, k_2; y) \equiv y^m_{(k_1)}y^n_{(k_2)} \Gamma^\alpha_{mn} + 
y^\sigma_{(k_1)}y^\gamma_{(k_2)} \Gamma^\alpha_{\sigma\gamma} + 
y^\sigma_{(k_1)}y^m_{(k_2)} \Gamma^\alpha_{\sigma m}, \nd
where the Christoffel symbols are again defined with respect to the un-warped metrics. In this form 
\eqref{mastronini} should be compared to $g_o$ in \eqref{kepnerA} which was defined using un-warped metric components also. We can also replace $\alpha$ by $m$ in \eqref{angelD}, but the form would remain unchanged. Therefore putting everything together, the functional form for ${\bf G}_{M0ij}$ becomes:

{\footnotesize
\bg\label{hokyra}
{\bf G}_{M0ij}  &=& {3g_{NM}\sqrt{-\gamma_{(2)}}\over \vert g_{00}\vert (1 + f_o)}\Bigg[\Gamma^N_{00} + 
4\Delta^2\Lambda \sum_{\{k_i\}} k_1 k_2 h^N_o
\left({g_s\over H}\right)^{2\Delta(k_1+k_2)} + {2\Lambda\over 9} 
\sum_{k_3} {k_3(2k_3 -7)}y^N_{(k_3)}
\left({g_s\over H}\right)^{2\Delta k_3} \nonumber\\
&& ~~~~~~~~~+ \vert g_{00}\vert (1 + f_o) g^{i'j'} \Gamma^N_{i'j'} -{8\Lambda^2\over 81}
\sum_{\{k_i\}}{k_1k_2k_3(k_1+k_2) g_o\over 1+f_o} y^N_{(k_3)}
\left({g_s\over H}\right)^{2\Delta\left(k_1+k_2+k_3\right)}\Bigg], \nd}
where everything is defined with respect to the un-warped metric except $\sqrt{-\gamma_{(2)}}$, which in turn is defined using the warped $2+1$ dimensional space-time metric, implying that the overall $g_s$ scaling of \eqref{hokyra} is $\left({g_s\over H}\right)^{-4}$. This {\it negative} $g_s$ scaling is important because other than that every term in \eqref{hokyra} scales as {\it positive} powers of $g_s$. Therefore with 
dynamical M2-branes, in the limit $g_s \to 0$, we can express ${\bf G}_{M0ij}$ alternatively as the following series:

{\footnotesize
\bg\label{kyratagra}
{\bf G}_{0ijM} &\equiv& \sum_{k \in {\mathbb{Z}\over 2}} 
{\cal G}^{(k)}_{0ijM}(y, k) \left({g_s\over H}\right)^{2\Delta(k - 2/\Delta)} \nonumber\\
&=& -\left({g_s\over H}\right)^{-4} \partial_M\left({\epsilon_{0ij}\over H^4}\right) + 
\sum_{k \in {\mathbb{Z}\over 2}} 
{\cal G}^{(k + 1/2)}_{0ijM}(y, k+1/2) \left({g_s\over H}\right)^{2\Delta(k + 1/2 - 2/\Delta)}, \nd}
which is somewhat similar to the expression for the other G-flux components in \eqref{ravali}.  However
similarities aside, the differences between \eqref{kyratagra} and \eqref{ravali} are important now. One of the main difference between these two expressions is that in \eqref{ravali}, $k \ge {3\over 2}$ for 
\eqref{olokhi} and $k \ge {9\over 2}$ for \eqref{ranjhita}. However for \eqref{kyratagra}, $k$ can be large or small: smaller $k$ implies, according to \eqref{lodger}, slowly moving M2-brane and for $k = 0$ it is completely static. Another difference is that even if we impose a {\it lower} bound on $k$, the $k$ independent piece
should always be there as one may infer from the exact expression in \eqref{hokyra}. It should also be clear from \eqref{hokyra}, when $k = 0$, 
${\cal G}^{(0)}_{0ijM}(y, 0) = -\partial_M\left({\epsilon_{0ij}\over H^4}\right)$.
  This is important, because it implies that no mater whether we allow dynamical M2-branes or not, the domination of the $k$ independent term in \eqref{kyratagra} over all other terms for $g_s < 1$ puts a strong confidence on our choice of the G-flux components ${\bf G}_{0ijm}$ and ${\bf G}_{0ij\alpha}$ for both cases \eqref{olokhi} and \eqref{ranjhita}. 

\vskip.1in

\noindent{\it Fluxes, seven-branes and additional dynamics}

\vskip.1in

\noindent The exact form of the G-flux components ${\bf G}_{0ijM}$ for $M = (m, \alpha)$ appearing in \eqref{hokyra} and \eqref{kyratagra}; as well as our ansatze for the other G-flux components in \eqref{ravali} pretty much summarize all the background fluxes that could be allowed in the set-up like ours. However, as the patient reader might have noticed, we did not express the G-flux components in terms of their three-form potentials except for the case studied in \eqref{hokyra}. In particular the three crucial G-flux components, namely 
${\bf G}_{mnab}, {\bf G}_{m\alpha ab}$ and ${\bf G}_{\alpha\beta ab}$, now require some explanations. It is of course clear that we do not want to express these three G-flux components in terms of the three-form potentials as ${\bf C}_{Mab}$ would create metric cross-terms ${\bf g}_{M3}$ in the type IIB side (other components can create un-necessary KK modes as in \eqref{montan}). This is not what we need so 
${\bf G}_{MNab}$ can only appear as {\it localized} fluxes in M-theory. In other words:
\bg\label{teskimey}
{\bf G}_{MNab}(y_1, y_2) = {\bf F}_{MN}(y_1) \otimes \Omega_{ab}(y_2), \nd
where we have divided the internal eight-dimensional coordinates $y$ as $y = (y_1, y_2)$, with $y_1$ parametrizing the coordinates of the four-dimensional base and $y_2$ parametrizing the coordinates of the remaining four-dimensional space. Such localized fluxes lead to gauge fields $-$ here we express them as ${\bf F}_{MN}$ $-$ on D7-branes. In other words, the orbifold points in M-theory lead to seven-branes in the type IIB side, in a charge-neutral scenario (much like in \cite{sendas}), wrapping appropriate four-manifolds that we shall specify below. 
As alluded to earlier, this set-up is then ripe for embedding the D3-D7 \cite{DHHK} or the KKLMMT  
\cite{kklmmt} inflationary model. The other factor in 
\eqref{teskimey}, namely $\Omega_{ab}(y_2)$, is the localized normalizable two-form near any of the orbifold singularities.           

In the time-independent case, \eqref{teskimey} is all that we need, but once time-dependences are switched on new subtleties arise. For example, the G-flux components 
${\bf G}_{MNab}$ have $g_s$ expansions as in 
\eqref{ravali}. Question then is how are the $g_s$ expansions for ${\bf F}_{MN}$ and 
$\Omega_{ab}$ defined here.  To analyze this, let us first consider the G-flux components ${\bf G}_{mnab}$. The flux quantization condition is described in \eqref{birjoani} on a four-cycle $\Sigma^{(1)}_4 \equiv
{\cal C}_2 \times {{\bf T}^2\over {\cal G}}$, where ${\cal C}_2$ is a two-cycle in ${\cal M}_4$. The gauge field
${\bf F}_{mn}$ will then have to be defined over this two-cycle, and we expect the corresponding D7-brane to wrap the four-cycle ${\cal M}_4$. 

Since all cycles in the internal eight-manifold is varying with respect to time, it would make sense to endow time-dependences on {\it both} the gauge flux components ${\bf F}_{mn}$ as well as the normalizable two-form $\Omega_{ab}$. The LHS of \eqref{birjoani} is where we introduce the split 
\eqref{teskimey}, and the RHS governs the quantization rule with seven-forms, which in turn may be divided into two sub-forms. Such a split doesn't have any new physics other than what we discussed in 
\eqref{birjoani}, but a new subtlety arises once we express the gauge field ${\bf F}_{mn}$ in terms of its potential ${\bf A}_m$ because of it's dependence on $g_s$ as well as on ($y^m, y^\alpha$). Similar subtlety will arise for the gauge potential ${\bf A}_{\alpha}$. Both these potentials will switch on:
\bg\label{guardian}
\partial_0{\bf A}_m(y^m, y^\alpha, g_s) \equiv H\sqrt{\Lambda} \left({\partial {\bf A}_m \over \partial g_s}\right), ~~~~~ 
\partial_0{\bf A}_\alpha(y^m, y^\alpha, g_s) \equiv H\sqrt{\Lambda} 
\left({\partial {\bf A}_\alpha \over \partial g_s}\right), \nd   
in addition to the existing field strengths. Clearly such components do not arise in the time-independent case and the split \eqref{teskimey} is all there is to it. The flux quantization conditions \eqref{harper1} and \eqref{harper2} tell us that the gauge field strengths 
${\bf F}_{\alpha\beta} = \partial_{[\alpha} {\bf A}_{\beta]}$ and 
${\bf F}_{m\alpha} = \partial_{[m} {\bf A}_{\alpha]}$ will have proper quantization schemes when defined over the two-cycles ${\cal M}_2$ and ${\bf S}_{(1)}^1 \times {\bf S}_{(2)}^1$ respectively where 
${\bf S}^1_{(1)} \in {\cal M}_4$ and ${\bf S}^1_{(2)} \in {\cal M}_2$. Both these one-cycles are allowed because neither ${\cal M}_4$, nor ${\cal M}_2$ are Calabi-Yau manifolds for the case \eqref{olokhi}. 
For the case \eqref{ranjhita}, {\bf Table \ref{blanchek}} will tell us that the latter is not well-defined. 
However now we need to deal with new components arising from temporal derivatives, that translate into $g_s$ derivatives, here. A way out this is to switch on 
electric potential ${\bf A}_0(y^m, y^\alpha, g_s)$ satisfying:
\bg\label{jodief}  
\partial_m{\bf A}_0 \equiv   H\sqrt{\Lambda} \left({\partial {\bf A}_m \over \partial g_s}\right), ~~~
\partial_\alpha{\bf A}_0 \equiv   H\sqrt{\Lambda} \left({\partial {\bf A}_\alpha \over \partial g_s}\right), \nd
which in turn will make ${\bf F}_{0m} = {\bf F}_{0\alpha} = 0$ and would not contribute to the energy-momentum tensors or the quantum terms \eqref{phingsha} and \eqref{phingsha2}. This could be generalized to the non-abelian case also but since we are only dealing with well-separated seven-branes in F-theory, 
\eqref{jodief} suffices. However the dependence of ${\bf A}_0$ on $g_s$ also switches on 
${\partial {\bf A}_0 \over \partial g_s}$, but this again does not contribute to the energy-momentum tensors or to the quantum terms \eqref{phingsha} and \eqref{phingsha2}.  

Interestingly, if we view {\it all} the G-flux components as localized fluxes of the form \eqref{teskimey}, then 
we are in principle dealing with only three gauge field components ${\bf F}_{mn}, {\bf F}_{m\alpha}$ and 
${\bf F}_{\alpha\beta}$ on D7-branes that are oriented along various directions in the internal space (they all do share the same $3+1$ dimensional space-time directions in the type IIB side).  This is an interesting scenario with only seven-brane gauge fluxes and no ${\bf H}_3$ and ${\bf F}_3$ three-form fluxes as these would require {\it global} ${\bf G}_{mnpa}, {\bf G}_{mn\alpha a}$ and ${\bf G}_{m\alpha\beta a}$ G-flux components. Such global G-flux components would in turn give rise to components 
${\bf G}_{0mnp}, {\bf G}_{0mn\alpha}$ and ${\bf G}_{0m\alpha\beta}$, which are not what we want here. 
Question then is whether it is possible to retain global {\it and} local G-flux components without encountering the issues mentioned above. 

It appears that there indeed exists a possible way out of this conundrum if we consider the modified Bianchi identity \eqref{uniprixtagra}, i.e the Bianchi identity with the full quantum corrections, carefully. In the absence of M5-branes, i.e when $N = 0$ in \eqref{uniprixtagra}, we can  rewrite \eqref{uniprixtagra}  as:
\bg\label{margoR}
d\left({\bf G}_4 - {c_2\over c_1} \hat{\mathbb Y}_4 + {c_3\over c_1} \ast_{11} \mathbb{Y}_7\right) = 0, \nd
where $c_i$ are constants, and $\mathbb{Y}_7$ and $\hat{\mathbb{Y}}_4$ are defined in 
\eqref{verasofmey} and \eqref{cambelle} respectively. Both of these have $g_s$ dependences and in fact 
$\mathbb{Y}_7$ features prominently in the flux quantization process as discussed earlier. The above equation allows us to introduce an exact form $d{\bf C}_3$, and so we can re-write \eqref{margoR} as:
\bg\label{fasit}
{\bf G}_4 = d{\bf C}_3 + {c_2\over c_1} \hat{\mathbb{Y}}_4 - {c_3\over c_1} \ast_{11} \mathbb{Y}_7, \nd
where all quantities are functions of $g_s$ as well as of ($y^m, y^\alpha$). The ${\bf C}_3$ could be understood as the potential, but ${\bf G}_4$ is not just $d{\bf C}_3$ because of the conpiracies of the quantum terms. Note that nothing actually depends explicitly on ${\bf C}_3$ (all quantum terms and the energy-momentum tensors, as well as the flux quantization rules and anomaly cancellation conditions, are expressed using ${\bf G}_4$), so we have some freedom in the choice of ${\bf C}_3$. We can use this freedom to set:
\bg\label{pashaclub}
{\bf G}_{0MNP}  \equiv \partial_{[0}{\bf C}_{MNP]} + {c_2\over c_1} (\hat{\mathbb{Y}}_4)_{0MNP} 
- {c_3\over c_1} \left(\ast_{11} \mathbb{Y}_7\right)_{0MNP} = 0, \nd
which amounts to putting ${\bf F}_{0M} = 0$ for the case ${\bf G}_{MNab}$, so they are still localized fluxes as \eqref{teskimey}, but the difference is now that we won't need to switch on an electric flux ${\bf A}_0$ on the world-volume of the D7-branes\footnote{In other words we can keep ${\bf C}_{0MN} = 0$ without loss of  generalities. Switching on ${\bf C}_{0MN}$ will be equivalent to switching on electric flux ${\bf A}_0$ on the D7-branes. Here the quantum terms help us {\it cancel} the $\partial_0{\bf C}_{MNP}$ piece without invoking, for example, pieces like $\partial_P{\bf C}_{0MN}$ in \eqref{pashaclub}. This is the leverage we get using the quantum terms in \eqref{pashaclub}.}. 
For the other G-flux components, we can now allow global fluxes so type IIB theory can have ${\bf H}_3$ and ${\bf F}_3$ three-form fluxes. However as discussed in 
\eqref{stewartk} the corresponding G-flux components ${\bf G}_{MNPa}$ do not have proper quantization schemes because of the absence of global four-cycles in the M-theory side. However in IIB global three-cycles do exist so these fluxes could be properly quantized in the IIB side. The quantization rule will however follow similar trend as in \eqref{stewartk}.  

\subsection{Stability, swampland criteria and the energy conditions \label{nala}}

Many questions could be raised at this point. For example how stable is our background? How do we overcome the swampland criteria? How do we satisfy the null-energy condition, the strong-energy condition and possibly the dominant-energy condition? In the following we will provide possible answers to the above set of questions.

\subsubsection{Stability of our background and quantum corrections \label{stabul}}

One of the important question now is the question of stability of our solution. Before going into this, let us answer a related question on what it means to introduce the series of quantum corrections to solve the EOMs. In other words, how do we interpret the quantum corrections here? 

\vskip.1in

\noindent {\it Stability and higher order quantum corrections}

\vskip.1in

\noindent To answer this, let us look at the metric components in the ($m, n$) i.e ${\cal M}_4$ direction. The EOM for $g_{mn}$ is given by \eqref{misslemon2}. The LHS of this equation has the Einstein tensor parts and the RHS is the sources, including the quantum terms. The quantum terms, i.e $\mathbb{C}_{mn}^{(0, 0)}$, are classified by $2/3 \le \theta'_k \le 8/3$ in \eqref{melamon2}, and the lower bound can at best renormalize the existing classical pieces as $\theta'_k = 2/3$ does not allow higher powers of G-flux or curvature components. Thus the RHS of \eqref{misslemon2} is classified by $\theta'_k \le 8/3$, and therefore knowing the G-flux components 
${\cal G}^{(3/2)}_{mnab}, {\cal G}^{(3/2)}_{m\alpha ab}$ and ${\cal G}^{(3/2)}_{\alpha\beta ab}$, and the curvature tensors, we can express the RHS of \eqref{misslemon2} in terms of the known quantities. 

Going to the next order should switch on the higher order quantum terms. How are they interpreted here? The G-flux components that we gather at the zeroth order in $g_s$, and the metric $g_{mn}$ that comes out of our zeroth order computation\footnote{The zeroth order actually mixes $g_{mn}, g_{\alpha\beta}$ as well as $g_{\mu\nu}$ together, so untangling them would require us to use {\it all} the zeroth order equations. We will avoid this subtlety for the sake of the present argument, but will become clearer as we go along.}, now serve as the {\it input} for the next order, i.e $g_s^{1/3}$, equations. What they do here is rather instructive. The next order equation is \eqref{rambha3}. The LHS of the equation is the $g_{mn}$ that we computed using all the zeroth order equations. The RHS is however made of quantum terms 
$\mathbb{C}_{mn}^{(1/2, 0)}$ as well as {\it new} G-flux components like ${\cal G}^{(2)}_{mnab},
{\cal G}^{(2)}_{m\alpha ab}$ and ${\cal G}^{(2)}_{\alpha\beta ab}$ generated at this level, including the higher order $C_k$ and $\widetilde{C}_k$ factors from the $F_i(t)$ functions. The quantum terms are now classified by $1 \le \theta'_k \le 3$ and appear as \eqref{melamon3}, thus clearly allowing at least to third order G-flux terms. All these new components and the quantum terms, with the background data at the zeroth order, balance each other in a precise way so as to preserve the zeroth order metric component $g_{mn}$. This is the meaning of \eqref{rambha3} (with the assumption that the non-perturbative corrections from section \ref{instachela} have been incorporated in).

The quantum terms are therefore computed on the zeroth order background, with additional new data from fluxes and the $(C_k, \widetilde{C}_k$) coefficients, to balance each other without changing the zeroth order metric and fluxes. Going to next order, i.e $g_s^{2/3}$, the equation is given by \eqref{clovbalti}. 
We see that the story is repeated in exactly the same fashion: the $g_s^{2/3}$ order switches on new quantum terms, i.e $\mathbb{C}^{(1, 0)}_{mn}$ classified by $\theta'_k = 4/3$ in \eqref{mameye} perturbatively and $\theta'_k \le 10/3$ non-perturbatively from section \ref{instachela}; new G-flux components and higher order $(C_k, \widetilde{C}_k$) coefficients; but they do not {\it de stabilize} the existing zeroth order metric $g_{mn}$ and the G-fluxes. The RHS of \eqref{clovbalti} is precisely the statement of balance: at the $g_s^{2/3}$ order the quantum terms use the data at the zeroth and next (i.e $g_s^{1/3}$) order including 
{\it new} G-flux components like ${\cal G}^{(5/2)}_{mnab},
{\cal G}^{(5/2)}_{m\alpha ab}$ and ${\cal G}^{(5/2)}_{\alpha\beta ab}$ to balance each other in such a way that LHS of \eqref{clovbalti} still remains $g_{mn}$. 

The story repeats in the same fashion as we go to higher powers of $g_s^{1/3}$. The quantum terms are  computed using the data generated at all lowers orders, including new G-flux components at this order along with the higher order $(C_k, \widetilde{C}_k$) coefficients. All these balance each other so as to keep the zeroth order data, that include metric $g_{mn}$ and G-flux components, unchanged. This delicate balancing act is responsible for keeping our background safe and stable.

Going to the ($\alpha, \beta$) directions, the zeroth order in $g_s$ reproduces the un-warped metric information $g_{\alpha\beta}$, once we have the full data on the G-flux components 
like ${\cal G}^{(3/2)}_{\alpha\beta ab},
{\cal G}^{(3/2)}_{\alpha m ab}$ and ${\cal G}^{(3/2)}_{mn ab}$, which are of course the same as 
before (see \eqref{uanaban}). On 
{\it this} background we now compute the quantum terms $\mathbb{C}_{\alpha\beta}^{(1/2, 0)}$ classified by $1 \le \theta'_k \le 3$ in \eqref{melamon2}. The balancing act starts again: new G-flux components like 
 ${\cal G}^{(2)}_{mnab},
{\cal G}^{(2)}_{m\alpha ab}$ and ${\cal G}^{(2)}_{\alpha\beta ab}$ that are required to this order in $g_s$
are added, to be pitted against the  quantum terms and the $F_i(t)$ coefficients, such that the metric 
$g_{\alpha\beta}$ doesn't change in \eqref{ajanta}. Going to order $g_s^{2/3}$, similar argument holds 
as seen from \eqref{vanandmey}. 

For the ($a, b$) directions, there are no zeroth order contributions. The first non-trivial order is  
$g_s^2$, and to this order the metric is flat i.e $\delta_{ab}$ from \eqref{buskaM}. This flat metric persists to all higher orders in $g_s$, as may be seen in \eqref{gargere} for $g_s^{7/3}$ and \eqref{hemra} for 
$g_s^{8/3}$ where for both cases the quantum terms computed from the lower order data plus new G-flux components to that order, balance against the fluxes and the $(C_k, \widetilde{C}_k)$ coefficients. 

The story takes a similar turn once we look at the space-time directions. The zeroth order in $g_s$ 
produces the space-time metric with full de Sitter isometries. The EOM is given by \eqref{fleuve}, and one may note that the flux components appear as before, and the quantum terms are classified by 
$\theta'_k \le 8/3$ in \eqref{melamon2} as shown in \eqref{oleport}. Such an equation has the following  implications. For $n_i = l_{34 + i} = 0$ in \eqref{oleport}, the $l_i$ can at best be bounded as $l_i \le 4$. Since 
$l_i$ for $i = 1,.., 27$ capture the curvature polynomials in \eqref{phingsha2}, this implies that at the 
{classical} level, the space-time EOM should have the fourth-order curvature terms. Not only that, \eqref{oleport} predicts that at the classical level all possible eighth-order\footnote{In derivatives.} polynomials with curvature, G-flux components
(classified by $l_{34+i}$) and derivatives (classified by $n_i$) are {\it necessary}. It was known for sometime 
in the literature that classically the fourth-order curvature polynomials (or eighth-order in derivatives) like:
\bg\label{tagmetmey}
J_0 \equiv t_8 t_8 {\bf R}^4, ~~~~~ E_8 \equiv \epsilon_{11} \epsilon_{11} {\bf R}^4, 
\nd 
should play a part, and now we not only can confirm this statement but also show that {\it all} eighth-order polynomials classified by \eqref{oleport} should play a part at the classical level. Of course the exact coefficients of these polynomials cannot be predicted from \eqref{phingsha2} or \eqref{oleport}, but the fact that this comes out naturally from our analysis should suggest that we are on the right track.

The quantum terms now do the same magic as before. To order $g_s^{1/3}$ the quantum terms, classified by $\theta'_k = 3$ in \eqref{melamon2} including the non-perturbative terms, balance each other as \eqref{urisis} in such a way that the four-dimensional de Sitter metric do not change. To next order in $g_s$, i.e $g_s^{2/3}$, the quantum terms, now classified by $\theta'_k = 10/3$, balance against the ($C_k, \widetilde{C}_k$) coefficients as in   
\eqref{ellemey} in a way as to again keep the zeroth order de Sitter metric invariant. The story progresses in the same way as we go to higher orders in $g_s$. 

From the above discussions we can now summarize our view of stability here. The classical EOMs, or the EOMs to the lowest order in 
$g_s$ (which for most cases are to zeroth order in $g_s$ with the exception of one where the lowest order is $g_s^2$), for all the components are \eqref{misslemon2}, \eqref{uanaban}, \eqref{buskaM} and 
\eqref{fleuve}. They involve the so-called quantum terms that not only renormalize the existing classical data but also add eighth-order (in derivatives) polynomials. Together with the G-flux components they determine the type IIB metric with four-dimensional de Sitter space-time and the un-warped internal six-dimensional non-K\"ahler metric. The quantum effects on this background, to order-by-order in powers of $g_s^{1/3}$, are balanced against the G-flux components and the coefficients ($C_k, \widetilde{C}_k$) coefficients, again to order by order in powers of $g_s^{1/3}$, in a way so as to preserve the form of the dual type IIB metric to the lowest order in $g_s$.  This is one important criteria of stability here.

\vskip.1in

\noindent {\it Stability, fluctuations and tachyonic modes}

\vskip.1in

\noindent Finally we turn our attention to the possible presence of tachyonic modes around our de Sitter background. This is an important question to determine the relationship between our background and the swampland criteria. The presence of tachyonic modes of sufficiently negative mass would be in agreement with the Hessian de Sitter criterion, while the absence of such would call for a re-examination of the criterion in the context of time-dependent backgrounds.

To determine the presence of tachyons we need to perturb our metric ansatze \eqref{vegamey} (and also the fluxes) and expand the quantum effective action to second order in the perturbations. Of course, the deciding factor is the sign of the various terms. Since we do not know the coefficients of all the quantum corrections, we can not hope to be completely sure of the absence of tachyonic modes using our approach. We do however have some information about the relative signs of some terms, from the requirement of positive four-dimensional curvature, so there may still be a consistency check available. The constraints on the curvature only manifest themselves in the metric equation of motion so we choose the following perturbations:
\bg\label{heigns}
\delta g_{MN}(x, y) = \phi^{(MN)}(x) g_{MN}(y),
\nd
where $x$ is the coordinate along the $2+1$ dimensional space-time directions and $y$ is the internal space coordinates.
For the internal components of the metric, $\phi^{(mn)}(x)$ are simply the scalars one obtains from dimensional reduction. For the space-time components these amount to the scalar modes of metric perturbations. The upside to using perturbations proportional to the ``background'' values of the fields is that the expansion of the quantum potential to second order in the perturbation is the same as calculating the second order variation of the quantum terms with respect to the original fields. The extra $x$ dependence can generate new contributions to the action, if derivatives along the space-time directions act on it. However this will not result in potential terms, but rather will contribute to the kinetic and higher-derivative terms for the scalar, which will have no bearing on the tachyon question. The downside of this choice of fluctuation is that it ignores the fields which are set to zero\footnote{We have assumed earlier that we have integrated such components out and that the effects of their fluctuations have thus already been incorporated into the quantum potential. This is strictly speaking only possible if their masses are above the scale at which we are studying the theory. Otherwise there are IR modes left over. Note that in either case, these modes are certainly not tachyonic in the ground state of our EFT, so the implicit hope here is simply that they also do not become tachyonic as we move to the coherent de Sitter state.}. Since terms involving these fields don't appear in our background quantum potential, their sign will not be constrained by the curvature conditions anyway. Other subtleties aside, 
the first variation of the action with respect to the metric is simply given by the equations of motion:
\bg\label{steelem}
\frac{\delta \mathbb{S}_{11}}{\delta {\bf g}^{MN}} &=& \int d^{11}x \sqrt{-{\bf g}_{11}} 
\left({\bf R}^{(11)}_{MN} - \frac{1}{2}{\bf R}^{(11)} {\bf g}_{MN} - \mathbb{T}^{G}_{MN} - \mathbb{T}^{Q}_{MN}   \right), \nd
where the metric components are all taken as the warped ones and the energy-momentum tensors, especially the quantum energy momentum tensor, take the form that we have used so far. For example the latter would appear from \eqref{phingsha2}, say if we consider only the case \eqref{olokhi}. In other words, we can use \eqref{phingsha2} to express the quantum energy-momentum tensor in the following way: 
\bg\label{milast}
\mathbb{T}^{Q}_{MN} = \frac{1}{2} {\bf g}_{MN} \mathcal{L}^{(Q)} - \frac{\delta \mathcal{L}^{(Q)}}
{\delta {\bf g}^{MN}}, 
\nd
where $\mathcal{L}^{(Q)}$ is the the sum of quantum terms in the action (i.e. without Lorenz indices). This is pretty much equivalent to \eqref{ducksoup}, with the quantum pieces expressed together as \eqref{chuachu}. Alternatively, we could also express it more directly as \eqref{neveC}. With these at hand, the second variation takes the form:

{\footnotesize
\bg\label{harmonir}
\frac{\delta^2 \mathbb{S}_{11}}{\delta {\bf g}^{PQ} \delta {\bf g}^{MN}} &=& \int d^{11} x 
\sqrt{-{\bf g}_{11}} \Bigg[\frac{\delta {\bf R}^{(11)}_{MN}}{\delta {\bf g}^{PQ}} - \frac{1}{2} 
\left({\bf R}^{(11)}_{PQ} 
{\bf g}_{MN} - {\bf R}^{(11)} {\bf g}_{M(P} {\bf g}_{Q)N} \right) - \frac{\delta \mathbb{T}^{G}_{MN}}
{\delta {\bf g}^{PQ}} \\
&+& \frac{1}{2} \mathcal{L}^{(Q)} {\bf g}_{M(P} {\bf g}_{Q)N} - \frac{1}{2} {\bf g}_{MN} 
\frac{\delta \mathcal{L}^{(Q)}}{\delta {\bf g}^{PQ} } + \frac{ \delta^2 \mathcal{L}^{(Q)} }{\delta {\bf g}^{PQ} \delta {\bf g}^{MN}}  \Bigg]  + \int d^{11} x\sqrt{-{\bf g}_{11}} ~{\bf g}_{PQ}(\rm{EOM})_{MN}. \nonumber
\nd}
Stable solutions to the equations of motion are local maxima of the action, so complete stability would require that the above expression is negative.

Note that the first variation of $\mathcal{L}^{(Q)}$ is still present in the expression, and can be re-expressed in terms of the quantum stress tensor $\mathbb{T}^{Q}_{MN}$, as in \eqref{neveC}, which contains the quantum corrections 
$\mathbb{C}^{(k_1,0)}_{MN}$ that appear in the lowest order equations of motion. From here, one approach could be to make a connection with the positivity of the cosmological constant by, for example, taking the same linear combination of diagonal components as was used to obtain \eqref{musemey}. However, there are still terms involving $\mathcal{L}^{Q}$ and more importantly its second variation, which does not appear in the equations of motion. These terms have signs that are not fixed by the trace of the metric equations of motion alone as they depend on all the components and fluxes. This means they would need to be determined by solving for all the metric and flux components. 

At this stage we could make some general observations. If we restrict the metric variations to be along the six-dimensional base 
${\cal M}_4 \times {\cal M}_2$, and only consider the case \eqref{olokhi} with quantum corrections, the second variation of 
${\cal L}^{(Q)}$ contains quantum terms classified by $\theta'_k - {4\over 3}$ perturbatively and at most
$\theta'_k - {10\over 3}$ non-perturbatively. This implies that to zeroth order in $g_s$, which we used to determine the EOMs, the contributions from the second variation of 
${\cal L}^{(Q)}$ come from the quantum terms\footnote{In other words, the first variations of the action i.e the EOMs, provide the background values of metric and G-flux components. These values enter inside the quantum terms classified by $\theta'_k$ in \eqref{melamon2} appearing from the second variations of the action.}  
classified by ${4\over 3} \le \theta'_k \le {10\over 3}$ in \eqref{melamon2}. In a similar vein, if one of the metric variation is along $\mathbb{T}^2/{\cal G}$ and the other along the six-dimensional base, or if both the variations are along $\mathbb{T}^2/{\cal G}$, then the second variations of ${\cal L}^{(Q)}$ come from the 
quantum terms classified {\it perturbatively} by $\theta'_k + {2\over 3}$ or $\theta'_k + {8 \over 3}$ respectively (non-perturbatively there are additional $g_s^{-2}$ suppressions as we saw from section \ref{instachela}). Clearly, non-perturbatively only the  first one can contribute to the zeroth order in $g_s$. On the other hand, if both the metric variations are along the $2+1$ dimensional space-time directions, the quantum terms contributing to the second variation of 
${\cal L}^{(Q)}$ are classified non-perturbatively by $\theta'_k = {22\over 3}$ in \eqref{melamon2}. In this way, one could go about finding other combinations, but the message should be clear. If all these contributions are such that  they make 
the RHS of \eqref{harmonir} negative definite, then there would be no tachyonic instability in our background.

Let us compare this to the first variation of ${\cal L}^{(Q)}$ contributing to the cosmological constant 
$\Lambda$ in \eqref{musemey}. The internal space quantum terms are classified by 
${2\over 3} \le \theta'_k \le {8\over 3}$ 
in \eqref{melamon2} and the $2+1$ dimensional space-time quantum terms are classified by $\theta'_k \le {8\over 3}$. Since the lower bounds on the internal space quantum terms simply renormalize the existing classical terms, the burden of getting {\it positive} cosmological constant rests solely on the quantum terms classified
by $\theta'_k \le {8\over 3}$. We want them to give positive contributions, so that the relative minus sign in \eqref{musemey} can make $\Lambda > 0$. Here, in \eqref{harmonir}, we want the opposite (assuming the contributions from the other terms are negligible). It is easy to see that, compared to the case 
\eqref{musemey}, there are now quantum terms classified by $\theta'_k \le {22\over 3}$ in \eqref{melamon2}, so we are no longer restricted only with the quantum terms classified by $\theta'_k \le {8\over 3}$. We now require these terms to make the RHS of \eqref{harmonir} negative definite to avoid the tachyonic instability. 

There are also second variations of the action with respect to the ${\bf C}_{MNP}$ fields, i.e 
$\frac{\delta^2 \mathbb{S}_{11}}{\delta {\bf C}_{MNP} \delta {\bf C}_{RSU}}$, that also need to be considered. Most of the three-form potentials scale in an identical way, so we expect the quantum terms 
contributing at the zeroth order being classified by $\theta'_k = 4\Delta k$ in \eqref{melamon2}
with $k \ge {3\over 2}$ for the case 
\eqref{olokhi}. We have put to zero components like ${\bf C}_{0MN}$ using \eqref{pashaclub}, and in fact the quantum term $\mathbb{Y}_7$ has enough degrees of freedom to keep these modes from contributing to the tachyonic instability. The space-time potentials ${\bf C}_{0ij}$ would contribute quantum terms classified by $\theta'_k + 8$, so they don't change the zeroth order equations. However now there could also be {\it mixed} variations like $\frac{\delta^2 \mathbb{S}_{11}}{\delta {\bf C}_{MNP} \delta {\bf g}^{RS}}$, and depending on the choice of $k$ and the orientations of the metric components, some of them would contribute to the zeroth order EOMs. Fortunately the quantum terms contributing to this order, or in general any orders, are finite in number so it is not a very difficult exercise to list all these terms appearing from the second variations of \eqref{phingsha2}, and see how the tachyonic instability, if any, could be 
removed\footnote{Analysis in terms of the {\it dual} six-form potentials are much more involved as every components scale in a  different way and may be extracted from {\bf Table \ref{dualforms}}.}.   
Similar arguments can be given for the case \eqref{ranjhita} but we will not pursue this here.

\subsubsection{Stability, landscape and the swampland criteria \label{jola}}

In the above discussions we summarized how the quantum corrections do not destabilize the 
background, and instead tend to stabilize it at every order in $g_s^{1/3}$. Our next exercise would be to see how the stability extends to keeping the background in the {\it landscape} and out of the {\it swampland}. In other words we want to show how the swampland criteria are averted by the time-dependences of the fluxes and the metric components and by our choice of the quantum potential.

\vskip.1in

\noindent {\it The exactness of the four-dimensional cosmological constant}

\vskip.1in

\noindent The quantum potential is given in \eqref{ducksoup} and basically incorporates the information of either 
\eqref{phingsha} and \eqref{phingsha2} for the two cases \eqref{ranjhita}  and \eqref{olokhi} respectively. 
However it is important to note that the cosmological constant $\Lambda$ appears almost exclusively from the $g_s$ independent, or time independent, parts of the potential (i.e most of the contribution to $\Lambda$ appears from the 
$g_s$ independent parts of $\mathbb{V}_Q$ in \eqref{ducksoup}), which goes without saying that it is truly a constant\footnote{In other words, and taking into account the time-independent Newton's constant from 
\eqref{olokhi}, the late-time cosmology will always be de Sitter in our set-up and {\it never} quintessence. There is of course the possibility that the Newton's constant may get {\it renormalized} while still remaining time-independent. Such a scenario is discussed in \cite{coherbeta}, wherein the renormalization effect comes from the zeroth order quantum terms and is therefore reflected in the expression of the cosmological constant.}. The exact form of which may be expressed as:

{\footnotesize
\bg\label{hathway}
\Lambda &=& 
{1\over 12 V_6} \left\langle [\mathbb{C}^i_{i}]^{(0, 0)} \right\rangle -
{1\over 24V_6 H^4} \left\langle\left[\mathbb{C}_a^a \right]^{(3, 0)} \right\rangle
-{1\over 48V_6H^4} \left\langle\left[\mathbb{C}_m^m \right]^{(0, 0)} \right\rangle
-{1\over 48 V_6H^4} \left\langle\left[\mathbb{C}_\alpha^\alpha\right]^{(0, 0)}\right\rangle 
\nonumber\\
&-& {\kappa^2 T_2 n_b \over 6 V_6 H^8}
-{5\over 384 V_6 H^8}\Big[\left\langle {\cal G}^{(3/2)}_{mnab}{\cal G}^{(3/2)mnab} \right\rangle 
+ \left\langle {\cal G}^{(3/2)}_{m\alpha ab}{\cal G}^{(3/2)m\alpha ab} \right\rangle +
\left\langle {\cal G}^{(3/2)}_{\alpha\beta ab}{\cal G}^{(3/2)\alpha\beta ab} \right\rangle\Big],
 \nd}
which may be easily inferred from \eqref{musemey}, and we have taken, just for simplicity, a very slowly varying function for $H$. Thus $H$ is essentially a constant and can come out of the integrals in 
\eqref{musemey}. $V_6$ is the volume of the six-dimensional base ${\cal M}_4 \times {\cal M}_2$, i.e the 
volume measured using un-warped metric components.
The other expectation values are defined in the standard way $-$ we take the functions and integrate over the volume element $-$ namely:

{\footnotesize
\bg\label{anneH}
\left\langle [\mathbb{C}^M_{M}]^{(a, 0)} \right\rangle \equiv 
\int d^6 y \sqrt{g_6}  [\mathbb{C}^M_{M}]^{(a, 0)}, ~
\left\langle {\cal G}^{(3/2)}_{MN ab}{\cal G}^{(3/2)MN ab} \right\rangle \equiv
\int d^6 y \sqrt{g_6} ~{\cal G}^{(3/2)}_{MN ab}{\cal G}^{(3/2)MN ab}, \nd}
where $g_6$ is the determinant of the un-warped metric of the six-dimensional base, ($M, N$) denote the coordinates of the base and the superscript $a = 0, 3$ depending on which quantum corrections we choose.  In fact as discussed earlier, the most {\it dominant} quantum terms are the ones classified by $\theta'_k = {8\over 3}$
or $\theta_k = {8\over 3}$ in \eqref{melamon2} and \eqref{miai} respectively. These mostly appear from the
non-perturbative contributions to  
$\left[\mathbb{C}^i_{i}\right]^{(0, 0)}$ and $\left[\mathbb{C}^M_{M}\right]^{(0, 0)}$, as we saw from section 
\ref{instachela}. The lower order perturbative contributions, i.e the ones classified by say $\theta'_k = 2/3$, simply renormalize the existing classical data. Since the fluxes are taken to be small everywhere and $n_b$ is 
small\footnote{Note that it doesn't matter whether we take M2 or anti-M2 branes in \eqref{hathway}. 
The {\it sign} of the cosmological constant $\Lambda$ {\it cannot} be changed from either of them $-$ a fact reminiscent of the no-go condition of \cite{gibbons, malnun}. In fact we can even go a step further. The presence or absence of M2 or anti-M2 branes $-$ because of the negative sign in \eqref{hathway} $-$ is a red herring in the problem because the positivity of the cosmological constant lies solely on the quantum terms,  especially on the $2+1$ dimensional space-time part of the quantum corrections, and not on branes or anti-branes. As such the back-reaction effect studied in \cite{bena} doesn't appear relevant here.}, the cosmological constant $\Lambda$ can be made positive, i.e $\Lambda > 0$, in the following way: consider all the quantum  terms in $\left[\mathbb{C}^i_{i}\right]^{(0, 0)}$. They appear from the perturbative and the non-perturbative interactions. The perturbative interactions only tend to renormalize the existing classical terms, and it's the non-perturbative corrections from section \ref{instachela} that dominate. In fact as long as the non-perturbative contributions dominate as a {\it positive-definite} quantity over any positive non-perturbative contributions to $\left[\mathbb{C}^M_{M}\right]^{(0, 0)}$ (because of the relative minus sign in \eqref{hathway}, any {\it negative} contributions would contribute positively to the cosmological constant), the cosmological constant would become positive. Additionally, the overall volume suppression in \eqref{hathway} tells us that for large enough $V_6$, $\Lambda$ could indeed be a 
tiny non-zero positive number. 
The crucial observation however is that the other parts of $\mathbb{V}_Q$ in \eqref{ducksoup} are used to {\it stabilize} the classical background in a way discussed earlier, but they do not contribute to the cosmological constant here\footnote{One might wonder why all the energies from $\mathbb{V}_Q$ do not contribute to the cosmological constant. The answer is simple to state. At the zeroth order in $g_s$, the energy contribution gives the cosmological constant $\Lambda$ as shown in \eqref{hathway}. As we go to the next order in $g_s$, new G-flux components are switched on, back-reacting on the geometry to create {\it negative} gravitational potentials. These negative potentials are exactly cancelled by the positive potential energies coming from $\mathbb{V}_Q$ to this order in such a way that the zeroth order energy, i.e $\Lambda$, does not change. The story repeats at every order in the same fashion. The net result is that the cosmological constant $\Lambda$ truly remains a {\it constant} at all orders in $g_s$. \label{expanda}}!

\vskip.1in

\noindent {\it Keeping the de Sitter vacua outside the swampland}

\vskip.1in

\noindent One may also ask how the swampland criteria are taken care of here. The fact that new degrees of freedom do not appear when we switch on time-dependences is easy to infer by looking at the $g_s$ scalings 
$\theta_k$ and $\theta'_k$ in \eqref{miai} and \eqref{melamon2} respectively. Putting $k =0$ is equivalent to switching-off the time-dependences, and we get $\theta'_0$ as in \eqref{kkkbkb2} which in-turn is defined with relative minus signs. Existence of such relative minus signs lead to an infinite number of states 
satisfying \eqref{evabmey2} for any given value of $\theta'_0$ in \eqref{evabmey2}. This proliferation of states is of course one sign of the breakdown of an EFT description, and therefore the theory is indeed in the swampland. Switching on time-dependences miraculously cure this problem as both $\theta_k > 0$ and $\theta'_k > 0$
for the cases \eqref{ranjhita} and \eqref{olokhi} respectively.

The above reasonings do provide a way to overcome the swampland {\it distance} criterion, namely, switching on time-dependences allows us to avoid inserting arbitrary number of degrees of freedom at any given point in the moduli space of the theory. The question now is how the original swampland 
criterion \cite{vafa1}, namely, $\partial_\phi V > c V$ is taken care of with $c = {\cal O}(1)$ number. To see this, let us consider the quantum terms \eqref{phingsha2} for the case \eqref{olokhi} (similar argument may be given for 
\eqref{phingsha} for the case \eqref{ranjhita}). The potential associated to this is 
\eqref{ducksoup}, and we can get scalars from the G-flux components as well as from the internal metric components. First let us take a simple example where the scalar fields appear from the G-flux components
in the following way:
\bg\label{dlane}
 {\bf C}_3({x}, y) &=& \langle{\bf C}_3(y)\rangle + \sum_i
\phi^{(i)}({x})  \Omega^{(i)}_{(3)}(y) +  \sum_j {\bf A}^{(j)}_1(x) \wedge \Omega^{(j)}_{(2)}(y) 
+ \sum_l {\bf B}^{(l)}_2(x) \wedge \Omega^{(l)}_{(1)}(y)\nonumber\\
{\bf G}_4({x}, y) &=& \langle{\bf G}_4(y)\rangle + \sum_i \phi^{(i)}({x}) d\Omega^{(i)}_{(3)} 
+ \sum_i d\phi^{(i)}({x}) \wedge \Omega^{(i)}_{(3)}(y) + \sum_j {\bf F}^{(j)}_2(x) \wedge \Omega^{(j)}_{(2)}(y) 
\nonumber\\
&-& \sum_j {\bf A}^{(j)}_1(x) \wedge d\Omega^{(j)}_{(2)}(y) 
+ \sum_l
 {\bf B}^{(l)}_2(x) \wedge d\Omega^{(l)}_{(1)}(y) + \sum_l {\bf H}_3^{(l)} \wedge \Omega^{(l)}_{(1)}, \nd
where $\Omega^{(j)}_{(k)}$ are the $k$-forms defined over the internal manifold (we can restrict them to the six-dimensional base ${\cal M}_4 \times {\cal M}_2$  with ($i, j$) representing 
the number of independent forms), and are not necessarily harmonic functions as 
the underlying background is non-supersymmetric and the six-dimensional base is non-K\"ahler. This also explains why we can allow one-forms like $\Omega^{(i)}_{(1)}$. The two-forms 
$\Omega^{(j)}_{(2)}$ should not be confused with the localized two-form $\Omega_{ab}$ in 
\eqref{teskimey}. Additionally, \eqref{teskimey} is the decomposition of the background data itself, whereas \eqref{dlane} is the decomposition of the {\it fluctuations} over our background \eqref{vegamey}\footnote{We expect ${\bf H}_3^{(l)} = 0$ because it has no dynamics in $2+1$ dimensions.}. 
We are also suppressing the $g_s$ dependences, and therefore both the $k$-forms and the $2+1$ dimensional 
space-time fields have $g_s$ dependences. In general, for a manifold whose geometry is varying with time, we expect:
\bg\label{boklikelly}
\int d\Omega_{(k)}^{(i)} \wedge \ast_6 ~d\Omega_{(k)}^{(j)} \equiv 
\sum_{\{l_i\}} \int d\Omega_{(k)}^{(l_1, i)} \wedge \ast_6~ d\Omega_{(k)}^{(l_2, j)}
\left({g_s\over H}\right)^{2\Delta(l_1 + l_2)}
\nd
over the six-dimensional base ${\cal M}_4 \times {\cal M}_2$ with the Hodge star defined over this base. 
Here $l_i$ denotes the mode expansion that we have used so far.  
In the standard time-independent supersymmetric case this would have vanished, but now we see 
explicit $g_s$ dependences complicating our analysis. Finally, 
the expectation values in \eqref{dlane} refer to the background values of the three- and the four-forms that we took earlier to solve the background EOMs (and thus they are functions of $y^M$). 
We have also given a small ${x}$ dependences to the {\it fluctuations} of the three- and the four-forms, and for computational efficiency, let us assume that we take the G-flux component ${\bf G}_{mnpq}$. 
For simplicity then, $i = 1$ in \eqref{dlane} with ${\bf A}_1^{(j)}(x) = {\bf B}_2^{(l)}(x) = 0$. Plugging 
\eqref{dlane} into \eqref{phingsha2} and \eqref{ducksoup}, we get the following form of the $2+1$ dimensional space-time potential:
\bg\label{serenity}
\mathbb{V}_Q({x}) \equiv \sum_{\{l_i\}, n}
 \int d^8 y \sqrt{{\bf g}_8} \left({\mathbb{Q}_T^{(\{l_i\}, n)} \over M_p^{\sigma(\{l_i\}, n)-8}}\right)= \sum_{l_{28}} \phi^{l_{28}}({x}) {\bf V}\left(\Phi({x})\right), \nd
where $\Phi({\bf x})$ are the set of all other scalars in the system, ${\bf V}(\Phi(x))$ is now dimensionless, 
and $l_{28}$ is a positive integer  that appears in \eqref{phingsha2}. We have ignored the non-perturbative corrections to \eqref{serenity}: they will simply make the story more involved without necessarily changing any of the outcomes that we are about to discuss in the following.
 For the purpose of our discussions we will take $l_{28} \ge 1$, and from the form of the G-flux components \eqref{ravali} it is clear that both $\phi(x)$ as well as $\Omega_{(3)}(y)$ should have $g_s$ dependences, confirming the $g_s$ dependence in \eqref{boklikelly}.  We can then assume:
\bg\label{katygorom}
\phi(x) \equiv \phi^{(1)} = \sum_p \phi^{(1, p)}({\bf x})\left({g_s\over H}\right)^{2\Delta p}, \nd
where $p$ has to be bounded below because the $k$ in G-flux components \eqref{ravali} are bounded below as $k \ge {3\over 2}$ or $k \ge {9\over 2}$ for \eqref{olokhi} and \eqref{ranjhita} respectively.
By construction \eqref{serenity} is derived from \eqref{ducksoup} and therefore ${\bf V}(\Phi(x))$ has the $g_s$ scaling given by $\theta'_k - 2\Delta p l_{28} - {2\over 3}$. 
 The swampland criterion then gives us:
\bg\label{lojjabot}
{\partial_\phi \mathbb{V}_Q \over \mathbb{V}_Q} = 
{\sum_{l_{28}}l_{28}\sum_{\{k_i\}}\phi^{(1, k_1)}...... \phi^{(1, k_{l_{28}})}\left({g_s\over H}\right)^{2\Delta(k_1 + ... k_{l_{28}})}
\over 
\sum_{\{r, q_i\}}\phi^{(1, r)}\phi^{(1, q_1)}...... \phi^{(1, q_{l_{28}})}\left({g_s\over H}\right)^{2\Delta(r+ q_1 + ... q_{l_{28}})}}
= {\cal O}\left({1\over g^n_s}\right) >> 1, \nd
where $n = {\cal O}(2\Delta r) \in \mathbb{Z}$ and $g_s < 1$. The above computation could be easily generalized to all scalar fields coming from the G-flux components in say \eqref{phingsha2}, provided of course the decomposition 
\eqref{dlane} is respected. For example taking all the components of $\phi^{(i)}$ in \eqref{dlane}, we get:
\bg\label{lina05}
{\left\vert\nabla \mathbb{V}_Q \right\vert \over \mathbb{V}_Q} = {\sqrt{g^{\phi^{(i)} \phi^{(j)}}
\partial_{\phi^{(i)}} \mathbb{V}_Q \partial_{\phi^{(i)}}\mathbb{V}_Q} \over \mathbb{V}_Q}  = 
{\cal O}\left(\sum_{k= 1}^{{\rm dim}\left({\cal M}_\phi\right)} {1\over g_s^{n_k}}\right) >> 1, \nd
where $g^{\phi^{(i)} \phi^{(j)}}$ is the metric on the moduli space ${\cal M}_\phi$
of all the scalars represented by 
$\phi^{(i)}$ which, in turn, could be decomposed as \eqref{katygorom}. The subscript $k$ in $n_k$ 
is summed from 1 to  ${\rm dim}\left({\cal M}_\phi\right)$, i.e 
dimension of the moduli space of the scalars. None of the scalars appearing from the G-fluxes are related to the inflaton, so the RHS being much bigger than identity is not unreasonable. 
Under these circumstances, clearly the swampland bound of \cite{vafa1} is easily 
satisfied. We can also confirm that any additional non-perturbative corrections to \eqref{serenity} cannot change the outcome.

On the other hand, the scalars coming from the metric components could in principle also be analyzed in a similar vein as \eqref{lina05}, but the analysis is complicated by the fact that the potentials for these scalars are not as simple as for the scalars from the G-flux components. In any case, the obvious redundancy in  indulging in such exercise should already be apparent from our earlier demonstration of the existence of
four-dimensional EFT descriptions with de Sitter isometries. Since these conclusions are derived from  
meticulously studying the $g_s$ scalings of the quantum terms, the swampland criteria are taken care of here, and these theories belong to the landscape of IIB vacua. 

\vskip.1in

\noindent {\it The null, strong and the dominant energy conditions}

\vskip.1in

\noindent Thus instead of getting involved in superfluous exercises to distinguish swampland versus landscape criteria, we can ask how the energy conditions are taken care of here. This is a meaningful question to ask because it brings us to the very foundation on which the no-go criteria of \cite{gibbons, malnun} are based.  
To proceed then we will make the assumption of a slowly varying warp-factor $H(y)$ so that the derivatives of the warp-factor do not un-necessarily complicate the ensuing analysis\footnote{In other words, the derivatives of the warp-factor $H(y)$ will add irrelevant functions to the traces that we perform below. We can absorb these functions in the quantum terms.}. 
To zeroth order in $g_s$ the trace of the energy-momentum tensor is defined as:
\bg\label{casin300}
\mathbb{T}_M^M \equiv \left[\mathbb{T}_M^M\right]^G + \left[\mathbb{T}_M^M\right]^Q, \nd
where the superscript $G$ and $Q$ correspond to the G-flux and the quantum energy-momentum tensors respectively. The traces of the individual pieces are taken with respect to the un-warped internal metric components. Restricting \eqref{casin300} to the ($m, n$), ($\alpha, \beta$) and ($a, b$) directions, yield the following traces:
\bg\label{montcross}
&&\mathbb{T}^\alpha_\alpha \equiv \left[\mathbb{C}_\alpha^\alpha\right]^{(0, 0)}  
+{1\over 8H^4} \Big({\cal G}^{(3/2)}_{\alpha\beta ab}{\cal G}^{(3/2)\alpha\beta ab} - 
{\cal G}^{(3/2)}_{mn ab}{\cal G}^{(3/2)mn ab}\Big)\\
&&\mathbb{T}^m_m \equiv 
\left[\mathbb{C}_m^m\right]^{(0, 0)}  
-{1\over 4H^4} \Big({\cal G}^{(3/2)}_{m\alpha ab}{\cal G}^{(3/2)m\alpha ab} + 
{\cal G}^{(3/2)}_{\alpha\beta ab}{\cal G}^{(3/2)\alpha\beta ab}\Big) \nonumber\\
&&\mathbb{T}^a_a \equiv 
\left[\mathbb{C}_a^a\right]^{(3, 0)}  
+ {1\over 8H^4} \Big(2{\cal G}^{(3/2)}_{m\alpha ab}{\cal G}^{(3/2)m\alpha ab} +
{\cal G}^{(3/2)}_{mn ab}{\cal G}^{(3/2)mn ab} 
+ {\cal G}^{(3/2)}_{\alpha\beta ab}{\cal G}^{(3/2)\alpha\beta ab} 
\Big), \nonumber \nd
where the individual energy-momentum tensors are defined in subsections \ref{kocu2}, \ref{kocu1} and 
\ref{kocu3} respectively for the case \eqref{olokhi}. We can easily insert in the non-perturbative corrections to the energy-momentum tensors from subsection \ref{instachela}.
A similar construction could be done for the case 
\eqref{ranjhita} too but we will not pursue this here. Note that, as an interesting fact, if we sum up all the three traces in \eqref{montcross}, we  will get:
\bg\label{pageT}
\mathbb{T}^m_m + \mathbb{T}^\alpha_\alpha + \mathbb{T}^a_a = \left[\mathbb{C}_m^m\right]^{(0, 0)} +
\left[\mathbb{C}_\alpha^\alpha\right]^{(0, 0)} + \left[\mathbb{C}_a^a\right]^{(3, 0)}, \nd
with no contributions from the G-flux components. Thus the total trace of the energy-momentum tensor in the internal space is only given by the quantum terms. These quantum terms are classified by 
${2\over 3} \le \theta'_k \le {8\over 3}$, and the lower bounds are related to the G-flux components as in \eqref{lulu}, thus renormalizing the existing classical data. The upper bounds give rise to the
eight-derivative terms that we discussed earlier. Similarly, the trace along the $2+1$ dimensional space-time direction yields:

{\footnotesize
\bg\label{crossmey}
&&\mathbb{T}^i_i  =   [\mathbb{C}^i_{i}]^{(0, 0)}  - \mathbb{A}^i_i, ~~~
\mathbb{T}^0_0  =   \left[\mathbb{C}^0_{0}\right]^{(0, 0)}  - \mathbb{A}^0_0\\
&&\mathbb{A}^i_i = \mathbb{A}^0_0 \equiv
{2 \kappa^2 T_2 n_b \over H^8 \sqrt{g_6}} \delta^8(y - Y)
+{1\over 8 H^8} \Big({\cal G}^{(3/2)}_{mnab}{\cal G}^{(3/2)mnab} +
2 {\cal G}^{(3/2)}_{m\alpha ab}{\cal G}^{(3/2)m\alpha ab} +  {\cal G}^{(3/2)}_{\alpha\beta ab}
{\cal G}^{(3/2)\alpha\beta ab}\Big), \nonumber \nd} 
where by construction $\mathbb{A}^i_i > 0$ and $\mathbb{A}^0_0 > 0$; and 
 both the quantum terms are classified by $\theta'_k \le {8\over 3}$ in \eqref{melamon2}. They therefore also involve eight-derivative terms as we saw in subsection \ref{kocu4} for the case \eqref{olokhi}. What we now need is:
 \bg\label{helen}
&& \mathbb{T}^i_i + \mathbb{T}^0_0 ~>~  \mathbb{T}^m_m + \mathbb{T}^\alpha_\alpha + \mathbb{T}^a_a\nonumber\\
&& [\mathbb{C}^i_{i}]^{(0, 0)} + \left[\mathbb{C}^0_{0}\right]^{(0, 0)} - \mathbb{A}^i_i - \mathbb{A}^0_0
~> ~ \left[\mathbb{C}_m^m\right]^{(0, 0)} +
\left[\mathbb{C}_\alpha^\alpha\right]^{(0, 0)} + \left[\mathbb{C}_a^a\right]^{(3, 0)}, \nd
which would be the null energy condition. Clearly when the quantum terms vanish, the inequality 
\eqref{helen} can never be satisfied, consistent with the no-go conditions of \cite{gibbons, malnun} and also \cite{nogo}. However once we allow the higher order quantum terms, and the very fact that the 
$[\mathbb{C}^\mu_\mu]^{(0, 0)}$ terms can be made to {\it dominate} over the other quantum terms (which was in fact necessary to reproduce $\Lambda > 0$ in the first place), the inequality \eqref{helen} can in principle be satisfied. To see this, let us recall that the lower bounds
$\theta'_k = {2\over 3}$ in \eqref{melamon2} for the internal quantum terms allow us to choose 
($l_{36}, l_{37}, l_{38}$) as (2, 0, 0), (0, 2, 0) or (0, 0, 2) in \eqref{phingsha2}, implying at most quadratic in these G-flux components. Additionally, the higher order quantum terms, classified by 
$\theta'_k \le {8\over 3}$, and appearing in the EOMs to zeroth order in $g_s$ are constrained as \eqref{lulu}. Combining these two, one possible solution could be the following:
\bg\label{sortega}
&& \left[\mathbb{C}_a^a\right]^{(3, 0)} = -{1\over 6 H^8} \Big({\cal G}^{(3/2)}_{mnab}{\cal G}^{(3/2)mnab} +
2 {\cal G}^{(3/2)}_{m\alpha ab}{\cal G}^{(3/2)m\alpha ab} +  {\cal G}^{(3/2)}_{\alpha\beta ab}
{\cal G}^{(3/2)\alpha\beta ab}\Big)\\
&&  \left[\mathbb{C}_m^m\right]^{(0, 0)} + \left[\mathbb{C}_\alpha^\alpha\right]^{(0, 0)} =
{1\over 24 H^8} \Big({\cal G}^{(3/2)}_{mnab}{\cal G}^{(3/2)mnab} +
2 {\cal G}^{(3/2)}_{m\alpha ab}{\cal G}^{(3/2)m\alpha ab} +  {\cal G}^{(3/2)}_{\alpha\beta ab}
{\cal G}^{(3/2)\alpha\beta ab}\Big), \nonumber \nd
which still leaves enough freedom to determine $\left[\mathbb{C}_m^m\right]^{(0, 0)}$ and 
$\left[\mathbb{C}_\alpha^\alpha\right]^{(0, 0)}$ individually. Such a choice would 
cancel the $\mathbb{A}^\mu_\mu$ terms in \eqref{helen}, yet satisfy \eqref{lulu}.
The viability of the choice \eqref{sortega} is guaranteed from the analysis of the EOMs in subsections \ref{kocu1}, \ref{kocu2} and \ref{kocu3}, where 
the input \eqref{sortega} could determine what kind of internal non-K\"ahler manifold we get. Note however that, in determining \eqref{sortega}, we have ignored the M2-brane contribution. Since $n_b \ne 0$ from \eqref{haroldR}, this can be justified from the fact that for $y^M \ne Y^M$ the M2-brane contributions vanish 
in $\mathbb{A}^\mu_\mu$ from \eqref{crossmey}. Therefore combining \eqref{sortega} with \eqref{helen}, we see that as long as:
\bg\label{kortega}
[\mathbb{C}^i_{i}]^{(0, 0)} + \left[\mathbb{C}^0_{0}\right]^{(0, 0)} ~ > ~ 0, \nd
the null energy condition may be easily satisfied. Since, and as mentioned repeatedly earlier, the 
$[\mathbb{C}^\mu_{\mu}]^{(0, 0)}$ are classified by eight derivative polynomials in G-flux and curvature tensors, \eqref{kortega} can be satisfied for our background, giving us a precise procedure to satisfy the null energy condition.
Under special choices of the higher order polynomials, we can even ask for stronger conditions like (see also \cite{russot}):
\bg\label{hanasuit}
\mathbb{T}^i_i + \mathbb{T}^0_0~ >~ 0 ~~~{\rm and/or}~~~ \mathbb{T}^0_0 ~ >~  0, \nd
leading to the strong and the dominant energy conditions respectively. Of course all our discussions have been on the M-theory side, but we could also construct similar criteria in the dual IIB side also as all M-theory ingredients have the corresponding IIB dual in our framework. Note that going beyond zeroth order in $g_s$ is not very meaningful here, at least in demonstrating the null, strong or dominant energy conditions,
because the Ricci curvature terms in the Einstein tensors \eqref{synecdoche} and \eqref{synecdoche2} only appear to the lowest order in $g_s$. Once we go to higher orders in $g_s$, the quantum terms, including higher order G-flux and metric terms, simply stabilize the zeroth order classical background in the way 
discussed in subsection \ref{stabul}.  

\vskip.1in

\noindent {\it An $\acute{\rm e}$tude on moduli stabilization}

\vskip.1in

\noindent Let us now discuss the issue of moduli stabilization both in the IIB and in the M-theory sides. One immediate question is the {\it meaning} of moduli {stabilization} in a time-dependent background when the metric of the internal space varies with respect to time. When the metric components vary, of course all the 
K\"ahler and the complex structure moduli will also vary with time\footnote{This is not generically true, but we can at least safely assume that there is a non-zero subset of the K\"ahler and the complex structure moduli that does vary with respect to time. For the case \eqref{olokhi}, the volume of the base 
${\cal M}_4 \times {\cal M}_2$ does not change with time, whereas this is not the case for \eqref{ranjhita}.}. 
Can  we give any meaning to the stabilization procedure $-$ that worked so well in the time-independent case $-$ in the time-dependent case now? Recall that in the time-independent case, the G-flux components ${\bf G}_{MNPa}$ are responsible for fixing the complex structure moduli in the IIB side because they lead to the RR and NS three-form fluxes \cite{GVW, DRS, GKP}. On the other hand, the K\"ahler structure moduli are fixed by the quantum terms (see for example \cite{Denef:2005mm}). In the time-dependent case we can then stabilize the moduli at every {\it instant}. In other words, at any instant of time, or alternatively, for any values of $g_s$,
the values of the G-flux components \eqref{ravali} and the quantum terms \eqref{phingsha2} (for the case \eqref{olokhi}) are fixed\footnote{In other words, fixed in time but have $y^m$ dependences.}. 
These values in turn fix the complex and the K\"ahler structure moduli for the given value of $g_s$. Once we change $g_s$, the moduli also change accordingly\footnote{The second moding scheme for the G-flux components, for both cases \eqref{olokhi} and \eqref{ranjhita}, allows $k \ge 0$ for the G-flux components 
${\bf G}_{mnpa}$, so we might expect fixed complex structure moduli at zeroth order in $g_s$.  However, as discussed in footnote \ref{plazamey} and also in sub-section \ref{maryse}, the anomaly cancellation condition \eqref{ramamey} as well as the flux EOMs
actually prefer the 
moding of $k \ge {3\over 2}$, at least for the case \eqref{olokhi}.}. 
In this sense the Dine-Seiberg runaway problem is avoided for any values of $g_s$. 

This is also the case when we view our de Sitter space as a coherent state. The moduli stabilization from such a viewpoint is even more logical as we showed recently in \cite{coherbeta}. Once we stabilize the moduli at the solitonic level, i.e for the background \eqref{betbab}, the coherent state automatically guarantees that moduli in the time-dependent background \eqref{pyncmey} vary in a controlled fashion so that there is never a Dine-Seiberg runaway problem in this system.

The puzzle however is when $g_s = 0$. In this limit one might worry that all the G-flux components and the quantum terms, since they are given in terms of series in $g_s$, would vanish, and therefore the moduli will remain unfixed. This is however not the case because precisely in this limit, the $3+1$ dimensional space-time part of the IIB metric \eqref{pyncmey} blows up and therefore one cannot construct the dual M-theory metric by T-dualizing the IIB configuration (this is signalled by the blowing-up of the M-theory metric
\eqref{vegamey} as well). This means $g_s = 0$ limit is not attainable in our set-up, and we will have to tread a bit more carefully to analyze the moduli stabilization now. 

The answer lies in re-parametrizing the temporal dependence (or alternatively the $g_s$ dependence) in a slightly different way. Let $t_o = \epsilon$ be the {\it smallest} time attainable by the system, i.e $t_o$ could be a very small number and its precise value does not concern us as long as it is non-zero. This could be related to graininess of time, much like the graininess of space encountered in the standard Wilsonian flow. We can then express the temporal coordinate $t$ as $t \equiv \hat{t} t_o$ such that $-\infty < \hat{t} \le 1$. The relation between $g_s$ and $t$ from \eqref{montse}, then allows us to express $g_s$ as 
$g_s \equiv \hat{g}_s g_{(o)}$ such that $1 \le \left({\hat{g}_s\over H}\right)^2 < {1\over g^2_{(o)}}$, with 
$g_{(o)}$ being a constant related to $t_o$ via \eqref{montse}. In this language, the G-flux components 
\eqref{ravali} can be re-written as:
\bg\label{sorserobbie}
{\bf G}_{MNPQ} = \sum_{k \in {\mathbb{Z}\over 2}} {\cal G}^{(k)}_{MNPQ}(y) g_{(o)}^{2\Delta k}
\left({\hat{g}_s\over H}\right)^{2\Delta k} \equiv \sum_{k \in {\mathbb{Z}\over 2}} \hat{\cal G}^{(k)}_{MNPQ}(y)
\left({\hat{g}_s\over H}\right)^{2\Delta k}, \nd
implying that when $\hat{g}_s = 1$, there would still be non-zero fluxes. We could even keep 
$\hat{\cal G}^{(k)}_{MNPQ}(y)$ finite, so that the complex structure moduli are fixed at finite values at any given values of $g_s$. Interestingly, in this limit, even the full quantum terms \eqref{phingsha2} (for the case 
\eqref{olokhi}) remain non-zero and finite and therefore the K\"ahler moduli could also be fixed at any given values of $g_s$. Once we view the de Sitter space as a coherent state, the moduli stabilization can be given a even simpler reasoning\footnote{For example, the moduli are already stabilized by the time-independent fluxes for the solitonic background \eqref{betbab}. The coherent state then simply tell us how the moduli evolve dynamically without any Dine-Seiberg runaway \cite{coherbeta}.} from \cite{coherbeta} as discussed above. 

\vskip.1in

\noindent {\it An explicit solution for the background EOMs}

\vskip.1in

\noindent Let us end this section by giving an {\it explicit} solution of the background de Sitter solution from our analysis. We will provide the solution from M-theory side, and the IIB result could be easily extracted by a duality transformation. 

\vskip.1in

\noindent $\bullet$  The type IIB metric that we want to reproduce takes the form \eqref{pyncmey}, whose M-theory uplift is given by \eqref{vegamey}.

\vskip.1in

\noindent $\bullet$ The unwarped base metric $g_{mn}$, where $(m, n) \in {\cal M}_4$ in \eqref{melisett}, can be expressed in terms of fluxes and quantum corrections as \eqref{rambhafin}. 

\vskip.1in

\noindent $\bullet$ The unwarped base metric $g_{\alpha\beta}$, where $(\alpha, \beta ) \in {\cal M}_2$ in \eqref{melisett}, can also be expressed in terms of fluxes and quantum corrections as \eqref{rambhafin2}. 

\vskip.1in

\noindent $\bullet$ The warp factor $H(y)$ that appears in our metric ansatze can be expressed in terms of the number of M2 and ${\overline{\rm M2}}$-branes as \eqref{evaB102}. 

\vskip.1in

\noindent $\bullet$ The G-flux components ${\bf G}_{MNPQ}$, where $(M, N) \in {\cal M}_4 \times {\cal M}_2$ in \eqref{melisett}, 
that are required to support a metric configuration like 
\eqref{vegamey} are given by \eqref{ravali}.

\vskip.1in

\noindent $\bullet$ The G-flux components of the form ${\bf G}_{MNab}$, where 
$(a, b) \in {\mathbb{T}^2\over {\cal G}}$, are the essential ingredients in the zeroth order EOMs. They are localized fluxes and are given by \eqref{teskimey}.

\vskip.1in

\noindent $\bullet$ Another set of G-flux components that are absolutely essential to switch on for consistency are the space-time fluxes of the form ${\bf G}_{0ijM}$ where $(0, i, j) \in {\bf R}^{2, 1}$. In the presence of dynamical M2-branes they are given by \eqref{kyratagra}. 

\vskip.1in

\noindent $\bullet$ The supersymmetry breaking condition can be expressed completely in terms of 
the localized fluxes ${\bf G}_{MNab}$ as a non-self-duality condition \eqref{evebe}. This condition breaks supersymmetry by giving masses to the fermions \cite{coherbeta}.
 
\vskip.1in

\noindent $\bullet$ The {\it exact} expression of the cosmological constant $\Lambda$ can be written in terms of the fluxes and a finite set of quantum corrections, and is given by \eqref{hathway}. It can be argued easily that it is possible to make it a {\it small} positive definite quantity.

\vskip.1in

\noindent $\bullet$ The time-period of the validity of our solution can be ascertained from the limit when the type IIA strong coupling sets in. This is detailed in \eqref{alibet} and is also related to the so-called {\it quantum break time} as advocated in \cite{dvali}. In our case, as discussed more recently in \cite{coherbeta}, this time interval may be associated with the time-interval where our de Sitter space remains as a coherent state. 

\vskip.1in

\noindent The consistent picture that evolves from our analysis is satisfying and puts a great deal of confidence on the fact that four-dimensional de Sitter vacua should be in the IIB string landscape and not in the swampland. The swampland criteria were developed, using the data of time-independent backgrounds,  to tackle backgrounds that only made sense with inherent time dependences. Clearly, as we showed here, this cannot work, and therefore the unsuitability of such an
approach is probably one of the main reason of its failure to predict backgrounds with positive cosmological constants and with four-dimensional de Sitter isometries.

\section{Discussions and conclusions}

In this paper we showed how it is possible for an ansatze \eqref{pyncmey} in IIB theory to be a solution to the string EOMs by lifting it to M-theory and taking all perturbative and non-perturbative as well as local and non-local quantum corrections into account. There are three main results of our paper:

\vskip.1in

\noindent $\bullet$ A IIB background with four-dimensional de Sitter isometries and time-independent six-dimensional internal space of the form \eqref{betta} along-with time-independent background fluxes 
{\it cannot} be a solution to the string EOMs no matter how many quantum corrections are added. In fact the 
$g_s$ scalings of the quantum terms, namely \eqref{kkkbkb2}, show that there are an infinite number of quantum terms that need to be inserted at any given order in $g_s$, ruining any EFT descriptions in four-dimensions. These theories then truly belong to the swampland \cite{vafa1} as shown in \cite{nogo, nodS}.

\vskip.1in

\noindent $\bullet$ Once time-dependences are allowed and we make the internal space and the  background fluxes time-dependent, the results change significantly.
 Generically this can make the four-dimensional Newton's constant time-dependent. The simplest example of this kind appears from \eqref{ranjhita}. In this case a IIB background of the form \eqref{pyncmey}, uplifted as \eqref{vegamey} to M-theory, with time-dependent G-flux components do appear to have an EFT description. This is evident from the $g_s$ scalings \eqref{miai} of 
the quantum terms 
\eqref{phingsha} that only allow a {\it finite} number of quantum terms at any given order in $g_s$. Unfortunately however there appears to be a late time singularity, amongst other issues, that prohibit such a configuration to be a viable model of late-time cosmology.

\vskip.1in

\noindent $\bullet$ Thus a IIB background with time-independent Newton's constant, again uplifted to M-theory as \eqref{vegamey} but now satisfying \eqref{olokhi}, with time-dependent fluxes, does allow an EFT description as evident from the $g_s$ scalings \eqref{melamon2} of the quantum terms \eqref{phingsha2}. Fortunately now there are no late-time singularities and the background also appears to overcome both the no-go and the swampland criteria to be a viable late-time cosmological model in the string landscape. All the issues plaguing the case 
\eqref{ranjhita} or the background \eqref{betta}, do not appear for this case.

\vskip.1in

\noindent The above conclusion justifies how time-dependences of metric and flux components are essential to generate a four-dimensional space-time with de Sitter isometries in the IIB landscape. The quantum terms are also equally important and time-dependences go hand in hand with the quantum corrections. Existence of $g_s$ and $M_p$ hierarchies then guarantee four-dimensional EFT descriptions as tabulated above. In addition to that, we also have many other results scattered throughout the paper that are derived from M-theory. In the following we list some of them.

\vskip.1in

\noindent $\bullet$ An exact expression for the cosmological constant $\Lambda$, completely in terms of the background fluxes and quantum corrections, can be expressed as \eqref{hathway}.

\vskip.1in

\noindent $\bullet$ An exact expression for the G-flux components, appearing from the back-reaction of a dynamical M2-brane, can be expressed as \eqref{hokyra}.

\vskip.1in

\noindent $\bullet$  Quantizations of the G-flux components and anomaly cancellations can be demonstrated even when time-dependences are switched on. The quantum corrections, like 
\eqref{phingsha2} for the case \eqref{olokhi}, play an important role here.

\vskip.1in

\noindent $\bullet$ The energy condition, for example the null-energy condition, can be shown to be satisfied with the choice of fluxes and quantum corrections. In fact it appears that the $2+1$ dimensional 
quantum corrections play a significant role in satisfying the null-energy condition as shown in 
\eqref{kortega}. For special choices of these quantum corrections, one could even satisfy the strong and the dominant energy conditions \eqref{hanasuit}.

\vskip.1in

\noindent Note that we haven't said anything about the fermions. We could introduce components of gravitino and their interactions with the bosonic degrees of freedom in M-theory. Giving a small mass to the gravitino components, one should be able to integrate out all the fermionic degrees of freedom in our model. This will result into the polynomial forms of the quantum terms. For very light fermions, we can express the G-flux ansatze \eqref{teskimey} in terms of fermionic terms as in \cite{petite} using eleven-dimensional Gamma matrices. One may then further decompose these fermions into space-time and internal fermionic degrees of freedom. Plugging this in  
the two set of quantum terms \eqref{phingsha} and \eqref{phingsha2}, for the two cases \eqref{ranjhita} and \eqref{olokhi}  respectively, and integrating out both the internal massive fermionic as well other bosonic 
degrees of freedom (those
that would have potentially ruined the de Sitter isometries), will provide the necessary fermionic quantum terms in $2+1$ dimensions. For more details see \cite{coherbeta}.

Thus it appears that our analysis may be generic enough, and therefore the fact that four-dimensional de Sitter vacua appear from such an approach, provides a strong indication that the landscape of string theory allows solutions with positive cosmological constants and time-independent Newton's constants to 
exist\footnote{Some recent works on generating de Sitter using different techniques than what we used here are in \cite{hardy}. It might be interesting to compare our results with those of \cite{hardy}.}.

\section*{Acknowledgements}

We would like to specifically thank Md. Muktadir Rahman for helping us with the Mathematica programme for all the Einstein's EOMs \cite{mathematica}, and to 
Savdeep Sethi for many helpful discussions. We would also like to thank Robert Brandenberger, Carlos Herdeiro, Onirban Islam, Dileep Jatkar, Gregory Moore, Sakura Schafer-Nameki, Cumrun Vafa and Edward Witten for useful correspondences.
The work of KD, ME and MMF is supported in part by the Natural Sciences and Engineering Research Council of Canada.


\end{document}